\documentclass[11pt]{book}
\usepackage[T1]{fontenc}
\usepackage[small]{caption}
\usepackage[dvips]{graphicx}
\usepackage{color,amsmath,amssymb,amsfonts,mathrsfs}
\usepackage{afterpage,fancyhdr,tabularx}
\usepackage{geometry}
\usepackage{cite}

\newcommand{\mc}{\mathcal}
\newcommand{\la}{\Lambda}
\newcommand{\tn}{\textnormal}

\begin{document}

\begin{titlepage}
\begin{center}
{\sffamily {\Large \bfseries 
~\\
Transport Through Correlated Quantum Dots\\
-A Functional Renormalization Group Approach-}\\[20ex]
\large
Diplomarbeit\\[20ex]
vorgelegt von \\[2ex]
{\bfseries Christoph Karrasch}\\[2ex]
aus Duderstadt\\[42ex]
angefertigt am\\[2ex]
Institut f\"ur Theoretische Physik\\der Georg-August Universit\"at G\"ottingen\\[2ex]
September 2006}
\end{center}

\newpage
\thispagestyle{empty}\setlength{\parindent}{0cm}{\bfseries
Contact Information:\\\\\\
Christoph Karrasch\\\\
Institut f\"ur Theoretische Physik\\
Friederich-Hund-Platz 1\\
37077 G\"ottingen\\\\
Tel. +49 551 399505\\
email: karrasch@theorie.physik.uni-goettingen.de\\
homepage: http://www.theorie.physik.uni-goettingen.de/$\sim$karrasch
}
\setlength{\parindent}{2ex}
\end{titlepage}

\tableofcontents

\chapter{Introduction}

Quantum dots have recently attracted experimental interest due to their `possible' application in future nano devices and for quantum information processing. From the theoretical point of view a many-particle method suited to account for the two-particle interactions between the electrons in the dot -- which strongly affect the physics -- is needed. In the literature, the numerical renormalization group (NRG) was frequently used as a very reliable method to study systems with local Coulomb correlations, but its applicability is limited due to the vast computational resources required. Practically, only simple geometries can be treated and even there exhaustive scans of the parameter space are impossible. Therefore other accurate methods that allow for a precise treatment of the two-particle interactions in quantum dots are needed, but none of the recent approaches succeeded in reaching the accuracy of the NRG calculations, most times not even on a qualitative level \cite{dotsystems}.

Here, we will present a recently-developed renormalization group (RG) scheme based on Wilson's general RG idea, the functional renormalization group (fRG). Its starting point is the replacement of the free propagator by one depending on an infrared cutoff $\la$ in the generating functional of the single-particle irreducible vertex functions $\gamma_m$. Differentiating the functional with respect to $\la$ one obtains an exact hierarchy of coupled flow equations for the functions $\gamma_m$. This infinite set has to be truncated in order to render it solvable, and within this thesis we will employ a truncation scheme that accounts for the flow of $\gamma_1$ (the self-energy) and the two-particle vertex $\gamma_2$ evaluated at zero external frequency (the effectice interaction), but neglects all higher-order functions. Hence we compute a frequency-independent approximation for $\gamma_1$ which can be viewed as a renormalization of the dot's single-particle energies.

An important transport property of a quantum dot is the linear-response conductance $G$. In a noninteracting model, it is frequently defined using single-particle scattering theory, which leads to the well-known Landauer-B\"uttiker formula. Since the aforementioned fRG approximation computes effective noninteracting parameters for the quantum dot system, one might be tempted to use the ordinary $U=0$ expression to calculate $G$. This is, however, conceptionally wrong since it neglects the fact that this quantity cannot be defined a priori using the ordinary scattering theory approach as one is confronted with a system of correlated electrons, whether the approximation we employ to describe it can be interpreted as an effective noninteracting one or not. Hence a better way to define the conductance is to consider the current-current response function within the full interacting problem and to show that the noninteracting expression follows consistently within our fRG framework without need to apply further approximations. 

Having an accurate method at hand to describe Coulomb correlations, we will first consider the low-energy physics (at zero temperature) of a spin-polarised dot containing several levels. This can be viewed as a suitable model for an experimentally realised single quantum dot, assuming that the experiment showed that the spin degree of freedom seems to play no role. A series of important experiments on such systems was performed in the past few years \cite{phase3,phase2,phase1}, and in addition to the conductance also the transmission phase and, in the very recent one, the average occupancy of the dot were measured. The conductance was found to exhibit resonances (Coulomb blockade peaks) each time another electron was added to the system. Surprisingly, the measurement of the phase yielded universal jumps by $\pi$ between consecutive peaks, in contradiction to the expected mesoscopic behaviour of jumping between some while evolving continuously between others, the actual realisation depending on the dot under consideration. Despite a large amount of theoretical work on this issue (for a review see \cite{sumtheoriephase1,sumtheoriephase2}; the most important works with electron-electron correlations as a key ingredient are \cite{theoriephase1,imry,theoriephase2,theoriephase3,GG}), none succeeded in giving a fully satisfactory explanation. Fortunately, the last experiment (\cite{phase1}) provided a clue to this phase lapse puzzle. For dots occupied by only a few electrons, the mesoscopic behaviour was indeed observed, and only if electrons were successively added to the system, the universally appearing phase lapses were recovered. As noted by \cite{phase1}, the most fundamental difference between both situations is the single-particle level spacing which one would expect to decrease for the electrons occupying the topmost states. Exploring this idea in a systematic study of the parameter space of our spinless quantum dot model with an accurate treatment of Coulomb correlations is the aim of the first part of this thesis.

Next, we describe a variety of spinful dot geometries where the physics is dominated by local correlations between the electrons of both spin directions. We will show how the Kondo energy scale can be extracted from the linear conductance in the zero temperature limit. We will study two geometries (a short Hubbard chain and two side-coupled dots) that have been tackled before by various authors using NRG, and we will demonstrate the power of the fRG by scanning a much larger (and in particular generic) region of the parameter space of these systems. For the Hubbard chain this will reveal new physics, while for the side-coupled geometry it will turn out that the behaviour observed within the NRG approach already covers all essential phenomena. We will finally investigate the situation of two parallel dots. This geometry allows for a large amount of parameters to be varied, and we will report on the physics that is observed in different regimes.

In the last chapter, we will establish the accuracy of the employed fRG approximation scheme by comparison to NRG data for every system that has previously been under consideration. We will show that the fRG is very reliable up to interaction strengths where the physics is dominated by correlations (in the sense that the behaviour of all quantities of interest is significantly altered with respect to the noninteracting case). Furthermore, we will prove that this method is far superior to simple mean-field approaches.

\chapter{The Functional Renormalization Group}

The functional renormalization group (fRG) is based on Wilson's general renormalization group (RG) idea. The RG is the appropriate framework to deal with quantum many-particle problems that contain a multitude of different energy scales as well as with problems that cannot be tackled by ordinary perturbation theory which breaks down if certain classes of (low-order) Feynman diagrams diverge. It starts from high energy scales (where infrared divergences are cut out) and gradually works its way down to the low energy region.

The fRG provides one way to implement the general renormalization group idea for interacting quantum many-particle systems \cite{salmhofer}. It is based on the functional integral approach to many-particle physics. Differentiating the generating functional of the single-particle irreducible $m$-particle vertex functions $\gamma_m$ (with the free propagator replaced by one that cuts out the low energy modes below a scale $\la$) with respect to a cutoff parameter yields an infinite hierarchy of flow equations for the $\gamma_m$. In practise, this set has to be truncated to make it solvable. A truncation scheme motivated by physical or practical reasons renders the fRG an approximate method, most times perturbative in the two-particle interaction. Integrating the resulting finite set of equations it turns out that in the limit $\la\to 0$, which corresponds to the original cutoff-free problem, one often obtains non-divergent results, and even in problems that are not plagued by infrared divergences (as those tackled within this thesis) it shows that the fRG produces results far superior to those of perturbation theory. An application of the method to the well-known one-dimensional single-impurity Anderson model can be found in \cite{ralf}. In \cite{draht} the fRG is used to successfully derive power laws in correlated one-dimensional quantum wires (Luttinger liquids).

This chapter is organised as follows. In the first section, we recall the functional integral approach to many-particle physics and how it is used to compute Green functions. In the second section, we replace the noninteracting propagator in the generating functional of the vertex functions by one containing a cutoff. Differentiating with respect to the cutoff parameter $\la$ then yields an infinite hierarchy of flow equations for the vertex functions. Since this hierarchy needs to be truncated in order to make it solvable, we will develop a truncation scheme that becomes exact in the limit of vanishing strength of the two-particle interaction. Finally, we will specify the cutoff to be an infrared cutoff in frequency space to write down coupled flow equations that perturbatively describe an interacting system.

\section{Many-particle Green Functions}

In this section we recall how functional integrals can be used to compute the grand canonical partition function of a system of interacting quantum mechanical particles as well as Green functions. This will be the appropriate framework to set up the fRG flow equations.

\subsection{The Partition Function as a Functional Integral}

\subsubsection{A Short Reminder of Functional Integrals in Many-Particle Physics}

The fundamental equation for the functional integral approach to many-particle physics reads
\begin{equation*}
e^{-\beta H} = \lim_{N\to\infty}\left(:e^{-\beta H/N}:\right)^N,
\end{equation*}
where $H$ is a normal-ordered but otherwise arbitrary many-particle Hamiltonian, and the colons denote normal-ordering. We insert the unity operator written in terms of the eigenstates of the annihilation operators between each of two factors of $:e^{-\beta H/N}:$. Since we are dealing with fermions, we have to introduce Grassmann variables, that is anti-commuting numbers, to write down these states. Because of the normal-ordering, the annihilation (creation) operators in the exponential can then be replaced by the corresponding eigenvalue and we are basically left with an `ordinary' (Grassmann) integral, that becomes a functional integral in the limit $N\to\infty$. In case that $H$ contains only single-particle terms, this integral is purely Gaussian and can be performed `almost trivially'. If $H$ contains more complicated terms (such as a two-particle interaction), carrying out the the integral analytically becomes impossible, but one can expand the interaction part of the exponential function into a Taylor series (if one notes that we are now only dealing with numbers and no longer with operators) to set up a perturbation theory. The contribution of each order can be evaluated using Wick's theorem, a property of Gaussian integrals.

A more detailed description of the functional integral approach to many-particle physics can be found in \cite{negeleorland}, or, very nicely, in \cite{schoenhammer}.

\subsubsection{The Partition Function}

As motivated above, the grand canonical partition function of a fermionic many-particle system can be expressed as a functional integral,
\begin{equation}\label{eq:FRG.zustandssumme}
\mathcal{Z}=\int{\mathcal{D}\bar{\psi}\psi\exp{\left\{-\int^{\beta}_0 d\tau~\left[\sum_l{\bar{\psi}_l(\tau+0)\frac{d\psi_l(\tau)}{d\tau} + \mathcal{H}(\left\{\bar{\psi}\right\},\left\{\psi\right\})} \right]\right\}} },
\end{equation}
where $\bar{\psi}$ and $\psi$ denote Grassman variables,
\begin{equation*}
\psi_l\stackrel{(-)}{\psi_{l'}}+\stackrel{(-)}{\psi_{l'}}\psi_l=0,
\end{equation*}
and $\mathcal{H}$ is obtained from a normal-ordered (but otherwise arbitrary) Hamilton operator $H$ by substituting
\begin{equation*}
\mathcal{H}\left(\left\{\bar{\psi}\right\},\left\{\psi\right\}\right) = H\left(c^{\dagger}_l \rightarrow \bar{\psi}_l(\tau+0), c_l \rightarrow \psi_l(\tau)\right).
\end{equation*}
The boundary conditions of the fields read
\begin{equation*}
\psi_l(\beta)=-\psi_l(0), \qquad \bar{\psi}_l(\beta)=-\bar{\psi}(0),
\end{equation*}
with $\beta$ being the inverse temperature. In the following, we will always choose $\mathcal{H}$ to contain a single-particle term as well as a two-particle interaction,
\begin{equation}
\mathcal{H}(\left\{\bar{\psi}\right\},\left\{\psi\right\}) = \sum_l \epsilon_l\bar{\psi}_l(\tau+0)\psi_l(\tau) + \frac{1}{4}\sum_{i,j,k,l} \bar{v}_{i,j,k,l} \bar{\psi}_i(\tau+0)\bar{\psi}_j(\tau+0)\psi_l(\tau)\psi_k(\tau).
\end{equation}
Here $l$ denotes a set of quantum numbers in which the single-particle part of $H$ is diagonal, $\epsilon_l$ the one-particle dispersion (including the chemical potential) and $\bar{v}$ the anti-symmetrised matrix elements of the two-particle interaction, respectively.

For the upcoming calculations it will prove useful to expand the Grassmann fields into a Fourier series
\begin{equation}\label{eq:FRG.fourier}
\stackrel{(-)}{\psi_l}(\tau) = \frac{1}{\sqrt{\beta}} \sum_{\omega_n} e^{\mp i\omega_n\tau} \stackrel{(-)}{\psi_l}(i\omega_n).
\end{equation}
The summation extends over all odd (fermionic) Matsubara frequencies. The inverse transformation is given by
\begin{equation}
\stackrel{(-)}{\psi_l}(i\omega_n) = \frac{1}{\sqrt{\beta}} \int_0^{\beta} d\tau~e^{\pm i\omega_n\tau} \stackrel{(-)}{\psi_l}(\tau).
\end{equation}
Inserting (\ref{eq:FRG.fourier}) into the imaginary-time functional integral (\ref{eq:FRG.zustandssumme}) yields
\begin{equation}\label{eq:FRG.zustandssumme2}\begin{split}
\frac{\mc{Z}}{\mc{Z}_0} = & \frac{1}{\mc{Z}_0} \int{\mc{D}\bar{\psi}\psi\exp{\Bigg\{ \sum_l\sum_{\omega_n} \bar{\psi}_l(i\omega_n)\left[\mc{G}_l^0(i\omega_n)\right]^{-1}\psi_l(i\omega_n)}} \\ & - \frac{1}{4}\frac{1}{\sqrt{\beta}}\sum_{i,j,k,l}\sum_{n,n',m,m'} \bar{v}_{i,j,k,l} \delta_{m+m',n+n'} \bar{\psi}_i(i\omega_m)\bar{\psi}_j(i\omega_{m'})\psi_l(i\omega_{n'})\psi_k(i\omega_n) \Bigg\},
\end{split}\end{equation}
with $\mc{G}_l^0(i\omega_n)$ being the noninteracting single-particle propagator.\footnote{We have chosen a basis where $\mc G^0$ is diagonal. This is, however, not a vital condition, since all properties of the noninteracting problem (like Wick's theorem) follow from the Gaussian nature of the functional integral (\ref{eq:FRG.zustandssumme2}) with $\bar v=0$, diagonal or not. The shorthand notation introduced in (\ref{eq:FRG.zustandssumme3}) therefore explicitly allows for non-diagonal terms in $\mc G^0$.} All convergence factors have been omitted to keep the notation short and the noninteracting partition function $\mc{Z}_0$ was introduced to cancel the functional determinant that appears due to the change of the integration variables $\psi(\tau)\rightarrow\psi(i\omega)$. Introducing the shorthand notation $(\bar{\psi},X\psi):=\sum_{k,k'}\bar{\psi}_kX_{k,k'}\psi_{k'}$ where $k=(\omega_n,l)$ allows for rewriting (\ref{eq:FRG.zustandssumme2}) in the simple form
\begin{equation}\label{eq:FRG.zustandssumme3}\begin{split}
\frac{\mc{Z}}{\mc{Z}_0} & = \frac{1}{\mc{Z}_0} \int\mc{D}\bar{\psi}\psi\exp{\Bigg\{ \left(\bar{\psi},[\mc{G}^0]^{-1}\psi\right) - \frac{1}{4}\sum_{k_1',k_2',k_1,k_2} \bar{v}_{k_1',k_2',k_1,k_2} \bar{\psi}_{k_1'}\bar{\psi}_{k_2'}\psi_{k_2}\psi_{k_1} \Bigg\}}
\\  &=: \frac{1}{\mc{Z}_0} \int\mc{D}\bar{\psi}\psi\exp{\Big\{S_0 - S_{\textnormal{int}} \Big\} }.
\end{split}\end{equation}
A factor of $\beta^{-1}$ as well as the frequency-conserving $\delta$-function have been absorbed into the anti-symmetrised two-particle interaction $\bar{v}$.

\subsection{Generating Functionals of Green Functions}

\subsubsection{Connected Green Functions}

We define the generating functional of the $m$-particle Green functions as
\begin{equation}\label{eq:FRG.greenfunktional}
\mathcal{W}(\left\{\bar{\eta}\right\},\left\{\eta\right\}) = \frac{1}{\mc{Z}} \int\mc{D}\bar{\psi}\psi\exp{\Big\{S_0 - S_{\textnormal{int}} - (\bar{\psi},\eta) - (\bar{\eta},\psi) \Big\} }.
\end{equation}
By construction, taking the functional derivative with respect to the external source fields $\eta$ and $\bar{\eta}$ yields the $m$-particle Green function in frequency space\footnote{From $\frac{\delta\mc{W}}{\delta\eta(i\omega_n)}
=\int \frac{\delta\eta(\tau')}{\delta\eta(i\omega_n)}\frac{\delta\mc{W}}{\delta\eta(\tau')}d\tau'
= \beta^{-1/2}\int e^{-i\omega_n\tau'}\frac{\delta\mc{W}}{\delta\eta(\tau')}d\tau'$ it follows that by differentiating the generating functional $\mc{W}$ with respect to the Fourier-transformed fields one obtains the Fourier transform of the imaginary-time Green function.}
\begin{equation}\begin{split}
G_m(k_1',\dots, k_m';k_1,\dots,k_m) :&= (-1)^m \left\langle\psi_{k_1'}\dots\psi_{k_m'}\bar{\psi}_{k_m}\dots\bar{\psi}_{k_1}\right\rangle
\\ & = \frac{\delta^m}{\delta\bar{\eta}_{k_1'}\dots\delta\bar{\eta}_{k_m'}}\frac{\delta^m}{\delta\eta_{k_m}\dots\delta\eta_{k_1}} \mathcal{W}(\left\{\bar{\eta}\right\},\left\{\eta\right\})\Bigg|_{\eta=\bar{\eta}=0}.
\end{split}\end{equation}
In the noninteracting case (\ref{eq:FRG.zustandssumme2}) is a Gaussian integral that can be solved easily leading to the well-known result for the one-particle Green function,
\begin{equation*}
G_1(l,i\omega_n) = \mc{G}_l^0(i\omega_n) = \frac{1}{i\omega_n-\epsilon_l},
\end{equation*}
and to Wick's theorem to calculate all higher-order functions diagrammatically using Feynman diagrams.

The functional $\mc{W}^c$ that generates the connected Green functions
\begin{equation}\begin{split}
G_m^c(k_1',\dots, k_m';k_1,\dots,k_m) :&= (-1)^m \left\langle\psi_{k_1'}\dots\psi_{k_m'}\bar{\psi}_{k_m}\dots\bar{\psi}_{k_1}\right\rangle_c
\\ & = \frac{\delta^m}{\delta\bar{\eta}_{k_1'}\dots\delta\bar{\eta}_{k_m'}}\frac{\delta^m}{\delta\eta_{k_m}\dots\delta\eta_{k_1}} \mathcal{W}^c(\left\{\bar{\eta}\right\},\left\{\eta\right\})\Bigg|_{\eta=\bar{\eta}=0}
\end{split}\end{equation}
is given by (for a proof see \cite{negeleorland})
\begin{equation}
\mathcal{W}^c(\left\{\bar{\eta}\right\},\left\{\eta\right\}) = \ln{\left[\mathcal{W}(\left\{\bar{\eta}\right\},\left\{\eta\right\})\right]}.
\end{equation}

\subsubsection{Vertex Functions}

Next, we consider so-called $m$-particle vertex functions which are (in a diagramatical fashion) defined to consist of all connected one-particle irreducible diagrams with the $m$ external legs amputated. In particular, this implies that they cannot be split into two pieces by cutting one single-particle line.

The generating functional $\Gamma$ for these vertex function is given by a Legendre transform of $\mc{W}^c$,
\begin{equation}\label{eq:FRG.vertexfunktional}
\Gamma(\left\{\bar{\phi}\right\},\left\{\phi\right\})=
-\mathcal{W}^c(\left\{\bar{\eta}\right\},\left\{\eta\right\}) - (\bar{\phi},\eta) - (\bar{\eta},\phi)  + (\bar{\phi},\left[\mc{G}^0\right]^{-1}\phi),
\end{equation}
where the new fields $\phi$ and $\bar{\phi}$ are defined in the standard way,
\begin{equation}\label{eq:FRG.felderphi}
\phi = -\frac{\delta}{\delta\bar{\eta}}\mathcal{W}^c(\left\{\bar{\eta}\right\},\left\{\eta\right\}),
\qquad \bar{\phi} = \frac{\delta}{\delta\eta}\mathcal{W}^c(\left\{\bar{\eta}\right\},\left\{\eta\right\}),
\end{equation}
and the last term in (\ref{eq:FRG.vertexfunktional}) was added for convenience. Of course it is far from obvious that the vertex functions can really be obtained from this functional by taking the derivative with respect to the external fields. Anyway, we define
\begin{equation}\label{eq:FRG.vertexfunktionen}
\gamma_m(k_1',\dots, k_m';k_1,\dots,k_m) := \frac{\delta^m}{\delta\bar{\phi}_{k_1'}\dots\delta\bar{\phi}_{k_m'}}\frac{\delta^m}{\delta\phi_{k_m}\dots\delta\phi_{k_1}} \Gamma(\left\{\bar{\phi}\right\},\left\{\phi\right\})\Bigg|_{\phi=\bar{\phi}=0}.
\end{equation}

\subsection{Relations between Vertex and Connected Green Functions}

We will now explicitly show that the one- and two-particle vertex functions are in fact obtained by (\ref{eq:FRG.vertexfunktionen}) (or, to say it the other way round, that the so-defined functions $\gamma_1$ and $\gamma_2$ consists only of connected one-particle irreducible diagrams with their two (four) external legs amputated). Therefore we first calculate
\begin{equation}\label{eq:FRG.ableitungvertexfuntional1}\begin{split}
\frac{\delta}{\delta\phi_k}\Gamma(\left\{\bar{\phi}\right\},\left\{\phi\right\}) & \stackrel{(\ref{eq:FRG.vertexfunktional})}{=}
  \sum_q\left[-\frac{\delta\mc{W}^c}{\delta\eta_q}\frac{\delta\eta_q}{\delta\phi_k}-
\frac{\delta\mc{W}^c}{\delta\bar{\eta}_q}\frac{\delta\bar{\eta}_q}{\delta\phi_k}  +
 \bar{\phi}_q \frac{\delta\eta_q}{\delta\phi_k} - \frac{\delta\bar{\eta}_q}{\delta\phi_k}\phi_q - \bar{\phi}_q\left[\mc{G}^0\right]^{-1}_{q,k}            \right] + \bar{\eta}_k
 \\ & \stackrel{(\ref{eq:FRG.felderphi})}{=} \bar{\eta}_k - \sum_q \bar{\phi}_q\left[\mc{G}^0\right]^{-1}_{q,k}
\end{split}\end{equation}
\begin{equation}\label{eq:FRG.ableitungvertexfuntional2}\begin{split}
\frac{\delta}{\delta\bar{\phi}_k}\Gamma(\left\{\bar{\phi}\right\},\left\{\phi\right\}) & \stackrel{(\ref{eq:FRG.vertexfunktional})}{=}
  \sum_q\left[-\frac{\delta\mc{W}^c}{\delta\eta_q}\frac{\delta\eta_q}{\delta\bar{\phi}_k}-
\frac{\delta\mc{W}^c}{\delta\bar{\eta}_q}\frac{\delta\bar{\eta}_q}{\delta\bar{\phi}_k}  +
 \bar{\phi}_q \frac{\delta\eta_q}{\delta\bar{\phi}_k} - \frac{\delta\bar{\eta}_q}{\delta\bar{\phi}_k}\phi_q + \left[\mc{G}^0\right]^{-1}_{k,q}\phi_q            \right]  + \bar{\eta}_k
 \\ & \stackrel{(\ref{eq:FRG.felderphi})}{=} \bar{\eta}_k + \sum_q \left[\mc{G}^0\right]^{-1}_{k,q}\phi_q.
\end{split}\end{equation}
The additional minus signs appear because we are taking the derivative with respect to Grassmann variables. One should be reminded that as we have turned to the Legendre transform $\Gamma$, $\phi$ and $\bar{\phi}$ are independent fields, implying $\frac{\delta\phi}{\delta\bar{\phi}}=0$ etc. Next, we consider
\begin{equation}\label{eq:FRG.zusammenhang1}\begin{split}
\delta_{k,k'}  = & \frac{\delta\phi_k}{\delta\phi_{k'}} \stackrel{(\ref{eq:FRG.felderphi})}{=} - \frac{\delta}{\delta\phi_{k'}}\frac{\delta\mc{W}^c}{\delta\bar{\eta}_k} = -\sum_q\left[\frac{\delta\eta_q}{\delta\phi_{k'}}\frac{\delta^2\mc{W}^c}{\delta\eta_q\delta\bar{\eta}_k} + \frac{\delta\bar{\eta}_q}{\delta\phi_{k'}}\frac{\delta^2\mc{W}^c}{\delta\bar{\eta}_q\delta\bar{\eta}_k}\right]   \\
 = & \sum_q\left[\left( \frac{\delta^2\Gamma}{\delta\phi_{k'}\delta\bar{\phi}_q} - \left[\mc{G}^0\right]^{-1}_{q,k'}\right)\frac{\delta^2\mc{W}^c}{\delta\eta_q\delta\bar{\eta}_k} - \frac{\delta^2\Gamma}{\delta\phi_{k'}\delta\phi_q}\frac{\delta^2\mc{W}^c}{\delta\bar{\eta}_q\delta\bar{\eta}_k}\right].
\end{split}\end{equation}
The last line follows from differentiating (\ref{eq:FRG.ableitungvertexfuntional1}) and (\ref{eq:FRG.ableitungvertexfuntional2}) by $\phi$ and $\bar{\phi}$, respectively. Performing similar calculations that start out with $\frac{\delta\bar{\phi}_k}{\delta\bar{\phi}_{k'}}=\delta_{k,k'}$ and $\frac{\delta\bar{\phi}_k}{\delta\phi_{k'}}=\frac{\delta\phi_k}{\delta\bar{\phi}_{k'}}=0$, and denoting the resulting equations in a compact form (by interpreting the derivatives $\frac{\delta^2A}{\delta x_{k_1}\delta y_{k_2}}$ as matrix indices $(k_1,k_2)$), we obtain
\begin{equation*}
\begin{pmatrix}
\frac{\delta^2\Gamma}{\delta\bar{\phi}\delta\phi} + \left[\mc{G}^0\right]^{-1} & \frac{\delta^2\Gamma}{\delta\bar{\phi}\delta\bar{\phi}} \\
\frac{\delta^2\Gamma}{\delta\phi\delta\phi} & \frac{\delta^2\Gamma}{\delta\phi\delta\bar{\phi}} - \left[\left[\mc{G}^0\right]^{-1}\right]^T
\end{pmatrix}
\cdot
\begin{pmatrix}
\frac{\delta^2\mc{W}^c}{\delta\bar{\eta}\delta\eta} & -\frac{\delta^2\mc{W}^c}{\delta\bar{\eta}\delta\bar{\eta}} \\
-\frac{\delta^2\mc{W}^c}{\delta\eta\delta\eta} & \frac{\delta^2\mc{W}^c}{\delta\eta\delta\bar{\eta}}
\end{pmatrix}
=1.
\end{equation*}
As we will need it later on, we furthermore define
\begin{equation}\label{eq:FRG.zusammenhang}
\mc{V}(\bar{\phi},\phi):=
\begin{pmatrix}
\frac{\delta^2\mc{W}^c}{\delta\bar{\eta}\delta\eta} & -\frac{\delta^2\mc{W}^c}{\delta\bar{\eta}\delta\bar{\eta}} \\
-\frac{\delta^2\mc{W}^c}{\delta\eta\delta\eta} & \frac{\delta^2\mc{W}^c}{\delta\eta\delta\bar{\eta}}
\end{pmatrix}
=
\begin{pmatrix}
\frac{\delta^2\Gamma}{\delta\bar{\phi}\delta\phi} + \left[\mc{G}^0\right]^{-1} & \frac{\delta^2\Gamma}{\delta\bar{\phi}\delta\bar{\phi}} \\
\frac{\delta^2\Gamma}{\delta\phi\delta\phi} & \frac{\delta^2\Gamma}{\delta\phi\delta\bar{\phi}} - \left[\left[\mc{G}^0\right]^{-1}\right]^T
\end{pmatrix}^{-1}.
\end{equation}

It is now possible to establish a relation between the connected Green functions and the vertex functions. Therefore we evaluate (\ref{eq:FRG.zusammenhang}) at vanishing external fields and look at the (1,1) -- element of the matrix equation which is now diagonal,\footnote{The off-diagonal elements are zero unless we are in a phase of broken symmetry.}
\begin{equation}\label{eq:FRG.dyson}
G_1(k';k)\stackrel{(*)}{=}G_1^c(k';k)
:=\frac{\delta^2\mc{W}^c}{\delta\bar{\eta}_{k'}\delta\eta_k}\Big|_{\eta=\bar{\eta}=0}
=\left[\frac{\delta^2\Gamma}{\delta\bar{\phi}\delta\phi}\Big|_{\phi=\bar{\phi}=0} + \left[\mc{G}^0\right]^{-1}\right]^{-1}_{k',k}.
\end{equation}
The identity $(*)$ is the so-called linked cluster theorem. Since $G_1$ is just the single-particle propagator $\mc{G}$ (calculated in presence of the interaction term $S_{\textnormal{int}}$), (\ref{eq:FRG.dyson}) is nothing else but the Dyson equation. Consequently, the one-particle vertex function is related to the self-energy $\Sigma$, which is well-known to be one-particle irreducible,\footnote{To be more precise, for a quantity $X$, being the sum of all connected one-particle irreducible diagrams with the two external legs amputated is equivalent to fullfilling the Dyson equation.} by
\begin{equation}
\gamma_1=\frac{\delta^2\Gamma}{\delta\bar{\phi}\delta\phi}\Big|_{\phi=\bar{\phi}=0}=-\Sigma.
\end{equation}
Ignoring the difference in sign, we will call $\gamma_1$ self-energy from now on.

To prove that the same (one-particle irreducibility) holds for the two-particle vertex, we differentiate (\ref{eq:FRG.zusammenhang1}) with respect to $\phi_l$, which leads to
\begin{equation*}\begin{split}
0=\sum_q\Bigg[\frac{\delta^3\Gamma}{\delta\phi_l\delta\phi_{k'}\delta\bar{\phi}_q}
\frac{\delta^2\mc{W}^c}{\delta\eta_q\delta\bar{\eta}_{k}} 
+&\left( \frac{\delta^2\Gamma}{\delta\phi_{k'}\delta\bar{\phi}_q} - \left[\mc{G}^0\right]^{-1}_{q,k'}\right)\times \\
&  \sum_s \left(\frac{\delta\eta_s}{\delta\phi_l}\frac{\delta^3\mc{W}^c}{\delta\eta_s\delta\eta_q\delta\bar{\eta}_{k}}
+\frac{\delta\bar{\eta}_s}{\delta\phi_l}\frac{\delta^3\mc{W}^c}{\delta\bar{\eta}_s\delta\eta_q\delta\bar{\eta}_{k}}\right)   \Bigg].
\end{split}\end{equation*}
Differentiating once again with respect to $\bar{\phi}_{l'}$ and setting all external fields to zero simplifies this equation enormously, since all terms that do not contain an equal number of field derivatives $\delta_{\phi}$ and $\delta_{\bar{\phi}}$ vanish. We obtain
\begin{equation}\label{eq:FRG.vertex2}\begin{split}
0=&\sum_q\Bigg[\frac{\delta^4\Gamma}{\delta\bar{\phi}_{l'}\delta\phi_l\delta\phi_{k'}\delta\bar{\phi}_q}
\frac{\delta^2\mc{W}^c}{\delta\eta_q\delta\bar{\eta}_{k}} + \left( \frac{\delta^2\Gamma}{\delta\phi_{k'}\delta\bar{\phi}_q} - \left[\mc{G}^0\right]^{-1}_{q,k'}\right)
 \sum_{s,s'} \left(\frac{\delta\eta_s}{\delta\phi_l}\frac{\delta\bar{\eta}_{s'}}{\delta\bar{\phi}_{l'}}\frac{\delta^4\mc{W}^c}{\delta\bar{\eta}_{s'}\delta\eta_s\delta\eta_q\delta\bar{\eta}_{k}}\right)\Bigg]\Bigg|_{\phi=\bar{\phi}=0}
  \\ =& -\sum_q\frac{\delta^4\Gamma}{\delta\bar{\phi}_{l'}\delta\phi_l\delta\phi_{k'}\delta\bar{\phi}_q}\Big|_{\phi=\bar{\phi}=0}
\mc{G}_{k,q} -  \sum_{q,s,s'} \left[ [\mc{G}]^{-1}_{q,k'}[\mc{G}]^{-1}_{s,l}[\mc{G}]^{-1}_{l',s'}\frac{\delta^4\mc{W}^c}{\delta\bar{\eta}_{s'}\delta\eta_s\delta\eta_q\delta\bar{\eta}_{k}}\Big|_{\phi=\bar{\phi}=0}\right] \\
= & -\sum_q\gamma_2(l',q;k',l)\mc{G}_{k,q} - 
 \sum_{q,s,s'} \left[ [\mc{G}]^{-1}_{q,k'}[\mc{G}]^{-1}_{s,l}[\mc{G}]^{-1}_{l',s'}G_2^c(s',k,q,s)\right],
\end{split}\end{equation}
where we have used
\begin{equation}\begin{split}
\frac{\delta\eta_s}{\delta\phi_k} &\stackrel{(\ref{eq:FRG.ableitungvertexfuntional1})}{=}
\left[\mc{G}^0\right]^{-1}_{s,k}-\frac{\delta^2\Gamma}{\delta\phi_k\delta\bar{\phi}_s}
\stackrel{(\phi=\bar{\phi}=0)}{=}[\mc{G}]^{-1}_{s,k}  \\
\frac{\delta\bar{\eta}_s}{\delta\bar{\phi}_k} &\stackrel{(\ref{eq:FRG.ableitungvertexfuntional2})}{=}
\left[\mc{G}^0\right]^{-1}_{k,s}+\frac{\delta^2\Gamma}{\delta\bar{\phi}_k\delta\phi_s}
\stackrel{(\phi=\bar{\phi}=0)}{=}[\mc{G}]^{-1}_{k,s}.
\end{split}\end{equation}
Solving (\ref{eq:FRG.vertex2}) for $\gamma_2$ yields
\begin{equation}\label{eq:FRG.dyson2}
\gamma_2(k_1',k_2';k_1,k_2) = -\sum_{q_1',q_2',q_1,q_2}[\mc{G}]^{-1}_{k_1',q_1'}[\mc{G}]^{-1}_{k_2',q_2'}[\mc{G}]^{-1}_{q_2,k_2}[\mc{G}]^{-1}_{q_1,k_1} G_2^c(q_1',q_2';q_1,q_2),
\end{equation}
which is the prove of the one-particle irreducibility of the two-particle vertex functions. Interpreted diagrammatically, (\ref{eq:FRG.dyson2}) just states that the two-particle vertex is obtained by cutting a full single-particle line at each external leg of every two-particle connected diagram. What is then left are those diagrams that cannot be further split by just cutting a single-particle line (since the split-off part would have been amputated before), which is precisely the definition of one-particle irreducibility.

\section{The fRG Flow Equations}

Having learnt how functional integrals can be used to compute the partition function as well as Green functions of a fermionic many-particle system, we have established the formalism to set up the fRG flow equations. In order to do so, we introduce a new independent variable $\la$ in the noninteracting propagator $\mc{G}^0$,
\begin{equation*}
\mc{G}^0 \rightarrow \mc{G}^0(\la).
\end{equation*}
Later on, we will think of $\la$ as an infrared cutoff in frequency space,
\begin{equation*}
\mc G^0(i\omega_n,\la) = \Theta(|\omega_n|-\la)\mc G^0(i\omega_n),
\end{equation*}
but for the derivation of the flow equations the actual form of the $\la$-dependence is completely irrelevant, so that we will not specify it right now.

In a very pragmatic sense, the fRG can be viewed as follows. The flow equations for any physical quantity $X[\mc{G}^0(\la)]$ (such as the generating functionals $\mc{W}^c$ and $\Gamma$) are obtained by taking the derivative with respect to $\la$,
\begin{equation}\label{eq:FRG.flow1}
\dot X(\la) = Y(\la),
\end{equation}
where $Y(\la)$ remains to be computed. If we now choose two values $\la^{\tn{initial}}$ and $\la^{\tn{final}}$ such that $X[\mc{G}^0(\la^{\tn{initial}})]$ is easy-to-calculate, while the propagator evaluated at $\la^{\tn{final}}$ is just the ordinary free propagator $\mc{G}^0(\la^{\tn{final}})=\mc{G}^0$, we can try to solve (\ref{eq:FRG.flow1}) with the initial condition $X(\la^{\tn{initial}})=X[\mc{G}^0(\la^{\tn{initial}})]$ to compute our desired quantity $X[\mc{G}^0(\la^{\tn{final}})]$. A `simple' $X[\mc{G}^0(\la^{\tn{initial}})]$ is in particular obtained by cutting out all degrees of freedom, that is choosing $\mc{G}^0(\la^{\tn{initial}})=0$, which will be our future choice.

If $X$ depends on other independent variables $\varphi_i$ (such as $\Gamma(\{\bar\phi\},\{\phi\})$), we can expand it into a Taylor series with coefficients $\mu_i$ describing our physical system (such as the self-energy), assuming that we have chosen an appropriate expansion point $\tilde\varphi_i$. (\ref{eq:FRG.flow1}) will then provide an infinite hierarchy of flow equations for the expansion coefficients. Solving this hierarchy is equivalent to solving the original flow equation for arbitrary values of the variables $\varphi_i$.

One could immediately argue that in general we are already unable to calculate the partition function of an interacting problem, and hence we will totally fail in solving infinitely many coupled differential equations whose solution would provide us with complete knowledge of our system of correlated fermions. This is certainly true. The fundamental point is that for a suitable choice of $X$ there is physical reasoning to truncate the infinite hierarchy of flow equations for the expansion coefficients, rendering their solution (at least numerically) easy. The generating functional of the vertex functions will turn out to be that suitable choice if we want to set up an approximation which treats the two-particle interaction perturbatively, justifying our extensive elaboration on it.

\subsection{Flow Equations of Connected Green Functions}

We will for the moment refrain from setting up a flow equation for the generating functional of the vertex functions and turn to the connected Green functions instead. The flow equations of the latter will be of no further interest, but they will facilitate the computation of the former.

For convenience, we replace the full partition function in the denominator of (\ref{eq:FRG.greenfunktional}) by the noninteracting one, so that we start out with the following functional
\begin{equation}\label{eq:FRG.greenfunktional2}
\mathcal{W}(\left\{\bar{\eta}\right\},\left\{\eta\right\},\la) = \frac{1}{\mc{Z}_0(\la)} \int\mc{D}\bar{\psi}\psi\exp{\Big\{S_0(\la) - S_{\textnormal{int}} - (\bar{\psi},\eta) - (\bar{\eta},\psi) \Big\} }.
\end{equation}
This replacement changes $\mc{W}^c$ and $\Gamma$ only by the trivial constant $\ln(\mc Z_0)$, while all higher-order functions $G_{m>0}^c$ and $\gamma_{m>0}$ remain unaffected. In order to derive the flow equation for the generating functional of the connected Green functions, we have to compute the derivative
\begin{equation}\label{eq:FRG.flow2}
\dot{\mc W}^c(\la) = \frac{1}{\mc{W}(\la)} \partial_\la\left[\frac{1}{\mc Z_0(\la)}\int\mc{D}\bar{\psi}\psi\exp{\Big\{S_0(\la) - S_{\textnormal{int}} - (\bar{\psi},\eta) - (\bar{\eta},\psi) \Big\} }\right].
\end{equation}
Therefore we first calculate
\begin{equation*}\begin{split}
\partial_\la \frac{e^{S_0(\la)}}{\mc Z_0(\la)} & =
\frac{1}{\mc Z_0(\la)}\dot{S_0}(\la)e^{S_0(\la)} - \left(\frac{1}{\mc Z_0(\la)}\right)^2e^{S_0(\la)}\partial_\la\mc Z^0(\la) \\ & = \frac{1}{\mc Z_0(\la)}\dot{S_0}(\la)e^{S_0(\la)} - \left(\frac{1}{\mc Z_0(\la)}\right)^2e^{S_0(\la)} \int\mc D \bar\psi\psi~\dot{S_0}(\la)e^{S_0(\la)} \\
& = \frac{e^{S_0(\la)}}{\mc Z_0(\la)} \left[\left(\bar\psi,\partial_\la[\mc G^0(\la)]^{-1}\psi\right) -  
\sum_{k,k'}\left\{\int \frac{\mc D\bar\psi\psi}{\mc Z_0(\la)}\bar\psi_k\psi_{k'}e^{S_0(\la)}\cdot\partial_\la [\mc G ^0(\la)]^{-1}_{k,k'} \right\}\right] \\
& = \frac{e^{S_0(\la)}}{\mc Z_0(\la)} \left[\left(\bar\psi,\partial_\la[\mc G^0(\la)]^{-1}\psi\right) -  \tn{Tr}\left(\mc G^0(\la)\partial_\la[\mc G^0(\la)]^{-1}\right)\right],
\end{split}\end{equation*}
which allows for rewriting (\ref{eq:FRG.flow2}) in the simple form
\begin{equation}\label{eq:FRG.flow3}
\dot{\mc W}^c(\la) = - \tn{Tr}\left(\mc G^0(\la)\partial_\la[\mc G^0(\la)]^{-1}\right) -
\underbrace{\frac{1}{\mc{W}(\la)}\left(\frac{\delta}{\delta\eta},\partial_\la[\mc G^0(\la)]^{-1}\frac{\delta}{\delta\bar\eta}\right)\mc W(\la)}_{=:X}.
\end{equation}
If we eliminate $\mc W$ in the second term,
\begin{equation*}\begin{split}
X & =
e^{-\mc{W}^c(\la)}\left(\frac{\delta}{\delta\eta},\partial_\la[\mc G^0(\la)]^{-1}\frac{\delta}{\delta\bar\eta}\right)e^{\mc W^c(\la)} \\
& = \left(\frac{\delta\mc W^c(\la)}{\delta\eta},\partial_\la[\mc G^0(\la)]^{-1}\frac{\delta\mc W^c(\la)}{\delta\bar\eta}\right)+\left(\frac{\delta}{\delta\eta},\partial_\la[\mc G^0(\la)]^{-1}\frac{\delta}{\delta\bar\eta}\right)\mc W^c(\la) \\
& = \left(\frac{\delta\mc W^c(\la)}{\delta\eta},\partial_\la[\mc G^0(\la)]^{-1}\frac{\delta\mc W^c(\la)}{\delta\bar\eta}\right) +
\sum_{k,k'}\left[ \frac{\delta^2\mc W^c(\la)}{\delta\eta_k\delta\bar\eta_{k'}}\partial_\la [\mc G^0(\la)]^{-1}_{k,k'} \right] \\
& = \left(\frac{\delta\mc W^c(\la)}{\delta\eta},\partial_\la[\mc G^0(\la)]^{-1}\frac{\delta\mc W^c(\la)}{\delta\bar\eta}\right) - \tn{Tr}\left(\partial_\la [\mc G^0(\la)]^{-1}\frac{\delta^2\mc W^c(\la)}{\delta\bar\eta\delta\eta}\right),
\end{split}\end{equation*}
we are finally able to write down an analytic expression for the flow equation of the generating functional of the connected Green functions,
\begin{equation}\label{eq:FRG.flowconnected}\begin{split}
\dot{\mc W}^c(\la) = & - \tn{Tr}\left(\mc G^0(\la)\partial_\la[\mc G^0(\la)]^{-1}\right) + \tn{Tr}\left(\partial_\la [\mc G^0(\la)]^{-1}\frac{\delta^2\mc W^c(\la)}{\delta\bar\eta\delta\eta}\right) \\
& -\left(\frac{\delta\mc W^c(\la)}{\delta\eta},\partial_\la[\mc G^0(\la)]^{-1}\frac{\delta\mc W^c(\la)}{\delta\bar\eta}\right).
\end{split}\end{equation}
Having solved this differential equation, we can calculate the grand canonical potential,\footnote{Remember our replacement $\mc Z\rightarrow\mc Z_0$ in the denominator of (\ref{eq:FRG.greenfunktional2}).}
\begin{equation*}
-T\ln\mc Z = -TG_0^c - T\ln\mc Z_0 = -TG_0^c(\la=\la_1) - T\ln\mc Z_0.
\end{equation*}

Since the flow equation for the generating functional of the Green functions follows directly from (\ref{eq:FRG.flow3}), we write it down for reasons of completeness
\begin{equation*}
\dot{\mc W}(\la) = - \left[\tn{Tr}\left(\mc G^0(\la)\partial_\la[\mc G^0(\la)]^{-1}\right) +
\left(\frac{\delta}{\delta\eta},\partial_\la[\mc G^0(\la)]^{-1}\frac{\delta}{\delta\bar\eta}\right)\right]\mc W(\la).
\end{equation*}

\subsection{Flow Equations of the Vertex Functions}

\subsubsection{Flow Equation of the Generating Functional}

We will now compute the flow equations for our primary quantity of interest, the vertex functions. We start out by simply differentiating (\ref{eq:FRG.vertexfunktional}) with respect to $\la$, bearing in mind that the fields $\eta$ and $\bar\eta$ have to be expressed in terms of $\phi$ and $\bar\phi$ via (\ref{eq:FRG.felderphi}) and therefore acquire a $\la$-dependence. We obtain
\begin{equation}\label{eq:FRG.flowvertex1}\begin{split}
\dot\Gamma(\left\{\bar{\phi}\right\},\left\{\phi\right\},\la)&=
-\frac{d}{d\la}\mathcal{W}^c(\left\{\bar{\eta}(\la)\right\},\left\{\eta(\la)\right\},\la) - (\bar{\phi},\dot\eta(\la)) - (\dot{\bar{\eta}}(\la),\phi)  + \left(\bar{\phi},\partial_\la[\mc{G}^0]^{-1}\phi\right) \\
&= -\dot{\mc W}^c-\sum_k\left[\dot\eta_k\frac{\delta\mc W^c}{\delta\eta_k}+\dot{\bar\eta}_k\frac{\delta\mc W^c}{\delta\bar\eta_k}\right]- (\bar{\phi},\dot\eta) - (\dot{\bar{\eta}},\phi)  + \left(\bar{\phi},\partial_\la[\mc{G}^0]^{-1}\phi\right) \\
& \hspace{-0.235cm} \stackrel{(\ref{eq:FRG.felderphi})}{=} -\dot{\mathcal{W}}^c(\left\{\bar{\eta}(\la)\right\},\left\{\eta(\la)\right\},\la)
+\left(\bar{\phi},\partial_\la[\mc{G}^0]^{-1}\phi\right) \\
& \hspace{-0.235cm} \stackrel{(\ref{eq:FRG.flowconnected})}{=} \tn{Tr}\left(\mc G^0(\la)\partial_\la[\mc G^0(\la)]^{-1}\right) - \tn{Tr}\left(\partial_\la [\mc G^0(\la)]^{-1}\frac{\delta^2\mc W^c(\la)}{\delta\bar\eta\delta\eta}\right) \\
& \hspace{-0.235cm} \stackrel{(\ref{eq:FRG.zusammenhang})}{=} \tn{Tr}\left(\mc G^0(\la)\partial_\la[\mc G^0(\la)]^{-1}\right) - \tn{Tr}\left(\partial_\la [\mc G^0(\la)]^{-1}\mc V^{(1,1)} (\bar\phi,\phi,\la)\right),
\end{split}\end{equation}
where the dot on top of $\mc W^c$ denotes the outer derivative, and we have defined $\mc V^{(1,1)}$ as the (1,1) - element of the matrix equation (\ref{eq:FRG.zusammenhang}) that relates the derivatives of $\mc W^c$ to those of $\Gamma$ (at arbitrary values of the external fields).

\subsubsection{Flow Equations of Vertex Functions}

In order to expand it into a Taylor series, we rewrite (\ref{eq:FRG.zusammenhang}) as
\begin{equation*}\begin{split}
\mc V(\la)=&
\begin{pmatrix}
\frac{\delta^2\Gamma}{\delta\bar{\phi}\delta\phi} + \left[\mc{G}^0\right]^{-1} & \frac{\delta^2\Gamma}{\delta\bar{\phi}\delta\bar{\phi}} \\
\frac{\delta^2\Gamma}{\delta\phi\delta\phi} & \frac{\delta^2\Gamma}{\delta\phi\delta\bar{\phi}} - \left[\left[\mc{G}^0\right]^{-1}\right]^T
\end{pmatrix}^{-1} \\
= & \left[
\begin{pmatrix}
[\mc G(\la)]^{-1} & 0 \\
0 & - \left[[\mc G(\la)]^{-1}\right]^{T}
\end{pmatrix}+
\begin{pmatrix}
\mc U(\la) & \frac{\delta^2\Gamma(\la)}{\delta\bar\phi\delta\bar\phi} \\
\frac{\delta^2\Gamma(\la)}{\delta\phi\delta\phi} & -\mc U(\la)
\end{pmatrix} \right]^{-1} \\
= & -\underbrace{\left[ 1- 
\begin{pmatrix}
-\mc G(\la) & 0 \\
0 &  [\mc G(\la)]^{T}
\end{pmatrix}\cdot
\begin{pmatrix}
\mc U(\la) & \frac{\delta^2\Gamma(\la)}{\delta\bar\phi\delta\bar\phi} \\
\frac{\delta^2\Gamma(\la)}{\delta\phi\delta\phi} & -\mc U(\la)
\end{pmatrix} \right]^{-1}}_{=:\hat{\mc V}}\cdot
\begin{pmatrix}
-\mc G(\la) & 0 \\
0 &  [\mc G(\la)]^{T}
\end{pmatrix},
\end{split}\end{equation*}
with $\mc U(\bar\phi,\phi,\la)$ being the difference of the one-particle vertex function and the second derivative of $\Gamma$ evaluated at arbitrary external fields,
\begin{equation*}\begin{split}
\mc U(\la) :&= \frac{\delta^2\Gamma(\la)}{\delta\bar\phi\delta\phi}-\frac{\delta^2\Gamma(\la)}{\delta\bar\phi\delta\phi}\Big|_{\phi=\bar\phi=0} \\
& = \frac{\delta^2\Gamma(\la)}{\delta\bar\phi\delta\phi} - \gamma_1(\la) \\
& = \frac{\delta^2\Gamma(\la)}{\delta\bar\phi\delta\phi} - [\mc G(\la)]^{-1} + [\mc G^0(\la)]^{-1}.
\end{split}\end{equation*}
Defining $\tilde{\mc V}:=\hat{\mc V}^{(1,1)}$, (\ref{eq:FRG.flowvertex1}) can be cast in the following form,
\begin{equation}\label{eq:FRG.flowvertex}
\dot\Gamma(\la) = \tn{Tr}\left(\mc G^0(\la)\partial_\la[\mc G^0(\la)]^{-1}\right) - 
\tn{Tr}\left(\mc G(\la)\partial_\la [\mc G^0(\la)]^{-1}\tilde{\mc V}(\la)\right).
\end{equation}

Of course there is no new physical insight in (\ref{eq:FRG.flowvertex}) as it is just a different formulation of (\ref{eq:FRG.flowvertex1}) with renamed variables. But these new variables allow for a Taylor expansion, which will be useful to derive ordinary differential equations from the functional differential equation (\ref{eq:FRG.flowvertex}). Namely,
\begin{equation}\label{eq:FRG.flowvertex2}
\tilde{\mc V}(\la) = 1 - \mc G\mc U +\mc G\mc U\mc G\mc U - \mc G \frac{\delta^2\Gamma}{\delta\bar\phi\delta\bar\phi}\mc G^T \frac{\delta^2\Gamma(\la)}{\delta\phi\delta\phi} + \ldots,
\end{equation}
where we have written down all terms up to second order of the geometric series.

Next, we expand $\Gamma(\bar\phi,\phi,\la)$ around $\bar\phi=\phi=0$,
\begin{equation*}
\Gamma(\bar\phi,\phi,\la)= \sum_{m=0}^{\infty} \frac{(-1)^m}{(m!)^2} \sum_{k'_1 \ldots k'_m}\sum_{k_1 \ldots k_m}
\gamma_m(k'_1,\ldots,k'_m;k_1,\ldots,k_m;\la)\bar\phi_{k'_1}\ldots\bar\phi_{k'_m}\phi_{k_m}\ldots\phi_{k_1},
\end{equation*}
which by substitution into (\ref{eq:FRG.flowvertex}) sets up flow equations for the physical meaningful expansion coefficients $\gamma_m$. For $\gamma_0$ we obtain
\begin{equation*}
\dot\gamma_0(\la) = \tn{Tr}\left(\mc G^0(\la)\partial_\la[\mc G^0(\la)]^{-1}\right) - 
\tn{Tr}\left(\mc G(\la)\partial_\la [\mc G^0(\la)]^{-1}\right),
\end{equation*}
which is the first term in (\ref{eq:FRG.flowvertex}) together with the zeroth order contribution of $\tilde{\mc V}$ to the second term, since $\mc U$ as well as $\frac{\delta^2\Gamma}{\delta\bar\phi\delta\bar\phi}$ and $\frac{\delta^2\Gamma}{\delta\phi\delta\phi}$ are at least of second order in the fields.\footnote{For $\frac{\delta^2\Gamma}{\delta\bar\phi\delta\bar\phi}$ and $\frac{\delta^2\Gamma}{\delta\phi\delta\phi}$ this follows because all terms in the Taylor expansion that not containing an equal number of fields $\bar\phi$ and $\phi$ vanish (unless we are in a phase of broken symmetry), while for $\mc U$ all zeroth order terms are cancelled by definition.}

\subsubsection{Flow of the Self-Energy}

To set up the flow equation for $\gamma_1$, we have to compute the part of $\tilde{\mc V}$ that is linear in $\bar\phi\phi$. It is given by
\begin{equation*}\begin{split}
\tilde{\mc V}^{\tn{lin}}  & = \frac{1}{(2!)^2}\frac{\delta^2}{\delta\bar\phi_{q'}\delta\phi_q}\sum_{k'_1,k'_2,k_1,k_2}\gamma_2(k'_1,k'_2;k_1,k_2;\la)\bar\phi_{k'_1}\bar\phi_{k'_2}\phi_{k_2}\phi_{k_1} \\
& = -\sum_{k',k}\gamma_2(k',q';k,q;\la)\bar\phi_{k'}\phi_k,
\end{split}\end{equation*}
where we have used that by construction it follows that
\begin{equation*}
\gamma_2(k'_1,k'_2;k_1,k_2;\la) = -\gamma_2(k'_2,k'_1;k_1,k_2;\la),
\end{equation*}
and likewise for the interchange of $k_1$ and $k_2$. Comparison of the coefficients of the term linear in $\bar\phi\phi$ on the left- and right hand side of (\ref{eq:FRG.flowvertex}) now yields the desired flow equation,
\begin{equation}\label{eq:FRG.flowse}\begin{split}
\dot\gamma_1(k';k;\la) & =  
\tn{Tr}\left(\mc G(\la)\partial_\la [\mc G^0(\la)]^{-1}\mc G(\la)\gamma_2(k',~;k,~;\la)\right),
\end{split}\end{equation}
with $\gamma_2(k',~;k,~)$ denoting a matrix with indices $[\gamma_2(k',~;k,~)]_{q',q}:=\gamma_2(k',q';k,q)$. The self-energy couples directly into this equation via the full propagator, $\mc G(\la)$.

\subsubsection{The Flow of $\gamma_2$ and Higher-Order Functions}

In order to derive a flow equation for the two-particle vertex, we have to find all terms on the right hand side of (\ref{eq:FRG.flowvertex}) (or rather of (\ref{eq:FRG.flowvertex2})) containing four external fields. Again, by comparison of coefficients we obtain
\begin{equation}\label{eq:FRG.flowww}\begin{split}
\dot\gamma_2(k'_1,k'_2;k_1,k_2;\la) = & \tn{Tr}\left(\mc G(\la)\partial_\la [\mc G^0(\la)]^{-1}\mc G(\la)\gamma_3(k'_1,k'_2,~;k_1,k_2,~;\la)\right) \\
- & \tn{Tr}\left(\mc G(\la)\partial_\la [\mc G^0(\la)]^{-1}\mc G(\la)\gamma_2(~,~;k_1,k_2;\la)[\mc G(\la)]^T\gamma_2(~,~;k_1,k_2;\la)\right) \\
- & \tn{Tr}\Big[\mc G(\la)\partial_\la [\mc G^0(\la)]^{-1}\mc G(\la)\gamma_2(k'_1,~;k_1,~;\la)\mc G(\la)\gamma_2(k'_2,~;k_2,~;\la) \\
  & \hspace{2cm} - [k'_1\leftrightarrow k'_2] - [k_1\leftrightarrow k_2] + [k'_1\leftrightarrow k'_2,k_1\leftrightarrow k_2] \Big].
\end{split}\end{equation}
The coupling to the three-particle vertex $\gamma_3$ emerges from the part that is proportional to $\bar\phi\bar\phi\phi\phi$ in the second term of (\ref{eq:FRG.flowvertex2}), while the second term results from the quadratic part of $\mc G \frac{\delta^2\Gamma}{\delta\bar\phi\delta\bar\phi}\mc G^T \frac{\delta^2\Gamma(\la)}{\delta\phi\delta\phi}$ (the fourth term of (\ref{eq:FRG.flowvertex2})). The last four terms arise from $\mc{GUGU}$ (third term of (\ref{eq:FRG.flowvertex2})) after anti-symmetrisation of the coefficient of the quadratic part (this is necessary since it is not anti-symmetric by itself; or put differently: anti-symmetrisation will naturally occur when writing down all possible ways of matching the coefficients on both sides of (\ref{eq:FRG.flowvertex})).

We could pursue this game forever and forever, writing down flow equations for the vertex functions of arbitrary order that would always turn out to couple to even higher-order functions. After having derived all these expressions (which is impossible) we would have to solve infinitely many coupled differential equations (which is impossible as well). Fortunately, having considered vertex functions will turn out to have been a very clever choice, because their flow becomes neglectible with increasing order if one assumes the two-particle interaction to be small. In fact, it will be sufficient to treat the infinite hierarchy only up to the second order (setting all other vertex functions to zero), and the corresponding flow equations were derived in this section.

The general structure of the higher-order equations, however, should have become clear by now. The flow equation for $\gamma_m$ contains one term involving $\gamma_{m+1}$ as well as contributions from all lower orders (except from $\gamma_0$). For example, the flow of $\gamma_1$ (\ref{eq:FRG.flowse}) is determined by $\gamma_2$ and by $\gamma_1$ itself (via $\mc G$).

\subsubsection{Feynman Diagrams}

\begin{figure}[t]	
	\centering
	\includegraphics[width=0.6\textwidth]{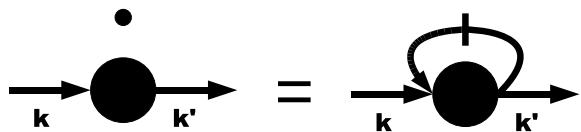}
        \caption{Diagrammatic representation of the flow equation for the self-energy (\ref{eq:FRG.flowse}).}
        \label{fig:FRG.feynman}
\end{figure}

It is often convenient to denote the matrix summations appearing on the right hand side of an arbitrary-order flow equation in analogy with Feynman diagrams. A $m$-particle vertex is symbolised by a dot with $2m$ external lines, and there are two (and only two, as should be clear from (\ref{eq:FRG.flowvertex2})) combinations that involve $\gamma_1$ on the right hand side of (\ref{eq:FRG.flowvertex}), namely $\mc G(\la)$ and $\mc G(\la)\partial_\la[\mc G(\la)]^{-1}\mc G(\la)$, and they are represented by a single line or a crossed-out line, respectively. The quantity $\mc G(\la)\partial_\la[\mc G(\la)]^{-1}\mc G(\la)=:\mc S(\la)$ is also called single-scale propagator. As for ordinary Feynman diagrams, summation over all internal lines is implicitly assumed. An example for the flow of the self-energy is shown in Fig.~\ref{fig:FRG.feynman}.

\subsubsection{Initial Conditions}

Up to now we have not thought of the initial conditions that are necessary for the solution of all the aforementioned differential equations to be well-defined. It is of course possible to derive these conditions analytically \cite{funrg}, but here we only give a very simple diagrammatic argument. Since at the beginning of the flow we want to cut out all degrees of freedom, that is $\mc G^0(\la^{\tn{initial}})=0$, all diagrams containing a finite number of free propagators are zero. The only nonvanishing diagram is the pure interaction $\bar v$ (its external legs $\mc G^0$ have been amputated by definition). Therefore the initial generating functional reads
\begin{equation*}
\Gamma(\la^{\tn{initial}}) = \frac{1}{4}\sum_{i,j,k,l}\bar v_{i,j,k,l}\bar\phi_i\bar\phi_j\phi_l\phi_k,
\end{equation*}
and the initial conditions for flow equations of the vertex functions are given by
\begin{equation}\label{eq:FRG.init1}
\gamma_2(k'_1,k'_2;k_1,k_2;\la^{\tn{initial}}) = \bar v_{k'_1,k'_2,k_1,k_2}
\end{equation}
for $m=2$, and
\begin{equation}\label{eq:FRG.init2}
\gamma_m(\la^{\tn{initial}})=0
\end{equation}
otherwise.\footnote{Sometimes one absorbs some additional one-particle terms $V_{k',k}$ into the interacting part of the action ($S_\tn{int}$). The initial condition for $\gamma_1$ in then reads $\gamma_1(k';k;\la^{\tn{initial}})=V_{k',k}$.}

\subsection{Truncation Schemes}\label{sec:FRG.truncation}

\subsubsection{General Arguments}

As mentioned above, we need to truncate the infinite hierarchy of flow equations for the vertex functions in order to render it solvable. In this section, we will set up a truncation scheme that is valid in the limit of a small two-particle interaction $\bar v$. It will turn out, however, that despite its perturbative nature the resulting finite set of equations will be sufficient in describing the effects of `fairly large' interactions even on a quantitative level in all situations considered in this thesis (we will discuss the limitations of this approximation in the last chapter).

The right hand side of the flow equation for each vertex function $\gamma_m$ contains contributions from $\gamma_{m+1}$ as well as from all lower-order vertices. At the beginning of the fRG flow, all functions $\gamma_m$ except for the two-particle interaction $\gamma_2(\la=\infty)=\bar v$ vanish, so that all vertices $\gamma_m$ with $m\ne 2$ are generated only by $\bar v$. Now, the fundamental point in setting up a truncation scheme perturbative in the interaction is the following. Since we are considering vertex functions, an $m$-particle vertex $\gamma_m(\la)$ has to be irreducible for arbitrary choice of $\la$. In particular, it can only be generated by irreducible diagrams on the right hand side of (\ref{eq:FRG.flowvertex2}). Since there are no such diagrams with $2m$ external (amputated) legs that contain less than $m$ terms $\bar v$,\footnote{Consider for example the three-particle vertex. There is one possibility to write down a diagram with two interaction vertices and six external (amputated) legs, but this diagrams contains one single-particle line connecting the interaction vertices, rendering it reducible.} all vertex functions $\gamma_{m\neq 2}$ are generated by terms that are at least of order $m$ in the interaction. If the latter is initially small and stays small for all $\la$ it is justified to cut the infinite hierarchy of flow equations at a certain order, neglecting the flow of all higher-order vertex functions.

\subsubsection{Approximation Schemes}

The very simplest approximation that emerges from these considerations is to neglect the flow of all vertices except the one for the self-energy. This leaves us with one single equation (remember that $\mc G = [(\mc G^0)^{-1} + \gamma_1]^{-1}$),
\begin{equation}\label{eq:FRG.flowse0}\begin{split}
\dot\gamma_1(k';k;\la)  & =  
\tn{Tr}\left(\mc G(\la)\partial_\la [\mc G^0(\la)]^{-1}\mc G(\la)\gamma_2(k',~;k,~;\la=0)\right) \\
& =  \sum_{q,q'}\left[\mc G(\la)\partial_\la [\mc G^0(\la)]^{-1}\mc G(\la)\right]_{q,q'}\bar v_{k',q',k,q},
\end{split}\end{equation}
which is easy to tackle numerically (and sometimes can even be solved analytically). Since the bare interaction $\bar v$ is frequency-independent, the self-energy will not acquire a frequency dependence during the flow, so that the effect of the interaction is just to renormalize the single-particle energies of our system by $\gamma_1(\la^{\tn{final}})$, allowing for an (effective) one-particle interpretation.

The next logical step would be to consider the coupled flow equations of the self-energy and the two-particle vertex. Unfortunately the resulting problem would not be frequency-independent any more, so that we would again have to solve infinitely many coupled equations (because we have infinitely many Matsubara frequencies). Therefore we implement another approximation: we neglect all frequency-dependencies of the two-particle vertex except for frequency conservation, which we have explicitly included in $\bar v$ and which is conserved by the flow equation (since $\mc G^0$ is diagonal in frequency space). As the bare interaction is frequency-independent, this will lead to errors only of second order for the self-energy, and of third order for the two-particle vertex. Alltogether, (\ref{eq:FRG.flowse}) and (\ref{eq:FRG.flowww}) can then be recast as
\begin{equation}\label{eq:FRG.flowse2}
\dot\gamma_1(k';k;\la)   =  \sum_{i\omega_n}\sum_{q,q'}[\underbrace{\mc G(i\omega_n,\la)\partial_\la [\mc G^0(i\omega_n,\la)]^{-1}\mc G(i\omega_n,\la)}_{=:\mc S(i\omega_n,\la)}]_{q,q'}\gamma_2(k',q';k,q;\la)
\end{equation}
for the flow of the self-energy, and
\begin{equation}\label{eq:FRG.flowww2}\begin{split}
\dot\gamma_2(k'_1,k'_2;k_1,k_2;\la) & = \\ \sum_{i\omega_n}\sum_{s',q',s,q}\Bigg[ 
 & - \mc S_{q,q'}(i\omega_n,\la)\gamma_2(q',s';k_1,k_2;\la)\mc G_{s,s'}(-i\omega_n,\la)\gamma_2(k'_1,k'_2;s,q;\la) \\
 & - \Big\{ \mc S_{q,q'}(i\omega_n,\la)\gamma_2(k'_1,q';k_1,s;\la)\mc G_{s,s'}(i\omega_n,\la)\gamma_2(k'_2,s';k_2,q;\la) \\
 & \hspace{1cm}- [k'_1\leftrightarrow k'_2] - [k_1\leftrightarrow k_2] + [k'_1\leftrightarrow k'_2, k_1\leftrightarrow k_2] \Big\} \Bigg]
\end{split}\end{equation}
for the flow of the effective interaction (as we will call the frequency-independent two-particle vertex from now on). Here all the labels $k_i$ only denote the single-particle quantum numbers, the frequency-dependence has been written out explicitly. It will turn out that the remaining summations over the Matsubara frequency $i\omega_n$ can be easily performed if we introduce a sharp cutoff in frequency space.

Also this more elaborate approximation scheme yields a self-energy $\gamma_1(\la^{\tn{final}})$ that is frequency-independent, which will again facilitate gaining further insights into our upcoming results as it allows for an interpretation in a simple effective single-particle picture.\footnote{A word of warning is in order. If the whole effect of the interaction is to renormalize the single-particle energy of our system, the properties of this effective system can of course be interpreted very easily due to its noninteracting nature. This does, however, not answer the question \textit{why} the interaction particularly renormalizes the levels the way it does.}

\subsubsection{Symmetries}

By applying the aforementioned approximations, we have boiled down the infinite hierarchy of flow equations to $N^2+N^4$ coupled ordinary differential equations, where $N$ denotes the number of single-particle quantum numbers that define our interacting system. For all $N$ considered in this thesis ($N\leq 8$), this set can be solved very easily by direct implementation on a state-of-the-art personal computer. However, if one wants to even further speed up the numerics, one should realise that many flow equations are still redundant since up to now we have not exploited any symmetries.

The most obvious symmetry to reduce the number of independent equations is the (anti)symmetry of the two-particle vertex. By definition, the bare interaction $\bar v$ is symmetric under the exchange of the first with the last two indices, and it is antisymmetric under the exchange of the first or the last indices, respectively. These symmetries are conserved by the flow equation (\ref{eq:FRG.flowww2}). For the first term on the right hand side conservation of the symmetry is seen if one considers
\begin{equation*}\begin{split}
&~ \mc S_{q,q'}(i\omega_n,\la)\gamma_2(q',s';k'_1,k'_2;\la)\mc G_{s,s'}(-i\omega_n,\la)\gamma_2(k_1,k_2;s,q;\la) \\
= &~ \mc S_{q,q'}(i\omega_n,\la)\gamma_2(s,q;k_1,k_2;\la)\mc G_{s,s'}(-i\omega_n,\la)\gamma_2(k'_1,k'_2;q',s';\la) \\
= &~ \mc S_{q',q}(-i\omega_n,\la)\gamma_2(q,s;k_1,k_2;\la)\mc G_{s',s}(i\omega_n,\la)\gamma_2(k'_1,k'_2;s',q';\la),
\end{split}\end{equation*}
where we have used $\mc G_{s,s'}(i\omega_n)=-\mc G_{s',s}(-i\omega_n)$ as well as (\ref{eq:FRG.morris1}). Renaming the summation indices $(q\leftrightarrow q', s\leftrightarrow s', i\omega_n\rightarrow -i\omega_n)$ then yields the desired symmetry under exchange of the first with the last indices. For the second term this follows directly from the definition, and it is obvious that (\ref{eq:FRG.flowww2}) preserves the anti-symmetry of the two-particle vertex as well.\footnote{The two-particle vertex is of course antisymmetric for arbitrary choice of $\la$. But since we are only computing an \textit{approximation} for the exact vertex it is, however, not from the beginning obvious that this approximate vertex obeys the same antisymmetry relations.}

The second important symmetry of the equations (\ref{eq:FRG.flowse2}) and (\ref{eq:FRG.flowww2}) is spin conservation. If the bare interaction and the free propagator are spin-conserving, no terms that are non-diagonal in spin space can be generated by the flow.

\subsubsection{Comparison to Ordinary Perturbation Theory}

In order to render the infinite hierarchy of flow equations solvable, we have applied a truncation scheme valid in the limit of vanishing strength of the two-particle interaction. One could now immediately ask why such a perturbative approach should be superior to ordinary perturbation theory (which does not require 20 pages of elaboration on functional integrals to be set up). This, however, can already be seen by considering only the very simplest approximation, the flow equation for the self-energy with the two-particle vertex set to its initial value $\bar v$ (\ref{eq:FRG.flowse0}),
\begin{equation*}\begin{split}
\dot\gamma_1(k';k;i\omega_n;\la)  & =  \sum_{i\omega_n}\sum_{q,q'}\left[\mc G(i\omega_n,\la)\partial_\la [\mc G^0(i\omega_n,\la)]^{-1}\mc G(i\omega_n,\la)\right]_{q,q'}\bar v_{k',q',k,q} \\
& \approx \sum_{i\omega_n}\sum_{q,q'}\left[-\partial_\la [\mc G^0(i\omega_n,\la)]^{-1}\right]_{q,q'}\bar v_{k',q',k,q}.
\end{split}\end{equation*}
In the second line we have furthermore replaced the full propagator $\mc G$ by the noninteracting one, neglecting the self-energy. The integration can then be performed trivially, leading to
\begin{equation*}
\gamma_1(k';k;i\omega_n;\la^{\tn{final}}) \approx -\sum_{i\omega_n}\sum_{q,q'}\left[\mc G^0(i\omega_n)\right]^{-1}_{q,q'}\bar v_{k',q',k,q},
\end{equation*}
which is precisely the expression that would have followed from first order perturbation theory, but it is only recovered because we have made a further approximation within our original flow equation. To say it the other way round, (\ref{eq:FRG.flowse0}) has to correspond to a summation of more than first order diagrams. In fact, it will turn out that in all situations considered in this thesis, the functional renormalization group will produce results far superior to those of perturbation theory.

\subsection{Specification of a Cutoff}

\subsubsection{Introduction of a Sharp Cutoff}

The only remaining step now is to specify a certain form of the $\la$-dependent propagator $\mc G^0(\la)$. As mentioned above, we choose $\la$ to be an infrared cutoff in frequency space, namely
\begin{equation}\label{eq:FRG.cutoff}
\mc G^0(i\omega_n,\la) = \Theta(|\omega_n|-\la)\mc G^0(i\omega_n).
\end{equation}
We have employed a `sharp' $\Theta$ - function cutoff to simplify our calculations and to speed up the numerics\footnote{Of course the special choice of the cutoff should not influence the underlying physics. It can nevertheless effect the quality of the results obtained. We will, however, not elaborate on this issue here.} (in particular to make it possible to carry out the summation over the Matsubara frequencies in (\ref{eq:FRG.flowse2}) and (\ref{eq:FRG.flowww2}) analytically). Since we want to cut out all degrees of freedom at the beginning of the flow, we choose $\la^{\tn{initial}}=\infty$, and we set $\la^{\tn{final}}=0$ to recover the original free propagator.

\subsubsection{Carrying Out the Freqency Integrals: Morris' Lemma}

In order to perform the remaining frequency summation in the flow equations for the self-energy and the effective interaction, we consider
\begin{equation}\label{eq:FRG.morris1}\begin{split}
\mc S(i\omega_n,\la) & = \mc G(i\omega_n,\la)\partial_\la [\mc G^0(i\omega_n,\la)]^{-1}\mc G(i\omega_n,\la) \\
& = \frac{\mc G^0\Theta}{1+\mc G^0\Theta\gamma_1}[\mc G^0\Theta]^{-1}\mc G^0\delta[\mc G^0\Theta]^{-1}\frac{\mc G^0\Theta}{1+\mc G^0\Theta\gamma_1} \\
& = \frac{\mc G^0\delta}{(1+\mc G^0\Theta\gamma_1)^2} \\&=\delta \partial_\Theta \left[\frac{\mc G^0\Theta}{1+\mc G^0\Theta\gamma_1}\right] \\[0.5ex]&= \delta(|\omega_n|-\la)\partial_\Theta \mc G(i\omega_n,\la).
\end{split}\end{equation}
In the zero-temperature limit, on which we will mainly focus within this thesis, we can write the summation over Matsubara frequencies as an integral, $\sum_{i\omega_n} \rightarrow (2\pi T)^{-1}\int d\omega$, and with the help of (\ref{eq:FRG.morris1}), (\ref{eq:FRG.flowse2}) becomes
\begin{equation}\label{eq:FRG.flowse3}\begin{split}
\dot\gamma_1(k';k;\la)   & =  \frac{1}{2\pi}\int d\omega \sum_{q,q'} \mc S_{q,q'}(i\omega,\la)\gamma_2(k',q';k,q;\la) \\
& = \frac{1}{2\pi}\int d\omega \sum_{q,q'} \delta(|\omega|-\la)\partial_\Theta \mc G_{q,q'}(i\omega,\la)\gamma_2(k',q';k,q;\la) \\
& \hspace{-0.05cm}\stackrel{(*)}{=} \frac{1}{2\pi}\int d\omega \sum_{q,q'} \delta(|\omega|-\la)\int_0^1 dt~\left\{\partial_t \mc G_{q,q'}(i\omega,\la;\Theta\rightarrow t)\right\}\gamma_2(k',q';k,q;\la) \\
& = \frac{1}{2\pi}\int d\omega \sum_{q,q'} \mc \delta(|\omega|-\la)\tilde{\mc G}_{q,q'}(i\omega,\la)\gamma_2(k',q';k,q;\la) \\
& = \frac{1}{2\pi}\sum_{\omega=\pm\la}\sum_{q,q'} \tilde{\mc G}_{q,q'}(i\omega,\la)\gamma_2(k',q';k,q;\la).
\end{split}\end{equation}
The factor of $T$ was cancelled by one implicitly contained in the two-particle vertex $\gamma_2$ via the initial condition (\ref{eq:FRG.init2}) (because we have initially absorbed a factor of $\beta^{-1}$ into the bare interaction which we will from now on consequently think of not containing the temperature any more), and we have introduced
\begin{equation*}
\tilde{\mc G}(\la) := \frac{1}{[\mc G^0]^{-1} + \gamma_1(\la)}.
\end{equation*}
The identity $(*)$ is the so-called Morris' lemma that allows for evaluating the at first sight ambiguous product of a $\delta$- with a $\Theta$-function, assuming that the sharp cutoff $\Theta$ is implemented as a limit of increasingly sharp broadened cutoff functions $\Theta_{\epsilon\rightarrow 0}$ \cite{morris},
\begin{equation*}
\delta_\epsilon(x-\la)f[\Theta_\epsilon(x-\la)] \rightarrow \delta(x-\la)\int_0^1 f(t) dt,
\end{equation*}
with $f$ being a continuous but otherwise arbitrary function. The functional form of $\Theta_\epsilon$ for finite $\epsilon$ does not affect the result in the limit $\epsilon\rightarrow 0$.

To show how the frequency summation in the flow equation for the effective interaction is performed, we examine
\begin{equation*}\begin{split}
 &  \hspace{-1cm} \sum_{i\omega_n} \mc S_{q,q'}(i\omega_n,\la) \mc G_{s,s'}(\mp i\omega_n,\la) \\
 = & \frac{\beta}{2\pi} \int d\omega~\delta(|\omega|-\la)\int_0^1 dt~\left\{\partial_t[\mc G_{q,q'}(i\omega,\la,\Theta\rightarrow t)]\mc G_{s,s'}(\mp i\omega,\la,\Theta\rightarrow t)\right\} \\
= & \frac{\beta}{2\pi} \int d\omega~\delta(|\omega|-\la)\int_0^1 dt~\left\{\frac{1}{2}\partial_t[\mc G_{q,q'}(i\omega,\la,\Theta\rightarrow t)\mc G_{s,s'}(\mp i\omega,\la,\Theta\rightarrow t)]\right\} \\
= & \frac{\beta}{4\pi} \int d\omega~\delta(|\omega|-\la)\tilde{\mc G}_{q,q'}(i\omega,\la)\tilde{\mc G}_{s,s'}(\mp i\omega,\la) \\
= & \frac{\beta}{4\pi} \sum_{\omega=\pm \la} \tilde{\mc G}_{q,q'}(i\omega,\la)\tilde{\mc G}_{s,s'}(\mp i\omega,\la).
\end{split}\end{equation*}
The third line follows because (\ref{eq:FRG.flowww2}) is symmetric under the exchange of the summation indices $(s\leftrightarrow q, s'\leftrightarrow q', i\omega_n\rightarrow\mp i\omega_n)$ (by precisely the same argument used to show the symmetry under the exchange of the first and the last two external indices). Altogether, the flow equation for the effective interaction can then be cast in the form
\begin{equation}\label{eq:FRG.flowww3}\begin{split}
\dot\gamma_2(k'_1,k'_2;k_1,k_2;\la) & = \\ \frac{1}{2\pi}\sum_{\omega=\pm\la}\sum_{s',q',s,q}\Bigg[ 
 & - \frac{1}{2}\tilde{\mc G}_{q,q'}(i\omega,\la)\gamma_2(q',s';k_1,k_2;\la)\tilde{\mc G}_{s,s'}(-i\omega,\la)\gamma_2(k'_1,k'_2;s,q;\la) \\
 & - \tilde{\mc G}_{q,q'}(i\omega,\la)\gamma_2(k'_1,q';k_1,s;\la)\tilde{\mc G}_{s,s'}(i\omega,\la)\gamma_2(k'_2,s';k_2,q;\la) \\
 & + \tilde{\mc G}_{q,q'}(i\omega,\la)\gamma_2(k'_2,q';k_1,s;\la)\tilde{\mc G}_{s,s'}(i\omega,\la)\gamma_2(k'_1,s';k_2,q;\la) \Bigg].
\end{split}\end{equation}
The special form of the cutoff has completely disappeared from both (\ref{eq:FRG.flowse3}) and (\ref{eq:FRG.flowww3}), ruling numerical problems due to the `discontinuous' $\delta$-function out from the beginning.

\subsubsection{Finite Temperatures}

At nonzero temperatures one cannot replace the frequency summation in (\ref{eq:FRG.flowse2}) and (\ref{eq:FRG.flowww2}) by an integral. However, using $\sum_{i\omega_n}=\int d\omega\sum_n\delta(\omega-\omega_n)$ we can recast the flow equations for $A=\{\gamma_1,\gamma_2\}$ as
\begin{equation*}
\dot A = \int\frac{d\omega}{2\pi} 2\pi T \sum_n\delta(\omega-\omega_n)\dot\Theta(|\omega|-\Lambda)B(\Theta,\omega),
\end{equation*}
with an appropriate choice of $B$. If we implement the sharp cutoff $\Theta$ as a limit of increasingly sharp cutoff functions we cannot apply Morris' lemma because the right hand side of the flow equations is now discontinuous. Choosing a smooth cutoff instead would though not giving rise to conceptional problems considerably slow down the numerics. One way out of the misery that was originally proposed by \cite{tilman2} is to take the limits in a different order. If we first replace $\delta(\omega-\omega_n)$ by a continuous $\delta$-function $\delta_x(\omega-\omega_n)$ sharply centred around $\omega=\omega_n$, we can implement the sharp cutoff and apply Morris' lemma,
\begin{equation*}\begin{split}
\dot A & = \int\frac{d\omega}{2\pi} 2\pi T \sum_n\delta_x(\omega-\omega_n)\delta(|\omega|-\Lambda)\tilde B(\omega) \\
& = \int\frac{d\omega}{2\pi} 2\pi T \sum_n\delta_x(\omega-\omega_n)\delta(|\omega|-\Lambda)\tilde B(\omega_n) \\
& = T \sum_n\delta_x(\Lambda-|\omega_n|)\tilde B(\omega_n),
\end{split}\end{equation*}
again assuming an appropriate $\tilde B$. If we choose $\delta_x$ to be a box of height $(2\pi T)^{-1}$ and width $2\pi T$, we obtain
\begin{equation}
\dot A = \frac{1}{2\pi}\sum_{|\omega_n|\approx\Lambda}\tilde B(\omega_n),
\end{equation}
which is exactly the zero temperature flow equation, but with the continuous $\omega=\pm\la$ replaced by the discrete Matsubara frequency nearest to $\la$, $\omega_n\approx\pm\la$.

\subsubsection{Again: Initial Conditions}

We have dropped all convergence factors $\exp(i\omega 0^+)$ when we initially replaced the imaginary-time functional integral for the partition function (\ref{eq:FRG.zustandssumme}) by its pendant in frequency space (\ref{eq:FRG.zustandssumme2}). For finite $\la$ this exponential factor is irrelevant in the above flow equations, but we need it to define their initial conditions at $\la^{\tn{initial}}=\infty$ properly. This can be seen if one integrates (\ref{eq:FRG.flowse3}) from infinity down to some arbitrary large $\la_0$, which yields a finite value,
\begin{equation}\label{eq:FRG.initlarge}\begin{split}
& \frac{1}{2\pi}\lim_{\la_0\rightarrow\infty}\int_\infty^{\la_0}d\la\sum_{\omega=\pm\la}\sum_{q,q'}e^{i\omega 0^+} \tilde{\mc G}(i\omega,\la)\gamma_2(k',q';k,q;\la) \\
= & \frac{1}{2\pi}\lim_{\la_0\rightarrow\infty}\int_\infty^{\la_0}d\la\sum_{\omega=\pm\la}\sum_{q,q'}e^{i\omega 0^+} \frac{\delta_{q,q'}}{i\omega}\bar v_{k',q',k,q} + O(\la_0^{-1})\\
= & \frac{1}{\pi}\lim_{\la_0\rightarrow\infty}\sum_q \bar v_{k',q,k,q}\int_\infty^{\la_0}d\la \left\{ \frac{\sin(\la 0^+)}{\la} \right\} + O(\la_0^{-1})  \\
= & \frac{1}{\pi}\lim_{\la_0\rightarrow\infty}\sum_q \bar v_{k',q,k,q}\Bigg[\underbrace{\int_{\la_0}^0 d\la \left\{ \frac{\sin(\la 0^+)}{\la} \right\}}_{= 0}-\underbrace{\int_0^\infty d\la \left\{ \frac{\sin(\la 0^+)}{\la} \right\}}_{=\frac{\pi}{2}}\Bigg] + O(\la_0^{-1}) \\
= & -\frac{1}{2}\sum_q \bar v_{k',q,k,q}.
\end{split}\end{equation}
In the second line we have replaced the two-particle vertex by its value at $\la=\infty$, and we have used that $\tilde{\mc G}_{q,q'}(i\omega) \stackrel{\omega\rightarrow\infty}{\longrightarrow} \frac{\delta_{q,q'}}{i\omega}$ \cite{negeleorland}. With these considerations, the initial conditions (\ref{eq:FRG.init1}) and (\ref{eq:FRG.init2}) read
\begin{equation}\label{eq:DOT.init}\begin{split}
& \gamma_1(k';k;\la^{\tn{initial}}\to\infty)  = - V_{k',k} - \frac{1}{2}\sum_q \bar v_{k',q,k,q}\\
& \gamma_2(k'_1,k'_2;k_1,k_2;\la^{\tn{initial}}\to\infty)  = \bar v_{k'_1,k'_2,k_1,k_2},
\end{split}\end{equation}
with $V_{k',k}$ being an additional one-particle potential not included in the free propagator.\footnote{As explained above, such a potential can be included in $S_\tn{int}$ on from the beginning and will affect the whole fRG scheme only via the initial conditions (\ref{eq:FRG.init1}) and (\ref{eq:FRG.init2}).} We can now start the integration of the flow equations at some large arbitrary $\la_0$ discarding the exponential convergence factor, which is especially important because in general this integration has to be carried out numerically.

\subsubsection{Is Everything Well-Defined?}

Finally, a few general words about potentially ill-defined expressions are in order. If we cut out all degrees of freedom at the beginning of the fRG flow, we have to take care of all those terms that contain a divsion by the free propagator $\mc G^0(\la=\infty)=0$. One such term is the first one in (\ref{eq:FRG.flowvertex}), but it is cancelled by a contribution from the second term if we put in the expansion $\mc G = \mc G^0 - \mc G^0\gamma_1\mc G^0 +\ldots$. Otherwise, $\partial_\la [\mc G^0]^{-1}$ appears only in the well-defined combination
\begin{equation*}
\mc G^0 \partial_\la [\mc G^0]^{-1} \mc G^0 = \delta(|\omega|-\la)\mc G^0,
\end{equation*}
so that there are no ill-defined expressions in the flow equation (\ref{eq:FRG.flowvertex}) if we choose a sharp cutoff.\footnote{At least if we implement it as the limit of increasingly sharp broadenend cutoff functions, as argued above.}

\chapter{Quantum Dots}

Quantum dots, which is the usual name for mesoscopically confined electrons, are recently of great experimental interest. This is mainly due to their `possible' application in nano-electron devices and for quantum information processing \cite{quantencomputer}.

The simplest realisation of a quantum dot system is the so-called single-electron transistor (SET), which contains a `droplet' of localised electrons coupled by tunnelling barriers to a sea of delocalised electrons (the `leads'). An external voltage can be applied by source and drain contacts. If the dot is small enough one would expect its energy level spacing to be large, so that only a few levels must be considered, at least if the temperature is sufficiently low. The simplest model to describe such a situation theoretically is the famous single-impurity Anderson model (SIAM). If one considers the simplest case of only one level containing spin up and spin down electrons, in the zero temperature limit one would intuitively expect a current to flow if the single-particle energy of this level $V_g$ crosses the Fermi energy of the leads, the lineshape of the linear-response conductance $G(V_g)$ only broadened due to the coupling of the confined electrons to the latter. This follows indeed if we solve the SIAM assuming that the spin up and down electrons do not interact. The lineshape observed in the experiment \cite{kondoexperiment} is, however, not Lorentzian but rather box-like consistent with the SIAM predictions if we take into account a sufficiently large local interaction. This enhancement of the conductance is in contrast to the simple picture that the usual effect of the interaction, the Coulomb repulsion, in combination with the spatial confinement of the electrons should lead to charge quantisation and Coulomb blockade transport properties. It is due to the Kondo effect which is generally active below the Kondo temperature if a dot with arbitrary many levels and the level spacing being much larger than the level broadening is occupied by an odd number of electrons. Thus Kondo correlation physics is the basis for the use of a single quantum dot as a nano-transistor, since the conductance properties can be manipulated by adding or removing single electrons.

In this chapter we will introduce a Hamiltonian that describes more complex geometries containing several dots and/or several levels with arbitrary level spacing. We will assume two-particle interactions to be present between the localised electrons in the dots (the `interaction region') while we model the source and drain as noninteracting semi-infinite tight-binding leads. For application of the fRG, the latter have to be projected out in order to get a finite set of flow equations. Finally we will show how the conductance through the dots can be computed in an approach beyond single-particle scattering theory which is used to describe transport if no correlations are present.

\section{General Setup}

\subsection{Experimental Realisation}
\begin{figure}[b]	
     \centering
     \begin{minipage}[t]{0.47\textwidth}
        \centering
	\includegraphics[height=5cm]{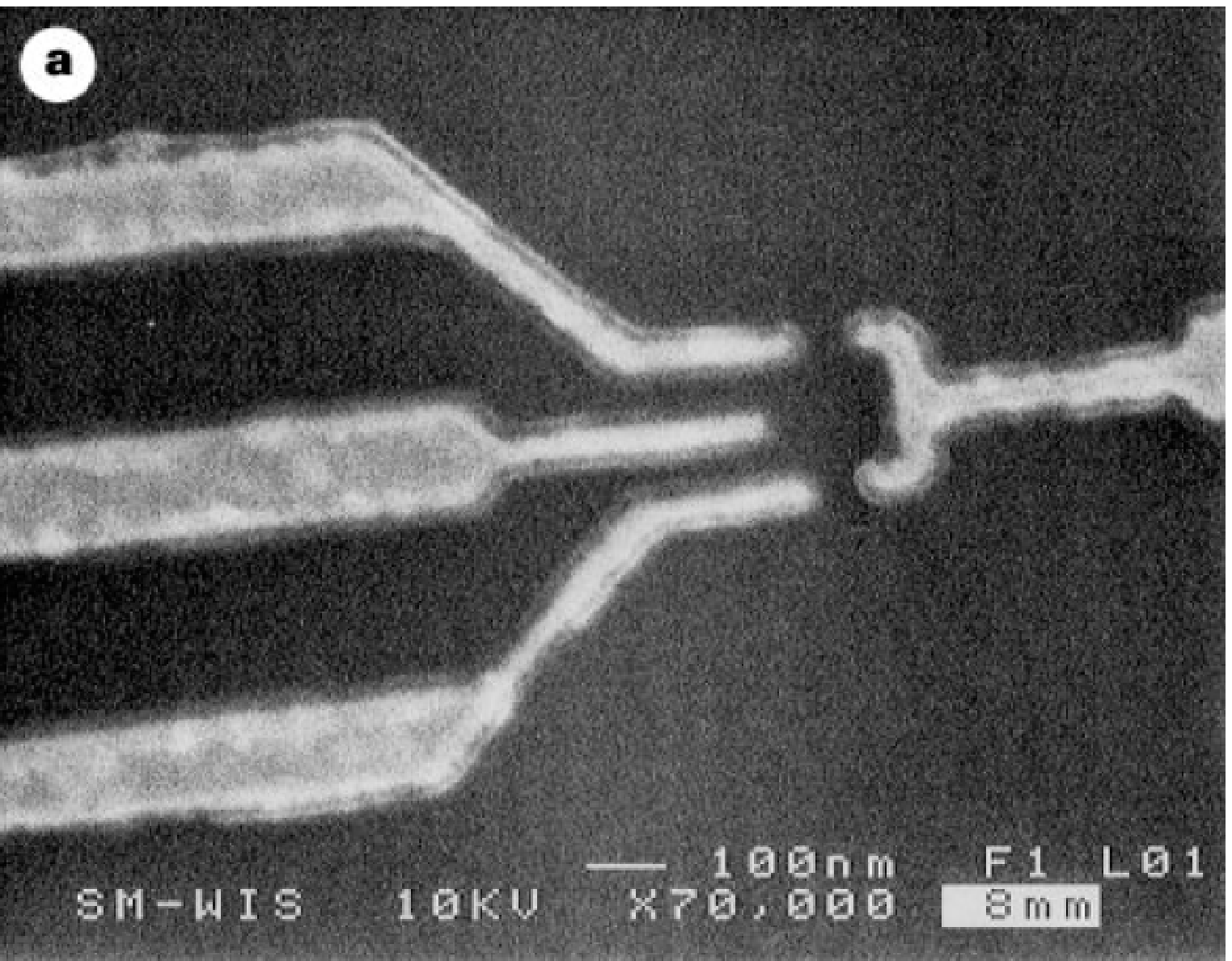}
        \caption{SEM picture of a GaAs/AlGaAs heterostructure and metallic gate electrodes. The electrode on the right and the upper and lower ones on the left define the quantum dot and control the strength of the tunnelling barriers to the rest of the electron gas (the `leads') by varying the magnitude of the applied voltage. The middle electrode on the left can be used to change the energy of the dot. Source and drain contacts are not shown. (Reprinted by permission from Macmillan Publishers Ltd: Nature {\bfseries 391}, 156, copyright 1998.)}
        \label{fig:DOT.exp1}
     \end{minipage} \begin{minipage}[t]{0.03\textwidth}~\end{minipage}
     \begin{minipage}[t]{0.47\textwidth}
        \centering
	\includegraphics[height=5cm]{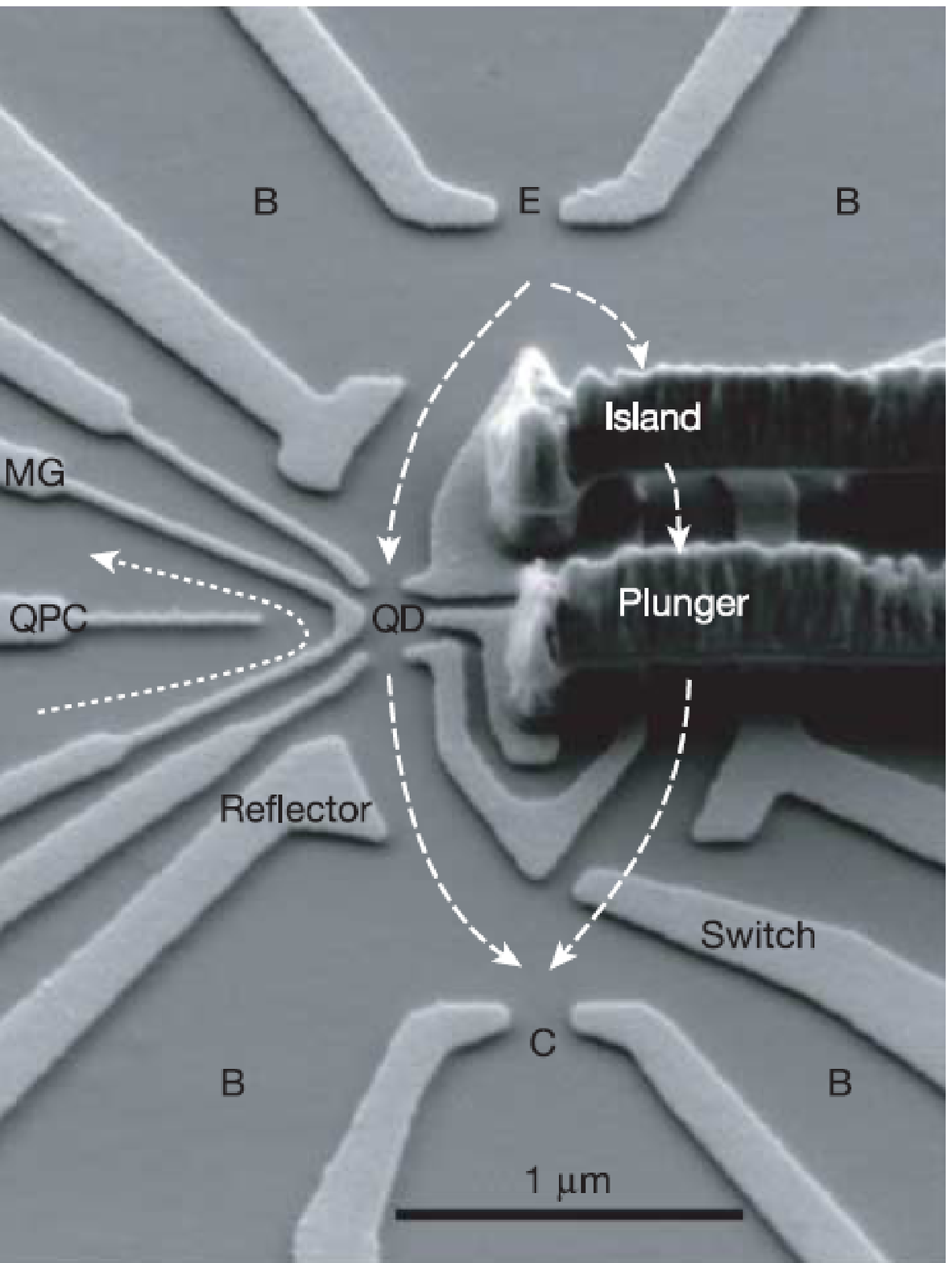}
        \caption{SEM image of a two-path interferometer formed on a GaAs/AlGaAs heterostructure. Electrons can travel from the emitter (E) to the collector (C) either through the reference arm (right path) or through the quantum dot (QD, left path) which are separated by an island. The ground regions (B) are based and serve as draining reservoirs, ensuring that interference is only between the two path. The energy of the QD is controlled via a plunger gate voltage. An additional quantum point contact (QPC) is capacitively coupled to the QD, allowing for measurement of its occupancy. All other electrodes are used for fine-tuning. (Reprinted by permission from Macmillan Publishers Ltd: Nature {\bfseries 436}, 529, copyright 2005.)}
        \label{fig:DOT.exp2}
     \end{minipage}
\end{figure}
A frequently used method to fabricate quantum dot systems (such as the aforementioned SET device) is to employ GaAs/AlGaAs heterostructures which contain a two-dimensional electron gas that results from the electronic properties of the different layers. Adding metallic gates and applying a negative voltage excludes the electrons from regions right below these electrodes, rendering it possible to separate a small region (the quantum dot) from the rest of the heterostructure by tunnelling barries which strength can be tuned by changing the applied voltage. The energy of the dot relative to the electron gas can be controlled by an additional gate electrode with potential $V_g$. By adding drain and source contacts one can then measure the linear response conductance $G$ as a function of $V_g$ by switching on a small bias voltage. A scanning electron microscope (SEM) picture that shows a SET used to measure the $G(V_g)$ dependency at very low temperatures exhibiting the typical Kondo box-like lineshape is shown in Fig.~\ref{fig:DOT.exp1}. Technical details on the fabrication of the quantum dot system can be found in \cite{dotherstellung}.

At $T=0$, which is a meaningful limit since in experiments the temperature can be tuned to be the smallest energy scale in the system (for example about one percent of the level-lead hybridisation strength in \cite{phase1}), the conductance $G$ is connected to the transmission probability $\hat T$ by a constant of proportionality. The transmission probability itself is the absolute square of the transmission amplitude $\hat T=|\hat t|^2$ that is related to matrix elements of the propagator by ordinary scattering theory. We will for the moment stick to this single-particle picture. Further comments on how to calculate the conductance of an interacting system and the connection to the ordinary scattering theory approach will be given in Sec.~\ref{sec:DOT.cond}. The transmission phase is now defined as the argument of the complex number $\hat t$, which is nothing else but the phase change of the wavefunction of the electron passing through the system. Thus to fully describe transport both the transmission probability and phase have to be determined. Theoretically the computation of the latter poses no problem as soon as the exact propagator of the system is known. Experimentally it can be extracted by placing the quantum dot in one arm of a double-slit interferometer and piercing the whole system by a magnetic flux $\Phi$. The transmission probability $\hat T_a$ through the whole system is then given by $\hat T_a=|\hat t+e^{i\Delta\varphi}\hat t_s|^2$, where $t_s$ refers to the amplitude through the second path of the interferometer and $\Delta\varphi$ is an additional phase difference induced by the flux. The interferometer has to be sufficiently open to avoid multi-path interference. Assuming fully coherent transport, as should be expected at sufficiently low temperatures, the interference term in this expression reads $|\hat t||\hat t_s|\cos (\Delta\varphi+\tn{arg}(\hat t_s)+\tn{arg}(\hat t))$. The transmission probability is therefore expected to oscillate as a function of the magnetic field strength, as it is indeed observed in the experiment \cite{phase1}. If we assume the transmission through the second arm $t_s$ to be constant (in particular independent of the energy of the quantum dot), a change in the phase of $\hat t$ will lead to a similar change in phase of the oscillations, which therefore allows for a direct measurement of the former. An SEM picture of an experimental setup that establishes an interferometer with a quantum dot placed in one arm is depicted in Fig.~\ref{fig:DOT.exp2}.

\subsection{Model Hamiltonian}

We will now specify a certain quantum-mechanical model to describe interacting quantum dots coupled to noninteracting semi-infinite leads. We assume that our general Hamiltonian consists of three parts, namely
\begin{equation}\label{eq:DOT.genham}
H = H_{\tn{lead}} + H_{\tn{dot}} + H_{\tn{lead-dot}}.
\end{equation}
Here $H_{\tn{lead}}$ describes the leads, $H_{\tn{dot}}$ the interaction region of the dots and $H_{\tn{lead-dot}}$ the coupling between the two. For simplicity, we assume the two leads to be equal and model them by a tight-binding approach,
\begin{equation}\label{eq:DOT.leadham}
H_{\tn{lead}} = - \sum_{s=L,R}\sum_\sigma\sum_{m=0}^\infty\left[\tau\left(c_{m,\sigma,s}^\dagger c_{m+1,\sigma,s} + c_{m,\sigma,s}c_{m+1,\sigma,s}^\dagger\right) +\mu_s c_{m,\sigma,s}^\dagger c_{m,\sigma,s}\right],
\end{equation}
with $c_{m,\sigma,s}^{(\dagger)}$ being the annihilation (creation) operator for an electron with spin direction $\sigma=\;\uparrow,\downarrow$ localised on lattice site $m$ of the left $(s=L)$ or right $(s=R)$ lead. $\tau$ denotes the hopping amplitude between two nearest-neighbour sites $m$ and $m+1$, and $\mu_s$ the chemical potential of the left or the right lead, respectively. For the time being, we will always assume our model to include spin degrees of freedom, since the Hamiltonian to describe spin-polarised situations just follows by dropping all spin indices in (\ref{eq:DOT.genham}).

The Hamiltonian that describes the quantum dots is made up of three terms,
\begin{equation}\label{eq:DOT.dotham}
H_{\tn{dot}} = H_{\tn{dot}}^{\tn{1P}} + H_{\tn{dot}}^{\tn{hop}} + H_{\tn{dot}}^{\tn{int}}.
\end{equation}
The first term denotes the on-site energies of the different dot levels,
\begin{equation}\label{eq:DOT.dotham1p}
H_{\tn{dot}}^{\tn{1P}} = \sum_\sigma\sum_l \epsilon_{l,\sigma}d_{l,\sigma}^\dagger d_{l,\sigma},
\end{equation}
with $d_{l,\sigma}^{(\dagger)}$ being the dot electron annihilation (creation) operators and $\epsilon_{l,\sigma}$ the one-particle dispersion\footnote{In fact, this name is slightly misleading because $H_{\tn{dot}}^{\tn{hop}}$ is of course a single-particle term as well which together with $H_{\tn{dot}}^{\tn{1P}}$ would determine the one-particle dispersion $\tilde{\epsilon}_{\tilde{l},\sigma}$.} of the states from the interaction region. In general, we will choose this dispersion to include a constant position (that can be different) for each level as well as a variable gate voltage $V_g$,
\begin{equation}
\epsilon_{l,\sigma} = \epsilon_{l,\sigma}^0 + V_g.
\end{equation}
Later on, we will mainly focus on studying the properties of the system described by (\ref{eq:DOT.genham}) as a function of $V_g$. If we apply a magnetic field $B$, the spin dependence of $\epsilon_{l,\sigma}^0$ reads
\begin{equation}
\epsilon_{l,\sigma}^0 = \epsilon_l^0 -\sigma\frac{B}{2},
\end{equation}
with the convention that $\sigma=\;\uparrow$ corresponds to $\sigma=+1$. The second term in (\ref{eq:DOT.dotham}) introduces a hopping between the different dots,
\begin{equation}\label{eq:DOT.dothamhop}
H_{\tn{dot}}^{\tn{hop}} = - \sum_\sigma\sum_{l>l'}\left[t_{l,l'}d_{l,\sigma}^\dagger d_{l',\sigma} + \tn{H.c.} \right],
\end{equation}
while the last one accounts for the two-particle interactions,
\begin{equation}\label{eq:DOT.dothamww}
H_{\tn{dot}}^{\tn{int}} = \frac{1}{2}\sum_{\sigma,\sigma'}\sum_{l,l'}U_{l,l'}^{\sigma,\sigma'}\left(d_{l,\sigma}^\dagger d_{l,\sigma}-\frac{1}{2}\right)\left(d_{l',\sigma'}^\dagger d_{l',\sigma'}-\frac{1}{2}\right).
\end{equation}
We choose $U$ to be symmetric, $U_{l,l'}^{\sigma,\sigma'}=U_{l',l}^{\sigma',\sigma}$, and of course we set $U_{l,l}^{\sigma,\sigma}=0$. The additional shift of the single-particle energies was chosen such that $V_g=0$ corresponds to the particle-hole symmetric point. As suggested by our notation, we assume this energy shift not to be included in the free propagator. It will then affect the fRG scheme via the initial condition (\ref{eq:DOT.init}), which now reads
\begin{equation*}\begin{split}
\gamma_1(l',\sigma';l,\sigma;\la=\infty)  & = \frac{1}{2} \sum_{l'',\sigma''} U_{l,l''}^{\sigma,\sigma''}\delta_{l,l'}\delta_{\sigma,\sigma'}  - \frac{1}{2}\sum_{l'',\sigma''} \bar v_{l'\sigma',l''\sigma'',l\sigma,l''\sigma''} \\
& = \frac{1}{2} \sum_{l'',\sigma''} U_{l,l''}^{\sigma,\sigma''}\delta_{l,l'}\delta_{\sigma,\sigma'}
- \frac{1}{2}\sum_{l'',\sigma''}U_{l,l''}^{\sigma,\sigma''}\delta_{l,l'}\delta_{\sigma,\sigma'} = 0.
\end{split}\end{equation*}
We have used that in order to obtain (\ref{eq:DOT.dothamww}), we have to define 
\begin{equation*}
\bar v_{l_1\sigma_1,l_2\sigma_2,l_3\sigma_3,l_4\sigma_4}:=-U_{l_1,l_3}^{\sigma_1,\sigma_3}	\delta_{l_1,l_4}\delta_{\sigma_1,\sigma_4}\delta_{l_2,l_3}\delta_{\sigma_2,\sigma_3} + U_{l_1,l_4}^{\sigma_1,\sigma_4} \delta_{l_1,l_3}\delta_{\sigma_1,\sigma_3}\delta_{l_2,l_4}\delta_{\sigma_2,\sigma_4}.
\end{equation*}

Up to now we were quite sloppy in using the terms `dot' and `level' when referring to the region in between the noninteracting leads. A more precise notion would be the following. If we want to use (\ref{eq:DOT.dotham}) to model several spatially separated quantum dots each containing several energy levels, it is meaningful to introduce hopping matrix elements $t_{l,l'}$ only between those $l$'s that belong to different dots. The choice of these hoppings therefore induces the spatial geometry of our system in the Hamiltonian. For each level there should be a local interaction $U_{l,l}^{\sigma,\bar\sigma}$ (at least if we want to include spin degrees of freedom), and we can introduce both inter-level and inter-dot interactions $U_{l,l'}^{\sigma,\sigma}$ and $U_{l,l'}^{\sigma,\bar\sigma}$. Having all this in mind, we will continue the aforementioned sloppiness and speak of `dots' when referring to the interacting region to keep the notation short.

\begin{figure}[t]	
	\centering
	\includegraphics[width=\textwidth,clip]{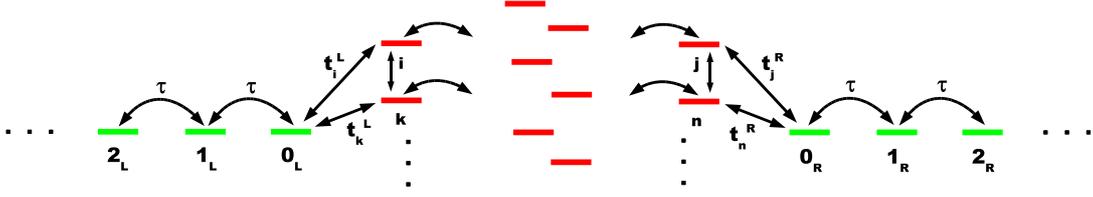}
        \caption{General setup described by the Hamiltonian (\ref{eq:DOT.genham}). The two noninteracting leads (green) are modeled by a tight-binding approach with a hopping matrix element $\tau$, and they are connected to the interaction region (red) via tunnelling barries $t_j^{s=L,R}$. The interaction region of the dots itself comprises levels with arbitrary onsite energies $\epsilon_{l,\sigma}$, two-particle interactions $U_{l,l'}^{\sigma,\sigma'}$, and inter-dot hoppings $t_{l,l'}$. The latter induce a spatial geometry into the system, and the whole Hamiltonian is qualified to model arbitrary many dots each with arbitrary many levels.}
\label{fig:DOT.dots}
\end{figure}

The last term in (\ref{eq:DOT.genham}) that needs to be discussed is the coupling between our quantum dots and the noninteracting leads. It is modeled by
\begin{equation}\label{eq:DOT.coupham}
H_{\tn{lead-dot}} = -\sum_\sigma\sum_{s=L,R}\sum_l\left( t_l^sc_{0,\sigma,s}^\dagger d_{l,\sigma} + \tn{H.c.}\right),
\end{equation}
where only the hoppings $t_l^s$ do not vanish if the corresponding dot $l$ is connected to the lead $s$. The general setup is again depicted in Fig.~\ref{fig:DOT.dots}.

\section{Theoretical Approach}

In this section we will show how the fRG flow equations derived in the previous chapter are applied to our dot system. We will map the infinite system described by the general Hamiltonian (\ref{eq:DOT.genham}) to a finite one by projecting out the noninteracting leads. Finally we will show how quantities that might be measurable in the experiment are calculated in an interacting system. For our main quantity of interest, the conductance, we will derive a generalisation of the Landauer-B\"uttiker formula, the latter describing transport through a noninteracting system.

\subsection{Application of the fRG}

\subsubsection{General Considerations}

The starting point in order to treat the interacting many-particle Hamiltonian (\ref{eq:DOT.genham}) are the coupled equations (\ref{eq:FRG.flowse3}) and (\ref{eq:FRG.flowww3}) for the flow of the self-energy and the two-particle vertex evaluated at zero frequency. But since our system is infinitely large due to the semi-infinite leads, solving these equations would mean that one would again have to tackle infinitely many coupled differential equations. Fortunately this set can be boiled down to one of size $N^2+N^4$ (not accounting for symmetries), where $N$ is the number of degrees of freedom within the dot region, by the following considerations. Since the interaction region is finite at the beginning, it will remain finite during the fRG flow. In particular, no additional single-particle or interaction terms with indices outside the dot region can be generated by (\ref{eq:FRG.flowse3}) and (\ref{eq:FRG.flowww3}), since all external indices on the left hand side of these equations are connected to a two-particle vertex on the right hand side, which by definition vanishes for $\la=\la^{\tn{initial}}$ if one index is chosen from the leads.\footnote{Of course this does not mean that the interaction in the dot region does not affect the leads. If, for example, we wanted to calculate the full propagator $\mc G_{m,m'}$ with lead indices $m$ and $m'$, there would always be a contribution from diagrams (if for the moment we think of the propagator exactly expanded in an infinite perturbation series) containing dot indices $l$ since both are coupled via (\ref{eq:DOT.coupham}). It only means that the single-particle energies in the leads are not renormalized by the flow.} Furthermore it follows from the usual expansion (with the implicit understanding of $\mc G$ referring to the interaction region only),
\begin{equation*}
\mc G = \frac{1}{(\mc G^0)^{-1}+\gamma_1} = \mc G^0 - \mc G^0\gamma_1\mc G^0 + \ldots,
\end{equation*}
that the free propagator entering the flow equations only needs to be evaluated at dot indices $l$, and hence we can replace $\mc G^0$ by its projection on the dot region, $P\mc G^0 P$. The computation of the latter will turn out to be very easy by applying a standard projection method presented below. Finally, $\tilde{\mc G}(\la)$ can be calculated as the inverse of an $N\times N$ -- matrix
\begin{equation}\label{eq:DOT.Gproj}
\tilde{\mc G}(\la) = \frac{1}{(P\mc G^0P)^{-1}+\gamma_1(\la)}.
\end{equation}
It is important to note that the projection technique that leads to (\ref{eq:DOT.Gproj}) is an exact procedure.

\subsubsection{Projecting Out the Leads}

The propagator of a noninteracting system\footnote{Later on, we will also need this projection technique when the interaction in the dot region is present. Generalising the results derived in this section will, however, turn out to be quite simple, so that we will stick to the noninteracting case for the time being.} described by the Hamiltonian $H$ reads
\begin{equation*}
\mc G(z) = \frac{1}{z-H}.
\end{equation*}
In our case, $H$ is obtained from the general Hamilton operator ($\ref{eq:DOT.genham})$ by setting the interaction $U$ to zero (and replacing the many-particle Hamiltonian by its pendant in the standard single-particle space). If we now define operators $P$ and $Q$ that split the Hilbert space by projecting on the dots ($P$) or the leads ($Q$), we can rewrite $\mc G$ as
\begin{equation*}
\mc G(z) =
\begin{pmatrix}
\mc G_{PP}(z) & \mc G_{PQ}(z) \\
\mc G_{QP}(z) & \mc G_{QQ}(z) 
\end{pmatrix}.
\end{equation*}
The projection on the dot's region $\mc G_{PP}:=P\mc GP$ can be computed by considering
\begin{equation*}\begin{split}
& (z-H)\mc G(z) = 1 \\
\Rightarrow	~ & (z-H)(P+Q) \mc G(z)P = P \\
\Rightarrow ~ & (zP - PHP)P\mc G(z)P - PHQQ\mc G(z)P = P \\ & \\
& (z-H)\mc G(z) = 1 \\
\Rightarrow ~ & (z-H)(Q+P) \mc G(z)P = P \\
\Rightarrow ~ & (zQ - QHQ)Q\mc G(z)P - QHPP\mc G(z)P = 0,
\end{split}\end{equation*}
where we have used that $P$ and $Q$ are projectors, that is $P^2=P$, $Q^2=Q$ and $P+Q=1$. By eliminating $\mc G_{QP}=Q\mc GP$ from these two equations, we obtain
\begin{equation}\label{eq:DOT.gpp}\begin{split}
\mc G_{PP}(z) = & \frac{1}{zP - H_{PP} - H_{PQ}\frac{1}{zQ - H_{QQ}}H_{QP}} \\
= & : \frac{1}{zP - H_{\tn{eff}}},
\end{split}\end{equation}
with the obvious definition $H_{PP}:=PHP,\ldots$, and likewise for the other components:
\begin{equation}\label{eq:DOT.gother}\begin{split}
\mc G_{PQ}(z) = & -\mc G_{PP}(z)H_{PQ}\frac{1}{zQ-H_{QQ}} \\
\mc G_{QP}(z) = & -\frac{1}{zQ-H_{QQ}}H_{QP}\mc G_{PP}(z) \\
\mc G_{QQ}(z) = & \frac{1}{zQ-H_{QQ}} + \frac{1}{zQ-H_{QQ}}H_{QP}\mc G_{PP}(z)H_{PQ}\frac{1}{zQ-H_{QQ}}.
\end{split}\end{equation}
Since only $H_{\tn{lead-dot}}$ will contribute when calculating $H_{PQ}$ and $H_{QP}$, we again write down its single-particle version $\tilde{H}_{\tn{lead-dot}}$ to avoid notational confusion:
\begin{equation}
\tilde{H}_{\tn{lead-dot}} = -\sum_\sigma\sum_{s=L,R}\sum_l\Big[ t_l^s|0,\sigma,s\rangle\langle l,\sigma| + (t_l^s)^*|l,\sigma\rangle\langle 0,\sigma,s|\Big].
\end{equation}
The states $|0,\sigma,s\rangle$ and $|l,\sigma\rangle$ denote the wavefunction with spin $\sigma$ localised at the last site ($m=0$) of the left ($s=L$) or right ($s=R$) lead and of the dot level $l$, respectively. We can then compute the effective Hamiltonian $H_{\tn{eff}}$ that determines the propagator $\mc G$ projected on the dots,
\begin{equation}\label{eq:DOT.heff}\begin{split}
H_{\tn{eff}} & =  H_{PP} + H_{PQ}\frac{1}{zQ - H_{QQ}}H_{QP} \\
& =  H_{PP} + \sum_{\sigma,\sigma'}\sum_{s,s'=L,R}\sum_{l,l'}\Big[(t_l^s)^*t_{l'}^{s'}|l,\sigma\rangle\langle 0,\sigma,s|(zQ-H_{QQ})^{-1}|0,\sigma',s'\rangle\langle l',\sigma'| \Big] \\
& =  H_{PP} + \sum_{\sigma}\sum_{s=L,R}\sum_{l,l'}\Big[(t_l^s)^*t_{l'}^s|l,\sigma\rangle\underbrace{\langle 0,\sigma,s|(zQ-H_{QQ})^{-1}|0,\sigma,s\rangle}_{=:g_s^\sigma(z)}\langle l',\sigma| \Big].
\end{split}\end{equation}
The last line follows because the left and the right lead do not couple directly. Except for the calculation of the function $g_s^\sigma(z)$ we have now succeeded in deriving a finite set of flow equations that describe the effects of the interaction in between the quantum dots.

As we will need them later on, we also calculate the following matrix elements:
\begin{equation}\label{eq:DOT.projg1}\begin{split}
\mc G_{l;0_s}^\sigma(z) : & = \langle l,\sigma|\mc G_{PQ}(z)|0,\sigma,s\rangle \\
& = -\sum_{\sigma'}\sum_{s'=L,R}\sum_{l'}\langle l,\sigma|\mc G_{PP}(z)|l',\sigma'\rangle(t_{l'}^{s'})^*\langle 0,\sigma',s'|(zQ-H_{QQ})^{-1}|0,\sigma,s\rangle \\
& = -\sum_{l'}\mc G_{l;l'}^\sigma(t_{l'}^s)^*g_s^\sigma(z), \\ & \\
\mc G_{0_s;l}^\sigma(z) : & = \langle 0,\sigma,s|\mc G_{QP}(z)|l,\sigma\rangle \\
& = -\sum_{\sigma'}\sum_{s'=L,R}\sum_{l'}\langle 0,\sigma,s|(zQ-H_{QQ})^{-1}|0,\sigma',s'\rangle t_{l'}^{s'}\langle l',\sigma'|\mc G_{PP}(z)|l,\sigma\rangle \\
& = -\sum_{l'}\mc G_{l';l}^\sigma t_{l'}^s g_s^\sigma(z),
\end{split}\end{equation}
for an arbitrary choice of $s=L,R$, and,
\begin{equation}\label{eq:DOT.projg2}\begin{split}
\mc G_{0_R;0_L}^\sigma(z) : & = \langle 0,\sigma,R|\mc G_{QQ}(z)|0,\sigma,L\rangle \\
& = \sum_{\sigma',\sigma''}\sum_{s',s''=L,R}\sum_{l',l''}\Bigg[\langle 0,\sigma,R|(zQ-H_{QQ})^{-1}|0,\sigma',s'\rangle t_{l'}^{s'}\langle l',\sigma'|\mc G_{PP}(z)|l'',\sigma''\rangle \\
& \hspace{6.5cm} (t_{l''}^{s''})^*\langle 0,\sigma'',s''|(zQ-H_{QQ})^{-1}|0,\sigma,L\rangle\Bigg] \\
& = \sum_{l',l''} \mc G_{l';l''}^\sigma (t_{l''}^L)^*t_{l'}^R g_L^\sigma(z)g_R^\sigma(z).
\end{split}\end{equation}
Here $l$ is always assumed to be a single-particle index from the dot region.

\subsubsection{Calculation of $g_s^\sigma(z)$}

Finally we have to calculate the propagator $g_s^\sigma(z)$ of a semi-infinite noninteracting lead described by (\ref{eq:DOT.leadham}) evaluated at the last lattice site. This is most easily achieved by a symmetry argument. Since both leads are assumed to be equally modeled by a tight-binding approach, their propagator $g_s^\sigma(z)$ will be of the same structure, so that without loss of generality we can consider the left semi-infinite lead with chemical potential $\mu_L$ ranging from $-\infty$ up to some arbitrary site $m$. The desired function $g_L(z)$ is then given by $g_L(z) = \langle m|(z-K)^{-1}|m\rangle$, where $K$ denotes the single-particle version of the standard tight-binding Hamiltonian (\ref{eq:DOT.leadham}), and we suppress the index $\sigma$ from now on since the leads are symmetric in the spin degree of freedom. Because the lead is semi-infinite and homogeneous, adding one additional site at the right end with the same hopping amplitude $\tau$ will not change the Green function at the last site, that is $\langle m+1|(z-K)^{-1}|m+1\rangle=\tilde{g}_L(z) \stackrel{!}{=}g_L(z)$. If we now apply the projection (\ref{eq:DOT.gpp}) with $P$ projecting on the site $m+1$ and $Q$ on the rest of the lead, we obtain
\begin{equation*}\begin{split}
[\tilde{g}_L(z)]^{-1} & = \langle m+1|(z-K)|m+1\rangle \\
& = \langle m+1|(z-K)|m+1\rangle-\langle m+1|K|m\rangle\langle m|(zQ-QKQ)^{-1}|m\rangle \langle m|K|m+1\rangle \\
& = z+\mu_L - \tau^2g_L(z)  \stackrel{!}{=}[g_L(z)]^{-1}.
\end{split}\end{equation*}
In the first line we could interchange the order of calculating the `inverse' and evaluating the scalar product since in this case $P$ is projecting on a one-dimensional space. The complex roots of the resulting quadratic equation,
\begin{equation*}
\tau^2g^2_L(z) - (z+\mu_L)g_L(z) + 1 = 0,
\end{equation*}
are given by
\begin{equation}
g_L(z) =
\begin{cases}
\frac{1}{2\tau^2}[z+\mu_L - i\sqrt{4\tau^2-(z+\mu_L)^2}] & \tn{if~~Im}(z)>0 \\
\frac{1}{2\tau^2}[z+\mu_L + i\sqrt{4\tau^2-(z+\mu_L)^2}] & \tn{if~~Im}(z)<0.
\end{cases}
\end{equation}
The sign was determined such that the imaginary part of $g_L(z)$ has a branch cut at the real axis and that $\lim_{\omega\to\pm\infty}g_L(i\omega)=0$ holds. The corresponding density of states at the last lattice site reads
\begin{equation}
\rho_{\tn{lead},L}(\omega) = -\frac{1}{\pi}\tn{ Im}g_L(\omega+i0) = \frac{1}{2\pi \tau^2}\sqrt{4\tau^2-(\omega+\mu_L)^2}.
\end{equation}
Later on, we will always take $\rho_{\tn{lead},L}(\omega)$ to be energy-independent, i.e. perform the so-called wide-band limit.

\subsubsection{Generalisation to the Interacting Problem}

Up to now, we have only developed the projection method for a completely noninteracting problem (arising from (\ref{eq:DOT.genham}) by setting $U_{l,l'}^{\sigma,\sigma'}=0$). In order to set up a finite set of fRG flow equations, this is all that we need, since these equations include only the \textit{free} propagator evaluated at dot indices. To calculate the conductance of an interacting system we will, however, also need to project out the leads when the interaction terms are present, that is in the \textit{full} propagator.

Fortunately, it is easy to see that (\ref{eq:DOT.projg1}) and (\ref{eq:DOT.projg2}) also hold in the interacting case. Therefore we expand the full propagator as usual,
\begin{equation*}
\mc G_{i;j}^\sigma = \mc (G^\sigma_{i;j})^0 - \sum_{l_1,l_2}\mc G^{\sigma}_{i;l_1}\gamma_1(l_1,\sigma;l_2,\sigma)(\mc G^\sigma_{l_2;j})^0,
\end{equation*}
where we have suppressed the frequency dependence. The summation extends only over the interaction region, since all self-energy diagrams vanish if one index is taken from the leads. If we now plug in the results from projecting the free propagator, we recover the noninteracting expression,
\begin{equation}\label{eq:DOT.fullgp1}\begin{split}
\mc G_{l;0_s}^\sigma & = -\sum_{l'}(\mc G_{l;l'}^\sigma)^0(t_{l'}^s)^*g_s + \sum_{l'}\sum_{l_1,l_2}\mc G^{\sigma}_{l;l_1}\gamma_1(l_1,\sigma;l_2,\sigma)   (\mc G_{l_2;l'}^\sigma)^0(t_{l'}^s)^*g_s \\
& = -\sum_{l'}\mc G_{l,l'}^\sigma(t_{l'}^s)^*g_s,
\end{split}\end{equation}
likewise for $\mc G_{0_s;l}^\sigma$, and
\begin{equation}\label{eq:DOT.fullgp2}\begin{split}
\mc G_{0_R;0_L}^\sigma & = \sum_{l',l''} (\mc G_{l';l''}^\sigma)^0 (t_{l''}^L)^*t_{l'}^R g_Lg_R
+  \sum_{l''}\sum_{l_1,l_2}\mc G^{\sigma}_{0_R;l_1}\gamma_1(l_1,\sigma;l_2,\sigma)   (\mc G_{l_2;l''}^\sigma)^0(t_{l''}^L)^*g_L \\
& = \sum_{l',l''} (t_{l''}^L)^*t_{l'}^R g_Lg_R\left[(\mc G_{l';l''}^\sigma)^0 - 
\sum_{l_1,l_2}\mc G^{\sigma}_{l';l_1}\gamma_1(l_1,\sigma;l_2,\sigma)   (\mc G_{l_2;l''}^\sigma)^0  \right] \\
& = \sum_{l',l''} \mc G_{l';l''}^\sigma(t_{l''}^L)^*t_{l'}^R g_Lg_R.
\end{split}\end{equation}

\subsection{The Conductance for an Interacting System}\label{sec:DOT.cond}

Transport through a noninteracting one-dimensional system can be described by the Landauer-B\"uttiker formalism \cite{pinkbook}, but since we are dealing with a system of \textit{interacting} fermions, it cannot be applied here. One could now argue that since we have set up an approximation scheme where the self-energy remains frequency-independent during the fRG flow, we map our interacting system to a noninteracting one, so that at the end all properties of the system (such as the conductance) can be calculated as in the well-known noninteracting model. For reasons of consistency it is, however, better to derive expressions for the desired quantities within the full interacting problem and then to verify that in our case these expressions correspond to the noninteracting ones (with renormalized parameters) without applying further approximations. Basically, this has been done before \cite{tilman}, but here we will present a slightly more general approach that allows for arbitrary couplings of the dots region with the noninteracting leads.

\subsubsection{Transport in Linear Response -- Kubo Formula}

If we apply a voltage $V=\mu_R-\mu_L$ between the `ends' of our two leads, we expect a current to flow. The linear conductance can in general be computed as \cite{oguri,pinkbook}
\begin{equation}\label{eq:DOT.cond}
G = \frac{e^2}{\hbar}\lim_{\omega\to 0}\frac{K(\omega)-K(0)}{i\omega},
\end{equation}
where the retarded current-current correlation function in frequency space $K(z)$ is given by the analytical continuation of
\begin{equation}\label{eq:DOT.K}
K(i\omega) = \int_0^\beta d\tau e^{i\omega\tau} \langle \mc T J_R(\tau)J_L(0)\rangle.
\end{equation}
Here $\omega$ is an even (bosonic) Matsubara frequency. $J_L$ and $J_R$ are the current operators at the left and right ends of the system. They are defined straight-forwardly as
\begin{equation}\begin{split}
J_L: & =-\dot{\mc N}_L = i\hbar [{\mc N}_L,H] = i\hbar [{\mc N}_L,H_{\tn{lead-dot}}] \\
& = -i\sum_\sigma\sum_l \left[ t_l^Lc_{0,\sigma,L}^\dagger d_{l,\sigma} - (t_l^L)^*d_{l,\sigma}^\dagger c_{0,\sigma,L} \right] \\ & \\
J_R: & =\dot{\mc N}_R = -i\hbar [{\mc N}_R,H] = -i\hbar [{\mc N}_R,H_{\tn{lead-dot}}] \\
& = i\sum_\sigma\sum_l \left[ t_l^Rc_{0,\sigma,R}^\dagger d_{l,\sigma} - (t_l^R)^*d_{l,\sigma}^\dagger c_{0,\sigma,R} \right],
\end{split}\end{equation}
with $\mc N_{L,R}$ being the particle number operator of the left or right lead, respectively.

\subsubsection{Calculation of $K^{\tn{c}}$}

We will begin the calculation of the correlation function (\ref{eq:DOT.K}) by expanding it into an infinite perturbation series. Due to the linked-cluster theorem, only connected diagrams will contribute. Importantly, it is not necessary that all four external legs are connected to an overall connected diagram. Put differently: we can split up the four-point function into two parts, the first one ($K^\tn{c}$) comprising all diagrams that consist of two unconnected parts each with two external legs, and the second one ($K^\tn{v}$) describing the overall connected diagrams (see also \cite[Eqn. (2.158b)]{negeleorland}). The contribution of the former to (\ref{eq:DOT.K}) reads
\begin{align*}
K^{\tn{c}}(i\omega) := \sum_{\sigma,\sigma'}\sum_{l,l'}\int_0^\beta d\tau e^{i\omega\tau} \Big\{ 
&-&&t_l^R t_{l'}^L&& \langle\mc T d_{l',\sigma'}(0)c_{0,\sigma,R}^\dagger(\tau)\rangle \langle\mc T d_{l,\sigma}(\tau)c_{0,\sigma',L}^\dagger(0)\rangle\\[-2ex]
&+&& t_l^R (t_{l'}^L)^*&&\langle\mc T c_{0,\sigma',L}(0)c_{0,\sigma,R}^\dagger(\tau)\rangle\langle\mc T d_{l,\sigma}(\tau)d_{l',\sigma'}^\dagger(0)\rangle\\[0.5ex]
&+&& (t_l^R)^*t_{l'}^L&&\langle\mc T d_{l',\sigma'}(0)d_{l,\sigma}^\dagger(\tau)\rangle\langle\mc T c_{0,\sigma,R}(\tau)c_{0,\sigma',L}^\dagger(0)\rangle\\
&-&& (t_l^R)^*(t_{l'}^L)^*&&\langle\mc T c_{0,\sigma',L}(0)d_{l,\sigma}^\dagger(\tau)\rangle\langle\mc T c_{0,\sigma,R}(\tau)d_{l',\sigma'}^\dagger(0)\rangle
\Big\}.
\end{align*}
Those terms where one creation is paired with one annihilation operator at equal times vanish. Since the Hamiltonian is time translational invariant, they do not depend on time at all, and the integral $\int_0^\beta d\tau \exp (i\omega\tau)$ with $\omega$ being an even Matsubara frequency is zero. If we now plug in the Fourier expansion of the Green functions and carry out the $\tau$-integration (which yields the inverse temperature times a delta function) as well as one spin summation (which is trivial because we assume spin conservation), we obtain
\begin{align*}
K^{\tn{c}}(i\omega) = T\sum_{\sigma}\sum_{l,l'}\sum_{i\epsilon} \Big [\hspace{0.5cm}
&-&&t_l^R t_{l'}^L && \mc G_{l';0_R}^\sigma(i\epsilon) \mc G_{l;0_L}^\sigma(i\omega+i\epsilon)\nonumber\\[-2ex]
&+&&t_l^R (t_{l'}^L)^* && \mc G_{0_L;0_R}^\sigma(i\epsilon) \mc G_{l;l'}^\sigma(i\omega+i\epsilon)\nonumber \\[0.5ex]
&+&&(t_l^R)^* t_{l'}^L && \mc G_{l';l}^\sigma(i\epsilon) \mc G_{0_R;0_L}^\sigma(i\omega+i\epsilon)\nonumber \\
&-&&(t_l^R)^* (t_{l'}^L)^* && \mc G_{0_L;l}^\sigma(i\epsilon) \mc G_{0_R;l'}^\sigma(i\omega+i\epsilon) \hspace{0.5cm}\Big].
\end{align*}
Next, we express the correlation function in terms of Green functions with indices only in the interaction region by applying the projection technique for the \textit{full} propagator (\ref{eq:DOT.fullgp1}, \ref{eq:DOT.fullgp2}),
\begin{align}\label{eq:DOT.condcalc1}
K^{\tn{c}}(i\omega) &= T\sum_{l,l',k,k';\sigma}\sum_{i\epsilon} \Big [
&&\hspace{-0.4cm}-t_l^R t_{l'}^L (t_k^R)^* (t_{k'}^L)^* g_R(i\epsilon) g_L(i\omega+i\epsilon) && \hspace{-0.3cm}\mc G_{l';k}^\sigma(i\epsilon) \mc G_{l;k'}^\sigma(i\omega+i\epsilon)\nonumber\\[-2ex]
& &&\hspace{-0.4cm}+t_l^R (t_{l'}^L)^*t_k^L (t_{k'}^R)^* g_R(i\epsilon) g_L(i\epsilon)&& \hspace{-0.3cm}\mc G_{k;k'}^\sigma(i\epsilon) \mc G_{l;l'}^\sigma(i\omega+i\epsilon)\nonumber \\[0.2ex]
& &&\hspace{-0.4cm}+(t_l^R)^* t_{l'}^L t_k^R (t_{k'}^L)^*g_R(i\omega+i\epsilon)g_L(i\omega+i\epsilon)&& \hspace{-0.3cm}\mc G_{l';l}^\sigma(i\epsilon) \mc G_{k;k'}^\sigma(i\omega+i\epsilon)\nonumber \\
& &&\hspace{-0.4cm}-(t_l^R)^* (t_{l'}^L)^* t_k^L t_{k'}^R g_R(i\omega+i\epsilon)g_L(i\epsilon)&& \hspace{-0.3cm}\mc G_{k;l}^\sigma(i\epsilon) \mc G_{k';l'}^\sigma(i\omega+i\epsilon)\Big]\nonumber \\ \nonumber \\
&\hspace{-1cm}= T\sum_{l,l',k,k';\sigma}\sum_{i\epsilon} \Big[&&\hspace{-1.4cm}-t_l^R t_{l'}^L (t_k^R)^* (t_{k'}^L)^*f_R(i\epsilon,i\epsilon+i\omega)f_L(i\epsilon,i\epsilon+i\omega) && \hspace{-0.3cm}\mc G_{l';k}^\sigma(i\epsilon) \mc G_{l;k'}^\sigma(i\omega+i\epsilon)\Big],
\end{align}
where we have defined
\begin{equation}\begin{split}
f_L(i\epsilon,i\epsilon+i\omega) &:= g_L(i\epsilon+i\omega) - g_L(i\epsilon) \\
f_R(i\epsilon,i\epsilon+i\omega) &:= g_R(i\epsilon) - g_R(i\epsilon+i\omega) \\
f(i\epsilon,i\epsilon+i\omega) & := f_L(i\epsilon,i\epsilon+i\omega)f_R(i\epsilon,i\epsilon+i\omega).
\end{split}\end{equation}
In order to carry out the $i\epsilon$-summation, we rewrite (\ref{eq:DOT.condcalc1}) as a a complex contour integral with an appropriate closed integration path $\mc C $ by substituting $T\sum_{i\epsilon}F(i\epsilon)\to -(2\pi i)^{-1} \int_{\mc C} dz F(z) g(z)$, where the simple poles at $z=i\epsilon$ of $g(z)$ should not coincide with those of $F(z)$.\footnote{This is the so-called Possion summation formula, for a complete review see e.g. \cite{mahan}.} Since $\epsilon$ is an odd Matsubara frequency in contrast to $\omega$, the Fermi function fullfills this condition. Hence we choose $g(\epsilon)=1/(\exp (\beta\epsilon)+1)$ from now on.\footnote{Be careful not to confuse this with the propagator $g_s(z)$.} In our case, $F(z)$ has a branch cut at the real axis $\tn{Im}(z)=0$ as well as at the line $\tn{Im}(z)=-\omega$ (because the propagators are nonanalytic at $z=i\omega$), so that the contour $\mc C$ consists of three parts. The Fermi function falls off exponentially for large positive arguments while it tends to one for large negative arguments. But since each propagator falls off linearly, we can deform the integration path into four lines slightly above and below the axes of nonanalyticity. This yields
\begin{equation}\label{eq:DOT.condcalc2}\begin{split}
K^{\tn{c}}(i\omega) = \hspace{-0.4cm}\sum_{l,l',k,k';\sigma}t_l^R t_{l'}^L (t_k^R)^* (t_{k'}^L)^*\int d\epsilon~\frac{g(\epsilon)}{2\pi i}\Big\{
\hspace{0.3cm}& f(\epsilon+i0,\epsilon+i\omega) \mc G_{l';k}^\sigma(\epsilon+i0) \mc G_{l;k'}^\sigma(\epsilon+i\omega) \\[-2ex]
\hspace{0.3cm}-& f(\epsilon-i0,\epsilon+i\omega) \mc G_{l';k}^\sigma(\epsilon-i0) \mc G_{l;k'}^\sigma(\epsilon+i\omega) \\[0.5ex]
\hspace{0.3cm}+& f(\epsilon-i\omega,\epsilon+i0) \mc G_{l';k}^\sigma(\epsilon-i\omega) \mc G_{l;k'}^\sigma(\epsilon+i0) \\
\hspace{0.3cm}-& f(\epsilon-i\omega,\epsilon-i0) \mc G_{l';k}^\sigma(\epsilon-i\omega) \mc G_{l;k'}^\sigma(\epsilon-i0) \hspace{0.1cm}\Big\}.
\end{split}\end{equation}
The first and fourth term are of $O(\omega^2)$, and hence they will vanish in the limit $\omega\to 0$ in (\ref{eq:DOT.cond}). The former can be seen if we consider
\begin{equation*}\begin{split}
g_s(\epsilon\pm i\delta) & = \frac{1}{2\tau^2}\left[\epsilon\pm i\delta+\mu \mp i\sqrt{4\tau^2-(\epsilon\pm i\delta+\mu)^2}\right] \\
&= \frac{1}{2\tau^2}\left[\epsilon+\mu \mp i\sqrt{4\tau^2-(\epsilon+\mu)^2}\right] + O(\delta),
\end{split}\end{equation*}
with $\mu$ being the chemical potential of both the left and right lead, which is now equal since we are calculating the conductance in linear response. Thus, for small $\omega$
\begin{equation*}\begin{split}
f_{L,R}(\epsilon\pm i\omega,\epsilon\pm i0) & = O(\omega) \\
f_{\stackrel{L}{R}}(\epsilon-i0,\epsilon+i\omega) & = \mp \frac{i}{\tau^2}\sqrt{4\tau^2-(\epsilon+\mu)^2} + O(\omega) \\
f_{\stackrel{L}{R}}(\epsilon-i\omega,\epsilon+i0) & = \mp \frac{i}{\tau^2}\sqrt{4\tau^2-(\epsilon+\mu)^2} + O(\omega).
\end{split}\end{equation*}
Performing the analytical continuation $i\omega\to\omega+i0$ and substituting $\epsilon\to\epsilon+\omega$ in the second term of (\ref{eq:DOT.condcalc2}), we obtain
\begin{equation}\begin{split}
K^{\tn{c}}(\omega) = &\sum_{l,l',k,k';\sigma}t_l^R t_{l'}^L (t_k^R)^* (t_{k'}^L)^* \int d\epsilon~\frac{g(\epsilon+\omega)-g(\epsilon)}{2\pi i}\\
& \hspace{1.5cm}\times \Big\{ f(\epsilon-i0,\epsilon+\omega+i0)\mc G_{l';k}^\sigma(\epsilon-i0) \mc G_{l;k'}^\sigma(\epsilon+\omega+i0)\Big\} + O(\omega^2).
\end{split}\end{equation}
The conductance that follows from this part of the correlation function then takes the simple form
\begin{equation}\label{eq:DOT.condKc}\begin{split}
G^{\tn{c}} & = \frac{e^2}{\hbar}\lim_{\omega\to 0}\frac{K^{\tn{c}}(\omega)-K^{\tn{c}}(0)}{i\omega} \\
& = -\frac{e^2}{h}\int d\epsilon~ g'(\epsilon) f(\epsilon-i0,\epsilon+i0)\sum_{l,l',k,k';\sigma}t_l^R t_{l'}^L (t_k^R)^* (t_{k'}^L)^*\mc G_{l';k}^\sigma(\epsilon-i0) \mc G_{l;k'}^\sigma(\epsilon+i0) \\
& = -\frac{e^2}{h}\sum_\sigma\int d\epsilon~ g'(\epsilon)\frac{4\tau^2-(\epsilon+\mu)^2}{\tau^4}\Big|\sum_{l,l'}t_l^R(t_{l'}^L)^*\mc G^\sigma_{l;l'}(\epsilon+i0)\Big|^2\\
& = -\frac{e^2}{h}\sum_\sigma\int d\epsilon~ g'(\epsilon)\Big|2\pi\rho_{\tn{lead}}(\epsilon)\sum_{l,l'}t_l^R(t_{l'}^L)^*\mc G^\sigma_{l;l'}(\epsilon+i0)\Big|^2,
\end{split}\end{equation}
where we have used that $\mc G^\sigma_{l;l'}(z)=\mc G^\sigma_{l';l}(z)$ due to time-reversal symmetry of the Hamiltonian,\footnote{Imagine the exact propagator $\mc G_{l;l'}$ expanded in an infinite perturbation series. If we now in each diagram interchange the direction of every internal line (which is possible if we assume that the single-particle dispersion is symmetric) and shift the time by $\tau\to-\tau$ (which is possible because the Hamiltonian was assumed to be time-reversal invariant), we will end up with a diagram of precisely the same structure that would appear in the expansion of $\mc G_{l';l}$.} and $\mc G(\omega+i\delta)=(\mc G(\omega-i\delta))^*$, and we have reintroduced the density of states at the last site $\rho_{\tn{lead}}(\omega)$.

\subsubsection{Vertex Corrections}

Before further commenting on (\ref{eq:DOT.condKc}), we have to calculate the second contribution to (\ref{eq:DOT.cond}) which we will call $K^{\tn{v}}$. It arises from those diagrams where all four external legs are connected to an overall connected part, i.e. the connected two-particle Green functions, which can be related to the two-particle vertex functions via (\ref{eq:FRG.dyson2}). This will prove meaningful since we are directly calculating this quantity within our fRG approach. If we then carry out the $\tau$ integration in (\ref{eq:DOT.K}) (which kills one frequency summation) and use that the two-particle vertex is frequency-conserving (which kills another frequency summation), we obtain
\begin{equation}\begin{split}
K^{\tn{v}} = -T^2\sum_{l,l'\atop\sigma,\sigma'}\sum_{l_1,\ldots,l_4}\sum_{i\epsilon,i\epsilon'}&\Big[
t_{l'}^L\mc G_{l_1;0_L}^{\sigma'}(i\epsilon+i\omega) \mc G_{l';l_2}^{\sigma'}(i\epsilon) - 
(t_{l'}^L)^*\mc G_{l_1;l'}^{\sigma'}(i\epsilon+i\omega) \mc G_{0_L;l_2}^{\sigma'}(i\epsilon) \Big]\\[-2ex]
&\hspace{1.7cm} \times\Gamma_{l_2,l_3;l_1,l_4}^{\sigma,\sigma'}(i\epsilon,i\epsilon'+i\omega;i\epsilon+i\omega,i\epsilon')\\[1ex]
& \hspace{-0.6cm}\times\Big[
t_l^R\mc G_{l;l_3}^\sigma(i\epsilon'+i\omega) \mc G_{l_4;0_R}^\sigma(i\epsilon') -
(t_l^R)^*\mc G_{0_R;l_3}^\sigma(i\epsilon'+i\omega) \mc G_{l_4;l}^\sigma(i\epsilon')\Big],
\end{split}\end{equation}
where the $l_i$ summations run only over the interaction region (since the two-particle vertex vanishes outside), and we have defined 
\begin{equation*}
\Gamma_{l_1,l_2;l_3,l_4}^{\sigma,\sigma'}(i\omega_1,i\omega_2;i\omega_3,i\omega_4):= 
\gamma_2\left(\{l_1,\sigma',i\omega_1\},\{l_2,\sigma,i\omega_2\};\{l_3,\sigma',i\omega_3\},\{l_4,\sigma,i\omega_4\}\right).
\end{equation*}
Applying the projection technique (\ref{eq:DOT.fullgp1}, \ref{eq:DOT.fullgp2}) yields
\begin{equation}\label{eq:DOT.condKv}\begin{split}
K^{\tn{v}} &= T^2\sum_{\sigma,\sigma'}\sum_{l,l',k,k'}\sum_{l_1,\ldots,l_4}\sum_{i\epsilon,i\epsilon'} \\ &\hspace{-0.3cm}\Big[
t_{l'}^L(t_k^L)^*\mc G_{l_1;k}^{\sigma'}(i\epsilon+i\omega) \mc G_{l';l_2}^{\sigma'}(i\epsilon)g_L(i\epsilon+i\omega) - 
(t_{l'}^L)^*t_k^L\mc G_{l_1;l'}^{\sigma'}(i\epsilon+i\omega) \mc G_{k;l_2}^{\sigma'}(i\epsilon)g_L(i\epsilon) \Big]\\[1ex]
&\hspace{3.4cm} \times\Gamma_{l_2,l_3;l_1,l_4}^{\sigma,\sigma'}(i\epsilon,i\epsilon'+i\omega;i\epsilon+i\omega,i\epsilon')\times\\[1ex]
&\hspace{-0.3cm}\Big[
t_l^R(t_{k'}^R)^*\mc G_{l;l_3}^\sigma(i\epsilon'+i\omega) \mc G_{l_4;k'}^\sigma(i\epsilon')g_R(i\epsilon') -
(t_l^R)^*t_{k'}^R\mc G_{k';l_3}^\sigma(i\epsilon'+i\omega) \mc G_{l_4;l}^\sigma(i\epsilon')g_R(i\epsilon'+i\omega)\Big] \\ & \\
& = T\sum_{\sigma'}\sum_{l',k}\sum_{l_1,l_2}\sum_{i\epsilon}
t_{l'}^L(t_k^L)^*\mc G_{l_1;k}^{\sigma'}(i\epsilon+i\omega) \mc G_{l';l_2}^{\sigma'}(i\epsilon)f_L(i\epsilon,i\epsilon+i\omega)
\Lambda_{l_3,l_4}^{\sigma'}(i\epsilon;i\omega)
\end{split}\end{equation}
with the definition
\begin{equation}\begin{split}
&\Lambda_{l_3,l_4}^{\sigma'}(i\epsilon;i\omega):=\\
&\hspace{0.5cm}\sum_{l,k';l_3,l_4\atop\sigma;i\epsilon'}\hspace{-0.2cm}\Gamma_{l_2,l_3;l_1,l_4}^{\sigma,\sigma'}(i\epsilon,i\epsilon'+i\omega;i\epsilon+i\omega,i\epsilon')
t_l^R(t_{k'}^R)^*\mc G_{l;l_3}^\sigma(i\epsilon'+i\omega) \mc G_{l_4;k'}^\sigma(i\epsilon')f_R(i\epsilon',i\epsilon'+i\omega).
\end{split}\end{equation}
Up to now, we have not applied any approximations except for the fact that we are describing the transport through our interacting system in \textit{linear} response. This means that (\ref{eq:DOT.condKc}) and (\ref{eq:DOT.condKv}) give an exact expression for the conductance, but in general it is impossible to evaluate at least the second one exactly even if we knew the exact propagator. In contrast, in the context of our fRG approximation scheme this becomes very simple which can be seen by an argument following \cite{tilman}. Since we have chosen a truncation scheme that keeps the two-particle vertex frequency-independent, it is easy to apply Poisson's summation formula again in complete analogy to (\ref{eq:DOT.condcalc2}) to write down
\begin{equation*}\begin{split}
\Lambda_{l_3,l_4}^{\sigma'}(i\epsilon;i\omega)=-\hspace{-0.2cm}\sum_{l,k';l_3,l_4\atop\sigma}\hat\Gamma_{l_2,l_3;l_1,l_4}^{\sigma,\sigma'}&t_l^R(t_{k'}^R)^*\int\frac{g(\epsilon')d\epsilon'}{2\pi i} \Big\{ \\
 &\mc G_{l;l_3}^\sigma(\epsilon'+i\omega) \mc G_{l_4;k'}^\sigma(\epsilon'+i0)f_R(\epsilon'+i0,\epsilon'+i\omega) \\
-&\mc G_{l;l_3}^\sigma(\epsilon'+i\omega) \mc G_{l_4;k'}^\sigma(\epsilon'-i0)f_R(\epsilon'-i0,\epsilon'+i\omega) \\
+&\mc G_{l;l_3}^\sigma(\epsilon'+i0) \mc G_{l_4;k'}^\sigma(\epsilon'-i\omega)f_R(\epsilon'-i\omega,\epsilon'+i0) \\
-&\mc G_{l;l_3}^\sigma(\epsilon'-i0) \mc G_{l_4;k'}^\sigma(\epsilon'-i\omega)f_R(\epsilon'-i\omega,\epsilon'-i0) ~~\Big\}.
\end{split}\end{equation*}
As before, in the first and the fourth term the function $f_R$ has frequency arguments at the same side of the branch cut and hence these terms are by an order of $\omega$ smaller than the other two. Performing the analytical continuation we can thus write
\begin{equation*}\begin{split}
\Lambda_{l_3,l_4}^{\sigma'}(i\epsilon;\omega+i0)&=-\hspace{-0.2cm}\sum_{l,k';l_3,l_4\atop\sigma}\hat\Gamma_{l_2,l_3;l_1,l_4}^{\sigma,\sigma'}t_l^R(t_{k'}^R)^*\int\frac{d\epsilon'}{2\pi i} \Big\{ \\
&\hspace{-0.5cm}[g(\epsilon'+\omega)-g(\epsilon')]\mc G_{l;l_3}^\sigma(\epsilon'+\omega+i0) \mc G_{l_4;k'}^\sigma(\epsilon'-i0)f_R(\epsilon'-i0,\epsilon'+\omega+i0)\Big\} \\[2ex]
&= O(\omega).
\end{split}\end{equation*}
This expression is independent of $i\epsilon$, hence we can perform the frequency summation in (\ref{eq:DOT.condKv}) as well. Since the structure of this equation is precisely the same, it will also be of $O(\omega)$, so that for the vertex corrections to the conductance we obtain
\begin{equation}
G^{\tn{v}} = \frac{e^2}{\hbar}\lim_{\omega\to 0}\frac{O(\omega^2)}{\omega}=0.
\end{equation}
Hence (\ref{eq:DOT.condKc}) is the expression for the conductance consistent with our approximation scheme. In the zero temperature limit, this result is much more general. Under quite weak assumptions for the two-particle vertex it is possible to show that the vertex corrections exactly vanish \cite{oguri}.

\subsubsection{Special Cases}

We will now discuss the form of (\ref{eq:DOT.condKc}) in some special cases. In this thesis, we will mainly focus on the zero temperature limit, in which minus the derivative of the Fermi function becomes a $\delta$-function, and the conductance reads
\begin{equation}\label{eq:DOT.leitwert}
G = \frac{e^2}{h}\sum_\sigma\Big|2\pi\rho_{\tn{lead}}(0)\sum_{l,l'}t_l^R(t_{l'}^L)^*\mc G^\sigma_{l;l'}(0)\Big|^2.
\end{equation}
If only two dots from the interaction region ($1$ and $N$) are connected to the left or right lead, respectively, we recover the familiar result
\begin{equation*}
G = \frac{e^2}{h}\sum_\sigma\Big|2\pi\rho_{\tn{lead}}(0)t_1^Lt_N^R\mc G^\sigma_{N;1}(0)\Big|^2.
\end{equation*}

The general form of (\ref{eq:DOT.condKc}) naturally allows for defining the partial conductance of the spin up and down electrons,
\begin{equation*}
G = G_\uparrow + G_\downarrow \stackrel{T=0}{=}
\frac{e^2}{h}\left\{\Big|2\pi\rho_{\tn{lead}}(0)\sum_{l,l'}t_l^R(t_{l'}^L)^*\mc G^\uparrow_{l;l'}(0)\Big|^2 +
\Big|2\pi\rho_{\tn{lead}}(0)\sum_{l,l'}t_l^R(t_{l'}^L)^*\mc G^\downarrow_{l;l'}(0)\Big|^2\right\}.
\end{equation*}
The name `partial conductance' is justified because the so-defined expression for $G_\sigma$ would follow if we calculate the current-current response function for the spin up and down electrons separately by replacing $J\to J_\sigma:=-\dot{\mc N}_\sigma$ in (\ref{eq:DOT.K}).

There is one case where one can obtain a very simple expression for the conductance valid at arbitrary temperature, and we will note it since it is frequently used in the literature to describe transport even at $T=0$ where (\ref{eq:DOT.condKc}) is exact. Namely, if the interaction region comprises only one single dot (with interacting spin up and down electrons), the conductance can be expressed as \cite{meirwingreen}
\begin{equation}\label{eq:DOT.leitwertsingle}
G = -\frac{e^2\pi^2}{h}\frac{4t_L^2t_R^2}{t_L^2+t_R^2}\sum_\sigma\int d\omega~g'(\omega)\rho(\omega)\tilde\rho_\sigma(\omega),
\end{equation}
with $t_L$ and $t_R$ being the couplings to the left and right lead, $g(\omega)$ the Fermi function, and $\rho(\omega)$ and $\tilde\rho_\sigma(\omega)$ the density of states at the last site of the leads and at the dot, respectively.

\subsubsection{Connection to Scattering Theory}

Everybody who is a bit confused by the form of the conductance if an arbitrary number of dots is connected to the leads or by its derivation should consider the following. In the noninteracting case (where (\ref{eq:DOT.condKc}) is exact) one usually defines the zero temperature conductance using ordinary single-particle scattering theory as a factor of $e^2/h$ times the absolute square of the transmission amplitude.\footnote{One should note, however, that although it frequently appears in the literature (see e.g. \cite{datta}), there is a lot of sloppiness buried in such a definition. An equally intuitive but much more stringent approach to derive the Landauer-B\"uttiker formula starts out from the time evolution of the current operator, $\lim_{t\to\infty}\langle \phi(t)|j|\phi(t)\rangle$ (implicitly weighted by the Fermi function), with $|\phi(0)\rangle$ being the ground state of the isolated leads (see \cite{schoenhammer2}). The subsequent computation, however, is crucially based on the single-particle nature of the Hamiltonian governing the system.}  The latter is most easily derived by looking at an eigenstate of the isolated left-lead Hamiltonian,
\begin{equation*}
\langle m|k,L\rangle =
\begin{cases}
\sqrt{\frac{2}{\pi}}\sin \left[ k(m-M) \right] & \tn{for $m$ from the left lead} \\
0 & \tn{otherwise}.
\end{cases}
\end{equation*}
$M$ is implicitly assumed to be chosen such that this linear combination of left- and right-moving waves vanishes at an imaginary site added to the right end of the lead. Since everything is assumed to be diagonal in spin space anyway, we suppress the $\sigma$ index. Scattering states are the defined as usual,
\begin{equation*}
|k,L,\pm\rangle := \lim\limits_{\eta\to 0} (\pm i\eta)\mc G (\epsilon_k\pm i\eta)|k,L\rangle.
\end{equation*}
The free propagator $\mc G$ of the system fullfills the Dyson equation,
\begin{equation*}
\mc G = \hat{\mc G} + \hat{\mc G}H_\tn{coup}\mc G = \hat{\mc G} + \mc G H_\tn{coup} \hat{\mc G},
\end{equation*}
with $\hat{\mc G}$ being the propagator of the full (noninteracting) system excluding the connection $H_\tn{coup}$ between the leads and the dot region, $\hat{\mc G}(z) = (z-H+H_\tn{coup})^{-1}$. Thus, for any $m$ from the right lead we can write 
\begin{equation*}\begin{split}
\langle m|k,L,+\rangle = & \langle m| \left[\frac{i0}{\epsilon_k-H+H_\tn{coup}+i0} + \mc G(\epsilon_k+i0)H_\tn{coup}\frac{i0}{\epsilon_k-H+H_\tn{coup}+i0}\right]|k,L\rangle \\
= & \underbrace{\langle m|k,L\rangle}_{=0} + \langle m|\mc G(\epsilon_k+i0)H_\tn{coup}|k,L\rangle \\
= & \langle m| \hat{\mc G}(\epsilon_k+i0)H_\tn{coup}\mc G(\epsilon_k+i0) H_\tn{coup}|k,L\rangle,
\end{split}\end{equation*}
which by plugging in the single-particle version of the level-lead coupling Hamiltonian (\ref{eq:DOT.coupham}),
\begin{equation*}
H_\tn{coup} = -\sum_{s=L,R}\sum_l\big[ t_l^s |0,s\rangle\langle l| + \tn{H.c.}\big],
\end{equation*}
simplifies to
\begin{equation*}\begin{split}
\langle m|k,L,+\rangle = & \sum_{l,l'}(t_l^L)^*t_{l'}^R\langle m|\hat{\mc G}(\epsilon_k+i0)|0,R\rangle \langle l|\mc G(\epsilon_k+i0)|l'\rangle \langle 0,L|k,L\rangle \\
= & \frac{1}{\sqrt{2\pi}}e^{ikm+i\varphi} \Big[\frac{2\sin(k)}{\tau}\sum_{l,l'}(t_{l'}^L)^*t_l^R\langle l|\mc G(\epsilon_k+i0)|l'\rangle\Big].
\end{split}\end{equation*}
Since the matrix element of $\hat{\mc G}$ is just an outgoing wave, $\langle m|\hat{\mc G}|0,R\rangle=\tau^{-1}\exp(ikm-i\varphi)$, we can now read off the transmission probability $T(\epsilon)$ as
\begin{equation}\label{eq:DOT.scattering}
T(\epsilon) = \left|\frac{2\sin(k)}{\tau}\sum_{l,l'}(t_{l'}^L)^*t_l^R\langle l|\mc G(\epsilon_k+i0)|l'\rangle\right|^2.
\end{equation}
If we furthermore introduce the density of states at the last site of the lead,
\begin{equation*}
\rho_\tn{lead}(\epsilon)=\frac{1}{2\pi\tau^2}\sqrt{4\tau^2-\epsilon^2}=\frac{1}{2\pi\tau}\sqrt{1-\cos^2(k)}=\frac{\sin(k)}{\pi\tau},
\end{equation*}
we precisely recover the expression (\ref{eq:DOT.leitwert}) derived above, showing that in the noninteracting case both approaches are equivalent. At nonzero temperatures, the distribution of the lead electrons is governed by the Fermi function, such that the obvious $T>0$ generalisation of the noninteracting scattering theory definition of the conductance is to introduce another weighted energy integration, which is again consistent with (\ref{eq:DOT.condKc}).

Our fRG approximation scheme maps the general interacting problem to an effective noninteracting one, and it turned out that we can compute the conductance using the $U=0$ expressions with renormalized parameters. However, since in the interacting case we have to define this quantity in a different way (as we cannot use single-particle scattering theory), this is not obvious.

From the scattering theory approach it is also immediately clear that since the transmission probability is bounded by one, the maximum conductance is given by $e^2/h$ per channel, that is $2e^2/h$ if take into account the spin, and $e^2/h$ in spin-polarised situations.

\subsubsection{Particle Numbers}

An important quantity that is also accessible in the experiment is the occupation number of each dot in the interaction region. For the dot with single-particle index $l$ it is given by
\begin{equation*}
\langle n_l^\sigma\rangle=\langle c_{l,\sigma}^\dagger c_{l,\sigma}\rangle=\langle \mc T c_{l,\sigma} c_{l,\sigma}^\dagger(-\delta)\rangle
=T\sum_{i\omega_n}e^{i\omega_n\delta}\mc G_{l;l}^\sigma(i\omega_n),
\end{equation*}
with $\delta$ going to zero. For $T\to 0$ the sum can be written as an integral,
\begin{equation}
\langle n_l^\sigma\rangle=\frac{1}{2\pi}\int d\omega~e^{i\omega\delta}\mc G_{l;l}^\sigma(i\omega),
\end{equation}
which is easy to tackle numerically, since the contribution from large $\omega$ can be computed similarly to (\ref{eq:FRG.initlarge}). At nonzero temperatures we use Poisson's summation formula to obtain
\begin{equation}\begin{split}
\langle n_l^\sigma\rangle=&~T\sum_{i\omega_n}e^{i\omega_n\delta}\mc G_{l;l}^\sigma(i\omega_n)\\
=& -\frac{1}{2\pi i} \int_{\mc C} dz~e^{z\delta}\mc G_{l;l}^\sigma(z)g(z)\\
=& -\frac{1}{2\pi i} \int d\omega~g(\omega)\left[\mc G_{l;l}^\sigma(\omega+i0)-\mc G_{l;l}^\sigma(\omega-i0)\right]\\
=& \int d\omega~g(\omega)\rho_{l;l}^\sigma(\omega),
\end{split}\end{equation}
which is easy to calculate numerically as well.

\chapter{Numerical Results}\label{sec:numericresults}

In this chapter we will present results from the application of the fRG to dot systems with a few levels for various geometries. Unless stated otherwise, we will always use the truncation scheme that includes the flow of the (frequency-independent) two-particle vertex. Despite that fact this approximation can strictly be justified only in the limit of small two-particle interactions, we will also apply it to fairly large $U$. Allover this chapter, we will only sporadically comment on what `fairly large' means and on the question which parameter regions are for sure out of reach within our approach. We will focus here on describing the physics that arises due to the presence of the interaction with the implicit understanding that the fRG produces reliable results for all situations shown. In the next chapter, by comparison to other data either obtained from a (complicated) exact solution or numerical methods known to be very precise we will verify that this is indeed in the case. We will furthermore specify in more detail which interaction strength can no longer be tackled by our simple fRG truncation scheme.

Since our main interest is in transport properties of dot systems, we will compute the linear-response conductance $G$ as well as the transmission phase $\alpha$ (at zero temperature, this is the argument of the transmission probability) as a function of the gate voltage $V_g$ that shifts the single-particle energies of each dot. For situations with spin degeneracy, in particular in absence of magnetic fields, the transmission phase of the spin up and down electrons is equal and will be denoted as $\alpha:=\alpha_\uparrow=\alpha_\downarrow$. Frequently, we will also calculate the occupancy of each level since this often facilitates the physical interpretation of the results. Unless stated otherwise (that is everywhere outside the section called `finite temperatures'), we will focus on the low-energy physics in the $T=0$ limit.

Every system under consideration exhibits three typical energy scales that determine the gate voltage dependence of the conductance, namely a single-particle energy $\Delta$ (which might be a level detuning between parallel dots or a nearest-neighbour hopping between sites of a chain), the two-particle interaction $U$, and the hybridisation strength with the leads $\Gamma$. To systematically study their influence on the physics that governs $G(V_g)$, we will pursue the following course of action which is guided by the limitations imposed by the perturbative nature of the fRG. First, we will discuss the noninteracting limit $U=0$, varying the single-particle spacing from $\Delta\ll\Gamma$ to $\Delta\gg\Gamma$. Next, we will turn on an interaction such that the physics is significantly influenced (which will mostly turn out to be $U\approx\Gamma$). If possible, we will finally increase $U$ such that all important limits can be captured ($\{\Delta\ll\Gamma,U\gg\Gamma\},\ldots$), allowing us to determine which quantity governs the physics in each case. This will, however, only turn out to be possible in the spinless two-level case and for the SIAM. Hence it is meaningful to focus on the effects of gradually turned on small interactions rather than concentrating on the extreme limits where one quantity is much larger (smaller) than the others.

It is very important to point out that everything shown in this chapter represents the generic behaviour of each particular system under consideration, and, even more, we will refrain from discussing special non-generic situations. In order to obtain a complete picture of what should be associated with the former and what with the latter, we have taken advantage of the small computational power required by fRG to scan wide regions of the parameter space. The consequences showing up if one carelessly sticks to situations of too high symmetry, which are a formidable candidate for yielding non-generic behaviour, can be quite drastic. In \cite{gefenwwfrei}, a noninteracting spinless parallel double dot was studied, and the authors focused on completely symmetric hybridisations. The paper aimed at giving a clue towards an explanation of the evolution of the transmission phase observed in quantum dot experiments (an issue that we will also tackle later on). Though giving many important insights, the results would have been more enlightening if generic hybridisations had been considered, as later on stated by the authors themselves \cite{GG}.

From now on, we set the chemical potential of the leads to zero. Furthermore, we perform the wide-band limit, which is achieved by substituting $\tau\to\eta\tau$ for the hopping matrix element of the leads and $t_l^s\to\sqrt{\eta}t_l^s$ for all hoppings from the last site of the latter into the dot region and taking the limit $\eta\to\infty$. The propagator $g_s^\sigma(i\omega)$ then reads
\begin{equation}
g_s^\sigma(i\omega) =
\begin{cases}
-i\tau^{-1} & \tn{Im}(\omega)>0 \\
+i\tau^{-1} & \tn{Im}(\omega)<0,
\end{cases}
\end{equation}
and the density of states $\rho_{\tn{lead}}$ becomes energy-independent, which justifies the name wide-band limit. It is important to note that we do not perform this approximation for computational reasons, but only because it usually appears in the literature. It is motivated by the fact that the details of the leads should not influence the transport properties of the system dramatically. For the system of two parallel spin-polarised dots (Sec.~\ref{sec:OS.dd}) we have verified that this is indeed the case. Since the density of states is independent of energy, the same will hold for the hybridisations
\begin{equation}
\Gamma_l^s:=\pi |t_l^s|^2\rho_{\tn{lead}}.
\end{equation}
Allover this chapter, the unit of energy is chosen to be\footnote{Choosing $\Gamma$ as the unit of energy might be questionable, especially if the system under consideration comprises a large number of dots $N$ where one would expect the physics to be governed by the ratio $U/\Gamma_\tn{typ}$ and $\Delta/\Gamma_\tn{typ}$ (with $\Gamma_\tn{typ}:=\Gamma/N$) rather than by $U/\Gamma$ and $\Delta/\Gamma$. Practically it turns out, however, that for all situations considered here $\Gamma$ is indeed the most suitable choice of the unit of energy.}
\begin{equation}
\Gamma:=\sum_l\Gamma_l:=\sum_{s=L,R}\Gamma_l^s,
\end{equation}
and we introduce the shorthand notation $\Gamma=\{\Gamma_A^L/\Gamma~\Gamma_A^R/\Gamma~\Gamma_B^L/\Gamma\ldots\}$.

\section{Spin-Polarised Dots}\label{sec:OS}

In this section we will apply the fRG to quantum dot systems where the spin degree of freedom is ignored. As mentioned above, the Hamiltonian to describe spin-polarised systems is obtained by dropping all spin indices in the general Hamiltonian (\ref{eq:DOT.genham}). We will focus on systems of up to six parallel dots each containing one level (or one dot containing up to six levels, or mixtures of both situations; we will keep the aforementioned linguistic sloppiness and do not distinguish between `dots' and `levels', ignoring the spatial geometry that is induced by the choice of the hoppings $t_{l,l'}$). The term `parallel' means that every dot is connected to the leads by nonzero tunnelling barriers $t_l^{s=L,R}$. 

On the one hand, the physical importance of a model neglecting the spin degree of freedom arises as it should reproduce the results of a model containing spin if a large magnetic field is applied. In the next section, we will exemplarily demonstrate that this is indeed true. On the other hand if it occurs in the experiment that the spin degree of freedom does not seem to play a role (signalised by the absence of Kondo physics), it might be justified to try to explain these experiments using a spin-polarised model.

This section is organised as follows. First, we will as an introduction describe transport through a single impurity in a homogeneous chain, which is by definition a noninteracting model and therefore exactly solvable. It will serve for a better understanding of transport through more than one dot if the singe-particle level spacing is assumed to be large so that the different levels do not overlap. This picture of transport occurring through each level individually will remain valid in presence of a two-particle interaction, only that the separation of the transmission resonances is enlarged due to Coulomb repulsion. In contrast, the effect of the interaction will be much more dramatic if the spacing between the levels decreases so that they overlap significantly. In a noninteracting picture, this would imply that they simultaneously contribute to the transport so that no well-separated resonances are to be expected. In presence of interactions, we will observe such peaks of good separation nevertheless due to Coulomb repulsion (`Coulomb blockade peaks'). Furthermore, we will frequently find the $G(V_g)$ curve to exhibits additional correlation induced resonances (CIRs) if the integration between the electrons exceeds a certain critical value depending on the dot parameters. These CIRs were first predicted for a two-level dot by \cite{cir} using the fRG.

As stated in the introduction, our main focus within this section will be on the number of resonance peaks and transmission zeros (and corresponding jumps in the phase $\alpha$) in the conductance $G(V_g)$ as a function of the dot's single-particle level spacing. If an experimentally realised quantum dot is successively filled with electrons, one would expect the latter to decrease for those electrons that occupy the states of highest energy. Taking this as an input from the experiment, our results might serve as an explanation of the puzzling behaviour of the phase evolution $\alpha(V_g)$ experimentally observed in quantum dots \cite{phase3, phase2, phase1}.

\subsection{A Single Impurity}
\begin{figure}[t]	
	\centering
	\includegraphics[width=0.495\textwidth,height=4.8cm,clip]{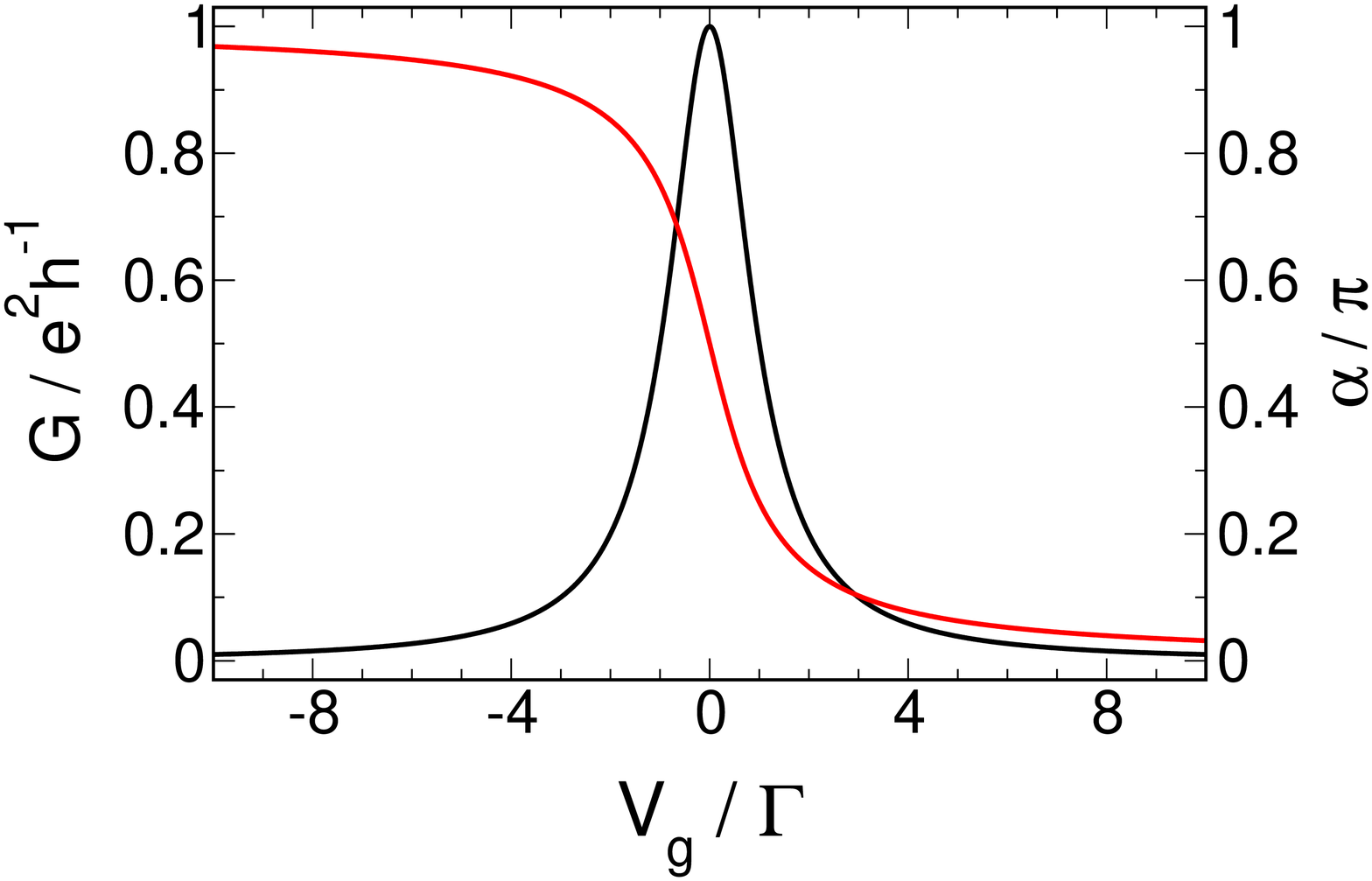}
        \caption{Conductance $G$ (black) and transmission phase $\alpha$ (red) as a function of the gate voltage for a single impurity with $\Gamma_L/\Gamma=\Gamma_R/\Gamma=0.5$.}
\label{fig:OS.singledot}
\end{figure}

For a single dot coupled to tight-binding leads the Hamiltonian (\ref{eq:DOT.genham}) reads
\begin{equation*}
H = V_g d^\dagger d - \left(t_L c_{0,L}^\dagger d + t_R c_{0,R}^\dagger d + \tn{H.c.}\right) + H_{\tn{lead}}.
\end{equation*}
The propagator projected on the dot (\ref{eq:DOT.gpp}) is then given by
\begin{equation}
\mc G(i\omega) = \frac{1}{i\omega - V_g + i~\tn{sgn}(\omega)\Gamma},
\end{equation}
and the linear-response conductance reads
\begin{equation}
G(V_g) = 4\frac{e^2}{h}\frac{\Gamma_L\Gamma_R}{V_g^2+\Gamma^2}.
\end{equation}
This is a Lorentzian of full width $2\Gamma$ and height $4\frac{e^2}{h}\frac{\Gamma_L\Gamma_R}{(\Gamma_L+\Gamma_R)^2}$ centred around $V_g=0$. In the left-right symmetric case ($\Gamma_L=\Gamma_R$) the conductance reaches the unitary limit, and the more asymmetric the hybridisations are chosen, the more it is suppressed. The transmission phase $\alpha$, which can be computed as
\begin{equation}
\alpha(V_g)=\arctan\left(\frac{\Gamma}{V_g}\right),
\end{equation}
changes by $\pi$ over the resonance. In fact, for a single dot at $T=0$ it is no independent quantity in the sense that it is directly connected to the conductance by a generalised Friedel sum rule \cite{hewson}. The same holds for the level occupancy, which equals the transmission phase in units of $\pi$. The conductance and the transmission phase are depicted in Fig.~\ref{fig:OS.singledot}.

\subsection{Parallel Double Dots}\label{sec:OS.dd}

\subsubsection{Application of the fRG}

We will now turn to a more complex system of two parallel spin-polarised dots. The free propagator with the leads projected out can be cast in the form
\begin{equation}\label{eq:OS.dd.g}
\left[\mc G^0(i\omega)\right]^{-1} =
\begin{pmatrix}
i\omega - V_g + \frac{\Delta}{2} + i~\tn{sgn}(\omega)~\Gamma_A & i~\tn{sgn}(\omega)\left[\sqrt{\Gamma_A^L\Gamma_B^L}+s\sqrt{\Gamma_A^R\Gamma_B^R}\right] \\
i~\tn{sgn}(\omega)\left[\sqrt{\Gamma_A^L\Gamma_B^L}+s\sqrt{\Gamma_A^R\Gamma_B^R}\right] & i\omega - V_g - \frac{\Delta}{2} + i~\tn{sgn}(\omega)~\Gamma_B
\end{pmatrix},
\end{equation}
where we have introduced the level spacing $\Delta$, and a vanishing direct hopping between both levels was assumed, since it can always be tuned away by a basis transformation of the dot states (leading to an additional level detuning). The notation used to derive the general expressions for the effective projected Hamiltonian (\ref{eq:DOT.gpp}) and the conductance (\ref{eq:DOT.condKc}) allows for all hopping matrix elements being arbitrary complex numbers. This is necessary if we for example want to describe the effect of a magnetic flux piercing the ring geometry. Here we will concentrate on real hoppings, in which case the off-diagonal element of the free propagator can be expressed by the square roots of the hybridisations $\Gamma_l^s$ and the relative sign $s$ of all hopping matrix elements, $s=\tn{sgn}(t_A^Lt_A^Rt_B^Lt_B^R)$. The fact that it is only the relative sign that enters the calculation of all quantities of interest is easily seen. Assuming an arbitrary choice for $\tilde s_l^s:=\tn{sgn}(t_l^s)$ and flipping $\tilde s_A^R\leftrightarrow\tilde s_B^R$ does not change the projected part of the free propagator (\ref{eq:DOT.gpp}) or the expression for the conductance (\ref{eq:DOT.leitwert}). The interchange of $\tilde s_A^L$ with $\tilde s_A^R$ changes the overall sign of the off-diagonal elements of $\mc G^0$, which implies that also the sign of the off-diagonal elements of the full propagator is flipped, while that of the diagonal terms is unaffected. This becomes clear if we expand the full propagator in an infinite perturbation series. Here this series only contains one single type of interaction vertices ($\bar v_{A,B,A,B}$), and hence each term in the expansion of $\mc G_{l;l}$ ($\mc G_{l;\bar l}$) has to contain an even (odd) number of factors $(\mc G_{l,\bar l})^0$. However, in the expression of the conductance every off-diagonal term acquires an additional minus sign as well, and all that enters the computation of the average level occupancies are the diagonal parts of the propagator. Hence all quantities of interest depend only on the relative sign of all couplings as claimed. By writing down (\ref{eq:OS.dd.g}) we have implicitly assumed that this relative sign is realised such that $t_l^L>0$, $t_A^R >0$, and $\tn{sgn}(t_B^R)=s$.

The fRG flow equations (\ref{eq:FRG.flowse3}) for the effective level position $V_{l=A,B}(\la):=-\gamma_1(l;l)$ and inter-level hopping $t(\la):=\gamma_1(A;B)$ read
\begin{equation}\begin{split}
\partial_\la V_l(\la) = & -\frac{U(\la)}{2\pi}\sum_{\omega=\pm\la}\tilde{\mc G}_{\bar l;\bar l}(i\omega;\la)\\
\partial_\la t(\la) = & -\frac{U(\la)}{2\pi}\sum_{\omega=\pm\la}\tilde{\mc G}_{A;B}(i\omega;\la),
\end{split}\end{equation}
with $\bar l$ denoting the complement, $\bar A = B$, and the flow of the effective interaction $U(\la):=\gamma_2(A,B;A,B)$ (\ref{eq:FRG.flowww3}) is given by
\begin{equation}\begin{split}
\partial_\la U(\la) = -\frac{U^2(\la)}{2\pi}\sum_{\omega=\pm\la}\Big[&
 ~~-\tilde{\mc G}_{A;A}(i\omega;\la)\tilde{\mc G}_{B;B}(-i\omega;\la)+\tilde{\mc G}_{A;B}(i\omega;\la)\tilde{\mc G}_{B;A}(-i\omega;\la)\\[-2ex]
&~~+\tilde{\mc G}_{A;B}(i\omega;\la)\tilde{\mc G}_{B;A}(i\omega;\la)- \tilde{\mc G}_{A;A}(i\omega;\la)\tilde{\mc G}_{B;B}(i\omega;\la)~~~~~~\Big].
\end{split}\end{equation}
We have chosen the matrix $\tilde{\mc G}$ as
\begin{equation*}\begin{split}
&\left[\tilde{\mc G}(i\omega;\la)\right]^{-1} = \\
&~~\begin{pmatrix}
i\omega - V_A(\la) + i~\tn{sgn}(\omega)~\Gamma_A & t(\la)+i~\tn{sgn}(\omega)\left[\sqrt{\Gamma_A^L\Gamma_B^L}+s\sqrt{\Gamma_A^R\Gamma_B^R}\right] \\
t(\la)+i~\tn{sgn}(\omega)\left[\sqrt{\Gamma_A^L\Gamma_B^L}+s\sqrt{\Gamma_A^R\Gamma_B^R}\right] & i\omega - V_B(\la) + i~\tn{sgn}(\omega)~\Gamma_B
\end{pmatrix},
\end{split}\end{equation*}
so that the initial conditions (\ref{eq:DOT.init}) read $V_A(\la\to\infty) = V_g-\frac{\Delta}{2}$, $V_B(\la\to\infty) = V_g+\frac{\Delta}{2}$, $t(\la\to\infty)=0$, and $U(\la\to\infty)=U$, with $U$ being the interaction between the electrons of both levels. Solving these differential equations provides us with the full propagator $\mc G = \tilde{\mc G}(\la=0)$ of the system and we can compute the zero temperature conductance using (\ref{eq:DOT.condKc}), which here takes the form
\begin{equation}\label{eq:OS.dd.cond}
G = 4\frac{e^2}{h}\left|\sqrt{\Gamma_A^L\Gamma_A^R}\mc G_{A;A}(0)+\sqrt{\Gamma_B^L\Gamma_A^R}\mc G_{A;B}(0)
+s\sqrt{\Gamma_A^L\Gamma_B^R}\mc G_{B;A}(0)+s\sqrt{\Gamma_B^L\Gamma_B^R}\mc G_{B;B}(0)\right|^2.
\end{equation}
Again, we have assumed that the relative sign of all couplings is realised in a way that $t_l^L>0$, $t_A^R >0$, and $\tn{sgn}(t_B^R)=s$.

\subsubsection{The Noninteracting Case}

Since the double dot system is already a quite complex geometry (in the sense that it has a fairly large number of parameters to be varied), we will first study  it without interaction. Replacing $\mc G$ by the free propagator (\ref{eq:OS.dd.g}) in (\ref{eq:OS.dd.cond}) then yields
\begin{equation}\label{eq:OS.dd.condwwfrei}
G(V_g) = 4\frac{e^2}{h}\frac{\Gamma_A^L\Gamma_A^R\epsilon_B^2+\Gamma_B^L\Gamma_B^R\epsilon_A^2
+2s\sqrt{\Gamma_A^L\Gamma_A^R\Gamma_B^L\Gamma_B^R}\epsilon_A\epsilon_B}
{\Big[\Gamma_A^L\Gamma_B^R+\Gamma_B^L\Gamma_A^R-2s\sqrt{\Gamma_A^L\Gamma_A^R\Gamma_B^L\Gamma_B^R}-\epsilon_A\epsilon_B\Big]^2
+ \Big[\epsilon_A\Gamma_B+\epsilon_B\Gamma_A\Big]^2 },
\end{equation}
with the obvious definition $\epsilon_A=V_g-\frac{\Delta}{2}$ and $\epsilon_B=V_g+\frac{\Delta}{2}$. As mentioned above, we will always focus on studying the \textit{generic} behaviour of the conductance and transmission phase as a function of the gate voltage. The term generic refers to that fact that we would expect the qualitative features and dependence on the dot parameters $U$, $\Delta$ and $s$ of both curves to be independent of the choice of the hybridisations. This is indeed true for all $\Gamma_l^s$ except for a three-dimensional manifold in the four-dimensional space of all couplings. In particular, it will prove that a left-right and $A$-$B$ asymmetry only influences the overall height of the conductance curve, the set-in of the crossover regime at scale $\Delta_\tn{cross}$, and the correlation induced structures which vanish for $\Delta>\Delta_\tn{CIR}$ (see below). An indication for this is already given by the noninteracting case if one analyses the exact expression (\ref{eq:OS.dd.condwwfrei}). For the interacting case, we have verified numerically by scanning a large part of the parameter space that the results presented in this section are indeed the generic ones. Many parameter sets $\{\Gamma_l^s\}$ included in the manifold of non-generic hybridisations are not of physical relevance anyway. For example, it shows that the qualitative behaviour of $G(V_g)$ for degenerate levels and $s=+$ is not generic in the left-right symmetric case. For a more detailed discussion of the manifold of non-generic hybridisations see \cite{cir}.
\begin{figure}[t]	
	\centering
	\includegraphics[width=0.485\textwidth,height=4.4cm,clip]{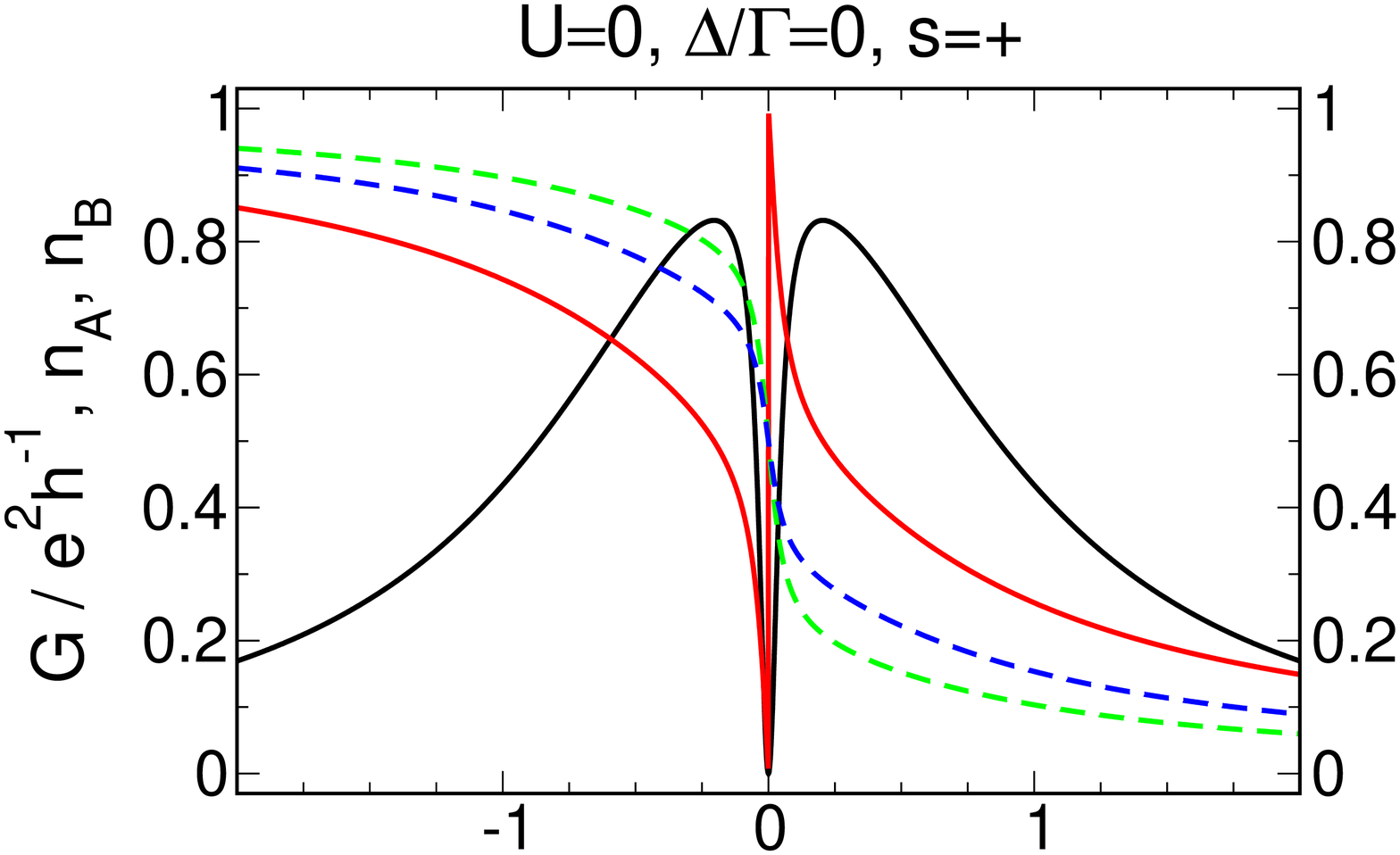}\hspace{0.015\textwidth}
        \includegraphics[width=0.485\textwidth,height=4.4cm,clip]{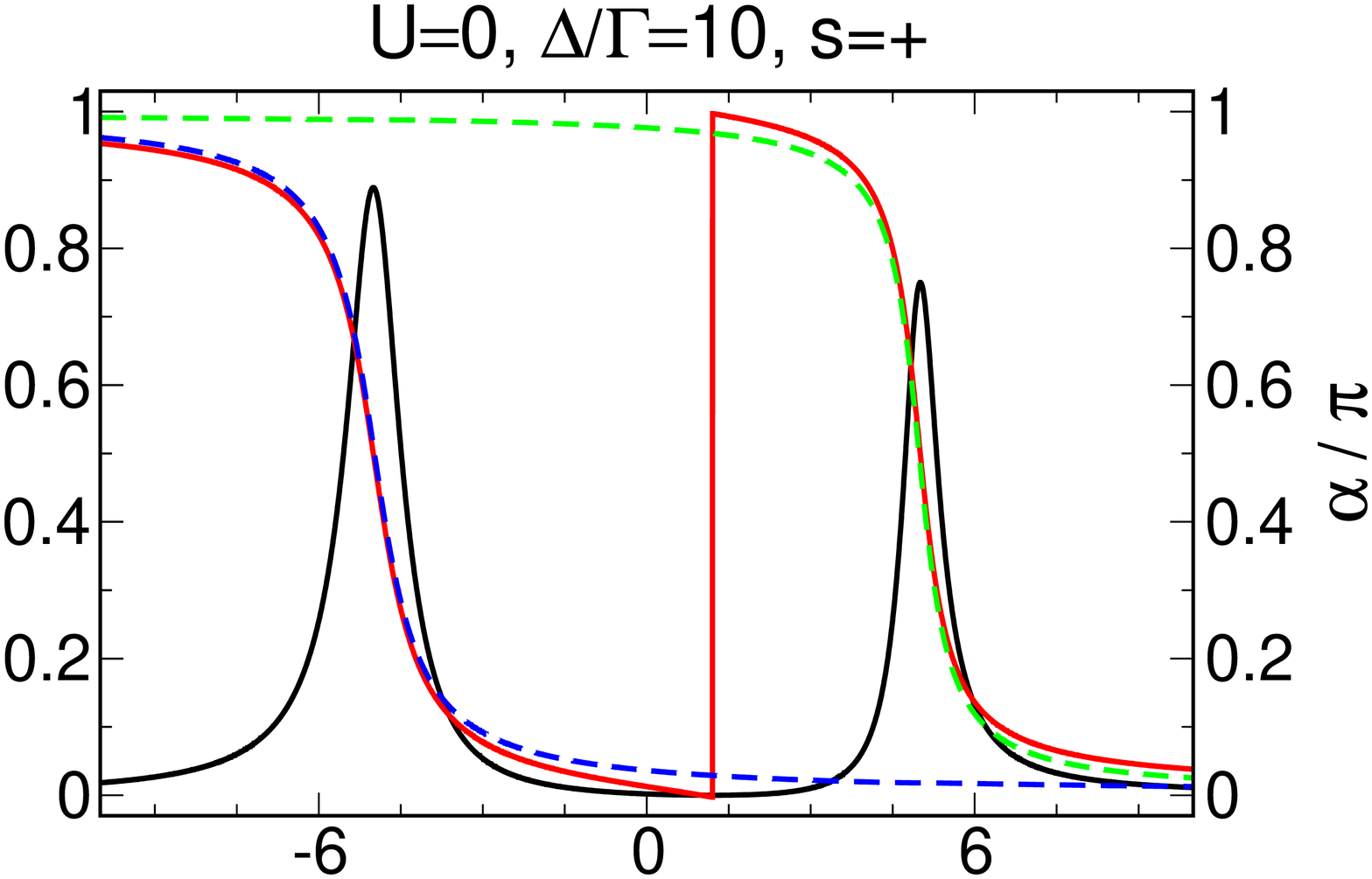}\vspace{0.3cm}
        \includegraphics[width=0.485\textwidth,height=5.2cm,clip]{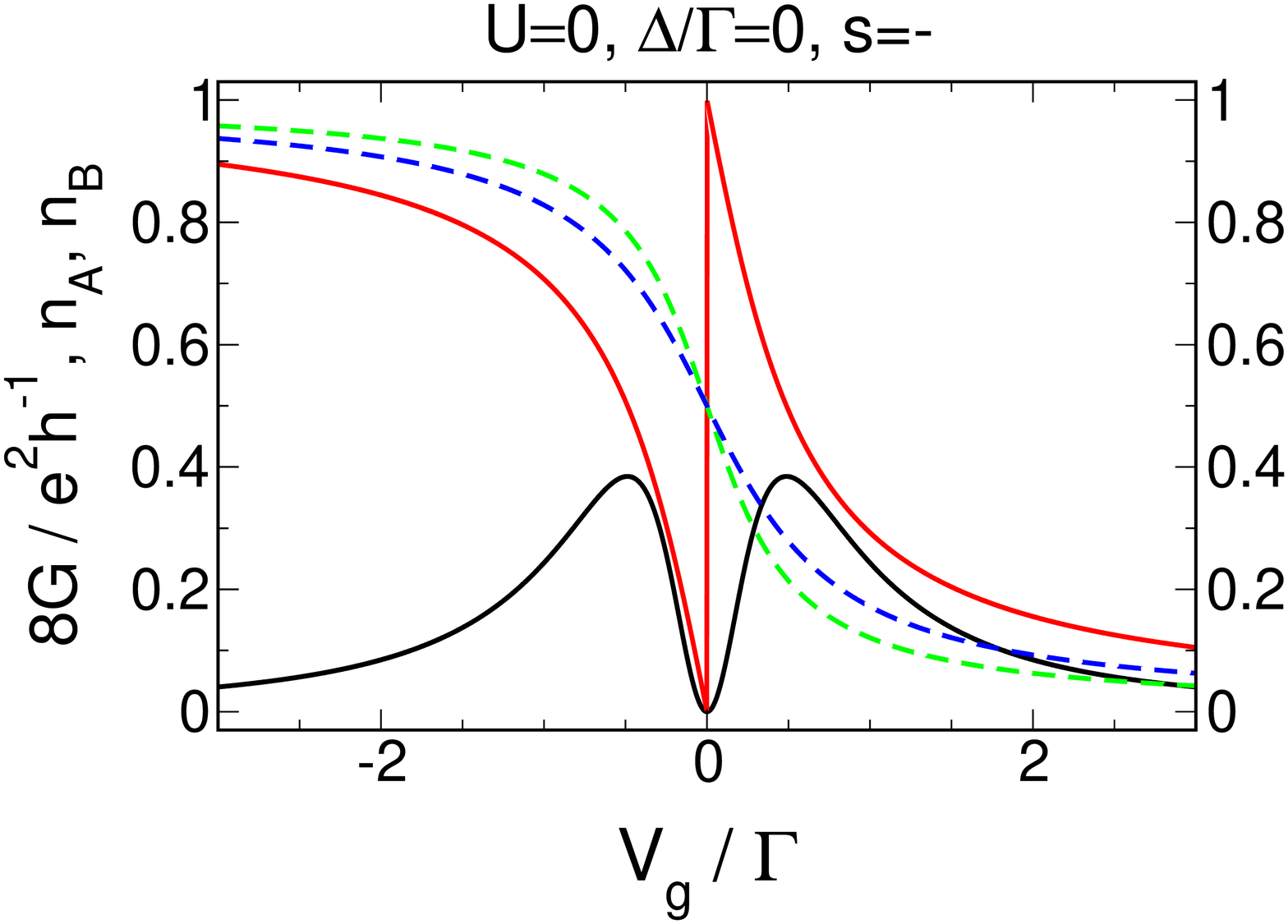}\hspace{0.015\textwidth}
        \includegraphics[width=0.485\textwidth,height=5.2cm,clip]{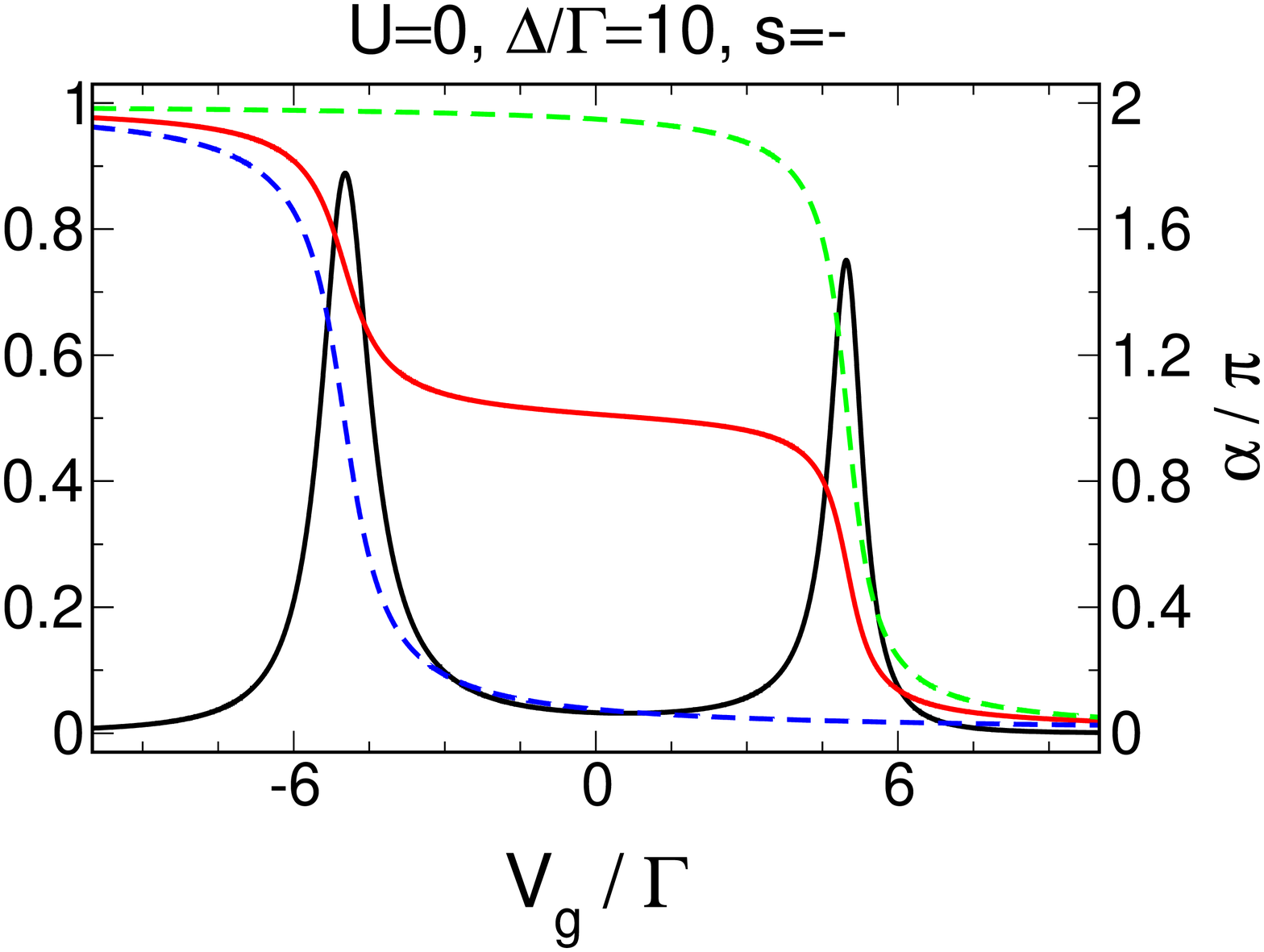}
        \caption{Gate voltage dependence of the conductance $G$ (black) and transmission phase $\alpha$ (red) of a noninteracting two-level dot with generic level-lead hybridisations $\Gamma=\{0.1~0.3~0.4~0.2\}$ for one small and one large level spacing. Level occupancies for dot A (green) and dot B (blue) are shown as well.}
\label{fig:OS.dd.wwfrei}
\end{figure}

For degenerate levels the curve described by (\ref{eq:OS.dd.condwwfrei}) is a Lorentzian of full width $2\Gamma$ for large gate voltages that has a dip and a transmission zero located at $V_g=0$. For a large $A$-$B$ asymmetry (that is for $\Gamma_A/\Gamma_B\gg1$, or vice versa), this can be viewed as a Fano anti-resonance resulting from destructive interference between the two different path through the system. The transmission phase changes by $\pi$ over each `peak' and shows a phase jump corresponding to the transmission zero in between. Increasing $V_g$ both levels are heightened in energy and therefore successively depleted.

For $s=+$ increasing $\Delta$ shifts the transmission peaks to larger $V_g$, the transmission zero always remaining in between. For large $\Delta\gg\Gamma$ their separation is given by $\Delta$, while the width and height of each peak is determined by the individual hybridisations $\Gamma_A^{L,R}$ and $\Gamma_B^{L,R}$. This is more or less obvious since in the limit of large $\Delta$, the dominant contributions to the conductance (\ref{eq:OS.dd.cond}) will come from the diagonal parts of the propagator which have support on $V_g$-regions separated by $\Delta$. The transmission phase changes by $\pi$ over each peak (and jumps in between), and again the form of $\alpha(V_g)$ is the same as in the single-level case. Both dots are depleted individually over their associated resonances, which finally manifests the intuitive picture that for nondegenerate levels with level spacing larger than the level broadening transport is observed each time the energy of one level crosses the Fermi energy of the leads. For $s=-$ the crossover regime from $\Delta\ll\Gamma$ to $\Delta\gg\Gamma$ is very different. One of the peaks present if the levels are degenerate splits up while the other becomes vanishingly small. For large $\Delta$ we recover the resonances corresponding to the individual levels, but now the transmission zero lies outside next to the (not observable) third resonance. Fig.~\ref{fig:OS.dd.wwfrei} again shows the described behaviour for the two regimes $\Delta\ll\Gamma$ and $\Delta\gg\Gamma$ for $s=+$ and $s=-$.

\subsubsection{Interacting Case, Nearly Degenerate Levels}
\begin{figure}[t]	
	\centering
	      \includegraphics[width=0.475\textwidth,height=5.2cm,clip]{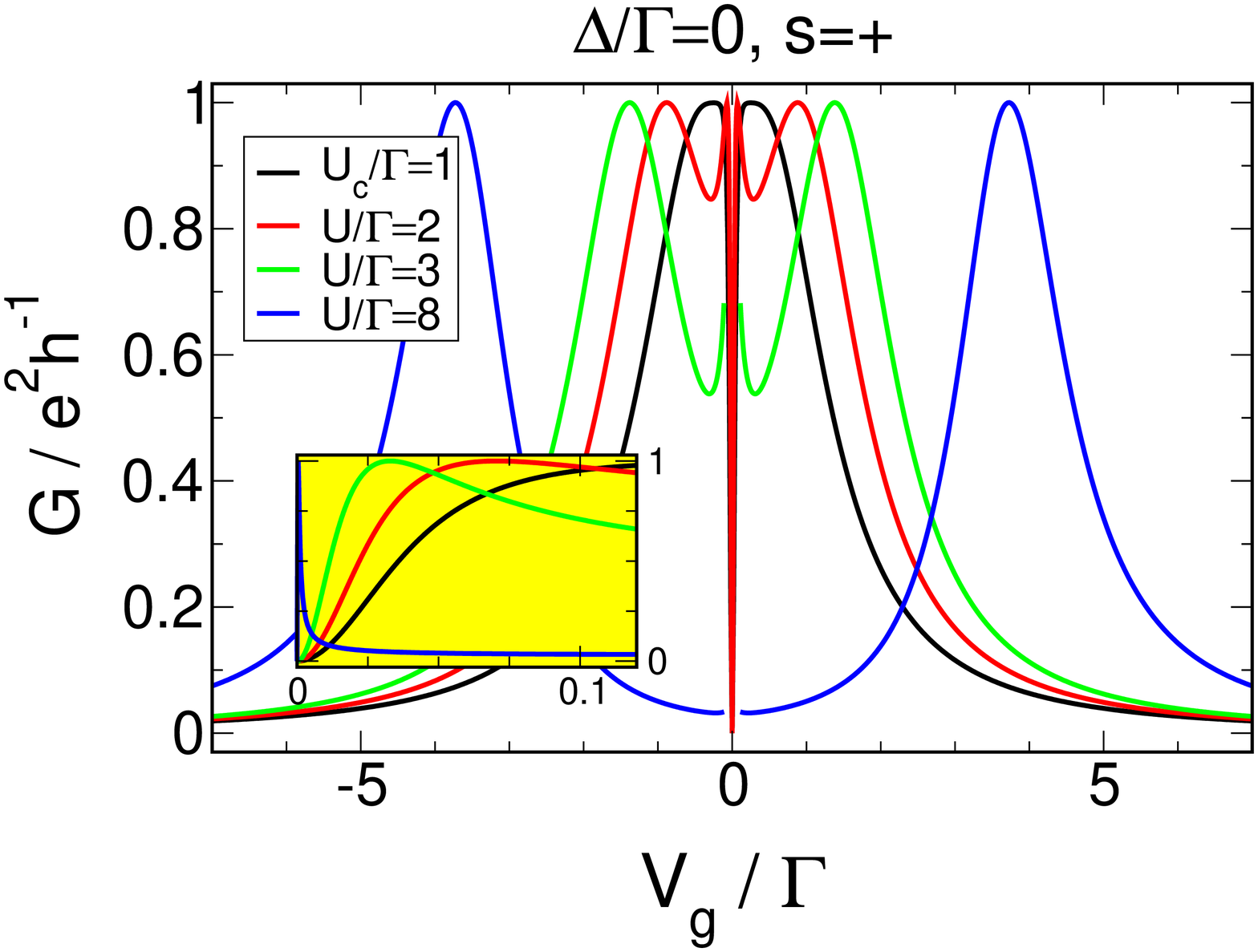}\hspace{0.035\textwidth}
        \includegraphics[width=0.475\textwidth,height=5.2cm,clip]{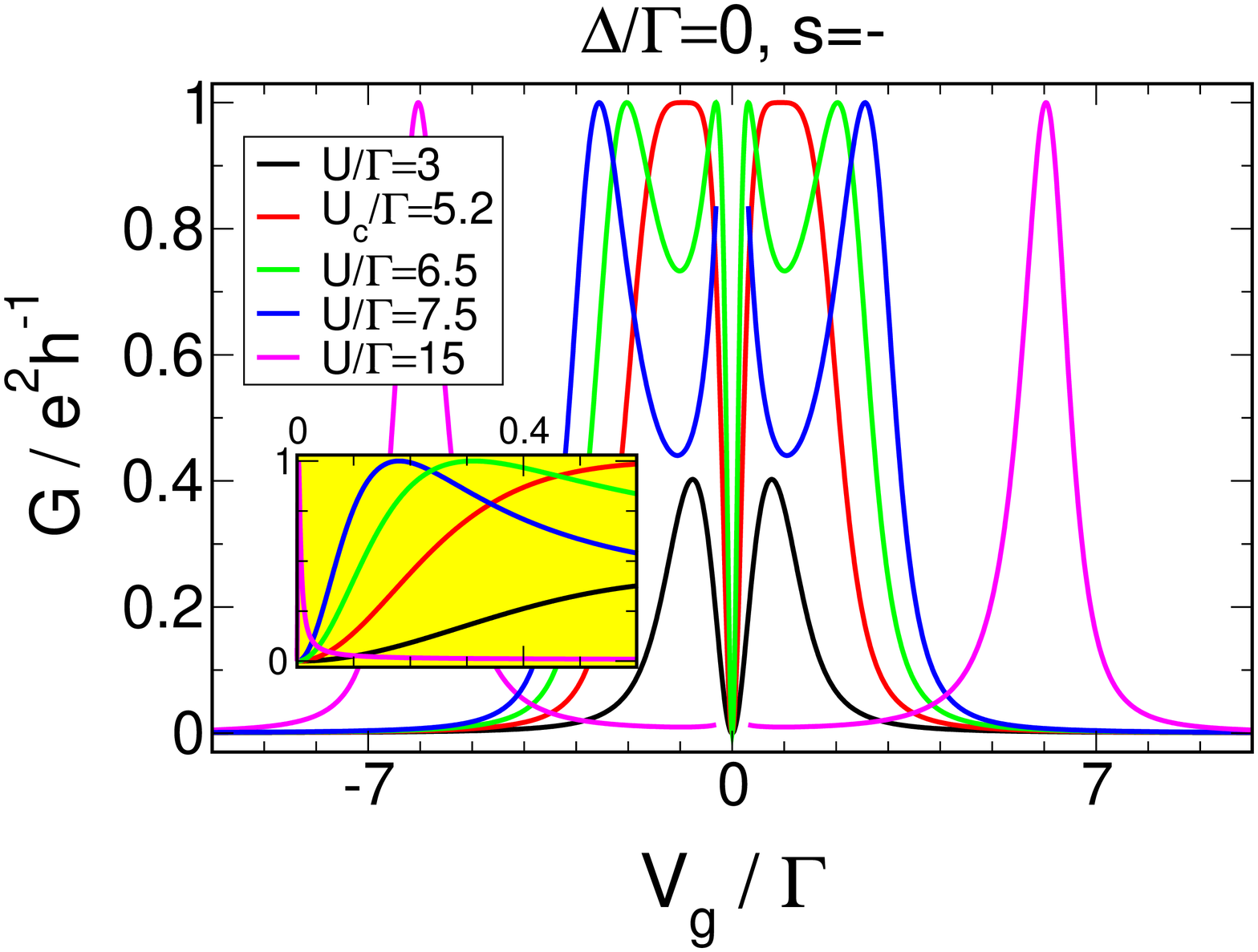}
        \caption{Gate voltage dependence of the conductance $G$ of a two-level dot with generic level-lead hybridisations $\Gamma=\{0.1~0.3~0.4~0.2\}$ at different local interactions. If the latter exceeds a critical strength depending on the dot parameters, additional sharp resonances split of the Coulomb blockade peaks located at $V_g=\pm U/2$. Note that the curves for the largest $U$ around $V_g=0$ are only shown in the insets.}
\label{fig:OS.dd.cir}
\end{figure}
If we add an interaction $U$ of order $\Gamma$ between the electrons on both levels, the physics changes. For nearly degenerate levels $\Delta\ll\Gamma$ increasing $U$ enlarges the separation of the two peaks (the wings of the Lorentzian with the dip at $V_g\approx 0$) and changes their lineshape. The height of the peaks increases as well as their separation which is given by $U$ and hence comparable to their width which is of order $\Gamma$ (for $U\approx\Gamma\leq U_c$). At a certain critical strength of the interaction (depending on the level-lead hybridisations) $U_c(\{\Gamma_l^s\},s)=O(\Gamma)$ each peak splits up into two, the height of all four resonances being equal and now independent of $U$. For large $U=O(\Gamma)$, the outer peaks are located approximately at $V_g\approx\pm U/2$, and their width is again of order $\Gamma$ and not of order $U$ or $\Gamma_l$. They are frequently interpreted in a usual Coulomb blockade picture. The inner resonances result from a subtle many-particle effect (justifying the name correlation induced resonances, CIRs) and cannot be explained in a simple way. Only recently they were predicted by \cite{cir} using the same fRG approach we employ here. Increasing $U$, the CIRs become very sharp and are located exponentially close to $V_g=0$. All this is shown in Fig.~\ref{fig:OS.dd.cir}.

\begin{figure}[t]	
	\centering
	      \includegraphics[width=0.5\textwidth,height=6cm,clip]{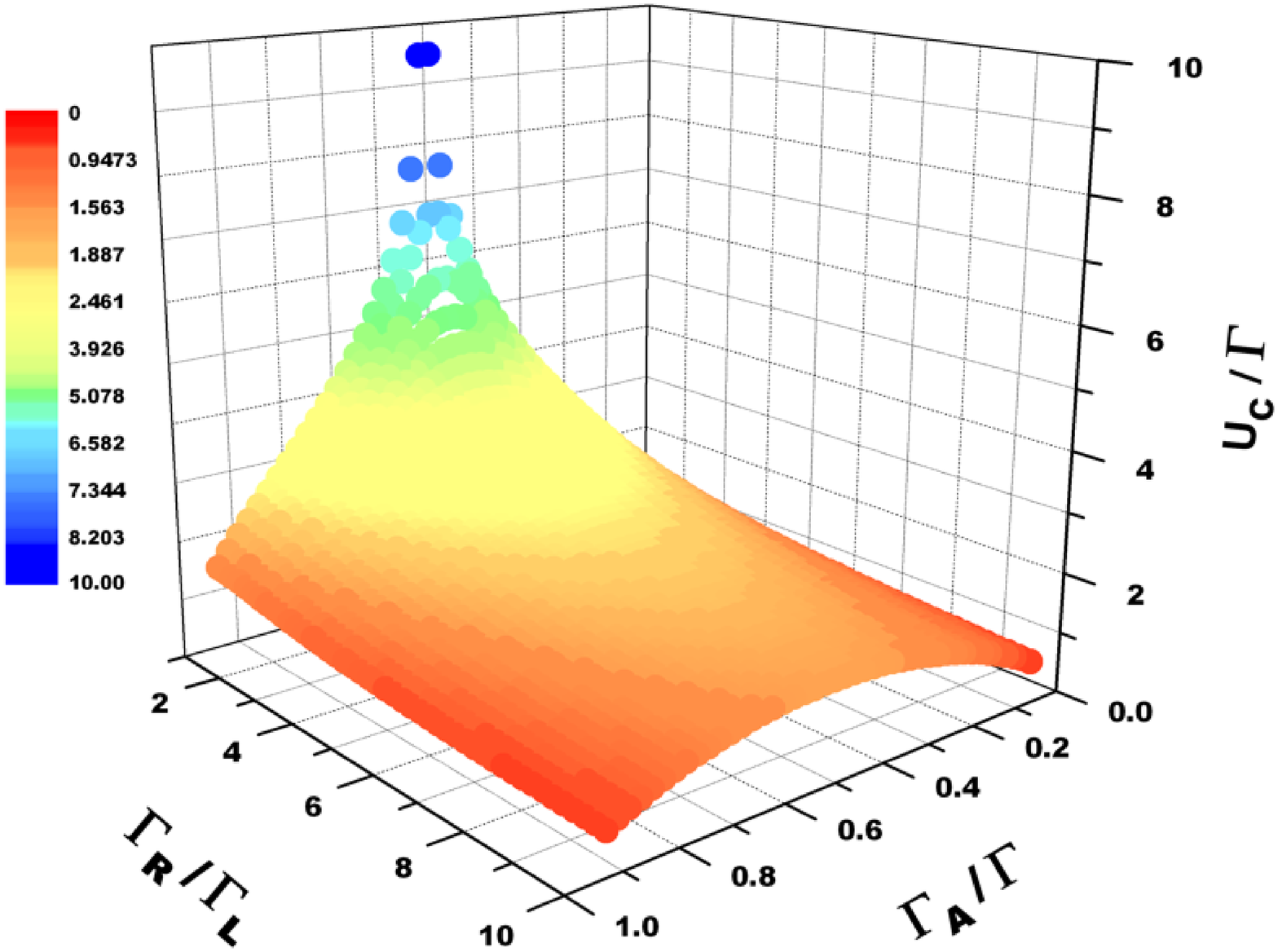}\hspace{0.015\textwidth}
        \includegraphics[width=0.45\textwidth,height=6.5cm,clip]{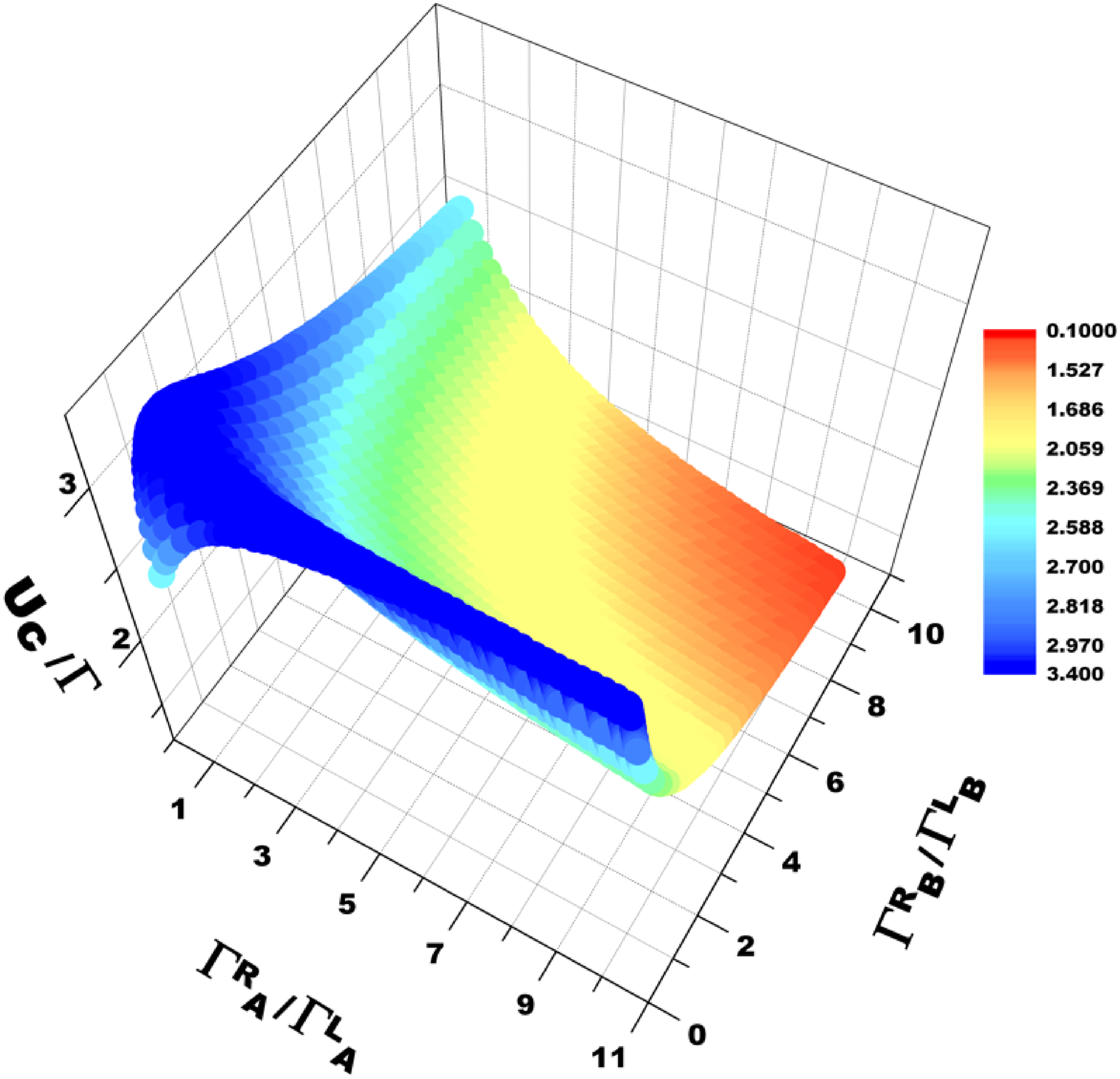}
        \caption{The critical $U_c$ that determines the appearance of the CIRs in the conductance of a double dot with degenerate levels and $s=-$ for fixed left-right asymmetry $\Gamma_A^R/\Gamma_A^L=\Gamma_B^R/\Gamma_B^L=\Gamma_R/\Gamma_L$ (left panel) and constant $A$-$B$ asymmetry $\Gamma_A/\Gamma=0.25$ (right panel).}
\label{fig:OS.dd.scancir}
\end{figure}

The critical $U_c$ that determines the interaction strength at which the additional resonances show up depends on the hybridisations $\Gamma_l^s$ as well as on the relative sign of the level-lead couplings $s$. If we fix $\Gamma$ as the unit of energy, we have to analyse a three-dimensional parameters space for each $s=+,-$ in order to obtain a complete picture of $U_c$ that might serve to better understand the nature of these structures. Since this is optically impossible, we stick to two-dimensional manifolds, and we start with $s=-$. One important choice for such a manifold is certainly that where the total coupling strength of dots $A$ and $B$ (governed by $\Gamma_A/\Gamma$) as well as the left-right asymmetry assumed to be equal for both levels, $\Gamma_A^R/\Gamma_A^L=\Gamma_B^R/\Gamma_B^L=\Gamma_R/\Gamma_L$, is varied (Fig.~\ref{fig:OS.dd.scancir}, left panel). We find that $U_c$ is always roughly of order $\Gamma$, and it decreases for decreasing symmetry ($A$-$B$ or left-right) of the system. Another important parameter manifold arises for fixed $\Gamma_A/\Gamma_B$ but arbitrary $\Gamma_A^L/\Gamma_A^R$ and $\Gamma_B^L/\Gamma_B^R$ (Fig.~\ref{fig:OS.dd.scancir}, right panel). If we assume dot $B$ being stronger coupled than dot $A$, we find that for constant $\Gamma_A^R/\Gamma_A^L$, $U_c$ decreases the more $\Gamma_B^R/\Gamma_B^L$ deviates from one. Furthermore, $U_c$ becomes larger (smaller) if for fixed $\Gamma_B^R/\Gamma_B^L<1$ ($\Gamma_B^R/\Gamma_B^L>1$) $\Gamma_A^R/\Gamma_A^L$ is increased. In the $s=+$ case, an equal left-right asymmetry does not represent generic parameters; computing $U_c$ as a function of $\Gamma_A/\Gamma_B$ and $\Gamma_A^R/\Gamma_A^L=2\Gamma_B^R/\Gamma_B^L$ yields the same picture as in the $s=-$ case with fixed asymmetry. We will refrain from a more detailed discussion here.

\begin{figure}[t]	
	\centering
	      \includegraphics[width=0.495\textwidth,height=4.4cm,clip]{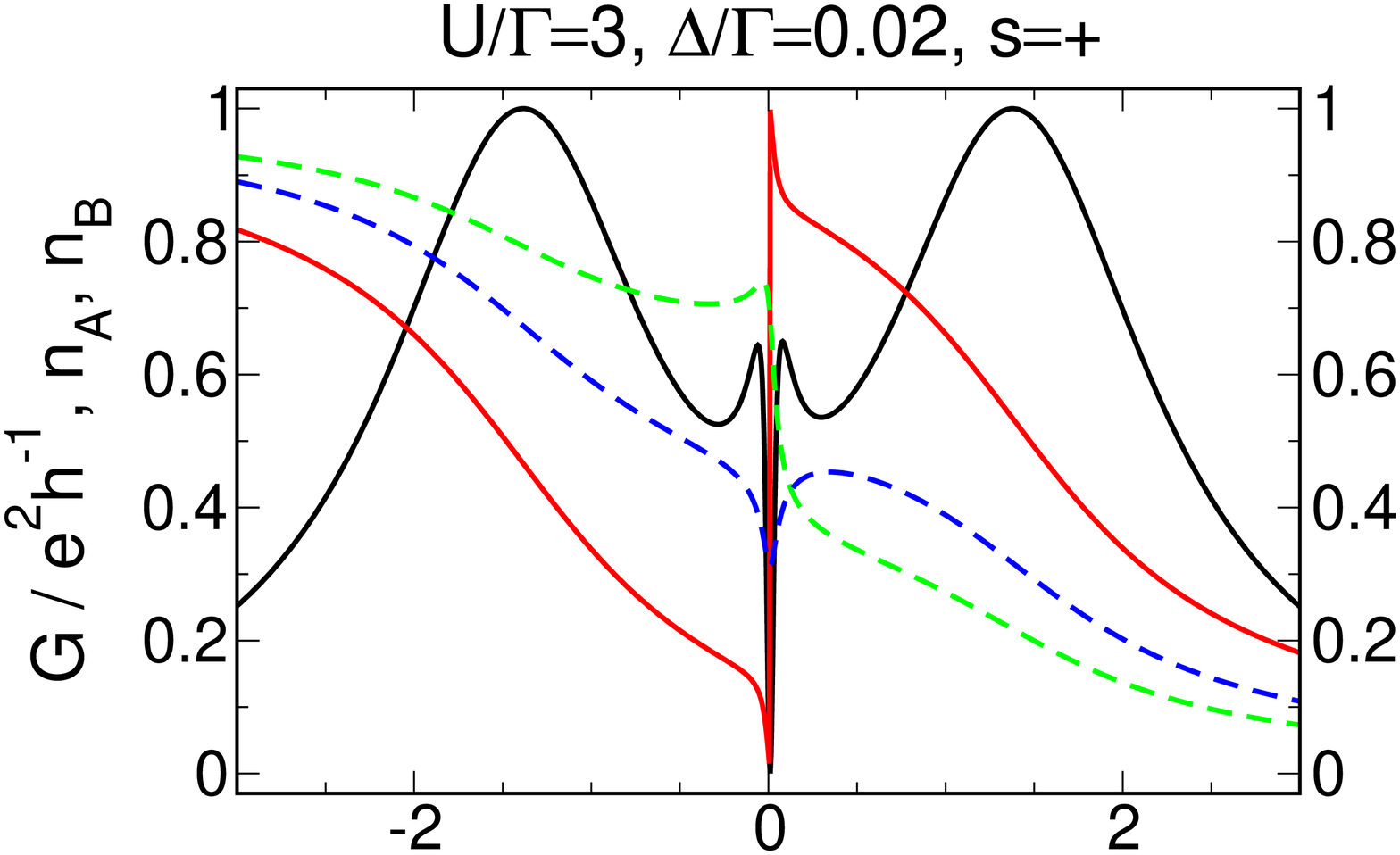}\hspace{0.015\textwidth}
        \includegraphics[width=0.475\textwidth,height=4.4cm,clip]{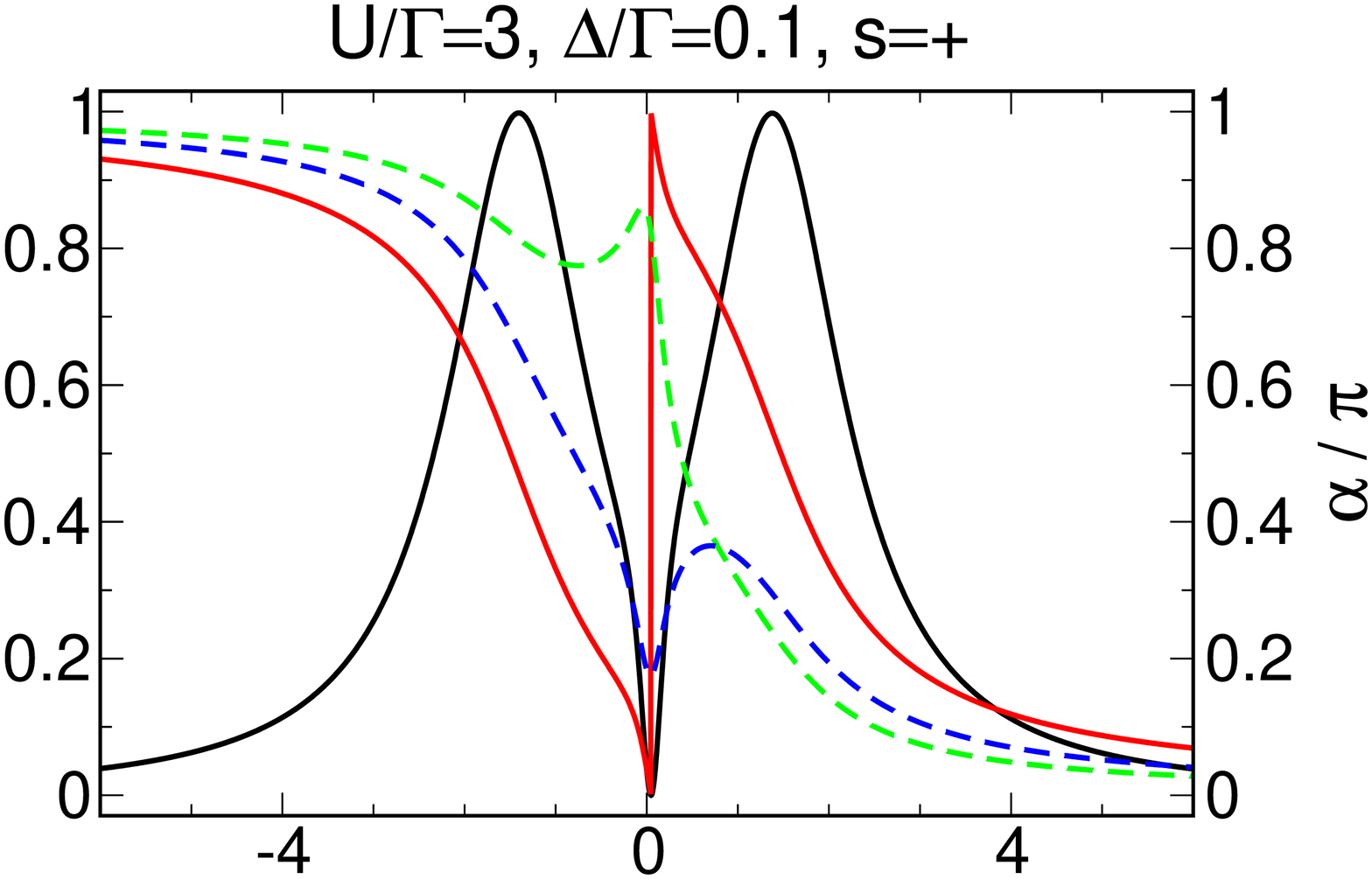}\vspace{0.3cm}
        \includegraphics[width=0.495\textwidth,height=5.2cm,clip]{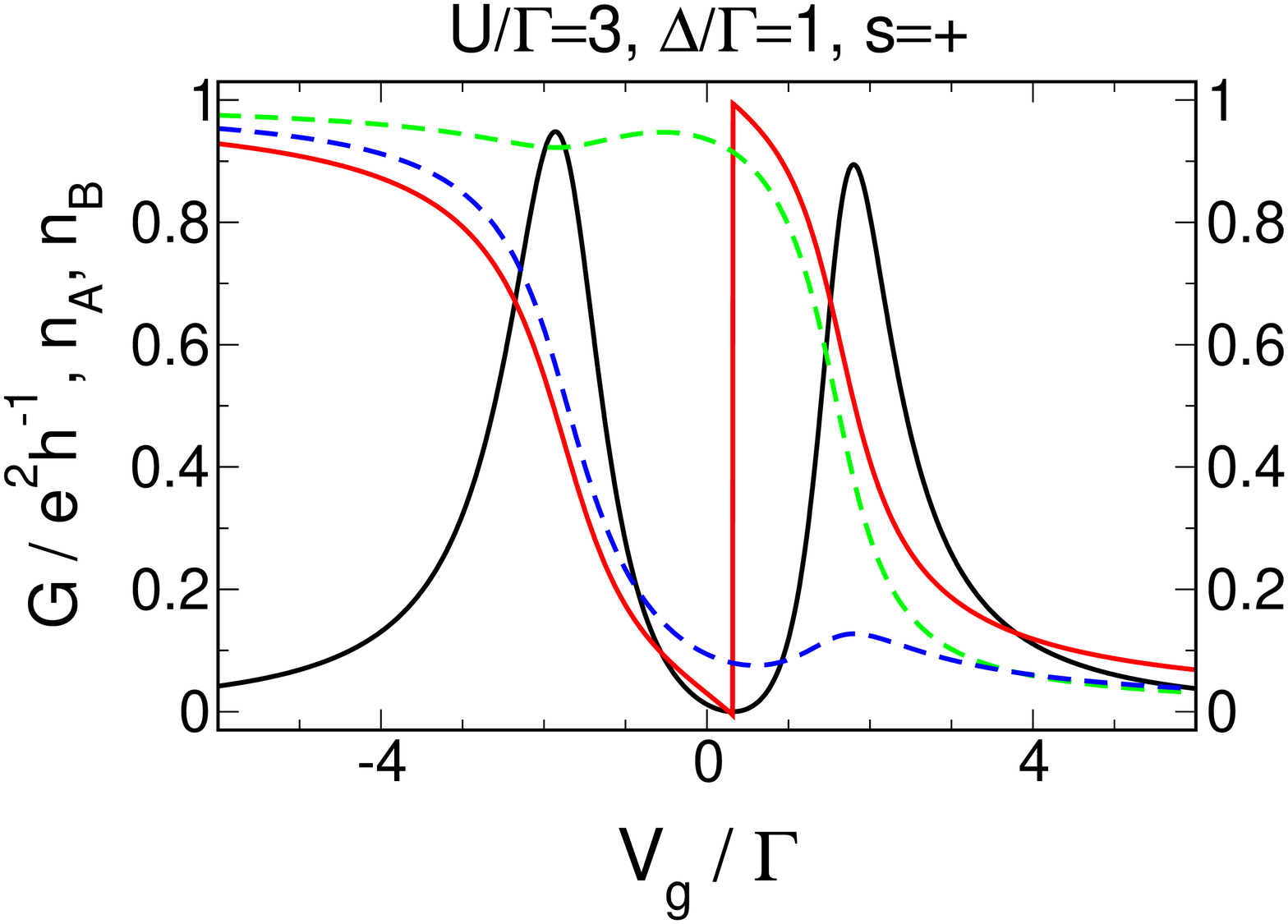}\hspace{0.015\textwidth}
        \includegraphics[width=0.475\textwidth,height=5.2cm,clip]{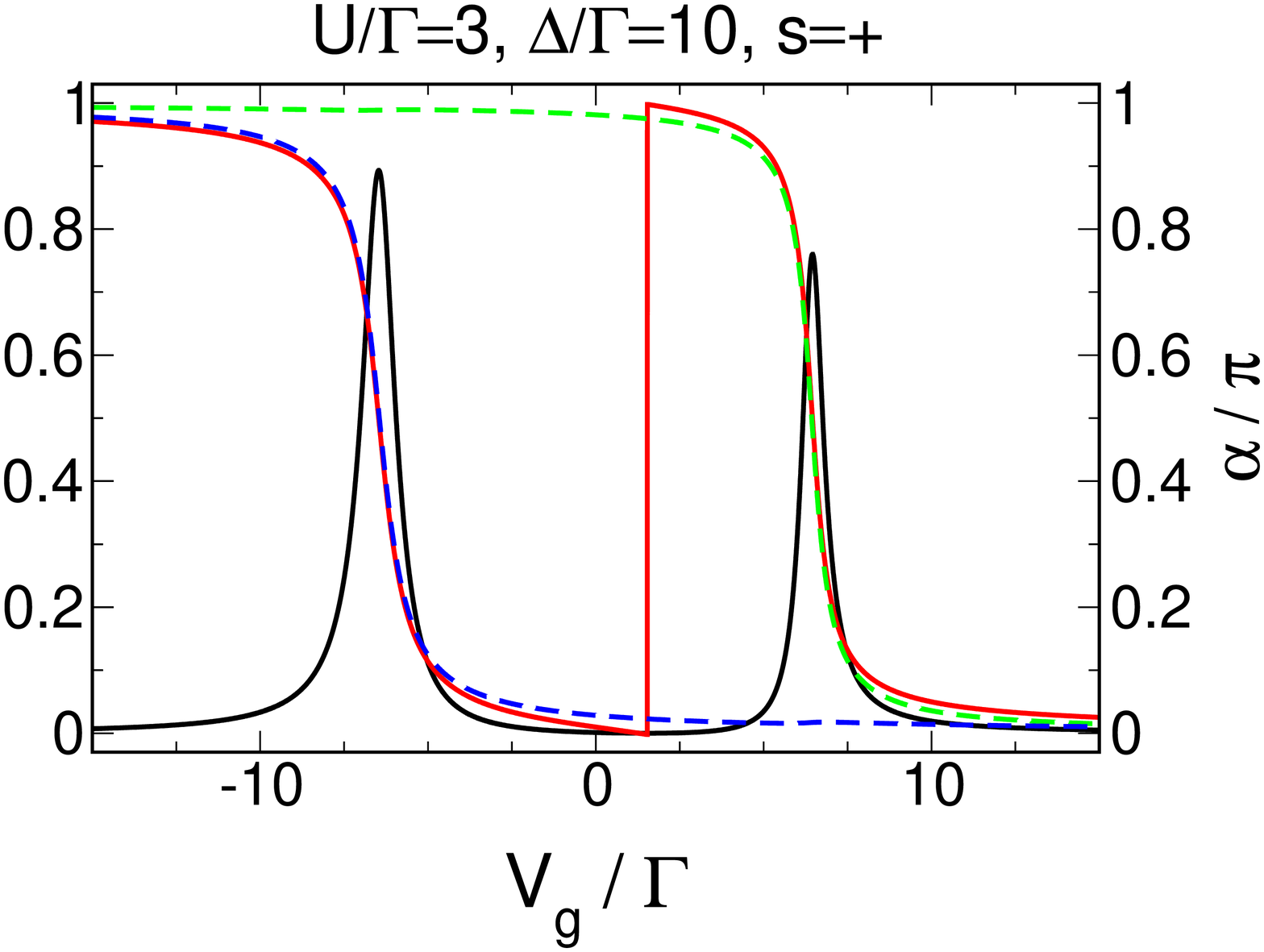}
        \caption{Gate voltage dependence of the conductance $G$ (black) and transmission phase $\alpha$ (red) of a two-level dot with generic level-lead hybridisations $\Gamma=\{0.1~0.3~0.4~0.2\}$ for four different level spacings $\Delta$. Level occupancies for dot A (green) and dot B (blue) are shown as well. For nearly degenerate levels, the phase changes rapidly because of the remnants of the CIRs. The effect of the population inversion is only weakly developed since the interaction is still too small. For large level spacings, the conductance is governed by transport through the individual levels.}
\label{fig:OS.dd.plusdelta}
\end{figure}

\begin{figure}[t]	
	\centering
	      \includegraphics[width=0.495\textwidth,height=4.4cm,clip]{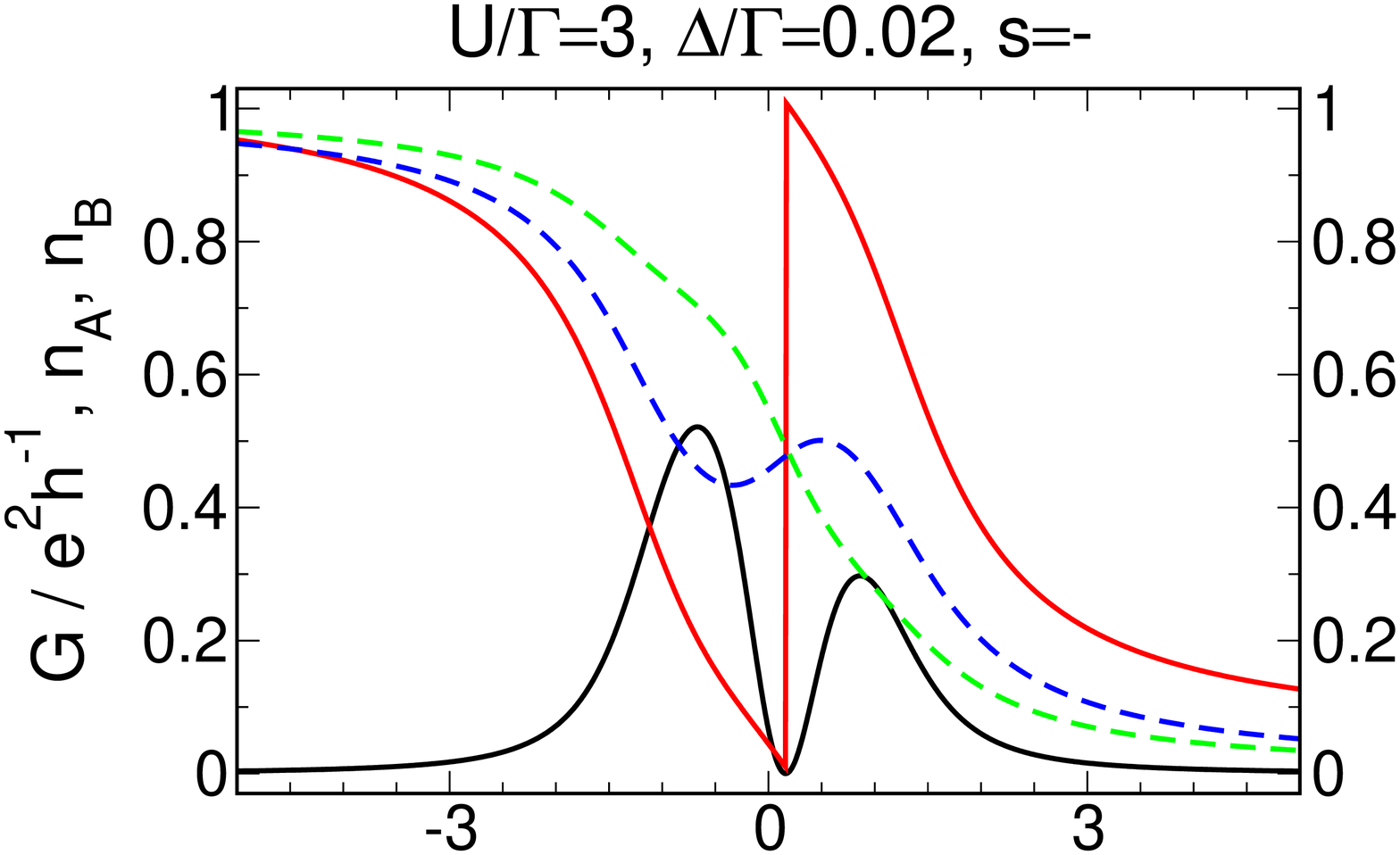}\hspace{0.015\textwidth}
        \includegraphics[width=0.475\textwidth,height=4.4cm,clip]{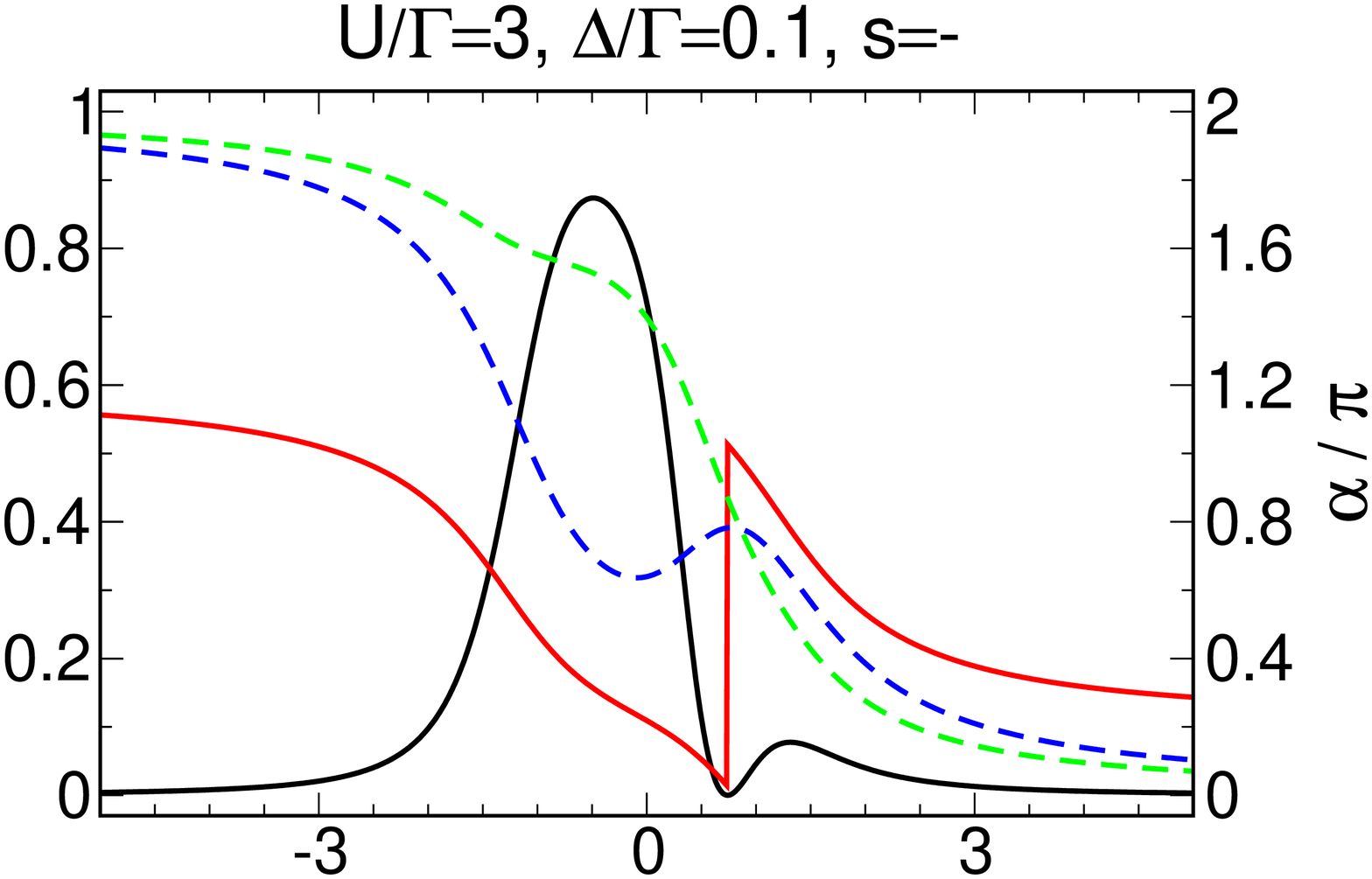}\vspace{0.3cm}
        \includegraphics[width=0.495\textwidth,height=5.2cm,clip]{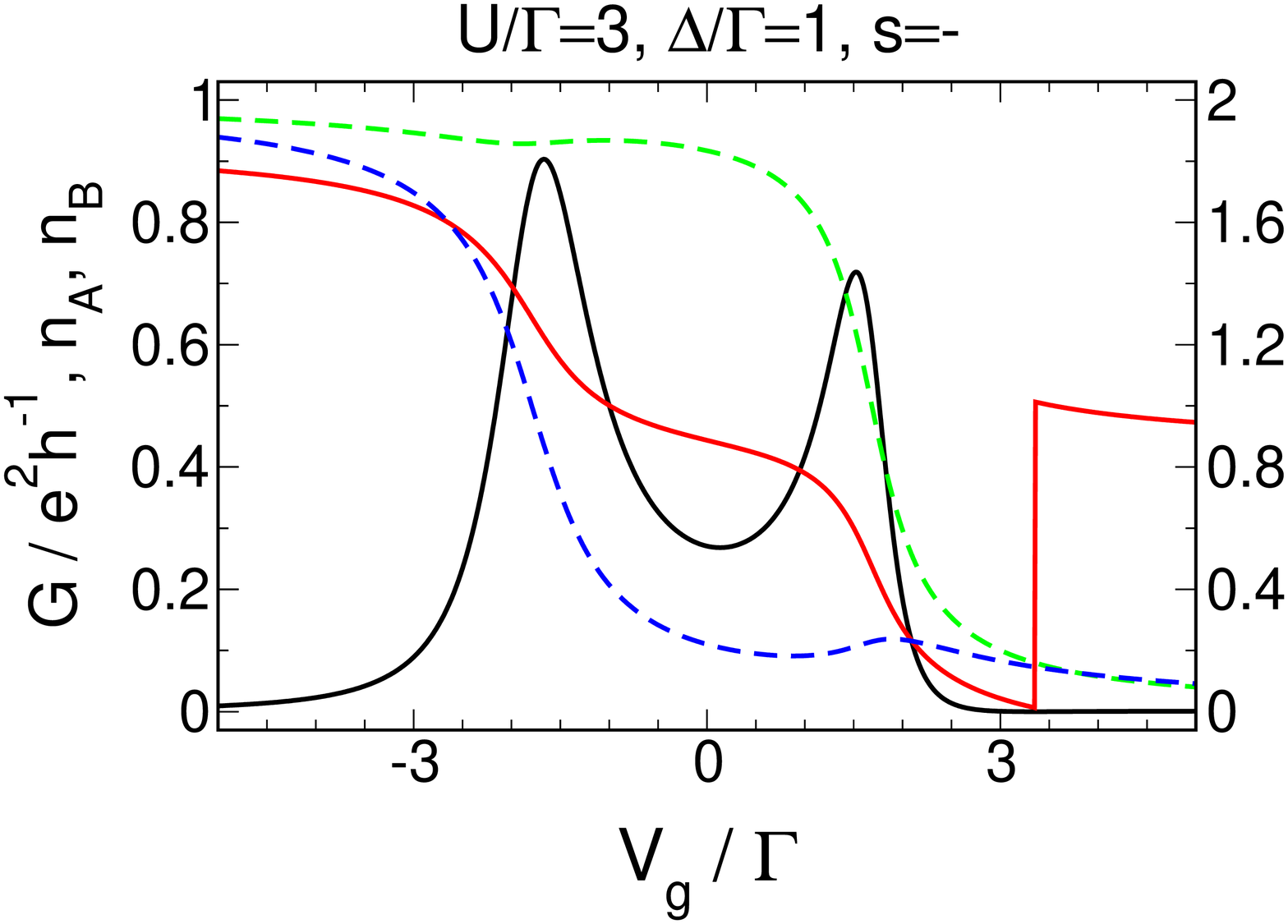}\hspace{0.015\textwidth}
        \includegraphics[width=0.475\textwidth,height=5.2cm,clip]{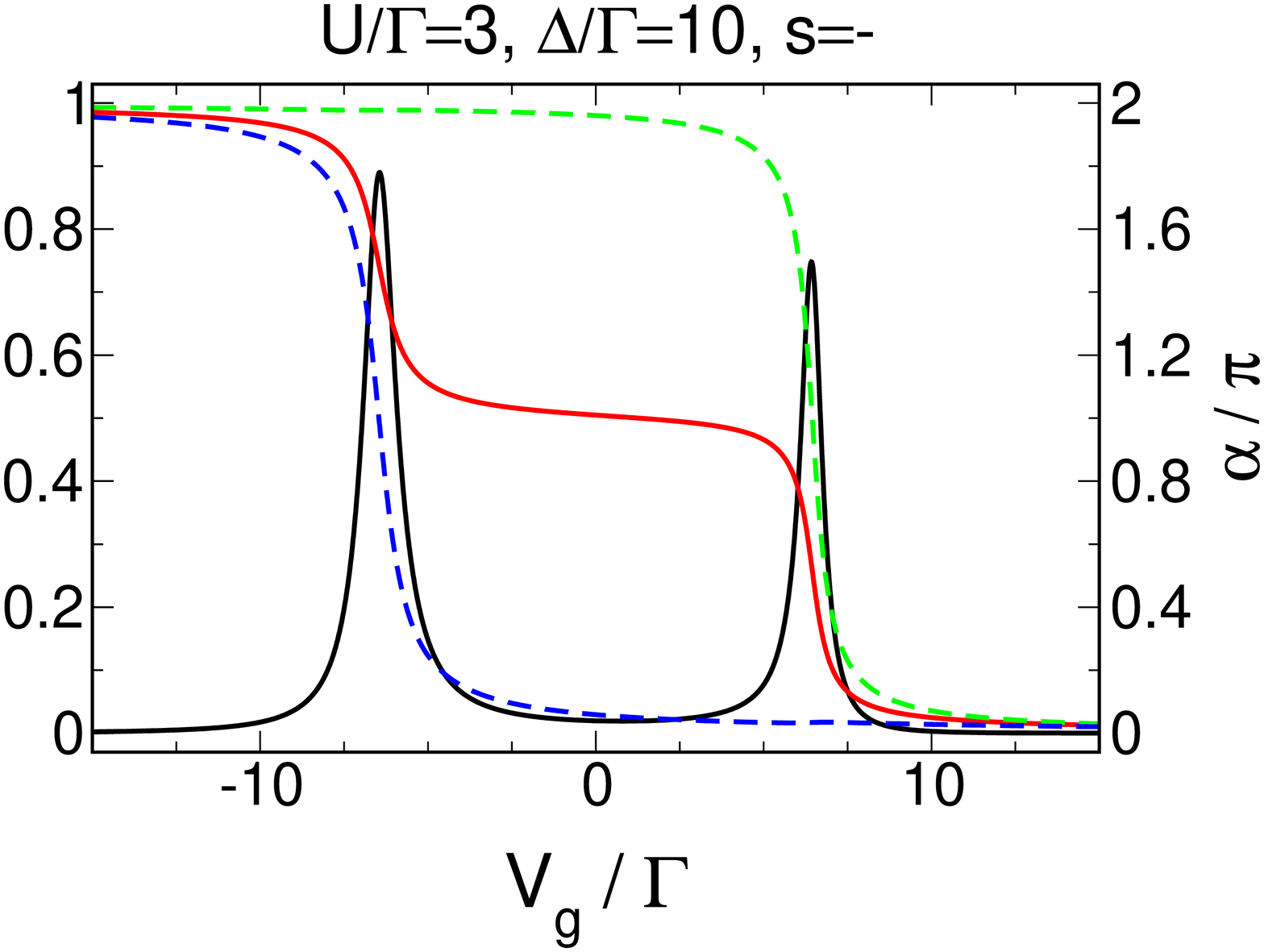}
        \caption{The same as Fig.~\ref{fig:OS.dd.plusdelta}, but for $s=-$. Here, the interaction does not exceed the critical $U_c$ so that the CIRs are not observed for nearly degenerate levels, while the population inversion is already visible. In the subsequent evolution, one peak splits up and the transmission zero moves outwards.}
\label{fig:OS.dd.minusdelta}
\end{figure}

The transmission phase $\alpha$ changes by $\pi$ over each Coulomb blockade peak and jumps by $\pi$ at the transmission zero which remains close to $V_g=0$ for arbitrary strength of the interaction. If present, the phase changes steeply when crossing the CIRs (see the upper left panel of Fig.~\ref{fig:OS.dd.plusdelta}), but basically it exhibits an S-like lineshape that becomes more pronounced the larger the interaction is chosen (compare the upper left panel of Fig.~\ref{fig:OS.dd.minusdelta} with $U/\Gamma=10.0$ in Fig.~\ref{fig:OS.dd.sum}). The level occupancies experience a more dramatic change if the interaction is switched on. In the limit of large $U$, they no longer depend monotonically on the level energy. Coming from large negative $V_g$, the dot that is more strongly coupled gets depleted over the first Coulomb blockade peak, while the occupation of the other stays close to one. At $V_g\approx 0$, the former is filled again while now the latter gets depleted, and a transmission zero instead of a resonance is observed. This inversion of the level populations if followed by another depletion of the more strongly coupled level over the second Coulomb blockade peak. In the limit of large $V_g$, both levels get depleted. It is important to note that the population inversion is not directly related to the appearance of the CIRs, since the strength of the interaction where the former sets in does not coincide with the critical $U_c$ (see for example Figs. \ref{fig:OS.dd.plusdelta} \& \ref{fig:OS.dd.minusdelta}). Furthermore, one should be aware that the total occupancy still depends monotonically on the gate voltage and changes by one over each resonance.

\subsubsection{Interacting Case, Large Level Spacing}

If we increase the level spacing $\Delta$, the CIRs gradually vanish. The largest $\Delta=:\tilde\Delta_{\tn{CIR}}$ where they are still observable is a function of the other parameters of the dot, but roughly it is given by $\Delta\approx\Gamma/50$. Applying a larger level detuning leaves only the Coulomb blockade peaks at position $V_g\approx\Delta\pm U/2$ which have width of order $\Gamma$ and the transmission zero in between. The subsequent evolution from $\Delta<\Gamma$ to $\Delta\gg\Gamma$ (the crossover regime) is completely similar to the noninteracting case. For $s=+$ the separation of the peaks increases, but the conductance always vanishes at some point close to $V_g=0$. For $s=-$ one of the peaks splits up (the one that one would associate with the more strongly coupled level\footnote{In that fashion that one would think of the level with higher energy to get depleted first; however, the universal peak width of $\Gamma$ shows that such an interpretation is incorrect. In fact, it would be best just to say that if dot A (B) is the more strongly coupled one, it is the right (left) peak that splits.}), while the other becomes vanishingly small. The scale $\tilde\Delta_{\tn{cross}}$ that separates the $\Delta\ll\Gamma$ regime where one observes two resonances with a transmission zero in between from the crossover regime where one of the peaks significantly decreases while the other begins to split depends on the parameters of the dot (and of course on its precise definition), but a very rough estimate should be $\tilde\Delta_{\tn{cross}}=\Gamma/10$. Surprisingly, $\Delta_\tn{cross}$ is always of order $\Gamma$ no matter how $U$ is chosen (Fig.~\ref{fig:OS.dd.sum}). If the CIRs are present, it might not be possible to distinguish their vanishing and the splitting up of one peak if for certain dot parameters it happens that $\tilde\Delta_{\tn{cir}}\approx\tilde\Delta_{\tn{cross}}$, as it is likely to happen if these structures are well-pronounced, that is for sufficiently large $U$. In both the $s=+$ and the $s=-$ case in the limit $\Delta\gg\Gamma$ one ends up with two peaks either with a transmission zero and associated phase jump of $\pi$ in between ($s=+$), or the conductance stays finite and the phase evolves continuously ($s=-$). As in the noninteracting case, this situation corresponds to separate transport through each of the original levels, but in contrast to $U=0$, the separation of both peaks is now given by $U+\Delta$. 

The latter follows from the intuitive picture that each time one dot gets depleted, all other levels are lowered in energy by $U$ due to the lack of the Coulomb repulsion. A more precise understanding can be gained if one diagonalises the interacting isolated dot system decoupled from the leads, which is meaningful if the energies of the former ($U$ and $\Delta$) are much larger than the energy scale that is determined by the coupling to the latter. In the occupation number basis $\{|0,0\rangle, |1,0\rangle, |0,1\rangle, |1,1\rangle\}$ the dot Hamiltonian reads
\begin{equation*}
H_{\tn{dot}} =
\begin{pmatrix}
0 & & & & \\
& V_g-\frac{U+\Delta}{2} & & \\
& & V_g-\frac{U-\Delta}{2} & \\
& & & 2V_g
\end{pmatrix}.
\end{equation*}
One would now expect transport to be possible when the states with zero and one or those with one and two electrons have the same energy. At $V_g=-(U+\Delta)/2$ this holds for $|0,0\rangle$ and the single-particle state with lowest energy, $|1,0\rangle$, and at $V_g=(\Delta+U)/2$ the latter and the two-particle state $|1,1\rangle$ are degenerate. These are approximately the positions of the resonance peaks for large level spacing. The same argumentation holds if the dot system comprises arbitrary many levels each with levels spacing $\Delta$ and two-particle interaction $U$ but no direct hopping. For arbitrary $V_g$, the $n+1$-particle state of lowest energy will then exceed the energy of that with $n$ particles by $V_g+n\Delta+nU$, and likewise the latter exceeds the energy of the $n-1$-particle state by $V_g+(n-1)\Delta+(n-1)U$. Hence the differences differ by $\Delta+U$ as claimed.

A final overview of the conductance and transmission phase of the double dot geometry that particularly accounts for the different energy scales within the system (though containing nothing that has not been mentioned before) is shown in Fig.~\ref{fig:OS.dd.sum}.

\subsubsection{Effective Level Interpretation}

To gain more insight into the interacting case it is often useful to exploit that the self-energy computed by our fRG scheme is frequency-independent. Hence the effect of the electron correlations is buried in the renormalized level positions and hoppings (which are not present in the free propagator, but in general they are generated by the flow) as a function of $V_g$. Equivalently, one can consider the eigenvalues $\omega_i$ and eigenvectors $\vec V^i$ of the dot system with these effective parameters decoupled from the leads. The eigenvectors $\{(\lambda_1,\lambda_2),(\mu_1,\mu_2)\}$ determine the hybridisations of new levels with energies $\omega_i(V_g)$ to the leads by expressing the coupling part of the Hamiltonian in terms of the new eigenstates of the effective isolated dot Hamiltonian and reading of the hopping matrix elements. More precisely, the old dot operators $d_{A,B}$ can be written as
\begin{equation*}\begin{split}
	d_A^{(\dagger)} =&~ \lambda_1 d_1^{(\dagger)} + \mu_1 d_2^{(\dagger)} \\
	d_B^{(\dagger)} =&~ \lambda_2 d_1^{(\dagger)} + \mu_2 d_2^{(\dagger)},
\end{split}\end{equation*}
hence the new couplings read
\begin{equation}\begin{split}
\tilde t_1^s = & ~\lambda_1 t_A^s + \lambda_2 t_B^s \\
\tilde t_2^s = & ~\mu_1 t_A^s + \mu_2 t_B^s,
\end{split}\end{equation}
and the hybridisations follow as $\gamma_l^s=\pi|\tilde t_i^s|^2\rho_\tn{lead}$. Transport through this new system is then equivalent to transport to the original one, which is evident because the Green function $\mc G^0_{1_L;1_R}$ is of course independent of the choice of a basis in the dot region, and in an (effective) noninteracting problem it is irrelevant which matrix element of the propagator we use to calculate the conductance, so that $\mc G^0_{1_L;1_R}$ determines the latter completely. If we have an arbitrary number $N$ of parallel dots, the new couplings follow similarly,
\begin{equation*}
\tilde t_i^s(V_g) = \sum_{j=1}^N V^i_j t_j^s.	
\end{equation*}
One should note that they are normalized because of the orthogonality of the $\vec V^i$,
\begin{equation*}
\gamma = \sum_{i,s}\gamma_i^s = \pi\rho_\tn{lead} \sum_{i,s}\sum_{j,j'} V^i_j V^i_{j'} t_j^s (t_{j'}^s)^* = \pi\rho_\tn{lead}\sum_{s,j} |t_j^s|^2 = \Gamma.
\end{equation*}

\begin{figure}[t]	
	\centering
	      \includegraphics[width=0.495\textwidth,height=5.2cm,clip]{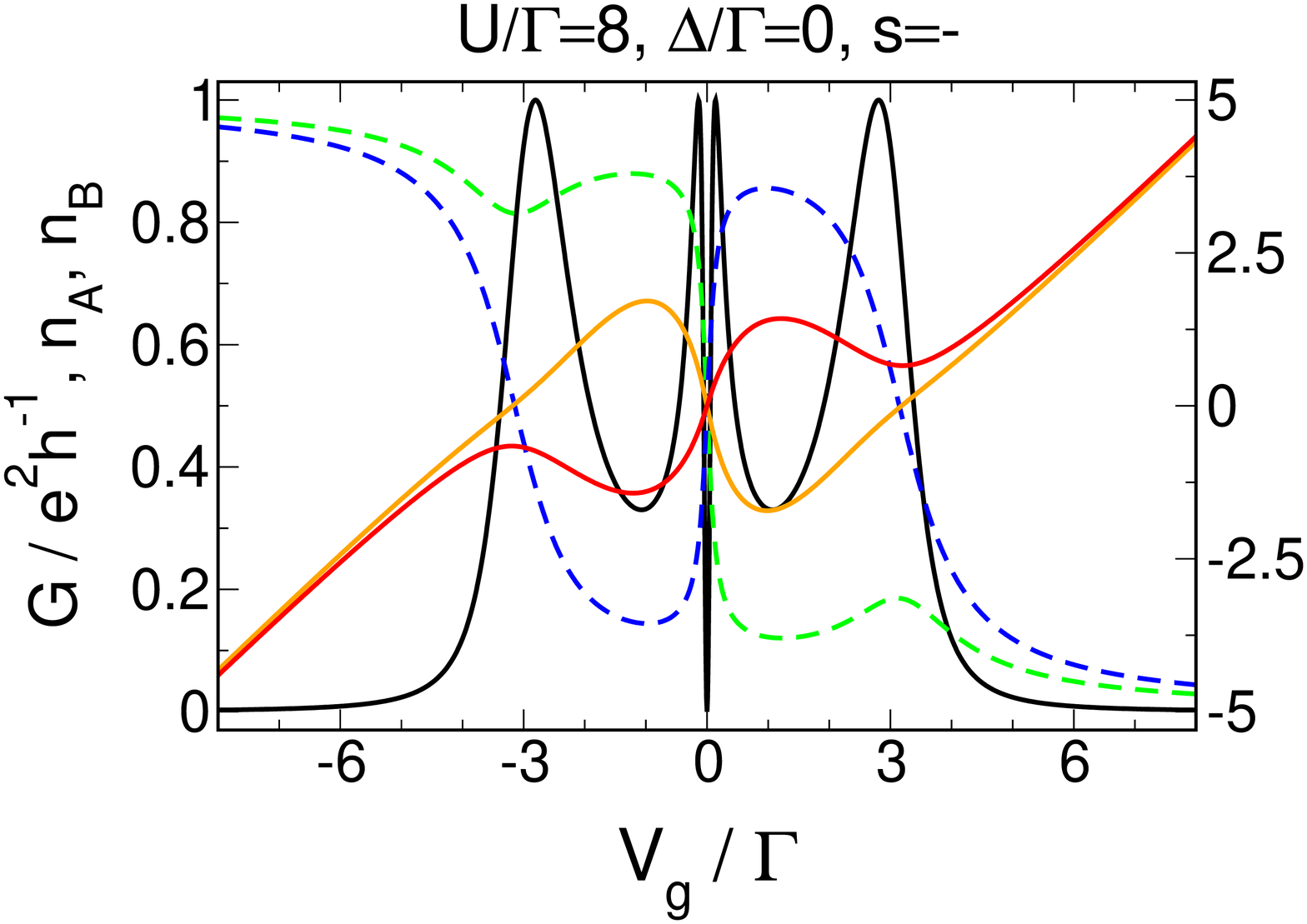}\hspace{0.015\textwidth}
        \includegraphics[width=0.475\textwidth,height=5.2cm,clip]{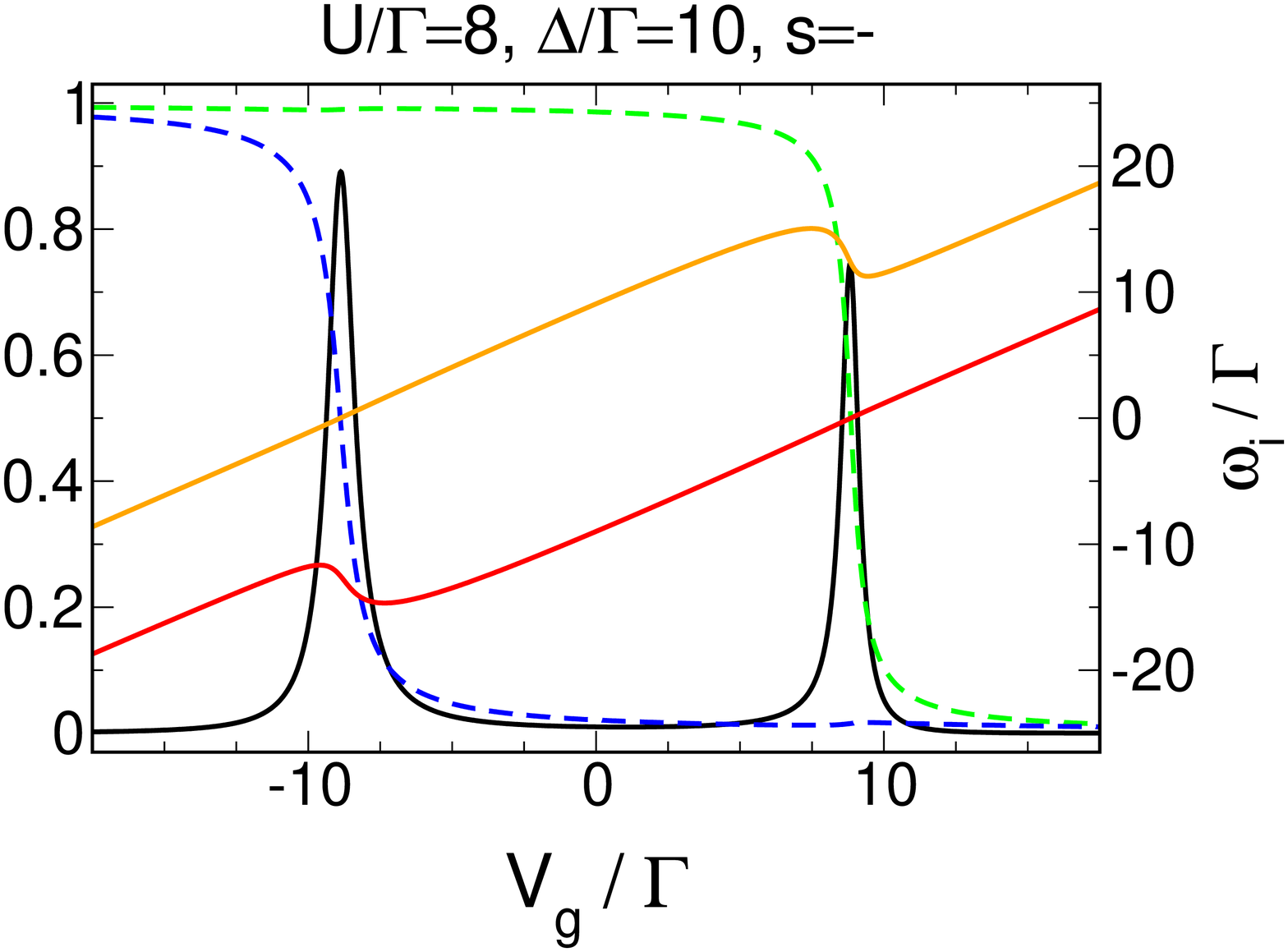}
        \caption{Conductance $G$ (black) of a two-level dot with generic level-lead hybridisations $\Gamma=\{0.1~0.3~0.4~0.2\}$ for two different level spacings. For almost degenerate levels, the level occupancies (dot A: green, dot B: blue) depend non-monotonically on the gate voltage $V_g$. The same holds for the effective eigenenergies $\omega_i$ (more weakly coupled dot: red, more strongly coupled dot: orange). Note that it is always the more strongly coupled effective level (original dot) that crosses the chemical potential (gets depleted) at each Coulomb peak. For large level spacings, $\omega_i$ crosses $\mu$ linearly at each peak. The new hybridisations almost coincide with the original ones and are therefore not shown (this does not longer hold for $N>2$  and $\Delta\ll\Gamma $ where correlations enhance the asymmetry between the dots).}
\label{fig:OS.dd.efflevel}
\end{figure}

In the case of nearly degenerate levels (Fig.~\ref{fig:OS.dd.efflevel}, left panel) and regardless of the relative sign $s$, we find that for $V_g\to\mp\infty$ the new effective levels $\omega_i$ (the eigenvalues of the isolated dot system at the end of the flow) are given by $\omega_i=V_g\pm U/2$. This is plausible since in the limit of large gate voltages, the latter is the dominant energy so that for $V_g\to\infty$ we would indeed expect $\omega_i=V_g-U/2$. If on the other hand both dots are filled ($V_g\to-\infty$) their energy is shifted by the constant $U$, which explains the observed behaviour $\omega_i=V_g+U/2$. In contrast, in the region around $V_g=0$, the interaction (assumed to be sufficiently large) has a more dramatic effect, in particular the eigenergies now no longer depend monotonically on the bare energy of the original levels. The more strongly coupled level (referring to the new couplings determined from the eigenvectors) crosses the Fermi energy of the leads at the first Coulomb blockade peak, but close to $V_g=0$ both levels cross (and cross $\mu$) so that a transmission zero (a type of Fano-antiresonance) instead of a peak is observed. At the second Coulomb resonance it is then again the more strongly coupled level that crosses the chemical potential. This is closely related to the aforementioned population inversion, since it turns out that the renormalized level positions are the dominant part of the effective Hamiltonian decoupled from the leads that one diagonalises in order to obtain the $\omega_i$, implying that the hybridisations barely change. The whole scenario of the most strongly coupled level crossing the chemical potential again and again at each peak is in close analogy to the one observed by \cite{imry} (SI) who studied chaotic quantum dots containing a few hundred levels. One should note, however, that the situation described here is more general since it does not require SI's assumption of strong asymmetry in the hybridisations. In particular, the behaviour of the level with largest $\gamma_i$ crossing $\mu$ at each resonance is observed generically even if not one of the original $\Gamma_i$ is much larger than the other.

\begin{figure}[t]	
	\centering
	      \vspace{-0.3cm}\includegraphics[width=0.352\textwidth,height=3.75cm,clip]{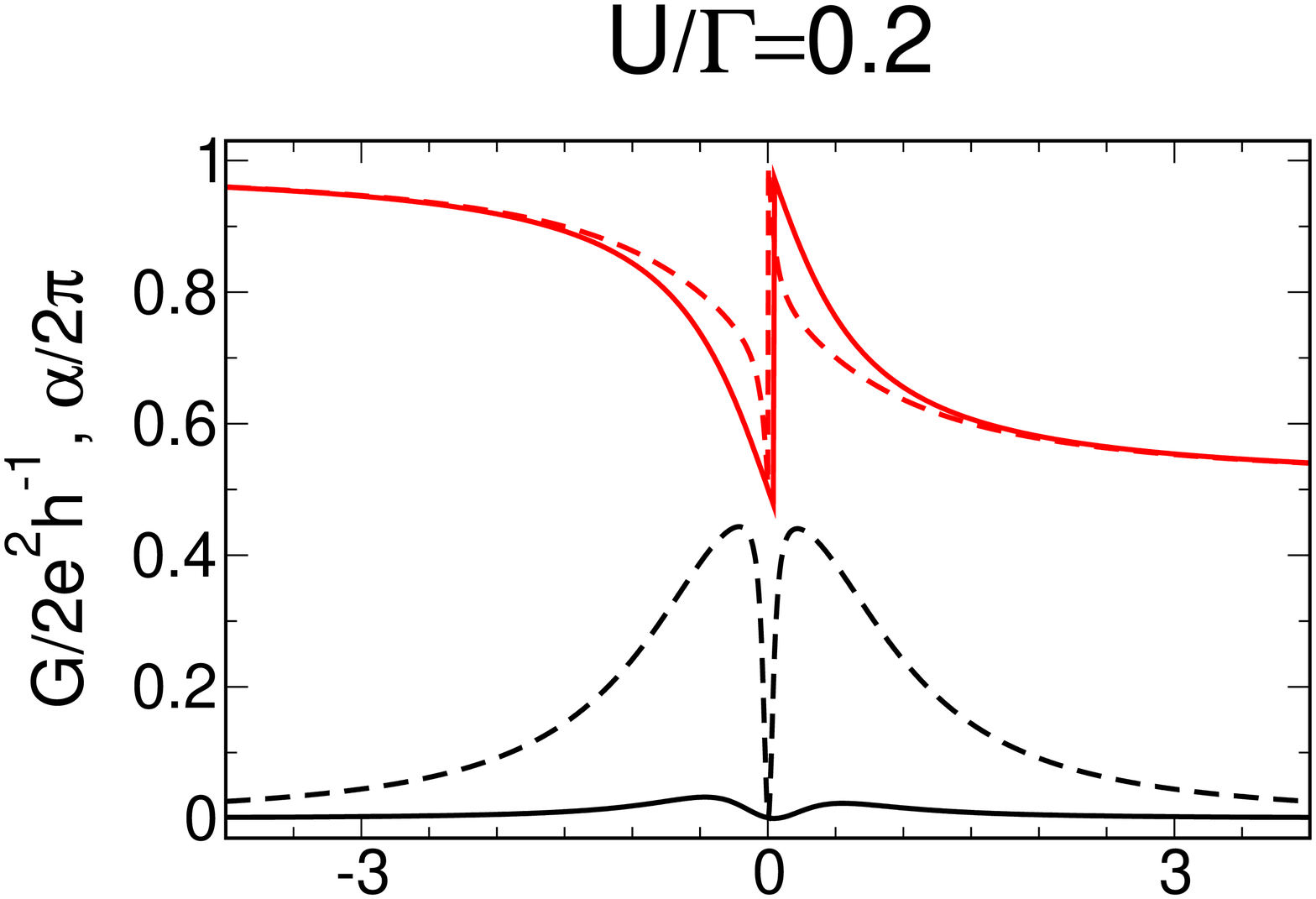}
	      \includegraphics[width=0.292\textwidth,height=3.75cm,clip]{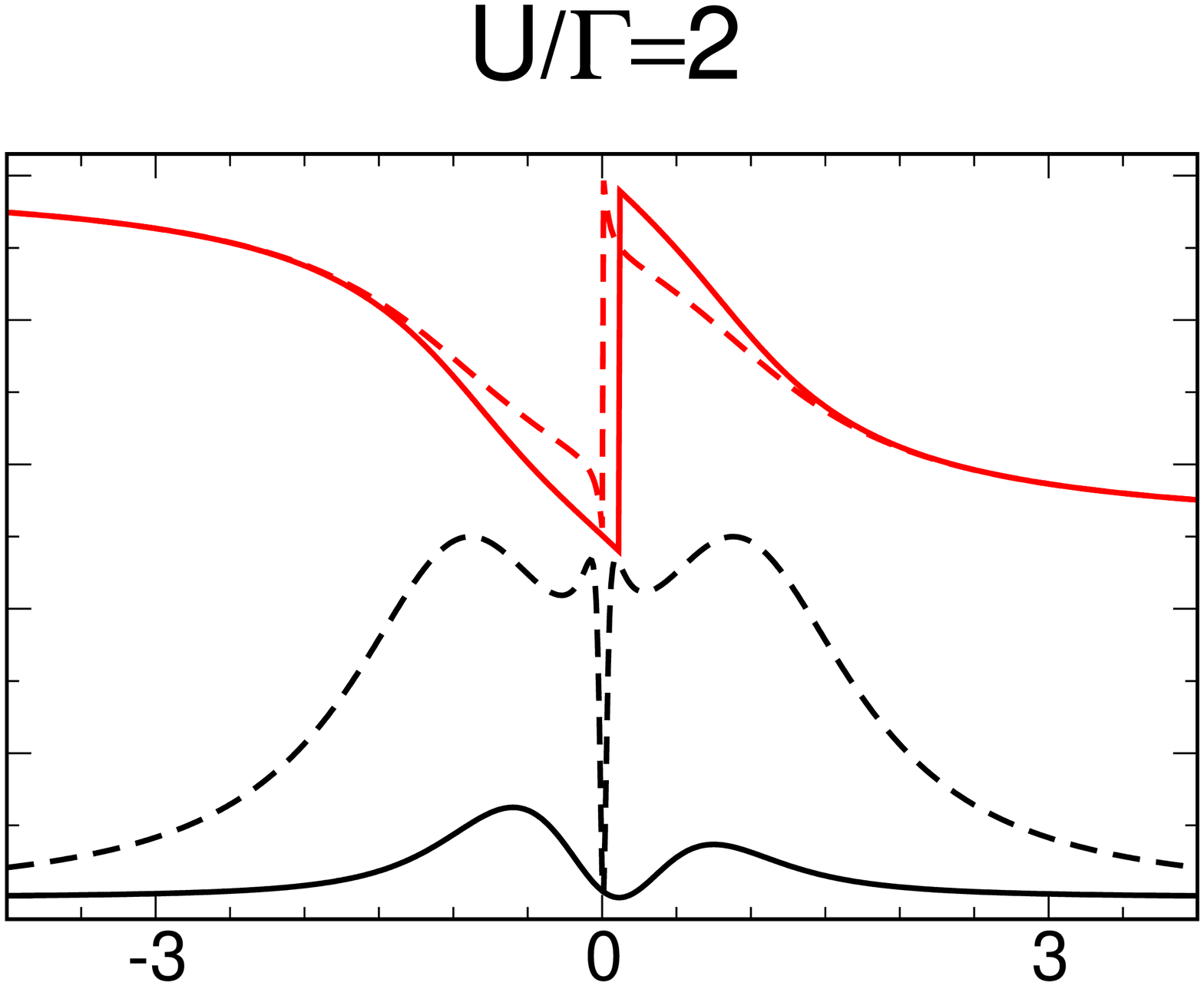}
        \includegraphics[width=0.33\textwidth,height=3.75cm,clip]{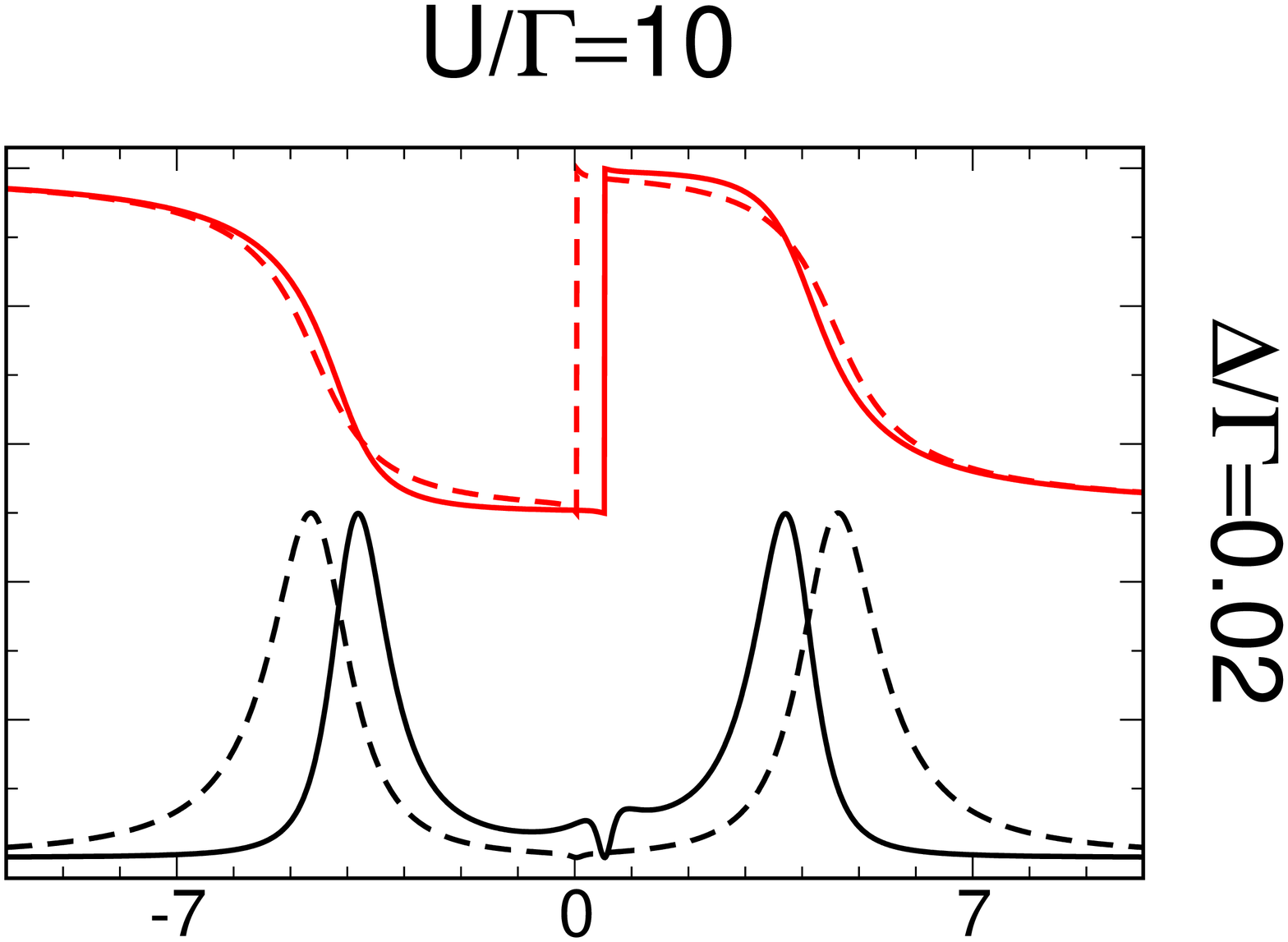}\vspace{0.3cm}
	      \includegraphics[width=0.352\textwidth,height=3.1cm,clip]{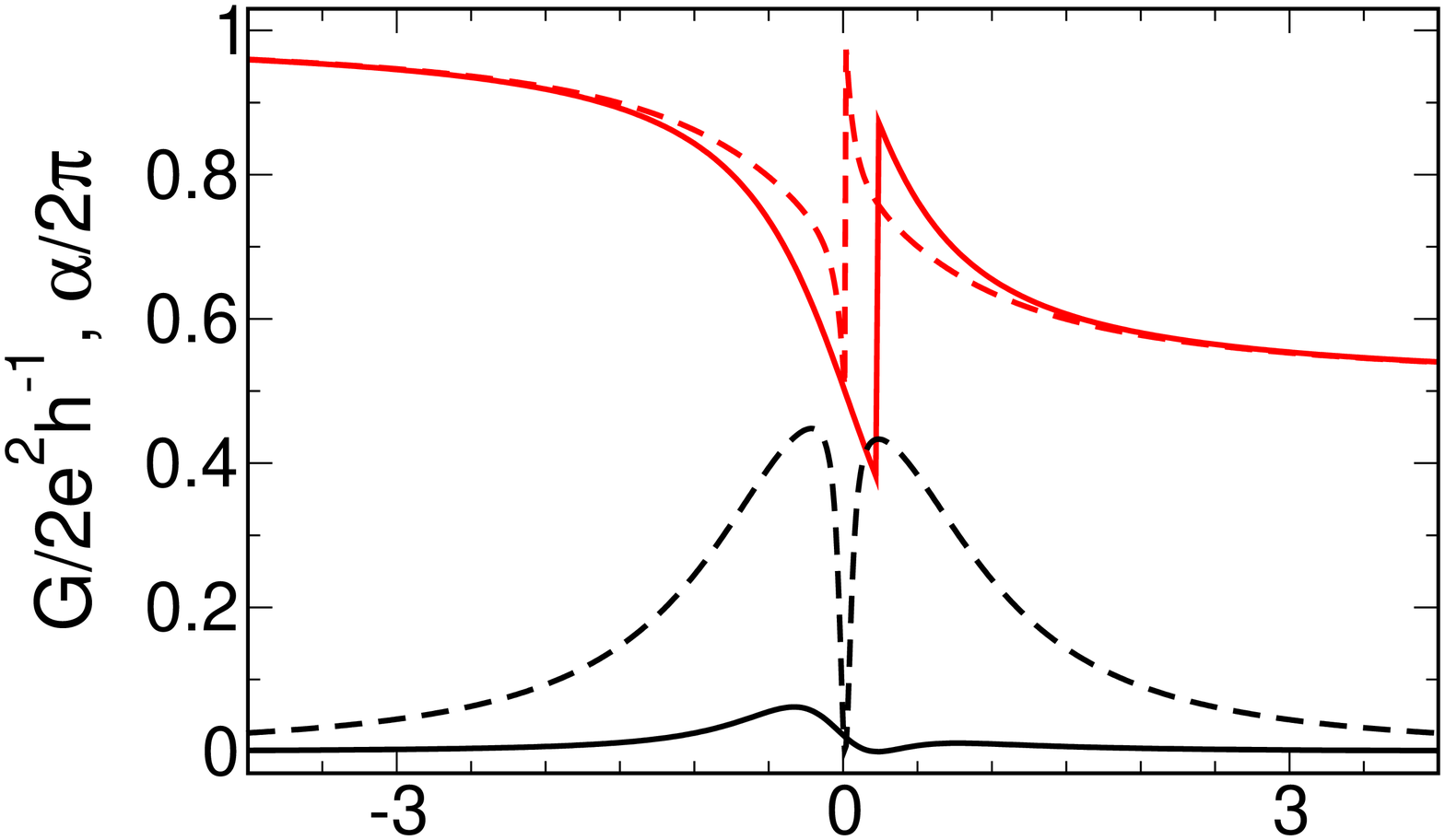}\hspace{0.001\textwidth}
	      \includegraphics[width=0.292\textwidth,height=3.1cm,clip]{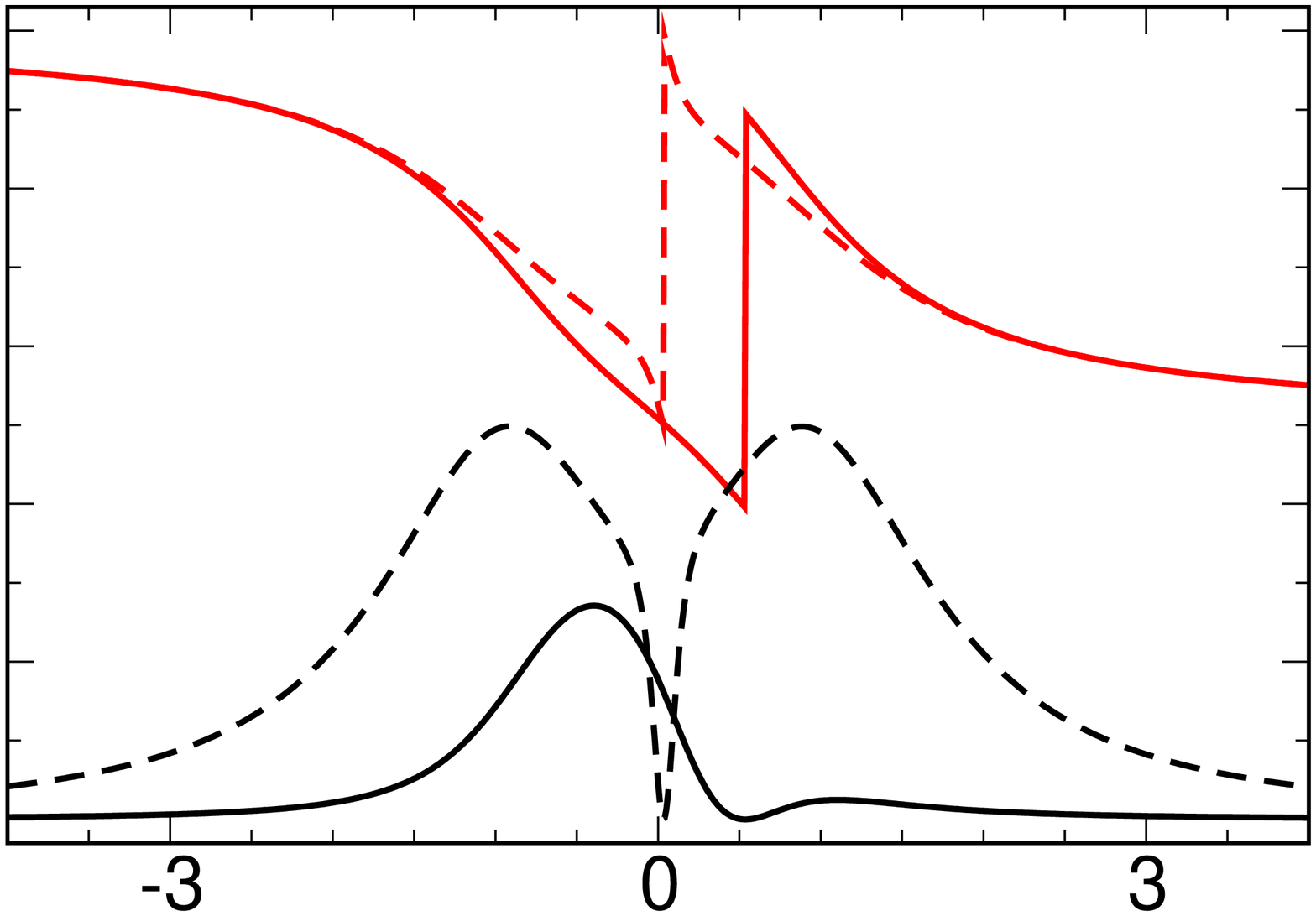}\hspace{0.001\textwidth}
        \includegraphics[width=0.33\textwidth,height=3.1cm,clip]{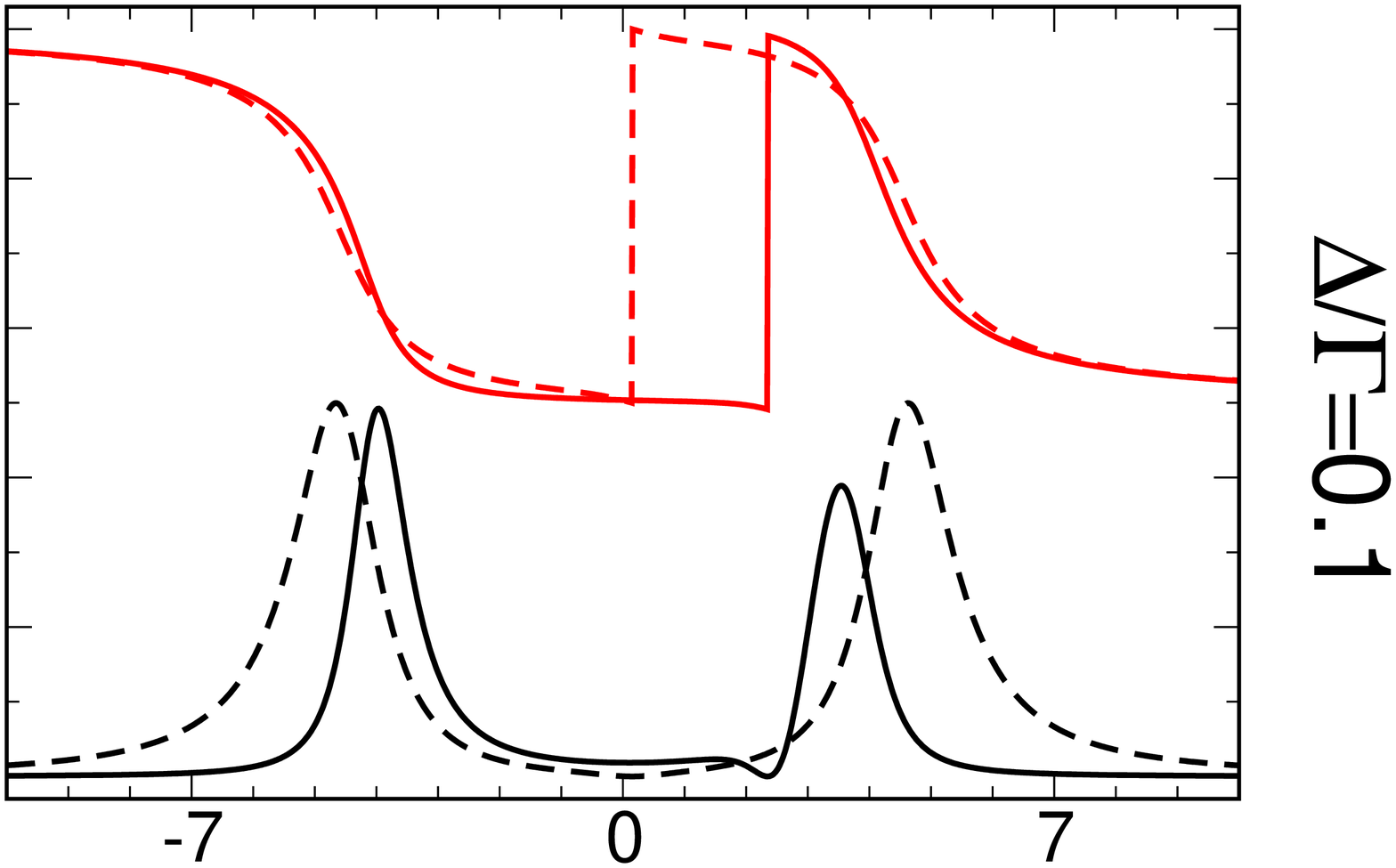}\vspace{0.3cm}
	      \includegraphics[width=0.352\textwidth,height=3.1cm,clip]{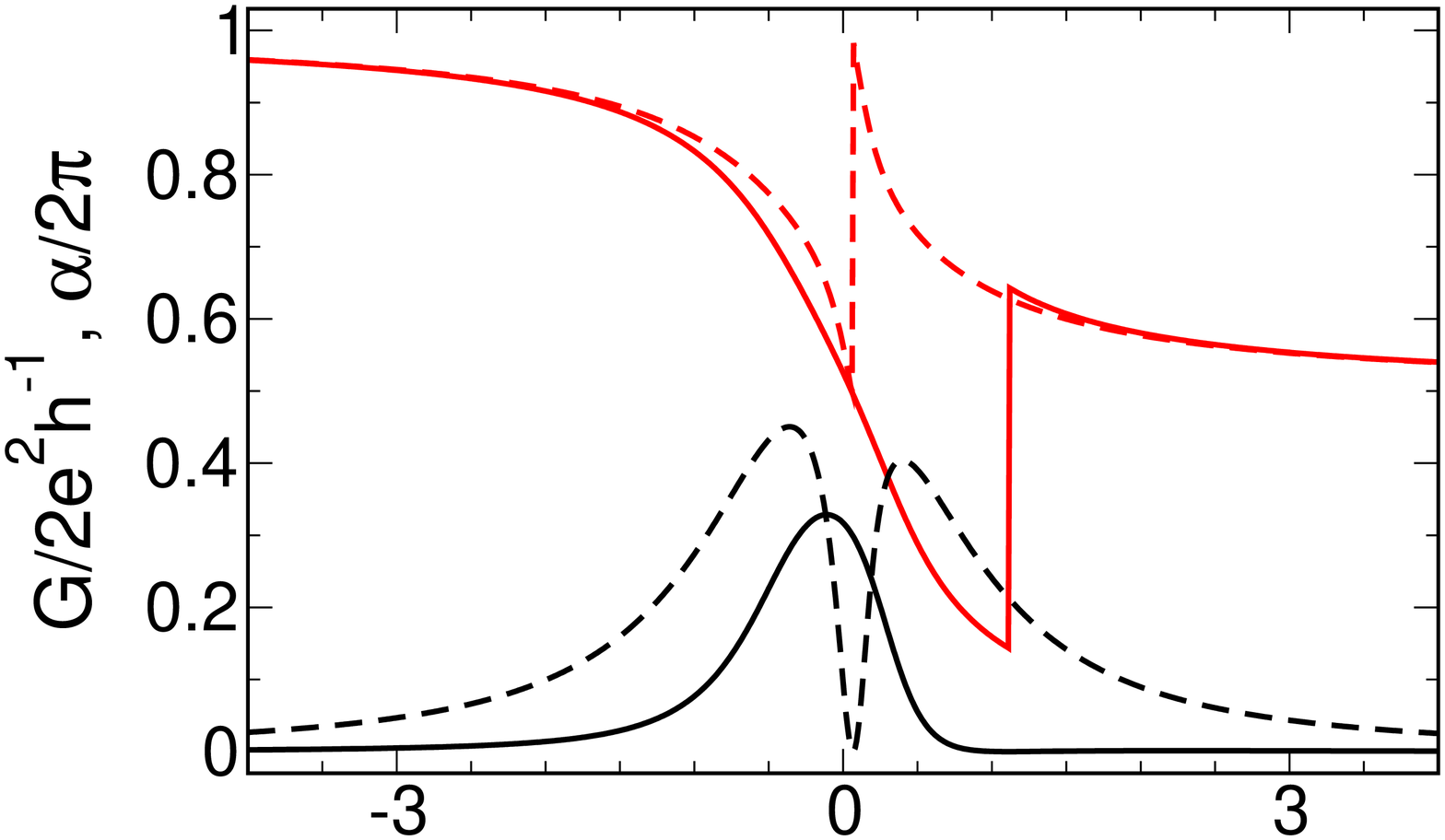}\hspace{0.001\textwidth}
	      \includegraphics[width=0.292\textwidth,height=3.1cm,clip]{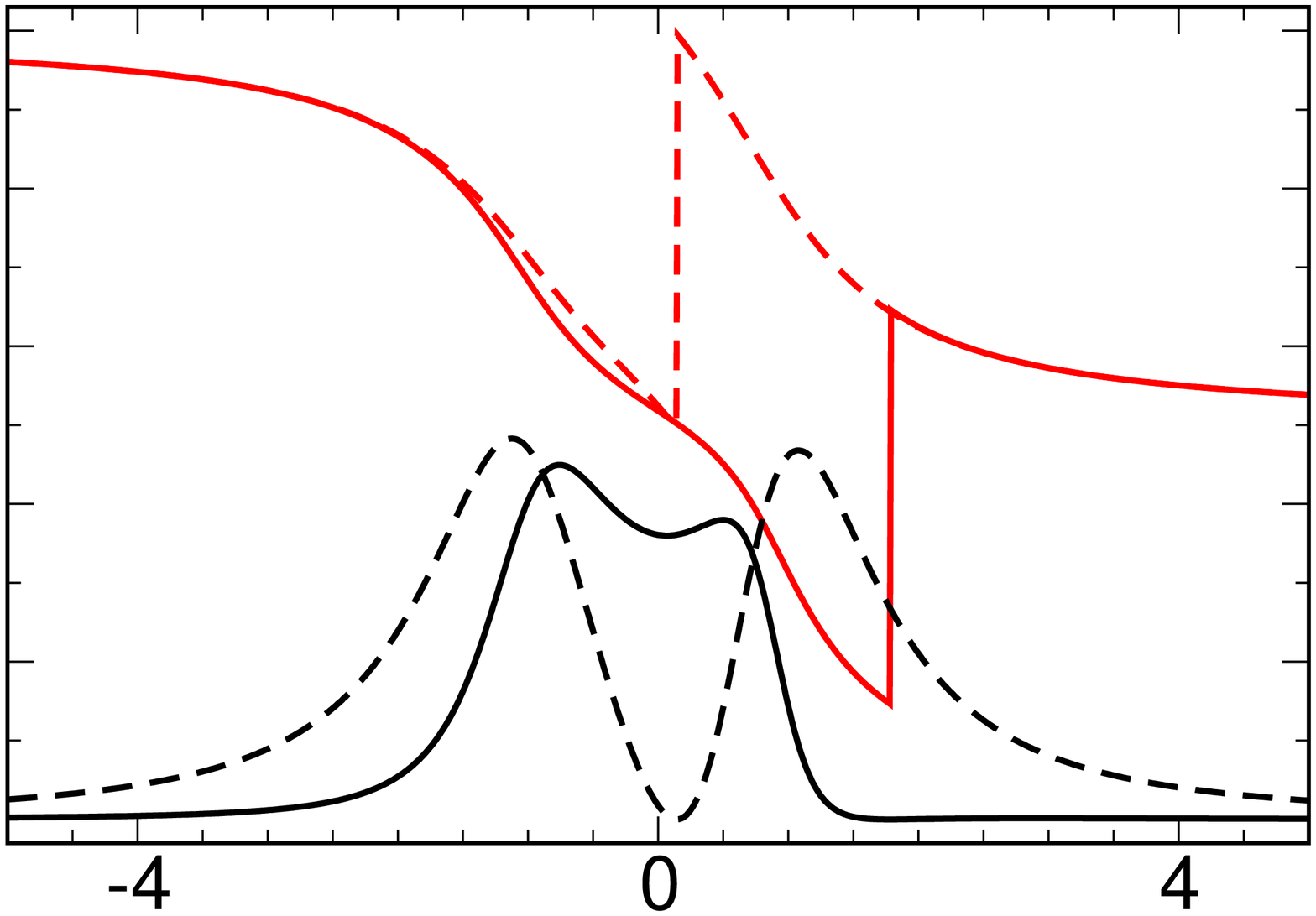}\hspace{0.001\textwidth}
        \includegraphics[width=0.333\textwidth,height=3.1cm,clip]{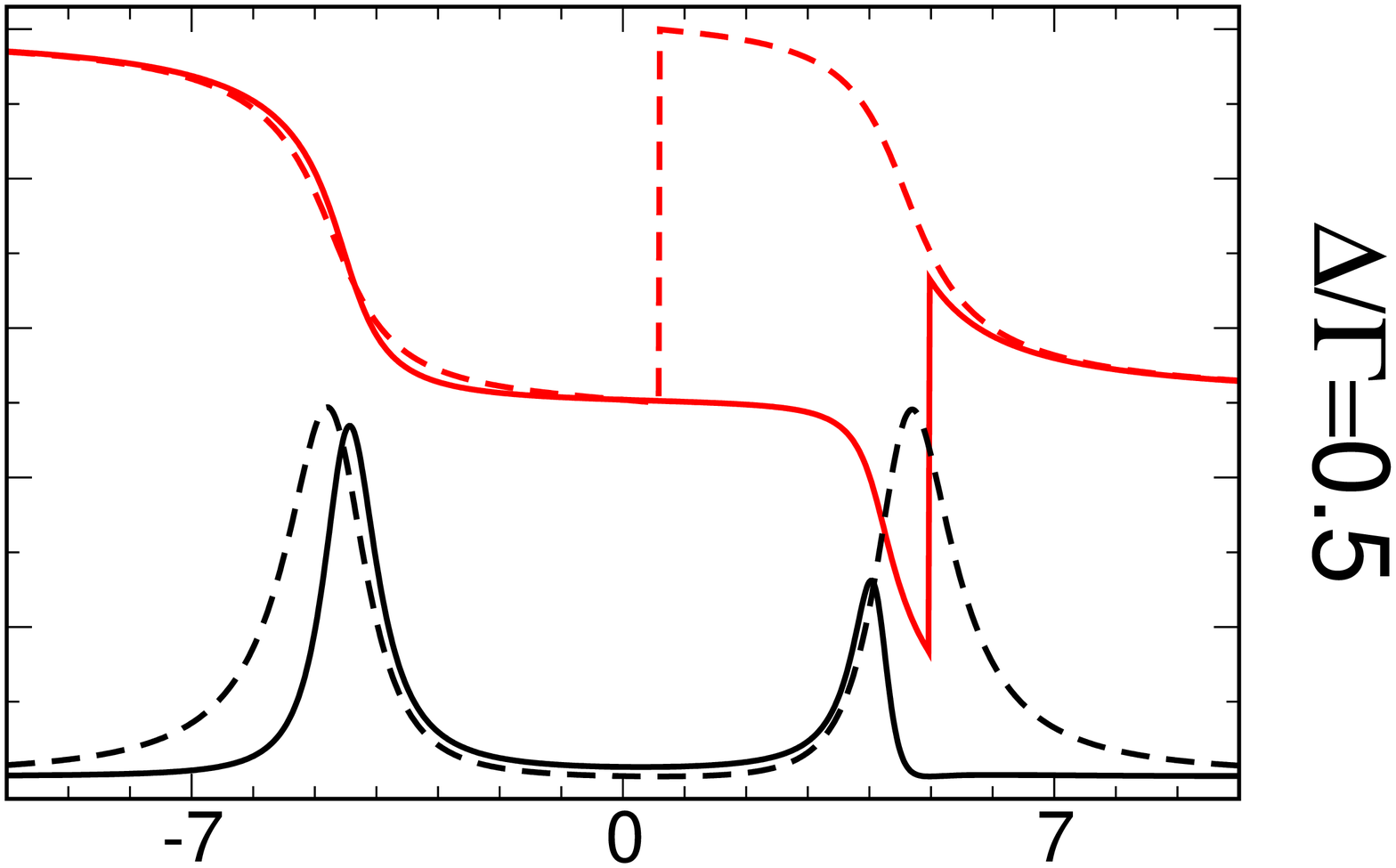}\vspace{0.3cm}
	      \includegraphics[width=0.352\textwidth,height=3.1cm,clip]{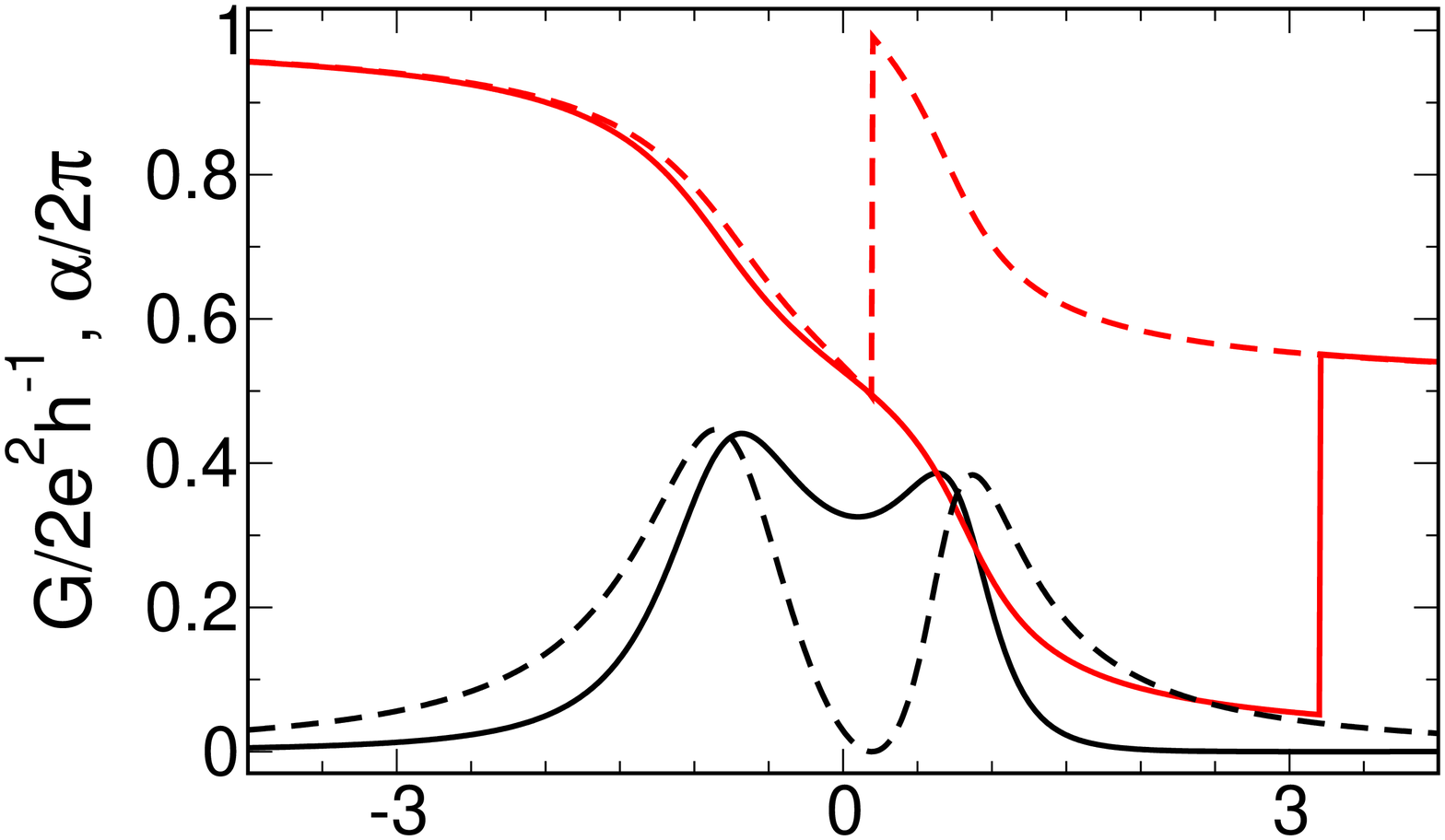}\hspace{0.001\textwidth}
	      \includegraphics[width=0.292\textwidth,height=3.1cm,clip]{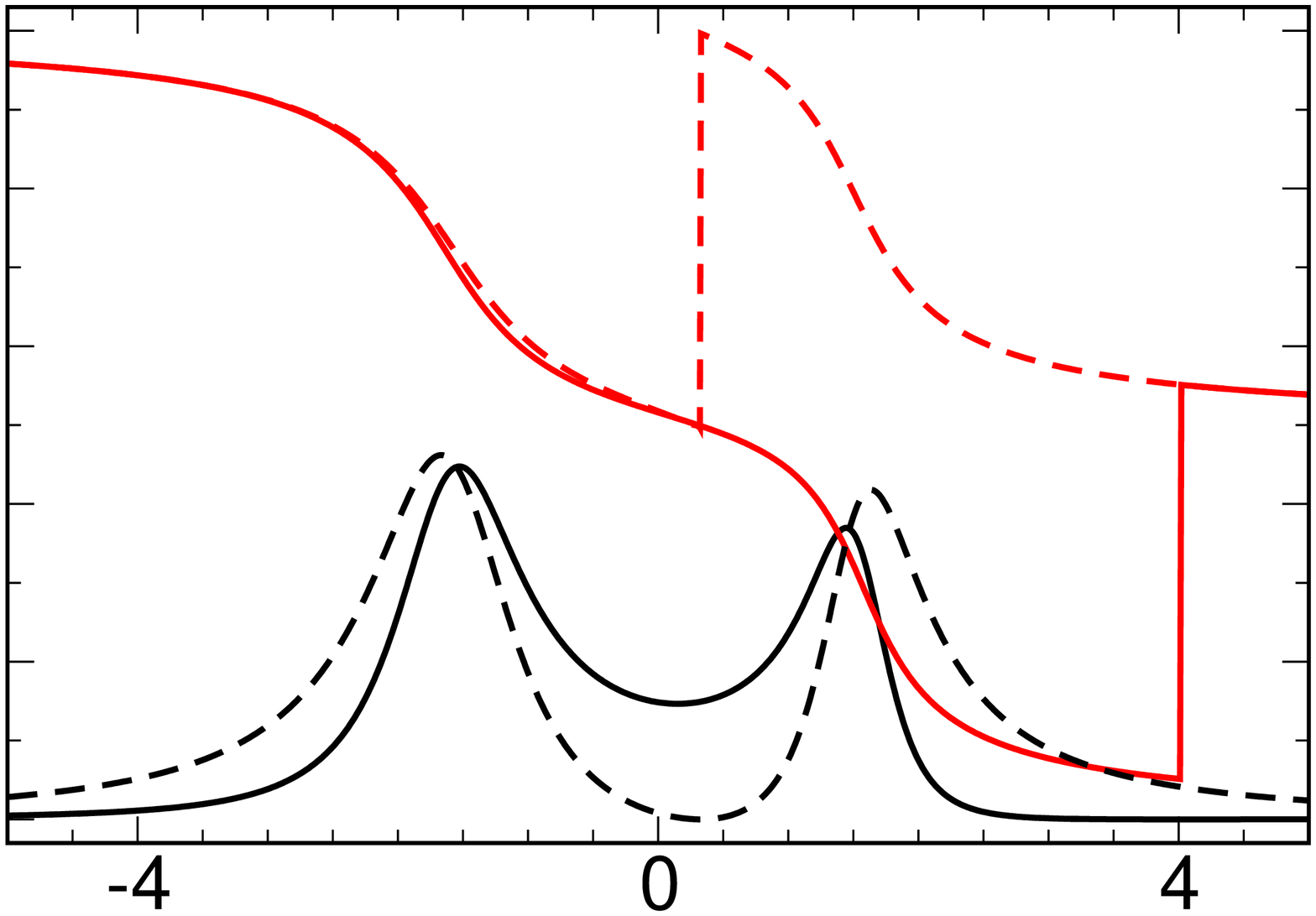}\hspace{0.001\textwidth}
        \includegraphics[width=0.33\textwidth,height=3.1cm,clip]{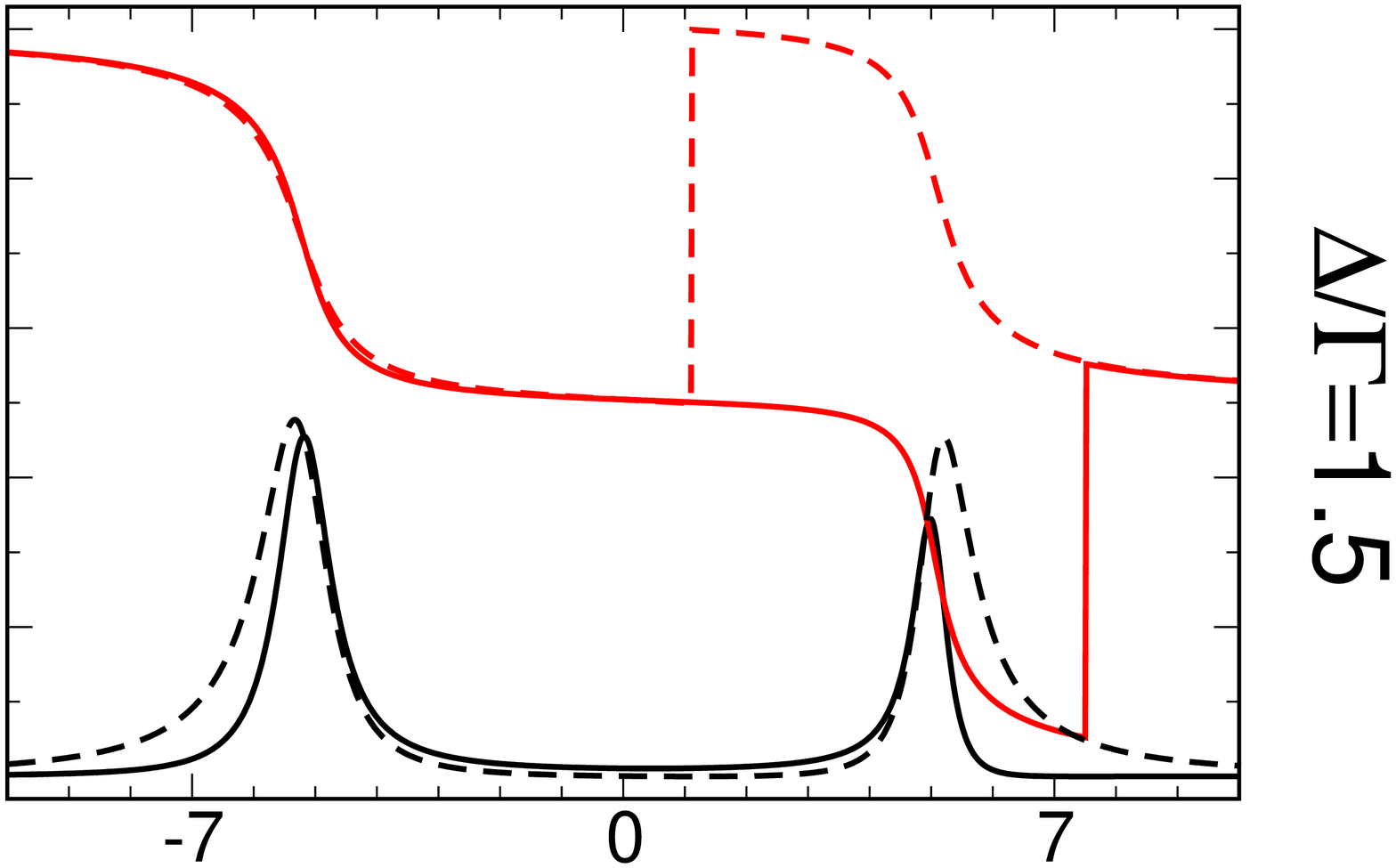}\vspace{0.3cm}
	      \includegraphics[width=0.352\textwidth,height=3.65cm,clip]{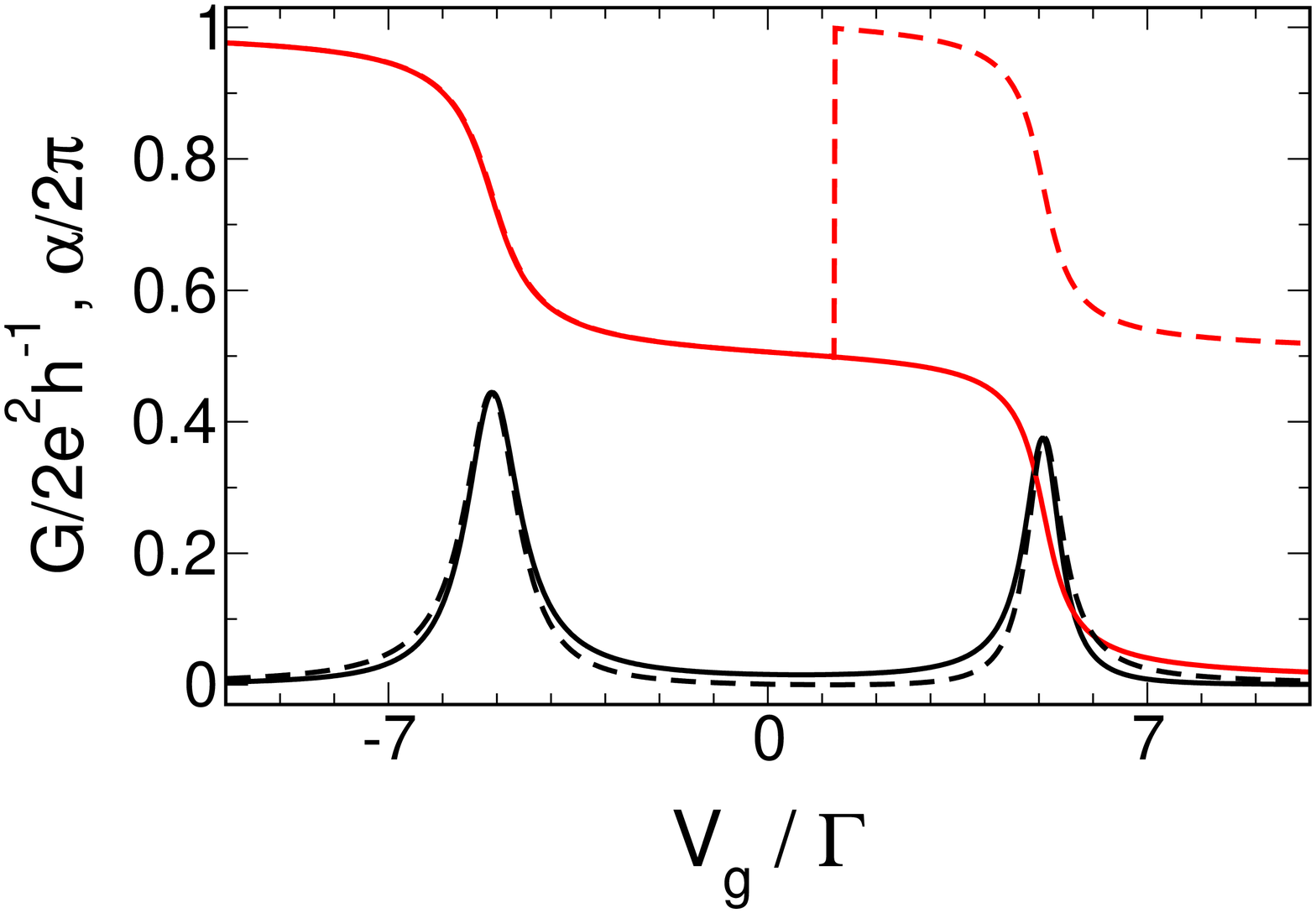}\hspace{0.001\textwidth}
	      \includegraphics[width=0.292\textwidth,height=3.65cm,clip]{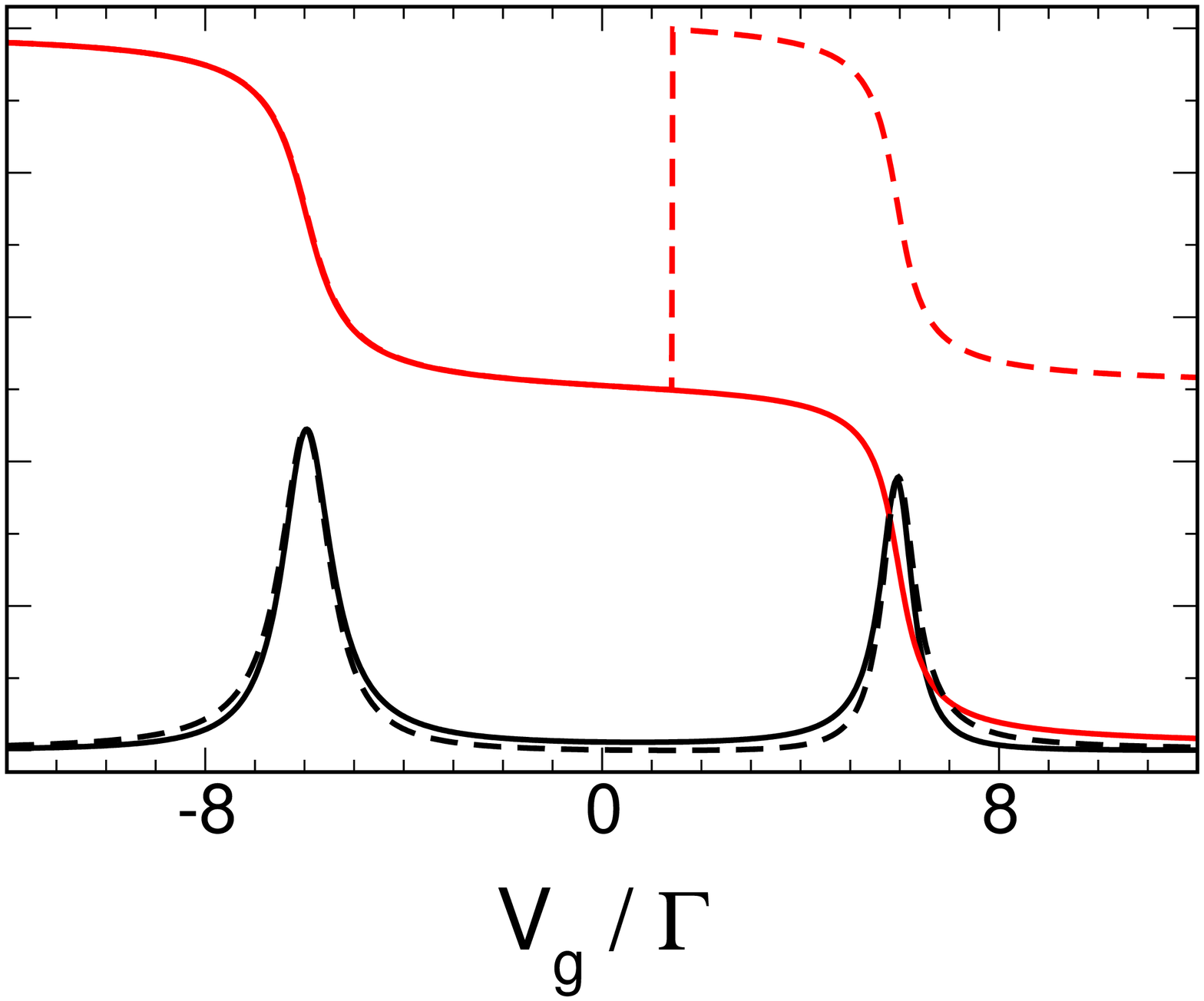}\hspace{0.001\textwidth}
        \includegraphics[width=0.33\textwidth,height=3.65cm,clip]{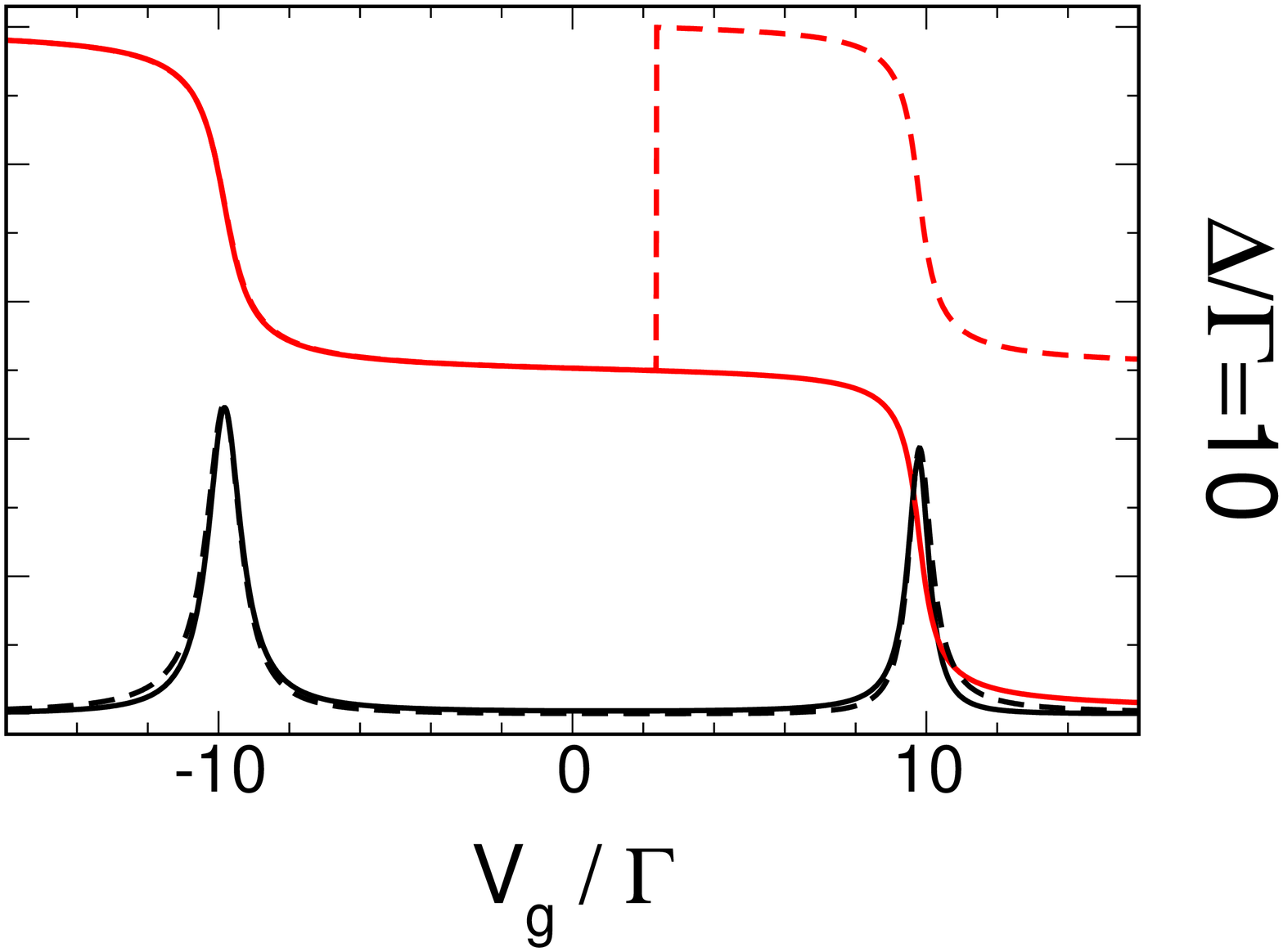}
        \caption{Systematic account for the three different energy scales (nearest-neighbour interaction $U$, level spacing $\Delta$, and hybridisation strength $\Gamma$) that govern the conductance $G$ (black) and transmission phase $\alpha$ (red) of a spinless two-level dot. The relative sign of the level-lead couplings is $s=-$ for the solid and $s=+$ for the dashed curves, and the hybridisations read $\Gamma=\{0.1~0.3~0.4~0.2\}$. Note that the depicted behaviour is the generic one and in particular qualitatively independent of the actual choice of $\Gamma=\{\ldots\}$. The most fundamental points of this figure are the following. For nearly degenerate levels, we observe two transmission resonances (Coulomb blockade peaks) with width of order $\Gamma$ separated by $U$. They are accompanied by additional correlation induced resonances for appropriate choices of $U$ and $\Delta\ll\Gamma$. The transmission phase shows an S-like change of $\pi$ over each peak and jumps by $\pi$ at the transmission zero located close to $V_g=0$. Increasing the level spacing, the behaviour of the curves begins to depend on $s$. In the limit $\Delta\gg\Gamma$, we find two resonances of width $\Gamma_l$ separated by $U+\Delta$. $\alpha$ drops by $\pi$ over each of them and jumps by $\pi$ at the zero which is still in between for $s=+$, while evolving continuously for $s=-$.}
\label{fig:OS.dd.sum}
\end{figure}
\afterpage{\clearpage}

For large level spacings (Fig.~\ref{fig:OS.dd.efflevel}, right panel) it shows that the renormalized level positions are not only dominant (as for nearly degenerate levels) but at least two orders of magnitude larger than the hoppings generated by the flow, and therefore the former are the eigenenergies of the effective dot system with unchanged $t_l^s$ and $s$. At each Coulomb blockade peak, one of them (say $\omega_A$) crosses the chemical potential. For all gate voltages except those close to the other Coulomb peak $\omega_A$ depends linearly on the gate voltage ($\omega_A=V_g+c$). This explains why the form of the resonances and the behaviour of the transmission phase are precisely the same as in the noninteracting case. In particular, the height and width of the peaks are determined by the parameters $\Gamma_l^s$ of the corresponding level through which transport takes place, and a phase jump is observed in between for $s=-$. At the other Coulomb peak (when $\omega_B$ crosses the chemical potential and the other dot is depleted), the energy $\omega_A$ is shifted by $U$ due to the interaction of both electrons, so that in the limit $V_g\to\mp\infty$ we recover $\omega_i=V_g\pm (U+\Delta)/2$.

A word of warning about the interpretation of the system in an effective single-particle picture is in order. Since at the end of the fRG flow we are left with a frequency-independent self-energy it is of course possible to describe transport from such a one-particle point of view, or, even simpler, to diagonalise the effective Hamiltonian to obtain new level energies $\omega_i$ with new hybridisations as a function of the gate voltage. However, it is important to note that in the first instance everything that can be extracted from this picture is that if we had a noninteraction dot system with energies $\omega_i$ and corresponding level-lead couplings $\tilde t_i^s$, then transport through such a system would coincide with transport through the effective interacting one and therefore allow for a more or less easy interpretation.\footnote{One should note, however, that even for $U=0$ and large $\Delta$ no simple explanation (besides the one buried in the exact expression for the conductance) of why $s$ governs the occurrence of a phase jump between the transmission resonances exists.} But what we a priori do not know is \textit{why} the interaction renormalizes the levels the way it does. The effective picture may therefore in some cases facilitate the interpretation of the results (as in the case of large level spacing, where the effective energies depend linearly on $V_g$ and are only shifted by $U$ if one level gets filled, whereas the new hybridisations coincide with the original ones and are in particular independent of $V_g$) while it proves to be of limited validity in others. This is most strikingly shown at nearly degenerate levels where the conductance exhibits Coulomb blockade peaks of width $\Gamma$. But since the hybridisations of the new levels basically coincide with the original ones (because the hopping generated by the interaction is small compared to renormalized level energy), it is impossible to interpret them as resonances arising from transport through separate levels. To put it the other way round, those peaks have to be caused by several overlapping effective levels, and it is of course far from evident why the interaction produces such a situation.

\subsection{Parallel Triple Dots}\label{sec:OS.td}

\subsubsection{General Considerations}

Next, we consider the parallel geometry that consists of three dots. The free propagator of this system can in complete analogy to the two-level case be cast in the form
\begin{equation}\label{eq:OS.td.g}
\left[\mc G^0(i\omega)\right]^{-1} =  
\begin{pmatrix}
i\omega-V_g+\Delta & & \\ & i\omega-V_g & \\ & & i\omega-V_g-\Delta
\end{pmatrix} - H_\tn{eff}+H_{PP},
\end{equation}
with the last term being the contribution from the projected leads,
\begin{equation*}\begin{split}
&-\frac{H_\tn{eff}-H_{PP}}{i~\tn{sgn}(\omega)}=\\[0.5ex]
&\hspace{0.5cm}\begin{pmatrix}
\Gamma_A &
\sqrt{\Gamma_A^L\Gamma_B^L}+s_1\sqrt{\Gamma_A^R\Gamma_B^R} &
\sqrt{\Gamma_A^L\Gamma_C^L}+s_2\sqrt{\Gamma_A^R\Gamma_C^R} \\
\sqrt{\Gamma_A^L\Gamma_B^L}+s_1\sqrt{\Gamma_A^R\Gamma_B^R} &
\Gamma_B &
\sqrt{\Gamma_B^L\Gamma_C^L}+s_1s_2\sqrt{\Gamma_B^R\Gamma_C^R} \\
\sqrt{\Gamma_A^L\Gamma_C^L}+s_2\sqrt{\Gamma_A^R\Gamma_C^R} &
\sqrt{\Gamma_B^L\Gamma_C^L}+s_1s_2\sqrt{\Gamma_B^R\Gamma_C^R} &
\Gamma_C
\end{pmatrix}.
\end{split}\end{equation*}
As usual, $V_g$ is a variable gate voltage that simultaneously shifts the on-site energies of the dots, $\Delta$ denotes a level spacing identical for all levels, and $s_1=\tn{sgn}(t_A^Lt_A^Rt_B^Lt_B^R)$ and $s_2=\tn{sgn}(t_B^Lt_B^Rt_C^Lt_C^R)$ the relative signs of the couplings, and $\Gamma_l^s$ the energy-independent hybridisations with the leads in the wide-band limit. As before, the relative signs are assumed to be realised such that $t_B^R=s_1$, $t_C^R=s_2$, and $t_l^s>0$ otherwise. Furthermore, we introduce the shorthand notation $s=\{s_1s_2\}$. The general expression for the zero temperature conductance (\ref{eq:DOT.leitwert}) then reads
\begin{equation}\label{eq:OS.td.cond}
G = \frac{e^2}{h}\Big|2\pi\rho_{\tn{lead}}(0)\sum_{l,l'}s_{l,l'}\sqrt{\Gamma_l^R\Gamma_{l'}^L}\mc G_{l;l'}(0)\Big|^2,
\end{equation}
with
\begin{equation*}
s_{l,l'}:=
\begin{cases}
s_1 & \tn{for } l=B \tn{ and arbitrary } l' \\
s_2 & \tn{for } l=C \tn{ and arbitrary } l' \\
1 & \tn{otherwise}.
\end{cases}
\end{equation*}
The full propagator in presence of interactions between all electrons,
\begin{equation*}
U:=\bar v_{A,B,A,B}=\bar v_{A,C,A,C}=\bar v_{B,C,B,C},	
\end{equation*}
is obtained by solving the flow equations for the self-energy and the effective interaction (\ref{eq:FRG.flowse3}, \ref{eq:FRG.flowww3}). Alltogether, they comprise twelve equations, six for the effective on-site energies $V_l(\la)$ and inter-dot hoppings $t_{l,l'}(\la)$ (which are in general of course generated by the flow), and six for all the independent components of the two-particle vertex, $\gamma_2(i,j;k,l;\la)$ with $(ijkl)=\{ABAB, ABAC, ABBC, ACAC, ACBC, BCBC\}$.

It is important to note that for otherwise arbitrary parameters relaxing the assumption of equal interactions between all electrons and introducing asymmetric level spacings only leads to quantitative changes but does not influence our results qualitatively.

In contrast to the previously studied double dot geometry, we will refrain from first describing the noninteracting case here. This is because for small level spacings $\Delta\ll\Gamma$ the $U=0$ case is qualitatively very different from that with $U>0$ and hence no additional insight into the interacting problem can be gained from considering the noninteracting case. Therefore we will delay the discussion of the latter until the end of this section. As before, we will focus on the generic behaviour of our system. In particular, we will ignore all choices of the hybridisations (and call them non-generic) where the conductance and transmission phase as a function of $V_g$, $\Delta$, $U$ and $s$ deviate from the behaviour observed for general $\Gamma_l^s$ if the dimension of the manifold of these hybridisations is smaller than six. By computing approximately 6500 data sets in the $(U,\Delta,s,\Gamma)$ -- parameter space ($\sim 5\cdot10^7$ data points $G(V_g)$) we have verified that the behaviour presented in this section is indeed the generic one.

\subsubsection{Interacting Case, Nearly Degenerate Levels}

\begin{figure}[t]	
	\centering
	      \includegraphics[width=0.495\textwidth,height=4.4cm,clip]{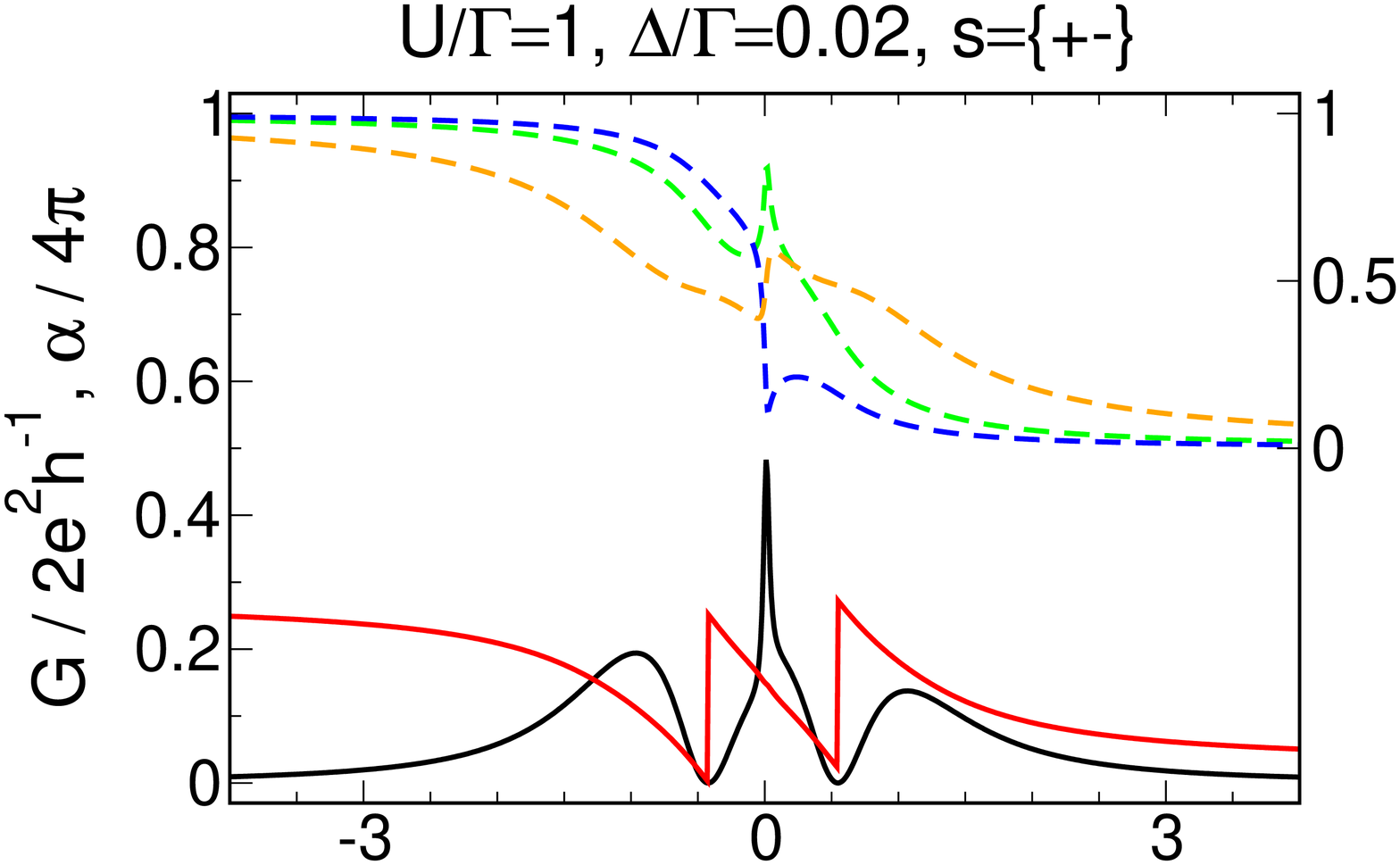}\hspace{0.015\textwidth}
        \includegraphics[width=0.475\textwidth,height=4.4cm,clip]{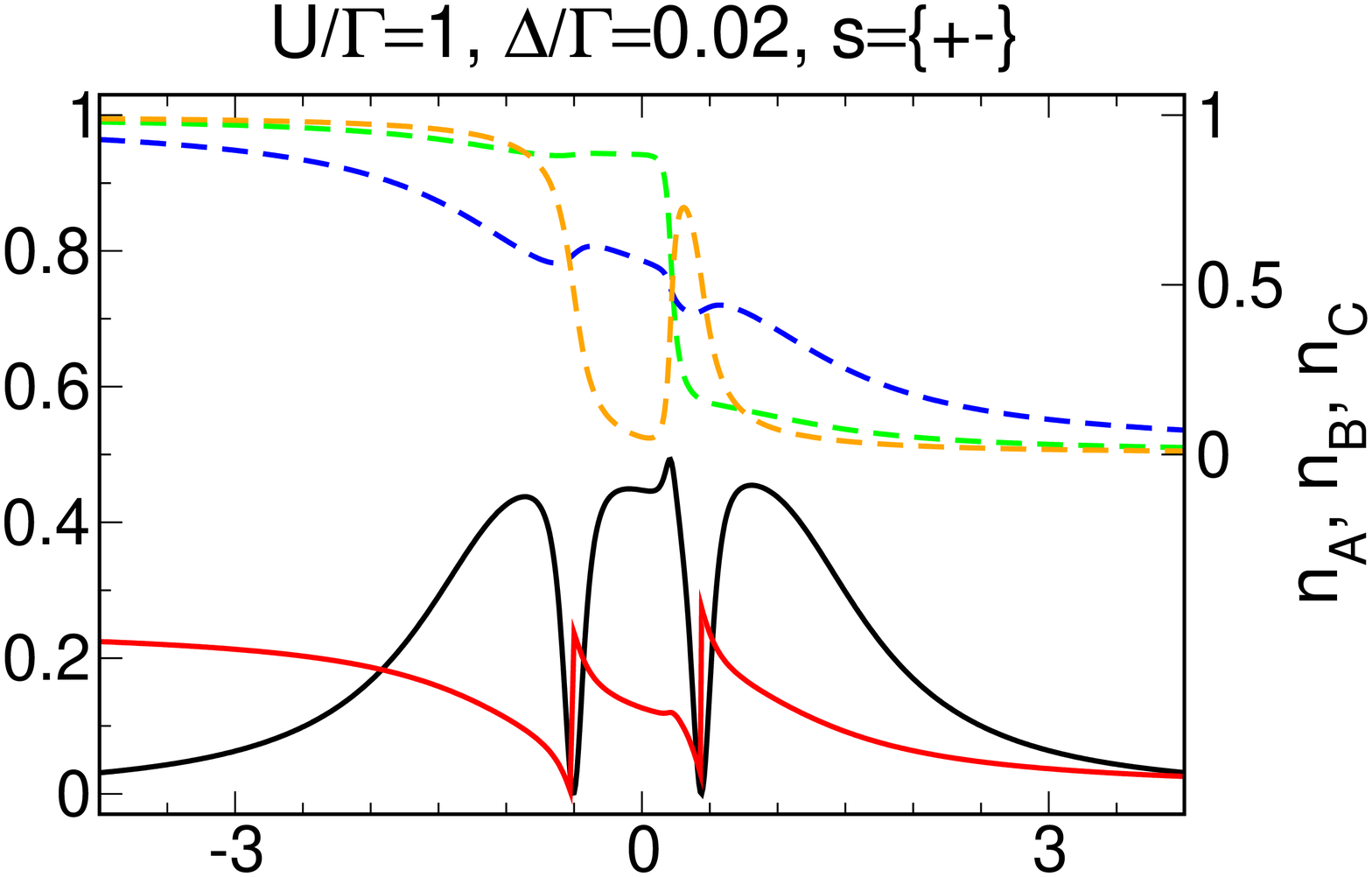}\vspace{0.3cm}
        \includegraphics[width=0.495\textwidth,height=5.2cm,clip]{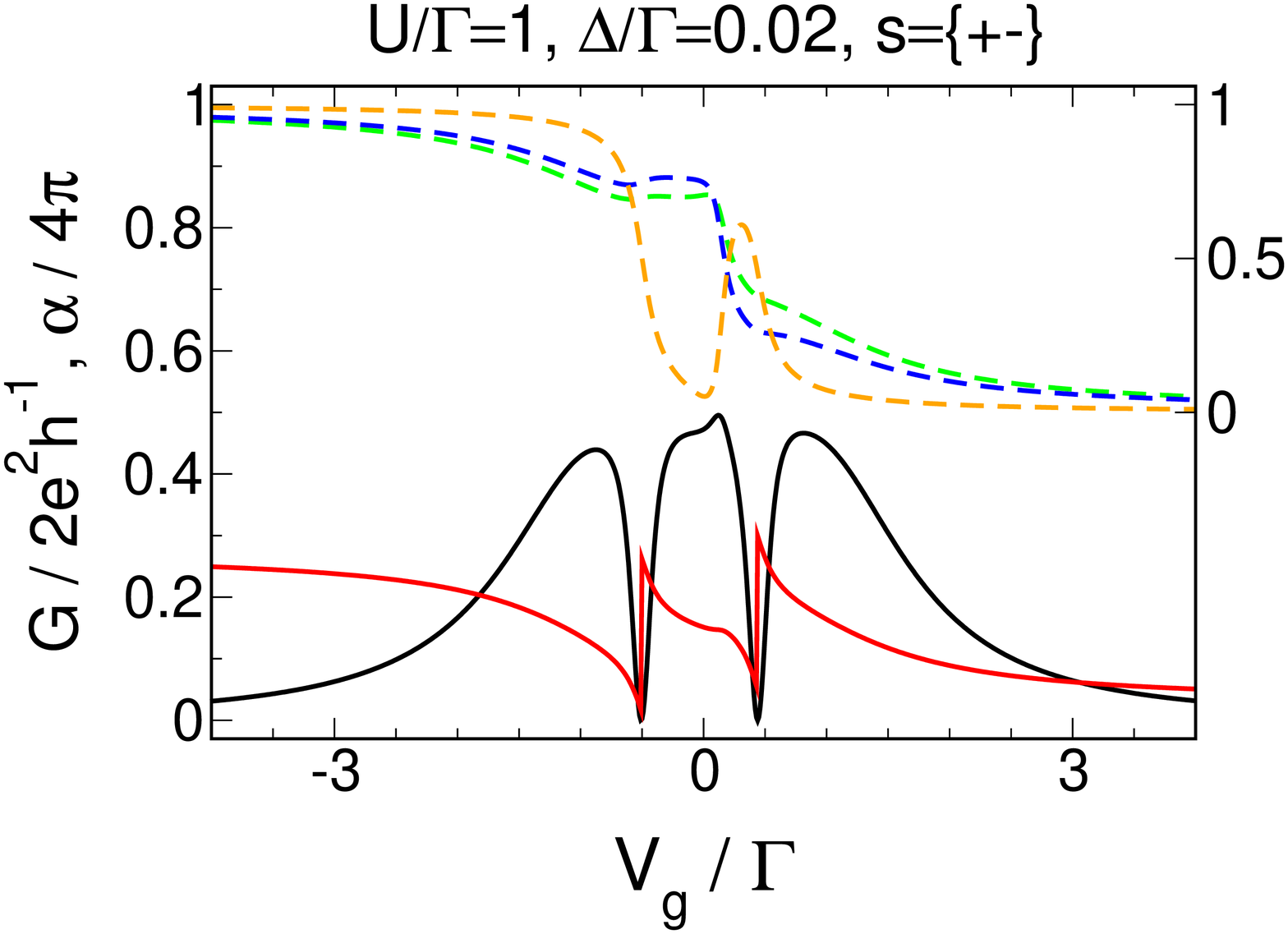}\hspace{0.015\textwidth}
        \includegraphics[width=0.475\textwidth,height=5.2cm,clip]{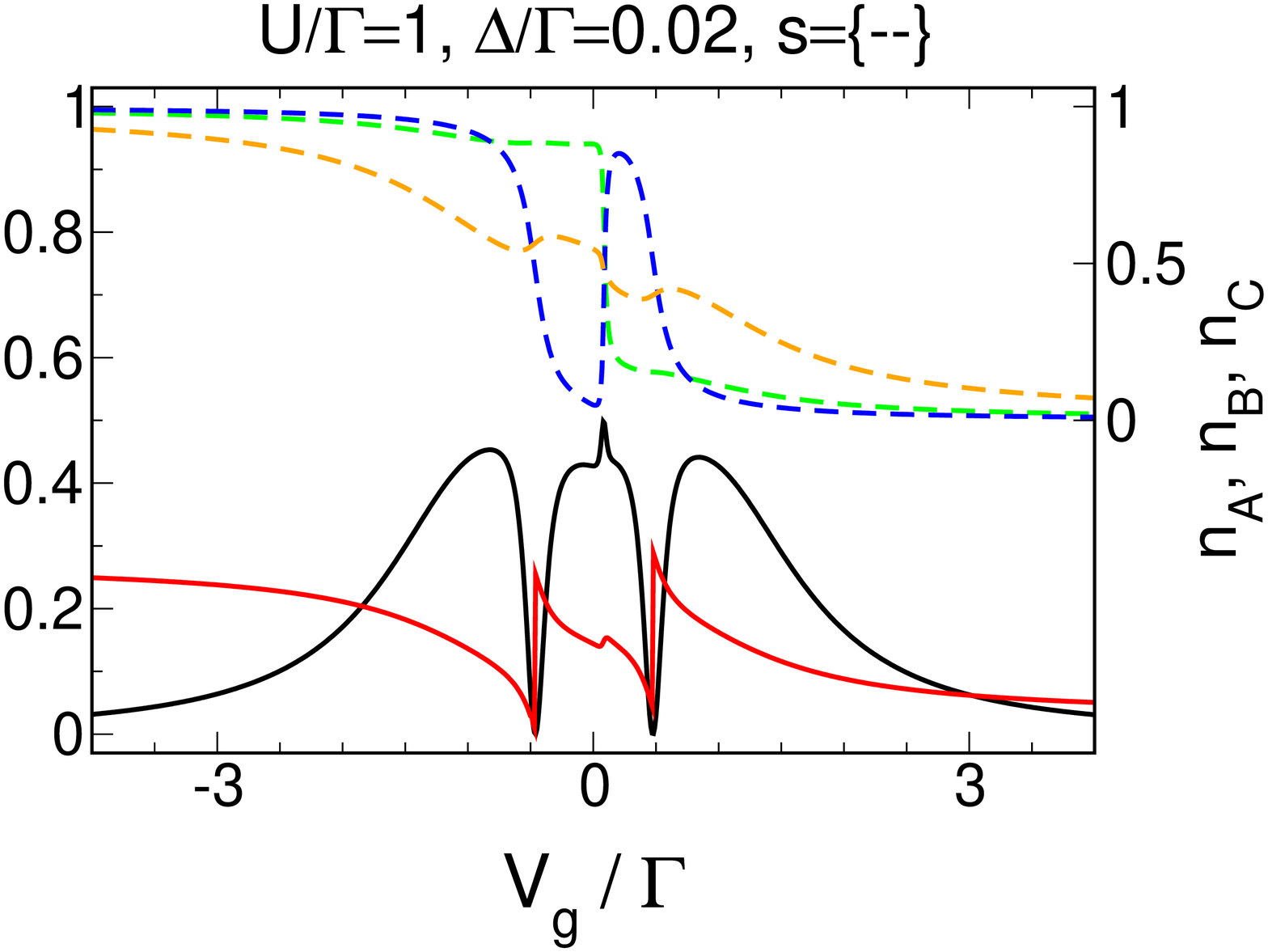}
        \caption{Conductance $G$ (black), transmission phase $\alpha$ (red), and average level occupancies (dot A: green, dot B: blue, dot C: orange) as a function of the gate voltage for parallel triple dots with nearly degenerate levels and different hybridisations $\Gamma=\{0.06~0.14~0.07~0.03~0.3~0.4\}$ (upper left and lower right), $\Gamma=\{0.06~0.14~0.3~0.4~0.07~0.03\}$ (upper right), $\Gamma=\{0.2~0.3~0.25~0.15~0.07~0.03\}$ (lower left). }
\label{fig:OS.td}
\end{figure}

For nearly degenerate levels, nonzero interactions of order $\Gamma$ and arbitrary $\{\Gamma,s\}$ we find three resonances separated by $U$ in the conductance if the gate voltage is varied. They are of almost equal height and width, the latter always being of the of the order of $\Gamma$ rather than of $\Gamma_l$ or $U$. The transmission phase $\alpha$ changes by $\pi$ over each of the peaks and jumps by $\pi$ at the transmission zeros located at $V_g\approx\pm U/2$ in between. The average level occupancies show a complicated behaviour. As for the double dot geometry, they depend nonmonotonically on the gate voltage and show population inversion, but here it is not always the most strongly coupled dot that is depleted each time $V_g$ is on-resonance. All this is shown in Fig.~\ref{fig:OS.td}.

\begin{figure}[t]	
	\centering
        \includegraphics[width=0.475\textwidth,height=5.2cm,clip]{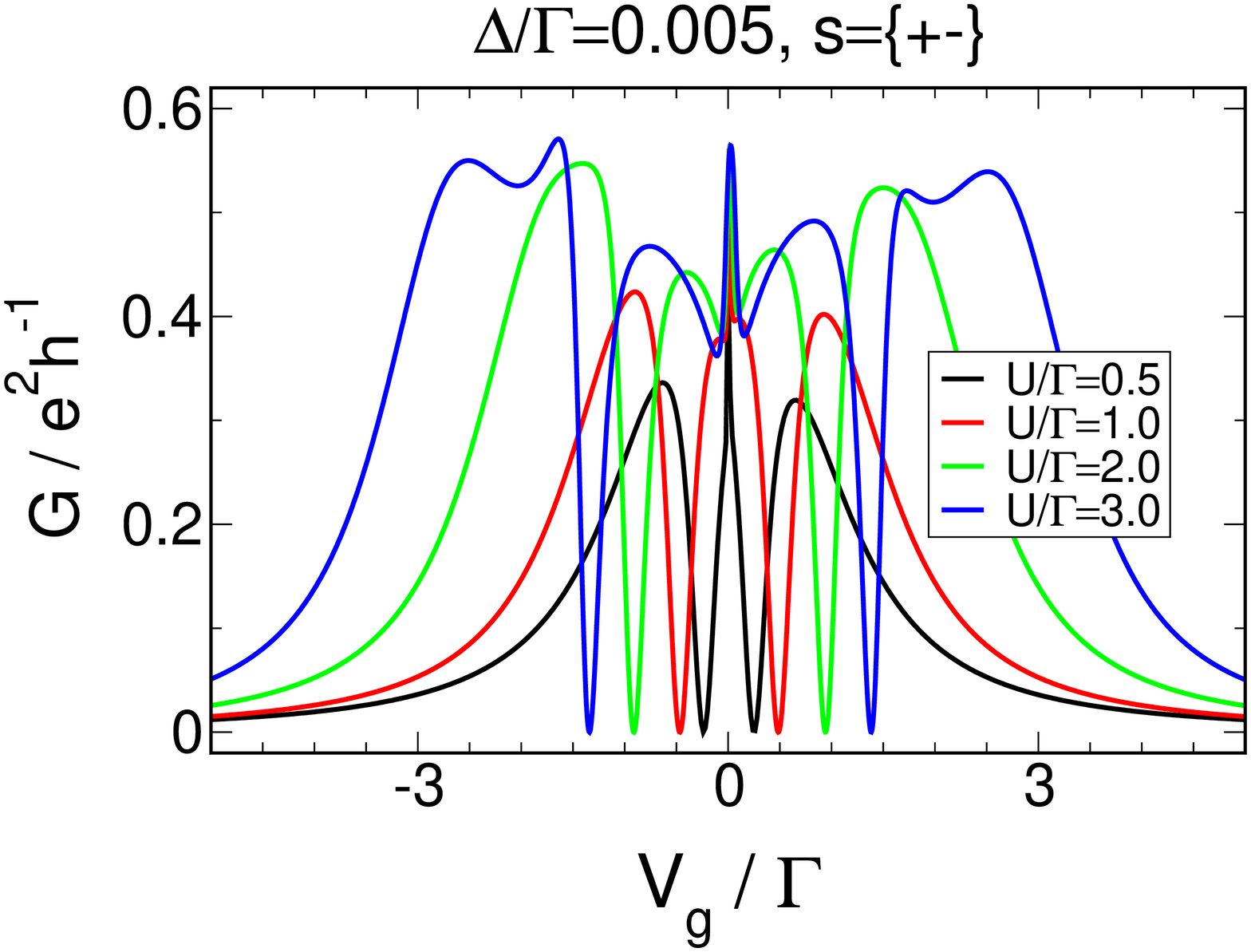}\hspace{0.035\textwidth}
        \includegraphics[width=0.475\textwidth,height=5.2cm,clip]{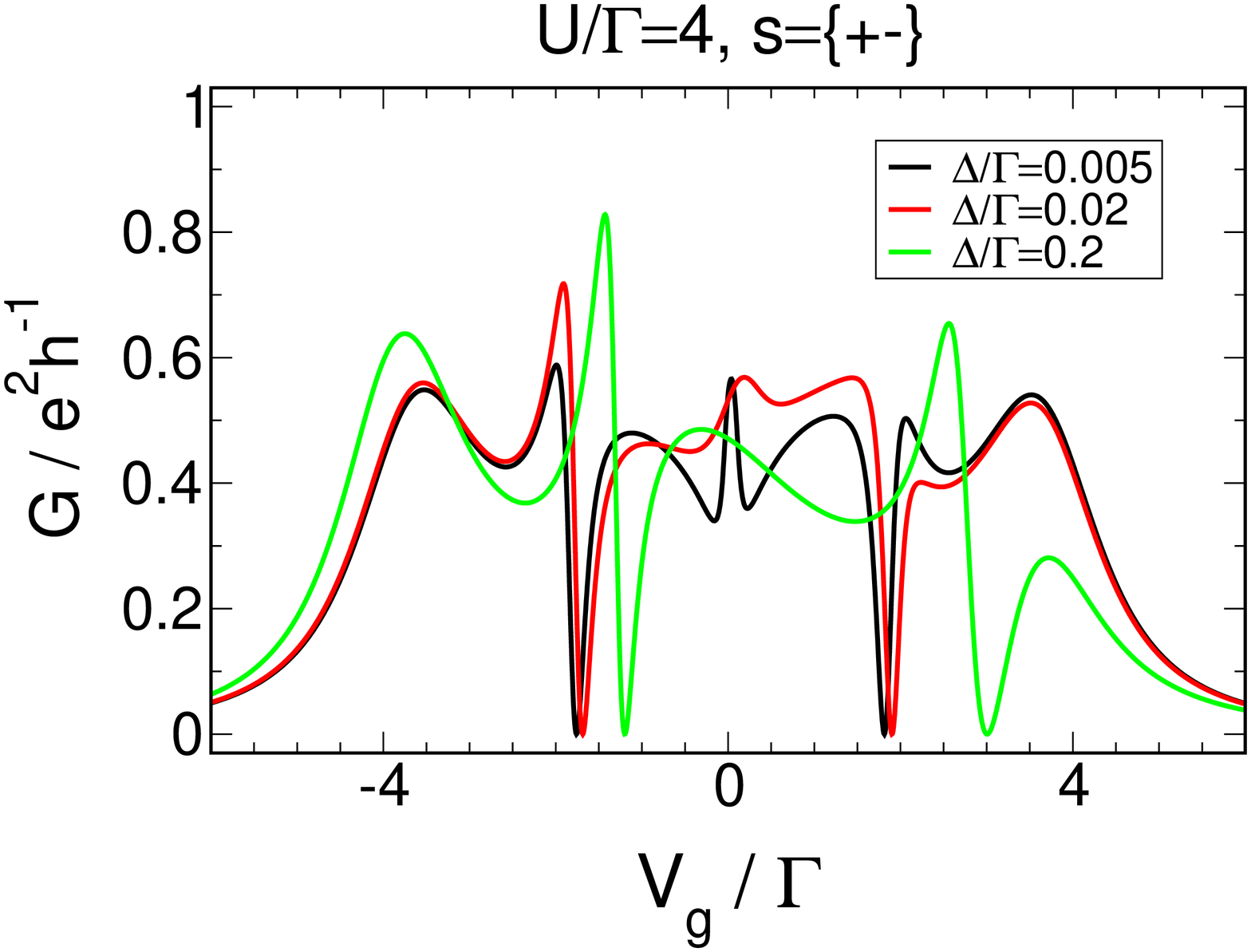}
        \caption{Appearance of additional correlation induced structures in the conductance of parallel triple dots with $\Gamma=\{0.03~0.07~0.04~0.06~0.2~0.6\}$.}
\label{fig:OS.td.u}
\end{figure}

Another similarity with the case of two dots is the gradual appearance of additional structures in the conductance if the interaction is increased (Fig.~\ref{fig:OS.td.u}). First, above a certain critical strength $U_c$ depending on the hybridisations $\Gamma_l^s$ and the relative signs $s$, the peaks located at $V_g\approx\pm U$ split up into two. For $U>U_c$ the outer of the new peaks are still located at $V_g\approx\pm U$, while the inner ones become more and more sharply centred at the transmission zeros at $V_g\approx\pm U/2$, such that one would expect only the outer peaks to remain observable in the limit of large $U$. Unfortunately, we are not able to tackle interactions large enough to confirm this within our fRG approximation scheme (see next chapter). The additional resonances vanish if one increases the level spacing above a scale $\Delta_\tn{CIR}\ll\Gamma$. This whole scenario is identical to the appearance of the correlation induced resonances for the double dot geometry.

\begin{figure}[t]	
        \centering
        \vspace{-0.25cm}\includegraphics[width=0.495\textwidth,height=4.4cm,clip]{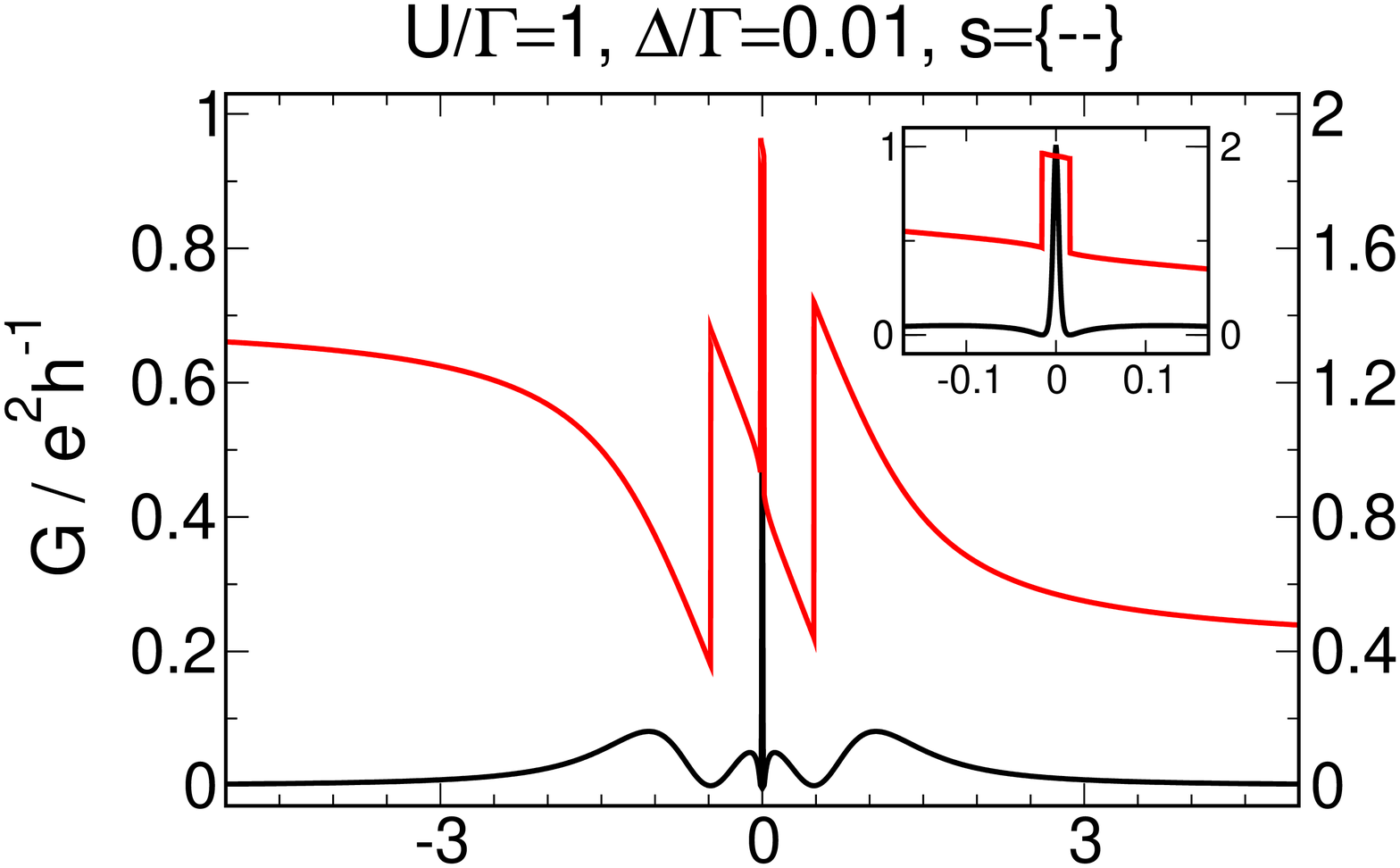}\hspace{0.015\textwidth}
        \includegraphics[width=0.475\textwidth,height=4.4cm,clip]{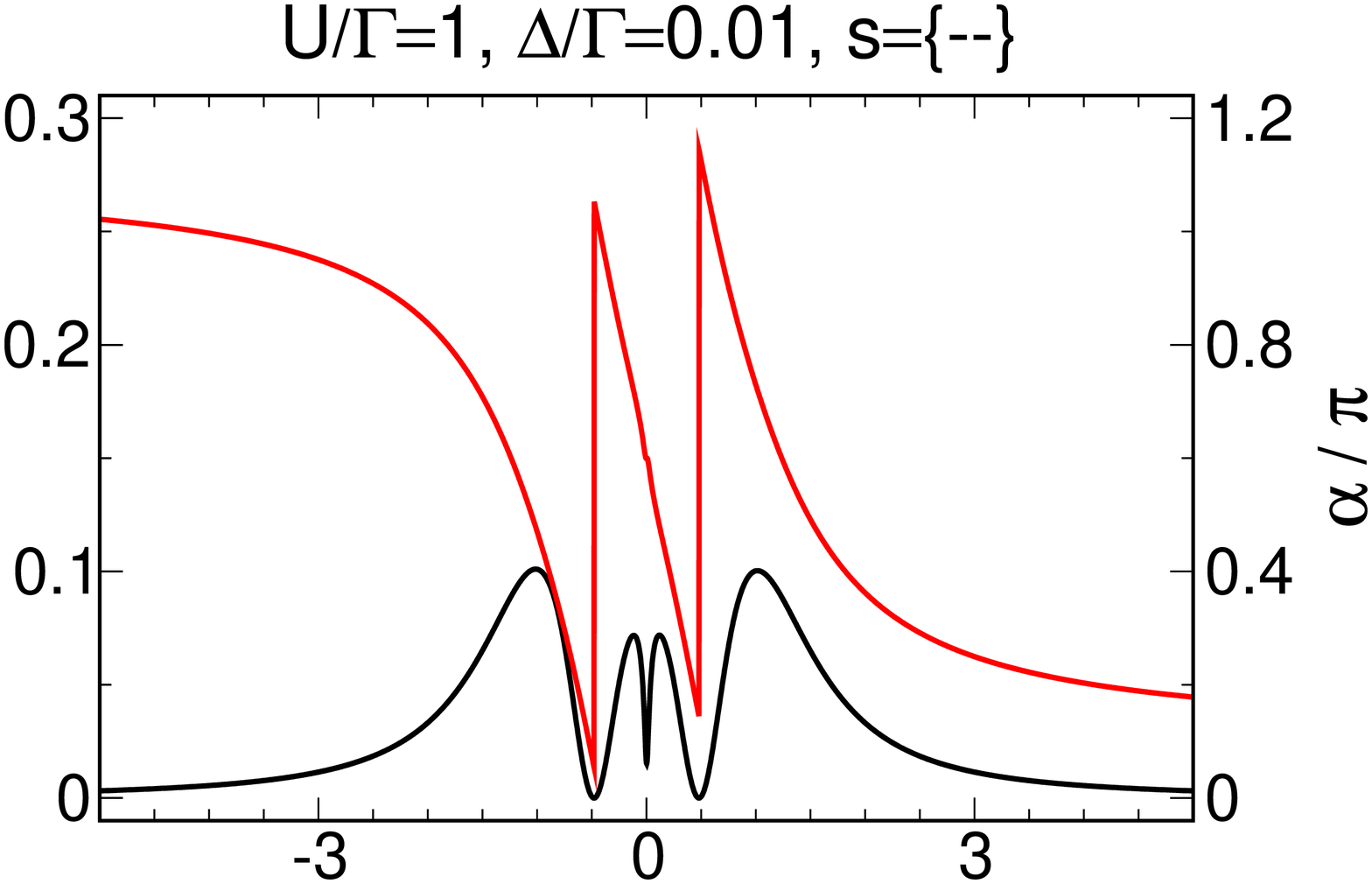}\vspace{0.3cm}
        \includegraphics[width=0.495\textwidth,height=5.2cm,clip]{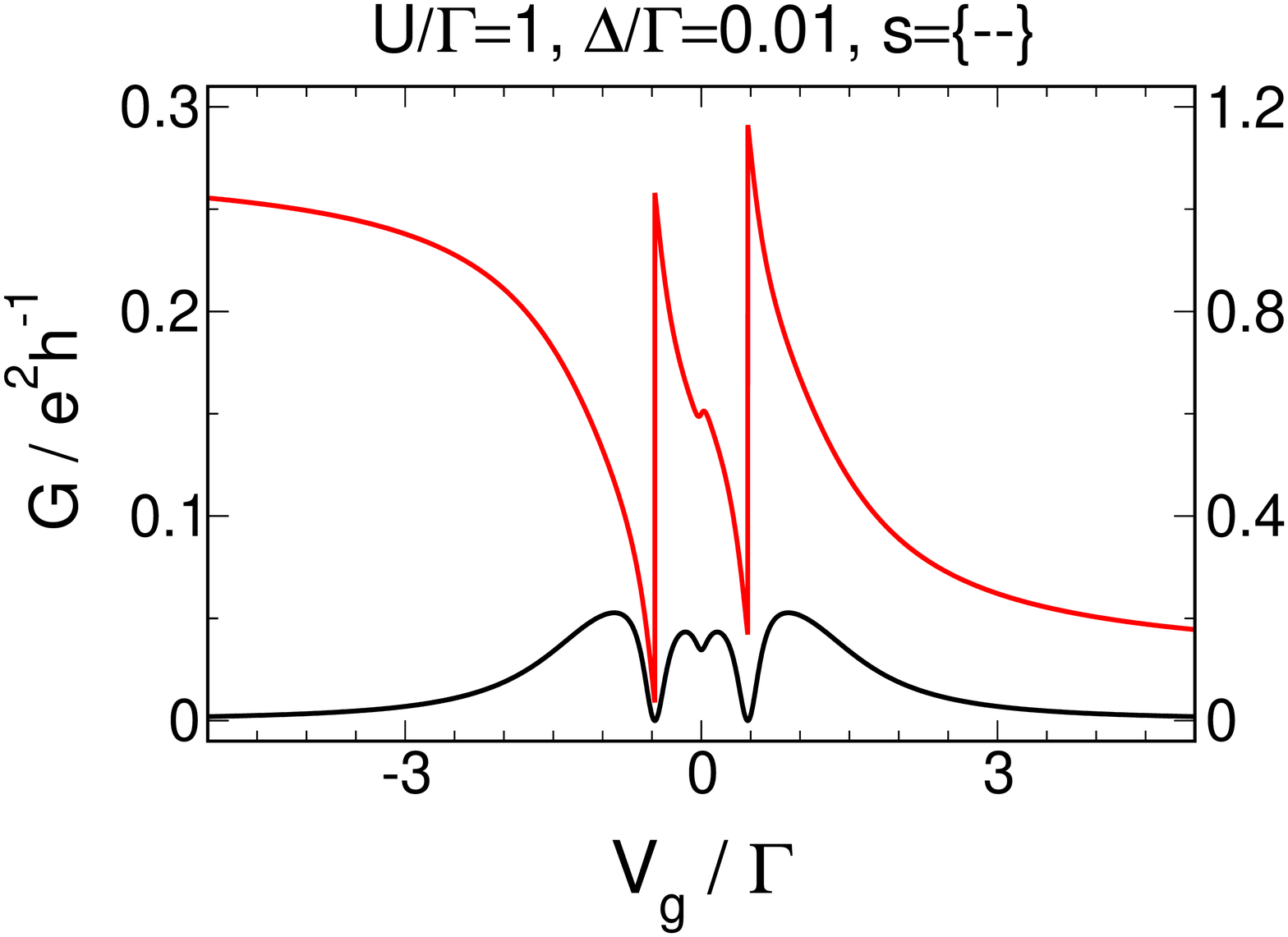}\hspace{0.015\textwidth}
        \includegraphics[width=0.475\textwidth,height=5.2cm,clip]{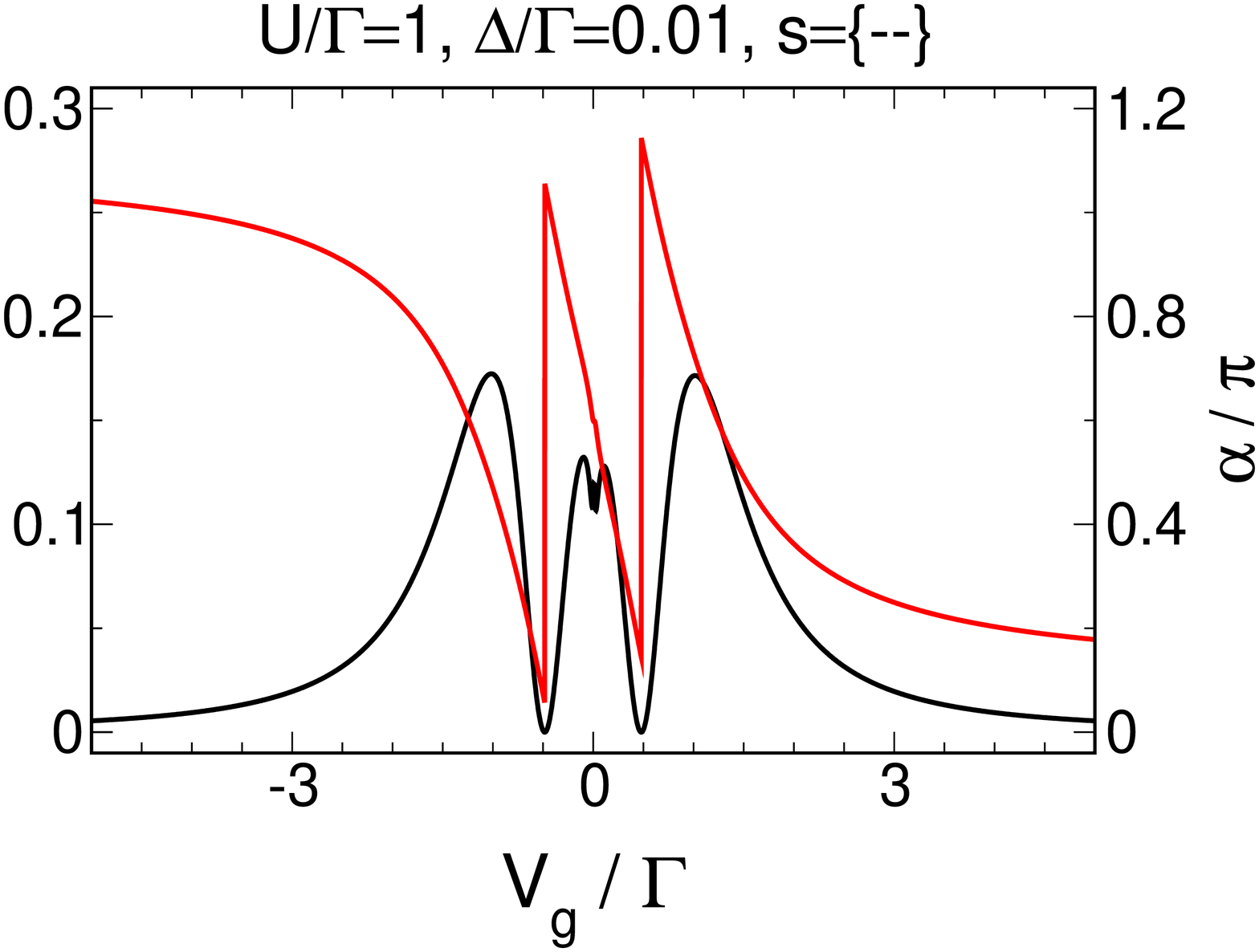}
        \caption{Illustration of the dependence of $\Delta_\tn{DPL}$ on left-right asymmetry $\Gamma_l^R/\Gamma_l^L$.  For level spacings within $\Delta_\tn{DPL}<\Delta<\Delta_\tn{cross}$, correlation induced double phase lapses (DPLs) are observed in the transmission phase $\alpha$ (red) of triple dots with $s=\{--\}$. The couplings read $\Gamma=\{0.15~0.15~0.2~0.2~0.15~0.15\}$, $\Gamma=\{0.1~0.2~0.1~0.3~0.12~0.18\}$, $\Gamma=\{0.05~0.25~0.05~0.35~0.05~0.25\}$, $\Gamma=\{0.1~0.2~0.05~0.35~0.2~0.1\}$ (from upper left to lower right). The conductance $G$ (black) is shown as well.}
\label{fig:OS.td.dpl_a}\vspace{0.6cm}
        \includegraphics[width=0.495\textwidth,height=4.4cm,clip]{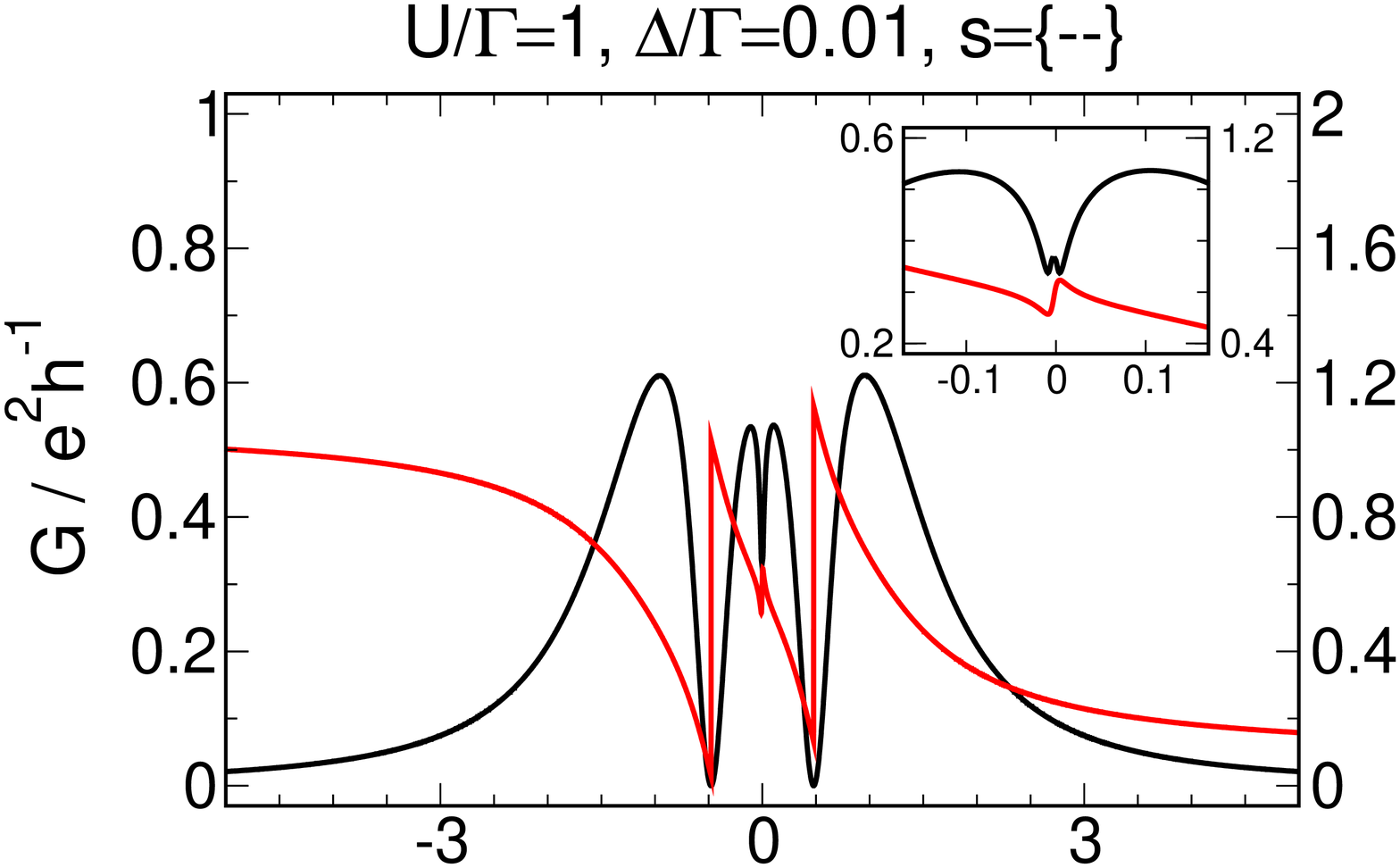}\hspace{0.015\textwidth}
        \includegraphics[width=0.475\textwidth,height=4.4cm,clip]{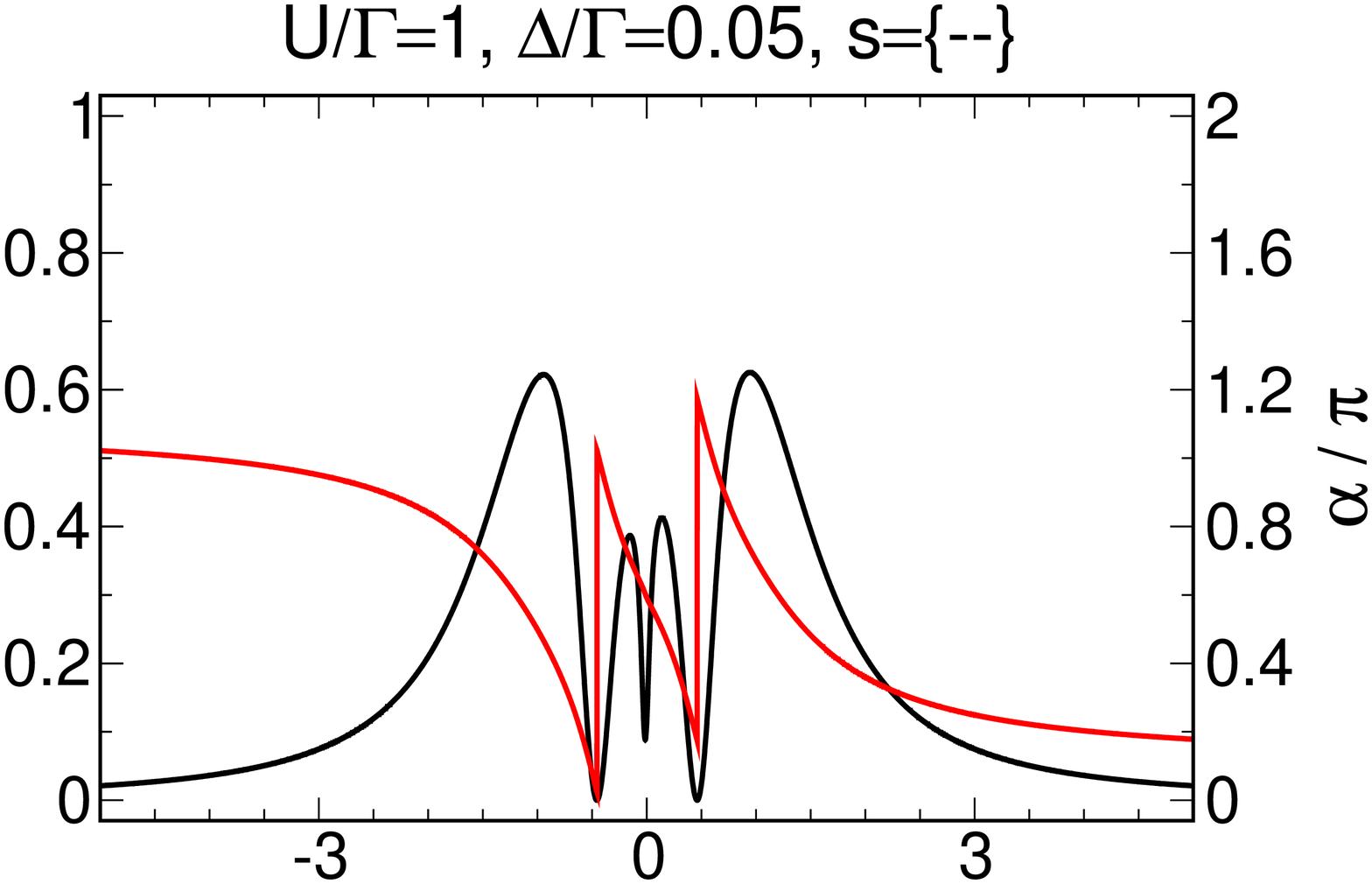}\vspace{0.3cm}
        \includegraphics[width=0.495\textwidth,height=5.2cm,clip]{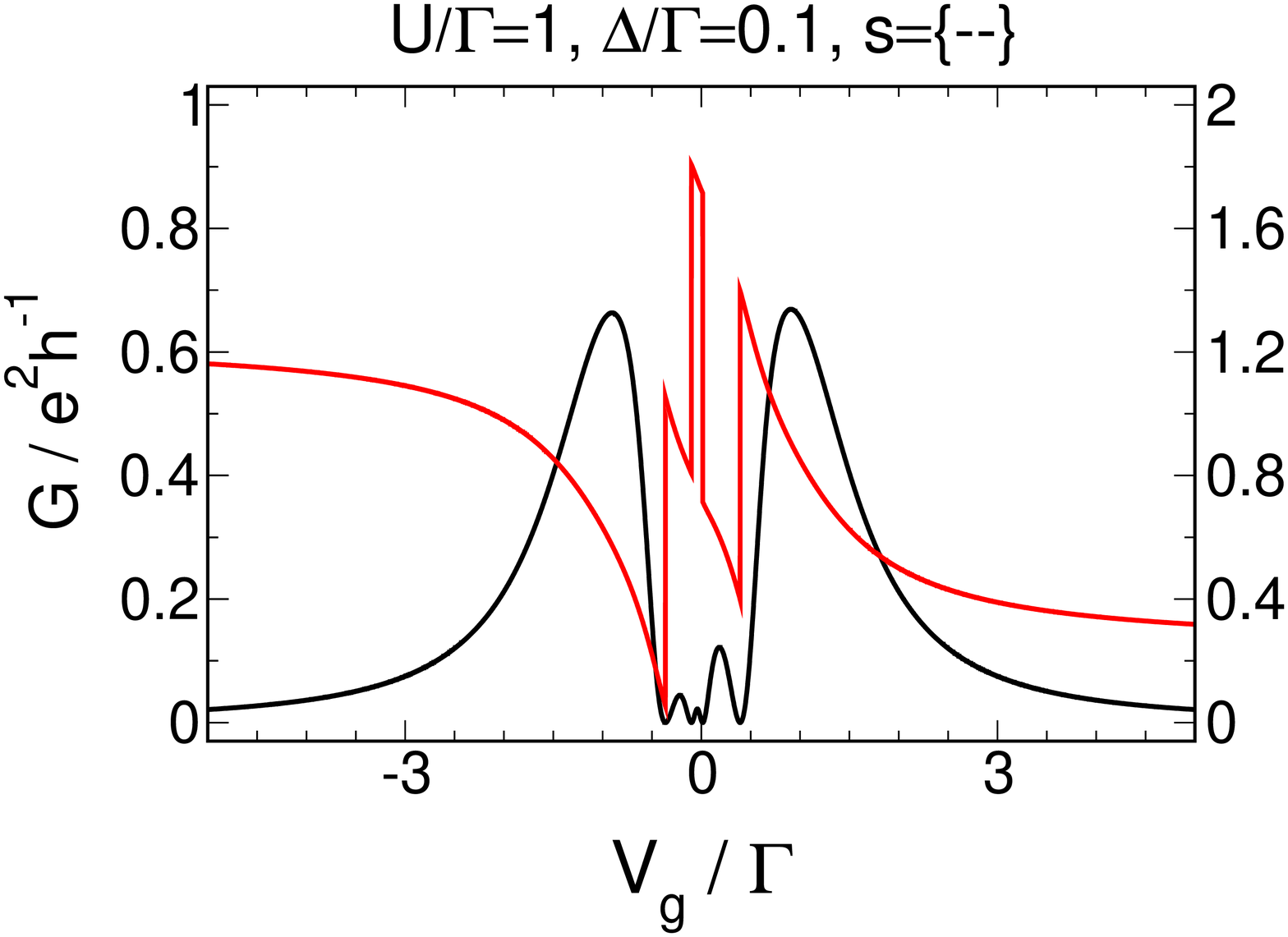}\hspace{0.015\textwidth}
        \includegraphics[width=0.475\textwidth,height=5.2cm,clip]{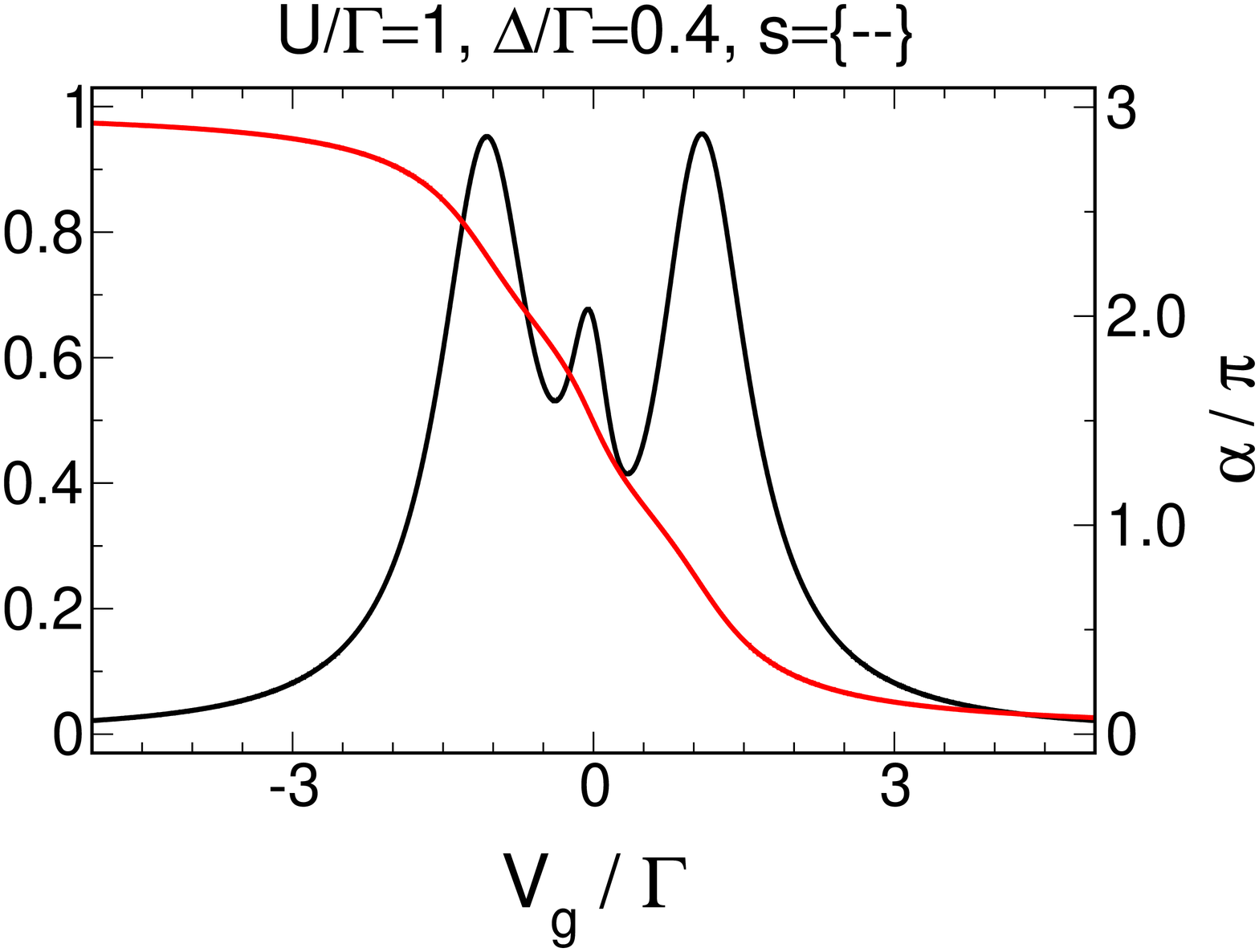}
        \caption{The same as Fig.~\ref{fig:OS.td.dpl_a}, but for $\Gamma=\{0.16~0.24~0.13~0.07~0.24~0.16\}$ and different level spacings. The DPLs appear for $\Delta_\tn{DPL}<\Delta<\Delta_\tn{cross}$.}
\label{fig:OS.td.dpl_b}
\end{figure}

Furthermore, the third peak centred at $V_g\approx 0$ shows very sharp structures that become further pronounced if the interaction is increased. Unfortunately, the precise shape of these `sharp structures' is not universal, but most frequently they take the form of an additional peak (sometimes accompanied by dips on its left and right side) of width much smaller than $\Gamma$ (Figs. \ref{fig:OS.td} \& \ref{fig:OS.td.u}). In general, the structures vanish for $\tilde\Delta_\tn{CIR}\ll\Gamma$. If (and only if) we choose $s_1=s_2=-$ and $\Gamma_B<\Gamma_A+\Gamma_C$, the dips in the conductance next to the additional resonance may even be transmission zeros for some special choices of the hybridisations (Fig.~\ref{fig:OS.td.dpl_a}, upper left panel). In particular, this is always the case for left-right symmetric $\Gamma_l^s$. Turning on slight asymmetries (which is the generic case) the zeros gradually disappear (Fig.~\ref{fig:OS.td.dpl_a}, other panels). Surprisingly, they do not vanish if the level spacing is increased but always appear above a scale $\Delta_\tn{DPL}$ which depends on the dot parameters, in particular on left-right asymmetry (Fig.~\ref{fig:OS.td.dpl_b}). For symmetric hybridisations, we have $\Delta_\tn{DPL}=0$, and this scale becomes larger and in particular no longer much smaller than $\Gamma$ the more asymmetric the $\Gamma_l^{L,R}$ are chosen. If it happens that $\Delta_\tn{DPL}$ and the level spacing where the crossover regime (see below) sets in become comparable, the additional zeros will never appear in between the limits $\Delta\ll\Gamma$ and $\Delta\gg\Gamma$.

As for the double dot geometry, the correlation induced structures are accompanied by a steep change of the transmission phase. If for $s_1=s_2=-$ the additional transmission zeros are present, $\alpha$ jumps by $\pi$ at each of them, so that we observe double phase lapses (DPLs) next to the sharp resonance.

Due to the very sharp structures close to $V_g=0$, one word about the reliability of the fRG results is in order. At first sight they might be interpreted as an artifact of our approximation, especially since they become more pronounced the larger the interaction is chosen. However, there are strong indications that these structures are indeed believable features of our interacting quantum dot model. First, they appear consistently within the different fRG truncation schemes, and second, they are confirmed in some special cases that can be tackled by an NRG calculation believed to give very precise results. All this will be explained in much more detail in the next chapter. Here, we should only keep in mind that everything presented should be taken to be reliable, as promised in the introduction.

\afterpage{\clearpage}

\subsubsection{From Small to Large Level Spacings: the Crossover Regime}

As mentioned above, the qualitative behaviour of the conductance as a function of $V_g$ is independent of the choice of the hybridisations (as long as they are generic) for nearly degenerate levels. In particular, similar to the double dot case the only influence of the left-right or the $A$-$B$-$C$ asymmetry is on the overall height of the conductance curves, on the scales $\Delta_\tn{CIR}$, $\tilde\Delta_\tn{CIR}$, and $\Delta_\tn{DPL}$ related to the correlation induced structures, and on the actual shape of the sharp feature close to $V_g=0$ (Figs.~\ref{fig:OS.td} \& \ref{fig:OS.td.dpl_a}). This does no longer hold (see below) when the level spacing exceeds $\Delta_\tn{cross}(\Gamma, s)$ and the dot merges into the crossover regime (as we will call the evolution from $\Delta\ll\Gamma$ to $\Delta\gg\Gamma$), which should be clear because similar to the double dot we would expect $\alpha(V_g)$ to be strongly dependend on $s$ for large level detunings.

In order to search for the different parameter regions that lead to distinct behaviour in the crossover regime, first, one has to convince oneself that there are only three independent choices of the two relative signs $s_1$ and $s_2$, namely $\{s_1s_2\}=\{++,+-,--\}$. This can be seen by considering the particle-hole transformation $d_l\to h_l^\dagger$ under which a Hamiltonian of the type
\begin{equation*}\begin{split}
H=&\sum_lV_l d_l^\dagger d_l + U\sum_{l>l'}\left(d_l^\dagger d_l-\frac{1}{2}\right)\left(d_{l'}^\dagger d_{l'}-\frac{1}{2}\right) \\
= & \sum_l\left(V_l-\frac{(N-1)U}{2}\right) d_l^\dagger d_l + U\sum_{l>l'}d_l^\dagger d_ld_{l'}^\dagger d_{l'}+\tn{const.}
\end{split}\end{equation*}
goes over to
\begin{equation*}\begin{split}
\tilde H = & \sum_l\left(V_l-\frac{(N-1)U}{2}\right) (1-h_l^\dagger h_l) + U\sum_{l>l'}(1-h_l^\dagger h_l)(1-h_{l'}^\dagger h_{l'})+\tn{const.} \\
= &\sum_l\left(-V_l-\frac{(N-1)U}{2}\right) h_l^\dagger h_l + U\sum_{l>l'}h_l^\dagger h_lh_{l'}^\dagger h_{l'}+\tn{const.},
\end{split}\end{equation*}
with $N$ being the number of levels in the system. Hence, this transformation implies that the choice $s_1=-,s_2=+$ is equivalent to $s_1=+$, $s_2=-$ if we additionally replace $V_g\to-V_g$. As argued above, the crossover regime will strongly depend on which of the three independent possibilities is realised. For $s=\{++\}$, it turns out to be independent of the size of the hybridisations, while for $s=\{+-,--\}$ it is the strength of the dot $\tilde A$ which couplings differ in sign from the other two, $\tn{sgn}(t_{\tilde A}^Lt_{\tilde A}^R)\neq \tn{sgn}(t_{\tilde B}^Lt_{\tilde B}^R)=\tn{sgn}(t_{\tilde C}^Lt_{\tilde C}^R)$, that determines the behaviour of the conductance in between $\Delta\ll\Gamma$ and $\Delta\gg\Gamma$. In particular, there are two distinct parameter regions in each case, namely $\Gamma_{\tilde A}<\Gamma_{\tilde B}+\Gamma_{\tilde C}$ (regime W), and $\Gamma_{\tilde A}>\Gamma_{\tilde B}+\Gamma_{\tilde C}$ (regime S).

\begin{figure}[t]	
        \centering
        \includegraphics[width=0.495\textwidth,height=5.2cm,clip]{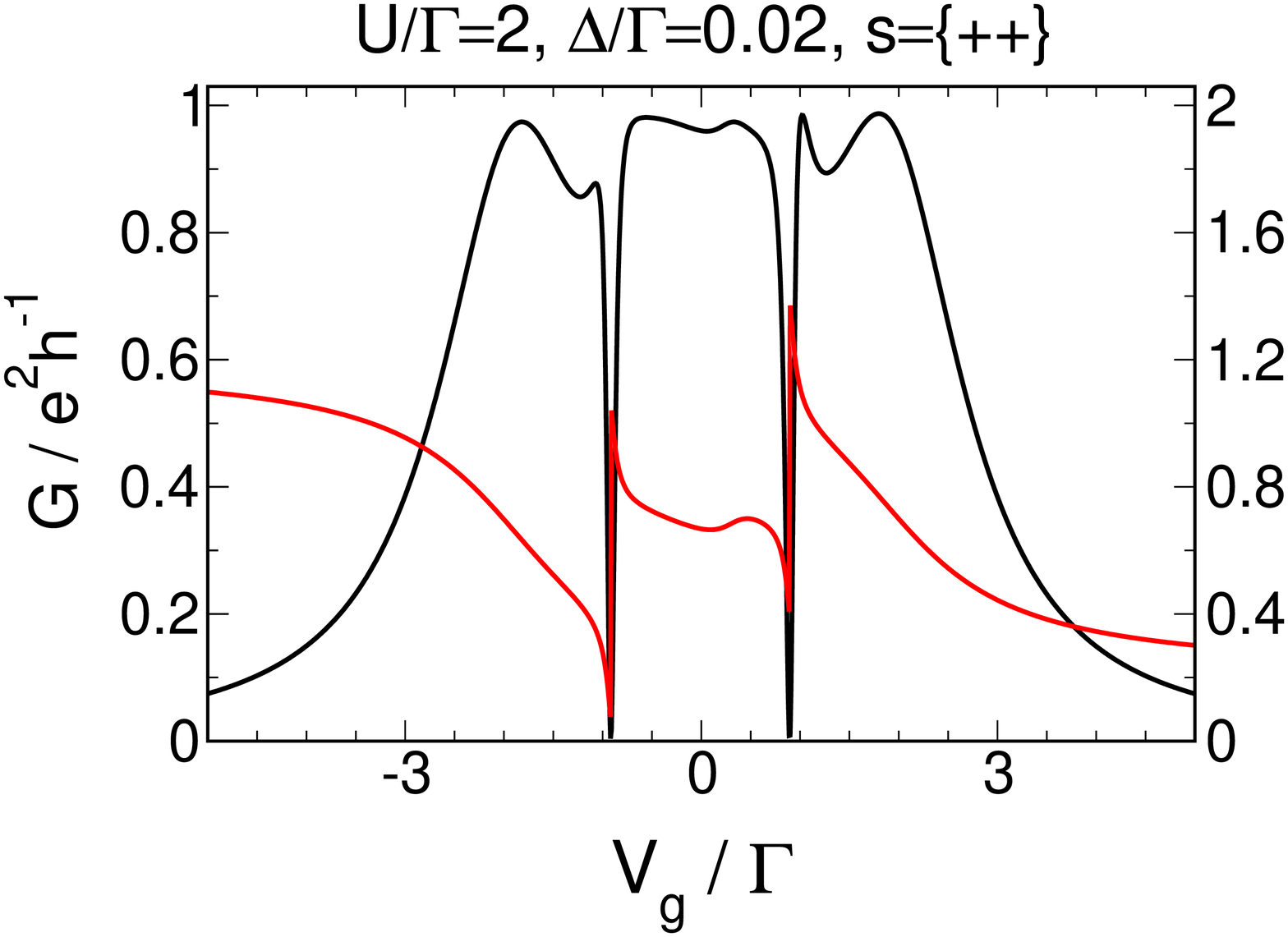}\hspace{0.015\textwidth}
        \includegraphics[width=0.475\textwidth,height=5.2cm,clip]{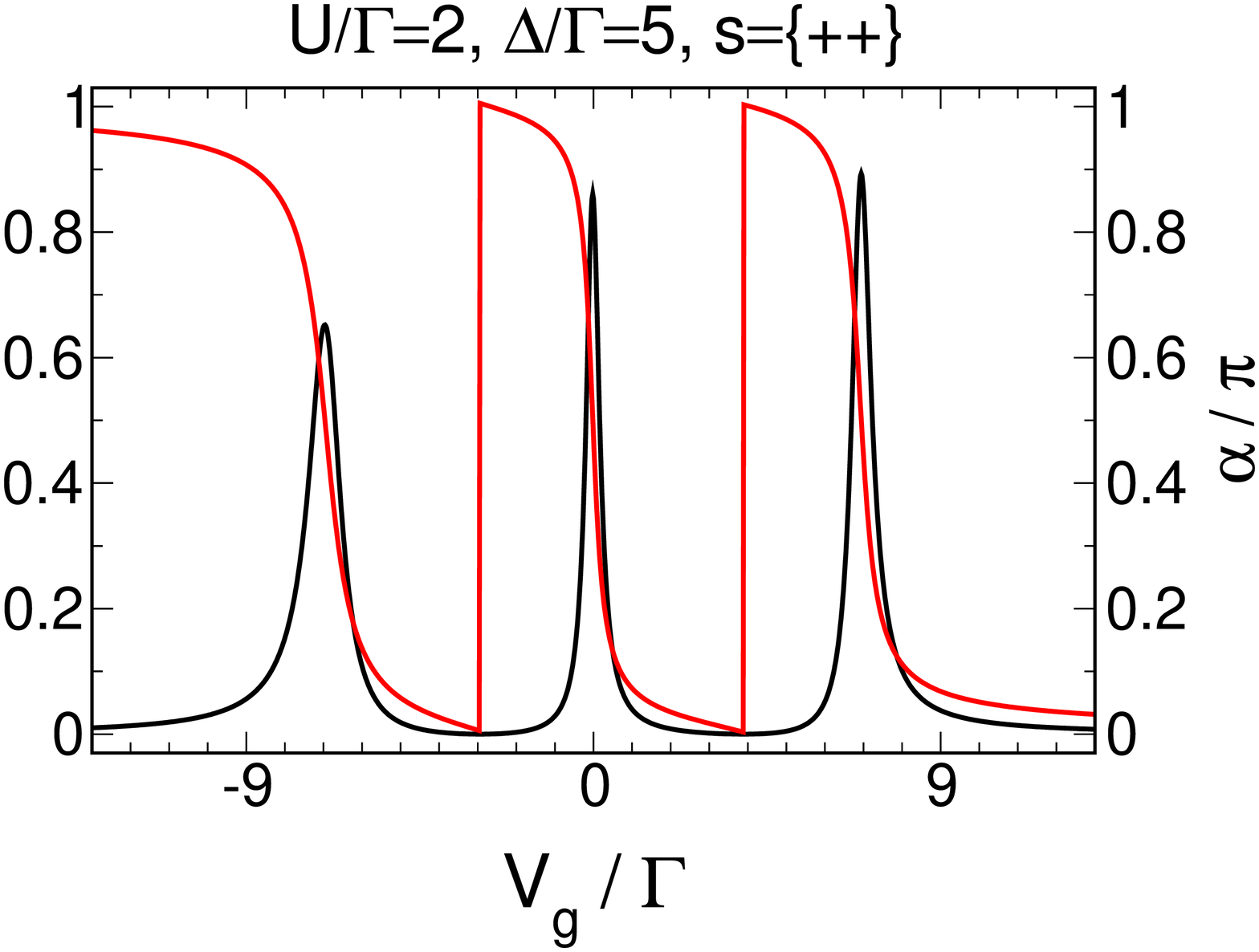}
        \caption{Conductance $G$ (black) and transmission phase $\alpha$ (red) as a function of the gate voltage for parallel triple dots with generic level-lead hybridisations $\Gamma=\{0.1~0.2~0.06~0.14~0.4~0.1\}$.}
\label{fig:OS.td.ppp}
\end{figure}

\begin{figure}[t]	
        \centering
        \includegraphics[width=0.495\textwidth,height=4.4cm,clip]{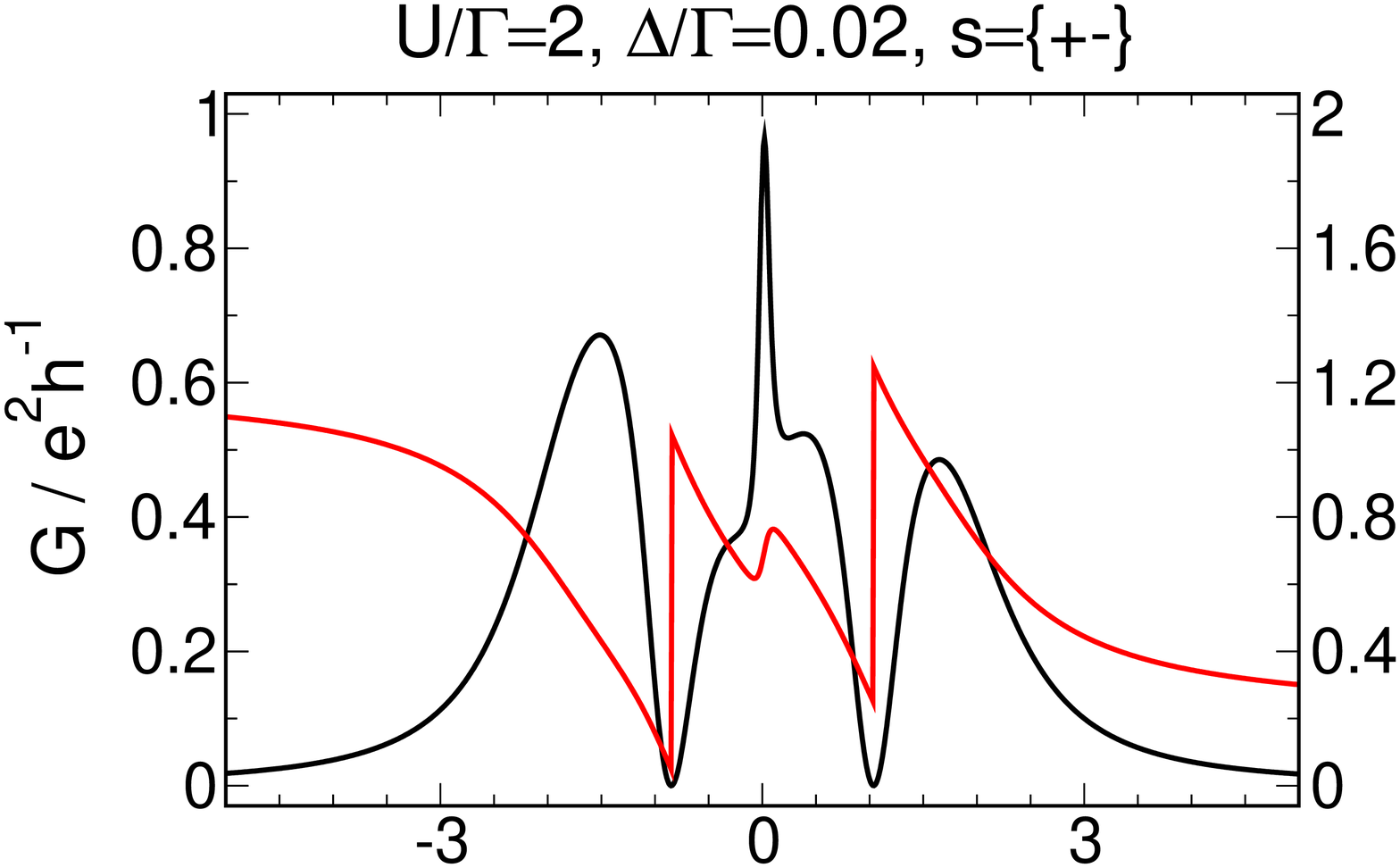}\hspace{0.015\textwidth}
        \includegraphics[width=0.475\textwidth,height=4.4cm,clip]{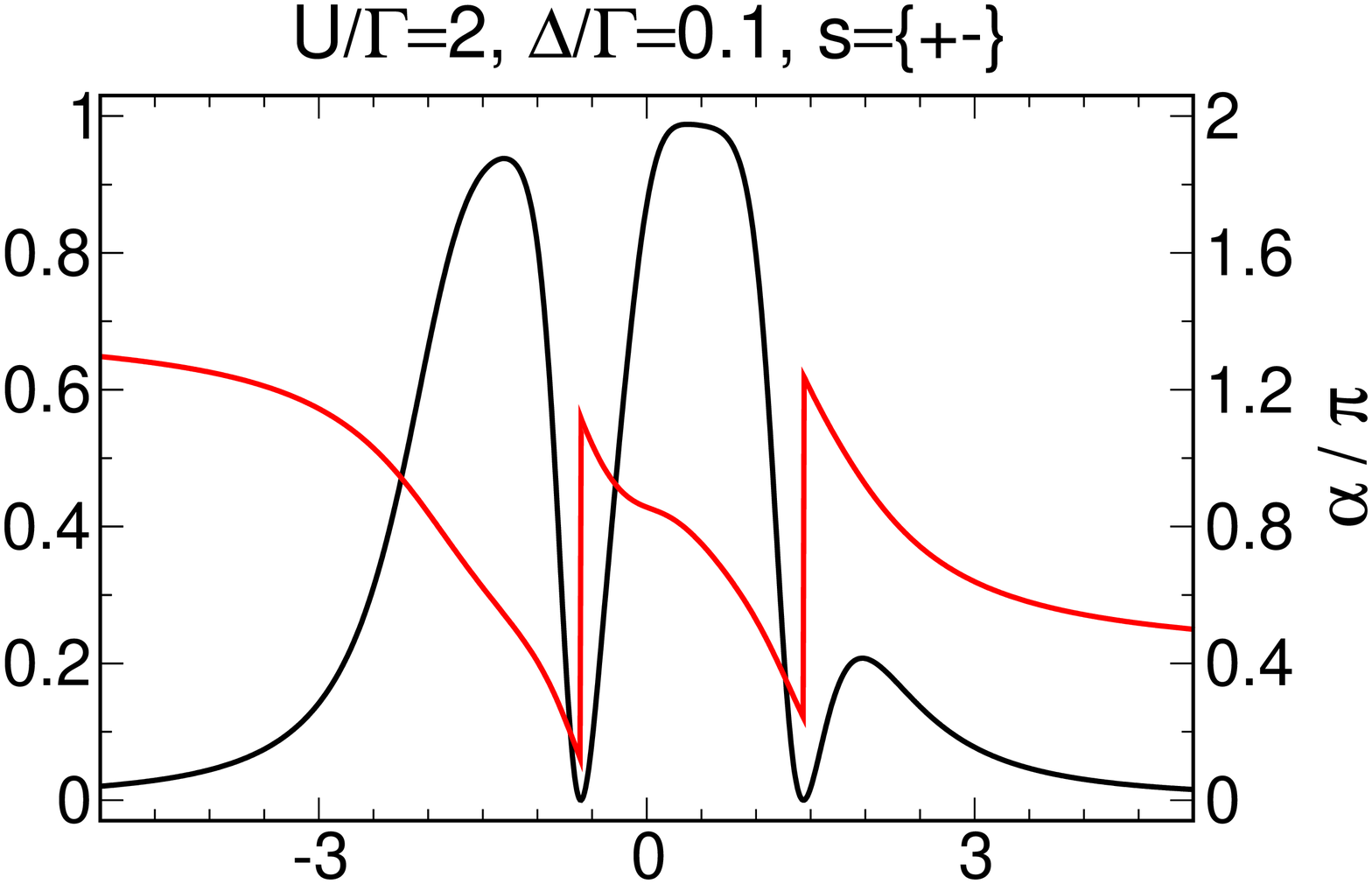}\vspace{0.3cm}
        \includegraphics[width=0.495\textwidth,height=5.2cm,clip]{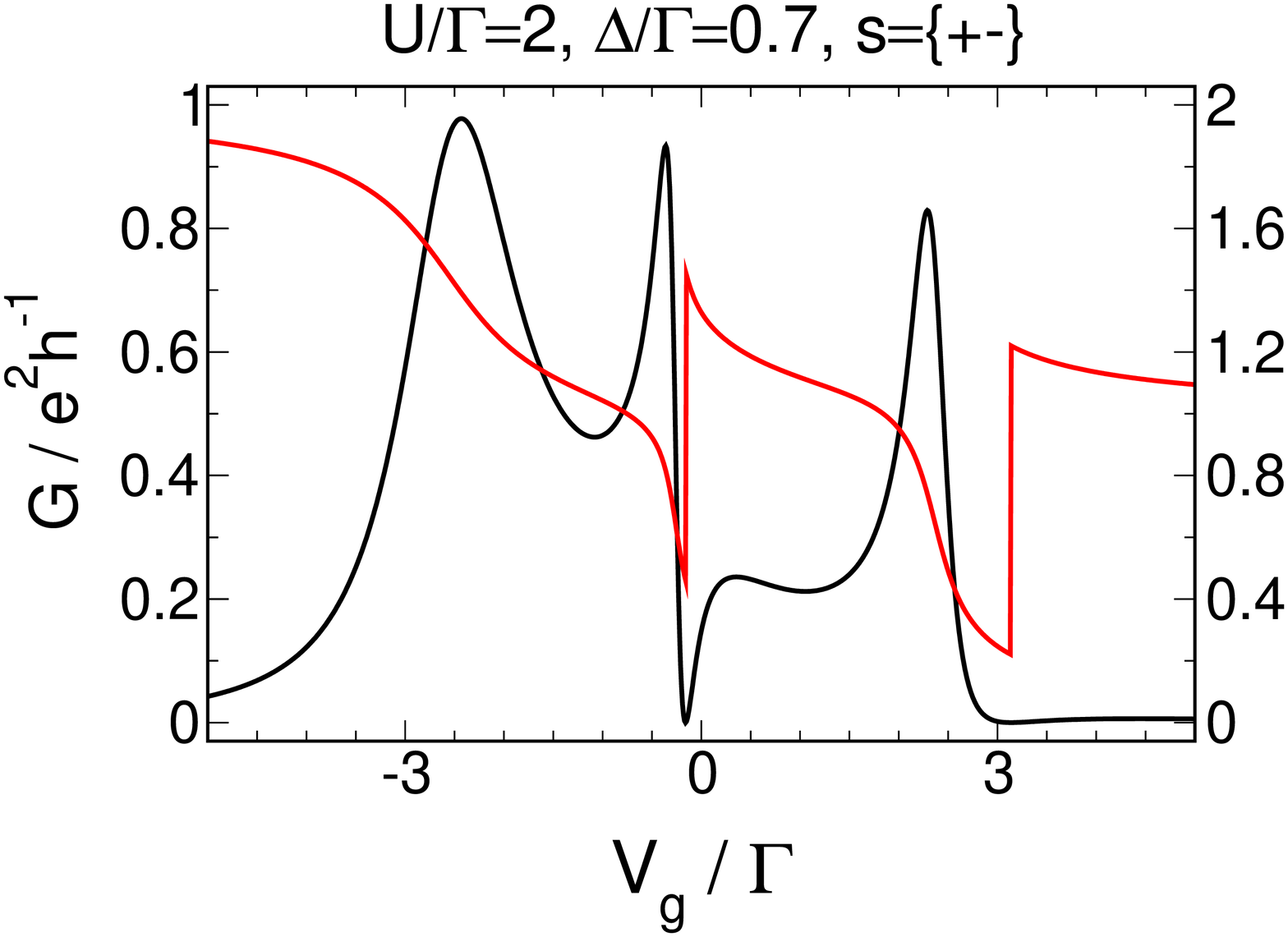}\hspace{0.015\textwidth}
        \includegraphics[width=0.475\textwidth,height=5.2cm,clip]{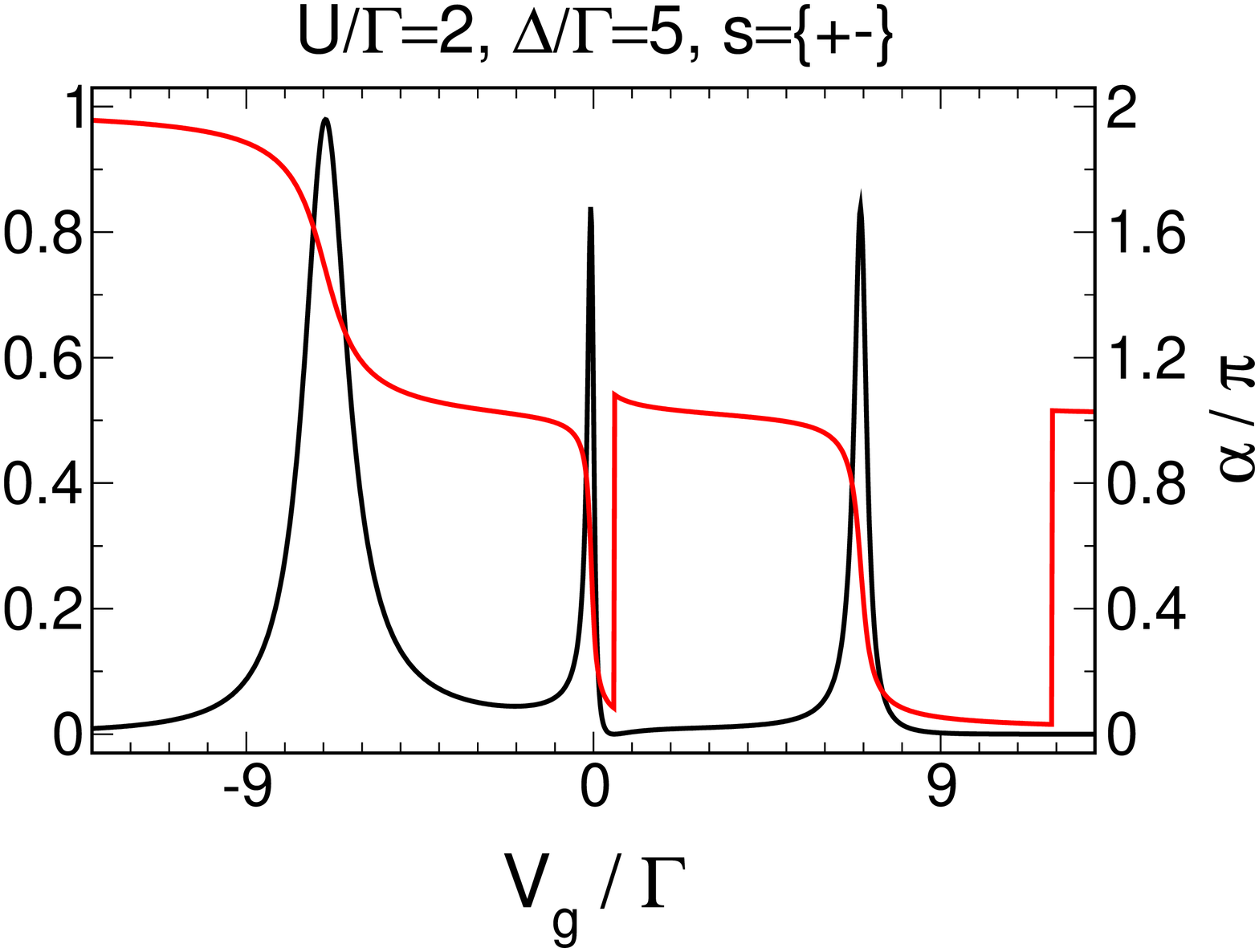}
        \caption{Gate voltage dependence of the conductance $G$ (black) and transmission phase $\alpha$ (red) for parallel triple dots with generic level-lead hybridisations $\Gamma=\{0.06~0.14~0.07~0.03~0.3~0.4\}$, illustrating the crossover from $\Delta\ll\Gamma$ to $\Delta\gg\Gamma$ in regime S.}
\label{fig:OS.td.ppmda}\vspace{0.6cm}
        \includegraphics[width=0.495\textwidth,height=4.4cm,clip]{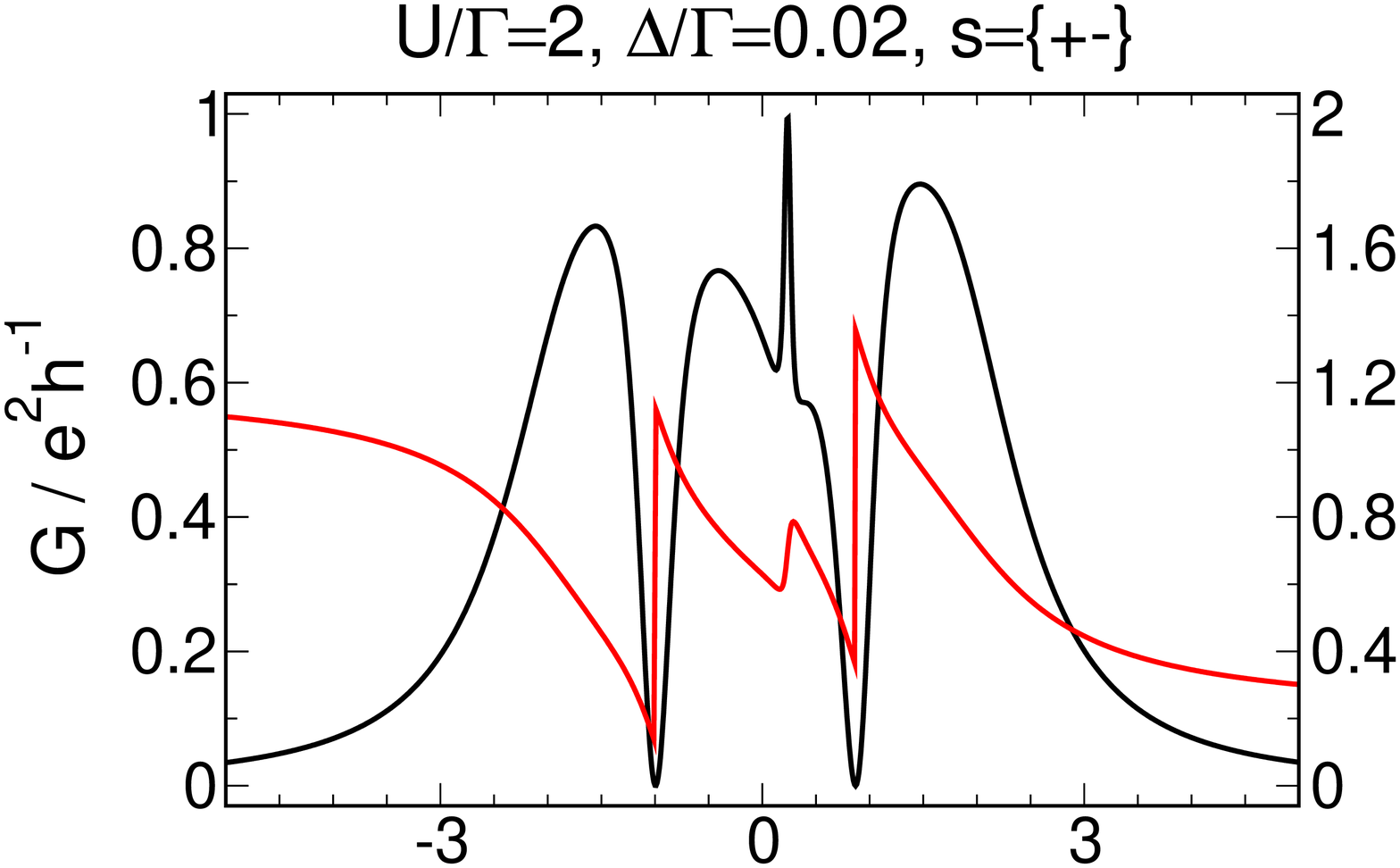}\hspace{0.015\textwidth}
        \includegraphics[width=0.475\textwidth,height=4.4cm,clip]{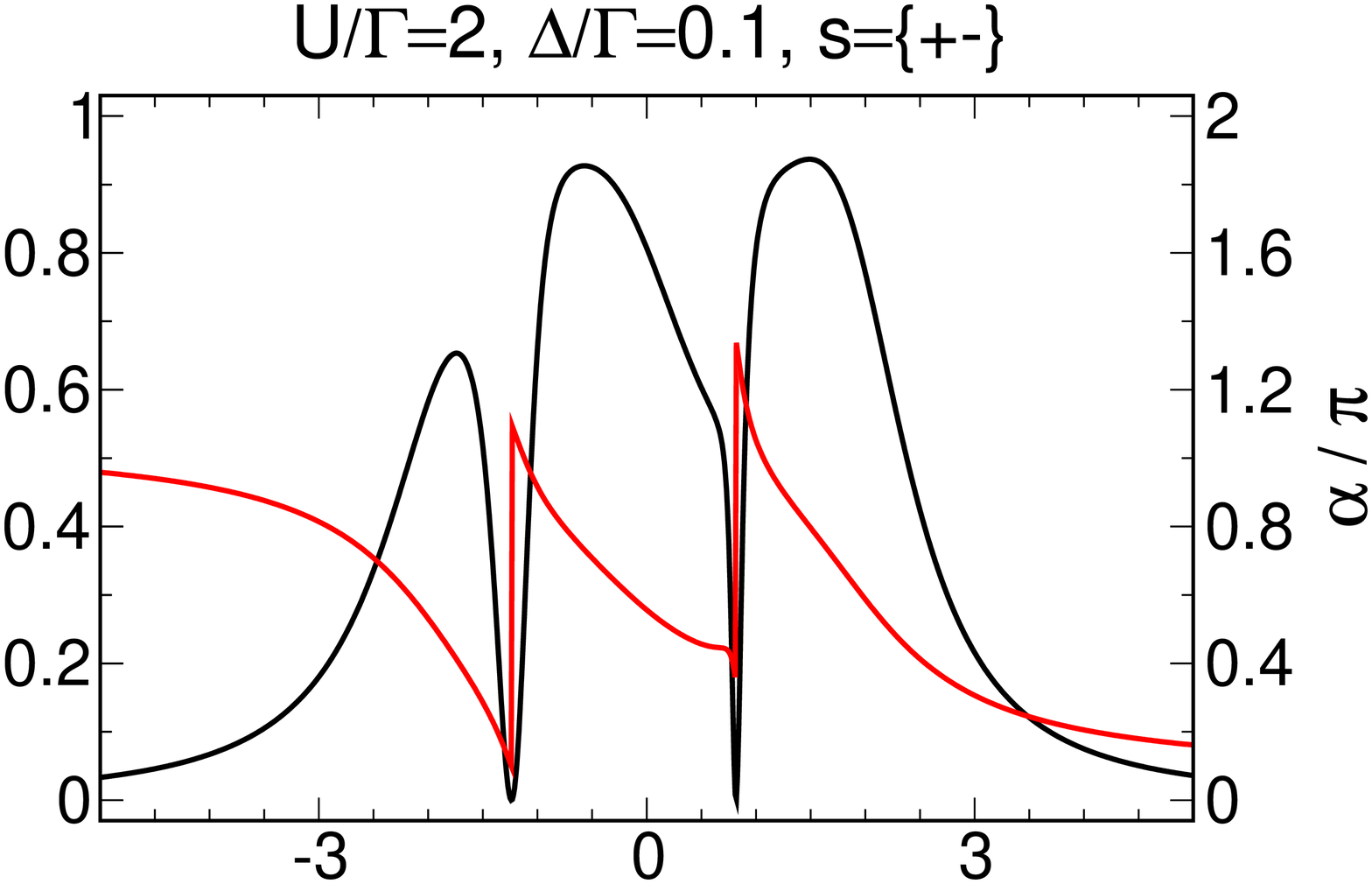}\vspace{0.3cm}
        \includegraphics[width=0.495\textwidth,height=5.2cm,clip]{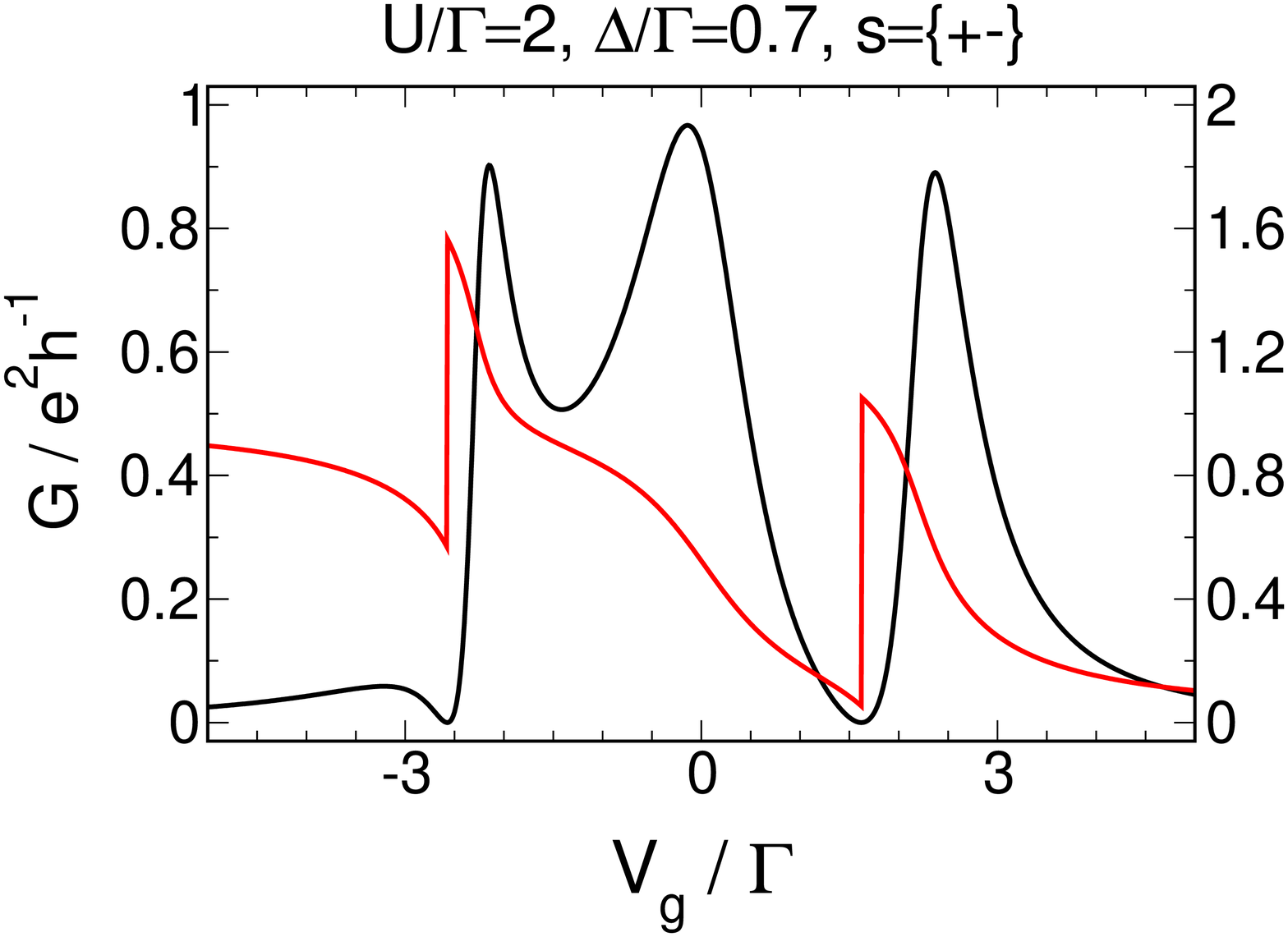}\hspace{0.015\textwidth}
        \includegraphics[width=0.475\textwidth,height=5.2cm,clip]{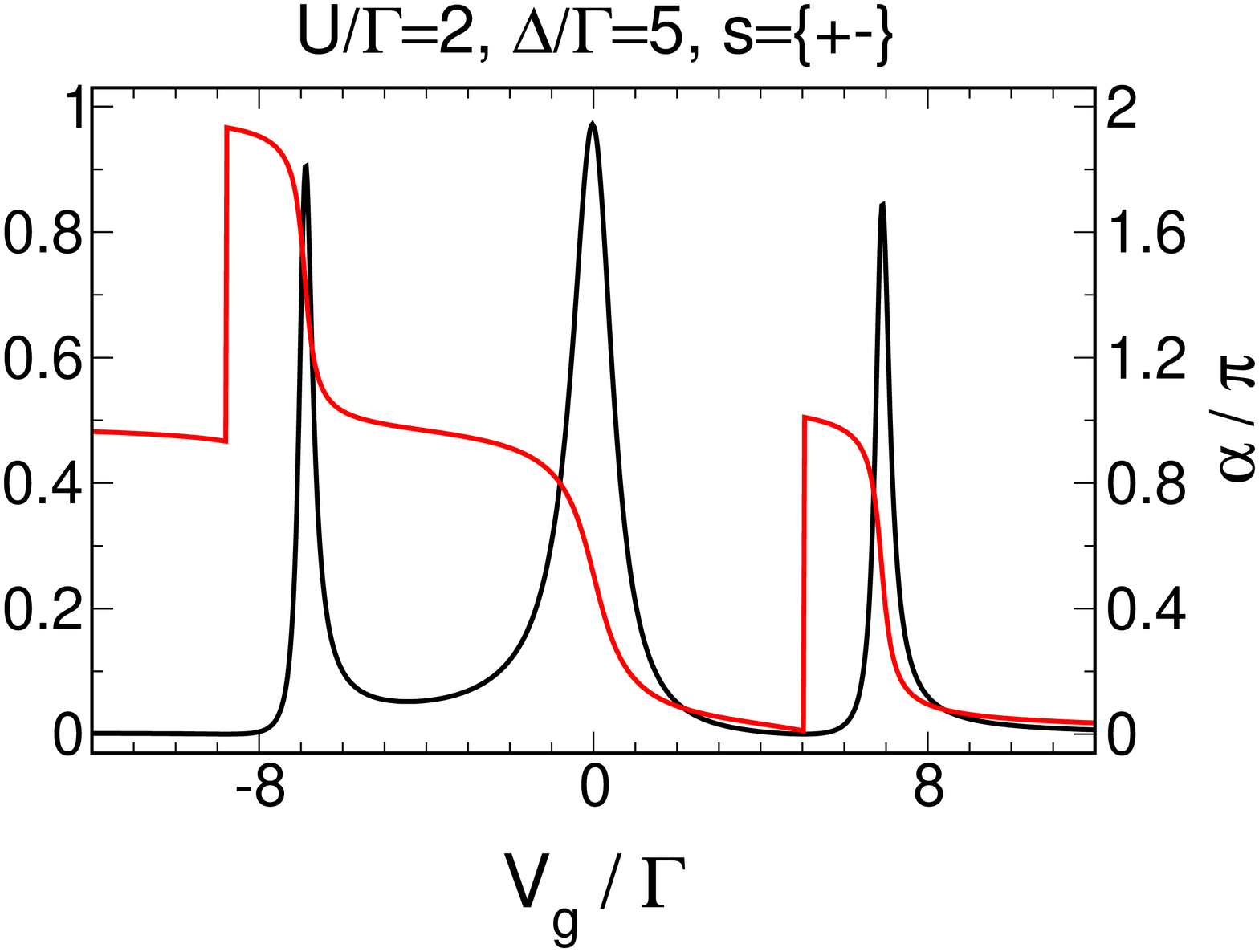}
        \caption{The same as Fig.~\ref{fig:OS.td.ppmda}, but for generic hybridisations from regime W, $\Gamma=\{0.06~0.14~0.25~0.35~0.13~0.07\}$.}
\label{fig:OS.td.ppmdb}
\end{figure}
\afterpage{\clearpage}

The most simple case is that with $s=\{++\}$. It is depicted in Fig.~\ref{fig:OS.td.ppp}. Starting from nearly degenerate levels, the correlations induced resonances close to $V_g=\pm U/2$ and the sharp feature at $V_g\approx 0$ vanish (if they are present at all) for $\Delta\approx\Delta_\tn{CIR}$ and $\tilde\Delta\approx\Delta_\tn{CIR}$, respectively. The separation of the three remaining resonances (the Coulomb blockade peaks) increases, their lineshape gradually changes and for $\Delta\gg\Gamma$ we finally observe three Lorentzian resonances separated by $U+\Delta$. They correspond to transport through the individual levels, and accordingly the height and width of the peak over which dot $l$ gets depleted is determined by the hybridisations $\Gamma_l^L$ and $\Gamma_l^R$. As for $\Delta\ll\Gamma$, the transmission phase changes by $\pi$ with the noninteracting single-level lineshape over each resonance and jumps by $\pi$ at the two transmission zeros in between.

For $s=\{+-\}$, the transition from small to large level spacings is more complicated. As before, the additional features observed in the three-peak structure gradually disappear if the level spacing is increased. As mentioned, the subsequent evolution in the crossover regime depends on the relative coupling strength of the different dots. Increasing $\Delta$, one of the Coulomb peaks splits up (the left one `corresponding' to dot $\tilde A$ in regime S; the central resonance in regime W) while another becomes increasingly small (regime S: right resonance; regime W: left resonance). In analogy to the parallel double dot, we define a scale $\Delta_\tn{cross}(\Gamma,s)$ characterising the set-in of the crossover regime (for example by the height of the resonance that almost disappears). This scale depends on the dot parameters, but it is roughly given by $\Delta_\tn{cross}\approx\Gamma/10>\Delta_\tn{CIR},\tilde\Delta_\tn{CIR}$. If for some parameters it happens that either $\Delta_\tn{CIR}\approx\Delta_\tn{cross}$ or $\tilde\Delta_\tn{CIR}\approx\Delta_\tn{cross}$, the vanishing of the correlation induced structures and the split-off of one of the peaks in the crossover regime might not be distinguishable. Finally, we obtain three Lorentzian resonances identical to the $s=\{++\}$ case for $\Delta\gg\Gamma$, only that now the phase that changes with the noninteracting lineshape over each resonance shows a jump by $\pi$ between those corresponding to level $A$ and $B$ while evolving continuously between $B$ and $C$. In particular, $\alpha$ only jumps in between the peaks related to those dots which couplings do not differ in sign ($\tn{sgn}(t_A^Lt_A^R)=\tn{sgn}(t_B^Lt_B^R)$). The transition from small to large level detunings for $s=\{+-\}$ is depicted in Fig.~\ref{fig:OS.td.ppmda} (regime S) and Fig.~\ref{fig:OS.td.ppmdb} (regime W).

\begin{figure}[t]	
        \centering
        \includegraphics[width=0.495\textwidth,height=4.4cm,clip]{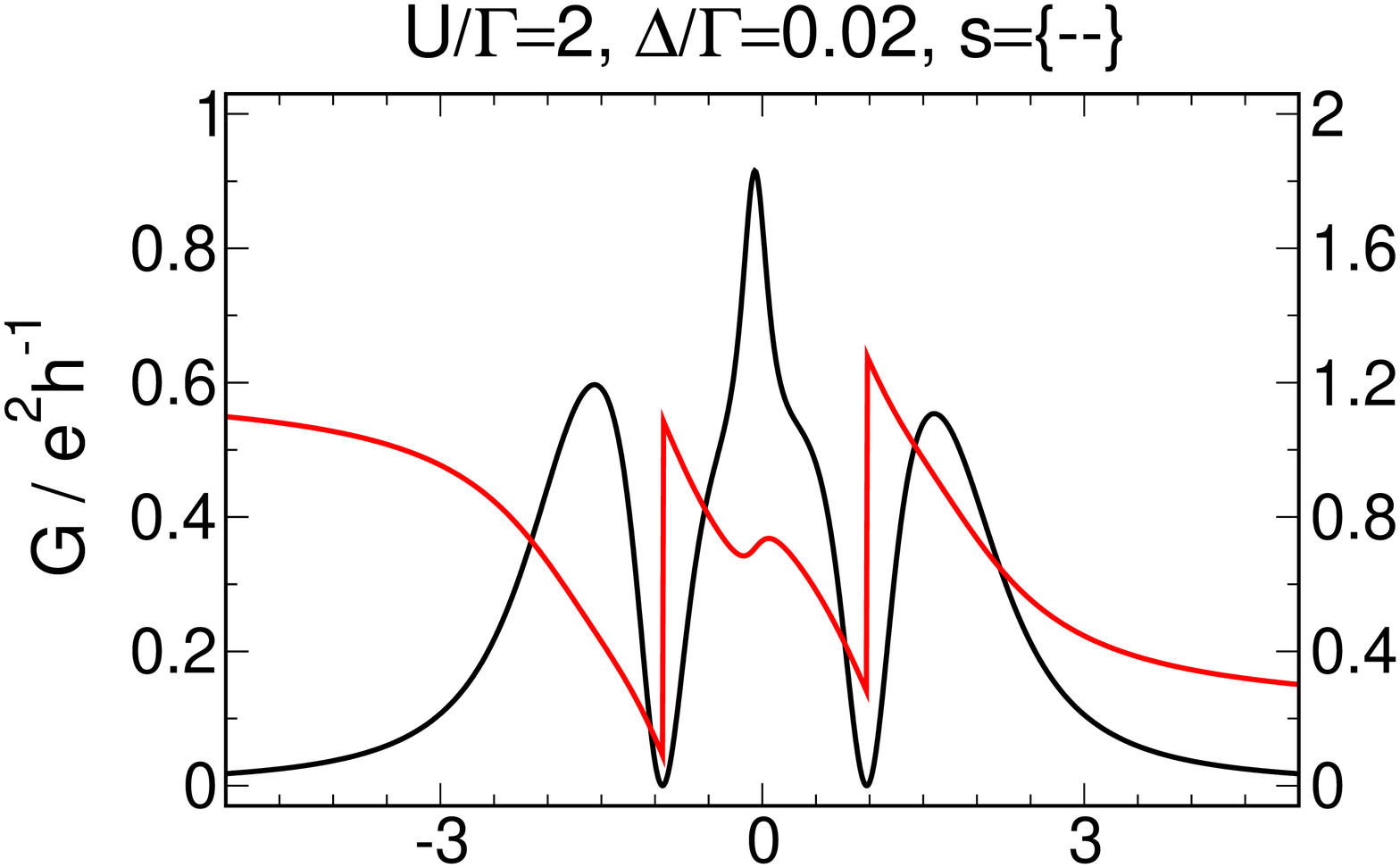}\hspace{0.015\textwidth}
        \includegraphics[width=0.475\textwidth,height=4.4cm,clip]{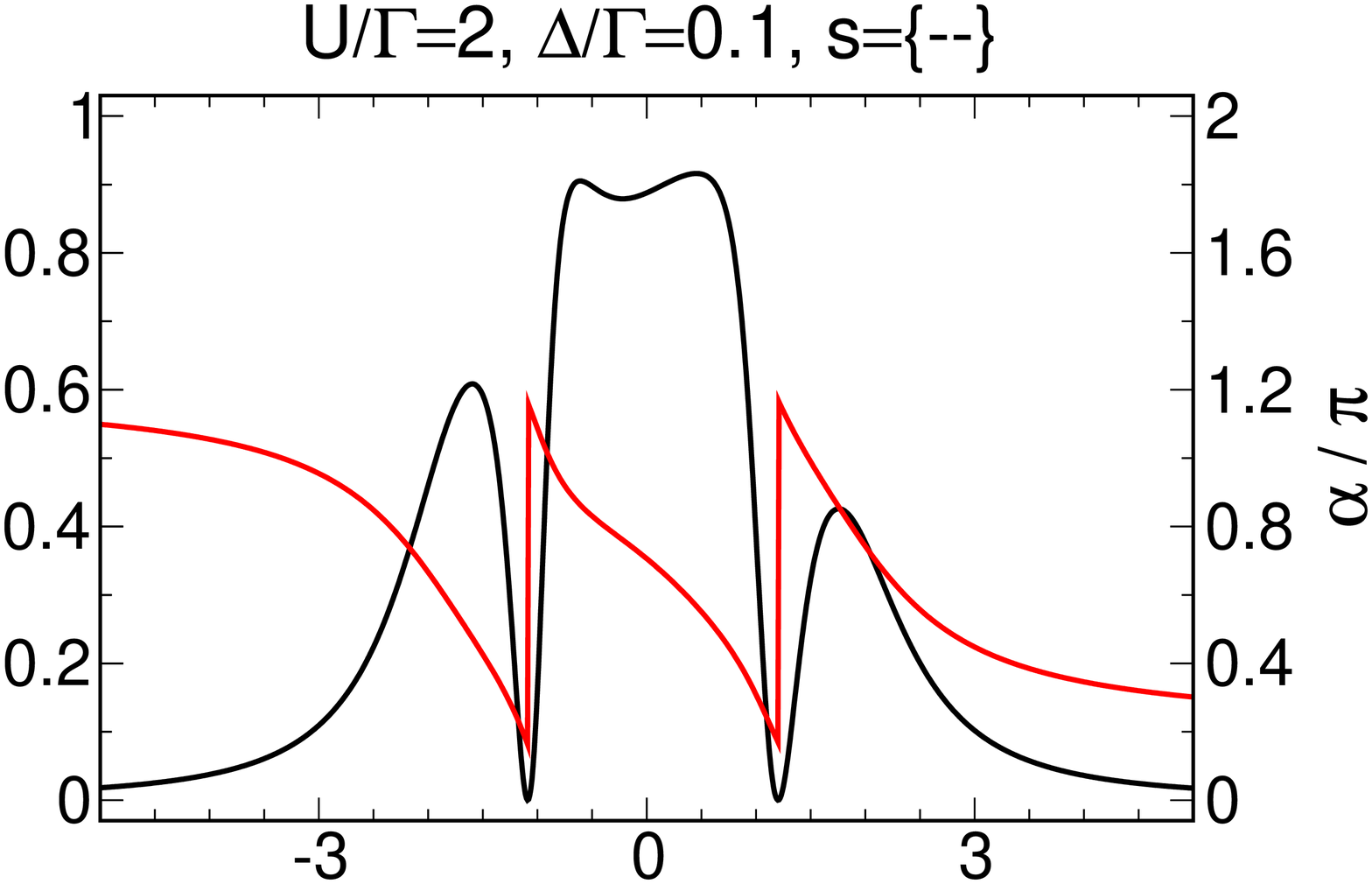}\vspace{0.3cm}
        \includegraphics[width=0.495\textwidth,height=5.2cm,clip]{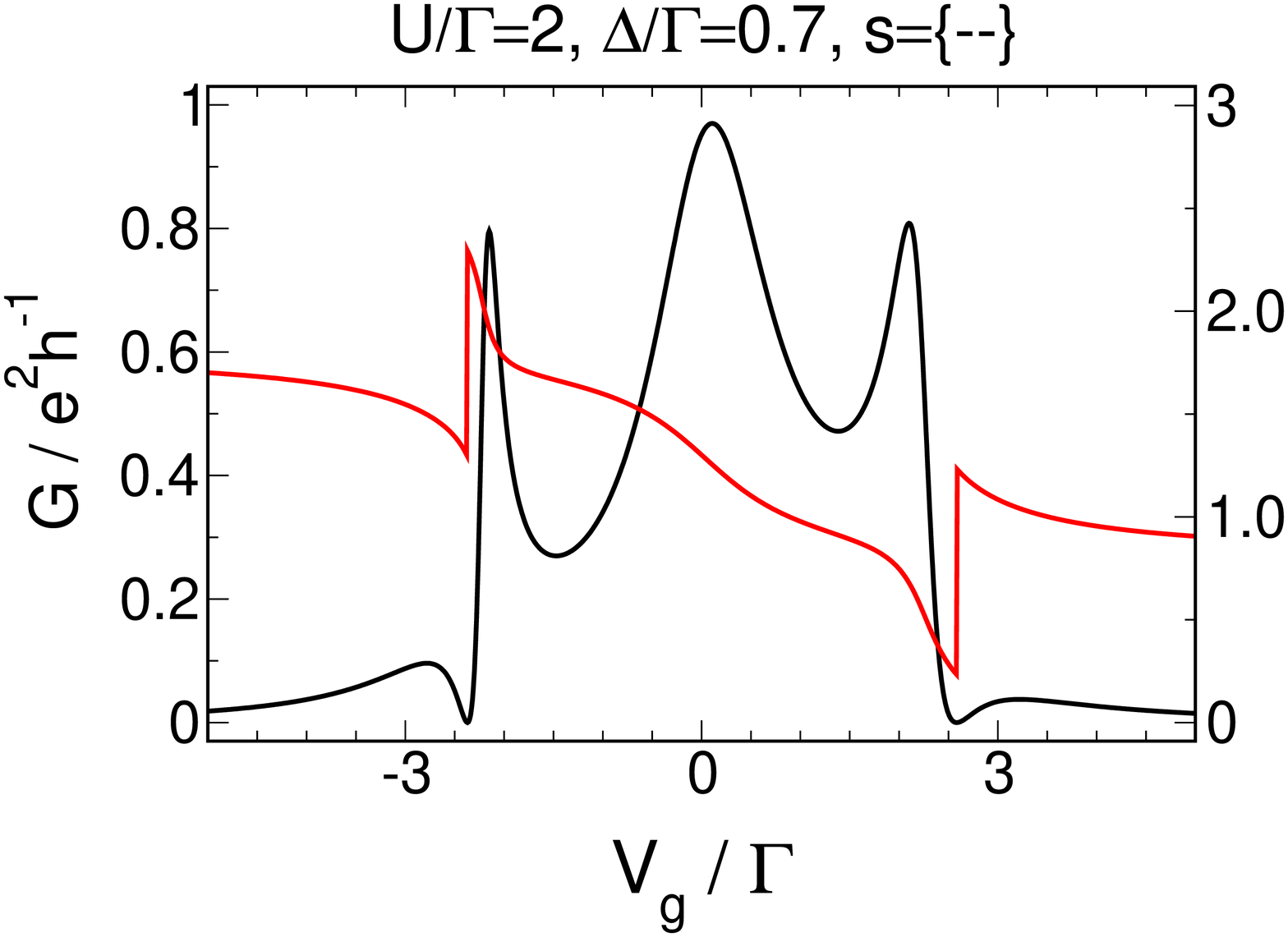}\hspace{0.015\textwidth}
        \includegraphics[width=0.475\textwidth,height=5.2cm,clip]{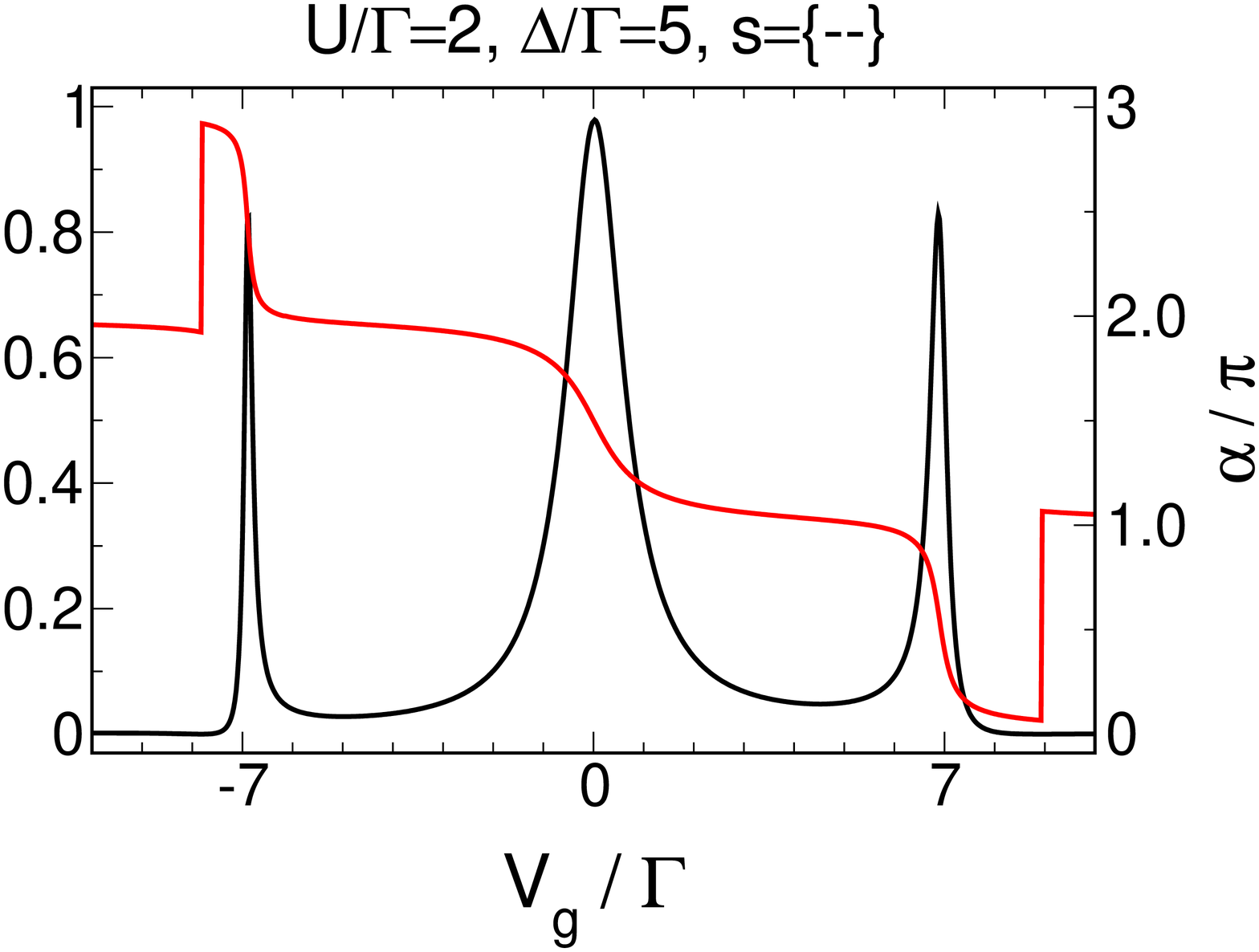}
        \caption{The same as Fig.~\ref{fig:OS.td.ppmda}, but for $s=\{--\}$, and $\Gamma=\{0.06~0.14~0.3~0.4~0.07~0.03\}$, illustrating the generic crossover from $\Delta\ll\Gamma$ to $\Delta\gg\Gamma$ for regime S.}
\label{fig:OS.td.pmpda}\vspace{0.6cm}
        \includegraphics[width=0.495\textwidth,height=4.4cm,clip]{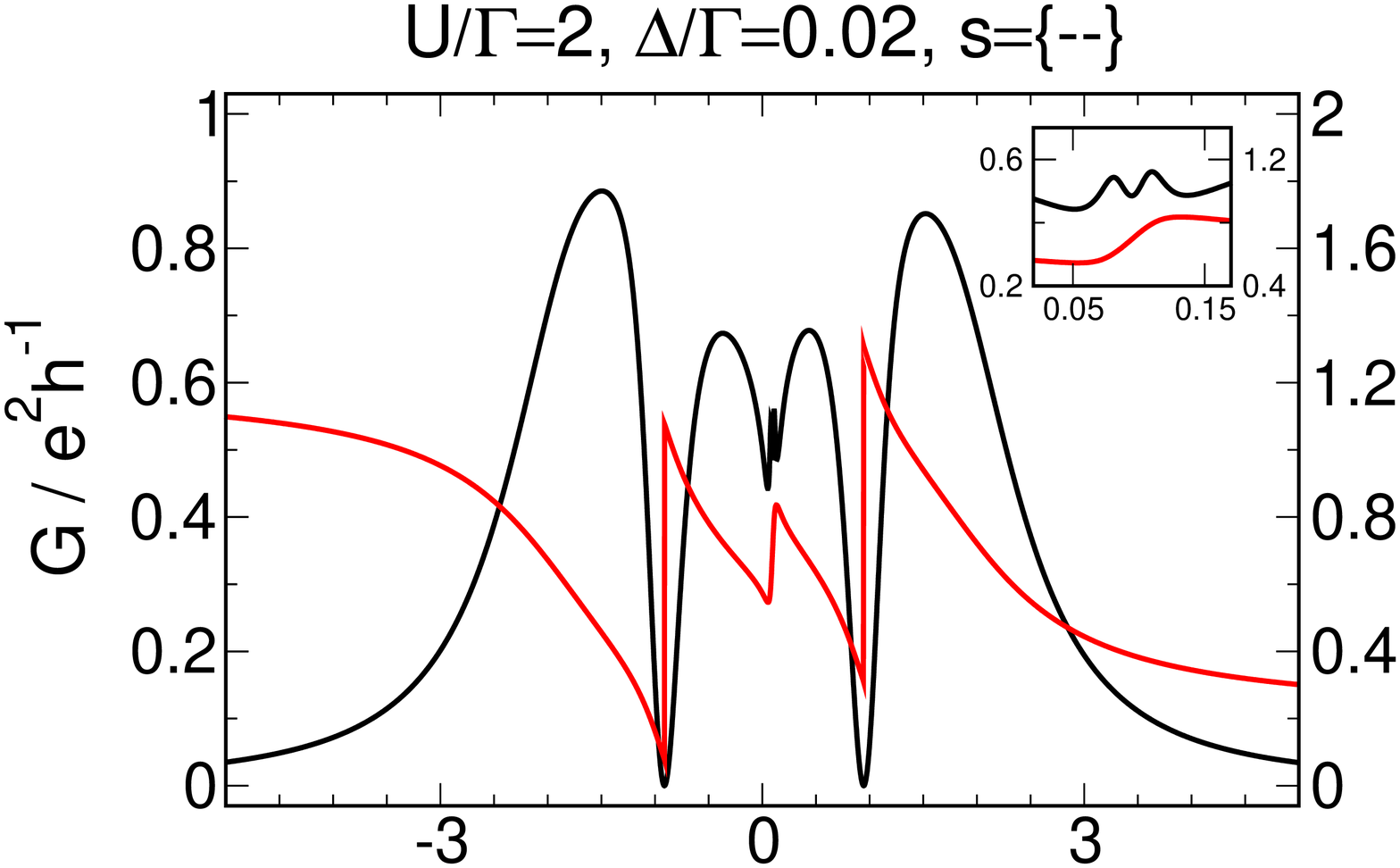}\hspace{0.015\textwidth}
        \includegraphics[width=0.475\textwidth,height=4.4cm,clip]{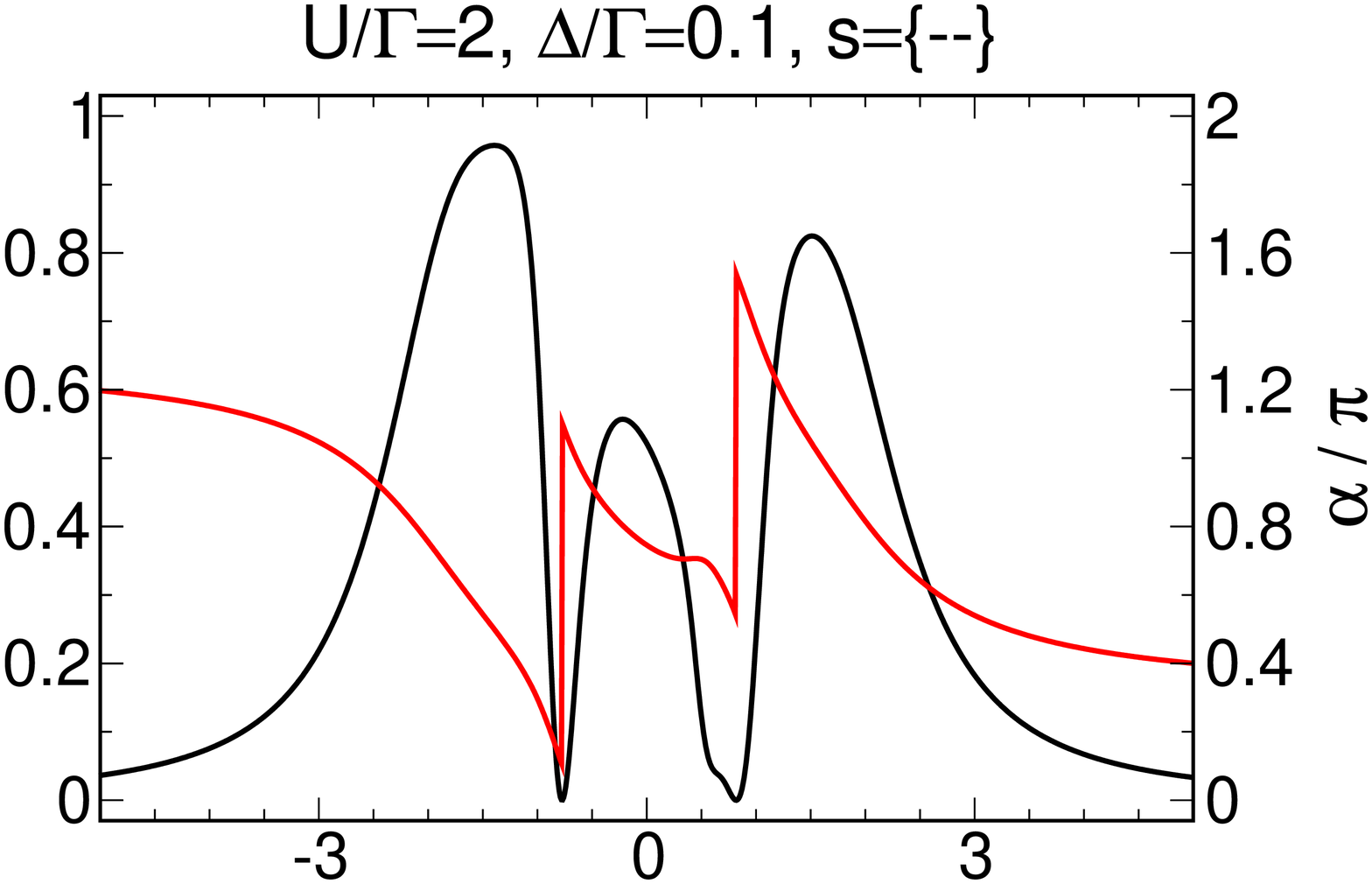}\vspace{0.3cm}
        \includegraphics[width=0.495\textwidth,height=5.2cm,clip]{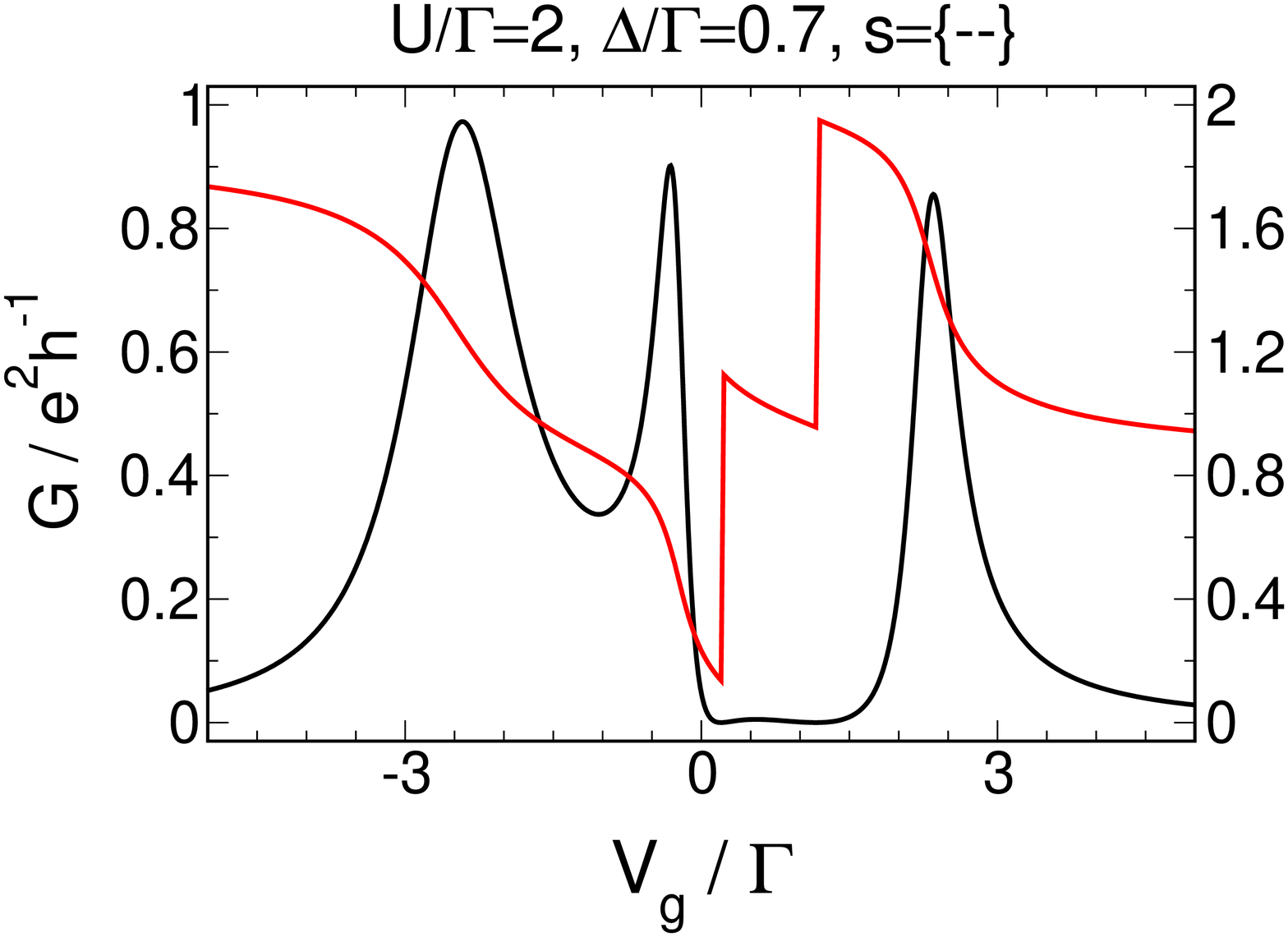}\hspace{0.015\textwidth}
        \includegraphics[width=0.475\textwidth,height=5.2cm,clip]{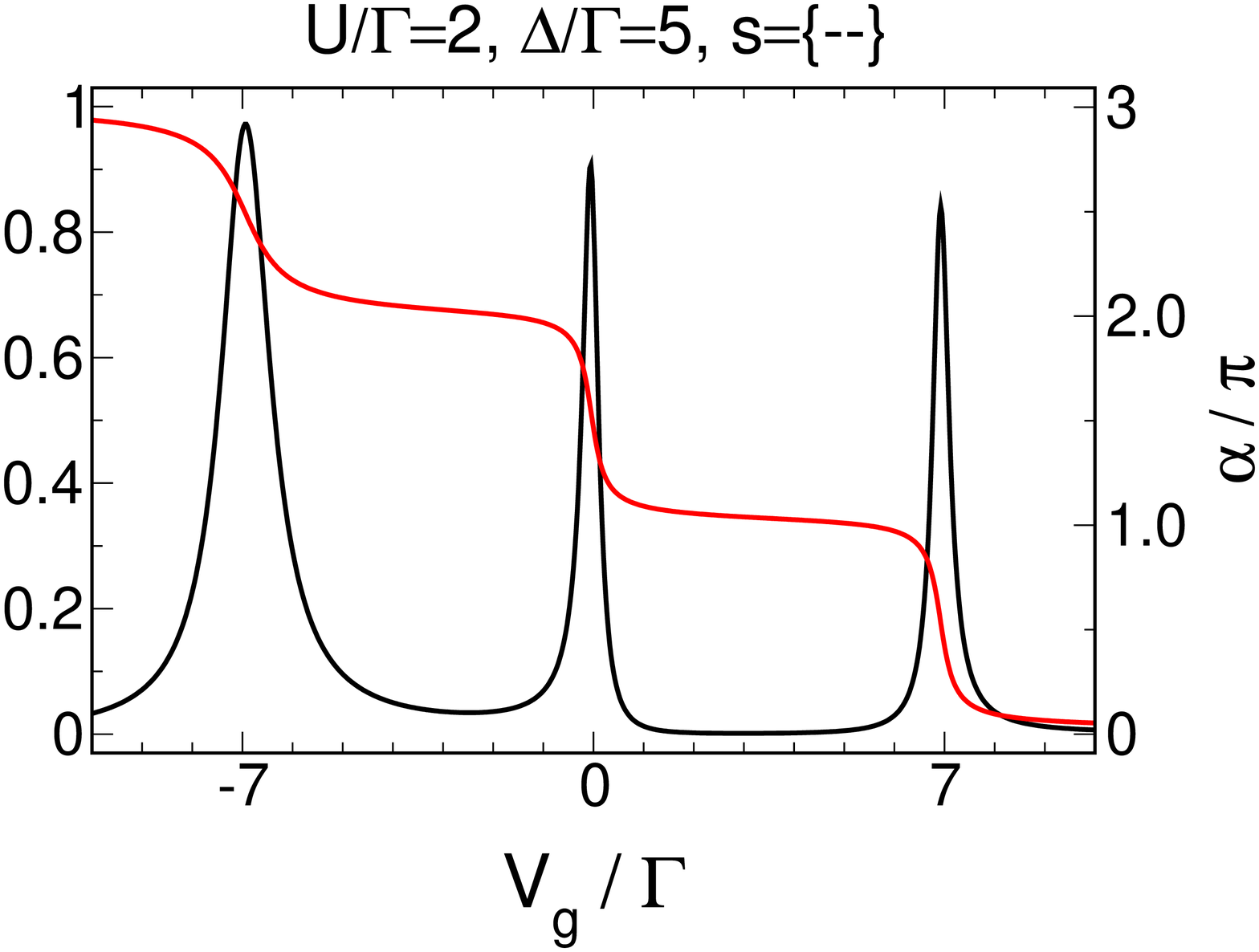}
        \caption{The same as Fig.~\ref{fig:OS.td.pmpda}, but with generic hybridisations for regime W, $\Gamma=\{0.06~0.14~0.13~0.07~0.25~0.35\}$.}
\label{fig:OS.td.pmpdb}
\end{figure}
\afterpage{\clearpage}

Finally, the behaviour of the conductance for $s=\{--\}$ in between the limits of small and large $\Delta$ can be described as follows. In regime S (shown in Fig.~\ref{fig:OS.td.pmpda}), the central peak (the one `corresponding' to the dot $\tilde A$) splits up into three at a scale $\Delta_\tn{cross}$ while the outer ones become vanishingly small. In contrast, the central resonance gradually disappears in regime W (depicted in Fig.~\ref{fig:OS.td.pmpdb}), and the outer one that one would relate to the more strongly coupled dot splits up into two. For extreme choices of the hybridisations, namely $\Gamma_B\ll\Gamma$ and $\Gamma_A\stackrel{\ll}{\gg}\Gamma_C$, the resonance close to $V_g=0$ in regime W always remains a maximum surrounded by transmission zeros and corresponding phase jumps though becoming unidentifiable on a scale $G$ of the order of the unitary limit. This similarly occurs in the noninteracting case and we will delay a more detailed characterisation of the affected parameter region. In general, the central resonance completely vanishes in the sense that it becomes a minimum and the transmission zeros disappear. As mentioned above, an additional correlation induced sharp peak and especially the double phase jumps (lapses) on both of its sides appear if $\Delta$ exceeds a scale $\Delta_\tn{DPL}$ depending on the hybridisations. Normally, this happens before the crossover regime is reached, but if in particular for large left-right asymmetries it happens that $\Delta_\tn{DPL}\approx\Delta_\tn{cross}$, these additional features will never be observed. Finally, in both regime S and W we end up with the well-known Lorentzian three-peak structure, only that now the phase does not exhibit any jumps between any of them. This is consistent with the cases considered before, since here the couplings of the levels corresponding to both pairs of subsequent resonances differ in sign, $\tn{sgn}(t_A^Lt_A^R)\neq\tn{sgn}(t_B^Lt_B^R)$, and $\tn{sgn}(t_B^Lt_B^R)\neq\tn{sgn}(t_C^Lt_C^R)$.

\subsubsection{Effective Level Interpretation}

\begin{figure}[t]	
\centering
      \includegraphics[width=0.495\textwidth,height=5.2cm,clip]{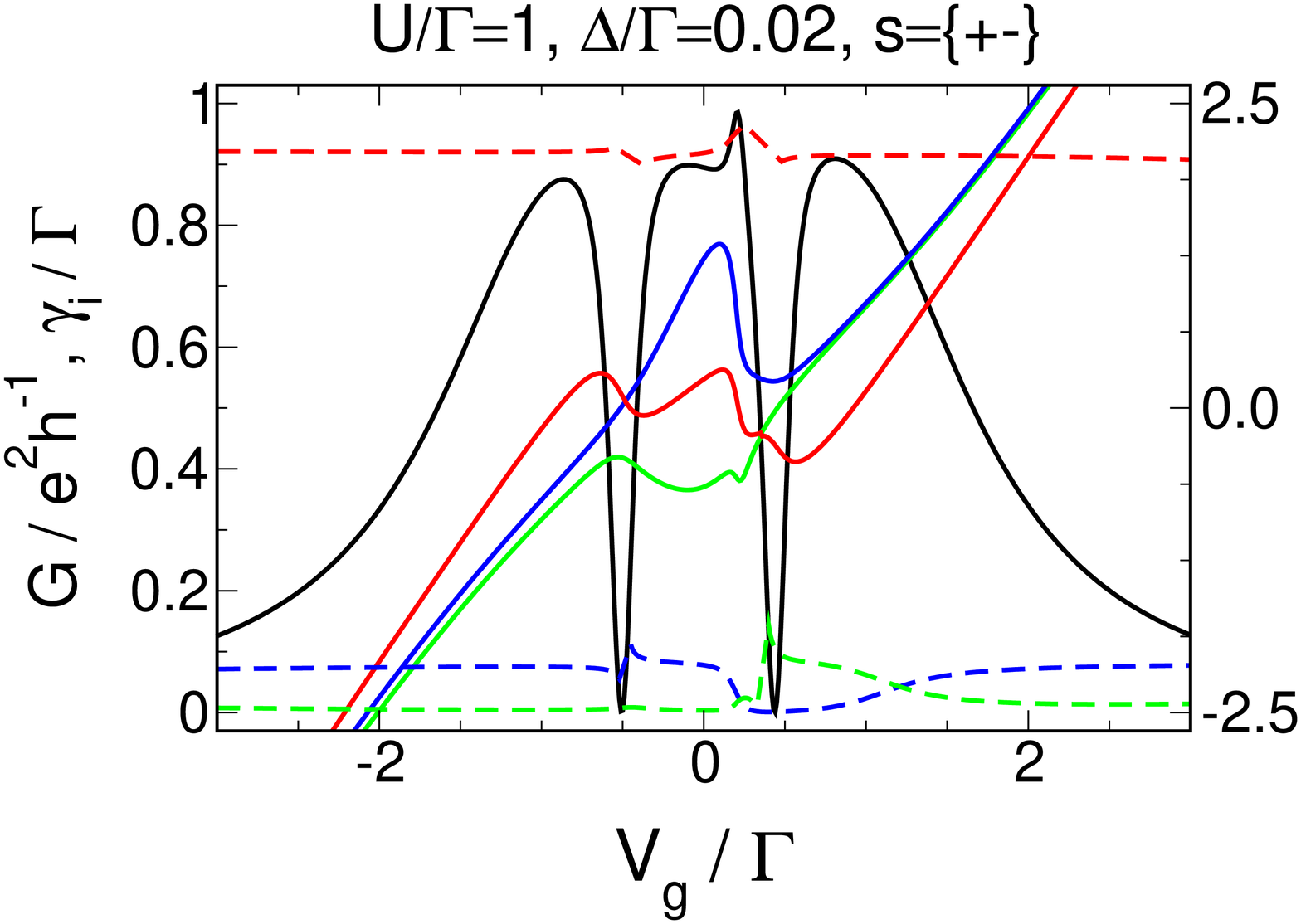}\hspace{0.015\textwidth}
        \includegraphics[width=0.475\textwidth,height=5.2cm,clip]{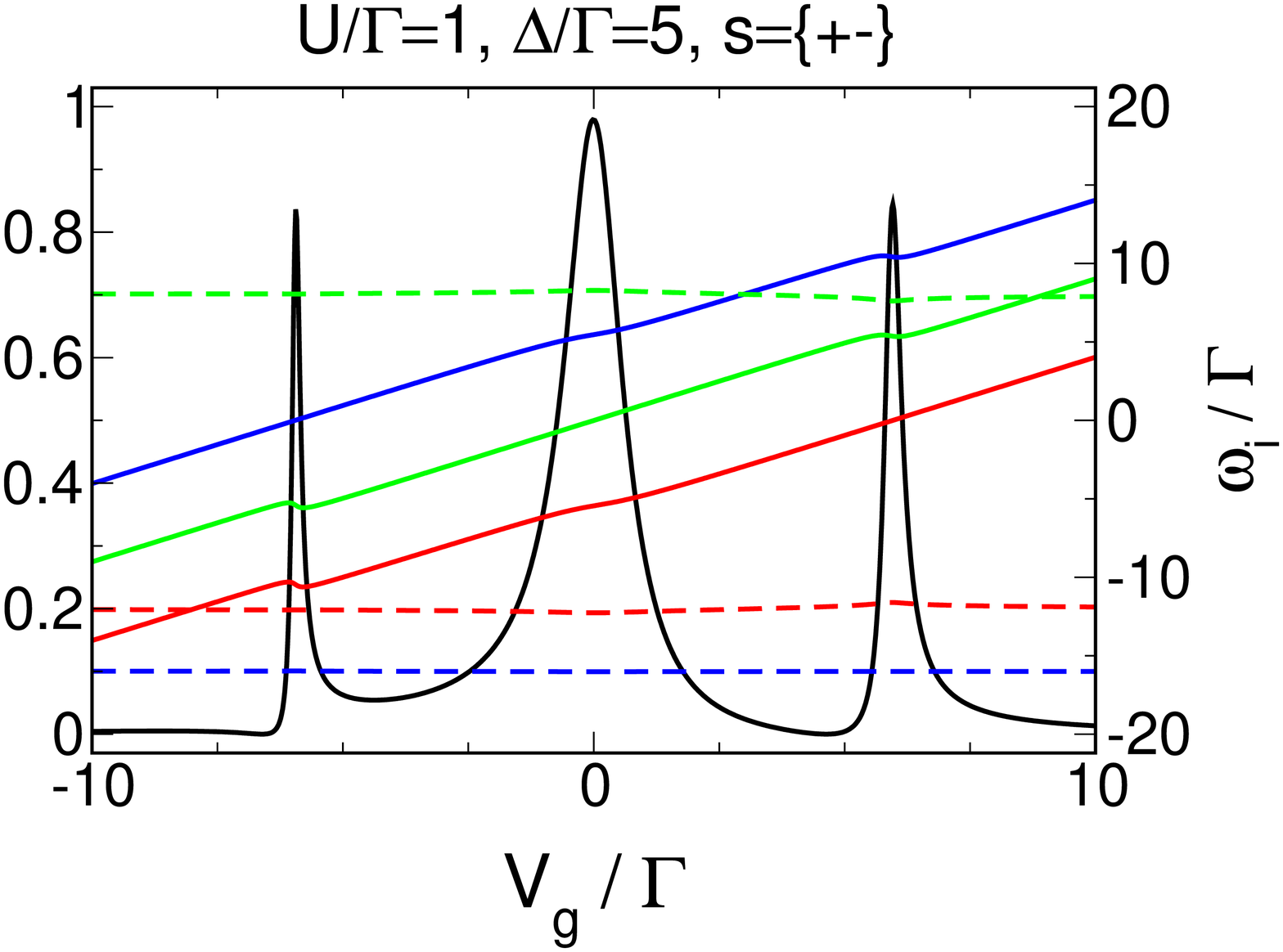}
        \caption{Conductance $G$ (black), effective eigenenergies $\omega_i$ (solid lines) and hybridisations $\gamma_i=\gamma_i^L+\gamma_i^R$ (dashed lines) for parallel triple dots with $\Gamma=\{0.06~0.14~0.3~0.4~0.07~0.03\}$ for two different level spacing. At $\Delta\ll\Gamma$, the most strongly coupled effective level (red) crosses the chemical potential upwards at each resonance, while close to the transmission zeros this level crosses $\mu$ downwards but at the same time it crosses another level, giving rise to a Fano-antiresonance. The interaction enhances the bare asymmetry of the hybridisations. Note that due to the degeneracy of the levels for $|V_g|\to\infty$, there is no need to recover the original $\Gamma_l$ in this limit. For $\Delta\gg\Gamma$, all $\omega_i$ cross $\mu$ linearly, hence $s$ governs the behaviour of $\alpha$ as for $U=0$.}
\label{fig:OS.td.efflevel}
\end{figure}

As in the double dot case, further insight in the underlying physics might be gained by considering the effective energies $\omega_i$ (the eigenvalues of the isolated dot Hamiltonian with renormalized parameters) and the new hybridisations $\gamma_i$ (obtained from the corresponding eigenvectors) of our system in the limits $\Delta\ll\Gamma$ and $\Delta\gg\Gamma$. For small level spacings, large $U$, and arbitrary $s$ we find (again in analogy to \cite{imry}) that at each of the three Coulomb peaks the most strongly coupled level crosses the chemical potential of the leads $\mu=0$ upwards leading to a resonance (Fig.~\ref{fig:OS.td.efflevel}, left panel). This is possible since the effective levels do not depend monotonically on the gate voltage (which is again related to population inversion). In particular, near the transmission zeros located in between the peaks, the level with the largest $\gamma_i$ crosses the chemical potential downwards, but roughly at the same point it crosses another level so that a zero instead of a peak is observed. This can be interpreted as a Fano-antiresonance, especially since it shows that the interaction enhances the bare asymmetry of the $\Gamma_l$ such that $\gamma_i$ of the most strongly coupled effective level is significantly larger than the largest $\Gamma_l$.

\renewcommand\tabularxcolumn[1]{m{#1}}
\begin{table}[t]
\begin{center}\begin{small}
\fbox{\begin{tabularx}{0.99\textwidth}{>{\centering}p{0.17\textwidth}|>{\centering}m{0.17\textwidth}|X}

Region of $\Delta$ & Typical values & \vspace{0.1cm}\centering Related phenomenon\vspace{0.1cm} \tabularnewline \hline

$\Delta\ll\Gamma$ & $\Delta/\Gamma=0.01$ & \vspace{0.1cm}Three `Coulomb' resonances of equal height, width of order $\Gamma$ and separation $U$ for arbitrary $\{s,\Gamma\}$; additional correlation induced structures if the interaction is large enough.\vspace{0.1cm}
\tabularnewline \hline

$\Delta_\tn{CIR}$ & $\Delta/\Gamma=0.05$ & \vspace{0.1cm}Vanishing of the CIRs located at $V_g\approx\pm U/2$.\vspace{0.1cm}
\tabularnewline \hline

$\tilde\Delta_\tn{CIR}$ & $\Delta/\Gamma=0.05$ & \vspace{0.1cm}Vanishing of the sharp features located close to half filling.\vspace{0.1cm}
\tabularnewline \hline

$\Delta_\tn{DPL}$ & \vspace{0.1cm}Depends on left-right asymmetry.\vspace{0.1cm} & Appearance of the additional transmission zeros and corresponding double phase jumps (lapses) for $s=\{--\}$ and $\Gamma_B<\Gamma_A+\Gamma_C$.
\tabularnewline \hline

$\Delta_\tn{cross}$ & $\Delta/\Gamma=0.1$ & \vspace{0.1cm}Set-in of the crossover regime where some peaks become increasingly small while others split up.\vspace{0.1cm}
\tabularnewline \hline

$\Delta\gg\Gamma$ & $\Delta/\Gamma=5.0$ & \vspace{0.1cm}Three Lorentzian resonances separated by $U+\Delta$ which height and width are determined by the corresponding hybridisations $\Gamma_l^L$ and $\Gamma_l^R$. The transmission phase jumps by $\pi$ between those peaks related to levels which couplings do not differ in sign.
\end{tabularx}}
\end{small}\end{center}
\caption{The different $\Delta$ - regimes of the triple dot geometry.}
\label{tbl:OS.td.scales}
\end{table}

For large level spacings every $\omega_i$ depends linearly on the gate voltage and is only shifted by $U$ when one of the other levels is filled (Fig.~\ref{fig:OS.td.efflevel}, right panel). The hybridisations $\gamma_i$ almost coincide with the original ones since virtually no hopping is generated by the flow. However, well-separated levels linearly crossing the chemical potential just lead to the noninteracting Lorentzian resonances which width and height are determined by the corresponding hybridisations $\Gamma_l^L$ and $\Gamma_l^R$, only that now their separation is $U+\Delta$ rather than $\Delta$. Furthermore it follows that the relation between the relative signs of the level-lead couplings $s$ and the behaviour of the transmission phase in between the peaks has to be the same as for $U=0$.

\subsubsection{The Noninteracting Case}

\begin{figure}[t]	
	\centering
        \includegraphics[width=0.475\textwidth,height=4.8cm,clip]{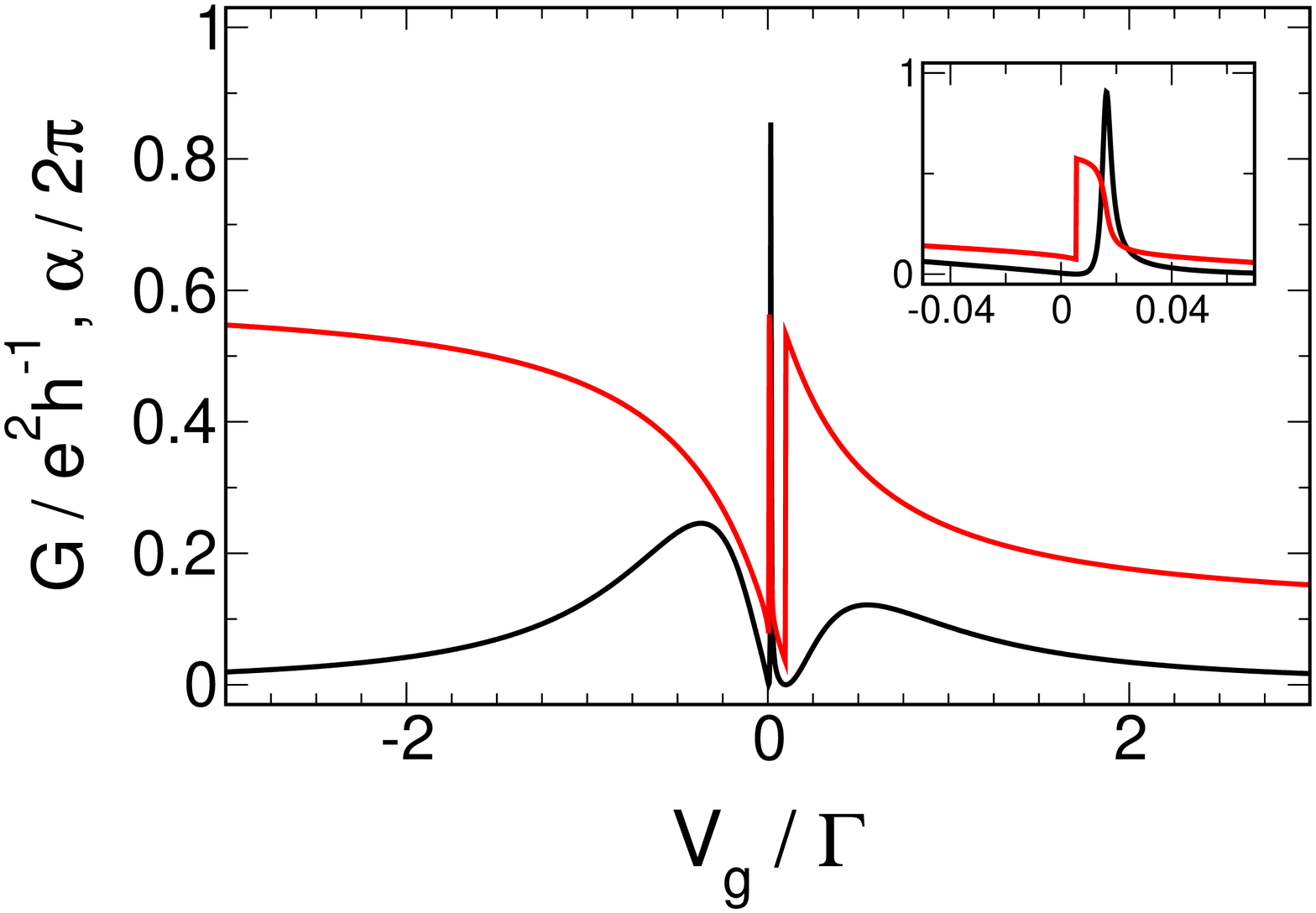}\hspace{0.035\textwidth}
        \includegraphics[width=0.475\textwidth,height=4.8cm,clip]{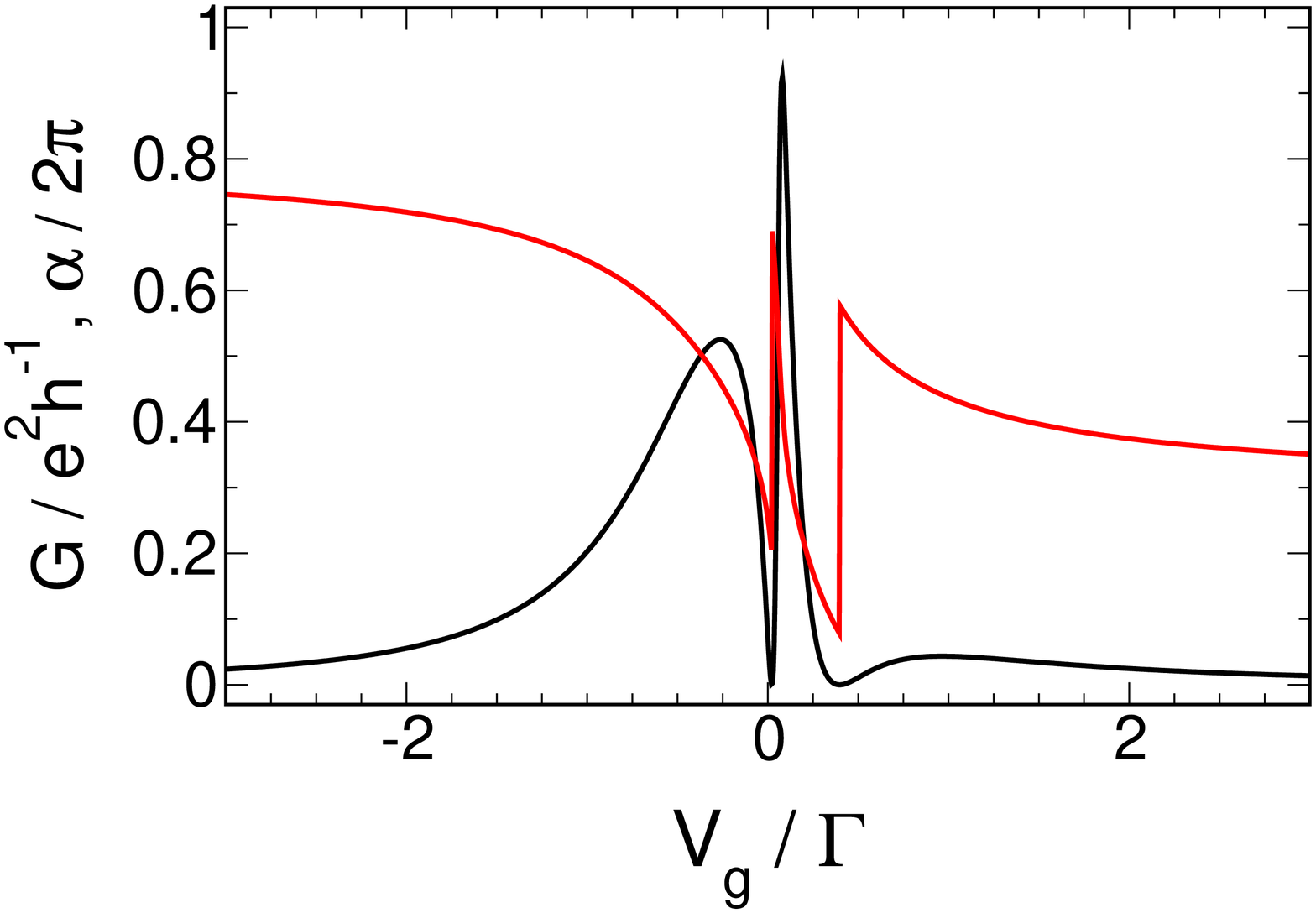}\vspace{0.3cm}
        \includegraphics[width=0.475\textwidth,height=4.8cm,clip]{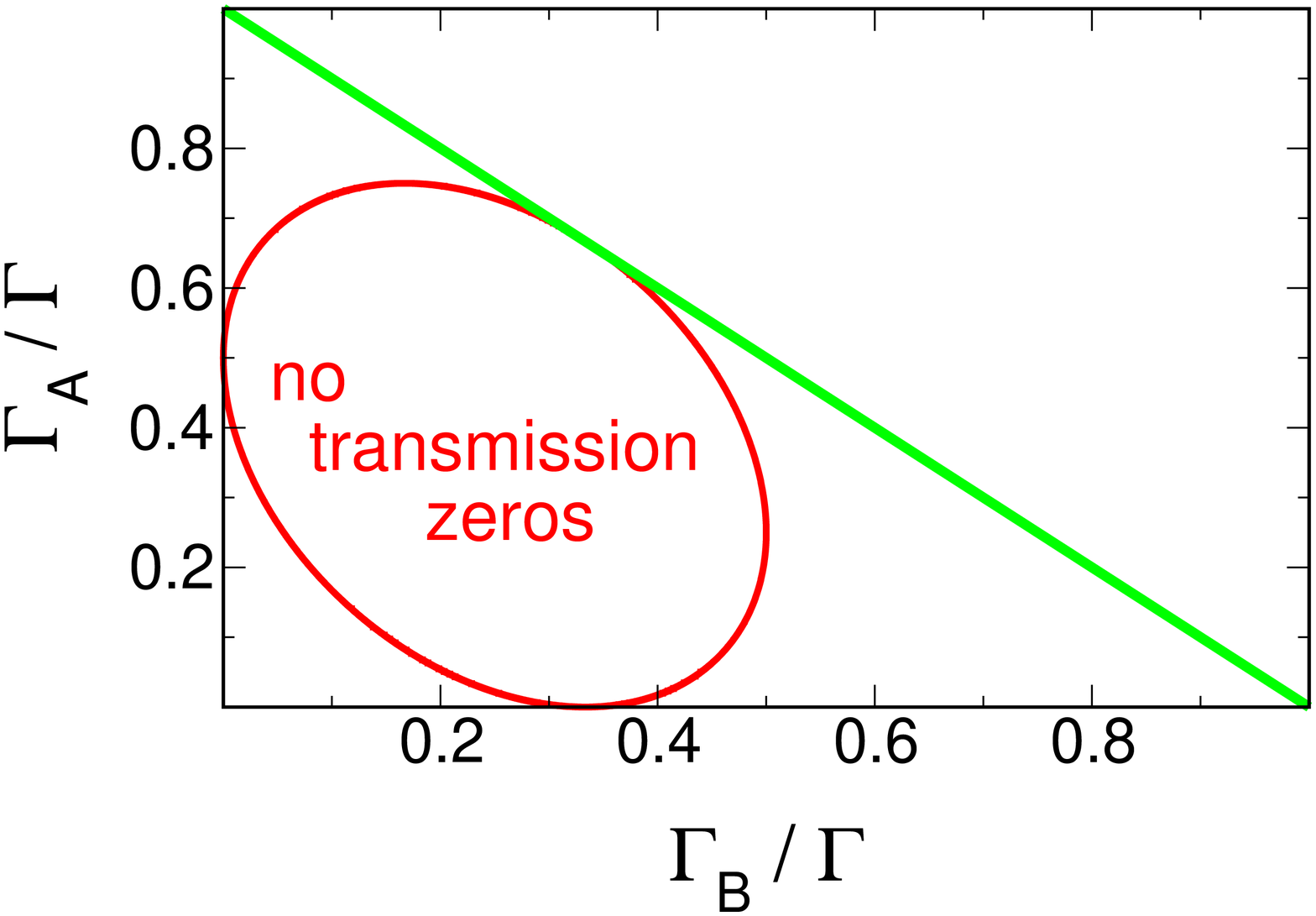}\hspace{0.035\textwidth}
        \includegraphics[width=0.475\textwidth,height=4.8cm,clip]{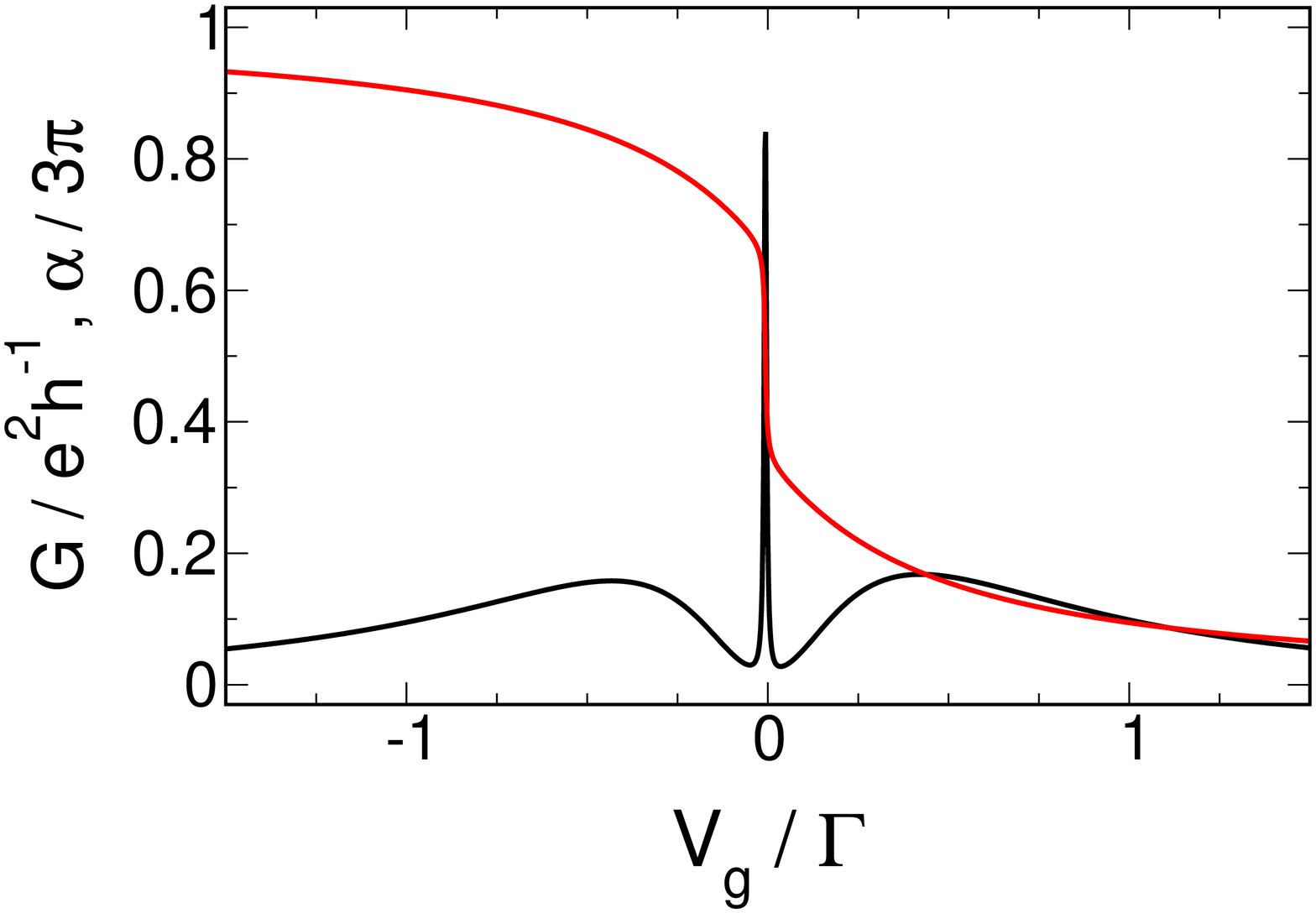}
        \caption{\textit{Upper row:} Gate voltage dependence of the conductance $G$ and transmission phase $\alpha$ (red) of noninteracting parallel triple dots with $s=\{+-\}$, $\Gamma=\{0.06~0.14~0.07~0.03~0.3~0.4\}$, $\Delta/\Gamma=0.05$ (left panel), and $\Delta/\Gamma=0.2$ (right panel). \textit{Lower row, left panel:} Appearance of transmission zeros as a function of the hybridisations for the left-right symmetric, noninteracting case with $s=\{--\}$ for arbitrary level spacing. \textit{Lower row, right panel:} The same as in the upper row, but for $s=\{--\}$, $\Gamma=\{0.15~0.25~0.1~0.2~0.12~0.18\}$, and $\Delta/\Gamma=0.05$.}
\label{fig:OS.td.wwfrei}
\end{figure}

Finally, we will consider the parallel triple dot geometry at $U=0$. Similar to the double dot, the behaviour of the conductance in the region of large level spacings and in the crossover regime does not change if the interaction is gradually switched off. However, one observes the width of the $V_g$-region where the conductance is significantly larger than zero to decrease when the level spacing is lowered. Hence in the limit $\Delta\ll\Gamma$ we would expect fairly sharp structures on a scale of $\Gamma$ arising from simultaneous transport through several levels in analogy to the sharp dip close to $V_g=0$ in $G(V_g)$ of the double dot with $U=0$. Therefore it is reasonable to seek for analytical statements that follow from the exact solution of the noninteracting problem. Unfortunately, evaluating the expression for the conductance (\ref{eq:OS.td.cond}) with $\mc G$ replaced by the free propagator yields a very lengthy expression that is hardly suited to reveal interesting information. Nevertheless, in the special case of left-right symmetry some insight in the number and position of the transmission zeros can be gained. For the double dot with finite level spacing, sticking to $\Gamma_l^L=\Gamma_l^R$ does not lead to non-generic behaviour and thus it seems reasonable that the same holds for the triple dot geometry. Regardless of that, we have numerically confirmed that everything stated here indeed represents the generic behaviour. Solving for the zeros of (\ref{eq:OS.td.cond}) then yields
\begin{equation}\label{eq:OS.td.wwfrei}
\frac{V_g}{\Delta} = f_1(\Gamma,s) \pm \frac{\sqrt{f_3(\Gamma,s)}}{f_2(\Gamma,s)},
\end{equation}
with
\begin{equation*}
f_1(\Gamma,s) =
\begin{cases}
-4\Gamma_A-2\Gamma_B+0.5\\[1ex]
\frac{4\Gamma_B-1}{16\Gamma_A+16\Gamma_B-2}\\[1ex]
\frac{8\Gamma_A+4\Gamma_B-1}{16\Gamma_B-2}
\end{cases}\hspace{-0.3cm},~~~
f_2(\Gamma,s)=
\begin{cases}
2 & \hspace{0.7cm}\tn{for } s={++} \\[1ex]
16\Gamma_A+16\Gamma_B-2 & \hspace{0.7cm} \tn{for } s={+-} \\[1ex]
16\Gamma_B-2 & \hspace{0.7cm}\tn{for } s={--}
\end{cases},
\end{equation*}
and, most important,
\begin{equation*}
f_3(\Gamma,s)=
\begin{cases}
64\Gamma_A^2-16\Gamma_A+64\Gamma_A\Gamma_B+8\Gamma_B+16\Gamma_B^2+1 & \tn{for } s={++} \\[1ex]
144\Gamma_B^2-48\Gamma_B+132\Gamma_A\Gamma_B+1 & \tn{for } s={+-} \\[1ex]
144\Gamma_B^2-48\Gamma_B+64\Gamma_A\Gamma_B+64\Gamma_A^2-16\Gamma_A+1 & \tn{for } s={--}
\end{cases}.
\end{equation*}
We have implicitly assumed that $\Gamma$ is chosen as the unit of energy, implying that we can substitute $\Gamma_C=1-\Gamma_A-\Gamma_B$. The first implication of (\ref{eq:OS.td.wwfrei}) is that the number of transmission zeros is independent of the level spacing, and it turns out (by multidimensional analysis or simply by performing a 3d plot) that for $s=\{++,+-\}$ the function $f_3(\{s,\Gamma\}$ is larger than zero for $[\Gamma_A,\Gamma_B]=[0\ldots1,0\ldots\Gamma_A]$ so that their number is indeed two. This is consistent with the general (numerical) observation that the region of large $\Delta$ as well as the crossover regime are identical for both $U=0$ and $U>0$ for arbitrary choice of $\{\Gamma\}$ if one keeps in mind that for $s=\{+-\}$ in both regime S and W the transmission zeros present for small level spacings lie outside the Lorentzian resonances but do not vanish in the limit $\Delta\gg\Gamma$. However, the second important insight that can be gained from the analytical solution is that the gate voltages with $G(V_g)=0$ depend linearly on the the level spacing with a prefactor that is roughly of the order of $\Gamma$. Hence for $\Delta\ll\Gamma$ though still having the correct number of transmission zeros we expect to find a very sharp three-peak structure on the usual scale of the hybridisations (Fig.~\ref{fig:OS.td.wwfrei}, upper left panel). The structures become broader when increasing $\Delta$, but for $s=\{+-\}$ the crossover regime sets in at scale $\Delta\lesssim\Gamma$, implying that the heights of the resonances begin to differ (Fig.~\ref{fig:OS.td.wwfrei}, upper right panel).

However, for $s=\{+-\}$ (and of course for $s=\{++\}$) the behaviour of the conductance is identical within both the interacting and the $U=0$ case if one is only concerned about the number of transmission zeros and the position of the resonances. This turns out to be different if we choose $s=\{--\}$, because here the function $f_3(s,\Gamma)$ is no longer positive for all hybridisations of interest. By solving $f_3(s,\Gamma)=0$ we can compute the region of $\Gamma$ where the conductance does not exhibit any zeros (Fig.~\ref{fig:OS.td.wwfrei}, lower left panel). We find that there are two solutions of $G(V_g)=0$ if dot B is strongly coupled, implying that the behaviour of the conductance in regime S is identical to the interacting case for all $\Delta$. On the other hand, in most parts of regime W we observe no transmission zeros which is still consistent with the claim that the crossover regime and region of large level spacings are identical for both $U\neq 0$ and $U=0$. However, since their number is independent of $\Delta$, additional zeros do not appear in the limit $\Delta\ll\Gamma$ in contrast to the interacting case. Hence for small level spacings, we observe a very sharp three-peak structure but the conductance stays finite in between them, and, more important, the phase evolves continuously (Fig.~\ref{fig:OS.td.wwfrei}, lower right panel). One exception to this is found for $\Gamma_A,\Gamma_B\ll\Gamma_C$ or $\Gamma_C,\Gamma_A\ll\Gamma_B$ (the upper and lower left corner in the lower left panel of Fig.~\ref{fig:OS.td.wwfrei}). Here, we find indeed two transmission zeros within the three-peak structure for $\Delta\ll\Gamma$. As the hybridisations are from regime W, one of these peaks splits up while the central one becomes small when the level spacing is increased. In contrast to the other parameters in regime W, the resonance close to $V_g=0$ always remains a maximum surrounded by zeros and phase jumps, but it becomes too small to identify on scale of the order of the unitary limit. As mentioned above, identical behaviour is observed in presence of the interactions.

\subsection{Parallel Geometries with More Than Three Dots}\label{sec:OS.more}

\subsubsection{Four Levels}

In this section, we will generalise the previously considered models by adding further levels. We will focus on the $\Delta$ regions of most interest, which are in particular the limits $\Delta\ll\Gamma$ and $\Delta\gg\Gamma$, and we refrain from providing a complete description of the crossover regime.

First, we will study a four-level dot where the projected free propagator reads
\begin{equation}\label{eq:OS.qd.g}
\left[\mc G^0(i\omega)\right]^{-1} = i\omega-V_g+
\begin{pmatrix}
1.5\Delta & & & \\ & 0.5\Delta & & \\
& & -0.5\Delta & \\
& & & -1.5\Delta
\end{pmatrix} - H_\tn{eff}+H_{PP},
\end{equation}
and the projected part $H_\tn{eff}-H_{PP}$ has the well-known form (\ref{eq:DOT.gpp}),
\begin{equation*}
H_\tn{eff}-H_{PP}=  
\sum_{s=L,R}\sum_{l,l'}\Big[-i~\tn{sgn}(\omega) S_{l,l'}(s)\sqrt{\Gamma_l^s\Gamma_{l'}^s}|l\rangle\langle l'| \Big],
\end{equation*}
with the definition
\begin{equation*}
S_{l,l'}(s):= \delta_{s,L}+\delta_{s,R}
\begin{pmatrix}
1 & s_1 & s_2 & s_3 \\
s_1 & 1 & s_1s_2 & s_1s_3 \\
s_3 & s_1s_2 & 1 & s_2s_3 \\
s_3 & s_1s_3 & s_2s_3 & 1
\end{pmatrix}.
\end{equation*}
The conductance is computed via (\ref{eq:DOT.leitwert}), which here reads
\begin{equation}\label{eq:OS.qd.cond}
G = \frac{e^2}{h}\Big|2\pi\rho_{\tn{lead}}(0)\sum_{l,l'}s_{l,l'}\sqrt{\Gamma_l^R\Gamma_{l'}^L}\mc G_{l;l'}(0)\Big|^2,
\end{equation}
where we have defined
\begin{equation*}
s_{l,l'}:=
\begin{cases}
s_1 & \tn{for } l=B \tn{ and arbitrary } l' \\
s_2 & \tn{for } l=C \tn{ and arbitrary } l' \\
s_3 & \tn{for } l=D \tn{ and arbitrary } l' \\
1 & \tn{otherwise}.
\end{cases}
\end{equation*}

\begin{figure}[t]	
        \centering
        \includegraphics[width=0.495\textwidth,height=4.4cm,clip]{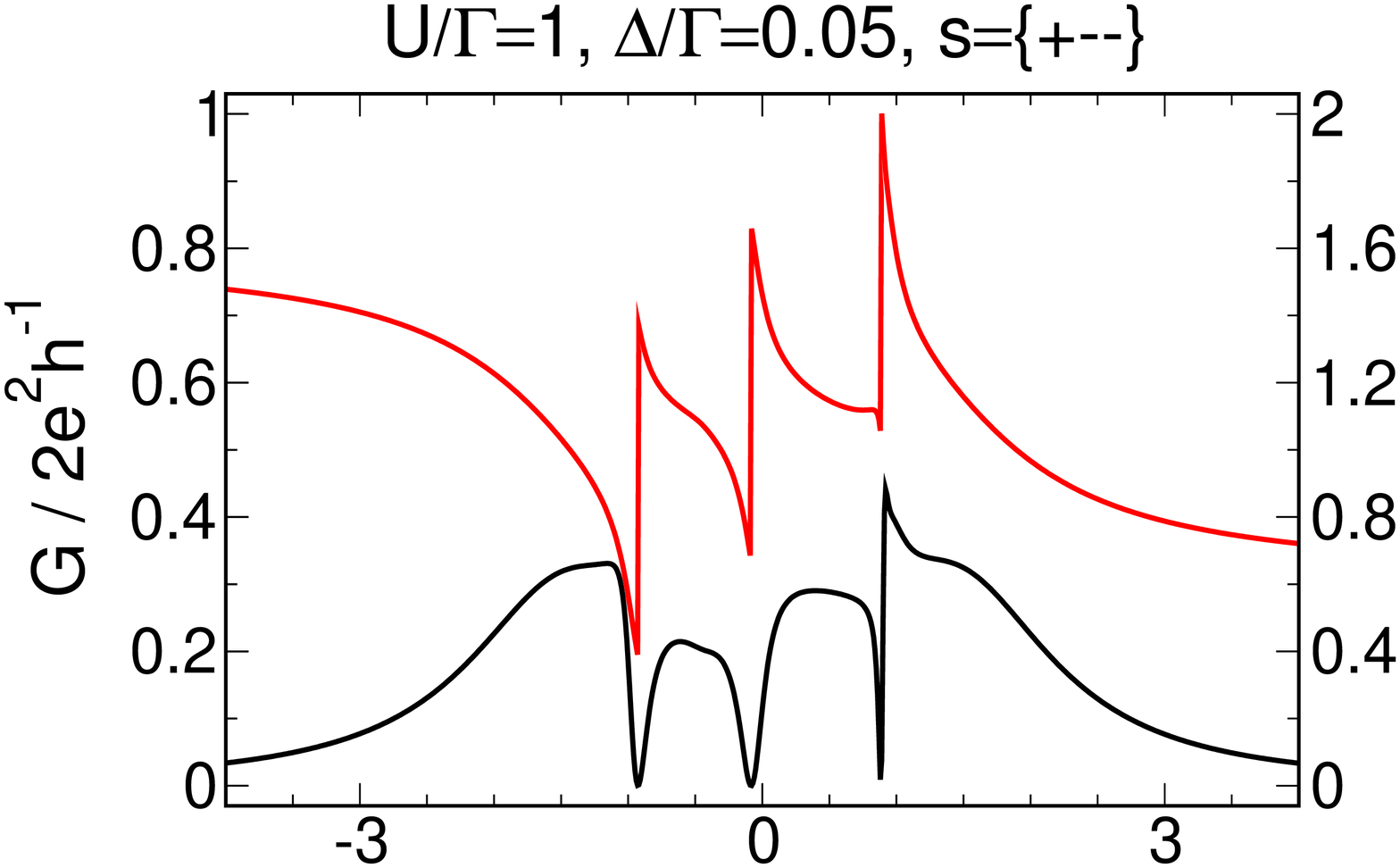}\hspace{0.015\textwidth}
        \includegraphics[width=0.475\textwidth,height=4.4cm,clip]{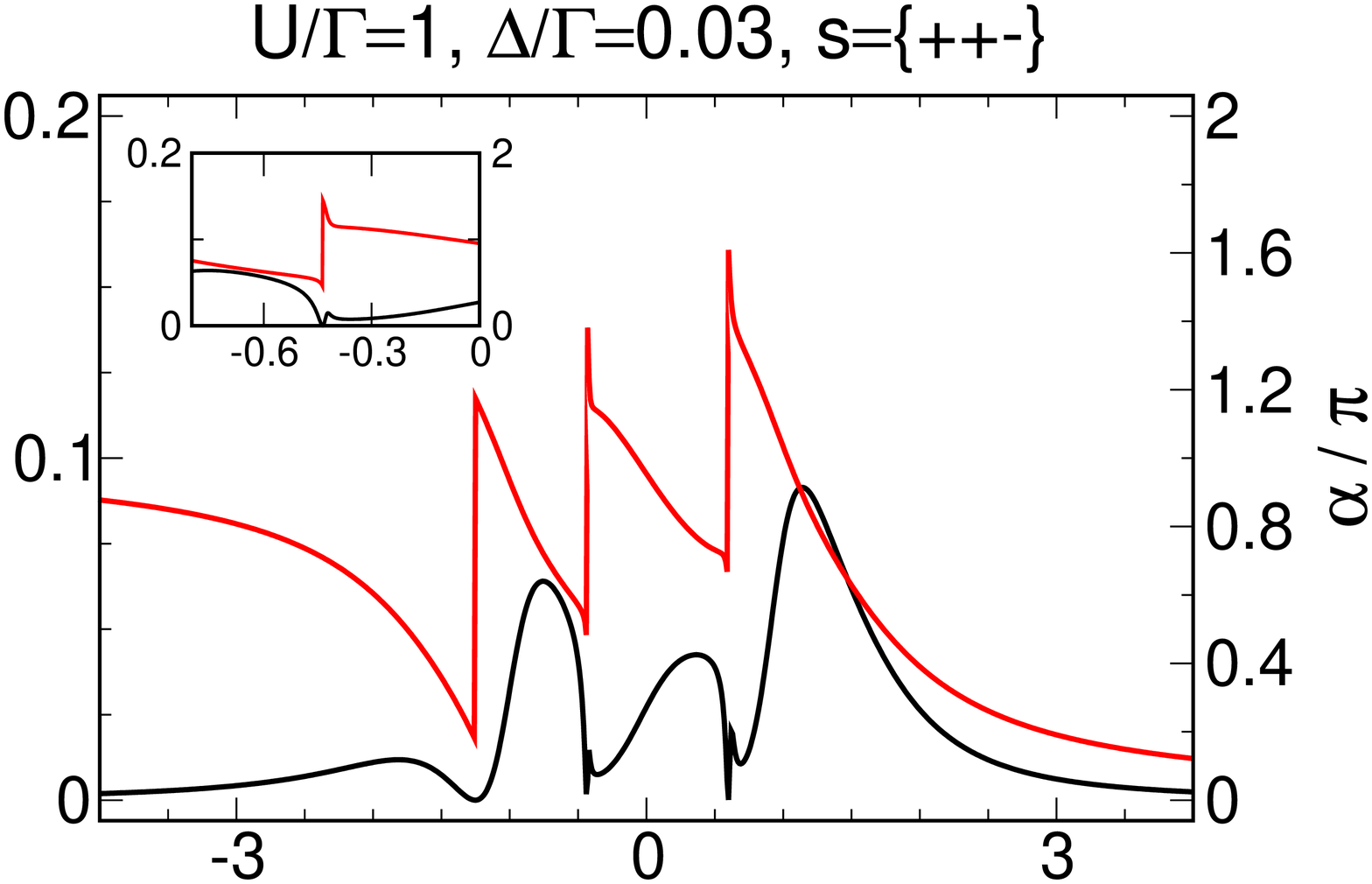}\vspace{0.3cm}
        \includegraphics[width=0.495\textwidth,height=5.2cm,clip]{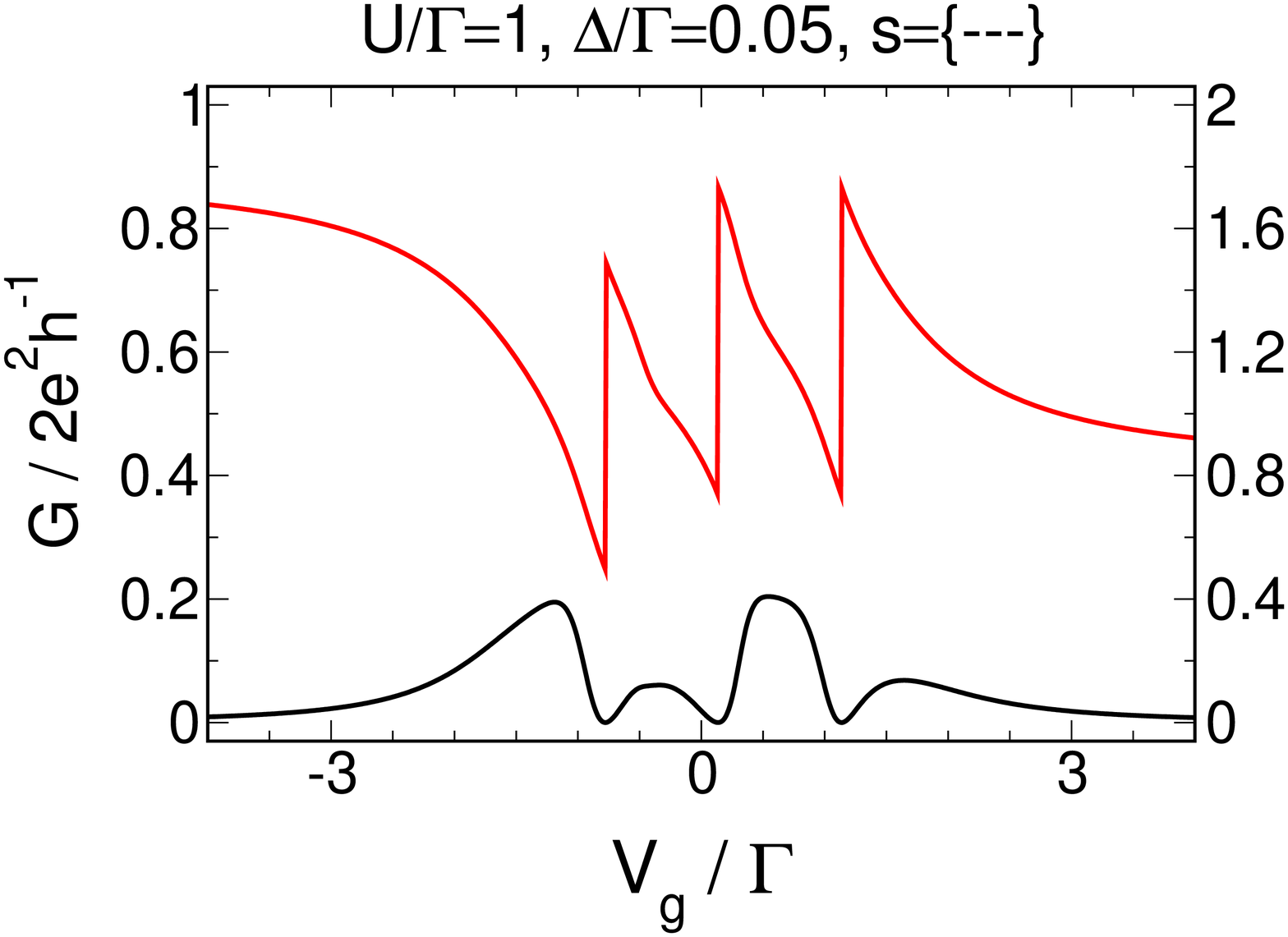}\hspace{0.015\textwidth}
        \includegraphics[width=0.475\textwidth,height=5.2cm,clip]{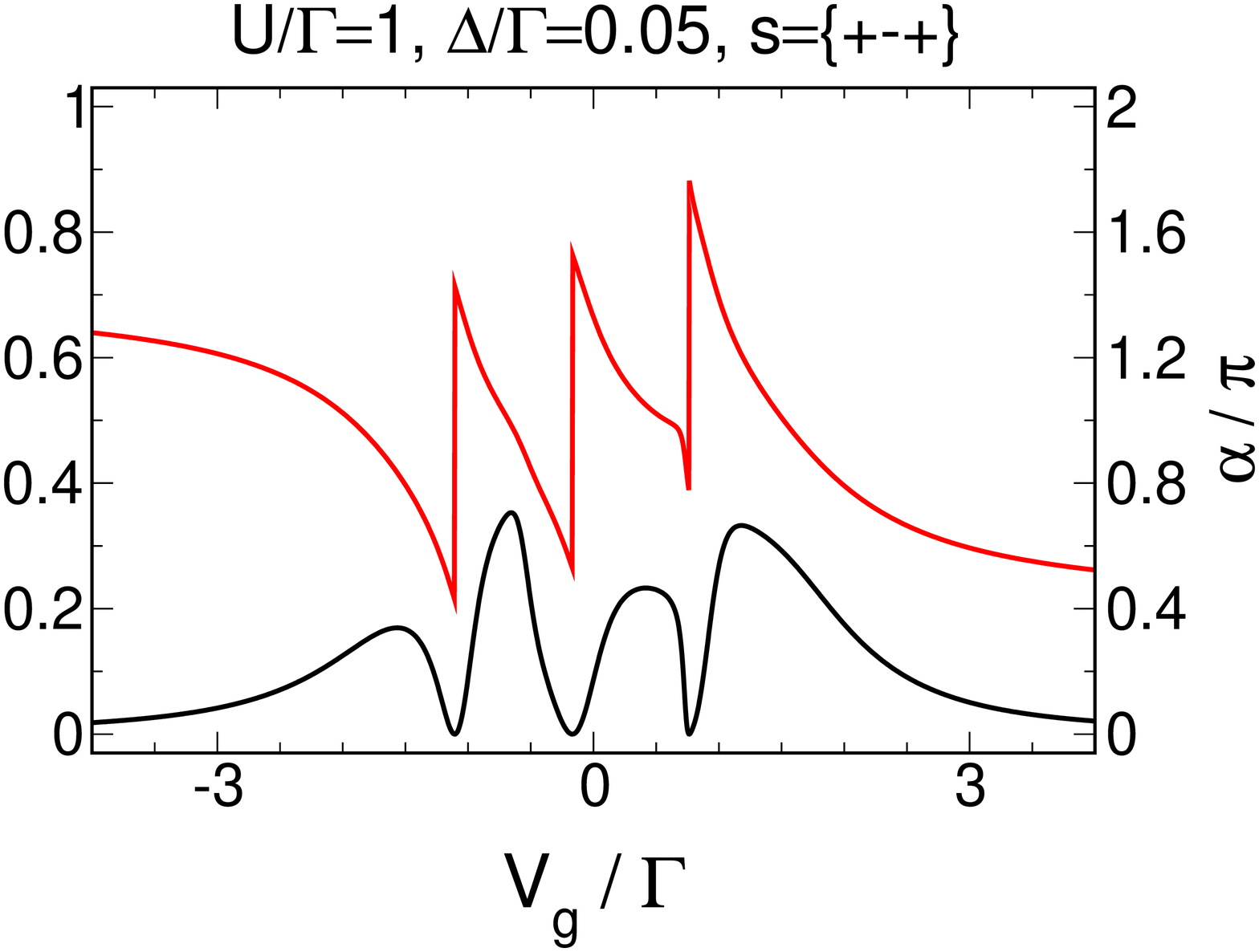}
        \caption{Gate voltage dependence of the conductance $G$ (black) and transmission phase $\alpha$ (red) for a four-level dot with nearly degenerate levels and generic level-lead hybridisations $\Gamma=\{0.05~0.15~0.075~0.225~0.025~0.075~0.1~0.3\}$ for various parameters. For $s=\{+-+\}$, these hybridisations are not generic, and hence we have chosen $\Gamma=\{0.12~0.28~0.1~0.3~0.04~0.06~0.03~0.07\}$.}
\label{fig:OS.qd.u}\vspace{0.6cm}
        \includegraphics[width=0.495\textwidth,height=4.4cm,clip]{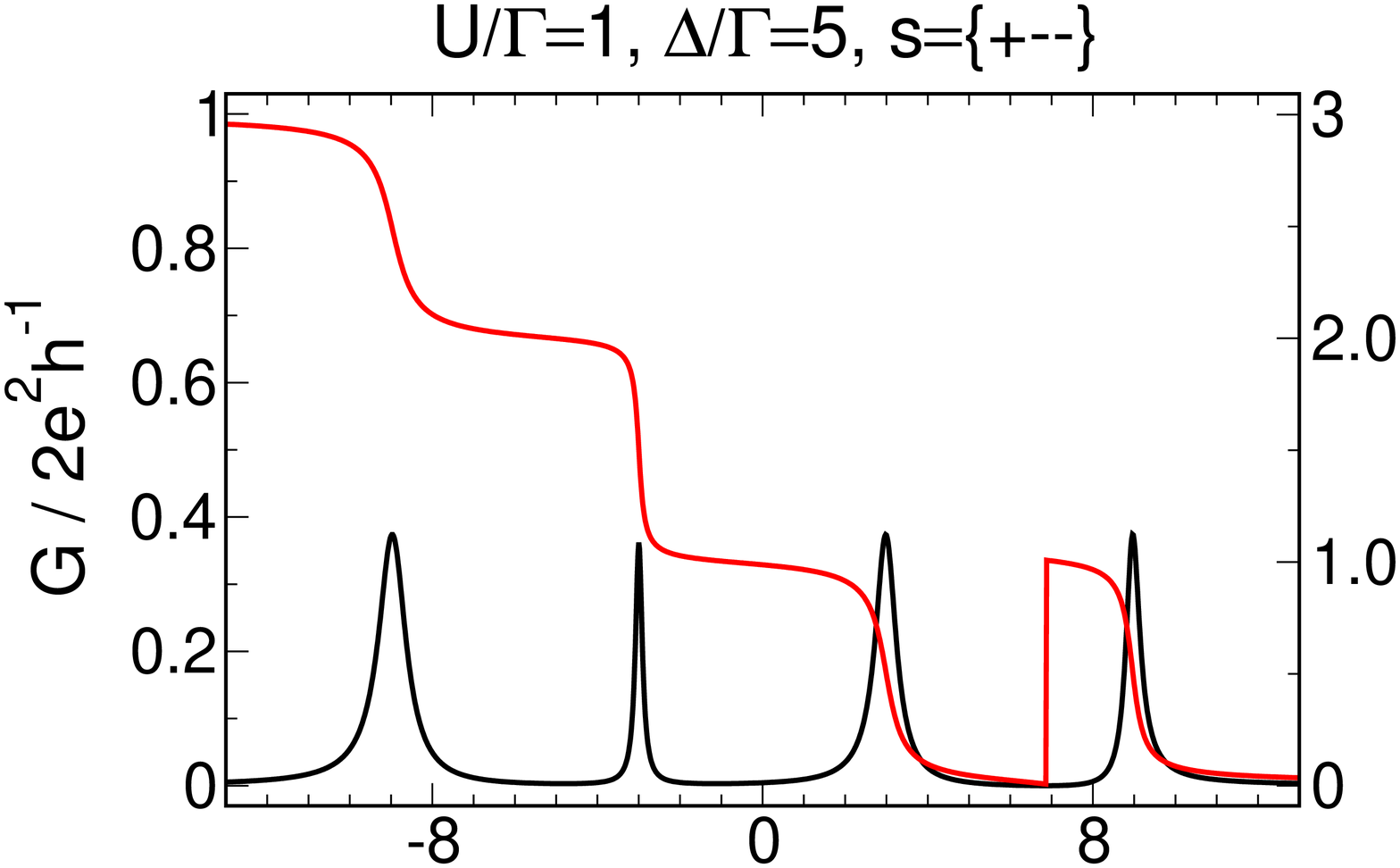}\hspace{0.015\textwidth}
        \includegraphics[width=0.475\textwidth,height=4.4cm,clip]{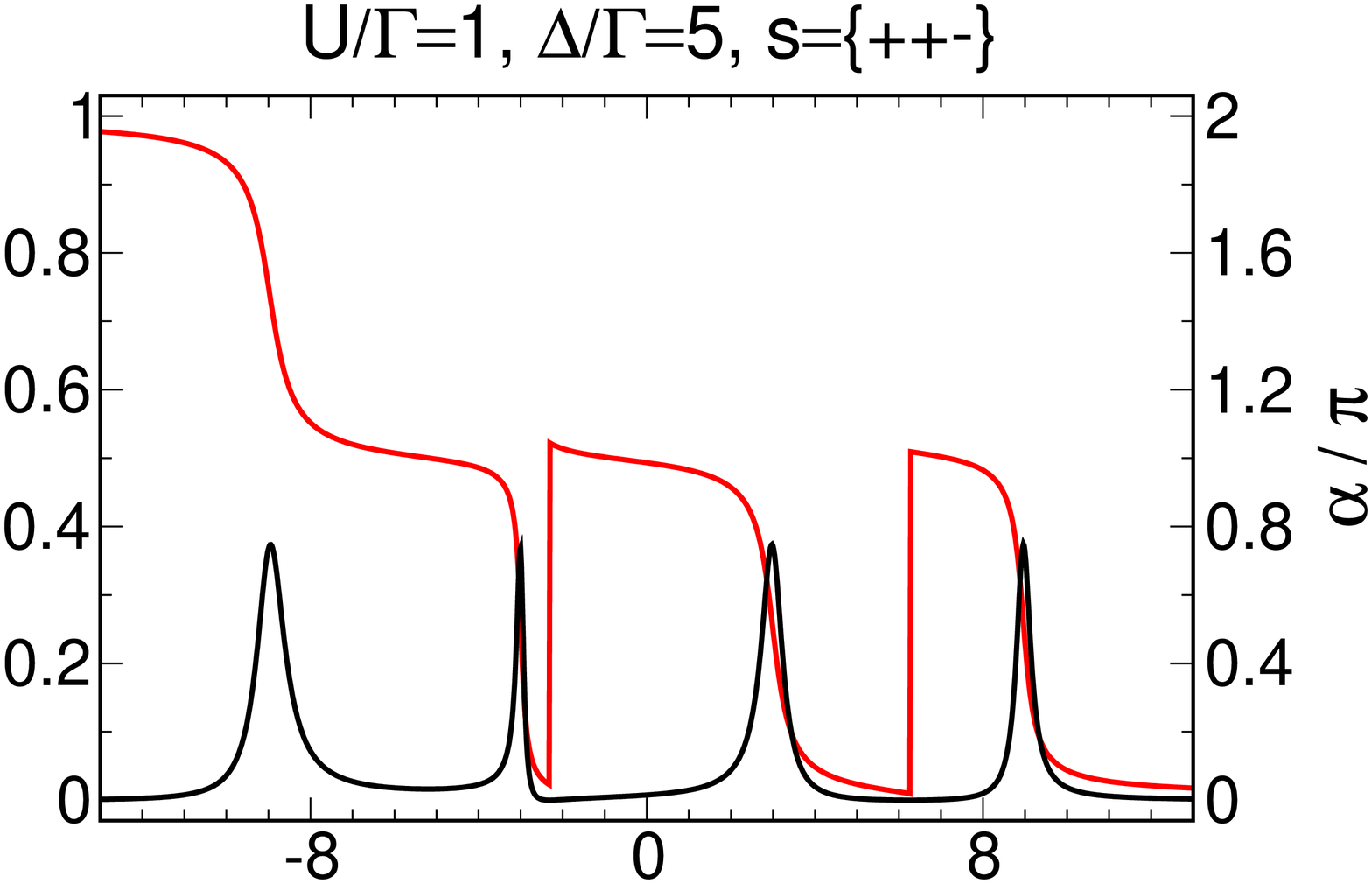}\vspace{0.3cm}
        \includegraphics[width=0.495\textwidth,height=5.2cm,clip]{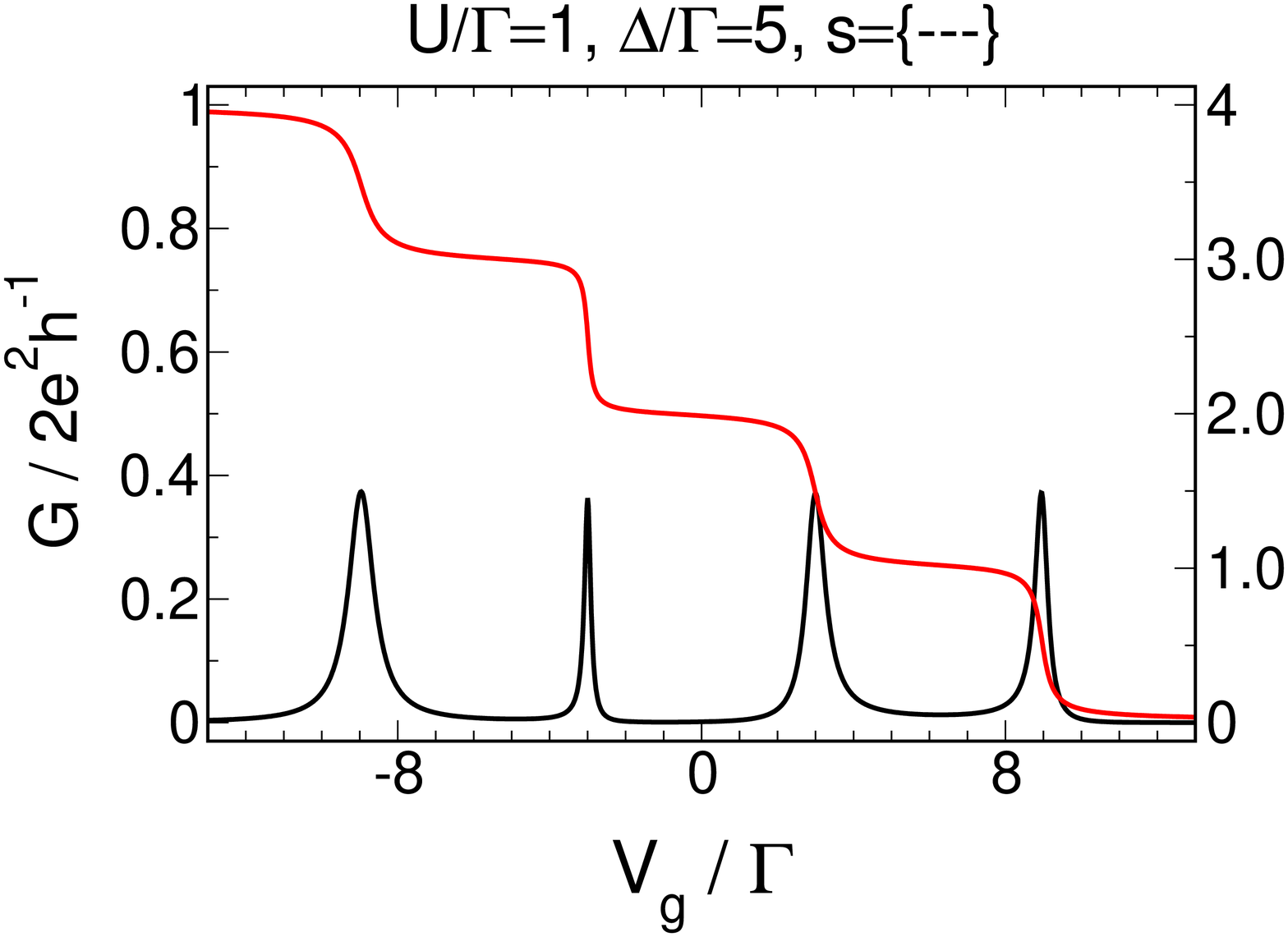}\hspace{0.015\textwidth}
        \includegraphics[width=0.475\textwidth,height=5.2cm,clip]{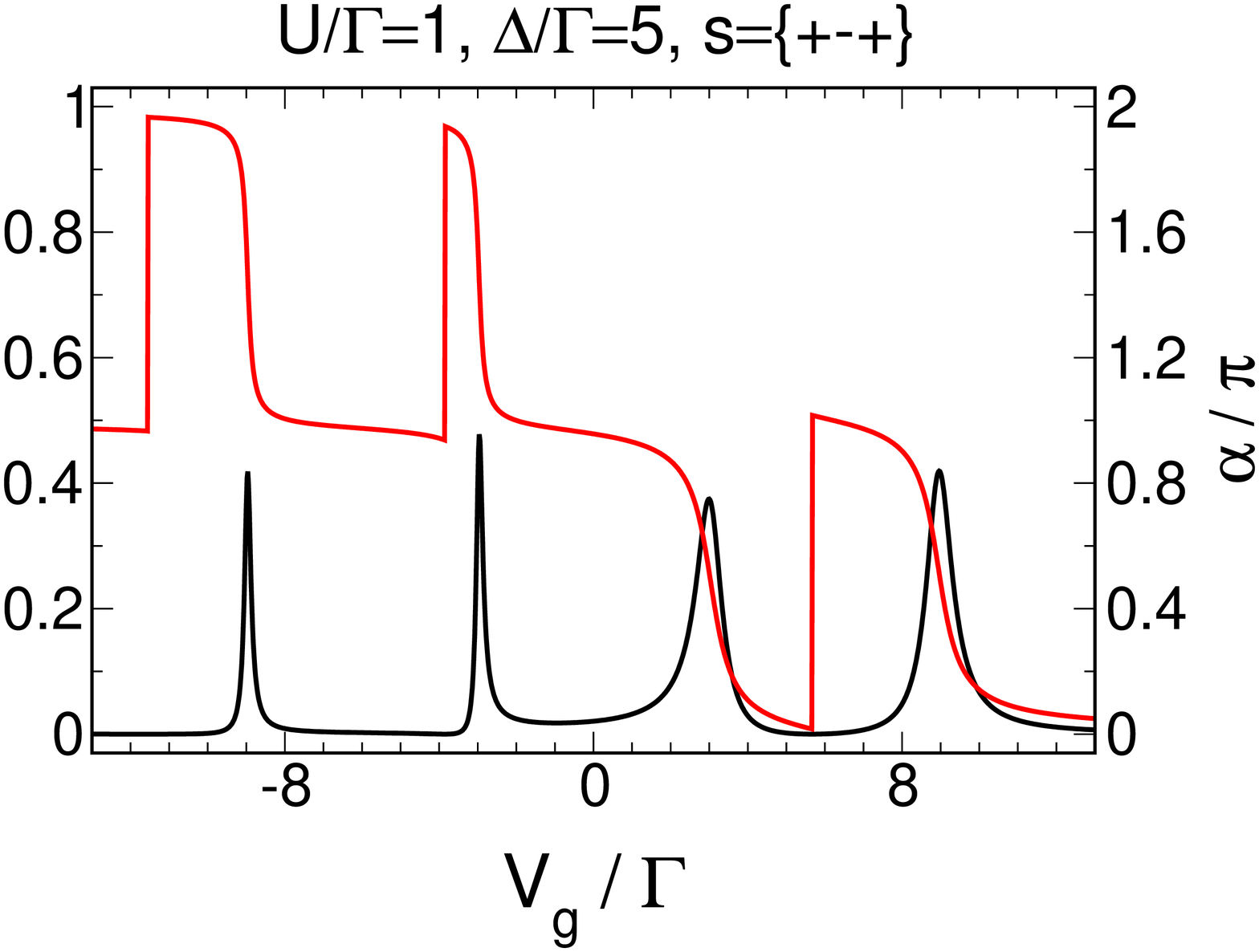}
        \caption{The same as Fig.~\ref{fig:OS.qd.u}, but for large level spacings.}
\label{fig:OS.qd.d}
\end{figure}
\afterpage{\clearpage}

It is straight-forward to convince oneself that the six independent choices for the relative signs of the level-lead couplings $s=\{s_1s_2s_3\}$ are given by $s=\{+++,++-,+--,+-+,-+-,---\}$. As before, we introduce $U$ as the strength of the two-particle interaction assumed to be equal between the electrons of all dots. The fRG flow equations comprise four for the on-site energies, six for the inter-dot hoppings and 21 for the independent components of the effective interaction.

For nearly degenerate levels, we observe four conductance resonances separated by $U$ of almost identical height and width (Fig.~\ref{fig:OS.qd.u}). The transmission phase changes approximately by $\pi$ when crossing each of them and jumps by $\pi$ at the three transmission zeros located in between. For increasing interaction strength, the peaks exhibit additional correlation induced structures at which the phase changes steeply. In particular, we find features similar to the CIRs of the parallel two-level dot (Fig.~\ref{fig:OS.qd.u}, upper right panel) as well as double phase lapses (jumps), the latter depending on the left-right symmetry of the hybridisations (compare the lower left panel of Fig.~\ref{fig:OS.qd.u} with the lower left panel of Fig.~\ref{fig:OS.qd.var}).

In the limit of large level spacings, we recover the usual structure of six Lorentzian resonances of separation $U+\Delta$ corresponding to transport through the individual levels (Fig.~\ref{fig:OS.qd.d}). The conductance vanishes for one gate voltage in between those peaks associated with levels which relative couplings do not differ in sign while remaining finite otherwise. Correspondingly, the transmission phase exhibits jumps by $\pi$ in the former case and evolves continuously in the latter. If the level spacing is chosen somewhere in between the limits $\Delta\ll\Gamma$ and $\Delta\gg\Gamma$, a crossover regime similar to the two- and three-level dot is observed, but we will refrain from giving a detailed description here.

Just like before, the noninteracting lineshape $G(V_g)$ differs strongly from that with $U>0$ in the limit of nearly degenerate levels. It shows very sharp structures on a scale $\Gamma$, and, more important, it does not exhibit four transmission resonances with a zero in between (Fig.~\ref{fig:OS.qd.var}, lower right panel) for arbitrary choice of the dot's parameters (the hybridisations and the relative signs of the couplings).

\subsubsection{Six Levels}

\begin{figure}[t]	
        \centering
        \includegraphics[width=0.495\textwidth,height=5.2cm,clip]{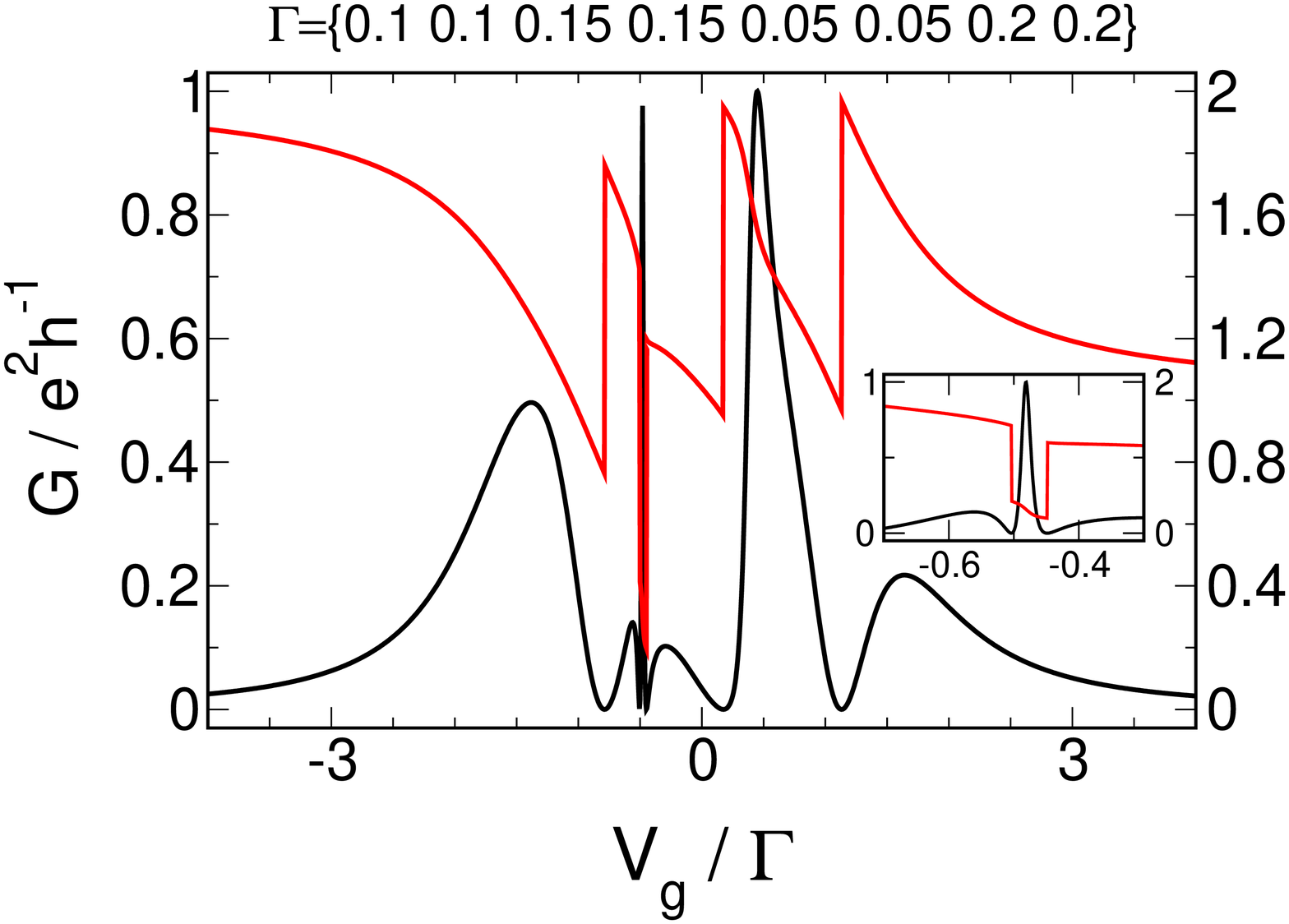}\hspace{0.015\textwidth}
        \includegraphics[width=0.475\textwidth,height=5.2cm,clip]{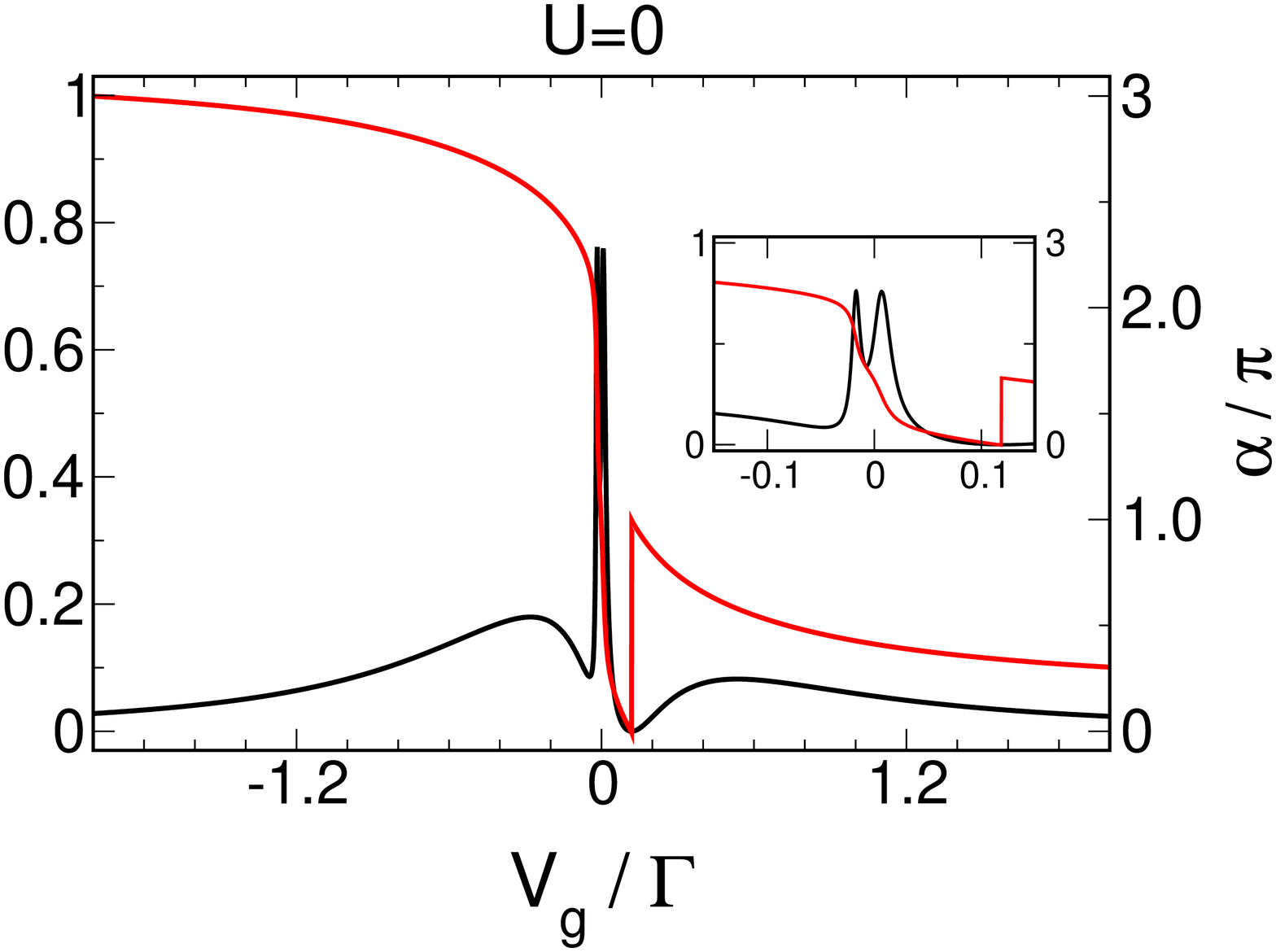}
        \caption{The same as the the lower left panel of Fig.~\ref{fig:OS.qd.u}, but for various combinations of parameters, illustrating the appearance of DPLs (left panel) and the lack of three transmission zeros in the noninteracting case (right panel).}
\label{fig:OS.qd.var}\vspace{0.3cm}
        \includegraphics[width=0.495\textwidth,height=5.2cm,clip]{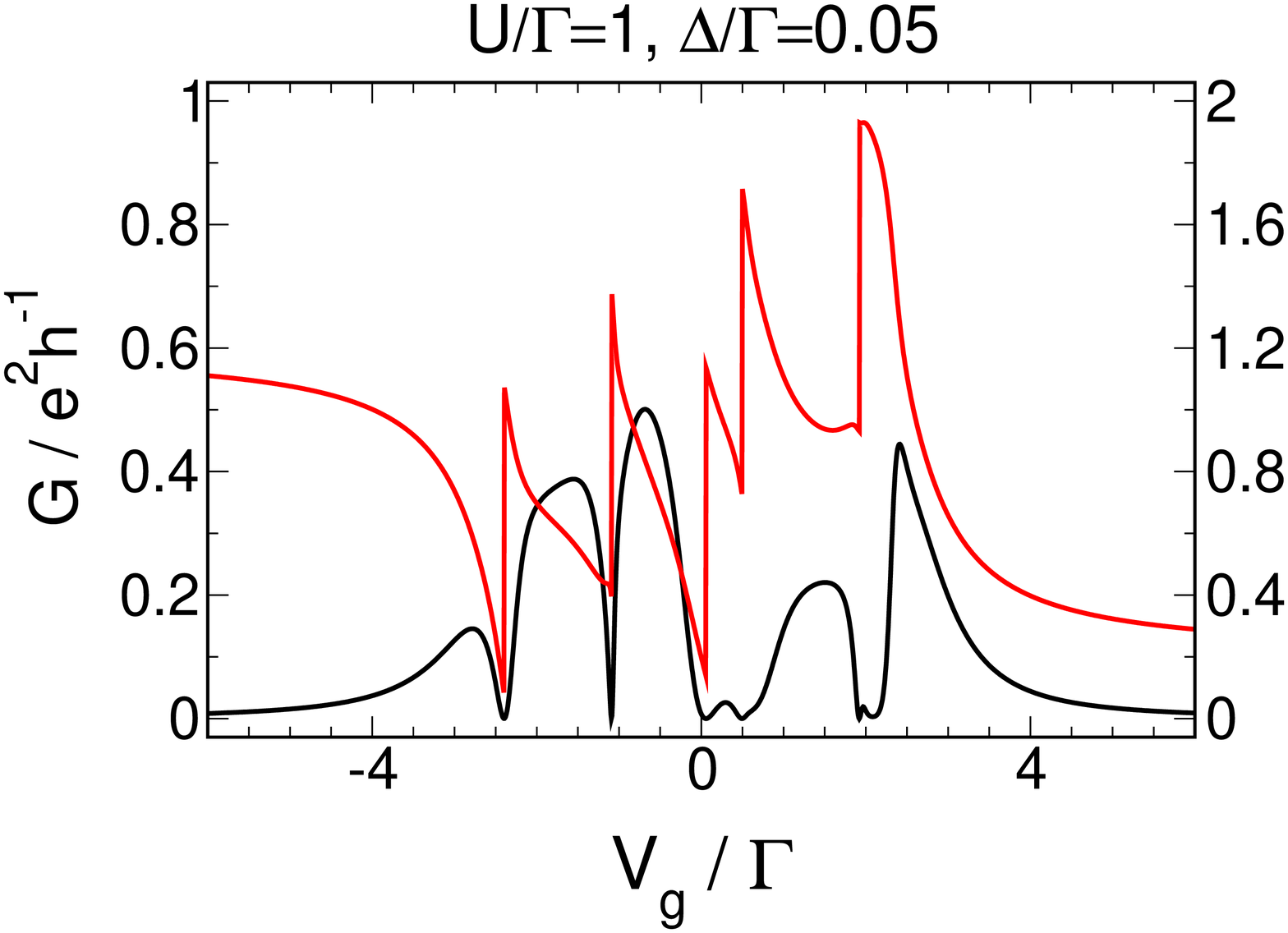}\hspace{0.015\textwidth}
        \includegraphics[width=0.475\textwidth,height=5.2cm,clip]{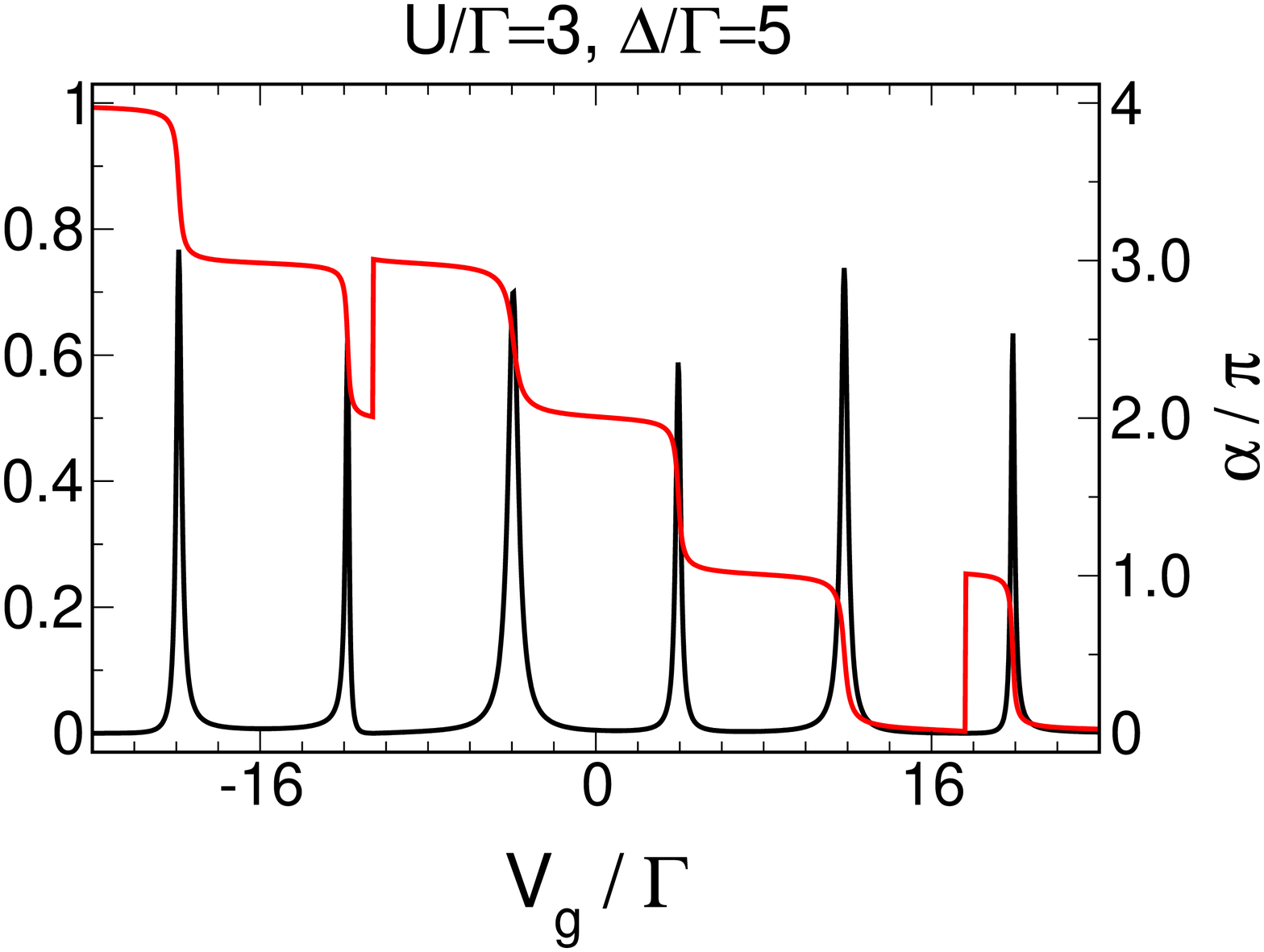}
        \caption{Conductance $G$ (black) and transmission phase $\alpha$ (red) for a six-level dot with $\Gamma=\{0.02~0.08~0.05~0.15~0.03~0.12~0.07~0.23~0.02~0.08~0.04~0.11\}$ and $s=\{+--+-\}$. For $\Delta\ll\Gamma$, one would expect the height and width of the resonances to become more comparable and the change of $\alpha$ in between the transmission zeros to become precisely $\pi$ when $\Delta$ is slightly decreased down to $\Delta/\Gamma=0.01$. Due to the limitations of the fRG, we cannot confirm this scenario explicitly (as we encounter serious problems if both $U\gg\Gamma$ and $\Delta\ll\Gamma$ and if many levels are considered; this will be explained in detail in the last chapter).}
\label{fig:OS.hd}
\end{figure}

We end this section with considering six parallel dots. The projected free propagator and the expression for the conductance can be written down in complete analogy with the four-level case, and also our results in the regimes $\Delta\ll\Gamma$ and $\Delta\gg\Gamma$ turn out to be the same. For nearly degenerate levels, we observe six resonances and five transmission zeros in between them in the $G(V_g)$ curve (Fig.~\ref{fig:OS.hd}, left panel). The transmission phase jumps by $\pi$ at each zero and approximately changes by $\pi$ over each resonance. Further correlation induced structures are observed if the strength of the interaction is increased. If on the other hand the level spacing is the dominant energy scale, we find six Lorentzian resonances of separation $U+\Delta$ and transmission zeros (and corresponding phase jumps) between those associated with transport through levels which couplings have the same relative sign (Fig.~\ref{fig:OS.hd}, right panel).

\subsection{Finite Temperatures}\label{sec:OS.T}

Up to now we have only considered the $T=0$ limit. This might be a meaningful simplification since transport through quantum dots is generally measured at very low temperatures. Nevertheless, it is important to investigate whether the qualitative results derived in this section are stable against very small but finite temperatures, especially when one is concerned with the very sharp structures (for example the CIRs or the DPLs) one would expect to be smeared out at finite $T$.

\begin{figure}[t]	
        \centering
        \includegraphics[width=0.495\textwidth,height=4.4cm,clip]{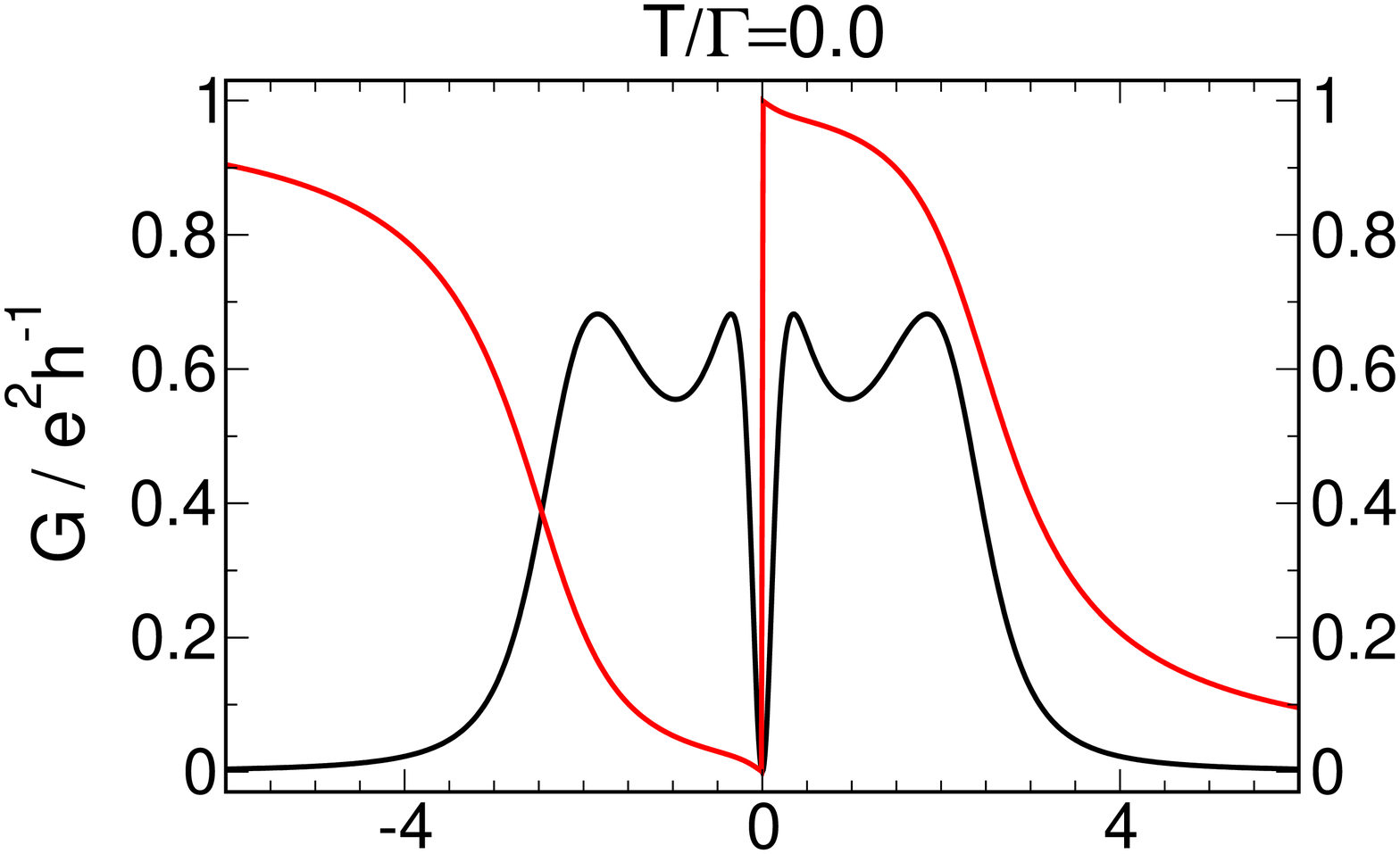}\hspace{0.015\textwidth}
        \includegraphics[width=0.475\textwidth,height=4.4cm,clip]{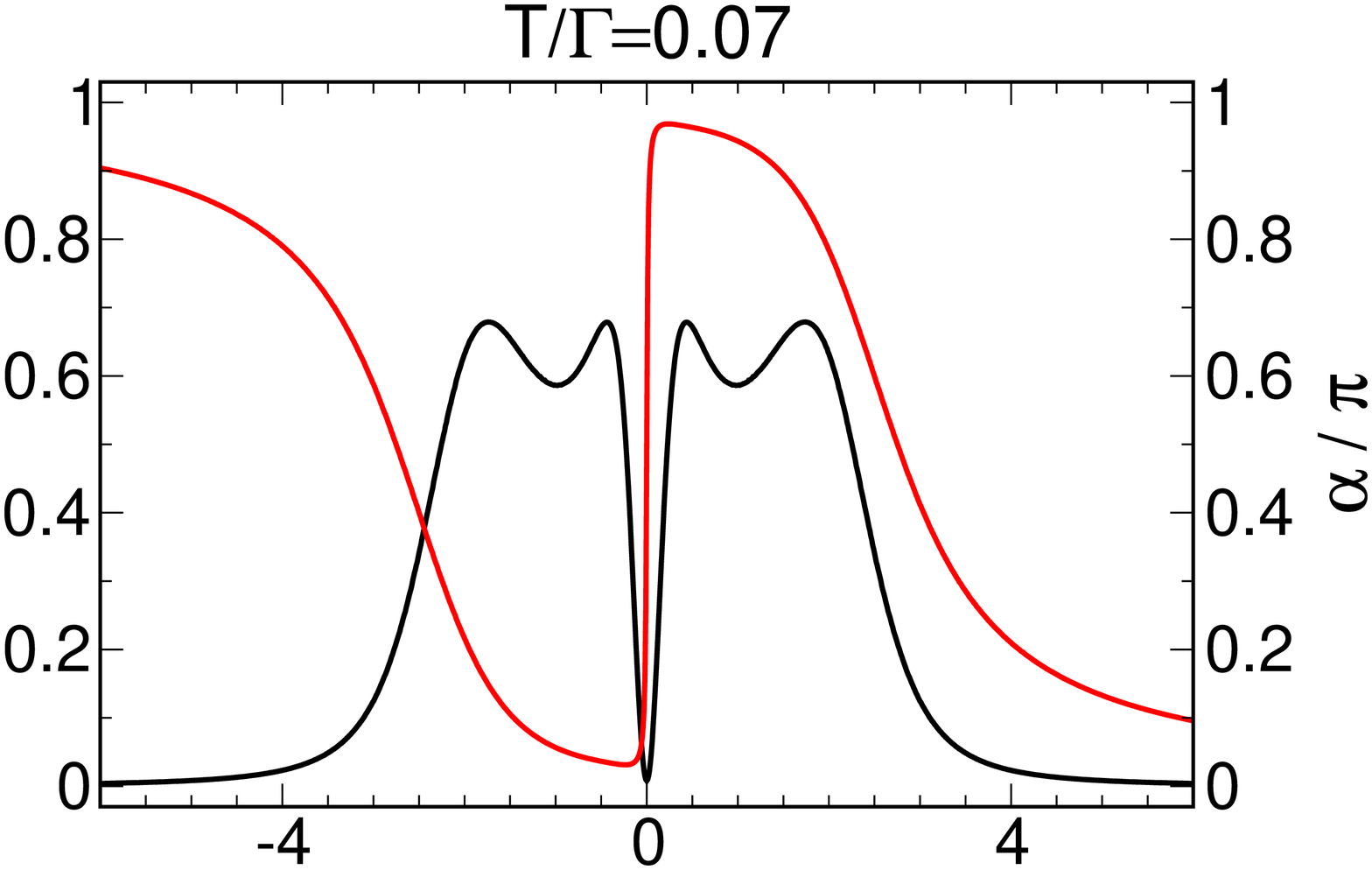}\vspace{0.3cm}
        \includegraphics[width=0.495\textwidth,height=5.2cm,clip]{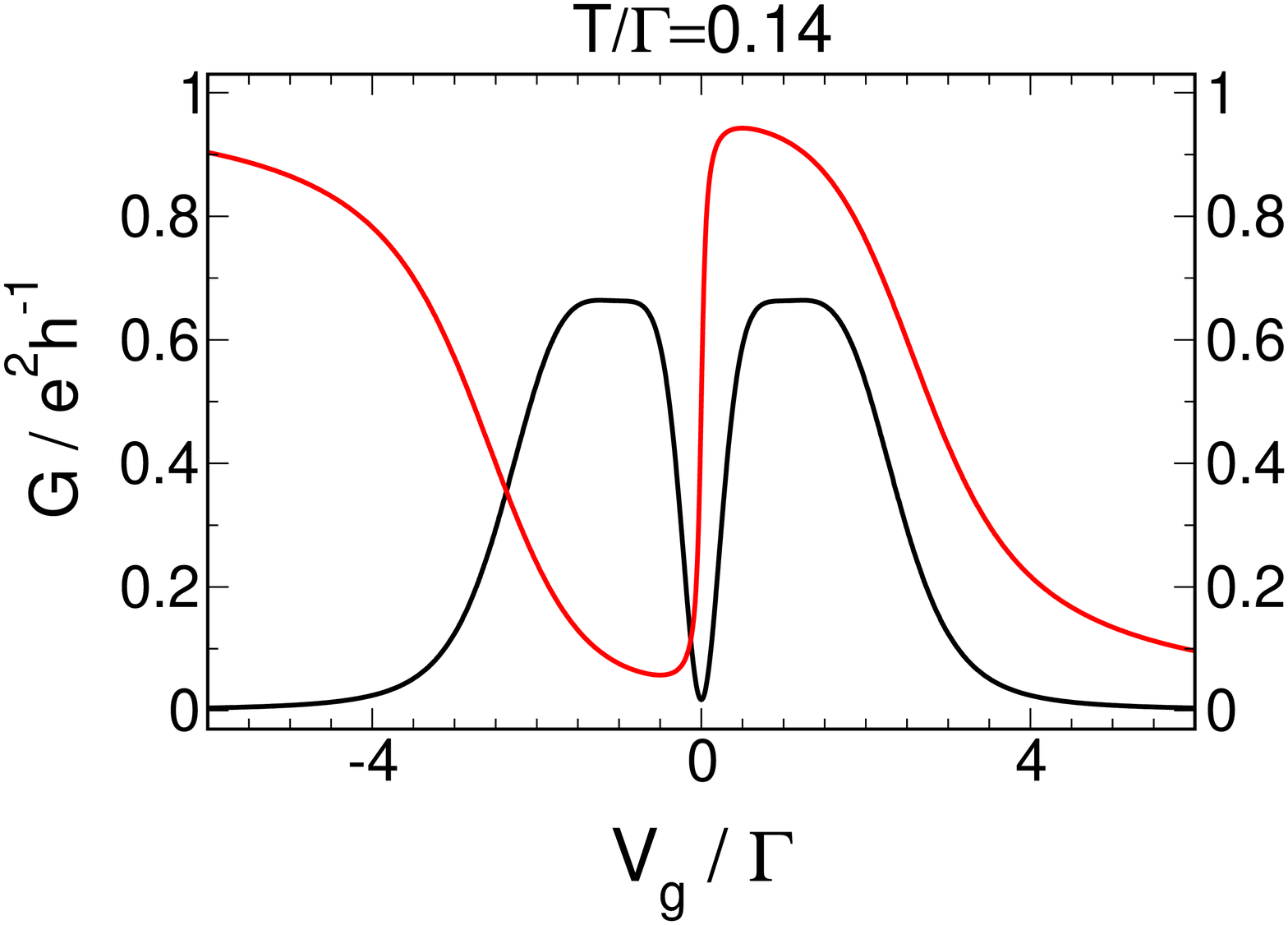}\hspace{0.015\textwidth}
        \includegraphics[width=0.475\textwidth,height=5.2cm,clip]{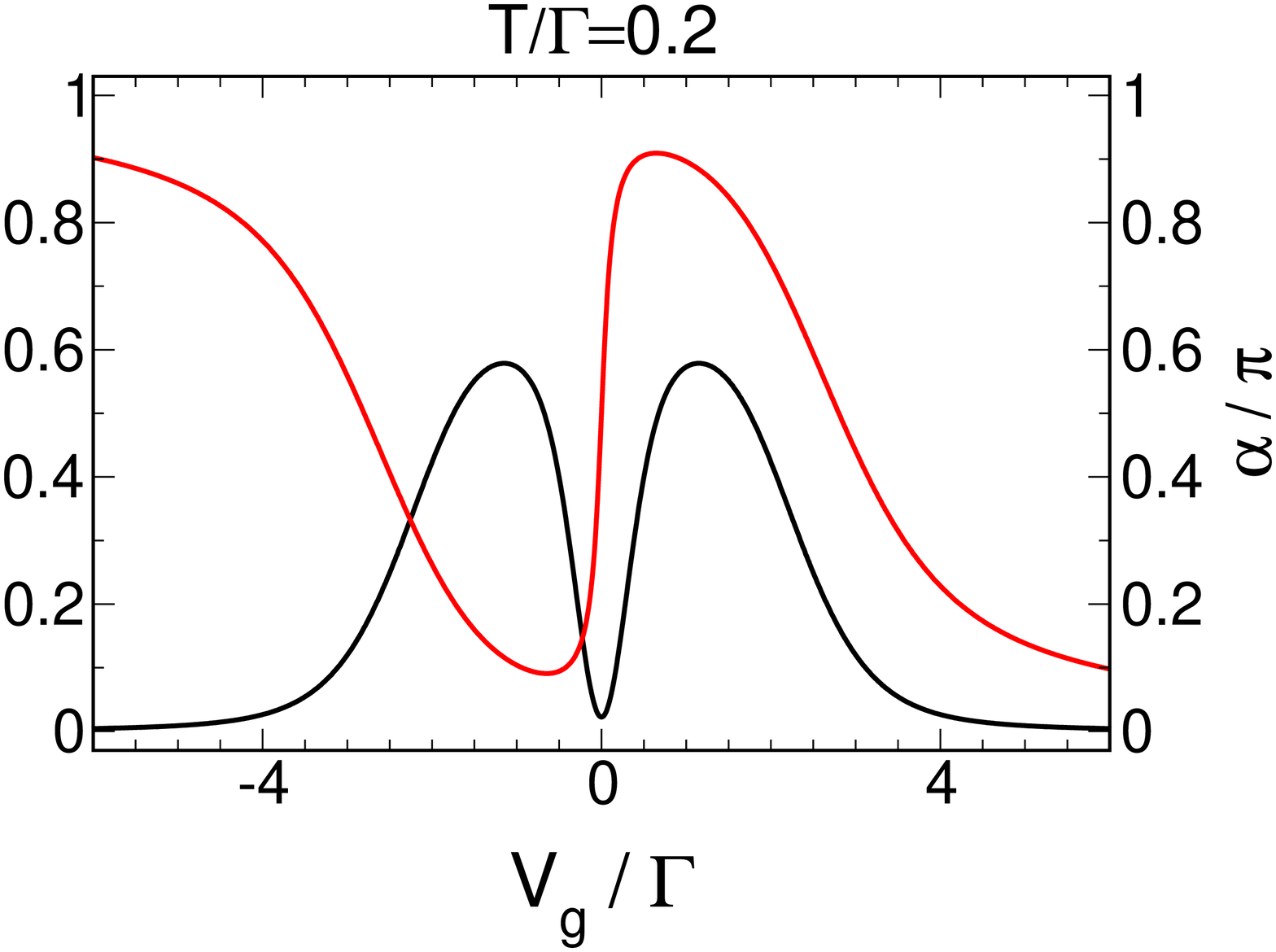}
        \caption{Gate voltage dependence of the conductance $G$ (black) and transmission phase $\alpha$ (red) for parallel double dots with $U/\Gamma=6.0$, degenerate levels, $s=-$, and generic level-lead hybridisations $\Gamma=\{0.27~0.33~0.16~0.24\}$ for finite temperatures. Increasing $T$, the CIRs gradually disappear.}
\label{fig:OS.T.dd}
\end{figure}

\begin{figure}[t]	
        \centering
        \includegraphics[width=0.475\textwidth,height=4.8cm,clip]{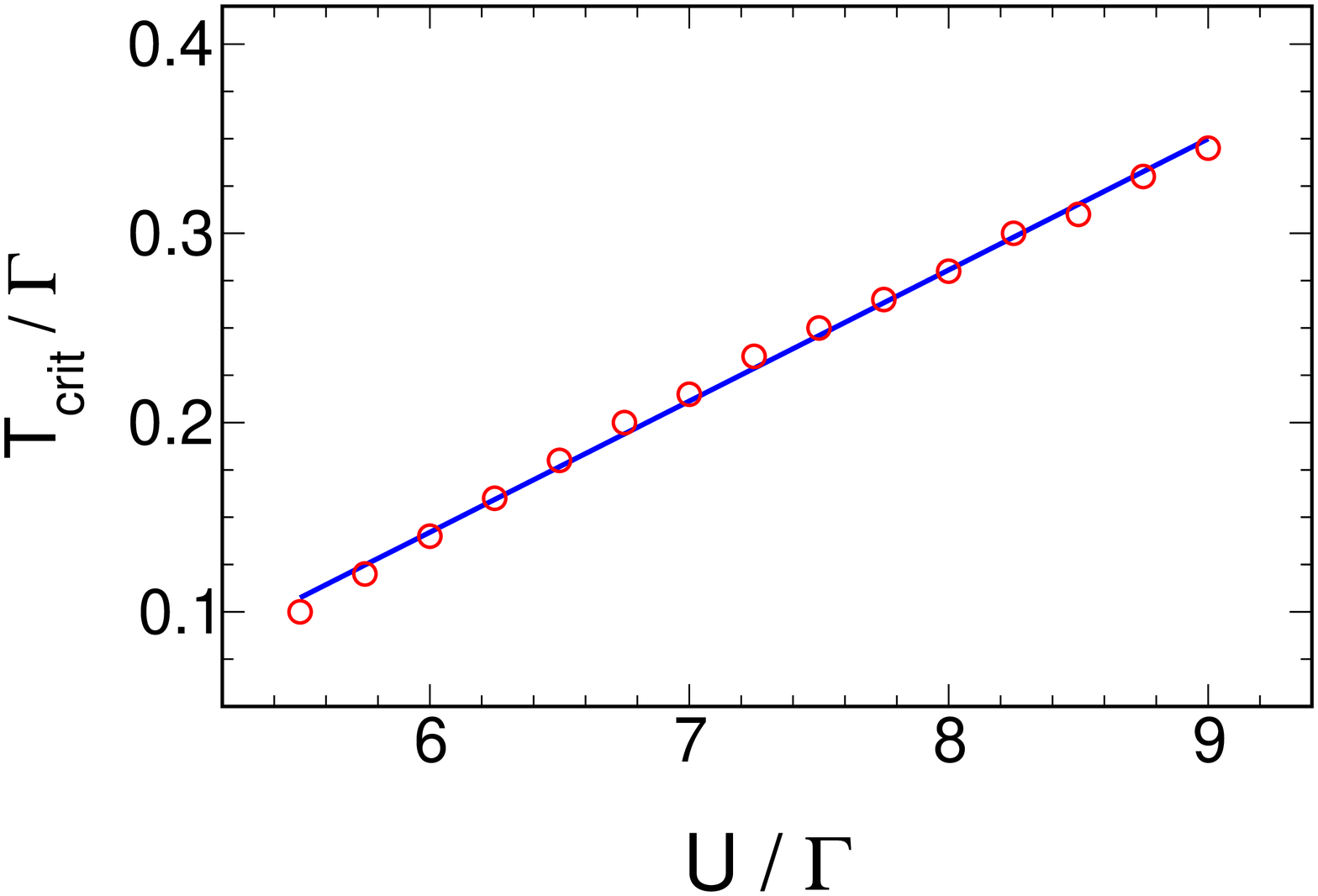}\hspace{0.035\textwidth}
        \includegraphics[width=0.475\textwidth,height=4.8cm,clip]{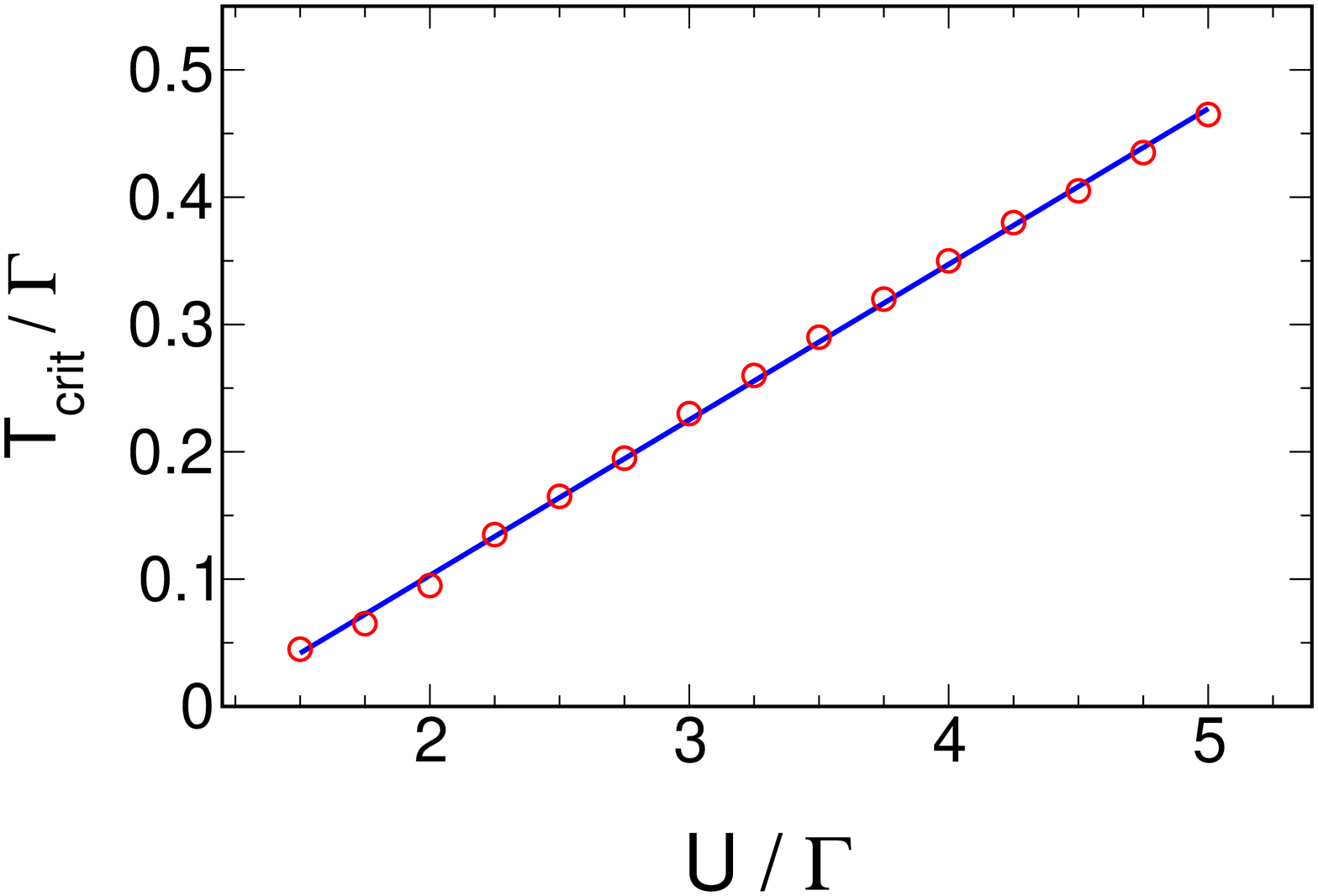}
        \caption{Critical temperature where the correlation induced resonances observed for parallel double dots with degenerate levels vanish as a function of the two-particle interaction $U$. \textit{Left panel:} Hybridisations $\Gamma=\{0.27~0.33~0.16~0.24\}$, and $s=-$. \textit{Right panel:} $\Gamma=\{0.5~0.25~0.07~0.18\}$, $s=+$.}
\label{fig:OS.T.dd_krit}
\end{figure}

For the geometry which we are going to discuss in detail in the next section, a single level containing interacting spin up and down electrons, we have already argued that our fRG truncation scheme which neglects all frequency dependencies yields a spectral function of Lorentzian lineshape very different from the exact one. Since for arbitrary $T$ the conductance is computed as an integral of the latter with the derivative of the Fermi function, we cannot expect to obtain reliable results beyond the limit of very small $T$ (and we have indeed checked that this is the case), and the same should hold for other spinful geometries. Nevertheless, we can hope that things are different for spin-polarised models. Unfortunately, no other calculations for such models at finite temperatures have been performed up to now, rendering it impossible to check the trustworthiness of our results. Therefore we will only employ the $T>0$ calculations to check the aforementioned stability of the sharp correlation induced features against small temperatures $T\lesssim\Gamma$ but refrain from a detailed study of arbitrary $T$.

The conductance for finite temperatures is computed using the general expression (\ref{eq:DOT.condKc}), and we define the $T\neq 0$ transmission phase as
\begin{equation}
\alpha_\sigma(T):=\tn{arg}\left(\int d\epsilon~ g'(\epsilon)\sum_{l,l'}t_l^R(t_{l'}^L)^*\mc G^\sigma_{l;l'}(\epsilon+i0)\right),
\end{equation}
with $g(\epsilon)$ being the Fermi function, and the expression for spin-polarised models follows by dropping all spin dependencies. First, we study the $T$ -- dependence of the CIRs observed for a parallel double dot with degenerate levels for sufficiently large interaction. It turns out that they are stable against temperatures $T\lesssim\Gamma$. To be more precise, the evolution of $G(V_g)$ with increasing temperature is similar to the one arising when $U$ is lowered in that sense that the four-peak structure observed for $T=0$ merges into two single resonances if the temperature exceeds a critical $T_\tn{crit}$ depending on all other dot parameters (Fig.~\ref{fig:OS.T.dd}). This temperature scale is roughly given by $T_\tn{crit}\approx\Gamma/4$, and it depends linearly on the interaction strength (Fig.~\ref{fig:OS.T.dd_krit}). On the other hand, the transmission phase which shows a jump by $\pi$ at $V_g=0$ for zero temperature evolves continuously for $T\neq 0$ with a slope that decreases if $T$ is increased. An interesting question is whether this `smoothing' of both $G(V_g)$ and $\alpha(V_g)$ is caused only by smearing with the Fermi function. In fact, we have tested that the $T$ -- dependence of $G(V_g)$ is qualitatively the same if one considers the finite temperature propagator at $z=i0$, or if one computes the latter at $T=0$ and calculates the conductance and transmission phase as an integral of this $\mc G(\epsilon+i0)$ with the Fermi function.

\begin{figure}[t]	
        \centering
        \includegraphics[width=0.495\textwidth,height=4.4cm,clip]{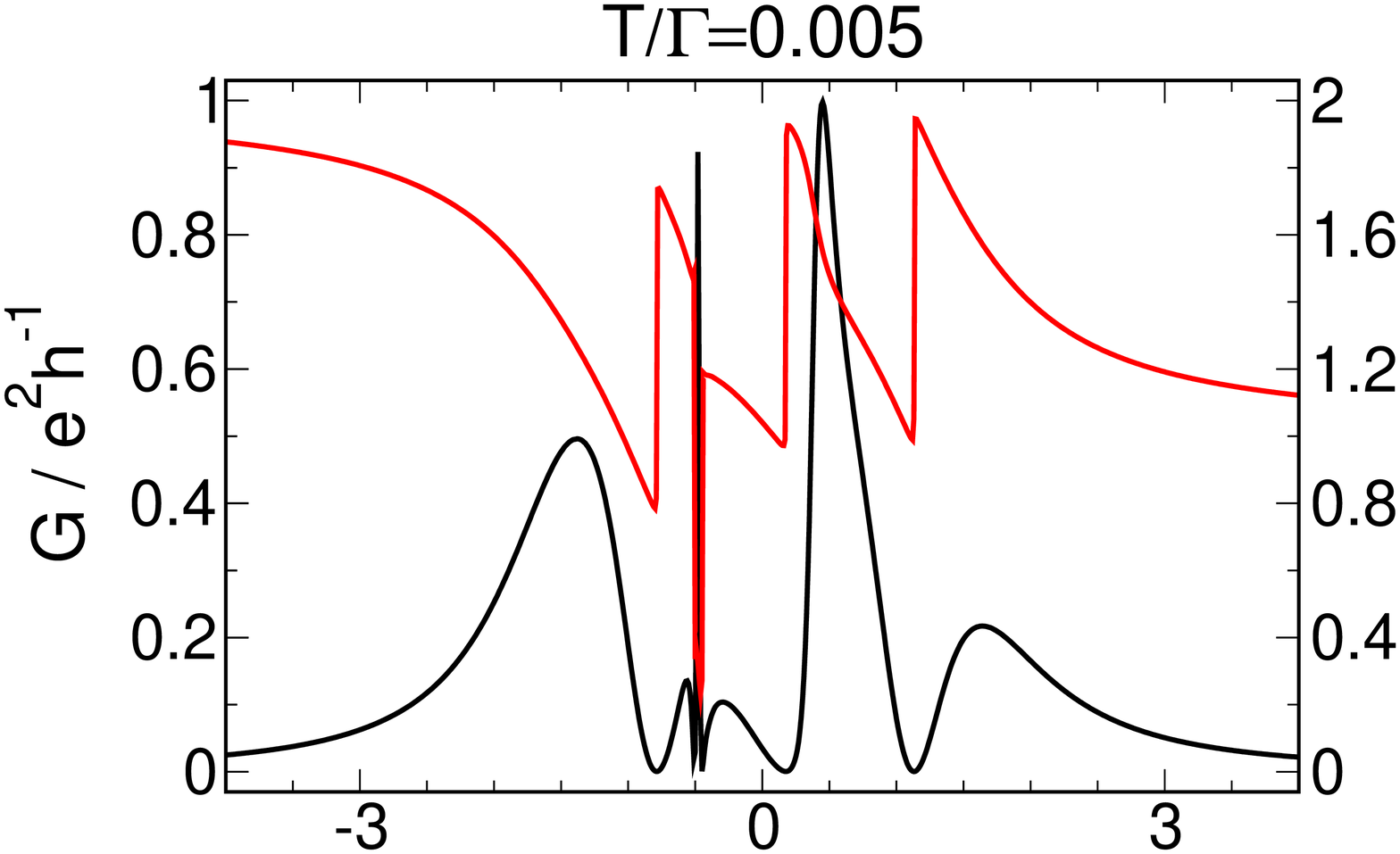}\hspace{0.015\textwidth}
        \includegraphics[width=0.475\textwidth,height=4.4cm,clip]{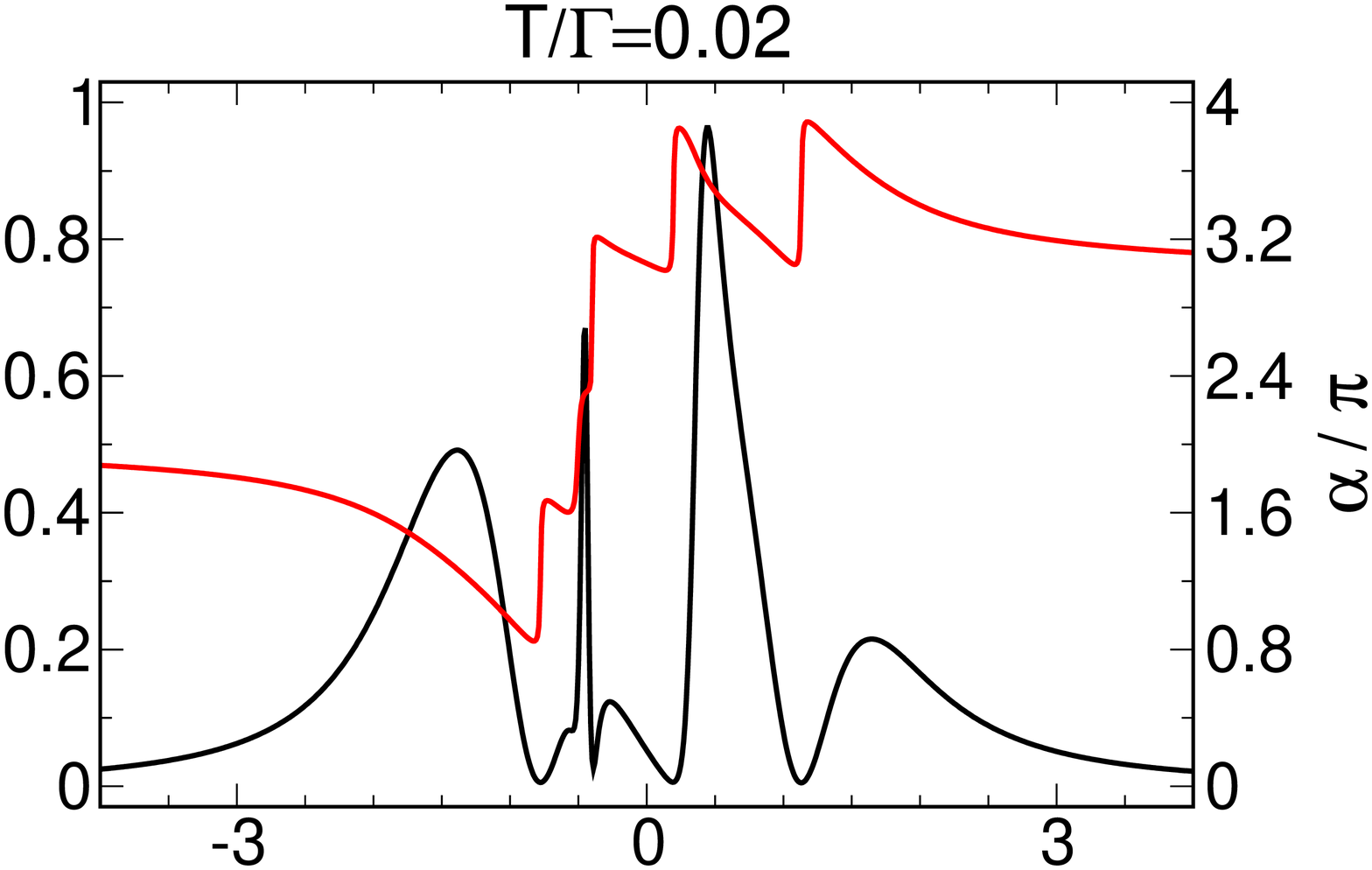}\vspace{0.3cm}
        \includegraphics[width=0.495\textwidth,height=5.2cm,clip]{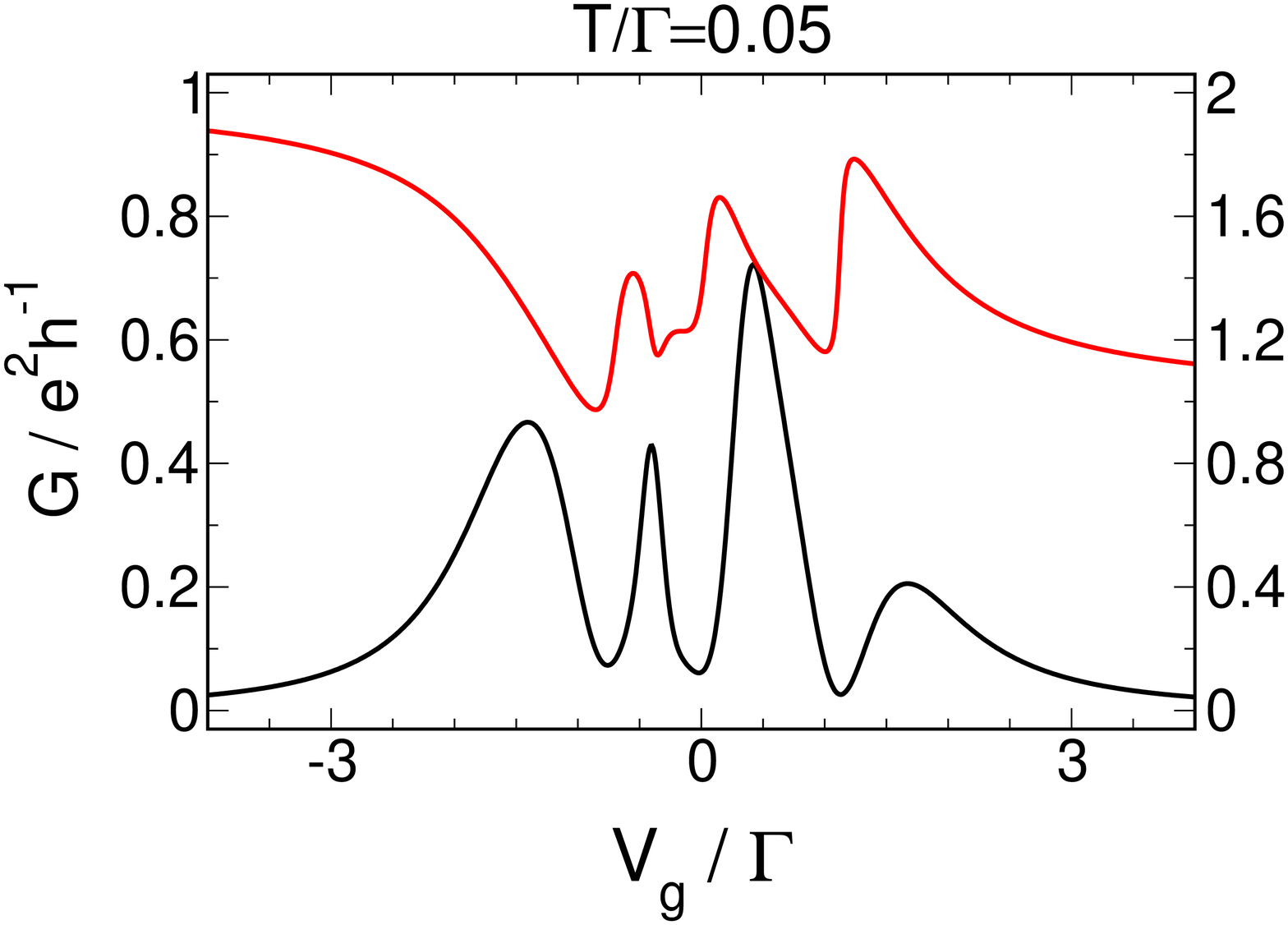}\hspace{0.015\textwidth}
        \includegraphics[width=0.475\textwidth,height=5.2cm,clip]{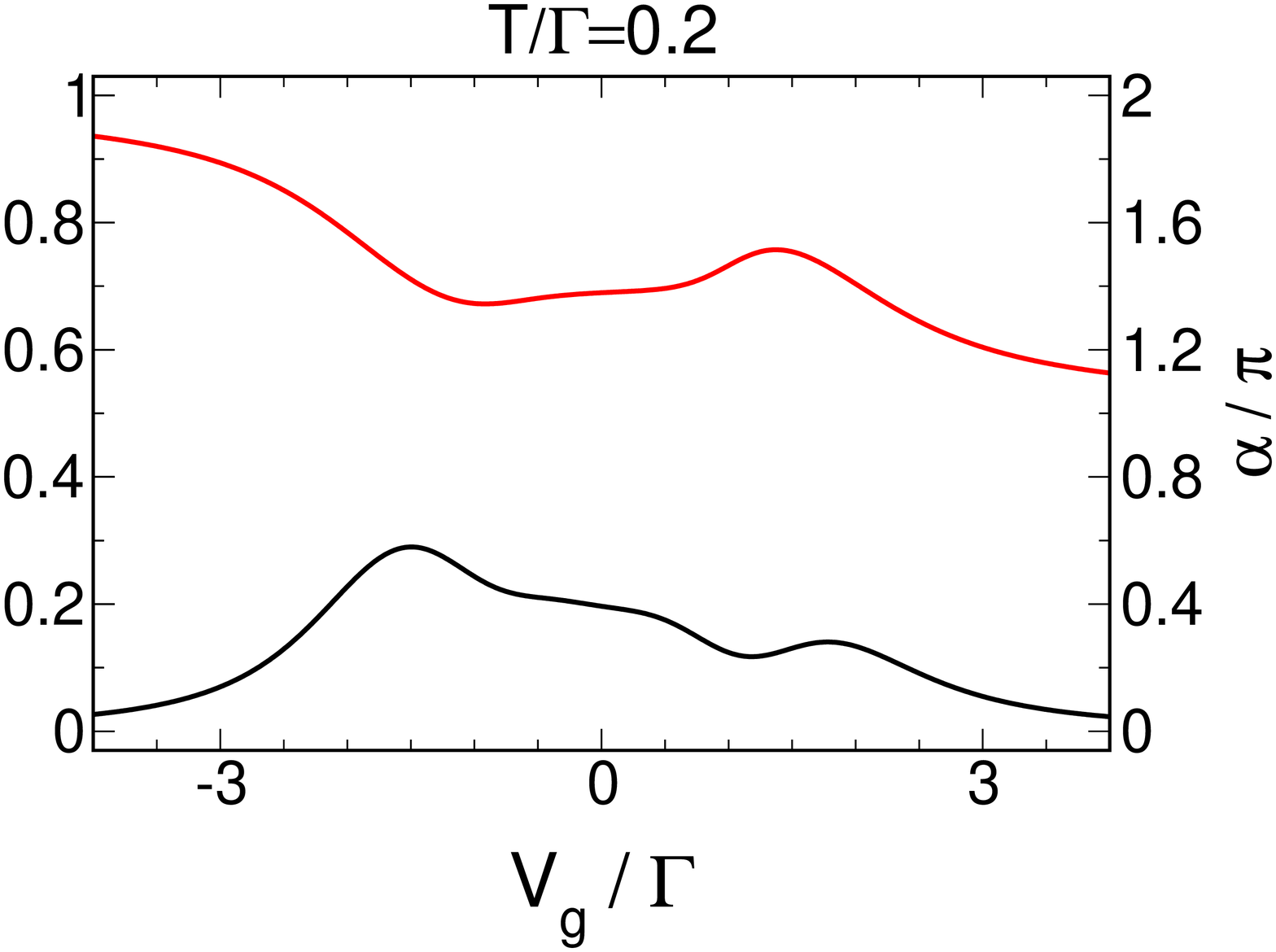}
        \caption{Conductance $G$ (black) and transmission phase $\alpha$ (red) as a function of the gate voltage for a four-level dot with $U/\Gamma=1.0$, $\Delta/\Gamma=0.05$, $\Gamma=\{0.1~0.1~0.15~0.15~0.05~0.05~0.2~0.2\}$, and $s=\{---\}$ for four finite temperatures.}
\label{fig:OS.T.dpl}
\end{figure}

Surprisingly, the sharp correlation induces features that appear for geometries with more than two dots (those similar to the ones observed for the triple dot at the central resonance) become more pronounced for $T\neq 0$. In particular, the double phase lapses do not vanish unless $T$ is so high that the whole structure of $G(V_g)$ is smeared. The generic evolution of the conductance and transmission phase curves with increased temperature is shown in Fig.~\ref{fig:OS.T.dpl}. The DPLs which show up because of the left-right symmetric hybridisations are more emphasised in the sense that they become less sharp.

\begin{figure}[t]	
        \centering
        \includegraphics[width=0.495\textwidth,height=4.8cm,clip]{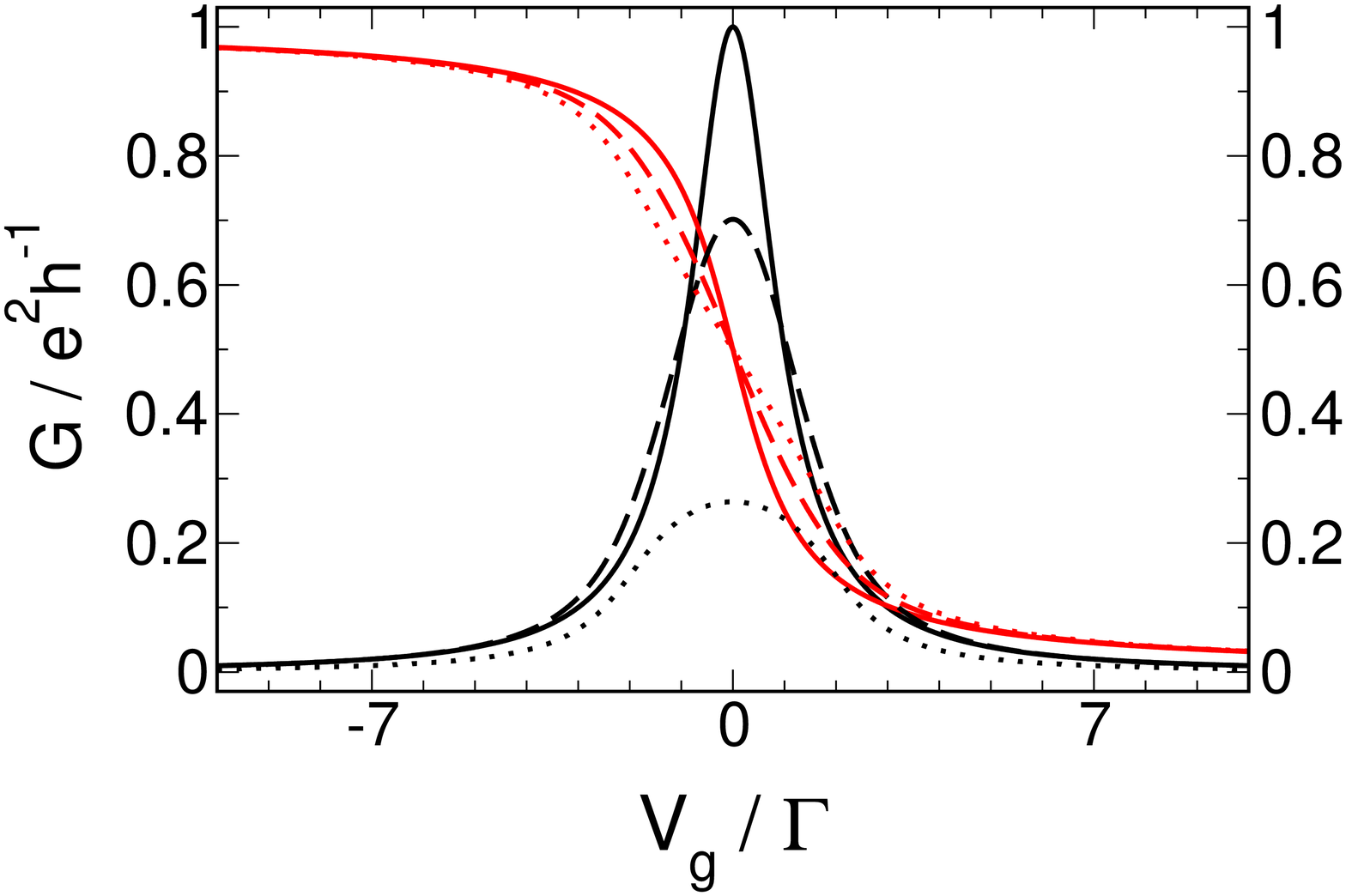}\hspace{0.015\textwidth}
        \includegraphics[width=0.475\textwidth,height=4.8cm,clip]{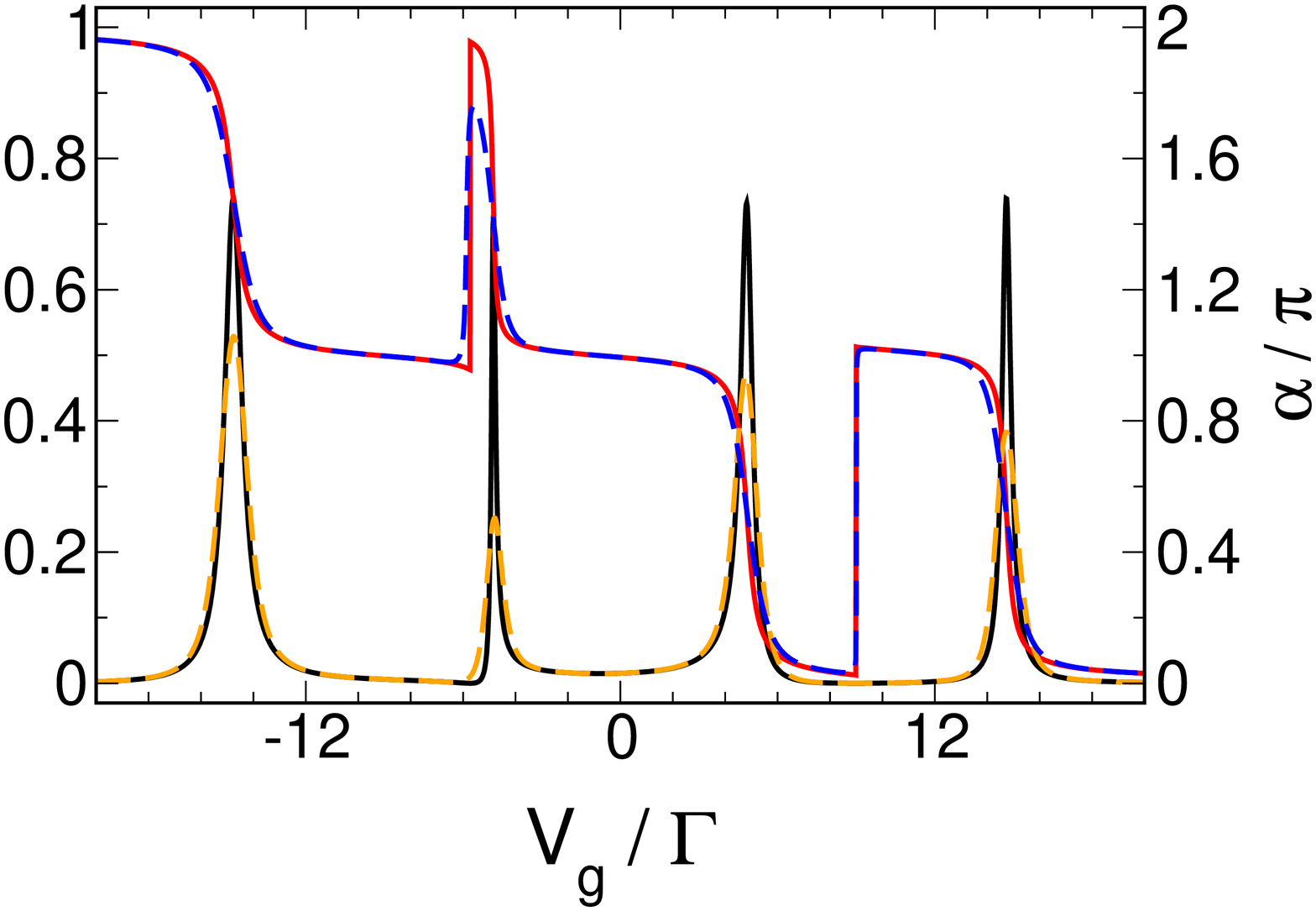}
        \caption{\textit{Left panel:} Conductance $G$ (black) and transmission phase $\alpha$ (red) for a single impurity with on-site energy $V_g$ for $T/\Gamma=0.0$ (solid lines), $T/\Gamma=0.5$ (dashed lines), and $T/\Gamma=2.0$ (dotted lines). Note that this is an exact result. \textit{Right panel:} The same from fRG for a four level dot with $U/\Gamma=5.0$, $\Delta/\Gamma=5.0$, $\Gamma=\{0.05~0.15~0.075~0.225~0.025~0.075~0.1~0.3\}$, and $s=\{+-+\}$ at zero temperature (black and red curves) and at $T/\Gamma=0.2$ (orange and blue curves).}
\label{fig:OS.T.delta}
\end{figure}

To understand the temperature dependence of $G(V_g)$ for large level spacings, it is meaningful to consider a single spin-polarised impurity first. The conductance can then be computed exactly as an integral of the derivative of the Fermi with the (exact) spectral function. As one would intuitively expect, the transmission resonance centred at $V_g=\mu$ becomes wider and its height decreases at temperatures $T\approx\Gamma$ (Fig.~\ref{fig:OS.T.delta}, left panel). The transmission phase evolves less steeply. Having this in mind, the temperature dependence of the peaks observed for interacting multi-level dots with large $\Delta$ is straight-forward to understand. The width (height) of each resonance increases (decreases) at a temperature scale given by the strength of the individual hybridisations, $T\approx\Gamma_l$. At the same time, the slope of the phase $\alpha$ becomes smaller and the jumps by $\pi$ at the transmission zeros are gradually smeared out, in particular if they are located close to a resonance. It is important to note that here it is no longer $T/\Gamma$ but rather $T/\Gamma_l$ which determines a typical temperature scale, implying that temperatures which are small for nearly degenerate levels have a considerable effect at large $\Delta$ if some levels are weakly coupled or if the number of dots $N$ becomes large such that $T/\Gamma\ll T/\Gamma_\tn{typ}=T/(\Gamma/N)$. This is shown in the right panel of Fig.~\ref{fig:OS.T.delta}.

\subsection{Summary \& Discussion}

In this section, we have discussed the \textit{generic} transport properties of $N$ quantum dots with level spacing $\Delta$ connected in parallel to source and drain leads. The latter were modelled by a noninteracting tight-binding approach while we allowed nearest-neighbour interactions $U$ between the electrons on the different dots. Using the fRG approach, we calculated the full propagator of the system which determines the linear-response conductance and the transmission phase. To fully describe transport, we computed $G$ and $\alpha$ as a function of a variable gate voltage that simultaneously shifts the energies of all dots.

For nearly degenerate levels $\Delta\ll\Gamma$ and nonzero interactions we found $N$ transmission resonances separated by $U$ of almost equal width and height, the former being of order $\Gamma$, with $N-1$ transmission zeros in between. The phase changes approximately by $\pi$ when crossing each peak and jumps by $\pi$ at the zeros. If the strength of the two-particle interactions exceeds a critical value depending on the dot parameters, additional correlation induced features in the $N$-peak structure are observed. They vanish for level detunings larger than $\Delta\left(\{s,\Gamma\}\right)\ll\Gamma$, only the double phase lapses observed for geometries with three and more dots become more pronounced if $\Delta$ is increased. However, the scale $\Delta_\tn{DPL}$ that determines their appearance strongly depends on the left-right asymmetry of the hybridisations $\Gamma_l^s$, and in particular it is no longer much smaller than $\Gamma$ for generic asymmetric $\Gamma_l^s$. Furthermore, the additional structures become exponentially sharp if the interaction strength is increased, and only the $N$ usual `Coulomb blockade peaks' remain visible.

If the level spacing is increased above $\Delta_\tn{cross}\left(\{s,\Gamma\}\right)\lesssim\Gamma$, the lineshape of the $N$ peaks changes as the quantum dot enters the crossover regime. Some of the resonances split up while others become vanishingly small. In the limit of $\Delta\gg\Gamma$, we end up with $N$ Lorentzian resonances separated by $\Delta+U$ corresponding to transport through the individual levels. The width and height of each peak is determined by the associated hybridisations $\Gamma_l^L$ and $\Gamma_l^R$. The transmission phase $\alpha$ drops by $\pi$ over each resonance. Transmission zeros and corresponding phase jumps are located between those peaks related to levels whose relative level-lead couplings do not differ in sign, while the conductance stays finite and $\alpha$ evolves continuously otherwise.

The regime of large level spacings is identical in both the $U=0$ and $U>0$ case, only that in the latter the separation of the resonances is increased due to Coulomb repulsion. For nearly degenerate levels, things are different. For some parameters, the conductance and transmission phase curves are similar in both cases if one is only concerned with the number of transmission zeros, but the observed structures are extremely sharp on a scale of $\Gamma$ and the width of the individual peaks differ by orders of magnitudes of $\Gamma$. For other parameters, there are even less than $N-1$ transmission zeros in the $U=0$ case. Thus the two-particle interactions are a vital ingredient in order to find $N$ almost equal resonances with $N-1$ phase jumps in between.
\begin{figure}[t]	
     \centering
 	\includegraphics[height=4cm]{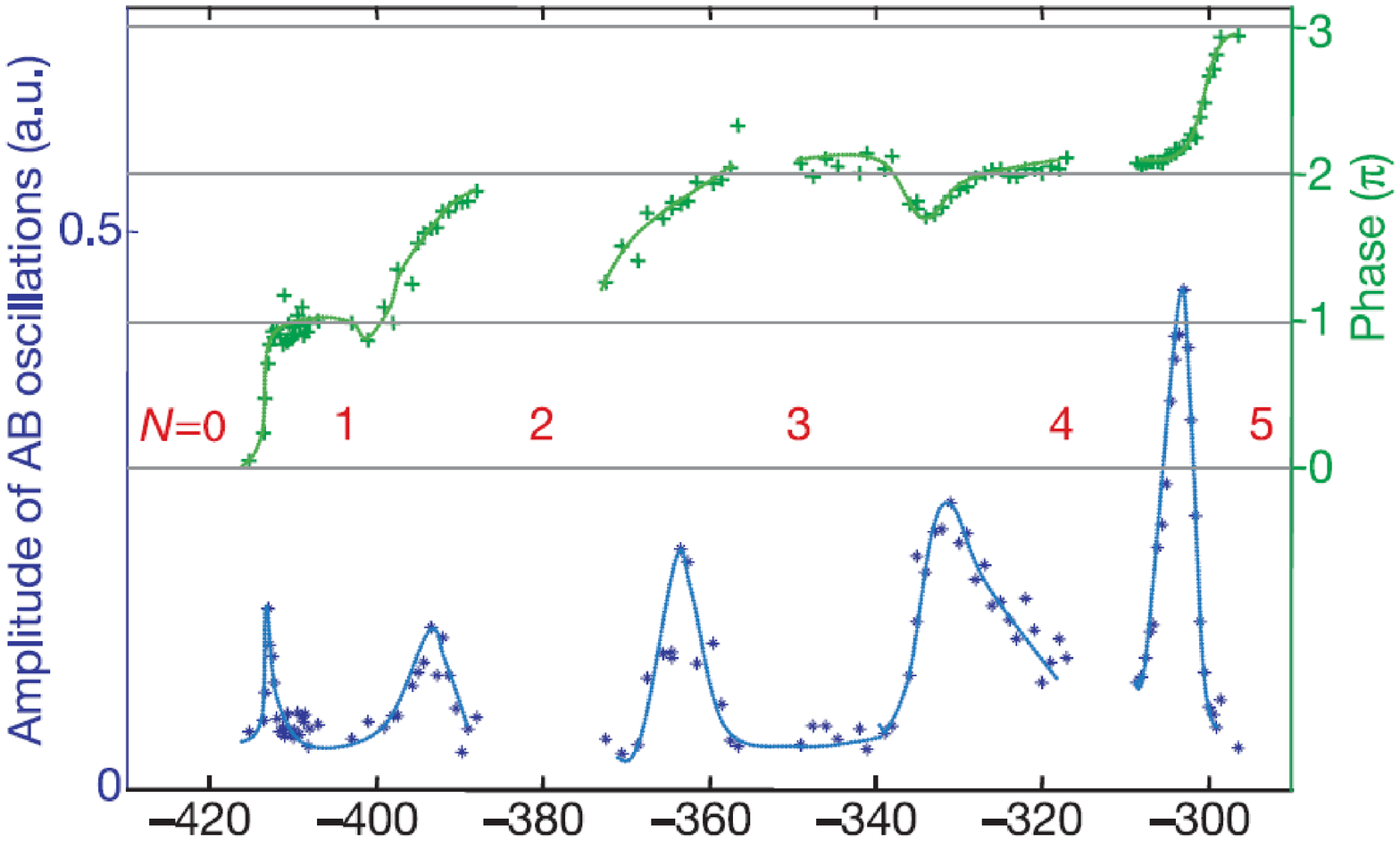}\hfill
 	\includegraphics[height=4cm]{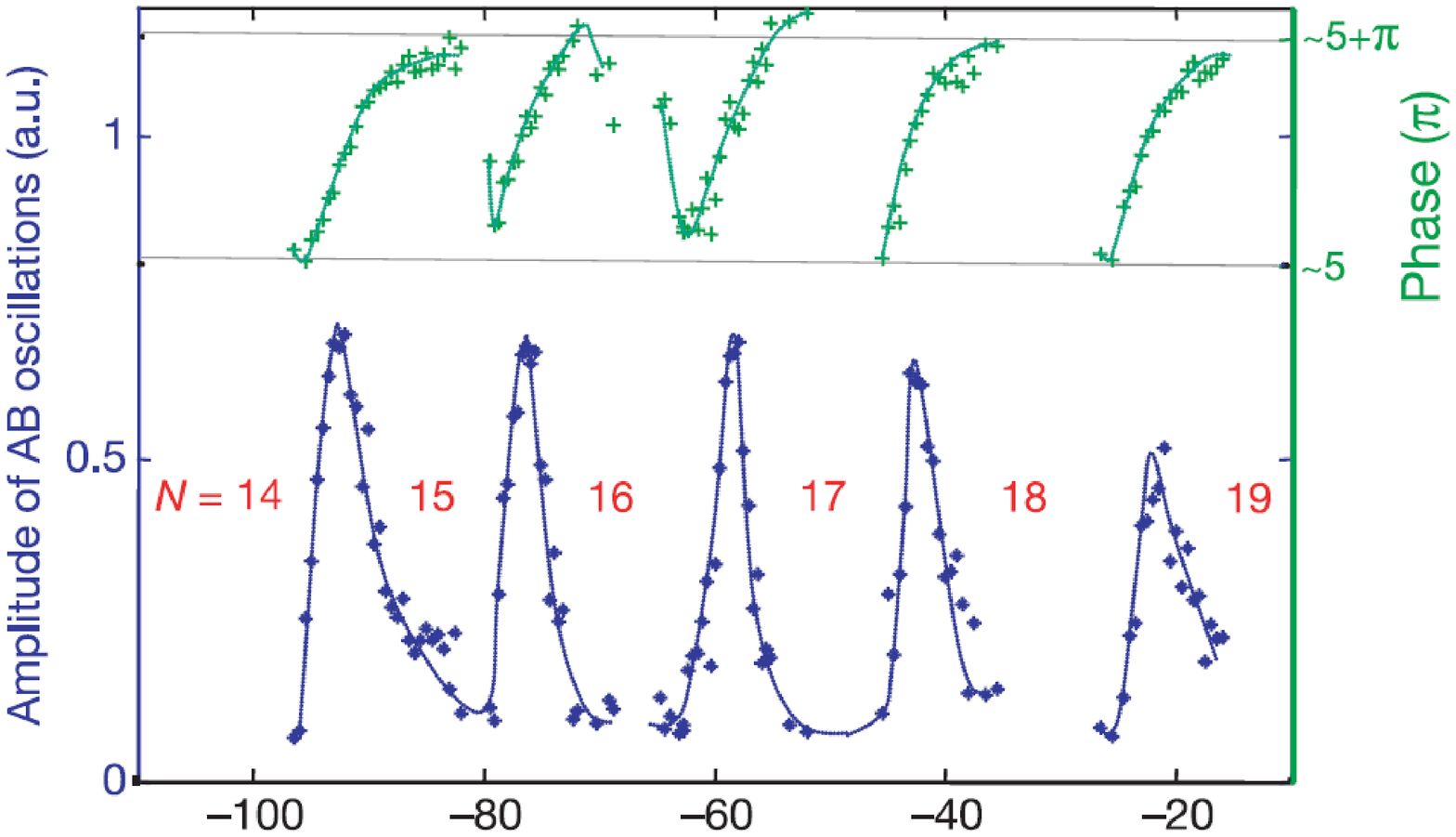}
        \caption{Measurement of the linear-response conductance (blue lines; the y axis is scaled in arbitrary units) and transmission phase (green lines) of a quantum dot experimentally realised as in Fig.~\ref{fig:DOT.exp1} as a function of a plunger gate voltage (x axis, scaled in $mV$) that lowers the energy of the dot region. The occupation $N$ of the dot was also measured and is show within the plots. (Reprinted by permission from Macmillan Publishers Ltd: Nature {\bfseries 436}, 529, copyright 2005.)}
        \label{fig:OS.expphase}
\end{figure}

As stated in the introduction, these results might serve as an explanation of the puzzling behaviour of the transmission phase observed in a series of transport experiments through single quantum dots \cite{phase1,phase2,phase3}, especially since in the very recent one (\cite{phase1}) in addition to the transmission probability and phase also the average occupation of the dot was measured. In the conductance, resonances were found each time another electron was added to the system. It showed that for dots that were only occupied by a few electrons ($\sim$5; mesoscopic regime) the phase either evolved continuously or jumped by $\pi$ in between,\footnote{Note that in the experimental papers the gate voltage is defined with opposite sign.} depending on the peak considered and on the dot measured (Fig.~\ref{fig:OS.expphase}, left panel), which is a typical `mesoscopic' behaviour. Surprisingly, $\alpha$ universally jumped by $\pi$ between all resonances for every experimental realised dot under consideration if the occupation of the latter was already `fairly large' ($\sim$15; universal regime, Fig.~\ref{fig:OS.expphase}, right panel). This is the behaviour which was observed in all previous experiments where the average occupation was not measured. As was already pointed out by \cite{phase1}, the fundamental difference between dots populated by a different number of electrons is the level spacing between the states that are occupied if further electrons are added to the system. For dots from the mesoscopic regime (those with a few electrons) one would expect the latter to be larger than for dots from the universal regime (which are occupied by a large number of electrons). Hence the size of the detuning between the levels should be taken as an input from the experiment if one seeks to set up a model to theoretically describe transport through such a single quantum dot. In this sense, our findings are consistent with the experimental observations. For $\Delta\ll\Gamma$ (universal regime), our calculations yield $N$ transmission resonances with $N-1$ zeros and corresponding phase jumps in between, with $N$ being the number of levels in the dot under consideration. This is precisely the behaviour found in the experiment for dots from the universal regime. On the other hand, for large $\Delta\gg\Gamma$ the evolution of the phase in between the $N$ conductance peaks depends on the actual parameters of the dot, in particular on the relative signs of the couplings of the levels associated with consecutive resonances. These signs are connected to the parity of the underlying wavefunctions and cannot be controlled in an experimentally realised system. Hence the theoretical prediction would just be that for dots occupied by a few electrons the phase should jump in between some peaks while evolving continuously between other and that the observed behaviour should change with the system under consideration. Again, this is precisely what was observed in the experiment.

A few things concerning our model calculations need to be pointed out. First, it is important to note that the behaviour in the universal regime follows from the presence of the two-particle interaction; it is not observed if one solves the noninteracting model. Second, the generic behaviour in both regimes is completely independent of all other dot parameters in the limits $\Delta\ll\Gamma$ and $\Delta\gg\Gamma$. The crossover regime and the scale $\Delta_\tn{cross}$ where the latter sets in depend on the level-lead hybridisations and the relative signs of the couplings, but in the limits of small and large level spacings one always ends up with the same behaviour characterising the universal and the mesoscopic regime. The typical ratio used in the calculations is $\Delta\gg\Gamma/\Delta\ll\Gamma\approx 200$, which corresponds to 15 levels if one assumes a quadratic dispersion of the dot level energies, a value which is in agreement with the experiment. Of course, one might argue that in the latter one probed the phase evolution for arbitrary many electrons in between both limits occupying the quantum dot, and hence the level spacing should decrease continuously. Our calculations, however, yield resonances accompanied with non-universal phase behaviour even for $\Delta$ from the crossover regime, consistent with what is measured for mesoscopic dots. Third, one has to contemplate about why the additional correlation induced structures were not observed in the experiment. This might be because the interaction is too small, the temperature too high, or because the spacing of the levels that are filled in the universal regime is still too large. However, this argumentation is flawed by the fact that it would lead to strongly non-universal behaviour depending on the dot under consideration. In particular, there are no universal values of $U$, $T$ or $\Delta$ such that the additional features do not appear (or are smeared) but at the same time the desired $N$-peak structure with $N-1$ fairly sharp phase jumps is still observed no matter how the couplings are chosen. For each individual dot, there are certainly choices of $\{U,T,\Delta\}$ such that no correlation induced phenomena but otherwise well-pronounced resonances and phase jumps are visible; however, another dot (with different couplings) might for the same choice $\{U,T,\Delta\}$ be in the crossover regime, or the temperature might be so high that the resonances or the jumps are already smeared in either the universal or the mesoscopic regime (remember that for the latter it is $T/\Gamma_l$ which determines a typical temperature scale). A much more likely explanation why the additional features are not observed that would not spoil the universality for $\Delta\ll\Gamma$ and $\Delta\gg\Gamma$ could be that the experimental resolution is much too small to observe these in general sharp structures which become even exponentially sharper with the interaction strength increased beyond $U_c$. Taking a look at the curves measured in the experiment (Fig.~\ref{fig:OS.expphase}), this does not seem to be too far-fetched. Fourth, one should note that it would of course be interesting to extend our model to include the spin degree of freedom (though the latter seems to play no role in the experiment (following \cite{imry,GG}); note that, however, our findings for spinful models in the next section do not contradict the interpretation presented here), and to perform calculations at finite temperatures (though $T$ is the smallest energy scale in the experiment; see \cite{phase1}). Finally, we want to point out very clearly that we do not seek to \textit{fit} the experimental curves by our calculations. It is of course important that the transmission resonances we computed are of comparable width and height and that the phase almost changes by $\pi$ over each of them. Nevertheless, the actual lineshape of the curves measured in the experiment might deviate from our findings, maybe because we use an oversimplified model, or maybe because the resolution in the experiment does not allow for a unique fit of the measured points. The most important observation, however, is that in the two-level case at large interactions $U\gg\Gamma$, our calculations yield curves which quantitatively resemble those observed in the experiment, in particular if one is concerned with the separation of the resonances and the S-like structure of the phase. Employing NRG calculations \cite{phaselapsespaper} one can show that the same holds for multi-level dots where we are unable to tackle large interactions $U\gg\Gamma$. However, all this is of minor importance since it does not influence or main argument: for small level spacings (universal regime), independent of the dot's parameters we find $N$ transmission resonances with $N-1$ phase jumps in between, while for large level spacings (mesoscopic regime) $\alpha$ evolves continuously or jumps by $\pi$ depending on the system under consideration. Thus we might hope that our calculations provide a natural explanation of the experiments.

\newpage
\section{Spinful Dots}\label{sec:MS}

In this section we will study transport through dots including the spin degree of freedom, where the local interaction of spin up and down electrons leads to Kondo physics \cite{hewson} at sufficiently low temperatures. Signatures of the Kondo effect were first observed in 1934 as an increase of the resistivity of gold when the temperature was sufficiently lowered. The set-in of the regime where the resistivity does no longer decrease with decreasing temperature was found to depend on the quality of the sample under consideration, and it was only observed at all if the latter contained local magnetic moments. In 1964 it was finally Y. Kondo who succeeded in fitting the experiments by quantum mechanical calculations based on a Hamiltonian that describes free electrons antiferromagnetically coupled to a spin one-half magnetic impurity. Below a characteristically energy scale, the Kondo temperature, this magnetic moment gets screened by the delocalised free electrons which leads to an increased scattering and hence the resistivity increases as well. If, on the other hand, a single quantum dot has an average occupation close to an odd integer number, it shows a resonance in the conductance due to quantum fluctuations between two spin-degenerate states. In fact, it can be shown that a quantum dot with a single electron in the state of highest energy modeled by an Anderson impurity type Hamiltonian is similar to a magnetic spin one-half impurity. In contrast to metals, an increased scattering leads to an increased conductance through the quantum dot (because additional transport channels are available).
\begin{figure}[b]	
	\centering
	      \includegraphics[width=\textwidth,clip]{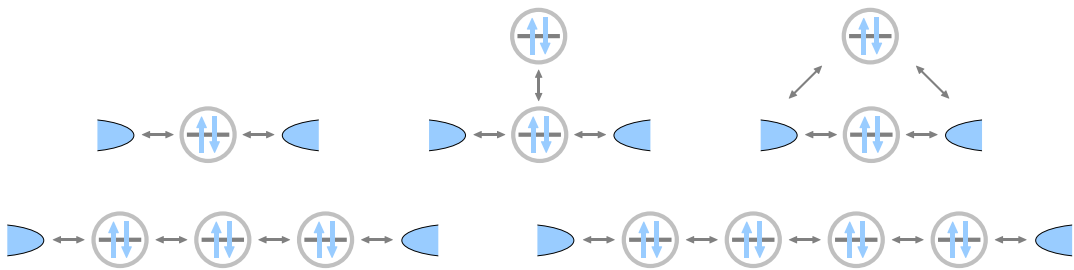}
        \caption{Dot geometries considered in this section.}
\label{fig:MS.geo}
\end{figure}

First, we will study the Kondo resonance in transport through a one-level quantum dot and especially show how the Kondo temperature, the typical energy scale that destroys this resonance, can be extracted even from the linear-response conductance at $T=0$ by the application of a magnetic field. We will then report on transport properties of up to four of these dots coupled in series, which can be viewed as a short Hubbard chain. Such systems have been studied recently using the numerical renormalization group (NRG) \cite{chain3a,chain3b,chain3c,chain4}. Here, we will demonstrate the power of the fRG to scan the parameter space for regions usually not covered by NRG. In particular, we will show that choosing non-symmetric couplings to the leads (which in general would correspond to the generic case) results in a vanishing of transmission resonances. Next, we will describe the so called two-stage Kondo effect that is characteristic for transport through a quantum dot to which another dot is attached by a tunnelling barrier, the so-called side-coupled geometry. This situation has been of recent interest as well \cite{side2,side3}, and we will again demonstrate the power of the fRG to scan the whole parameter space, this time verifying that the results obtained by the aforementioned authors using the NRG indeed represent the generic situation. Finally we will again turn to the system of parallel double and triple dots studied in the previous section, but here we will describe the physics that arises if we include the spin degree of freedom. These geometries have not been investigated in the literature so far.

\subsection{A Single Impurity}\label{sec:MS.sd}

\subsubsection{Application of the fRG - the Kondo Resonance}

The free propagator with the leads projected out (\ref{eq:DOT.gpp}) reads
\begin{equation}\label{eq:MS.singledot.g}
\mc G_\sigma^0(i\omega) = \frac{1}{i\omega - V_g + \sigma B/2 +i~\tn{sgn}(\omega)~\Gamma},
\end{equation}
with $B$ being a magnetic field applied to the system, $V_g$ the gate voltage that shifts the on-site energy of the dot, $\Gamma$ the sum of left and right hybridisations, $\Gamma=\Gamma_L+\Gamma_R$, and we have (as usual) performed the wide band limit. The fRG flow equations for the self-energy and the effective interaction (\ref{eq:FRG.flowse3}) and (\ref{eq:FRG.flowww3}) take the form
\begin{equation*}\begin{split}
\partial_\la V_\sigma(\la) = & -\frac{U(\la)}{2\pi}\sum_{\omega=\pm\la}\tilde{\mc G}_{\bar\sigma}(i\omega)
= \frac{1}{\pi}\frac{U(\la)V_{\bar\sigma}(\la)}{(\la+\Gamma)^2+V_{\bar\sigma}^2(\la)} \\[2ex]
\partial_\la U(\la) = & \frac{U^2(\la)}{2\pi}\sum_{\omega=\pm\la}\left[
\tilde{\mc G}_\uparrow(i\omega)\tilde{\mc G}_\downarrow(-i\omega) + 
\tilde{\mc G}_\uparrow(i\omega)\tilde{\mc G}_\downarrow(i\omega) \right] \\
= & \frac{2}{\pi}\frac{U^2(\la)V_\uparrow(\la)V_\downarrow(\la)}
{[(\la+\Gamma)^2+V^2_\uparrow(\la)][(\la+\Gamma)^2+V^2_\downarrow(\la)]},
\end{split}\end{equation*}
where $\tilde{\mc G}$ is obtained from (\ref{eq:MS.singledot.g}) by the replacement $V_g-\sigma B/2\to V_\sigma$, and the initial conditions read $V_\sigma(\la\to\infty)=V_g-\sigma B/2$ and $U(\la\to\infty)=U$, with $U:=U^{\sigma,\bar\sigma}$ being the local interaction of spin up and down electrons.
\begin{figure}[t]	
	\centering
	\includegraphics[width=0.475\textwidth,height=5.2cm,clip]{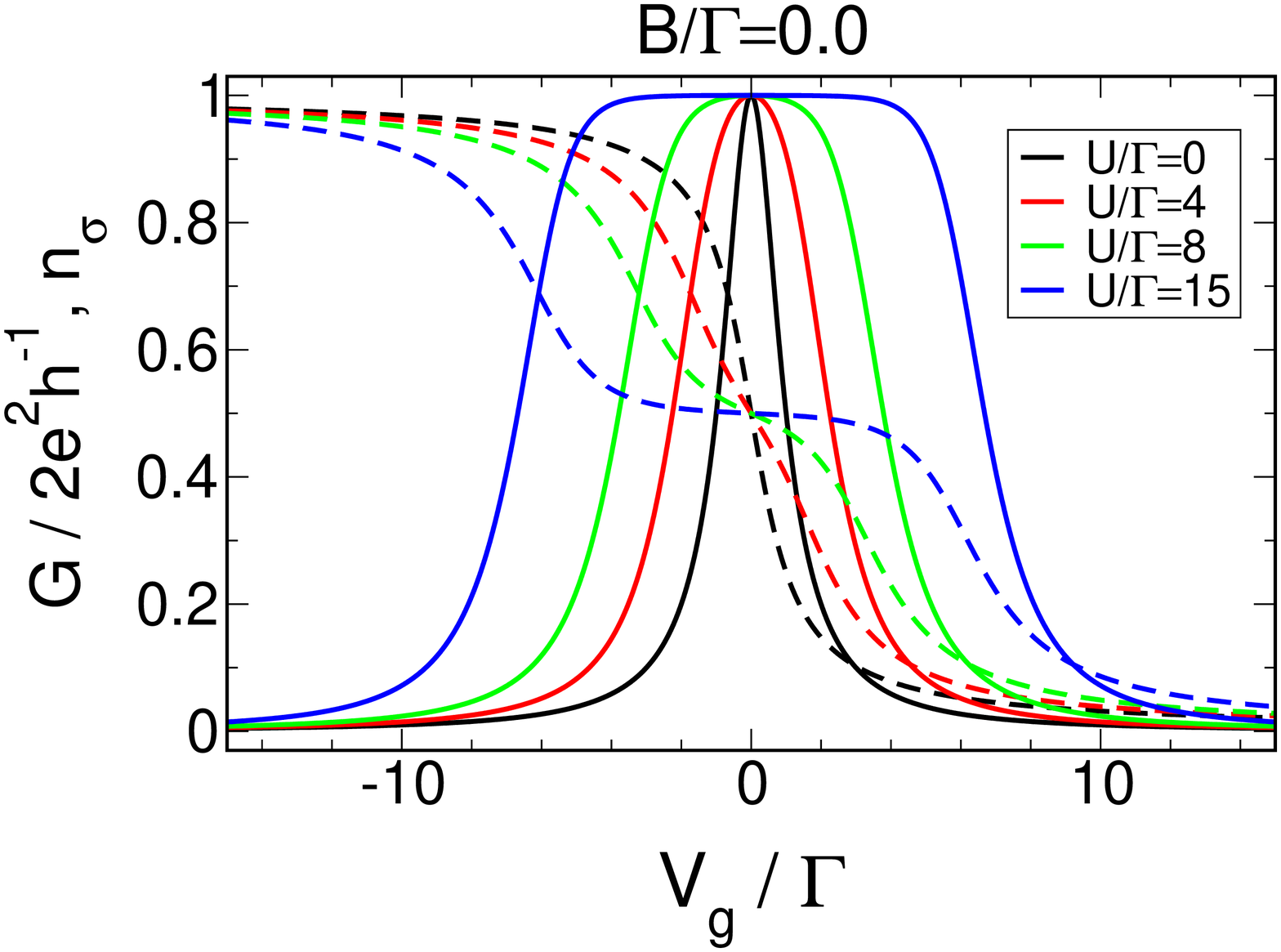}\hspace{0.035\textwidth}
        \includegraphics[width=0.475\textwidth,height=5.2cm,clip]{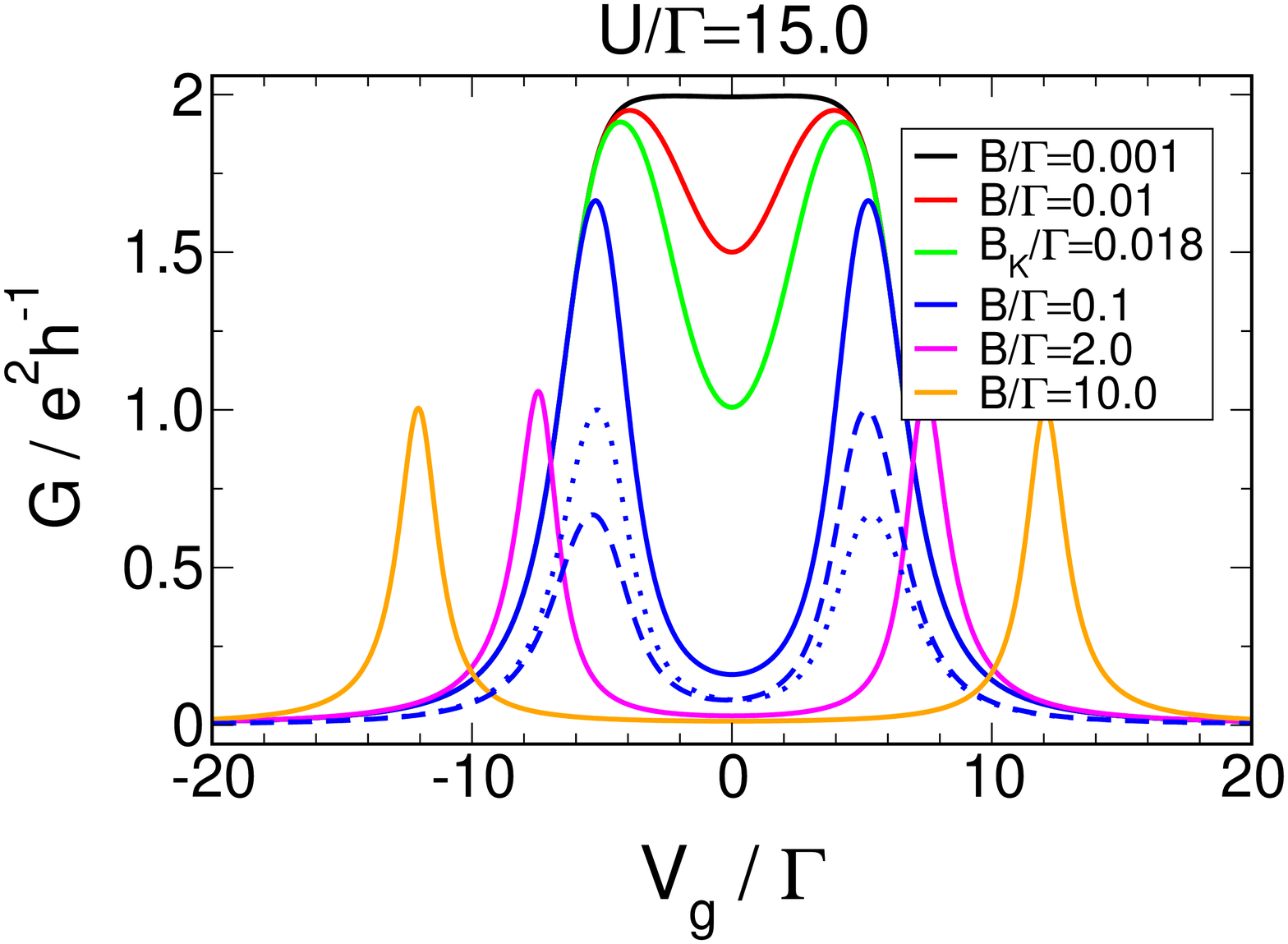}
        \caption{\textit{Left panel:} Gate voltage dependence of the conductance $G$ (solid lines) and average level occupation (dashed lines) for a single-level dot with $\Gamma_L/\Gamma=\Gamma_R/\Gamma=0.5$ and zero magnetic field for different local interactions. If the interaction is large enough ($U/\Gamma\gg 1$), the Lorentzian lineshape of $G(V_g)$ observed in the noninteraction case is altered and the Kondo effect leads to a wide plateau of almost perfect conductance in a region of width $U$ centred around $V_g=0$. Within this plateau, the average occupation of the dot is constant (and in particular odd). \textit{Right panel:} The same as the left panel, but for fixed $U/\Gamma=15.0$ and various magnetic fields. The spin degeneracy is lifted and the Kondo resonance gets destroyed at $B\approx T_K$. This implies that the Kondo temperature can be extracted from the linear-response conductance at $T=0$. For $B/\Gamma=0.1$, the partial conductance of the spin up (dashed) and spin down (dotted) electrons is shown as well.}
\label{fig:MS.singledot.ub}
\end{figure}

If we solve these differential equations numerically and compute the linear-response conductance,
\begin{equation}
G=G_\uparrow+G_\downarrow = 4\frac{e^2}{h}\frac{\Gamma_L\Gamma_R}{V_\uparrow^2(0)+\Gamma^2} + 4\frac{e^2}{h}\frac{\Gamma_L\Gamma_R}{V_\downarrow^2(0)+\Gamma^2},
\end{equation} 
we recover precisely what was claimed in the introduction. The Lorentzian lineshape that corresponds to the noninteracting case is gradually transformed into a box of width $U$ where the conductance reaches the unitary limit if the interaction is increased (this is shown in the left panel of Fig.~\ref{fig:MS.singledot.ub} for the left-right symmetric case; choosing $\Gamma_L\ne\Gamma_R$ changes only the height of the resonance). For gate voltages within this Kondo resonance the mean occupation of the dot is odd and the spin degeneracy of the states with one electron give rise to an increased scattering amplitude and can therefore be understood as the reason for the perfect conductance. Despite the fact that the transmission phase $\alpha$ and level occupation $n_\sigma$ are directly related to $G$ by a generalised Friedel sum rule \cite{hewson} (which holds exactly in our approximation since we are deriving an effective noninteracting model), the latter is shown as well in Fig.~\ref{fig:MS.singledot.ub}.

Since we have dropped all frequency dependencies in the fRG flow equations and mapped the interacting system to an effective noninteracting one, for arbitrary $V_g$ the spectral function will always be a Lorentzian centred around $V_\sigma(0)$. This lineshape differs strongly from the exact one (obtained from the analytic Bethe ansatz solution \cite{hewson}, or very precise NRG data \cite{kondonrg}). In particular, neither the resonance at $\omega=0$ is very sharp, nor the spectral function exhibits the signatures of atomic physics for large $U$, the Hubbard bands. Hence, we cannot expect to obtain reliable results for the conductance at high temperatures. For $T=0$, however, all that enters in the computation of the latter is the spectral function at zero frequency, and it turns out that the fRG successfully captures its pinning at $\omega=0$ for gate voltages close to $V_g=0$. In particular, the renormalized level energy $V_\sigma(0)$ is approximately zero in a region of size $U\gg\Gamma$ around $V_g=0$. For the fRG truncation scheme that keeps only the flow of the self-energy an exponential pinning of the spectral weight at $\omega=0$ can even be shown analytically \cite{dotsystems}. If on the other hand one keeps the whole frequency dependence of the two-particle vertex, it is possible to compute a spectral function that shows a the sharp Kondo resonance as well as the Hubbard bands, although with much higher numerical effort \cite{ralf}.

Adding a magnetic field $B$ to the system lifts the spin degeneracy at the particle-hole symmetric point and thus destroys the Kondo resonance. The wide plateau in $G(V_g)$ present for $B=0$ splits up and for large fields (in particular when $B$ is the dominant energy scale) we recover two Lorentz-like resonances separated approximately by $U+B$ which correspond to individual transport of the spin up and spin down electrons (Fig.~\ref{fig:MS.singledot.ub}, right panel). This is similar to the spinless two-level dot with large level spacing.
\begin{figure}[t]	
	\centering
        \includegraphics[width=0.475\textwidth,height=4.8cm,clip]{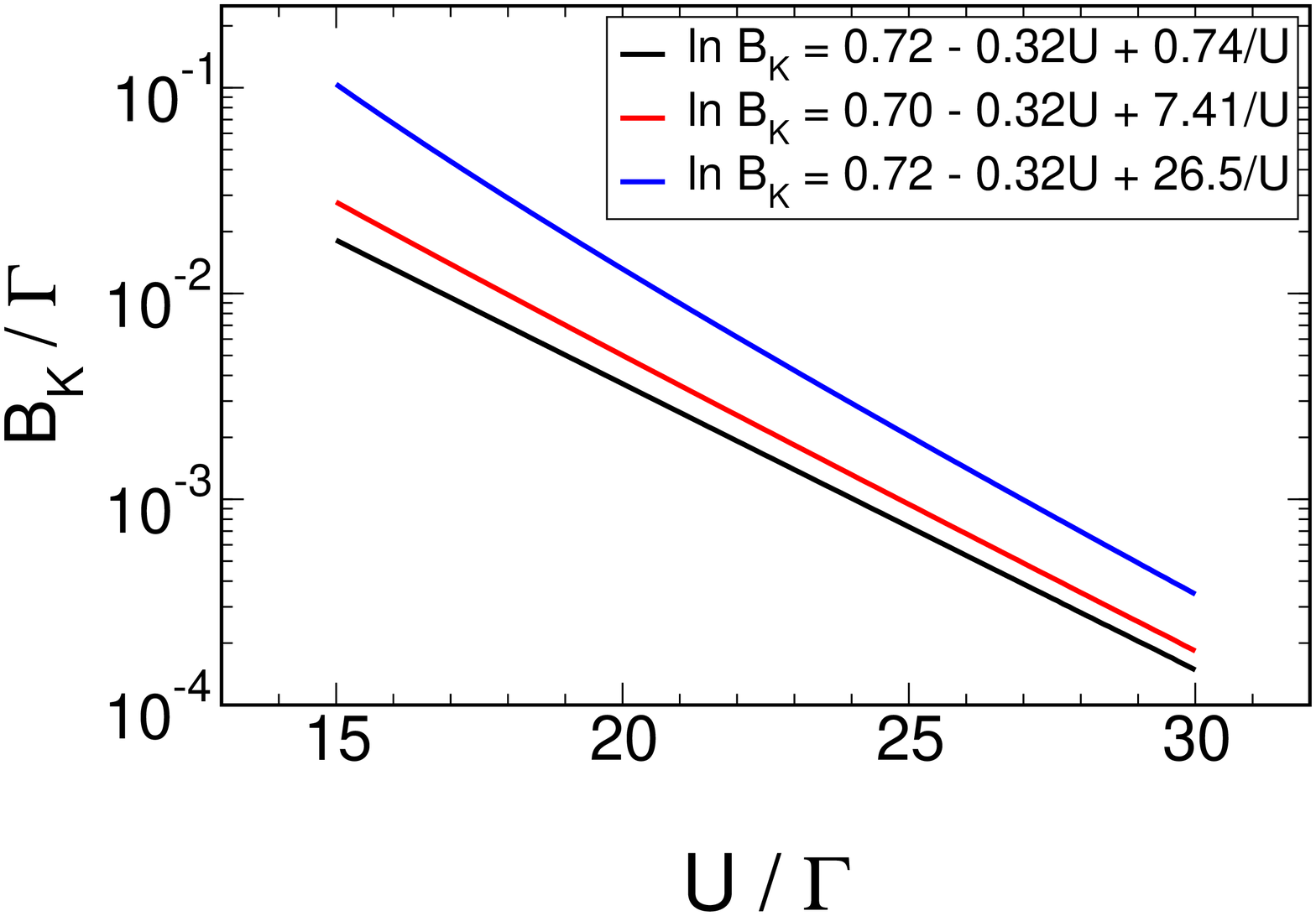}\hspace{0.035\textwidth}
        \includegraphics[width=0.475\textwidth,height=4.92cm,clip]{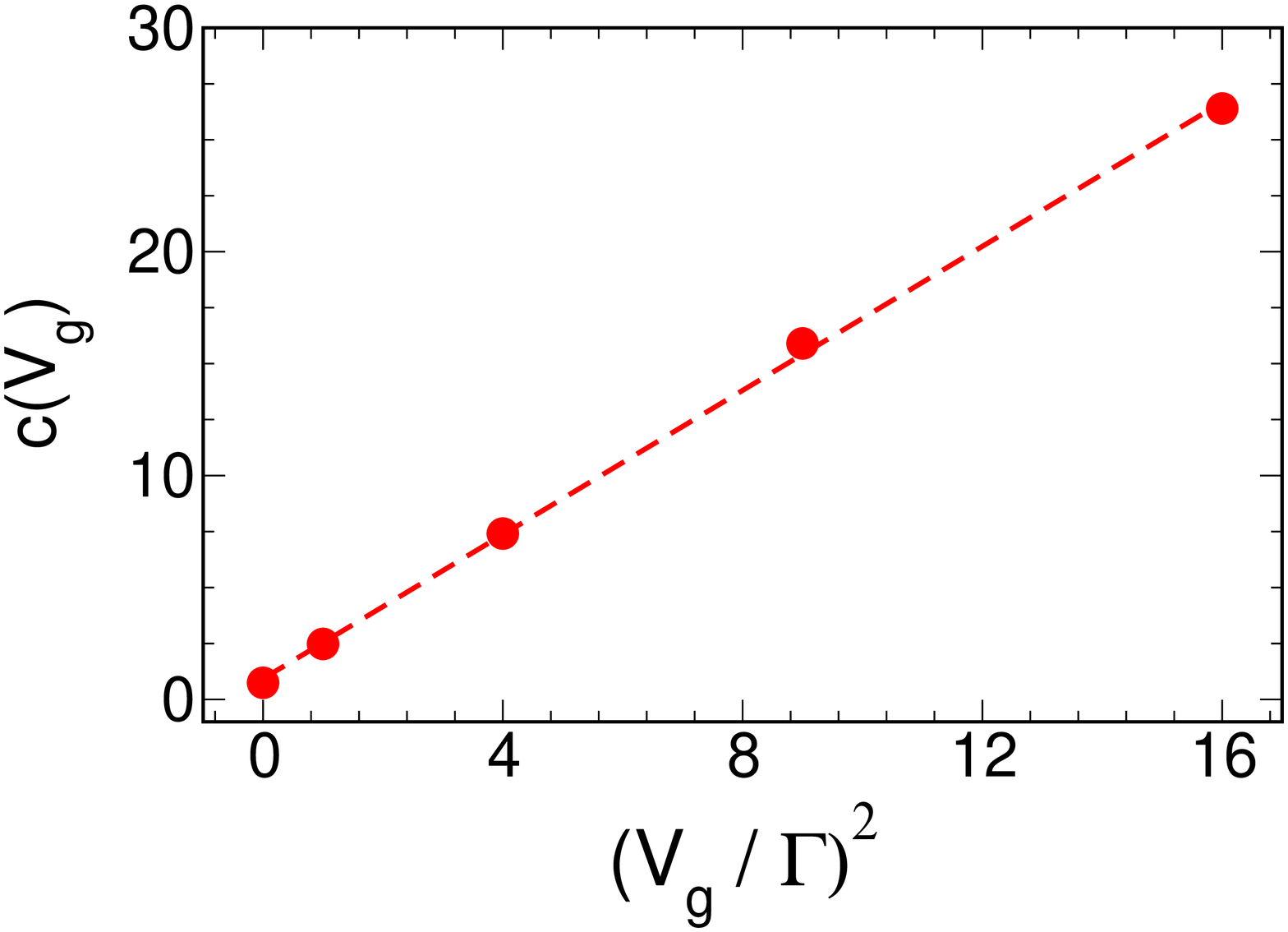}
        \caption{\textit{Left panel:} The Kondo energy scale $B_K$ of a single-level dot as a function of $U$ for different gate voltages within the Kondo resonance, $V_g=0.0$ (black), $V_g=2.0$ (red), and $V_g=4.0$ (blue). The data points obtained from fRG where fitted as $B_K=a\exp\left(-\left|bU/\Gamma-c\Gamma/U\right|\right)$. The values computed for the fitting parameters are shown in the inset. Their error is less than one percent, and the fitting curves cannot be distinguished from the original data on the scale of this plot. \textit{Right panel:} Gate voltage dependence of the fitting paramater $c$.}
\label{fig:MS.singledot.kondo}
\end{figure}

\subsubsection{Determination of the Kondo Energy Scale}

A meaningful definition of the energy scale on which the Kondo resonance survives could be the magnetic field strength $B_K$ required to reduce the conductance down to half the unitary limit (with $V_g$ implicitly assumed to be chosen such that $G(V_g)$ is of almost unitary height, i.e. within the conductance plateau). If on the other hand one tackles the system by an NRG approach that gives the correct spectral function, one usually defines this energy scale, which is then called Kondo temperature $T_K$, as the width of the Kondo resonance at zero frequency. In the exact solution, its dependence on the dot paramaters is given by
\begin{equation}
T_K = C(U,V_g) \exp \left[-\left|\frac{\pi}{8}\frac{U}{\Gamma}-\frac{\pi}{2}\frac{V_g^2}{\Gamma^2}\frac{\Gamma}{U}\right|\right],
\end{equation} 
where the prefactor $C(U,V_g)$ is constant to leading order in $V_g$ and $U$ \cite{hewson}. Hence it is meaningful to probe whether the fRG results for the Kondo energy scale defined by the magnetic field dependence of the conductance give reliable results by computing $B_K$ for several values of $U$ and $V_g$ and fit the resulting curves $B_K(U)$ by a function of the form
\begin{equation}
f(U/\Gamma) = a(V_g)\exp\left[-\left|b(V_g)\frac{U}{\Gamma}-c(V_g)\frac{\Gamma}{U}\right|\right].
\end{equation} 
This is done in Fig.~\ref{fig:MS.singledot.kondo}. It turns out that the fitting parameter $b\approx 0.32$ is almost independent of $V_g$ and indeed close to the exact value $\pi/8\approx 0.39$. Furthermore, the factor $c$ depends approximately quadratically on the gate voltage (Fig.~\ref{fig:MS.singledot.kondo}, right panel), while the $V_g$ dependence of prefactor $a$ is only weak. Altogether, this shows that the exponentially small Kondo energy scale can indeed by extracted from the linear-response conductance computed by our fRG approximation scheme at zero temperature.

\subsection{Linear Chains of Dots}\label{sec:MS.chain}

\subsubsection{General Considerations}

In this section we study transport through a system that contains $N$ single-level dots connected in series, a short Hubbard chain. The projected free propagator for this geometry reads
\begin{equation}\label{eq:MS.chain.g}
\left[\mc G^0_\sigma(i\omega)\right]^{-1} = i\omega - V_g + \sigma\frac{B}{2}+
\begin{pmatrix}
i~\tn{sgn}(\omega)~\Gamma_L & t_{12}& \cdots & t_{1N} \\
t_{21} & -\Delta_1 & \cdots &\vdots\\
\vdots & & & \\
t_{N1} & \cdots & &- \Delta_{N-1} + i~\tn{sgn}(\omega)~\Gamma_R
\end{pmatrix},
\end{equation}
with real hoppings $t_{ij}=t_{ji}$, arbitrary level spacings $\Delta_i$, a magnetic field $B$, the gate voltage $V_g$ that simultaneously shifts the on-site energies of the different levels, and hybridisations with the left and right lead $\Gamma_L:=\pi|t_1^L|^2\rho_{\tn{lead}}$ and $\Gamma_R:=\pi|t_N^R|^2\rho_{\tn{lead}}$, respectively. In the following we will mainly focus on the special case where only equal nearest-neighbour hoppings $t:=t_{i,i+1}$ are present and all on-site energies are the same, implying $\Delta_i=0$. Furthermore, we will assume that there is only a local interaction $U:=U_{i,i}^{\sigma,\bar\sigma}$ between spin up and down electrons, which we will also choose to be independent of the level considered. Later on, we will relax all these assumptions and show that all our main results remain unchanged. It is important to note again that within our fRG approach this neither causes conceptional problems nor extreme numerical effort since all parameters enter the flow equations only via the free propagator (or the initial conditions), leaving at maximum $((2N)^2+(2N)^4 - \tn{symmetries})$ ordinary differential equations to solve. We will not consider the effects of a magnetic field here.

If we solve the fRG flow equations (\ref{eq:FRG.flowse3}, \ref{eq:FRG.flowww3}), we can as usual calculate the full propagator of the system $\mc G=\tilde{\mc G}(\la=0)$ and compute the linear-response conductance at zero temperature (\ref{eq:DOT.leitwert}) as
\begin{equation}
G = 4\frac{e^2}{h}\Gamma_L\Gamma_R\left\{\left|\mc G_{1;N}^\uparrow(0)\right|^2+\left|\mc G_{1;N}^\downarrow(0)\right|^2\right\}.
\end{equation} 
Even in the case that we start out only with nearest-neighbour hoppings and local interactions, it turns out that in general all elements of the self-energy and two-particle vertex (except those that violate spin-conservation and anti-symmetry) such as hoppings between arbitrary sites and long-range interactions are generated by the flow.

\subsubsection{A Chain of Three Dots}
\begin{figure}[t]	
	\centering
	      \includegraphics[width=0.485\textwidth,height=4.4cm,clip]{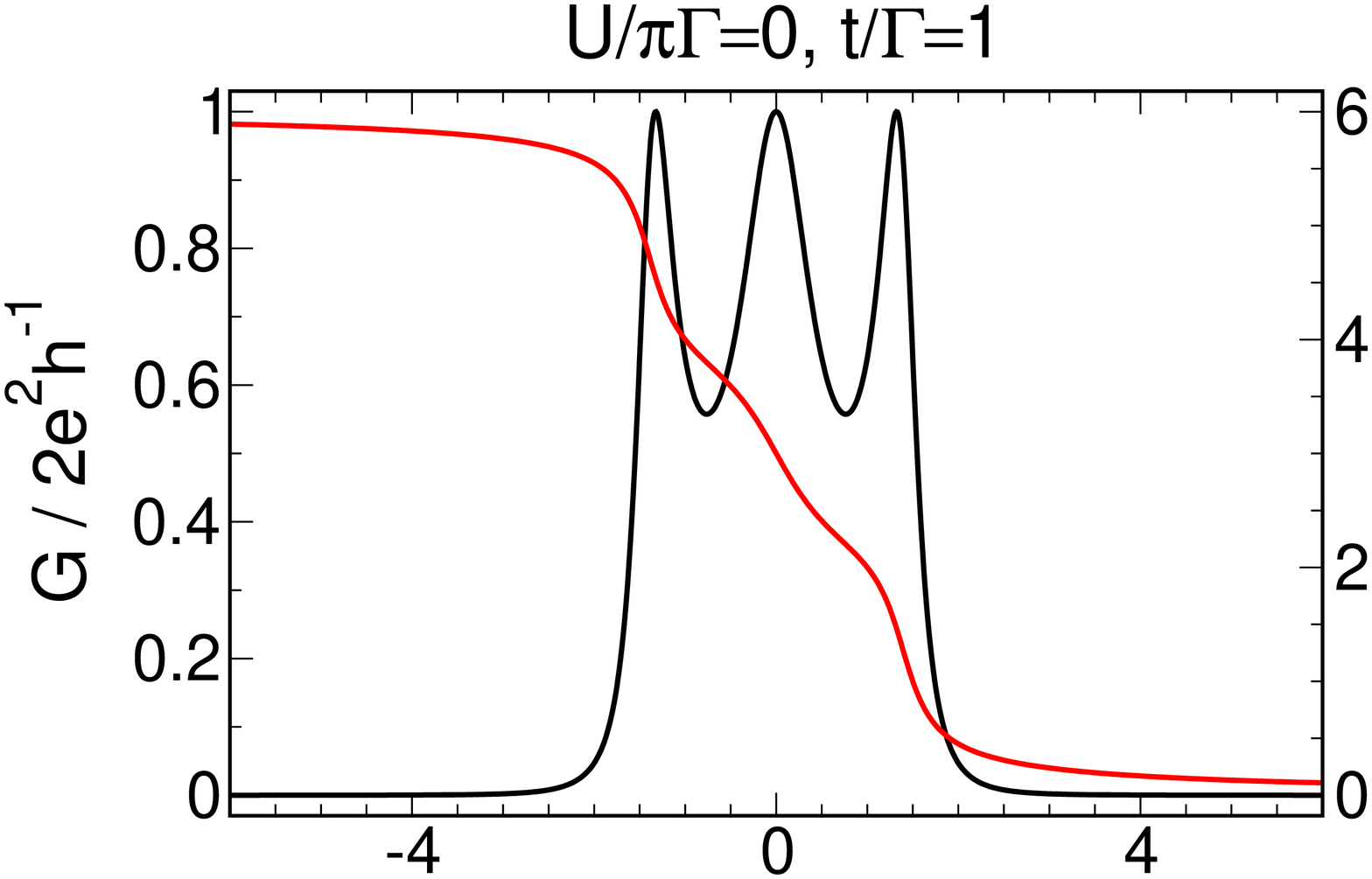}\hspace{0.015\textwidth}
        \includegraphics[width=0.485\textwidth,height=4.4cm,clip]{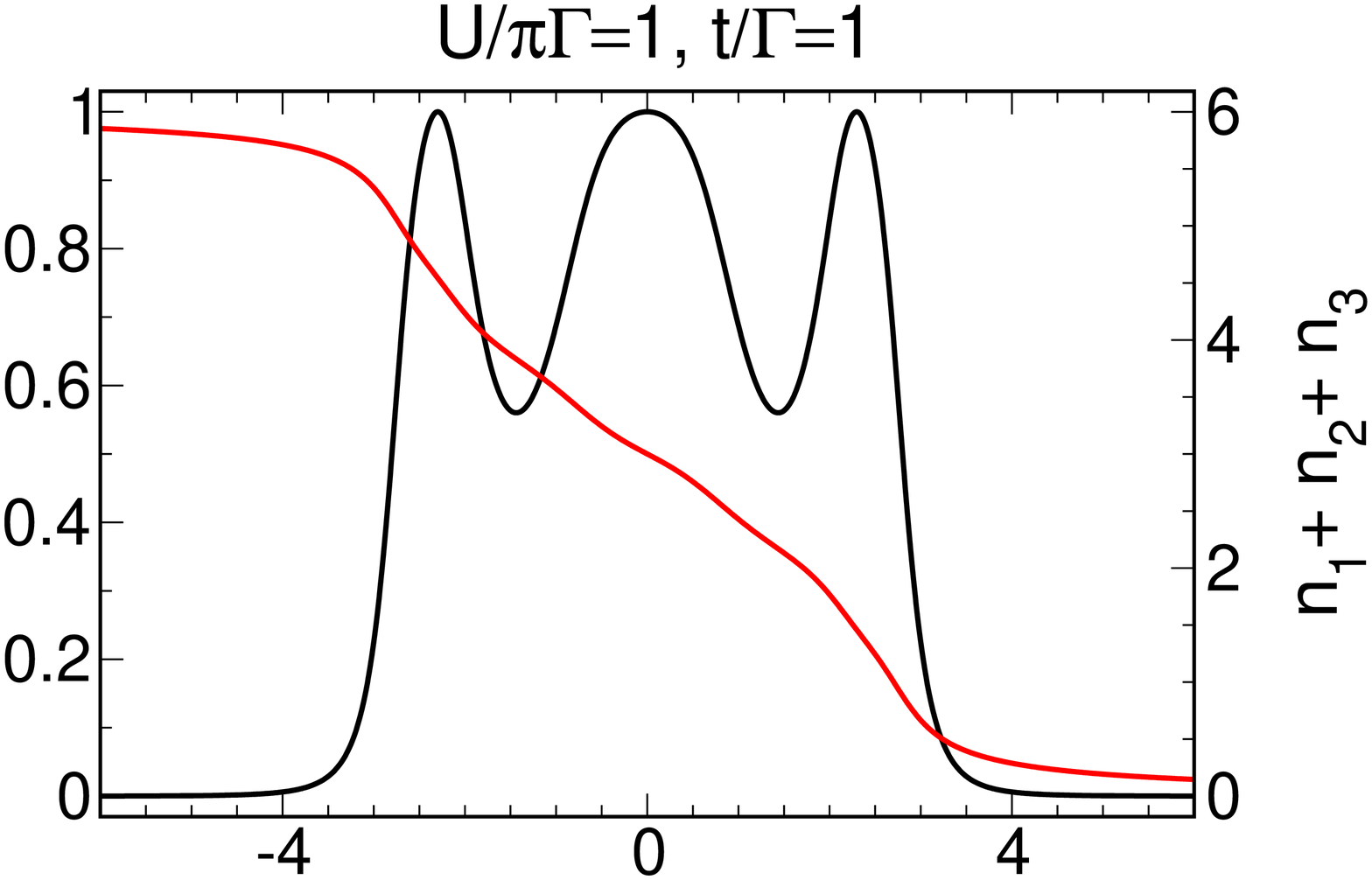}\vspace{0.3cm}
        \includegraphics[width=0.485\textwidth,height=5.2cm,clip]{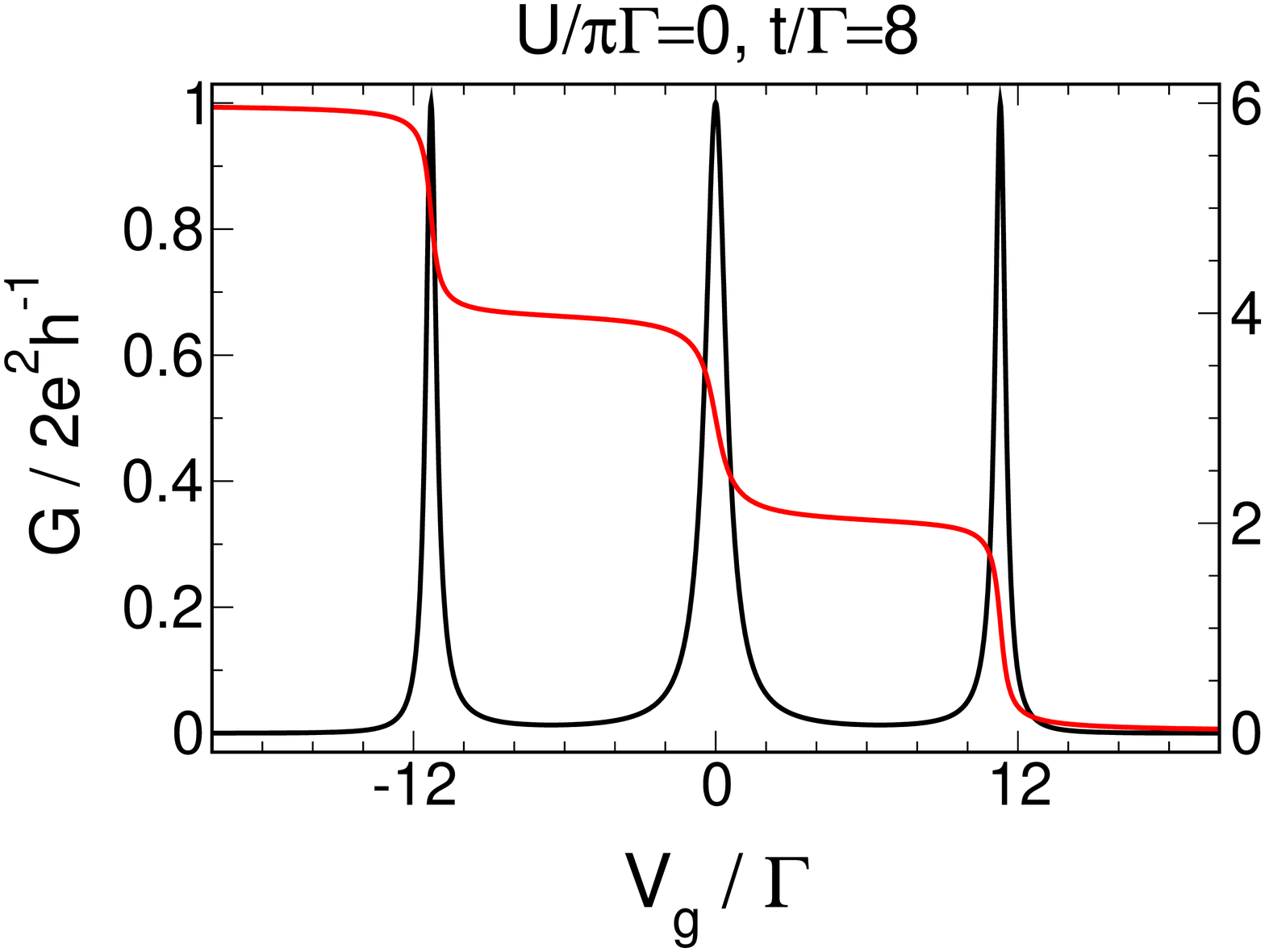}\hspace{0.015\textwidth}
        \includegraphics[width=0.485\textwidth,height=5.2cm,clip]{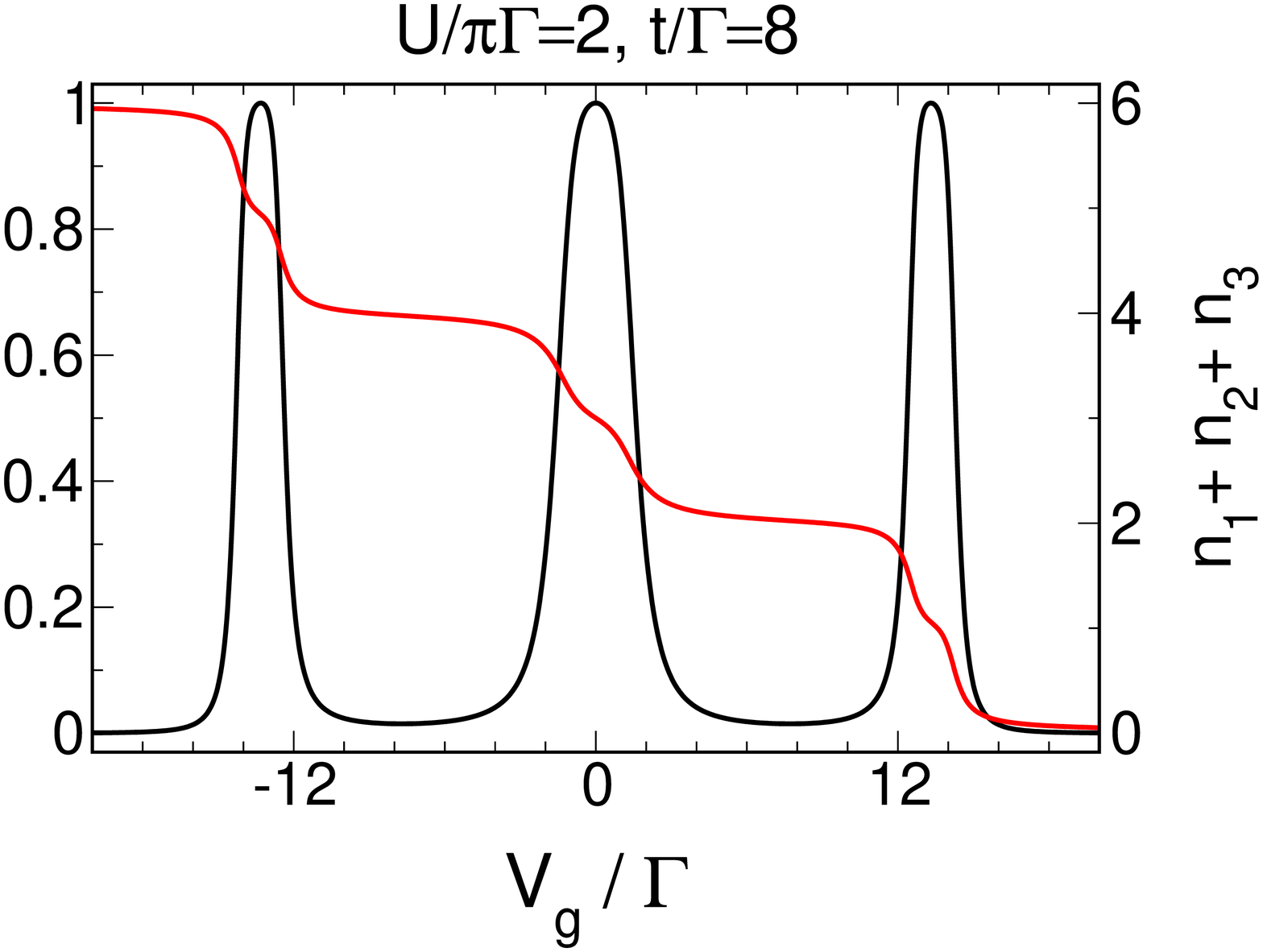}
        \caption{Gate voltage dependence of the conductance $G$ (black) and total average occupation (red) of a linear chain of three dots with $\Gamma_L/\Gamma=\Gamma_R/\Gamma=0.5$ for two different nearest-neighbour hoppings $t$ for the noninteracting case and for the largest $U/\Gamma$ that can be tackled by our fRG approximation scheme. Well-pronounced plateaus of unitary conductance and especially the wide plateau at $V_g=0$ caused by the suppression of charge fluctuations are only observed for larger interactions (see for example $t/\Gamma\approx 8$, $U/\Gamma\approx 16\pi$ in \cite{chain3a}).}
\label{fig:MS.chain.n3delta}
\end{figure}

We will start with three dots with nearest-neighbour hoppings $t$ and equal hybridisations with the left and right lead, respectively. It is again instructive to consider the noninteracting case first, which is very simple if we choose a basis where the dot Hamiltonian is diagonal. Since the eigenvalues of the latter are given by $\omega_2=V_g$ and $\omega_{1,3}=V_g\mp\sqrt{2}\hspace{0.2ex}t$ and the corresponding normalized eigenvectors read $(-\frac{1}{\sqrt{2}},0,\frac{1}{\sqrt{2}})$ and $(\frac{1}{2},\pm\frac{1}{\sqrt{2}},\frac{1}{2})$, transport through the system can be understood by three parallel levels with level spacing $\sqrt{2}\hspace{0.2ex}t$, hybridisations $\Gamma_2^L/\Gamma=\Gamma_2^R/\Gamma=0.25$ and $\Gamma_{1,3}^L/\Gamma=\Gamma_{1,3}^R/\Gamma=0.125$, and relative signs of the level-lead couplings $s_1=-$, and $s_2=-$. Hence for large nearest-neighbour hoppings $t$ compared to the strength of the level-lead hybridisation $\Gamma=\Gamma_L+\Gamma_R$ (the molecular orbital regime) we expect to find three Lorentzian conductance resonances of width $2(\Gamma_i^L+\Gamma_i^R)$ separated by $\sqrt{2}\hspace{0.2ex}t$, and the total average occupancy $n$ of the dot system to decrease by two (due to the two spin directions) over each peak. The transmission phase $\alpha_\sigma$ is no independent quantity since it is connected to $n$ by $\alpha_\sigma=\pi n/2$ \cite{chain4}, implying that it changes by $\pi$ over each peak and evolves continuously in between, as required by the two relative signs $s_{1,2}=-$. If the nearest-neighbour hopping is decreased and becomes comparable to $\Gamma$, the three levels overlap and contribute simultaneously to the transport. The resonances merge and for $t\ll\Gamma$ only one peak at $V_g=0$ is observed.

If we turn on an local interaction $U$ between the spin up and down electrons, the Kondo effect alters the $G(V_g)$ lineshape. For a chain comprising $N$ dots, the sharp Lorentzian resonances away from the particle-hole symmetric point in the molecular orbital regime are transformed into box-like peaks of width $U/N$, while the resonance at $V_g=0$ becomes a box even wider than the others. \cite{chain4} suggested to interpret this as a strong suppression of charge fluctuations at half-filling of the chain. The average number of electrons in the dots is consistent with the single-level Kondo behaviour. For gate voltages on-resonance, the occupation is odd and almost constant which leads to a half-integer total spin and hence a conductance plateau at the unitary limit. Between the resonances the occupation is even and transport is suppressed. Since for a chain with an odd number of sites half-filling corresponds to an odd number of electrons, the suppression of charge fluctuations around $V_g=0$ leads to a wide region of almost perfect conductance.

If the interaction becomes much larger than the separation of the peaks (which is still of order of the nearest-neighbour hopping $t$), or $t$ is smaller than the hybridisations with the leads such that already the noninteracting chain is no longer in the molecular orbital regime, the resonances begin to merge, and only one wide plateau of perfect conductance around $V_g=0$ is observed (the local regime). Within our approximation scheme, however, this parameter region is out of reach. A much more detailed discussion of the limitations of the fRG truncation scheme that neglects the frequency dependence of the two-particle vertex as well as the flow of all higher-order vertex functions will be given in the next chapter. In particular, by explicit comparison to very accurate NRG data we will establish upper boundaries for the interaction strength that can still be tackled within this approach for every system under consideration. A comparison between the noninteracting case and the largest $U$ that can be treated for a chain with three sites is shown in Fig.~\ref{fig:MS.chain.n3delta}.
\begin{figure}[t]	
	\centering
        \includegraphics[width=0.485\textwidth,height=4.8cm,clip]{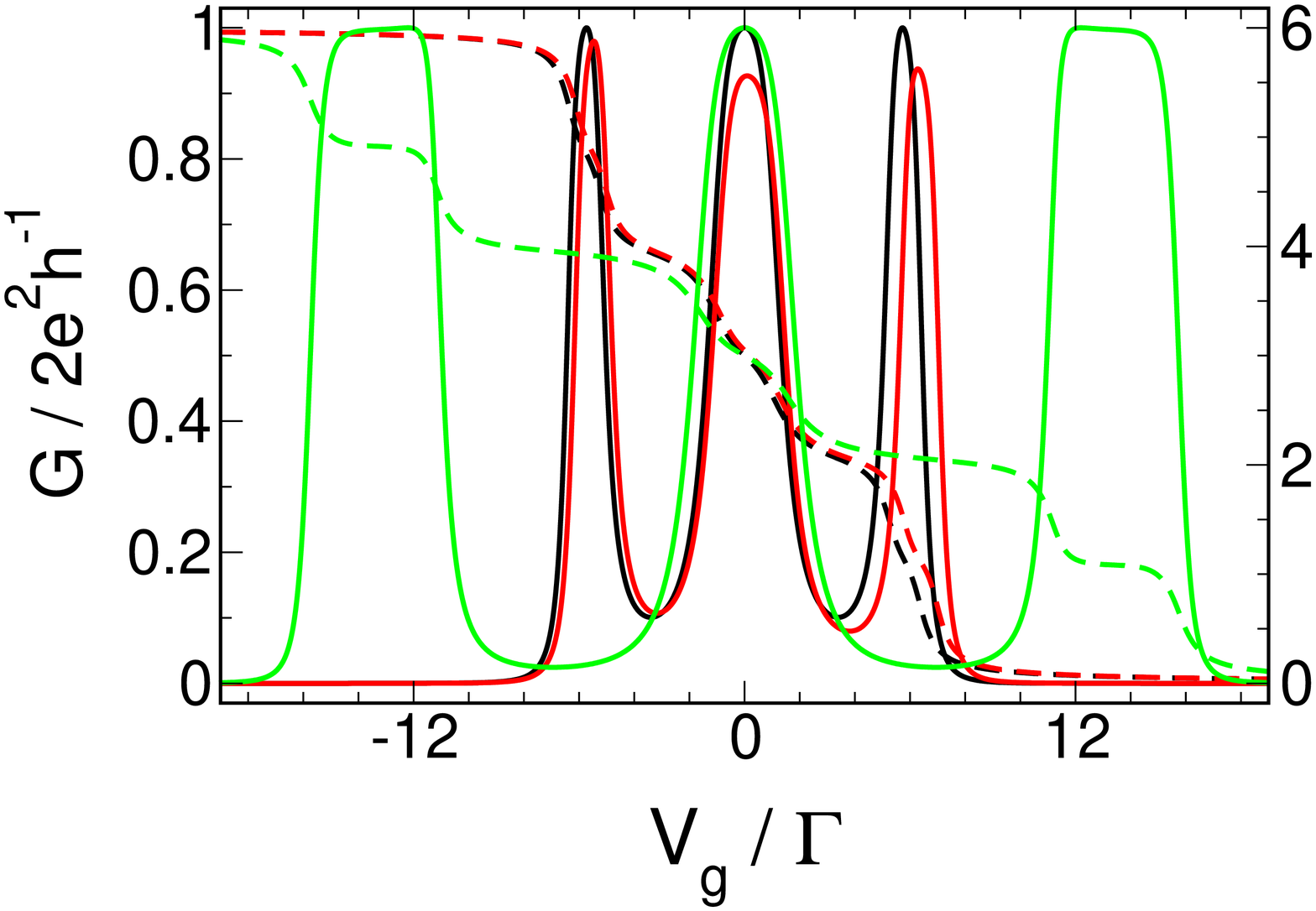}\hspace{0.015\textwidth}
        \includegraphics[width=0.485\textwidth,height=4.8cm,clip]{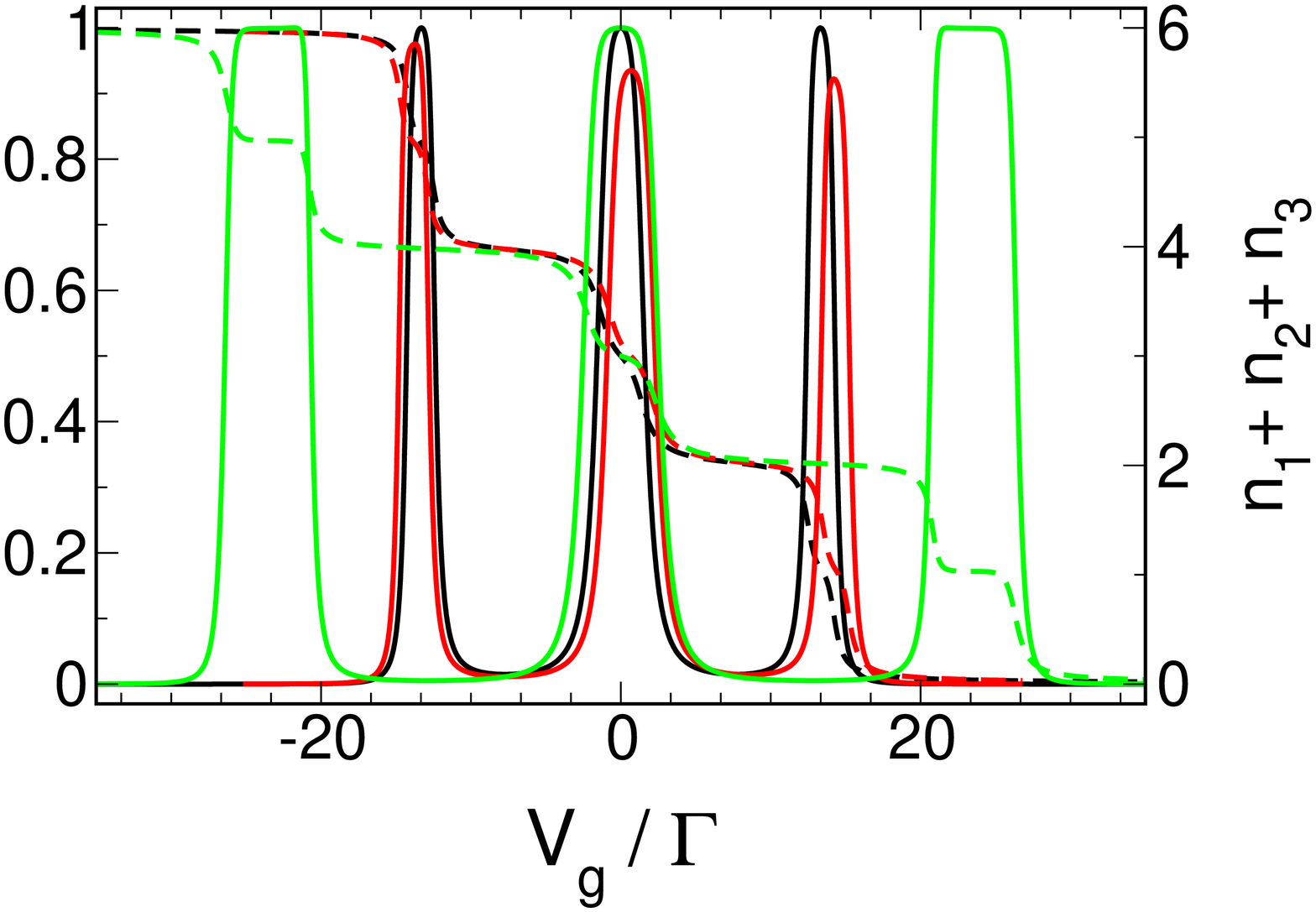}
        \caption{Gate voltage dependence of the conductance $G$ (solid lines) and total average occupation (dashed lines) of a linear chain of three dots with $\Gamma_L/\Gamma=\Gamma_R/\Gamma=0.5$ for six different choices of the nearest-neighbour hoppings $t_{ij}$, level spacings $\Delta_i$, local interaction $U_{\tn{loc}}$ and inter-level interaction $U_N:=U_{i,j}^{\sigma,\sigma}=U_{i,j}^{\sigma,\bar\sigma}$ present between all electrons of different levels. \textit{Left panel:} $t_{12}/\Gamma=t_{23}/\Gamma=3.0$, $\Delta_i/\Gamma=0.0$, $U_\tn{loc}/\Gamma=1.5\pi$, $U_N/\Gamma=0.0$ (black curves), $t_{12}/\Gamma=2.7$, $t_{23}/\Gamma=3.5$, $\Delta_1/\Gamma=-0.7$, $\Delta_2/\Gamma=-0.3$, $U_\tn{loc}/\Gamma=1.5\pi$, $U_N/\Gamma=0.0$ (red curves), $t_{12}/\Gamma=t_{23}/\Gamma=3.0$, $\Delta_i/\Gamma=0.0$, $U_\tn{loc}/\Gamma=U_N/\Gamma=1.5\pi$ (green curves). \textit{Right panel:} $t_{12}/\Gamma=t_{23}/\Gamma=8.0$, $\Delta_i/\Gamma=0.0$, $U_\tn{loc}/\Gamma=2.0\pi$, $U_N/\Gamma=0.0$ (black curves), $t_{12}/\Gamma=7.4$, $t_{23}/\Gamma=9.5$, $\Delta_1/\Gamma=0.9$, $\Delta_2/\Gamma=-0.9$, $U_\tn{loc}/\Gamma=2.0\pi$, $U_N/\Gamma=0.0$ (red curves), $t_{12}/\Gamma=t_{23}/\Gamma=8.0$, $\Delta_i/\Gamma=0.0$, $U_\tn{loc}/\Gamma=U_N/\Gamma=2.0\pi$ (green curves).  }
\label{fig:MS.chain.n3asymut}
\end{figure}

Relaxing the assumption of direct hoppings being present only between nearest-neighbours and that all levels have the same on-site energy does not lead to a qualitative change of the results. Adding an inter-level interaction between all electrons of different levels shifts the resonances further outwards, as should be expected due to the additional Coulomb repulsion. Surprisingly, now the outer peaks become wider and their box-like structure evolves earlier than for the resonance at half-filling, at least for the interactions we can treat here. This is shown in Fig.~\ref{fig:MS.chain.n3asymut}.
\begin{figure}[t]	
	\centering
	\includegraphics[width=0.475\textwidth,height=4.4cm,clip]{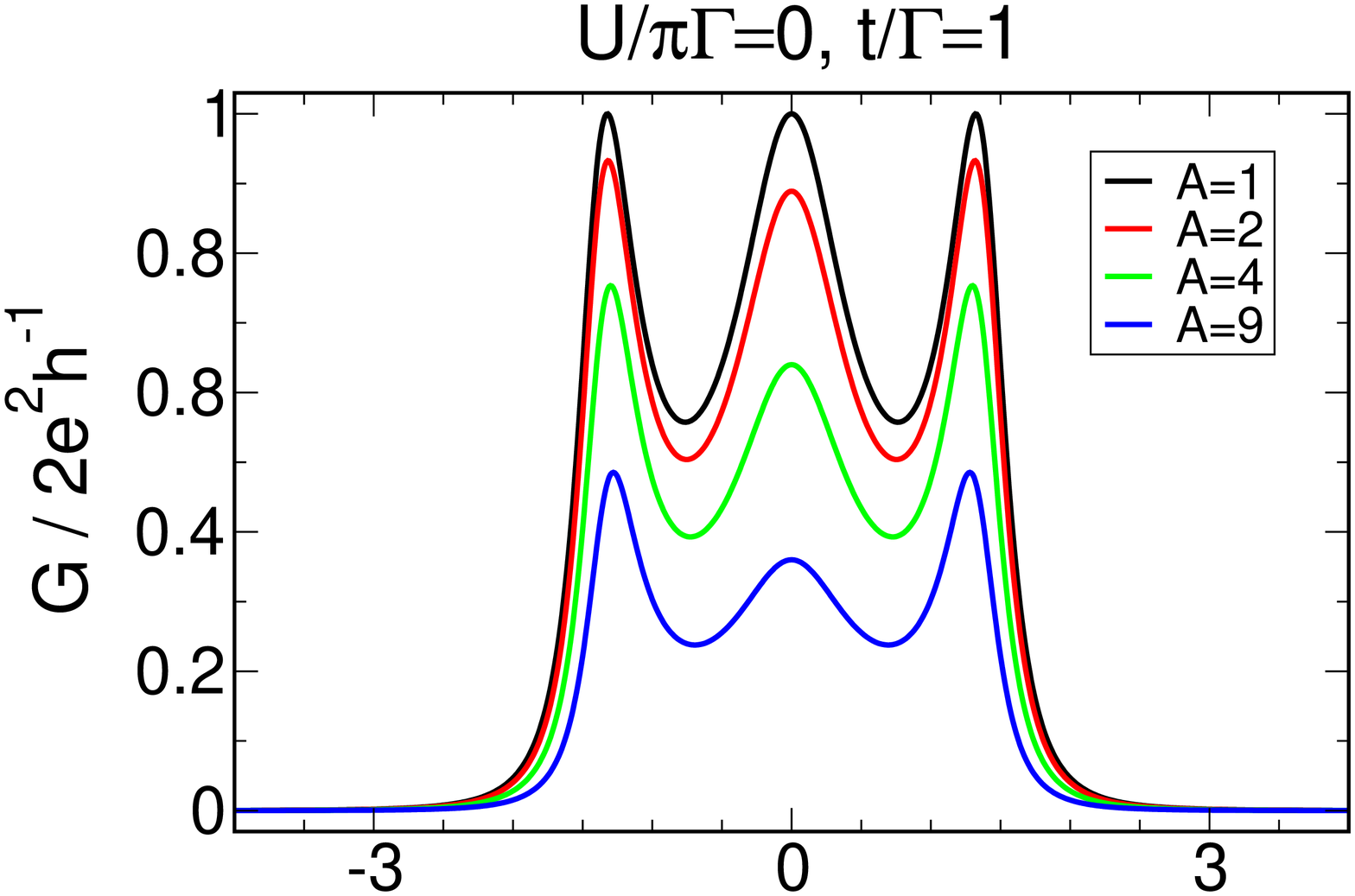}\hspace{0.035\textwidth}
        \includegraphics[width=0.475\textwidth,height=4.4cm,clip]{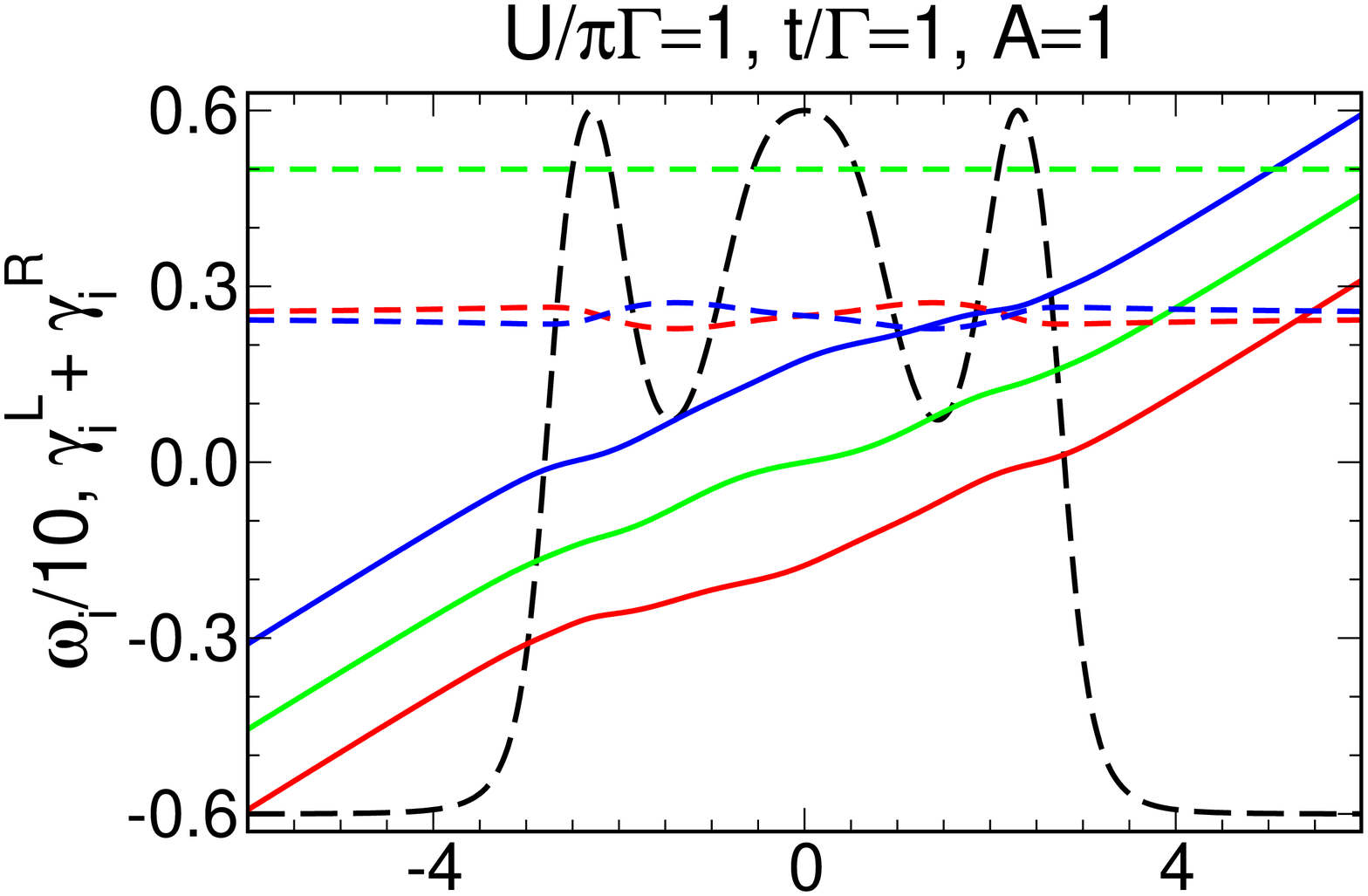}\vspace{0.3cm}
        \includegraphics[width=0.475\textwidth,height=5.2cm,clip]{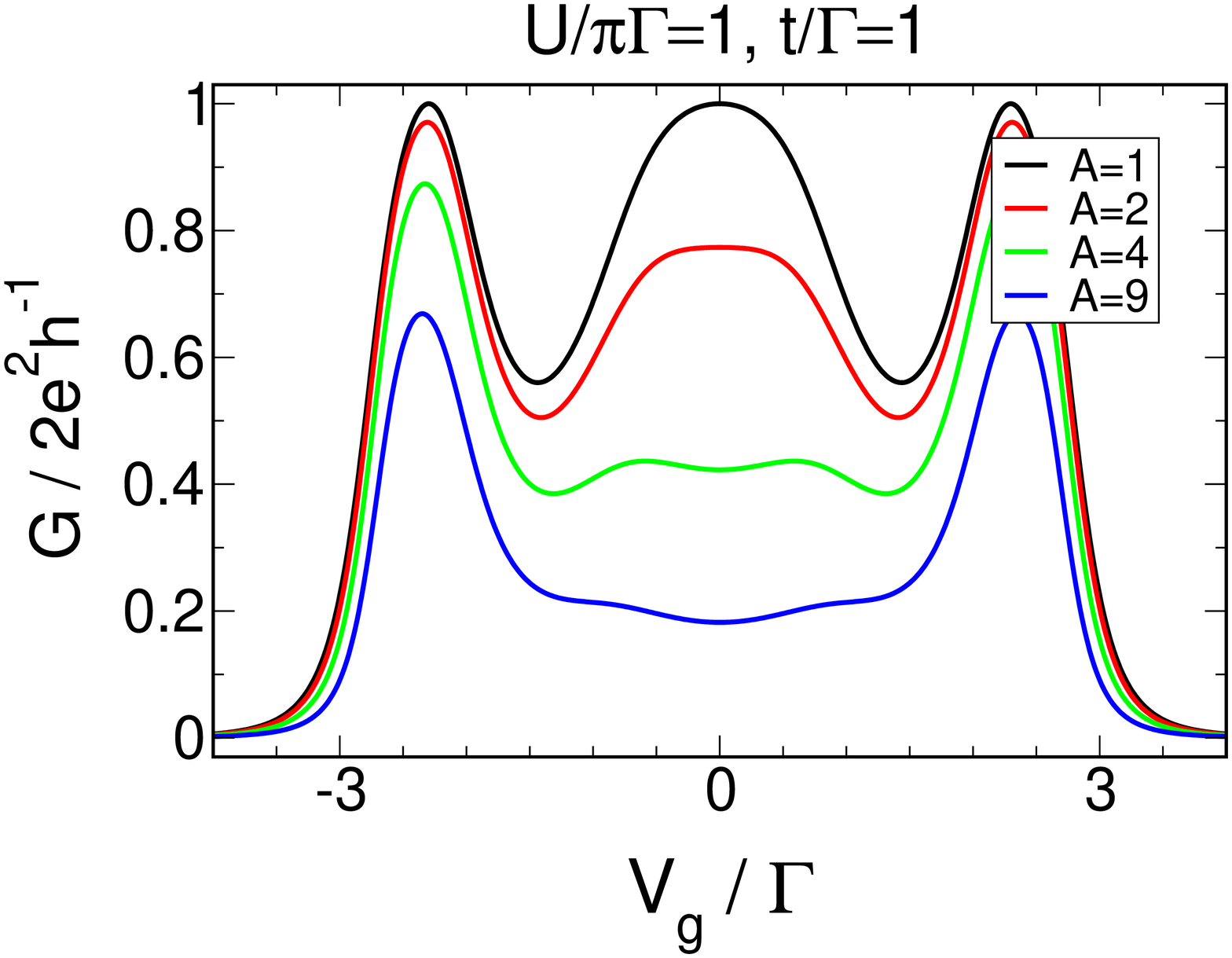}\hspace{0.035\textwidth}
        \includegraphics[width=0.475\textwidth,height=5.2cm,clip]{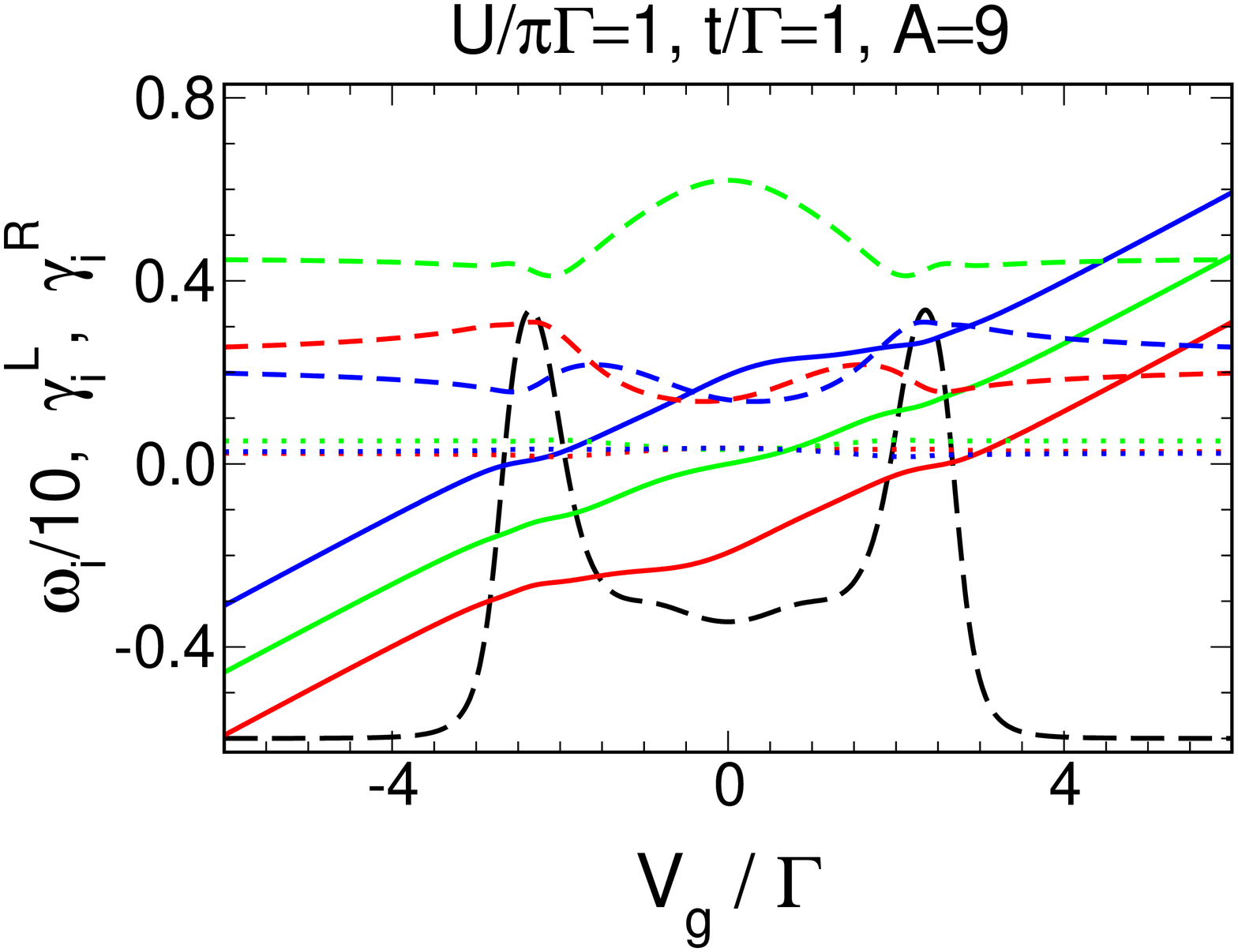}
        \caption{\textit{Left panels:} Gate voltage dependence of the conductance through a linear chain of three dots for the noninteracting case and for $U/\Gamma=1.0$ for four different left-right asymmetries $A:=\Gamma_R/\Gamma_L$. For $U>0$, the central resonance vanishes if $A$ is increased. \textit{Right panels:} Eigenenergies $\omega_i$ (solid lines) and corresponding hybridisations with the leads $\gamma_i^L$ and $\gamma_i^R$ of the isolated dot system with renormalized parameters for $U/\pi\Gamma=1.0$ as a function of $V_g$ for two different asymmetries $A=1$ (dashed lines: $\Gamma_i^L+\Gamma_i^R$) and $A=9$ (dotted lines: $\Gamma_i^L$, dashed lines: $\Gamma_i^R$). For reference, the conductance is shown as well (black long-dashed line; the y-axes is scaled to one).}
\label{fig:MS.chain.n3asymg}
\end{figure}

Applying a left-right asymmetry $A:=\Gamma_R/\Gamma_L$ between the hybridisations with the left and right lead has a more dramatic effect if the hopping $t$ is comparable to the hybridisation strength, $t\approx\Gamma$. While for the noninteracting case increasing $A$ basically leads to an overall decrease of the conductance (which is more or less obvious for not too small $t$ since the height of each peak arising from one level crossing the chemical potential decreases with increased $A$), the peak at $V_g=0$ completely vanishes in presence of a local interaction if one moves away from the left-right symmetric case that was treated before (Fig.~\ref{fig:MS.chain.n3asymg}, left panel). This might be understood as follows. Due to spin degeneracy, we can describe our system by three effective energies $\omega_i(V_g)$ (the eigenenergies of the dot system decoupled from the leads at the end of the flow), which for the noninteracting case cross the chemical potential separately, leading to a transmission resonance no matter how $A$ is chosen. One would not expect this picture to break down if a small interaction is turned on, and indeed it turns out that also in this case for $A=1$ one effective level crosses the chemical potential approximately at the position of each peak (Fig.~\ref{fig:MS.chain.n3asymg}, upper right panel). Surprisingly, the same holds for very large asymmetries (Fig.~\ref{fig:MS.chain.n3asymg}, lower right panel). In fact, the effective energies for this case nearly coincide with the ones for $A=1$ (which implies that the total average occupation of the chain barely changes as well). What causes the resonance at $V_g=0$ to vanish is that the new hybridisations $\gamma_i^L(V_g)$ and $\gamma_i^R(V_g)$, which almost coincide with the noninteracting ones in the left-right symmetric case, are strongly altered by the interaction for $A\gg 1$. In particular, the asymmetry and coupling strength of the level that crosses the chemical potential at half-filling are both increased in those regions of $V_g$ where it significantly contributes to the transport. In contrast, the left-right asymmetry of the levels that cross at the outer peaks is slightly reduced while their hybridisation strength remains almost unchanged compared to the $A=1$ case, at least close to the point where they cross $\mu$. The combined effect of the increased asymmetry and level broadening explains the vanishing of the resonance at half-filling. For a nearest-neighbour hopping from the molecular orbital regime, this effect is no longer observed.

\subsubsection{A Chain of Four Dots}
\begin{figure}[t]	
	\centering
	      \includegraphics[width=0.485\textwidth,height=4.4cm,clip]{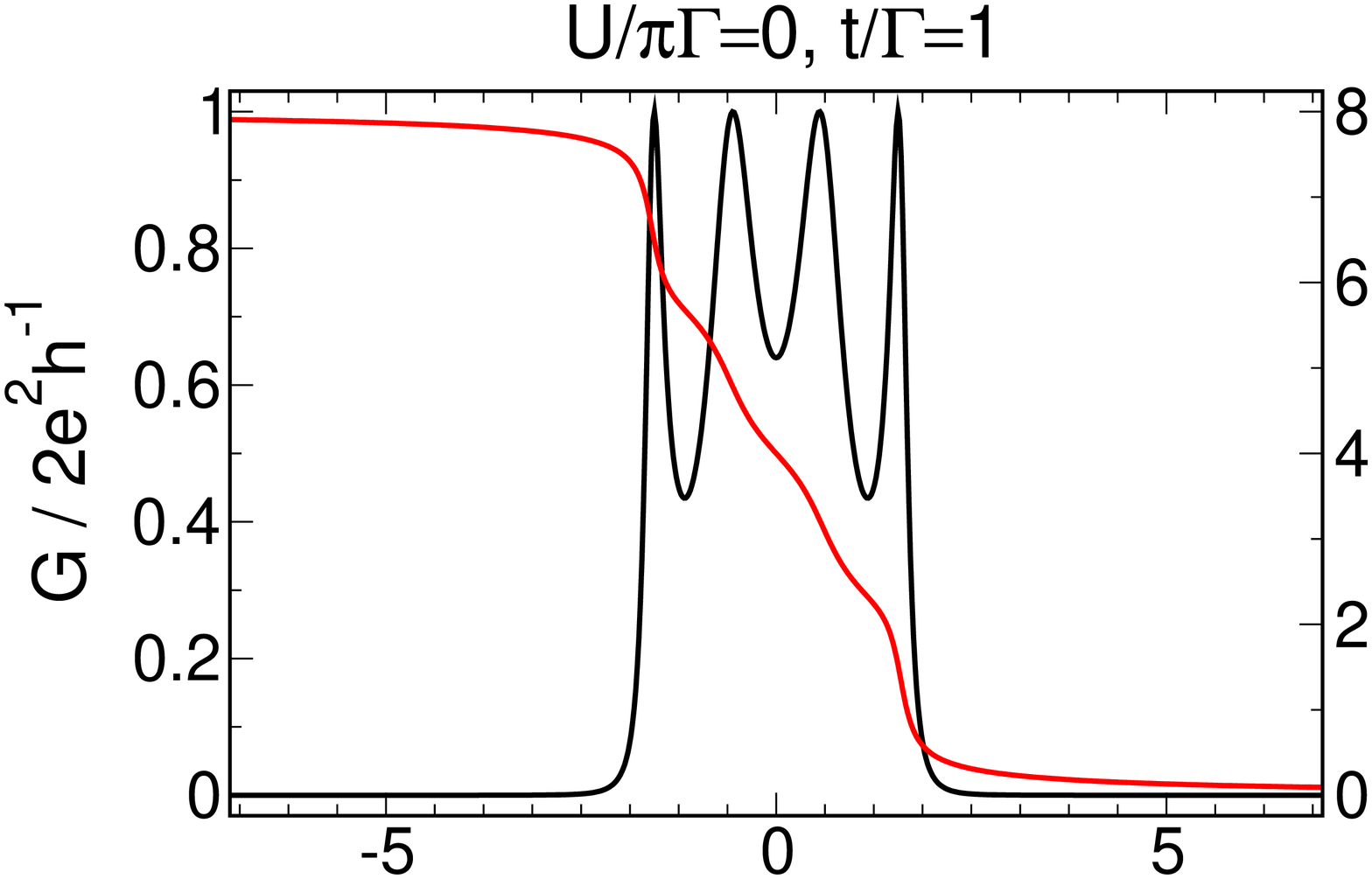}\hspace{0.015\textwidth}
        \includegraphics[width=0.485\textwidth,height=4.4cm,clip]{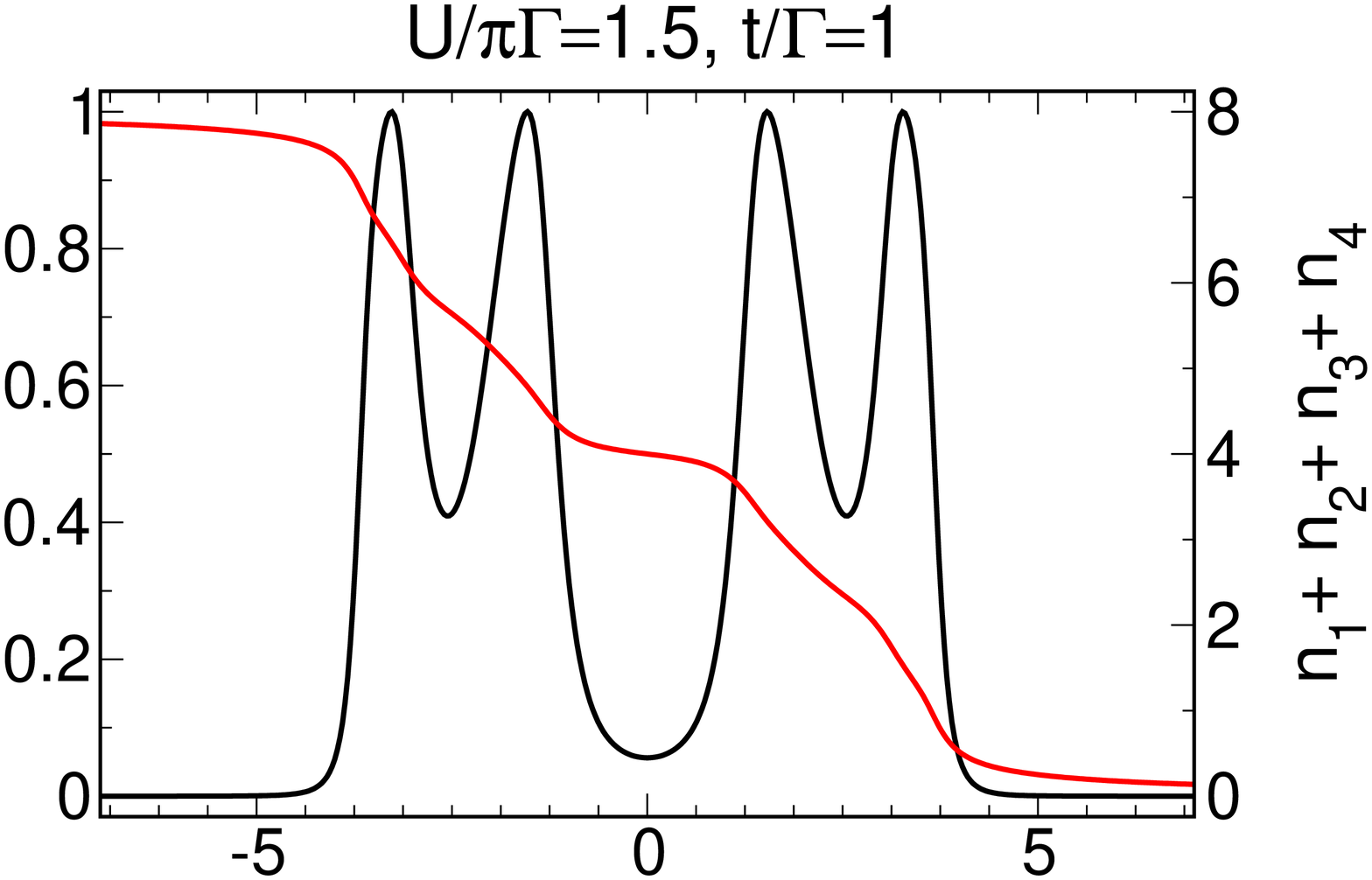}\vspace{0.3cm}
        \includegraphics[width=0.485\textwidth,height=5.2cm,clip]{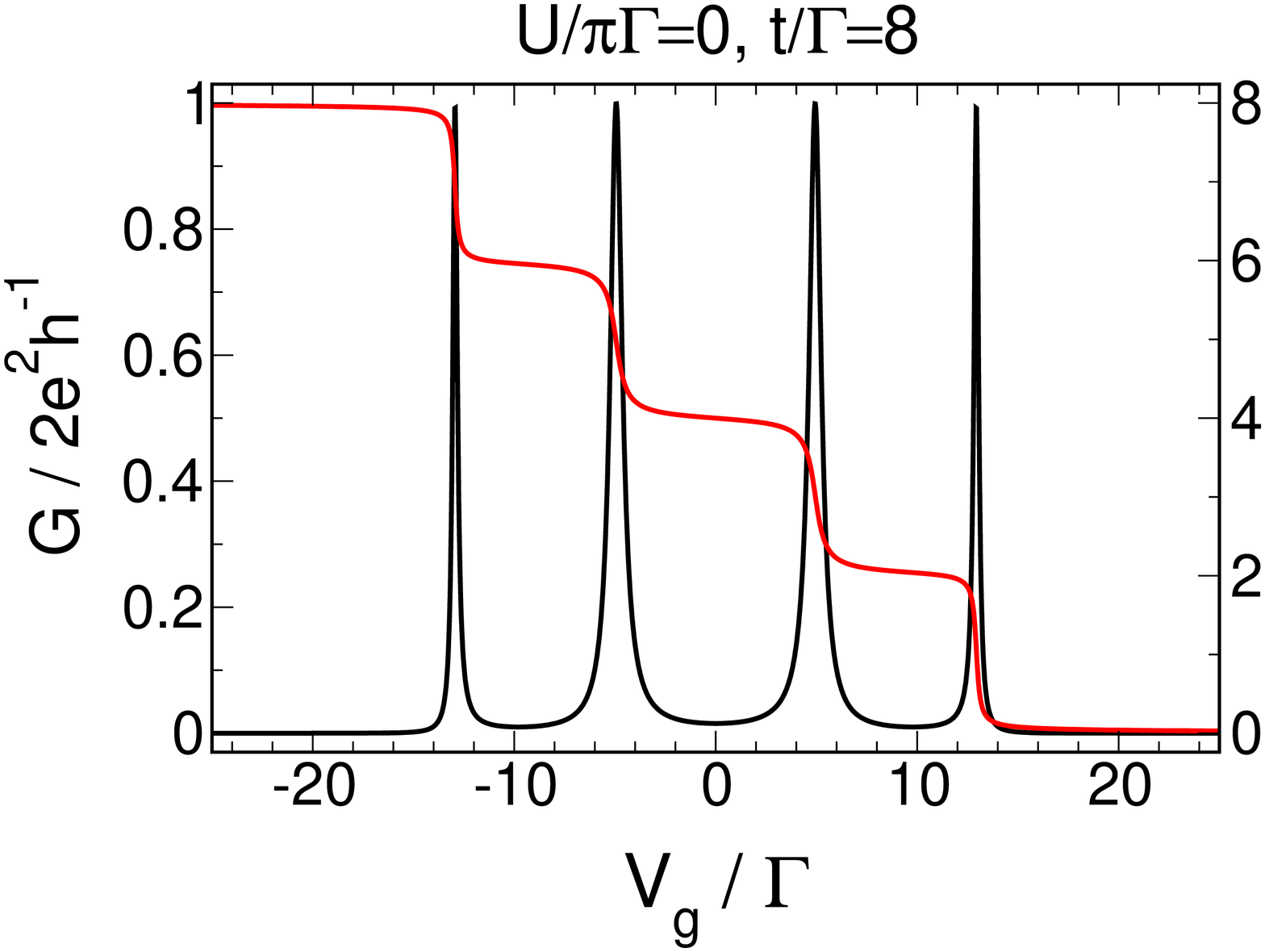}\hspace{0.015\textwidth}
        \includegraphics[width=0.485\textwidth,height=5.2cm,clip]{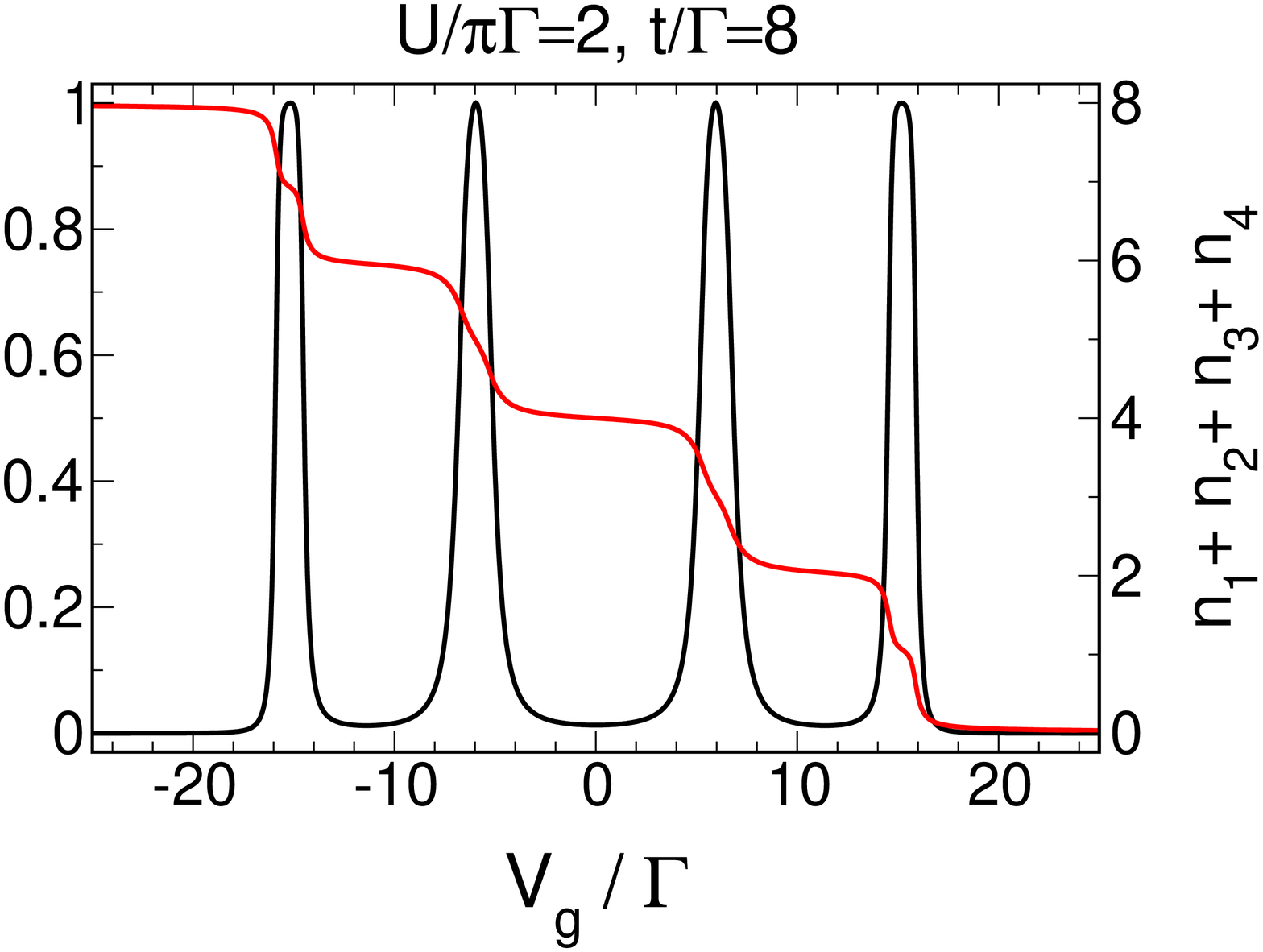}
        \caption{Gate voltage dependence of the conductance $G$ (black) and total average occupation (red) of a linear chain of four dots with $\Gamma_L/\Gamma=\Gamma_R/\Gamma=0.5$ for two different nearest-neighbour hoppings $t$ for the noninteracting case and for the largest $U/\Gamma$ that can be tackled by our fRG approximation scheme. To observe well-pronounced plateaus of unitary conductance, the interaction needs to be even larger (see for example $t/\Gamma\approx 8$, $U/\Gamma\approx 16\pi$ in \cite{chain4}), while the `Mott-insulating' type of behaviour at half-filling is already visible.}
\label{fig:MS.chain.n4delta}
\end{figure}

The situation of a chain containing four dots is very similar. For $U=0$ and large hoppings $t\gg\Gamma$ we observe four resonances that merge if the nearest-neighbour hopping $t$ becomes comparable to the hybridisation with the leads $\Gamma=\Gamma_L+\Gamma_R$. In the molecular orbital regime (large $t$) a local interaction $U$ transforms the Lorentzian lineshape of the peaks into boxes of width $U/N$ due to the set-in of the Kondo effect. The total occupation, which in the noninteracting case changes by two at each resonance and remains constant in between, gradually becomes step-like, in particular odd for gate voltages within the resonances and even in between. Again, charge fluctuations are strongly suppressed close to $V_g=0$, but this time half-filling corresponds to an even number of electrons in the chain so that a conductance valley is observed instead of a plateau of unitary height. Even if $t$ is lowered such that in the noninteracting case the conductance is not small at $V_g=0$ due to the overlap and decreased separation of the two nearest levels, the interaction leads to an exponential reduction of the conductance. Hence due to the suppression of charge fluctuations at half-filling the four-site chain tends towards a Mott insulator.

One should note that the behaviour of $G(V_g)$ at $V_g=0$ observed here holds for arbitrary long chains. If the latter comprises an odd number of sites, there is a wide conduction plateau of unitary height at the particle-hole symmetric point, while for an even number of dots half-filling corresponds to an even number of electrons and transport is strongly suppressed.

The effects of the interaction in a four-site chain were computed by \cite{chain4} employing an NRG approach. Our fRG approximation scheme allows to treat values of $U$ as large such that the strong suppression of charge fluctuations at $V_g=0$ shows up, but we are unable to treat $U$ large enough to observe strongly-developed box-like resonances (see next chapter). The conductance $G(V_g)$ for the noninteracting case and the largest interaction that can still be tackled by our approach is depicted in Fig.~\ref{fig:MS.chain.n4delta}. Introducing arbitrary hoppings, level spacings, left-right asymmetric hybridisations, or an additional inter-level interaction between the dots does not alter the results qualitatively.

\subsection{The Side-Coupled Geometry}\label{sec:MS.side}

\subsubsection{General Considerations}

\begin{figure}[t]	
	\centering
	      \includegraphics[width=0.4\textwidth,clip]{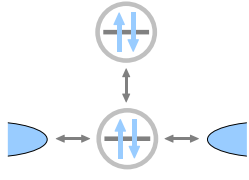}
        \caption{The side-coupled geometry.}
\label{fig:MS.geo_side}
\end{figure}

The side-coupled geometry consists of a single-level dot connected to the leads (the embedded dot) to which another dot with one level (the side-coupled dot) is attached by a tunnelling barrier $t$ (see Fig.~\ref{fig:MS.geo_side}). The projected free propagator reads
\begin{equation}\label{eq:MS.side.g}
\left[\mc G^0_\sigma(i\omega)\right]^{-1} = i\omega - V_g + \sigma\frac{B}{2}+
\begin{pmatrix}
i~\tn{sgn}(\omega)~[\Gamma_L+\Gamma_R] & t \\
t & 0
\end{pmatrix},
\end{equation}
where $B$ is a magnetic field applied to the system, $V_g$ denotes the gate voltage that controls the on-site energies of both dots, and $\Gamma_{L,R}$ are the hybridisations with the left and right lead in the wide-band limit. As usual, we choose the total hybridisation strength $\Gamma=\Gamma_L+\Gamma_R$ as the unit of energy. The flow equations (\ref{eq:FRG.flowse3}) and (\ref{eq:FRG.flowww3}) yield the full propagator $\mc G = \tilde{\mc G}(\la=0)$ of the system in presence of the interactions, which in general will be a local interaction $U:=U_{l,l}^{\sigma,\bar\sigma}$ between spin up and spin down electrons on both dots and a nearest-neighbour interaction $U_N:=U_{l,\bar l}^{\sigma,\sigma}=U_{l,\bar l}^{\sigma,\bar\sigma}$ between electrons on the two different dots. The zero temperature conductance can then be computed as (\ref{eq:DOT.leitwertsingle}),
\begin{equation}
G=\frac{4\pi e^2}{h}\frac{\Gamma_L\Gamma_R}{\Gamma_L+\Gamma_R}\sum_\sigma\rho_\sigma(0).
\end{equation}
Here $\rho_\sigma(i\omega)$ denotes the spectral function of the embedded dot.

When considering the parallel double dot in the next section, it will prove helpful to relate this problem back to the side-coupled one. In fact, it is possible to map the side-coupled dot onto a parallel double dot by a basis transformation to bonding and antibonding states,
\begin{equation*}
\{|e,\sigma\rangle, |s,\sigma\rangle\}\to \left\{|A,\sigma\rangle:=\frac{1}{\sqrt{2}}(|e,\sigma\rangle+|s,\sigma\rangle), |B,\sigma\rangle:=\frac{1}{\sqrt{2}}(|e,\sigma\rangle-|s,\sigma\rangle)\right\},
\end{equation*} 
with $|e(s),\sigma\rangle$ being the single-particle wavefunction of the embedded (side-coupled) dot. The noninteracting part of the Hamiltonian $H_\tn{1P}$ describing the side-coupled geometry,
\begin{equation*}\begin{split}
H_{\tn{1P}}= \sum_\sigma\Bigg[\sum_{l=e,s}(V_g+\sigma B/2)n_{l,\sigma} -  t&\left(d_{e,\sigma}^\dagger d_{s,\sigma} + \tn{H.c.}\right) \\
-&\left(t_L d_{e,\sigma}^\dagger c_{0,\sigma,L} + t_R d_{e,\sigma}^\dagger c_{0,\sigma,R} + \tn{H.c.}\right)\Bigg],
\end{split}\end{equation*}
then maps onto
\begin{equation*}
\tilde H_{\tn{1P}} =\sum_\sigma\Bigg[(V_g+\sigma B/2+t)n_{A,\sigma} + (V_g+\sigma B/2-t)n_{B,\sigma}
-\hspace{-0.2cm}\sum_{l=A,B\atop s=L,R}\left[\frac{t_s}{\sqrt{2}}d_{l,\sigma}^\dagger c_{0,\sigma,s}+\tn{H.c.}\right]\Bigg].
\end{equation*}
This is just the noninteracting part of a Hamiltonian describing a parallel system of single-level dots with level spacing $2t$ and level-lead couplings $t^A_{L,R}=t^B_{L,R}=\frac{t_{L,R}}{\sqrt{2}}$. The mapping of the interaction part of the side-coupled Hamiltonian,
\begin{equation*}
H_\tn{int} = U\left( n_{e,\uparrow}n_{e,\downarrow}+n_{s,\uparrow}n_{s,\downarrow}\right) + U_N \left(n_{e,\uparrow}n_{s,\uparrow}+n_{e,\uparrow}n_{s,\downarrow}+n_{e,\downarrow}n_{s,\uparrow}+n_{e,\downarrow}n_{s,\downarrow}\right),
\end{equation*}
is very simple for equal local and nearest-neighbour interactions $U$ and $U_N$. Namely,
\begin{equation*}\begin{split}
H_\tn{int}/U & =  n_{e,\uparrow} (n_{e,\downarrow}+n_{s,\uparrow}+n_{s,\downarrow}) + n_{s,\uparrow}(n_{s,\downarrow}+n_{e,\downarrow}) +n_{e,\downarrow}n_{s,\downarrow} \\
& = n_{e,\uparrow} (n_{A,\downarrow}+n_{B,\downarrow}+n_{s,\uparrow}) + n_{s,\uparrow}(n_{A,\downarrow}+n_{B,\downarrow}) +n_{e,\downarrow}n_{s,\downarrow} \\
& = (n_{A,\uparrow}+n_{B,\uparrow})(n_{A,\downarrow}+n_{B,\downarrow}) + n_{e,\uparrow}n_{s,\uparrow}+n_{e,\downarrow}n_{s,\downarrow} \\
& = (n_{A,\uparrow}+n_{B,\uparrow})(n_{A,\downarrow}+n_{B,\downarrow}) + n_{A,\uparrow}n_{B,\uparrow}+n_{A,\downarrow}n_{B,\downarrow}\\
& =\tilde H_\tn{int}/U,
\end{split}\end{equation*}
where we have used that for both spin directions the following identity holds,
\begin{equation*}\begin{split}
4n_en_s & = 4c_e^\dagger c_ec_s^\dagger c_s \\
& = (c_A^\dagger+c_B^\dagger)(c_A+c_B)(c_A^\dagger-c_B^\dagger)(c_A-c_B) \\
& = -(c_A^\dagger+c_B^\dagger)(c_A^\dagger-c_B^\dagger)(c_A+c_B)(c_A-c_B) \\
& = (c_A^\dagger c_B^\dagger - c_B^\dagger c_A^\dagger) (c_Bc_A-c_Ac_B) \\
& = 4n_An_B.
\end{split}\end{equation*} 
Alltogether this shows that an equal integration between all electrons of the side-coupled dot maps to the same kind of interaction of the parallel two-level dot. Unfortunately, this becomes more complicated if initially only a local interaction $U$ in the side-coupled geometry is present. In this case, also correlated hoppings are generated by the transformation,
\begin{equation*}\begin{split}
U(n_{e,\uparrow}n_{e,\downarrow} + n_{s,\uparrow}n_{s,\downarrow}) = 
\frac{U}{2}\Big[~n_{A,\uparrow}n_{A,\downarrow}&+n_{B,\uparrow}n_{B,\downarrow}+n_{A,\uparrow}n_{B,\downarrow}+n_{B,\uparrow}n_{A,\downarrow}\\
-&c_{A,\uparrow}^\dagger c_{A,\downarrow}^\dagger c_{B,\uparrow}c_{B,\downarrow}
-c_{A,\uparrow}^\dagger c_{B,\downarrow}^\dagger c_{B,\uparrow}c_{A,\downarrow}\\
-&c_{B,\uparrow}^\dagger c_{A,\downarrow}^\dagger c_{A,\uparrow}c_{B,\downarrow}
-c_{B,\uparrow}^\dagger c_{B,\downarrow}^\dagger c_{A,\uparrow}c_{A,\downarrow} ~~\Big].
\end{split}\end{equation*}

\subsubsection{Small Inter-Dot Hoppings: Two-Stage Kondo Regime}

\begin{figure}[t]	
	\centering
        \includegraphics[width=0.495\textwidth,height=5.2cm,clip]{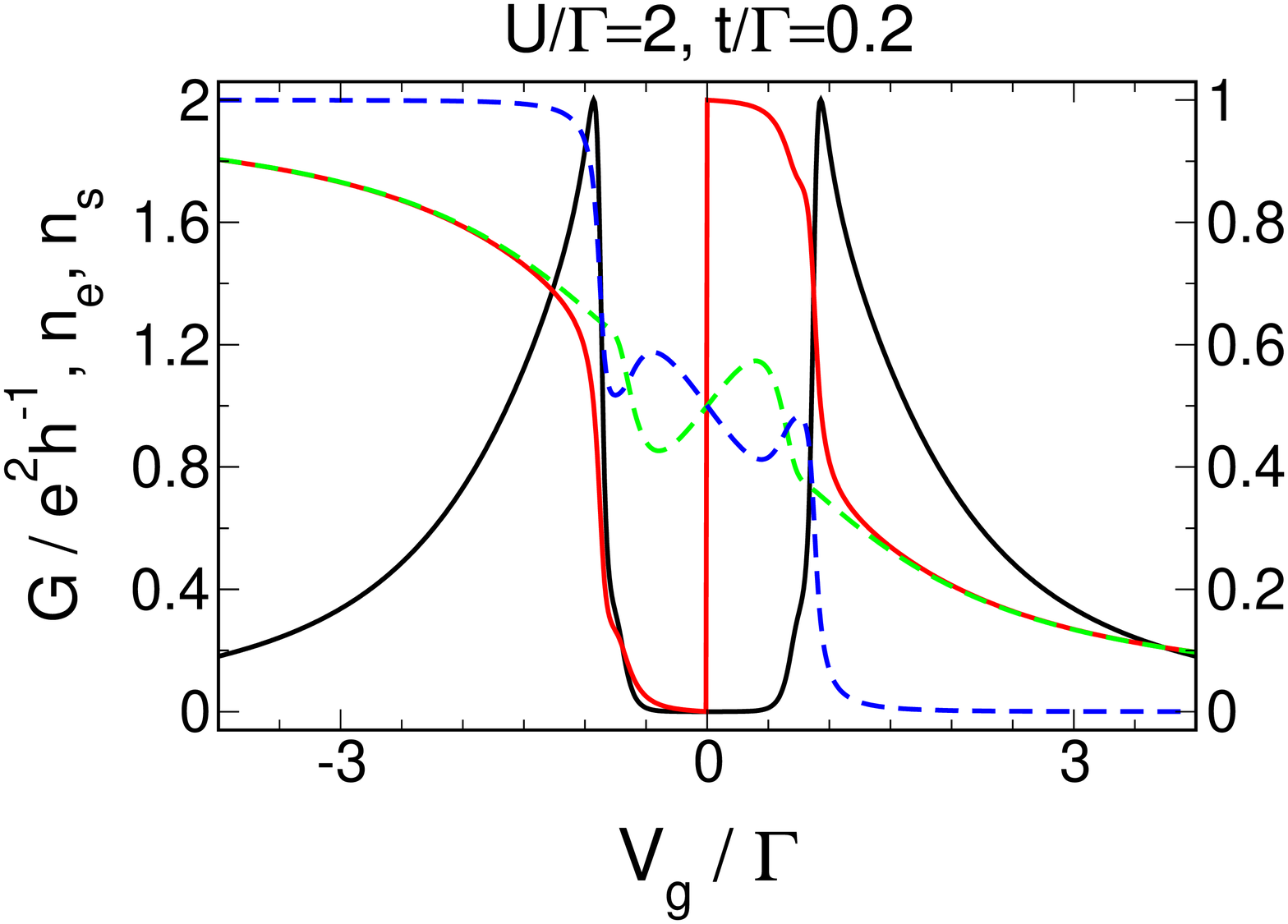}\hspace{0.015\textwidth}
        \includegraphics[width=0.475\textwidth,height=5.2cm,clip]{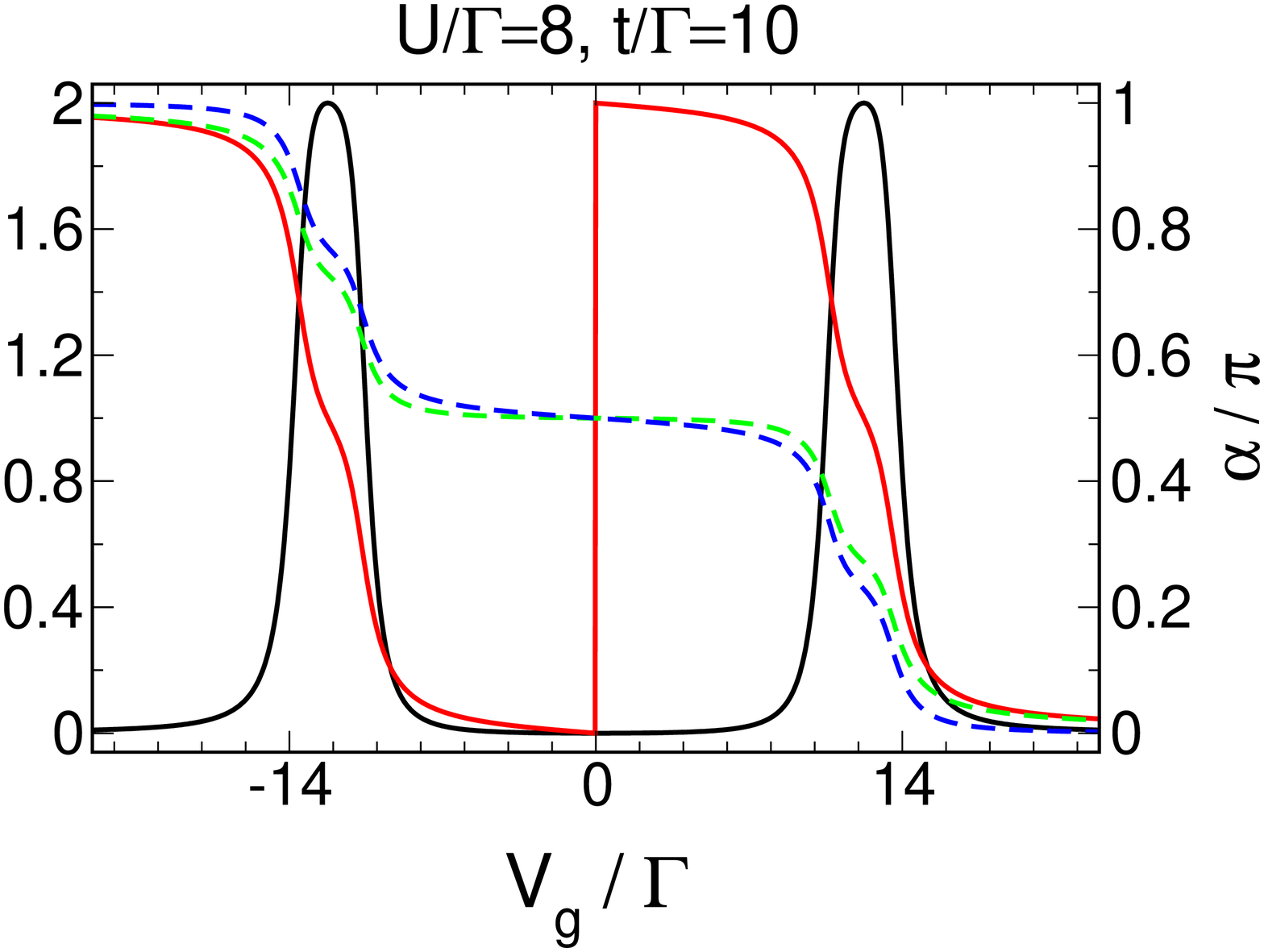}
        \caption{Gate voltage dependence of the conductance $G$ (black), transmission phase $\alpha$ (red) and average level occupancies (green: embedded dot, blue: side-coupled dot) of the side-coupled geometry with $\Gamma_L/\Gamma=\Gamma_R/\Gamma=0.5$ and zero magnetic field for parameters from the two-stage Kondo (left panel) and the local spin-singlet (right panel) regime. For the former, the two-stage Kondo effect causes a valley of width $U$ where transport is strongly suppressed. In contrast, the local spin-singlet phase can be understood as individual transport through the bonding and anti-bonding states of the isolated molecule.}
\label{fig:MS.side.t}
\end{figure}

\begin{figure}[t]	
	\centering
	\includegraphics[width=0.475\textwidth,height=5.2cm,clip]{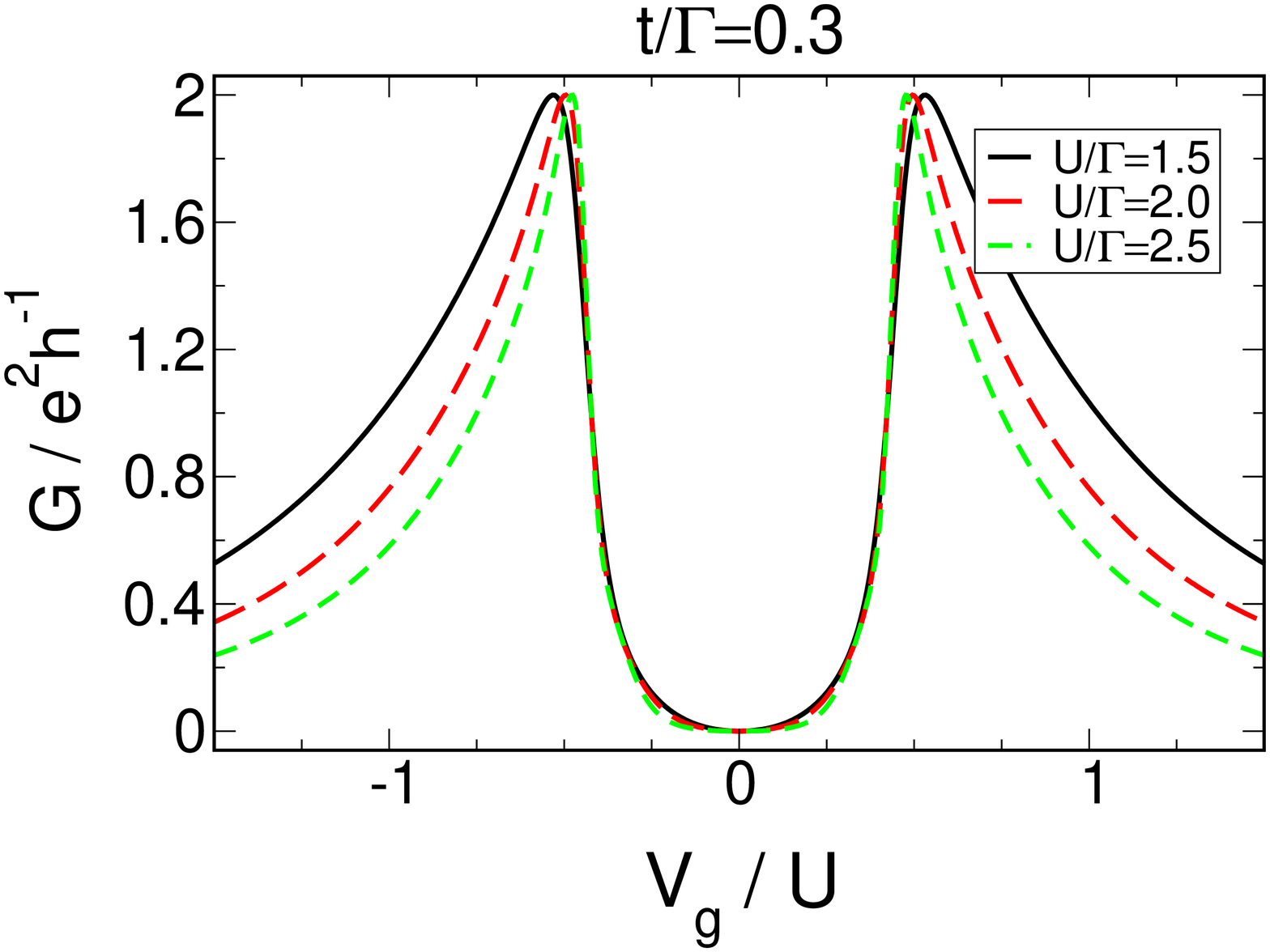}\hspace{0.035\textwidth}
        \includegraphics[width=0.475\textwidth,height=5.2cm,clip]{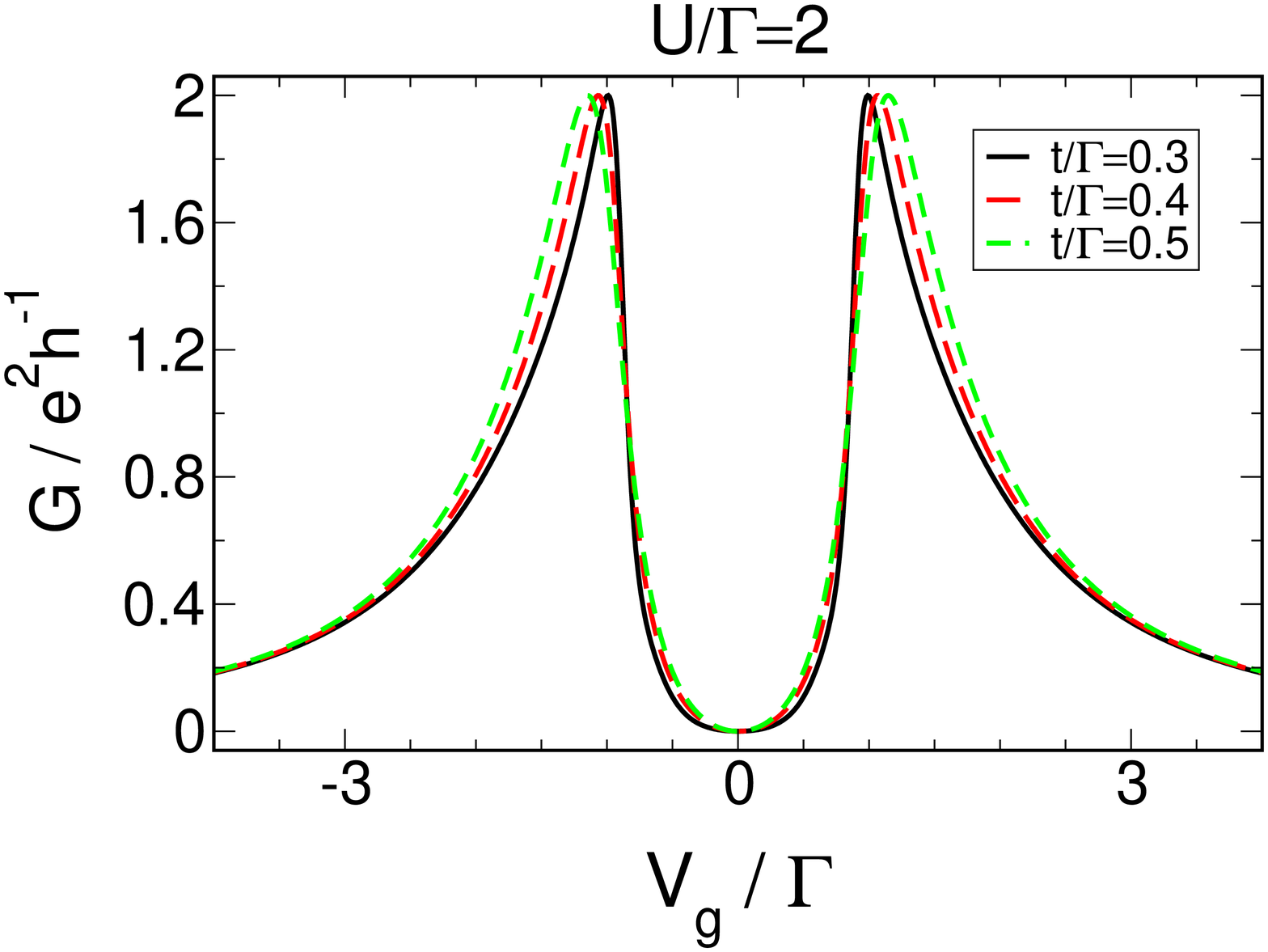}
        \caption{Gate voltage dependence of the conductance $G$ of the side-coupled geometry with left-right symmetric hybridisations and zero magnetic field for various parameters from the two-stage Kondo regime.}
\label{fig:MS.side.varut}
\end{figure}

The mapping between both geometries will be helpful to gain further insight into the interacting double dot studied in the next section. In the noninteracting case, however, the transport through the parallel two-level dot with spin is by definition equivalent to transport through the spin-polarised one which has been studied in Sec.~\ref{sec:OS.dd}. This immediately gives insight into the behaviour of the side-coupled geometry if no interaction is present. In particular, if the inter-dot hopping $t$ is small compared to the hybridisation strength $\Gamma$, the conductance as a function of the gate voltage $V_g$ has a Lorentzian lineshape with a dip close to the particle-hole symmetric point. Since for $t=0$ the side-coupled geometry maps onto a non-generic situation of the double dot which in particular does not exhibit a dip in $G(V_g)$ at $V_g=0$, the separation of the `resonances' for small but finite $t$ is very poor (for example compared with the situation shown in Fig.~\ref{fig:OS.dd.wwfrei} for degenerate levels). This dramatically changes if we add a local interaction $U$ of the spin up and spin down electrons on the embedded and side-coupled dot to the system. At zero temperature, in absence of the side-coupled dot the spin on the embedded dot gets screened by the lead electrons which gives rise to the Kondo resonance in the conductance (the first stage Kondo effect). Hence, coming from large negative gate voltages, $G$ increases with a slope determined by $U/\Gamma$ similar to the single-level case (Fig.~\ref{fig:MS.side.varut}, left panel). If the side-coupled dot is present its spin gets screened as well (the second stage Kondo effect), so that at some point the conductance falls off and merges into a valley where transport is strongly suppressed (Fig.~\ref{fig:MS.side.t}, left panel). Again in analogy to the single-level case, it is roughly $U/t^2$ that determines how steeply $G(V_g)$ falls off (as slightly indicated in Fig.~\ref{fig:MS.side.varut}). Thus, if the inter-dot hopping is small compared to the hybridisation strength, already for $U\approx\Gamma$ we would expect a well-pronounced valley around $V_g=0$, while the Kondo resonance on the embedded dot has not evolved yet and hence the conductance should vanish slowly when increasing $|V_g|$. If the interaction is large compared to the hopping $t$, the width of the region where transport is suppressed is given by $U$ (Fig.~\ref{fig:MS.side.varut}, left panel) and is independent of the actual choice of $t$ (Fig.~\ref{fig:MS.side.varut}, right panel), which is consistent with the single-level case.

The behaviour of the average level occupancies is in agreement with this picture. For $U\gg t^2$, the side-coupled dot is either occupied by two electrons or unoccupied outside a region of width $U$ around $V_g=0$, while at $V_g=\pm U/2$ the occupation abruptly changes. The occupation of the embedded dot falls off much more continuously from $V_g=-\infty$ to $V_g=+\infty$ because $U/\Gamma$ is small. All this is similar to the single-level case, but the behaviour within the Kondo valley is different. The occupancy of the side-coupled dot does not show a plateau but depends non-monotonically on the gate voltage. This also holds for the embedded dot. In contrast, the total occupation depends monotonically on $V_g$ which is necessary because of the behaviour of the conductance to which it can be directly related by a generalised Friedel sum rule \cite{side2}. For the side-coupled geometry, this relation reads $G/e^2h^{-1}=2\sin^2(n\pi/2)$, which implies that $G$ almost reaches the unitary limit for an occupation close to an odd integer number while it vanishes if there is an even number of electrons in the dot region. This again justifies why transport is strongly suppressed if the Kondo effect is active on both the embedded and the side-coupled dot. The transmission phase $\alpha$, which is also related to the conductance by a Friedel sum rule, changes by $\pi$ over each of the two `resonances' and jumps by $\pi$ in between, which is similar to the noninteracting case. Both the average level occupancies and the phase for a small hopping from the two-stage Kondo regime are also shown in Fig.~\ref{fig:MS.side.t}. 

Choosing asymmetric couplings to the left and right lead only lowers the overall height of the conductance curve $G(V_g)$ but does not lead to new physics (Fig.~\ref{fig:MS.side.var}, upper left panel), which is similar to the single-level case. Relaxing the assumption of equal local interactions on both dots only leads to quantitative changes. An important observation, however, is that the strength of the interaction on the side-coupled dot determines the width of the Kondo valley for small $U/\Gamma$, whereas its dependence on the the local interaction strength on the embedded dot is only weak (Fig.~\ref{fig:MS.side.var}, upper right panel). This manifests the interpretation that at $T=0$ the Kondo effect on the side-coupled dot is active and causes the wide valley of small conductance. Adding a nearest-neighbour interaction to the system spreads up the whole curve but does not lead to new physics.

The energy scale that determines the breakdown of Kondo physics for this geometry can be determined from the (numerically) exact $T=0$ spectral function obtained by NRG, which shows a sharp resonance at zero frequency characteristic for the (first stage) Kondo effect of the embedded dot and an even sharper dip at $\omega=0$ within this resonance due to the (second stage) Kondo effect on the side-coupled dot. The width of these two resonances then define the energy scales $T_K^1$ and $T_K^2$ at which the corresponding Kondo resonance gets destroyed. As indicated, it turns out that $T_K^1\gg T_K^2$. This is consistent with the single-level dot where the Kondo temperature is proportional to $\exp (-U/\Gamma)$, and hence for small inter-level hoppings we would expect the energy scale $T_K^1$ determined by $U/\Gamma$ to much larger than $T_K^2$ determined by $U/t^2$. Thus, if we move away from $T=0$, the conductance at $V_g=0$ should increase because the spin on the side-coupled dot is no longer screened above a temperature $T_K^2$. If the $T$ exceeds the second characteristic energy scale, $T_K^1$, also the electron on the embedded dot is no longer screened and thus one would expect the conductance to decrease again. All this was shown in an NRG calculation by \cite{side2}. As explained above, since the fRG truncation scheme we use neglects all frequency dependencies we cannot expect to obtain reliable results for finite temperatures. In contrast, it is no problem to tackle another quantity that destroys the Kondo resonances, a magnetic field, to extract the two Kondo energy scales. In analogy to the single-level dot, we define the Kondo `temperature' $T_K^2$ of the second stage of the Kondo effect as the field required to raise the conductance at $V_g=0$ to half the unitary limit, while the Kondo scale $T_K^1$ is just that of the single-level dot obtained from the side-coupled geometry if we set $t=0$. The fRG results obtained from the magnetic field definition of $T_K^1$ and $T_K^2$ are consistent with those obtained from the spectral function within the NRG approach. A more detailed comparison will be given in the next chapter.

The magnetic field dependence of the conductance within the two-stage Kondo regime is shown in the left panel of Fig.~\ref{fig:MS.side.b}. As explained above, for $B>0$ the conductance increases and the valley around $V_g=0$ evolves into a maximum since the second stage of the Kondo effect is no longer active above an energy scale $T_K^2$. For small magnetic fields $B\approx 0\ldots10T_K^2$, the conductance curves are qualitatively identical with those obtained by \cite{side2} for $B=0$ but finite $T$. This shows again that a magnetic field destroys the Kondo resonance similar to the temperature, justifying our definition of the Kondo energy scales. Further increase of the magnetic field strength lets the conductance around $V_g=0$ decrease again since also the Kondo resonance on the embedded dot gets destroyed for $B\approx T_K^e$. In the limit of large fields we finally recover twice the spin-polarised structure in the $G(V_g)$ curve, which corresponds to the noninteracting one if only a local interaction is present.

\subsubsection{Large Inter-Dot Hoppings: Local Spin-Singlet Phase}
\begin{figure}[t]	
	\centering
	      \includegraphics[width=0.475\textwidth,height=4.4cm,clip]{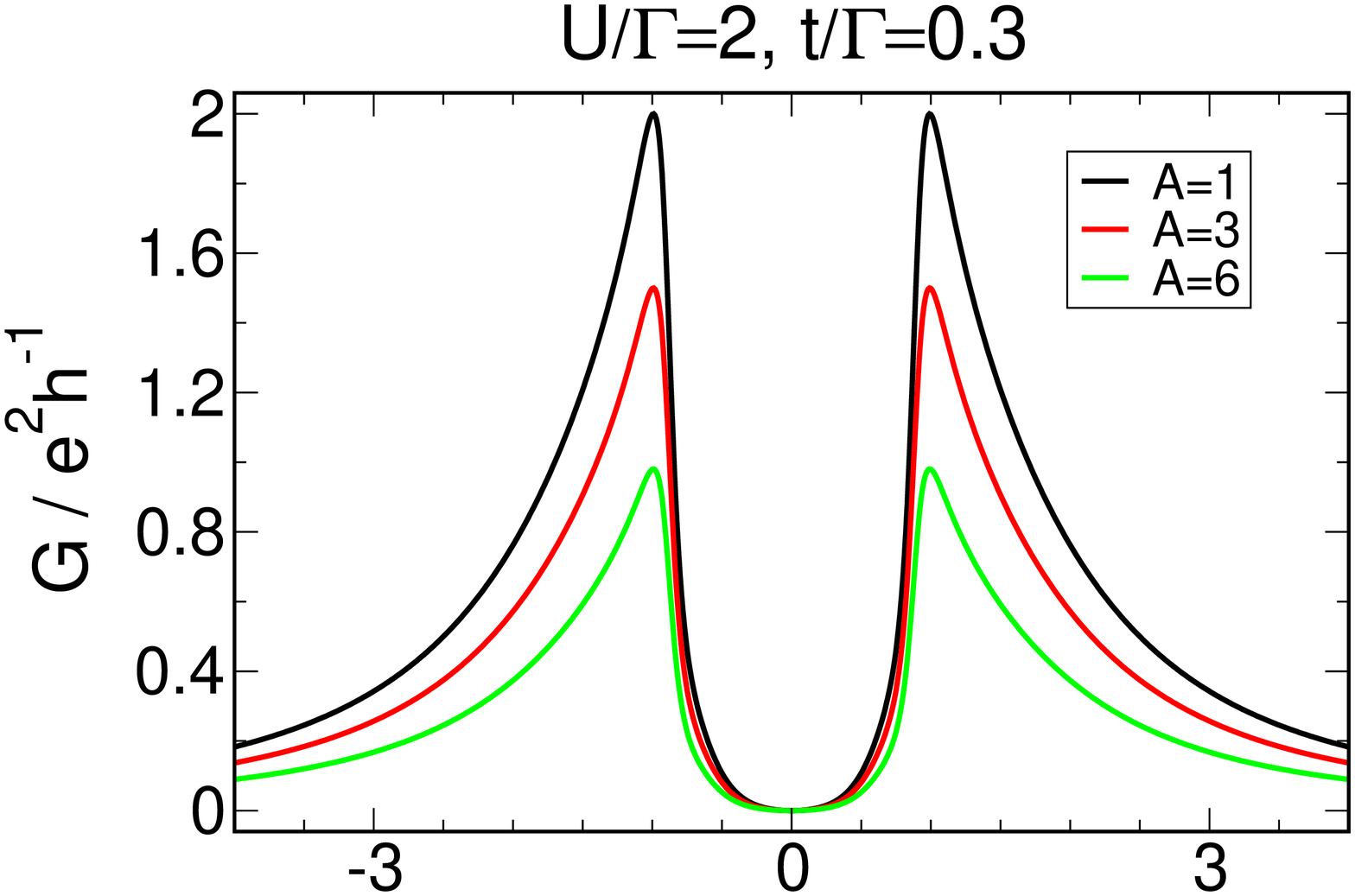}\hspace{0.035\textwidth}
        \includegraphics[width=0.475\textwidth,height=4.4cm,clip]{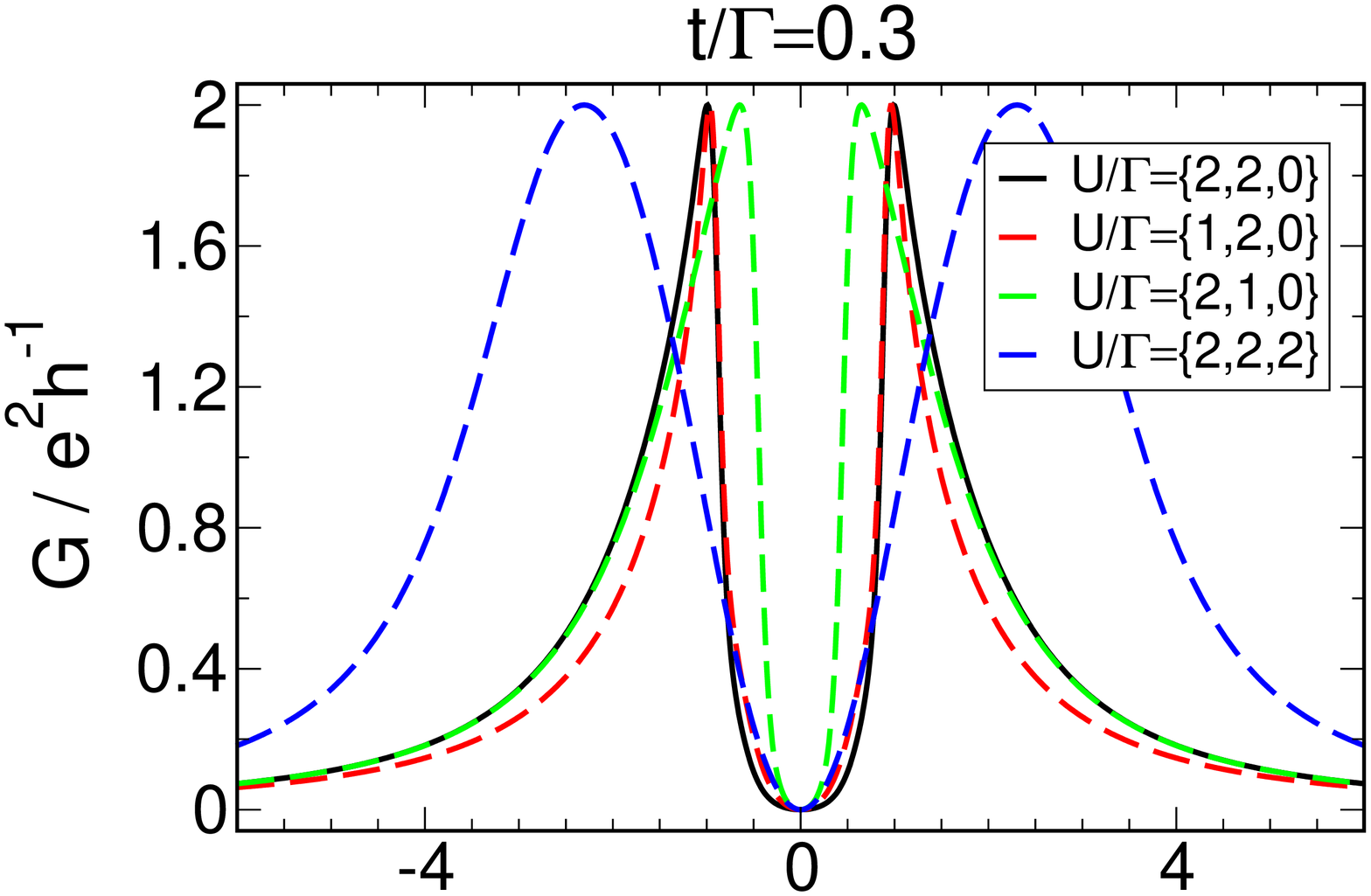}\vspace{0.3cm}
        \includegraphics[width=0.475\textwidth,height=5.2cm,clip]{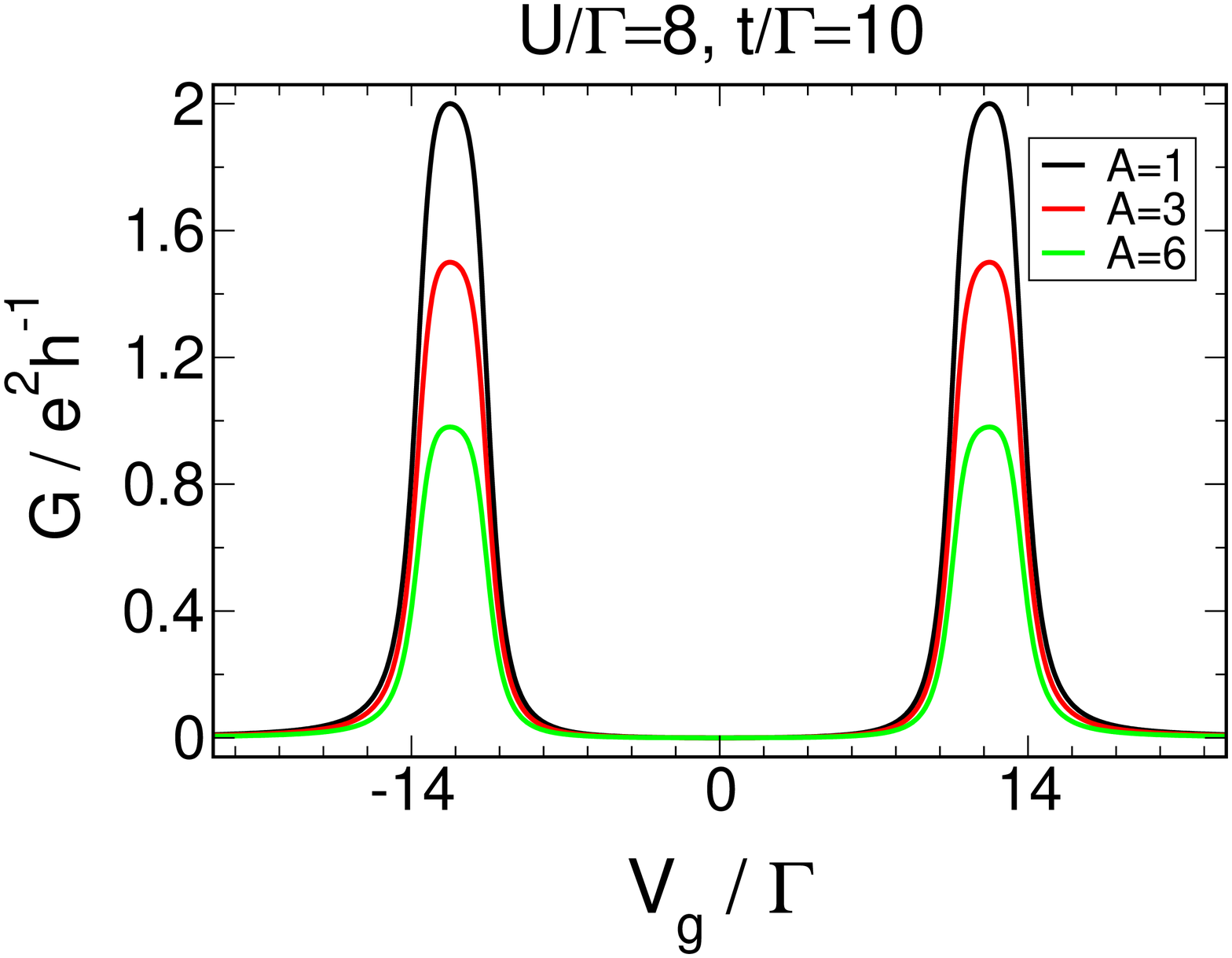}\hspace{0.035\textwidth}
        \includegraphics[width=0.475\textwidth,height=5.2cm,clip]{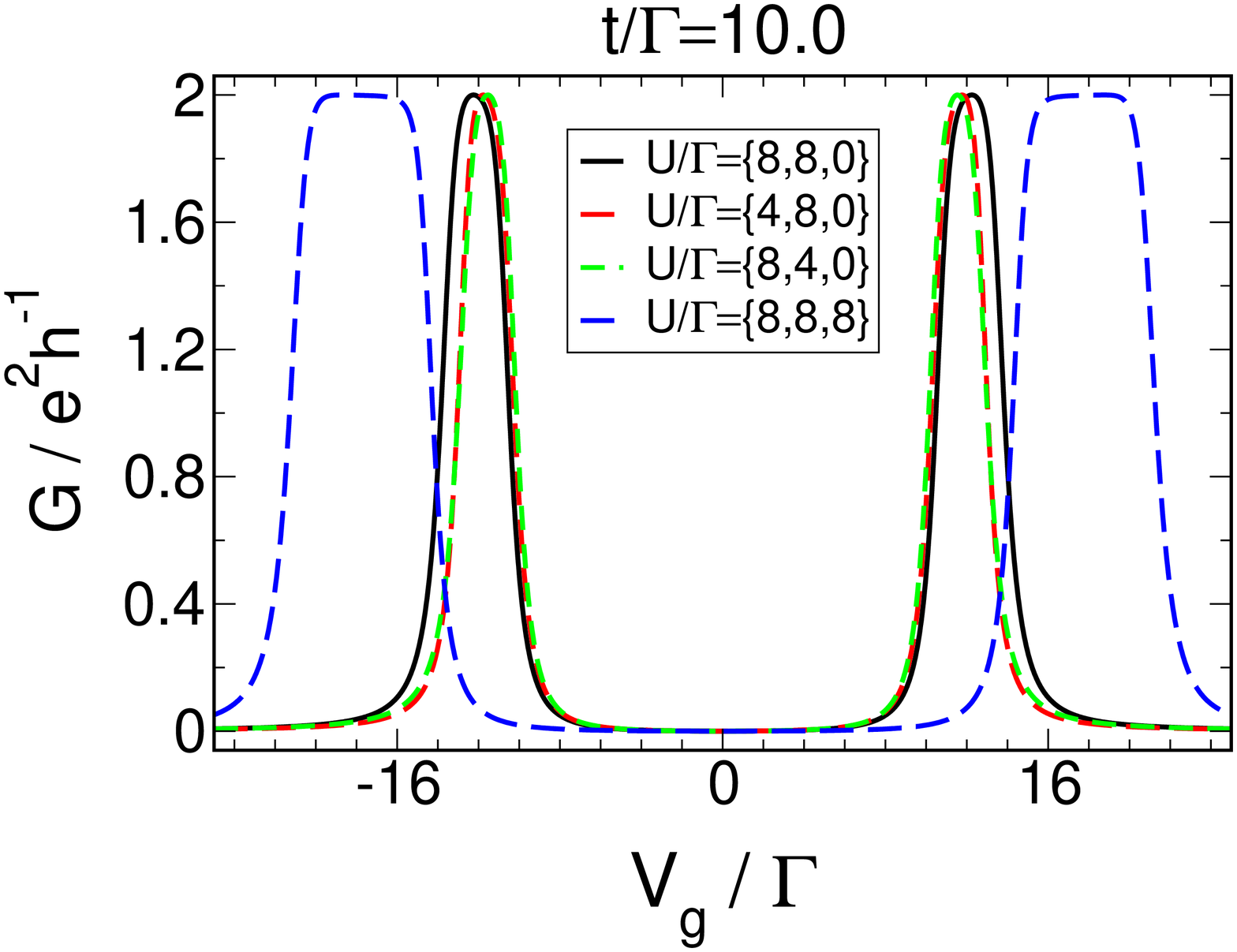}
        \caption{\textit{Left panel:} Conductance of the side-coupled geometry as a function of $V_g$ for $B=0$ and different left-right asymmetric hybridisations with the leads $A:=\Gamma_R/\Gamma_L$ for one parameter set from the two-stage Kondo regime (upper row) and one from the local spin-singlet phase (lower row). \textit{Right panel:} The same for symmetric hybridisations, but different local interactions on the embedded and side-coupled dot $U_e$ and $U_s$ and an additional nearest-neighbour interaction $U_N$ (denoted as $U=\{U_e,U_s,U_N\}$).}
\label{fig:MS.side.var}
\end{figure}

The physics in the limit of large hopping between the embedded and the side-coupled dot is very different from the two-stage Kondo regime. In the noninteracting case, which can again easily be understood by the mapping on the double dot geometry, and at zero magnetic field the conductance shows two Lorentzian peaks of width $\Gamma$ separated by $2t$. These resonances correspond to transport through the bonding and antibonding level of the isolated two-dot `molecule'. The average occupancy drops by two at the two peaks. The transmission phase changes by $\pi$ over each of them and jumps by $\pi$ in between since side-coupled dot maps onto a double dot with relative sign of the hybridisations $s=+$. Turning on a local interaction on both dots gradually transforms the Lorentzian resonances into box-like structures and enlarges their separation (Fig.~\ref{fig:MS.side.t}, right panel) due to the set-in of the Kondo effect on the bonding and antibonding level. For large $U$, the average occupation within each resonance is constant and close to an odd integer.

For large interactions, it is again meaningful to treat the isolated two-dot `molecule' decoupled from the leads to understand the width and the separation of the resonances. The single-particle part of the Hamiltonian of the single-level double dot with parameters arising from the mapping of the side-coupled geometry is diagonal with entries $V_g-U/2+t$ and $V_g-U/2-t$. Hence the vacuum and the single-particle state of lowest energy are degenerate if $V_g=t+U/2$. For higher gate voltages, the isolated molecule is unoccupied. In the occupation number basis,
\begin{equation*}\begin{split}
\{|n_{A,\uparrow},n_{A,\downarrow},&n_{B,\uparrow},n_{B,\downarrow}\rangle\}=\\[1ex] &\{|1,1,0,0\rangle, |0,0,1,1\rangle, |0,1,0,1\rangle, |1,0,0,1\rangle, |0,1,1,0\rangle, |1,0,1,0\rangle\}, 
\end{split}\end{equation*}
the two-particle part of the Hamiltonian reads
\begin{equation*}
2V_g-U + 
\begin{pmatrix}
2t+U/2 & +U/2 & & & & \\
+U/2 & -2t+U/2 & & & & \\
& & 0 & & & \\
& & & +U/2 & -U/2 & \\
& & & -U/2 &+U/2 & \\
& & & & & 0
\end{pmatrix}.
\end{equation*}
The smallest eigenvalue, which is obtained from diagonalising the first block of this matrix, is given by $2V_g-0.5U-0.5\sqrt{U^2+16t^2}$. This eigenvalues and the lowest of the single-particle part of the Hamiltonian are degenerate if $V_g=-t+0.5\sqrt{U^2+16t^2}$, hence for gate voltages in the interval $(t+U/2, -t+0.5\sqrt{U^2+16t^2})$ the single-particle state is that of lowest energy and thus at $T=0$ the molecule is occupied by one electron. Alltogether this means that we would expect a transmission resonance of width $2t+U/2-0.5\sqrt{U^2+16t^2}$ centred at $0.25(\sqrt{U^2+16t^2}+U)$ (and likewise for negative gate voltages due to particle-hole symmetry), which is indeed observed for large interactions \cite{side3}. As for the chain of dots studied in the previous section, we are unable to reach regions of $U$ where the conductance shows completely box-like structures which width and position is given by these values within our fRG approach.
\begin{figure}[t]	
	\centering
        \includegraphics[width=0.475\textwidth,height=5.2cm,clip]{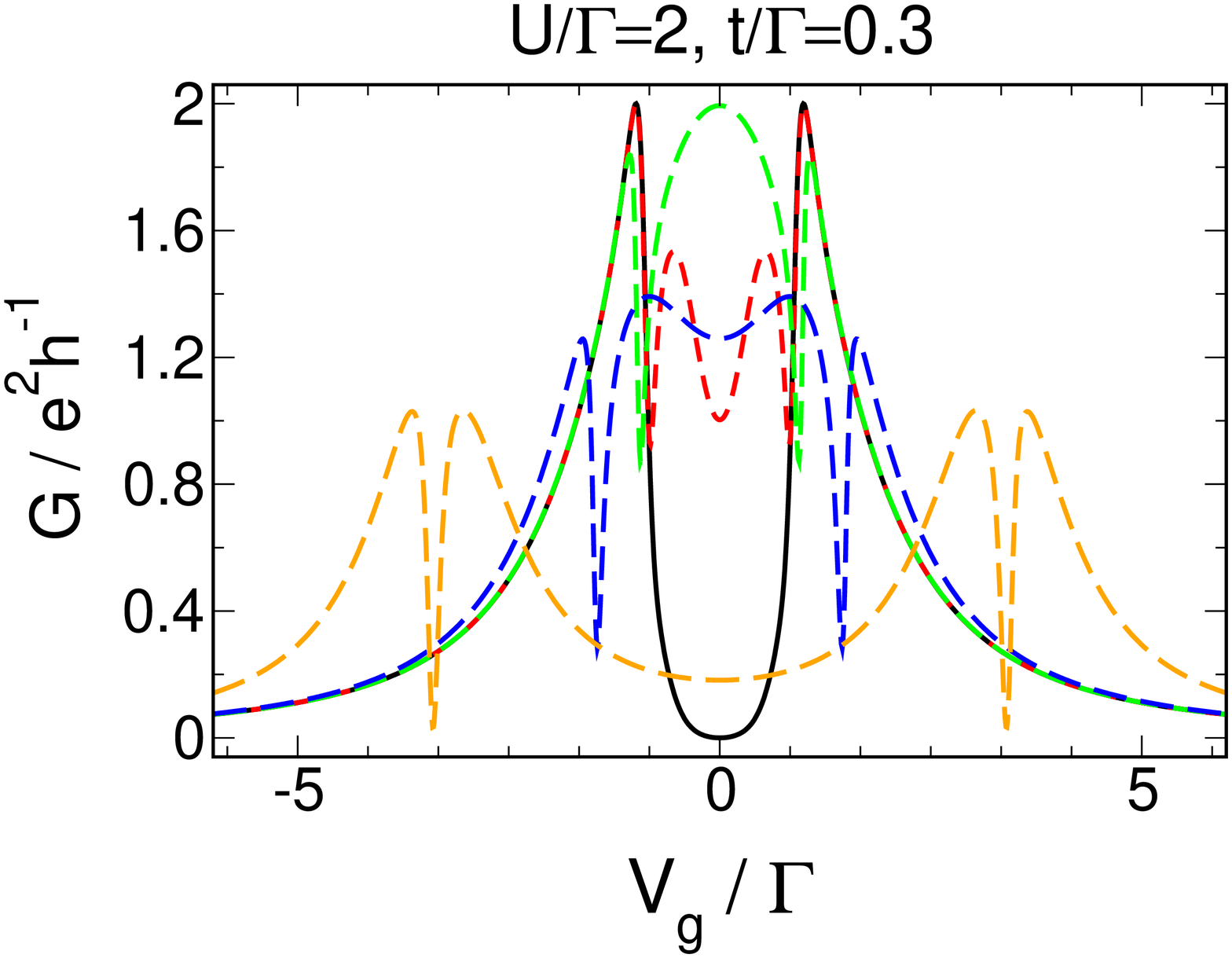}\hspace{0.035\textwidth}
        \includegraphics[width=0.475\textwidth,height=5.2cm,clip]{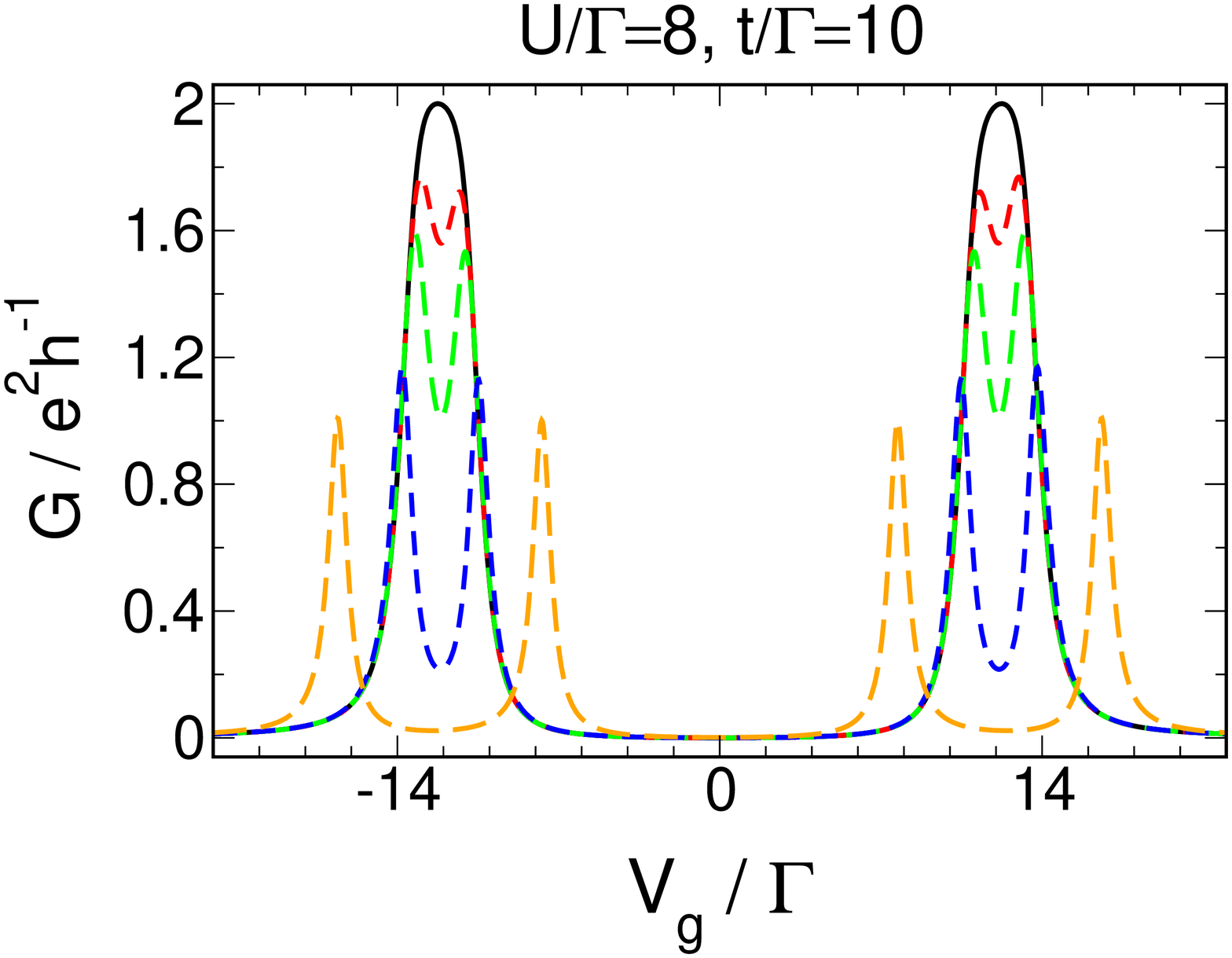}
        \caption{Conductance $G$ as a function of $V_g$ of the side-coupled geometry with $\Gamma_L/\Gamma=\Gamma_R/\Gamma=0.5$ for various zero magnetic fields. \textit{Left panel:} Two-stage Kondo regime with $T_K^2/\Gamma=0.0096$ and $T_K^1/\Gamma=1.16$ for $B/T_K^2=0.0$ (black), $B/T_K^2=1.0$ (red), $B/T_K^2=10.0$ (green), $B/T_K^2=100\approx 1.2T_K^1/T_K^2$ (blue), and $B/T_K^2=500\approx 6T_K^1/T_K^2$ (orange). \textit{Right panel:} Local spin-singlet phase with $T_K^1/\Gamma=0.11$ for $B/T_K^1=0.0$ (black), $B/T_K^1=0.5$ (red), $B/T_K^1=1.0$ (green), $B/T_K^1=5.0$ (blue), and $B/T_K^1=50.0$ (orange).}
\label{fig:MS.side.b}
\end{figure}

The eigenvector that corresponds to the two-particle state of lowest energy at half-filing is of the form
\begin{equation*}\begin{split}
\lambda |1,1,0,0\rangle + \mu |0,0,1,1\rangle &= (\lambda |A\rangle+\mu |B\rangle)\otimes(|\hspace{-0.7ex}\uparrow\downarrow\rangle-|\hspace{-0.7ex}\downarrow\uparrow\rangle) \\
 &= (\tilde\lambda |e\rangle+\tilde\mu |s\rangle)\otimes(|\hspace{-0.7ex}\uparrow\downarrow\rangle-|\hspace{-0.7ex}\downarrow\uparrow\rangle),
\end{split}\end{equation*}
which is a spin-singlet state. Therefore the `molecular' regime for large inter-dot hoppings is also called local spin-singlet phase.

An asymmetry in the hybridisations with the left and right lead only results in an overall decrease of the conductance (Fig.~\ref{fig:MS.side.var}, lower left panel). Relaxing the assumption of equal local interactions or introducing an additional nearest-neighbour interaction to the system leads only to qualitative changes but no new physics (Fig.~\ref{fig:MS.side.var}, lower right panel).

Applying a magnetic field $B$ to the system lifts the spin degeneracy and hence the Kondo resonances of the bonding and antibonding levels get destroyed. Similar to the single-level case the box-like resonances split up, whereas the conductance at half-filling always remains a minimum. As before, the Kondo energy scale $T_K$ can be defined by the magnetic field required to lower $G$ to half the unitary limit within the peaks, and again the results obtained from this definition yield values consistent with those calculated in an NRG approach \cite{side2} (see the next chapter for more details). Furthermore, the $G(V_g)$ curves for small magnetic fields $B\in [0,T_K]$ are similar to those computed by the aforementioned authors for finite temperatures, which again illustrates that a small magnetic field and small nonzero temperatures indeed similarly destroy the Kondo resonance, justifying our definition of the energy scale $T_K$. For large magnetic fields we recover twice the spin-polarised structure in $G(V_g)$. Hence, if only local correlations are present we recover four Lorentzian peaks. The influence of a magnetic field in the molecular-orbital regime is again shown in the right panel of Fig.~\ref{fig:MS.side.b}.

As the evolution of the conductance curves for small magnetic fields at $T=0$ and finite temperatures in both regimes is similar, the former provides a zero temperature criterion whether a certain choice of parameters corresponds to the two-stage Kondo regime or the local spin-singlet phase. The distinction between the two is not unambiguously possible if one only considers the gate voltage dependence of the conductance since the latter is strongly suppressed at half-filling in both cases. For all parameters shown in this section, we have tested the evolution with finite magnetic fields of $G(V_g)$ to justify that they indeed correspond to the regime claimed.

\subsection{Parallel Double Dots}\label{sec:MS.dd}

\begin{figure}[t]	
	\centering
	      \includegraphics[width=0.4\textwidth,clip]{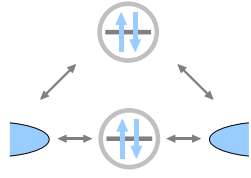}
        \caption{The double dot geometry.}
\label{fig:MS.geo_dd}
\end{figure}

In this section, we will generalise the results derived for the parallel double dot geometry in Sec.~\ref{sec:OS.dd} to the case including the spin degree of freedom. The geometry is again depicted in Fig.~\ref{fig:MS.geo_dd}. The free propagator, which is by definition diagonal in spin space, reads
\begin{equation}\label{eq:MS.dd.g}
\left[(\mc G^\sigma(i\omega))^0\right]^{-1} =
\begin{pmatrix}
i\omega - V_g + \frac{\Delta+\sigma B}{2} + i~\tn{sgn}(\omega)~\Gamma_A & i~\tn{sgn}(\omega)\left[\sqrt{\Gamma_A^L\Gamma_B^L}+s\sqrt{\Gamma_A^R\Gamma_B^R}\right] \\
i~\tn{sgn}(\omega)\left[\sqrt{\Gamma_A^L\Gamma_B^L}+s\sqrt{\Gamma_A^R\Gamma_B^R}\right] & i\omega - V_g - \frac{\Delta-\sigma B}{2} + i~\tn{sgn}(\omega)~\Gamma_B
\end{pmatrix},
\end{equation}
where $V_g$ is the gate voltage that shifts the on-site energies of the dots, $\Delta$ a level detuning, and $B$ is a magnetic field applied to the system. The hybridisations with the left and right leads are denoted by $\Gamma_{A,B}^{L,R}$, the relative sign of the level-lead couplings by $s$, and as usual we define $\Gamma_l:=\Gamma_l^L+\Gamma_l^R$, and $\Gamma:=\Gamma_A+\Gamma_B$. If we introduce correlations to the systems, which will in general be a local interaction $U_{l,l}^{\sigma,\bar\sigma}$ between the spin up and spin down electrons on each dot as well as nearest-neighbour interactions $U_{l,\bar l}^{\sigma,\sigma}$ and $U_{l,\bar l}^{\sigma,\bar\sigma}$, we have to solve the flow equations (\ref{eq:FRG.flowse3}) and (\ref{eq:FRG.flowww3}) to obtain (an approximation to) the full propagator $\mc G = \tilde{\mc G} (\la=0)$. The conductance for each spin direction can than be calculated in analogy to the spin-polarised case, (\ref{eq:OS.dd.cond}),
\begin{equation}\label{eq:MS.dd.cond}
G_\sigma = 4\frac{e^2}{h}\left|\sqrt{\Gamma_A^L\Gamma_A^R}\mc G_{A;A}^\sigma(0)+\sqrt{\Gamma_B^L\Gamma_A^R}\mc G_{A;B}^\sigma(0)
+s\sqrt{\Gamma_A^L\Gamma_B^R}\mc G_{B;A}^\sigma(0)+s\sqrt{\Gamma_B^L\Gamma_B^R}\mc G_{B;B}^\sigma(0)\right|^2.
\end{equation}

The parallel double dot geometry allows for a large amount of parameters to be varied. Fortunately, the noninteracting case is well understood since it is by definition equivalent to the spin-polarised case which was discussed in detail in Sec.~\ref{sec:OS.dd}. In presence of correlations, it seems reasonable to stick to two special choices out of all possible interactions, namely the one where only local correlations on each of the dots are present and that with equal interactions between all electrons. The first situation can be interpreted as two spatially well-separated dots with one level each, while the second would be a reasonable model for a single dot containing two levels. Despite the fact that it might be rational to introduce a direct hopping $t$ between two separated dots, we will consider $t=0$ here since such a hopping could always be tuned away by a basis transformation of the dot states. As for spin-polarised dots, we will focus on studying the generic behaviour of all quantities of interest. In particular, we would expect that all the $G(V_g)$ curves in dependence of the other parameters are qualitatively similar for all level-lead hybridisations up to a few pathological cases. As in the spin-polarised case we have checked that this is indeed true by computing roughly 5000 points within the parameter space $\{U, \Delta, B, \Gamma_{A,B}^{L,R}, s\}$.

\subsubsection{Purely Local Interactions, $s=+$}

\begin{figure}[t]	
	\centering
	      \includegraphics[width=0.475\textwidth,height=4.0cm,clip]{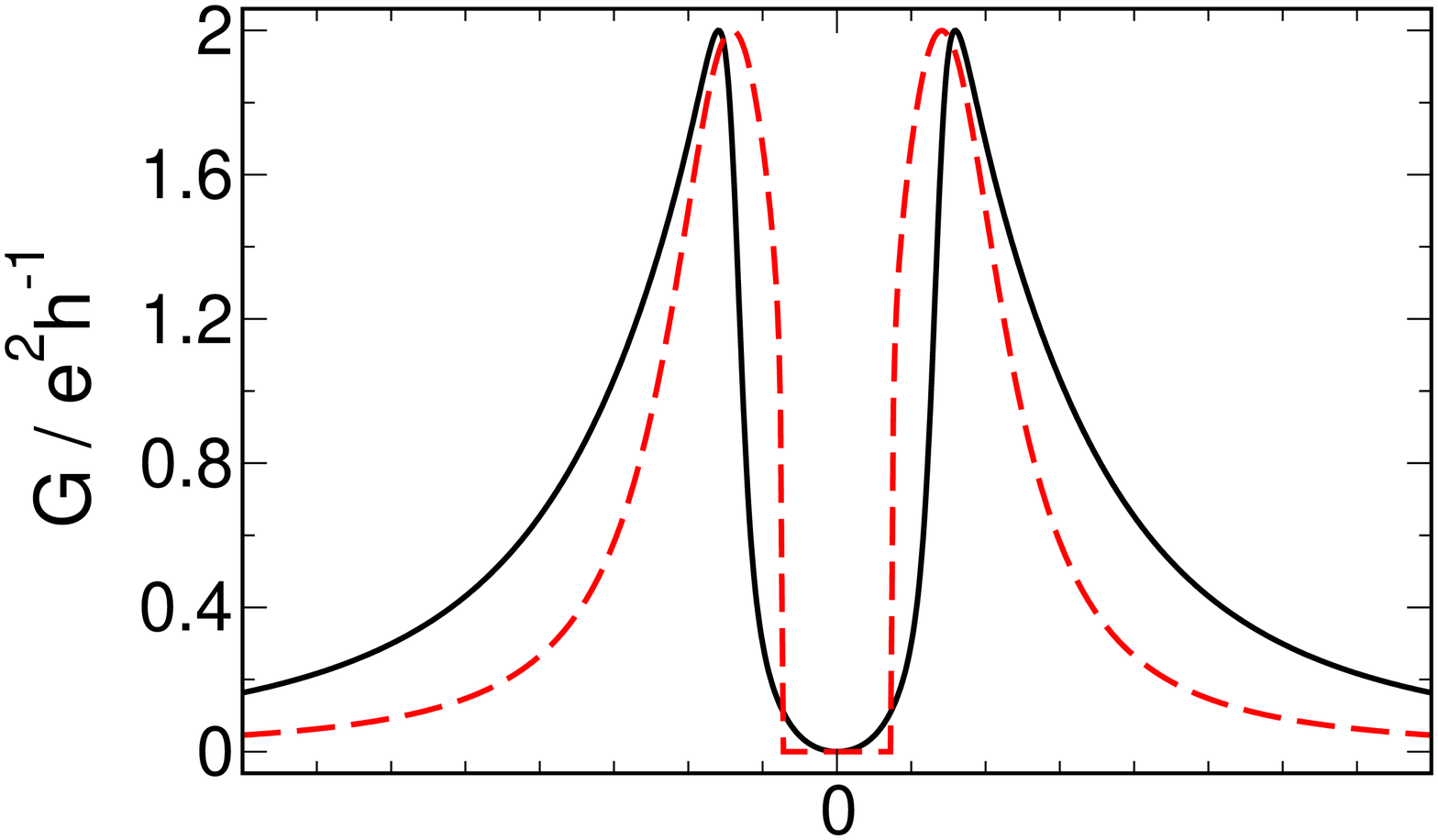}\hspace{0.035\textwidth}
        \includegraphics[width=0.475\textwidth,height=4.0cm,clip]{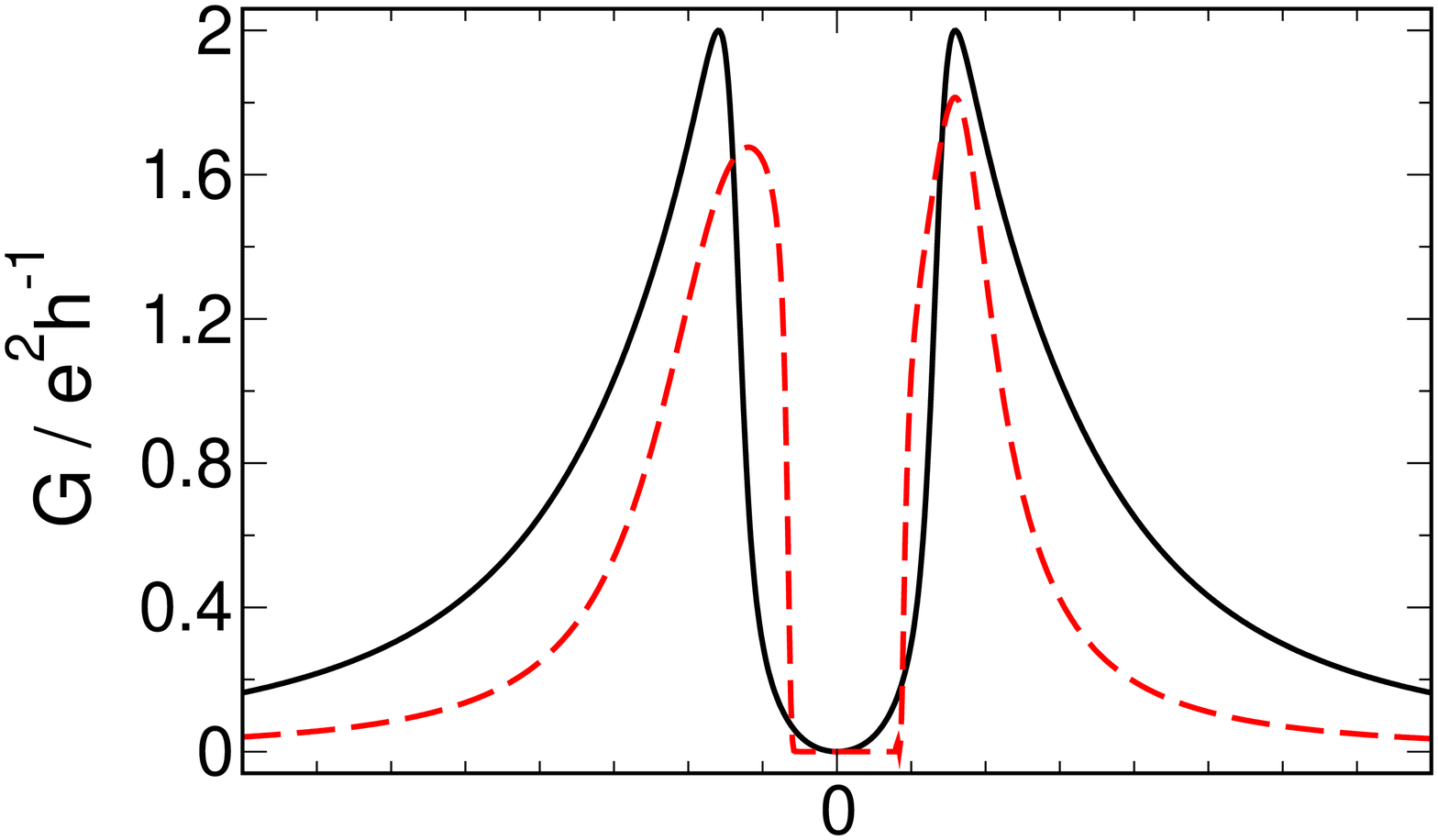}\vspace{0.3cm}
        \includegraphics[width=0.475\textwidth,height=4.8cm,clip]{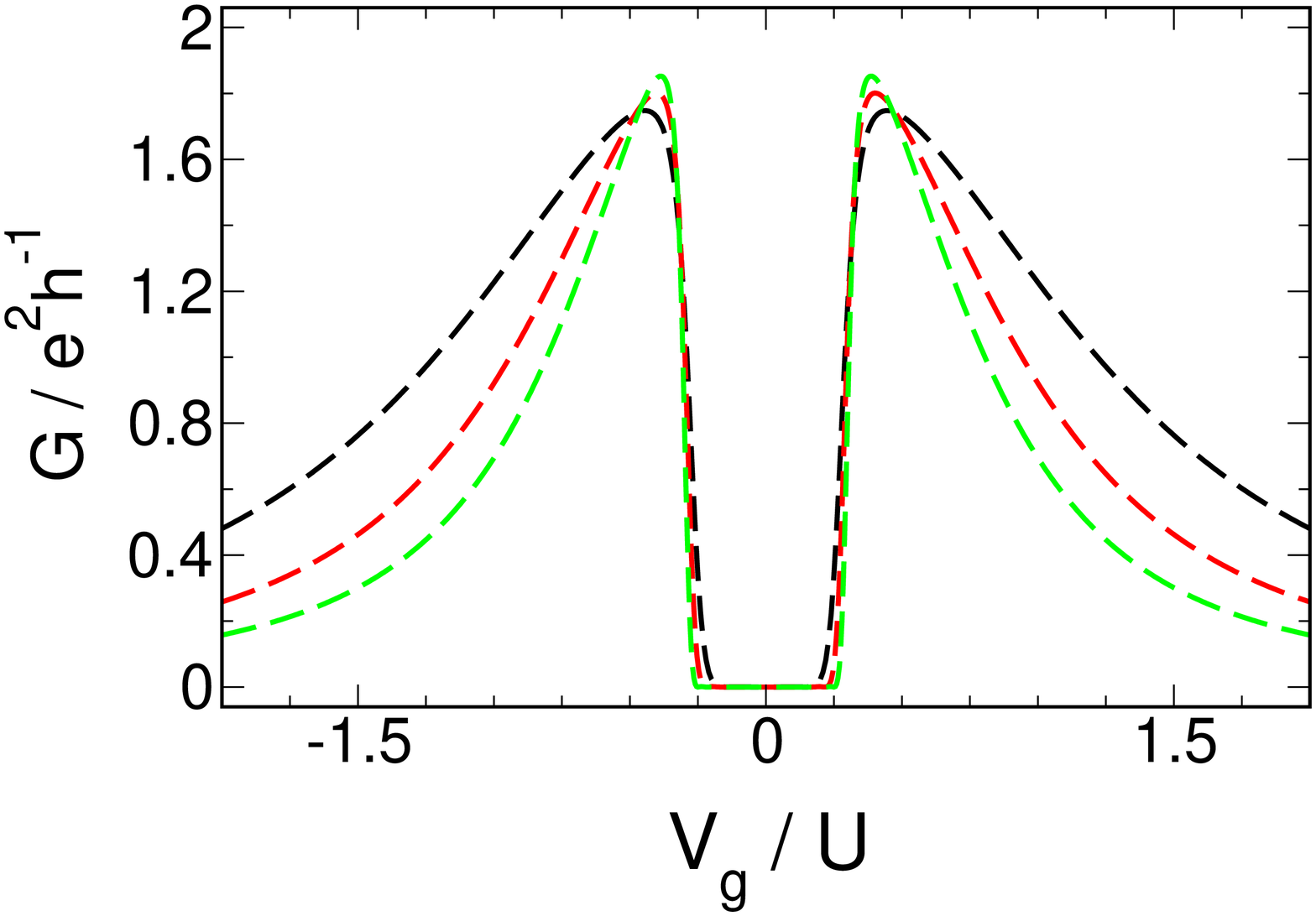}\hspace{0.035\textwidth}
        \includegraphics[width=0.475\textwidth,height=4.8cm,clip]{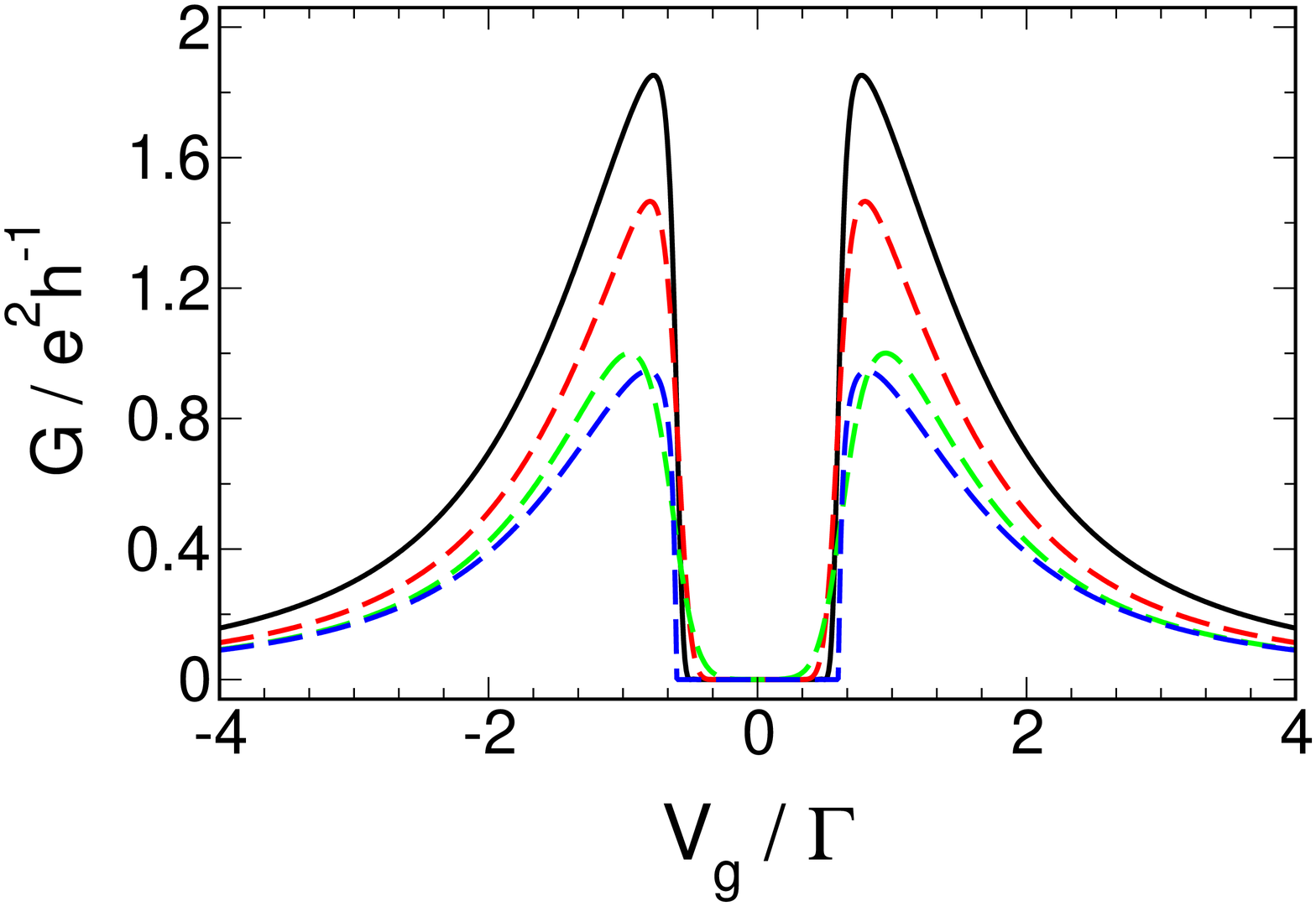}
        \caption{Comparison of the double dot at $s=+$ and purely local interactions $U$ with the side-coupled geometry. \textit{Upper row:} Gate voltage dependence of the conductance for the side-coupled dots with $\Gamma_L=\Gamma_R=0.5$, $U=1.5$, $t=0.3$ (black) and of the parallel double dot (red) with $U=1.5$, $\Delta=0.6$ for $\Gamma=\{0.125~0.125~0.125~0.125\}$ (left panel), and $\Gamma=\{0.05~0.1~0.25~0.1\}$ (right panel). For clarity, all energies are given in arbitrary units. \textit{Lower left panel:} Conductance through the double dot geometry with degenerate levels, $\Gamma=\{0.1~0.2~0.5~0.2\}$ for $U/\Gamma=1.0$ (black), $U/\Gamma=1.5$ (red), and $U/\Gamma=2.0$ (green). \textit{Lower right panel:} The same but for $U/\Gamma=2.0$ and $\Gamma=\{0.1~0.2~0.5~0.2\}$ (black), $\Gamma=\{0.1~0.2~0.6~0.1\}$ (red), $\Gamma=\{0.04~0.26~0.6~0.1\}$ (green), and $\Gamma=\{0.26~0.04~0.6~0.1\}$ (blue).}
\label{fig:MS.dd.loc_pcomp}
\end{figure}

We will start with $s=+$, small level detunings $\Delta\ll\Gamma$ and the case where only local correlations $U$ between the electrons on each of the two dots are present. Already for small interactions $U\approx\Gamma$, the `Fano-antiresonance' observed in the noninteracting case is transformed into a wide valley of width $U$ where the conductance is strongly suppressed. At $V_g\approx\pm U/2$ two resonances show up whose wings become smaller for $|V_g|>U/2$ if the interaction is increased (Fig.~\ref{fig:MS.dd.loc_pcomp}, lower left panel). An increased left-right asymmetry in the hybridisations leads to an overall decrease of $G$ (Fig.~\ref{fig:MS.dd.loc_pcomp}, lower right panel). Not surprisingly, all this is in complete analogy to the side-coupled dots in the two-stage Kondo regime. As explained in the previous section, the latter can be mapped onto the parallel double dot geometry with $A$-$B$-symmetric hybridisations and additional interaction terms. However, by going over to arbitrary hybridisations and slowly changing the interaction to a purely local one we convinced ourselves that the physics does not change. The conductance as a function of the gate voltage has qualitatively the same structure (Fig.~\ref{fig:MS.dd.loc_pcomp}, upper row), and, more important, the valley at half filling disappears when turning on small magnetic fields (see below). Alltogether, this means that the physics of the double dot geometry with $s=+$, small level spacings and purely local interactions can be understood from the side-coupled dot in the two-stage Kondo regime.

Since we know that for the side-coupled geometry with direct hopping $t$ it is $U/t^2$ that determines whether the second stage of the Kondo effect is active or not, for small $t/\Gamma$ the interaction already becomes important for fairly small $U/\Gamma$ and hence we should expect the fRG only to give reliable results for $U\lesssim\Gamma$. At the beginning of this chapter, we have claimed that we were only going to show data for interaction strength where the fRG still produces assured results. In all situations under consideration up to now this was indeed fullfilled, as will be shown in the next chapter by explicit comparison to very accurate NRG data. However, for the parallel double dot geometry an interesting parameter regime is certainly that with degenerate levels, where by the aforementioned argumentation the fRG results should be questionable already for fairly small interactions. This indeed turns out to be the case, but again we will delay an appropriate discussion to the next chapter, since it proves that the fRG still covers the essential physics. Here, we should only keep in mind, that for $s=+$ and small level detunings close to (away from) half filling we will capture the behaviour of $G(V_g)$ and all other quantities of interest qualitatively (quantitatively).

\begin{figure}[t]	
	\centering
	      \includegraphics[width=0.495\textwidth,height=4.4cm,clip]{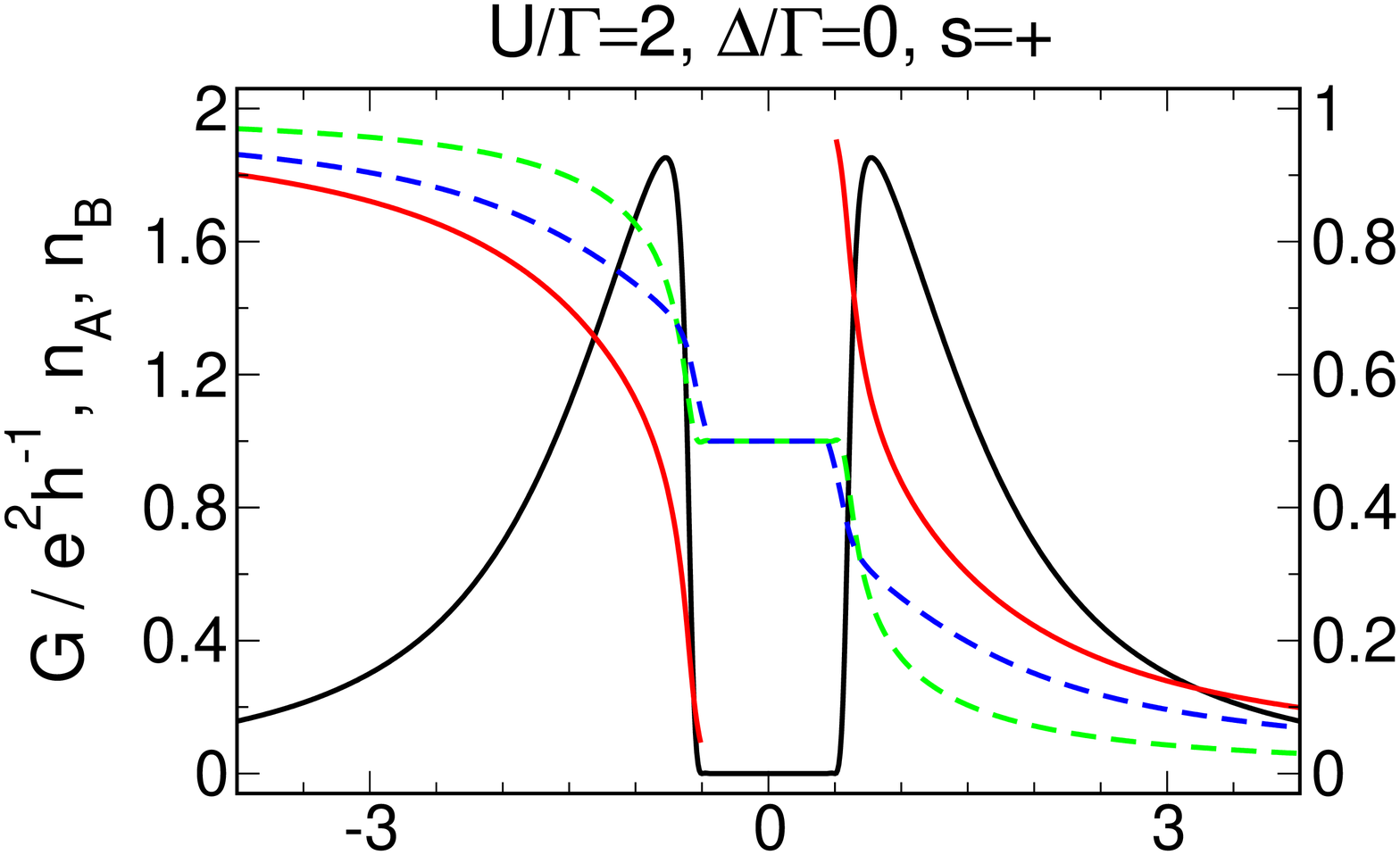}\hspace{0.015\textwidth}
        \includegraphics[width=0.475\textwidth,height=4.4cm,clip]{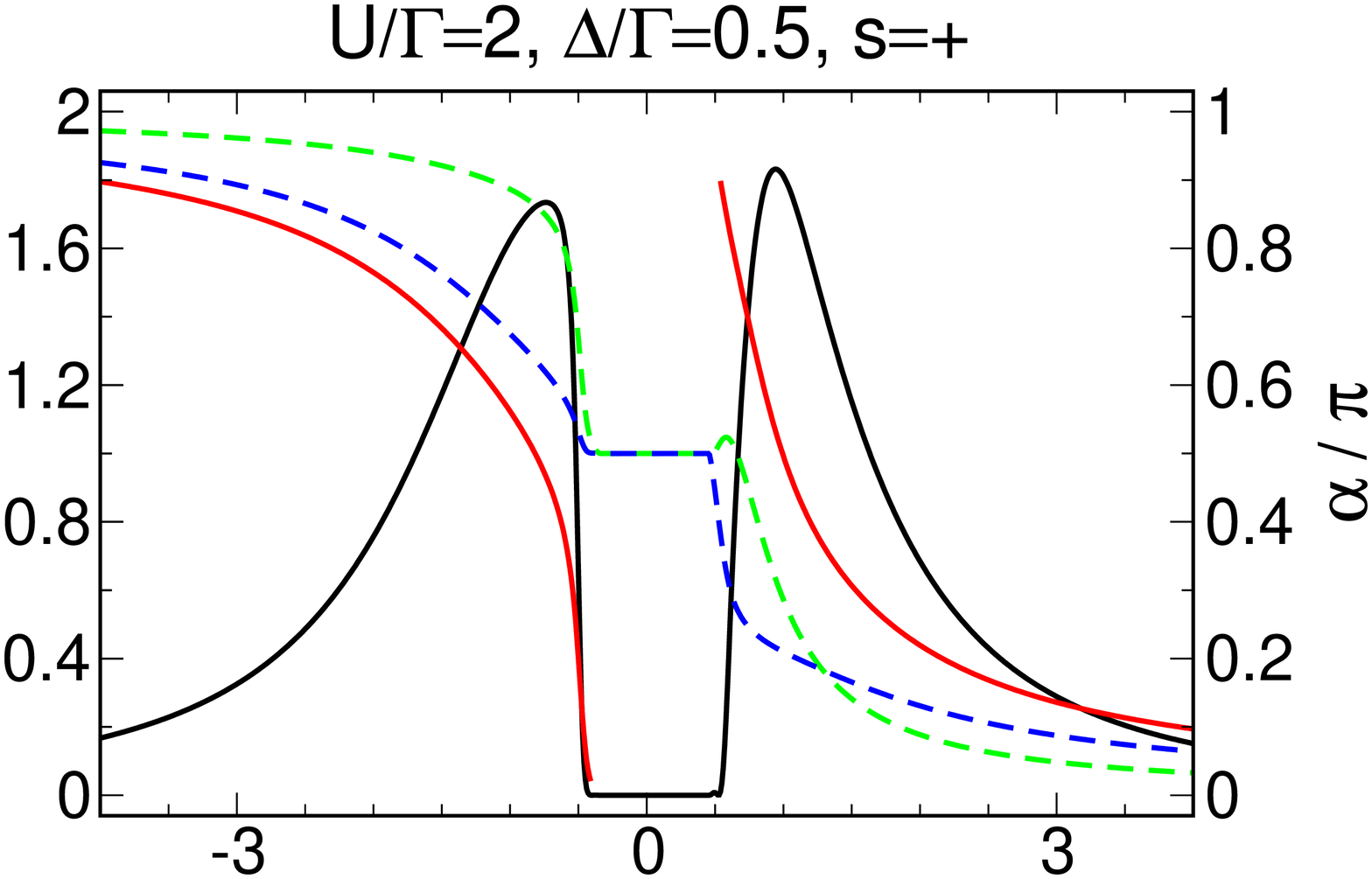}\vspace{0.3cm}
        \includegraphics[width=0.495\textwidth,height=5.2cm,clip]{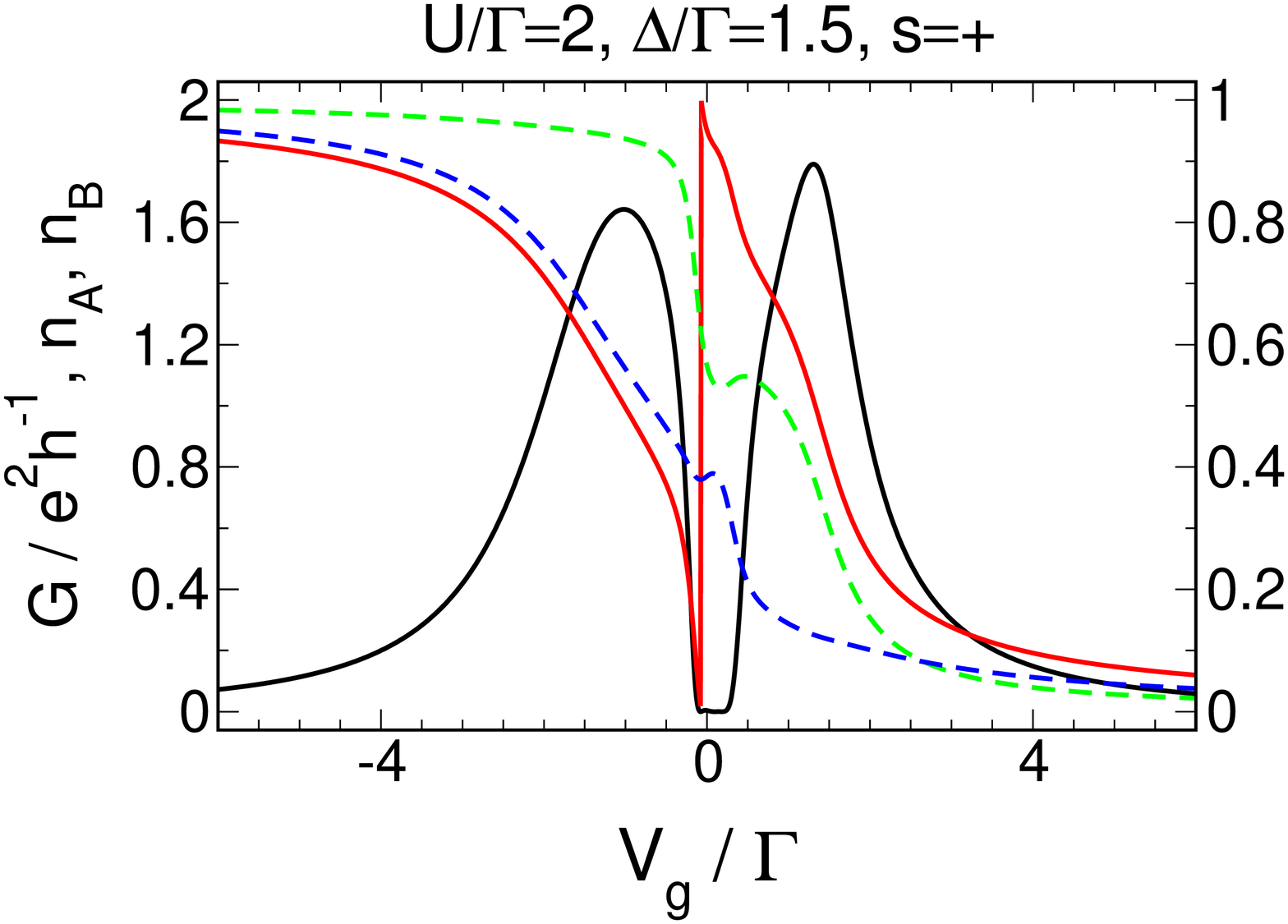}\hspace{0.015\textwidth}
        \includegraphics[width=0.475\textwidth,height=5.2cm,clip]{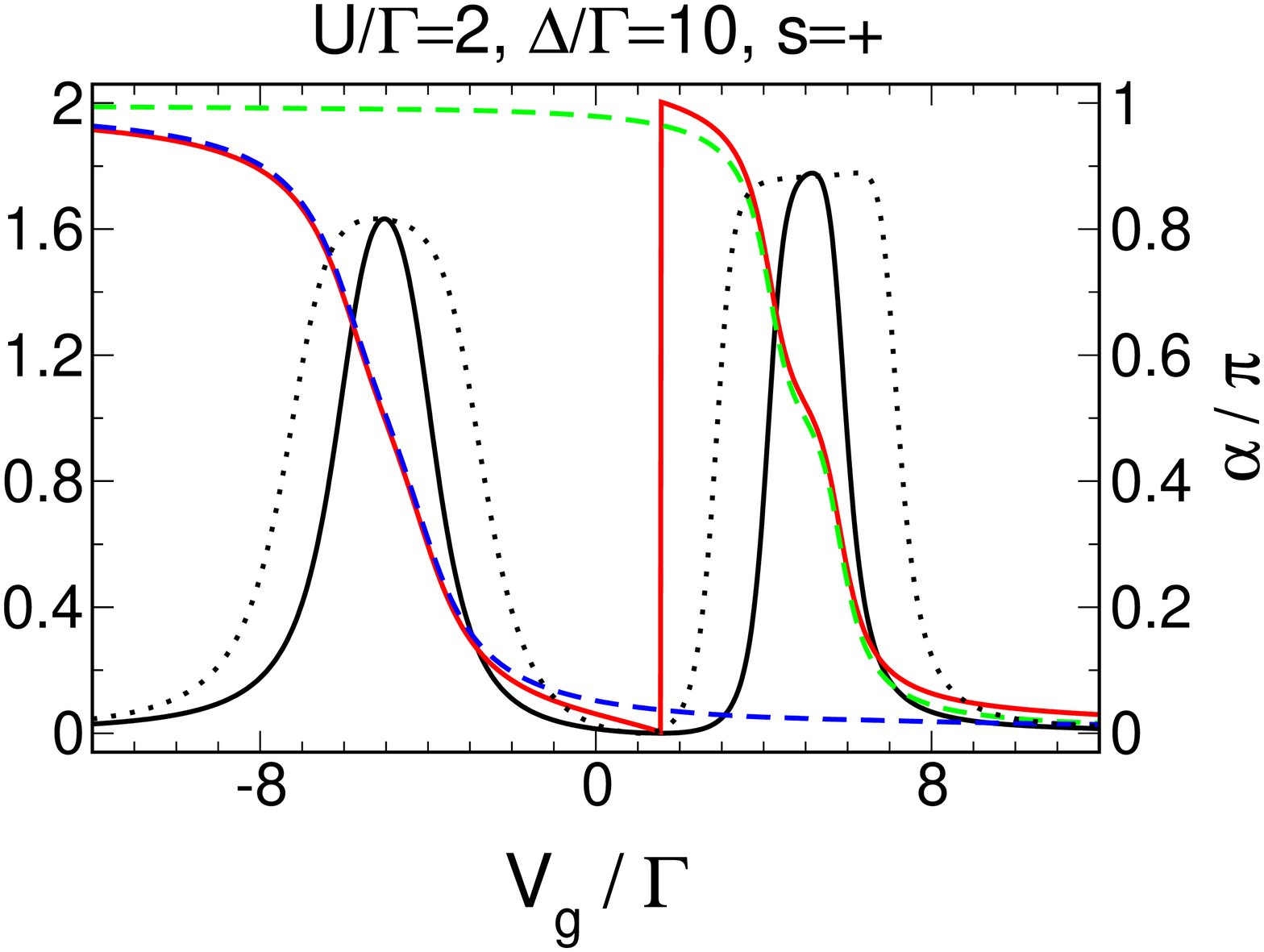}
        \caption{Gate voltage dependence of the conductance $G$ (black), transmission phase $\alpha$ (red) and average level occupancies (dot A: green, dot B: blue) for the parallel double dot with generic level-lead hybridisations $\Gamma=\{0.1~0.2~0.5~0.2\}$ and purely local interactions $U$ for four different level spacings $\Delta$. For small level detunings, the phase was hidden within the region where the conductance is strongly suppressed (see text). For $\Delta/\Gamma=10.0$, the conductance for $U/\Gamma=5.0$ is shown as well (black dotted line).}
\label{fig:MS.dd.loc_pd}
\end{figure}

Similar to the noninteracting case, the transmission phase $\alpha$ changes by $\pi$ over each of the resonances. Within the valley, it is almost constant up to several jumps by $\pi$. They are caused by additional transmission zeros numerically appearing because the conductance is far underestimated. Hence these phase jumps are almost certainly an artifact caused by the application of the fRG at too large $U$. The same holds for the average level occupancies. For large negative gate voltages, both dots are occupied by two electrons. The more weakly a level is coupled, the more steeply its occupation falls off close to $V_g=-U/2$ and merges into a plateau within the valley, which also seems to be overestimated. All this is depicted in the upper left panel of Fig.~\ref{fig:MS.dd.loc_pd}.

Increasing the level spacing up to $\Delta\approx\Gamma$ does not change the lineshape of $G(V_g)$ substantially (Fig.~\ref{fig:MS.dd.loc_pd}), which is plausible since even for $t/\Gamma\approx 0.5$ we found the side-coupled geometry to be within the two-stage Kondo regime (Fig.~\ref{fig:MS.side.varut}). An interesting observation, however, is that for such level detunings (where the fRG is known to give reliable results) the average occupancies depend non-monotonically on the gate voltage within the conductance valley (Fig.~\ref{fig:MS.dd.loc_pd}, lower left panel), a scenario similar to the side-coupled geometry. For large $\Delta\gg\Gamma$, we finally recover two separated resonances. Their Lorentzian lineshape exhibited at $U=0$ is transformed into box-like structures for large interactions $U/\Gamma_{A,B}\gg 1$ because of the Kondo effect that is active on one dot at each peak. This picture is supplemented by the fact that each level occupancy shows the typical Kondo behaviour at one of the resonances while being constant otherwise. The same holds for the transmission phase (up to a phase jump of $\pi$ in between the resonances). In addition, the height of the individual peaks is determined by the left-right asymmetry of the hybridisations $\Gamma_A^L/\Gamma_A^R$ and $\Gamma_B^L/\Gamma_B^R$. For large $U$ and $\Delta$ (and $U<\Delta$) the position of the two resonances can be understood as usual by diagonalising the isolated dot system. The lowest energies of states with one, two, three and four electrons read $V_g-\Delta/2-U/2$, $2V_g-\Delta+U/2$, $3V_g-\Delta/2+U/2$, and $4V_g+3U/2$, and hence at zero temperature the dot is occupied by an odd number of electrons in the $V_g$-regions $[\Delta/2,\Delta/2+U]$ and $[-\Delta/2-U,-\Delta/2]$. Assuming that in analogy to the single-level case transport is possible in these situations, it follows that we would expect two peaks of width $U$ separated by $\Delta$.
\begin{figure}[t]	
	\centering
	      \includegraphics[width=0.49\textwidth,height=5.2cm,clip]{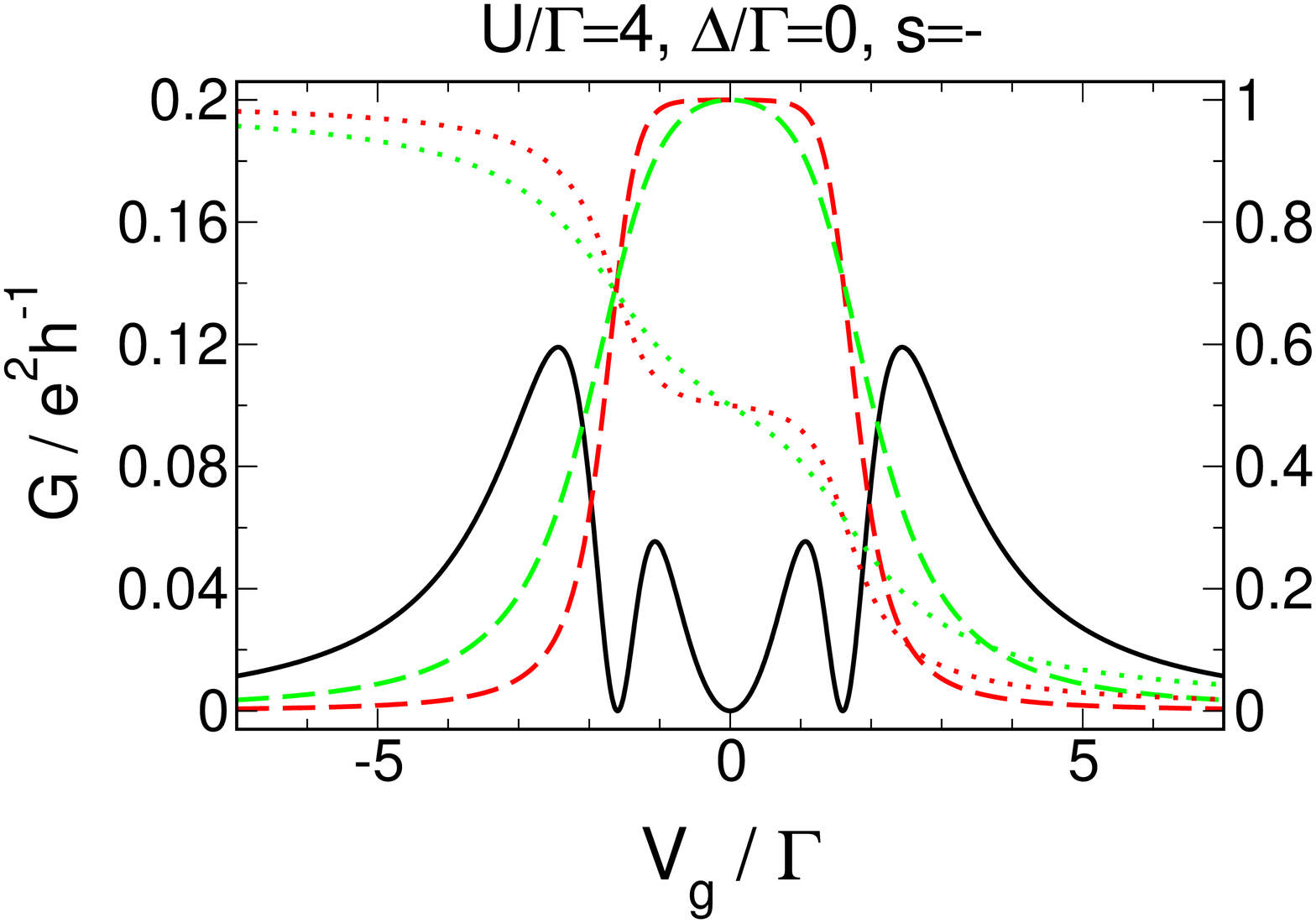}\hspace{0.015\textwidth}
        \includegraphics[width=0.48\textwidth,height=5.2cm,clip]{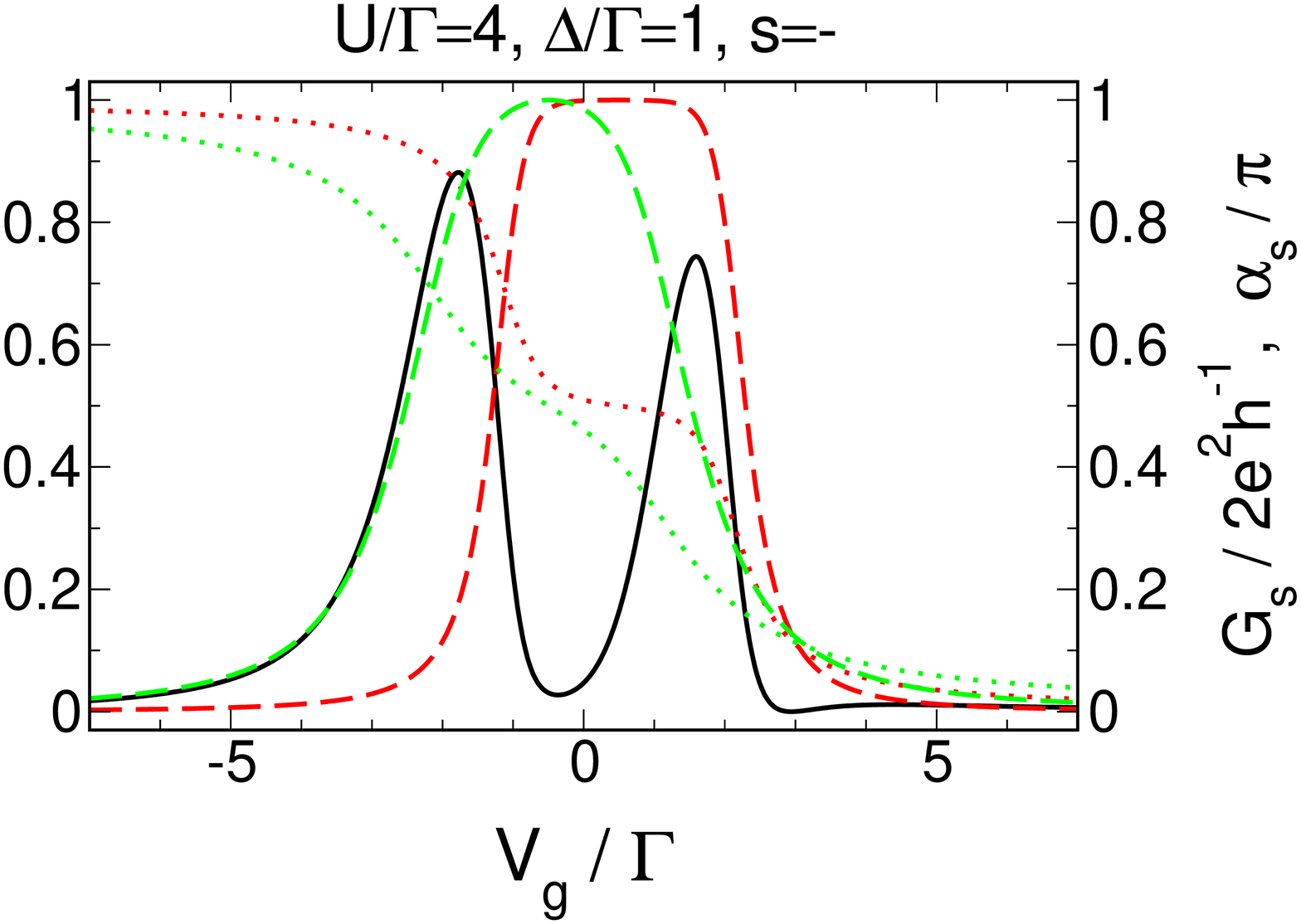}
        \caption{Gate voltage dependence of the conductance $G$ (black) for the double dot geometry with purely local interactions, left-right symmetric hybridisations $\Gamma=\{0.15~0.15~0.35~0.35\}$, and $s=-$. For these parameters, $G(V_g)$ (and in particular its four-peak structure) can be understood from the behaviour of the conductance $G_s$ (dashed lines) and transmission phase $\alpha_s$ (dotted lines) of two single-level dots with $\Gamma_L/\Gamma=\Gamma_R/\Gamma=0.15$ (red) and $\Gamma_L/\Gamma=\Gamma_R/\Gamma=0.35$ (green) separated by $\Delta$ (note that $\Gamma$ still refers to the hybridisation strength of the double dot).}
\label{fig:MS.dd.loc_mc}
\end{figure}

Applying a magnetic field to the system closes the analogy between the side-coupled geometry and the parallel double dot. For small $\Delta$, the conductance close to half filling is first increased for finite fields $B\ll\Gamma$ due to the suppression of the second stage Kondo effect. At some $B\lesssim\Gamma$, it starts to decrease again and for large $B\gg\Gamma$ we recover twice the spin-polarised structure in the lineshape of $G(V_g)$, which is here identical with the noninteracting one (Fig.~\ref{fig:MS.dd.loc_b}, upper left panel). For large $\Delta$, the situation for finite magnetic fields is again completely similar to the single-level case at each resonance. The box-like structures split up on an energy scale determined by the corresponding Kondo temperatures proportional to $\exp(-U/\Gamma_{A,B})$ (Fig.~\ref{fig:MS.dd.loc_b}, lower left panel). For large $\Delta$, we recover the noninteracting lineshape for the partial conductance and spin-dependent transmission phase $\alpha_\sigma$ of the spin up and down electrons (Fig.~\ref{fig:MS.dd.loc_b}, lower right panel).

\subsubsection{Purely Local Interactions, $s=-$}

\begin{figure}[t]	
	\centering
        \includegraphics[width=0.475\textwidth,height=5.2cm,clip]{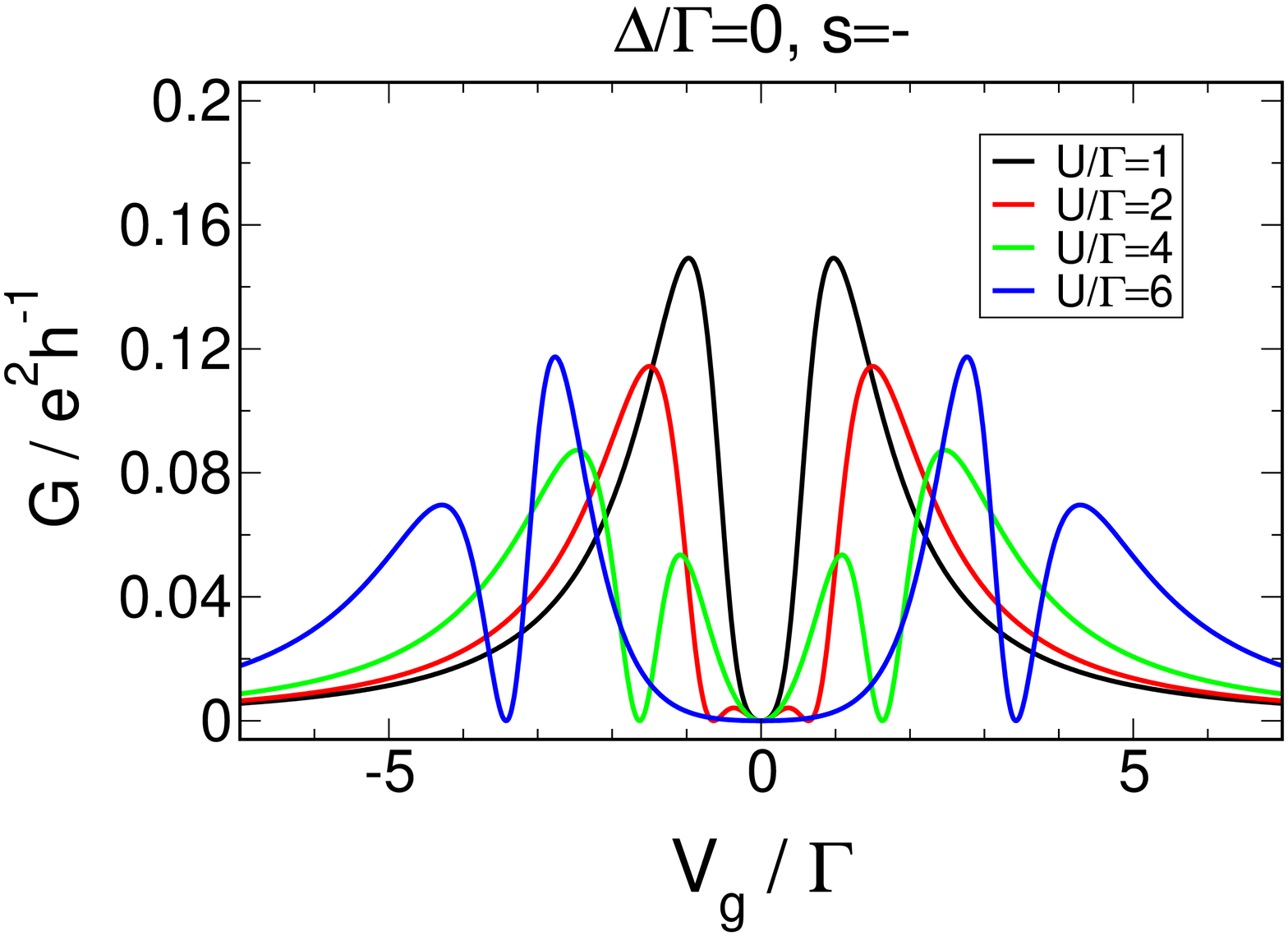}\hspace{0.035\textwidth}
        \includegraphics[width=0.475\textwidth,height=5.2cm,clip]{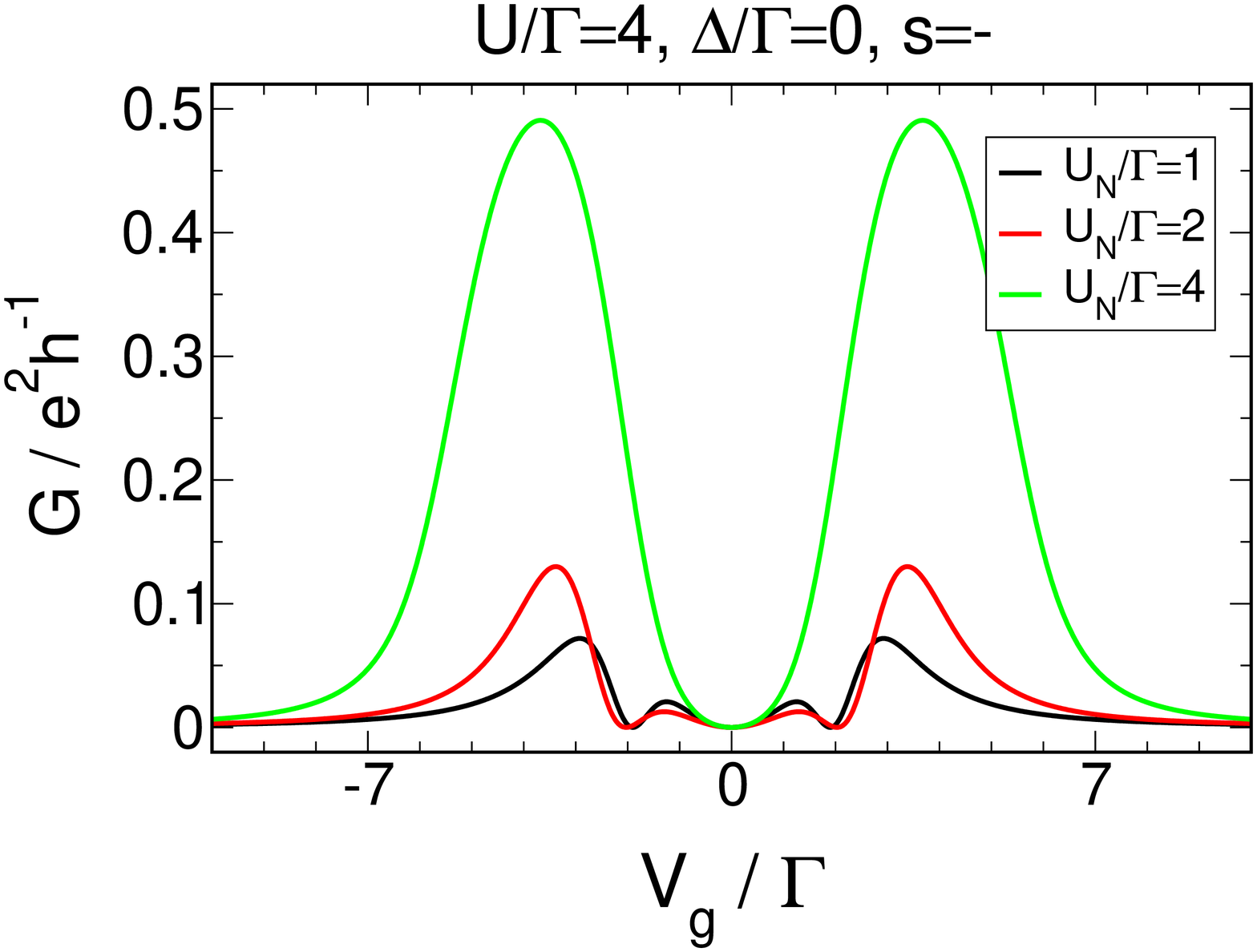}
        \caption{Gate voltage dependence of the conductance $G$ for the double dot geometry with degenerate levels, $s=-$, and generic hybridisations $\Gamma=\{0.1~0.2~0.5~0.2\}$ for purely local interactions $U$ (left panel) and in presence of additional nearest-neighbour correlations $U_N$ (right panel). The evolution of four resonances in the left panel can be easily understood in the left-right symmetric case (Fig.~\ref{fig:MS.dd.loc_mc}).}
\label{fig:MS.dd.loc_m}
\end{figure}

If the relative sign of the level-lead couplings is $s=-$, we can no longer map the parallel double dot geometry to the side-coupled one. Fortunately, for the special case of left-right symmetric hybridisations the flow equations for the self-energy and the effective interaction can be related to those of a single-level, allowing for an analogy between both situations within our approximation. In particular, if $\Gamma_l^L=\Gamma_l^R$, the free propagator (\ref{eq:MS.dd.g}) is diagonal, and if only local correlations are present, no off-diagonal terms are generated by the flow. This means that the full propagator of the system is diagonal with entries that correspond to two individual single-dot problems each with interaction between spin up and down electrons $U$ and level-lead hybridisations $\Gamma_A$ and $\Gamma_B$. The conductance $G$ of the system is then proportional to the absolute square of the difference of both single-level propagators evaluated at $z=i0$, implying that transport through the system can be interpreted as quantum interference between two independent channels each corresponding to a single spinful impurity.

In Fig.~\ref{fig:MS.dd.loc_mc} we show the conductance through the double dot system with purely local interactions and $s=-$. For degenerate levels and $U$ exceeding a critical value depending on the hybridisations, we observe four transmission resonances instead of two. They gradually disappear when $\Delta$ is increased. As stated above, this can be understood by considering the $G(V_g)$ curves of the individual levels. In general, the box-like structure of these curves is developed with different strength because of the different $U/\Gamma_l$, and hence both curves will intersect somewhere around $V_g\approx\pm U/2$, at least if the interaction is large enough for the width of both peaks to be determined by $U$ rather than by $\Gamma_{A,B}$. Since the same holds for the transmission phase, the conductance of the double dot system shows three transmission zeros at $V_g=0$ and $V_g\approx\pm U/2$, which implies a total of four resonances whose separation should increase with $U$ (see also Fig.~\ref{fig:MS.dd.loc_m}, left panel). A nonzero level spacing shifts two of the zeros closer together and they finally disappear for $\Delta=O(U)$. At first sight this might sound surprisingly since the $G(V_g)$ curves of the individual levels shifted by $\Delta$ against each other still intersect twice for arbitrary large level level spacings. However, this does not hold for the transmission phase. In fact, the latter determines the conductance by a Friedel sum rule, $G=2\sin^2\alpha$, but this does not imply that on the other hand the conductance fixes the phase uniquely. Hence, an intersection of the two $G(V_g)$ curves is not necessarily accompanied by equal transmission phases, so that it is no contradiction that for $\Delta=O(U)$ only one of the zeros in the conductance remains.

\begin{figure}[t]	
	\centering
	      \includegraphics[width=0.495\textwidth,height=4.4cm,clip]{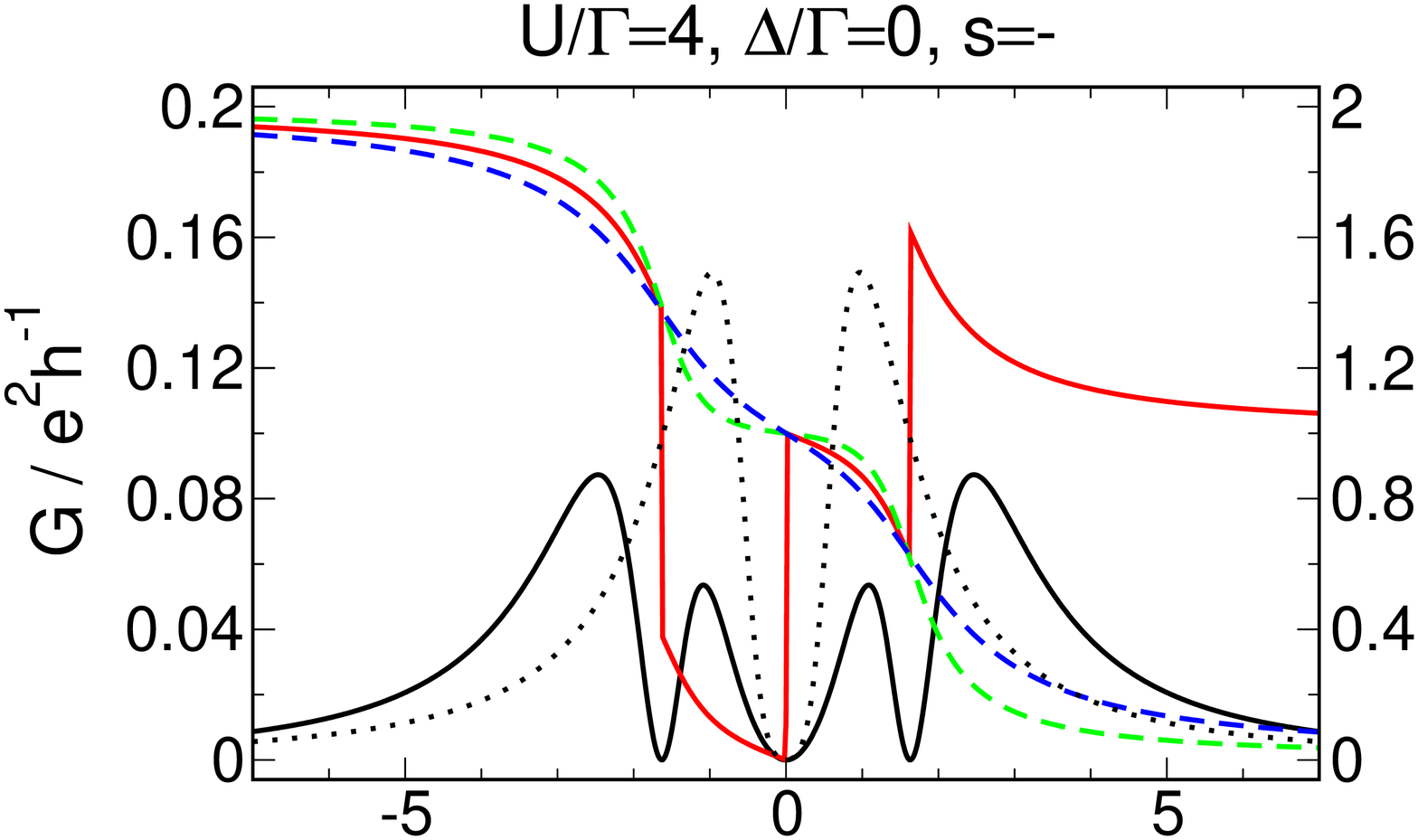}\hspace{0.015\textwidth}
        \includegraphics[width=0.475\textwidth,height=4.4cm,clip]{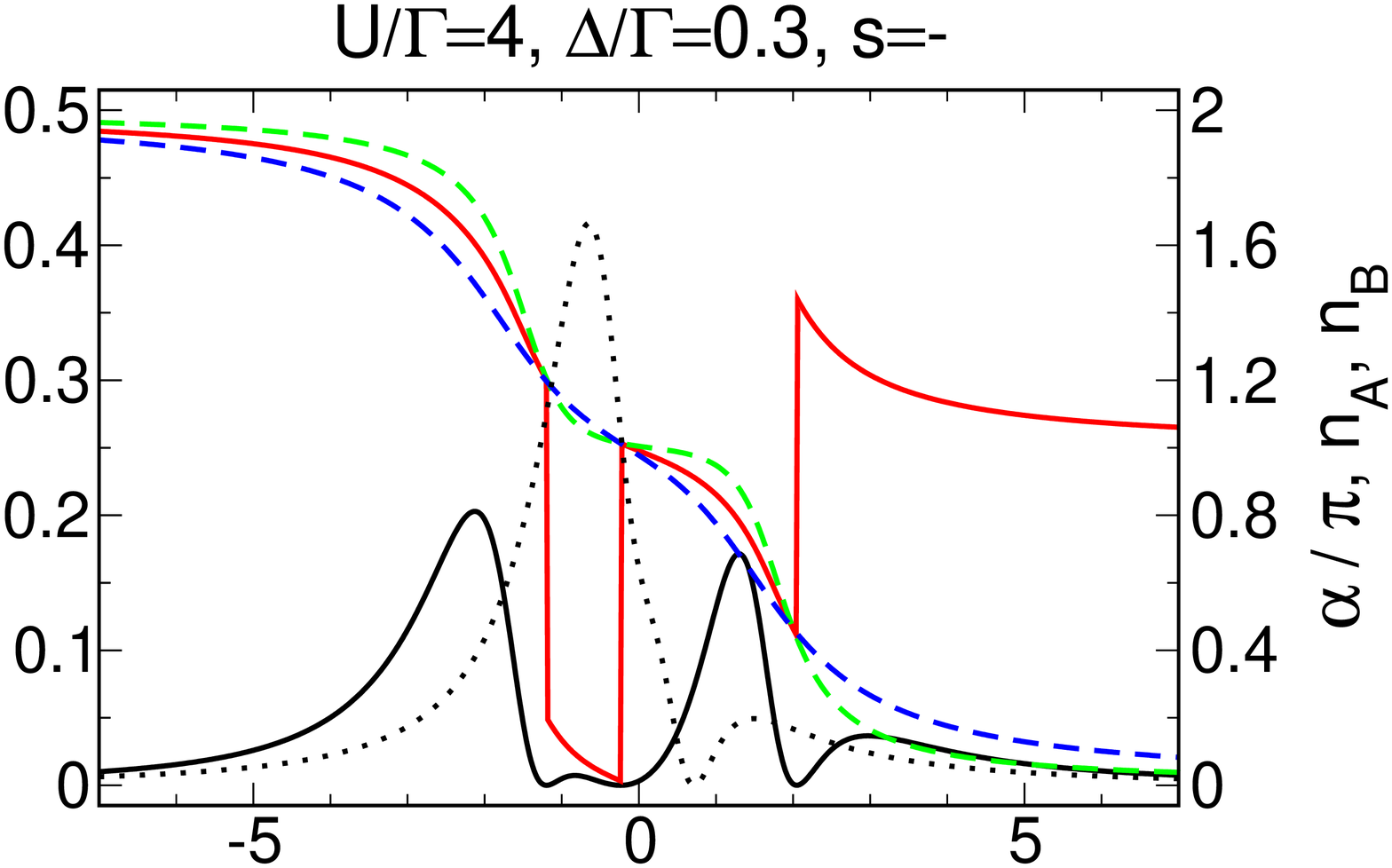}\vspace{0.3cm}
        \includegraphics[width=0.495\textwidth,height=5.2cm,clip]{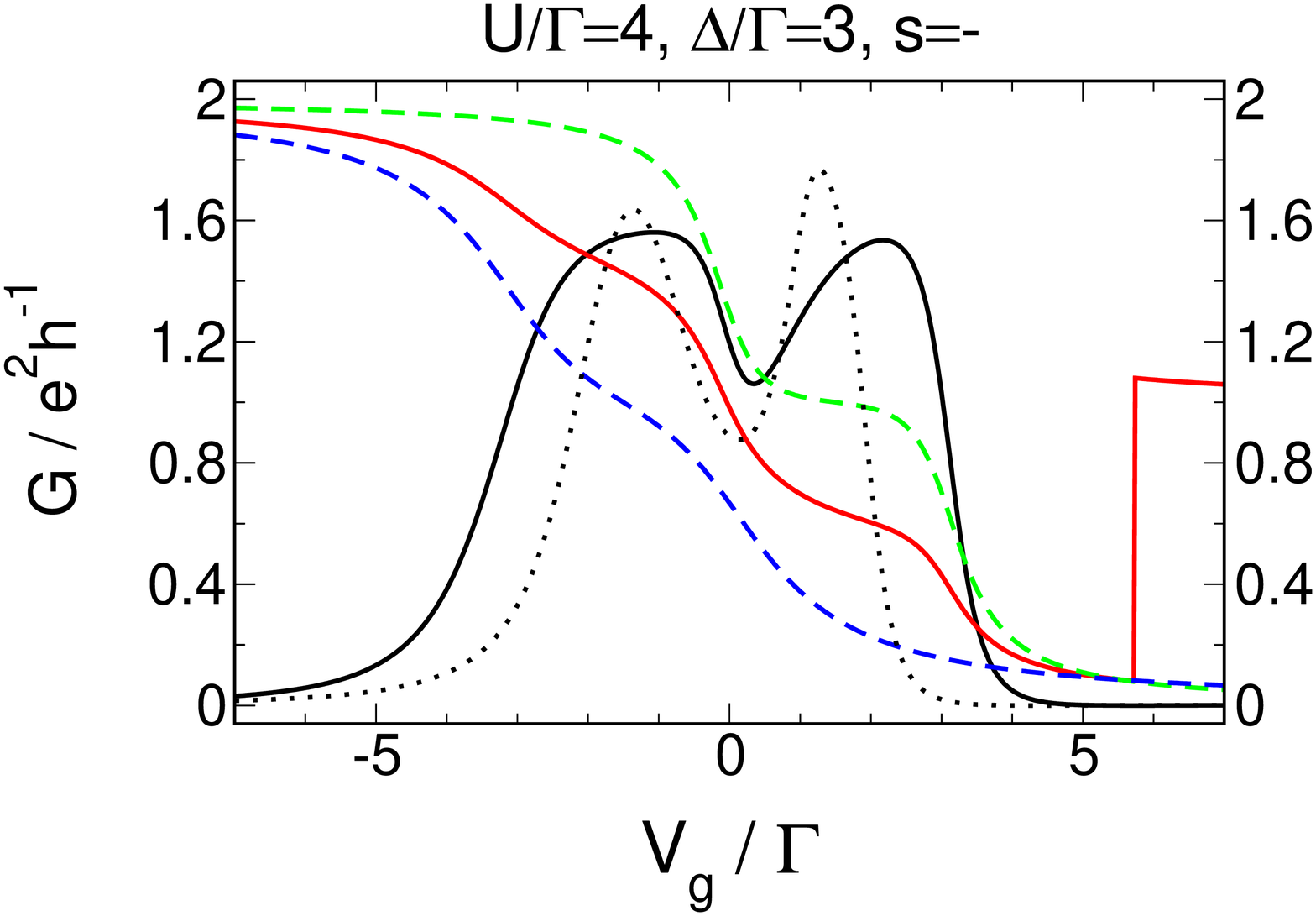}\hspace{0.015\textwidth}
        \includegraphics[width=0.475\textwidth,height=5.2cm,clip]{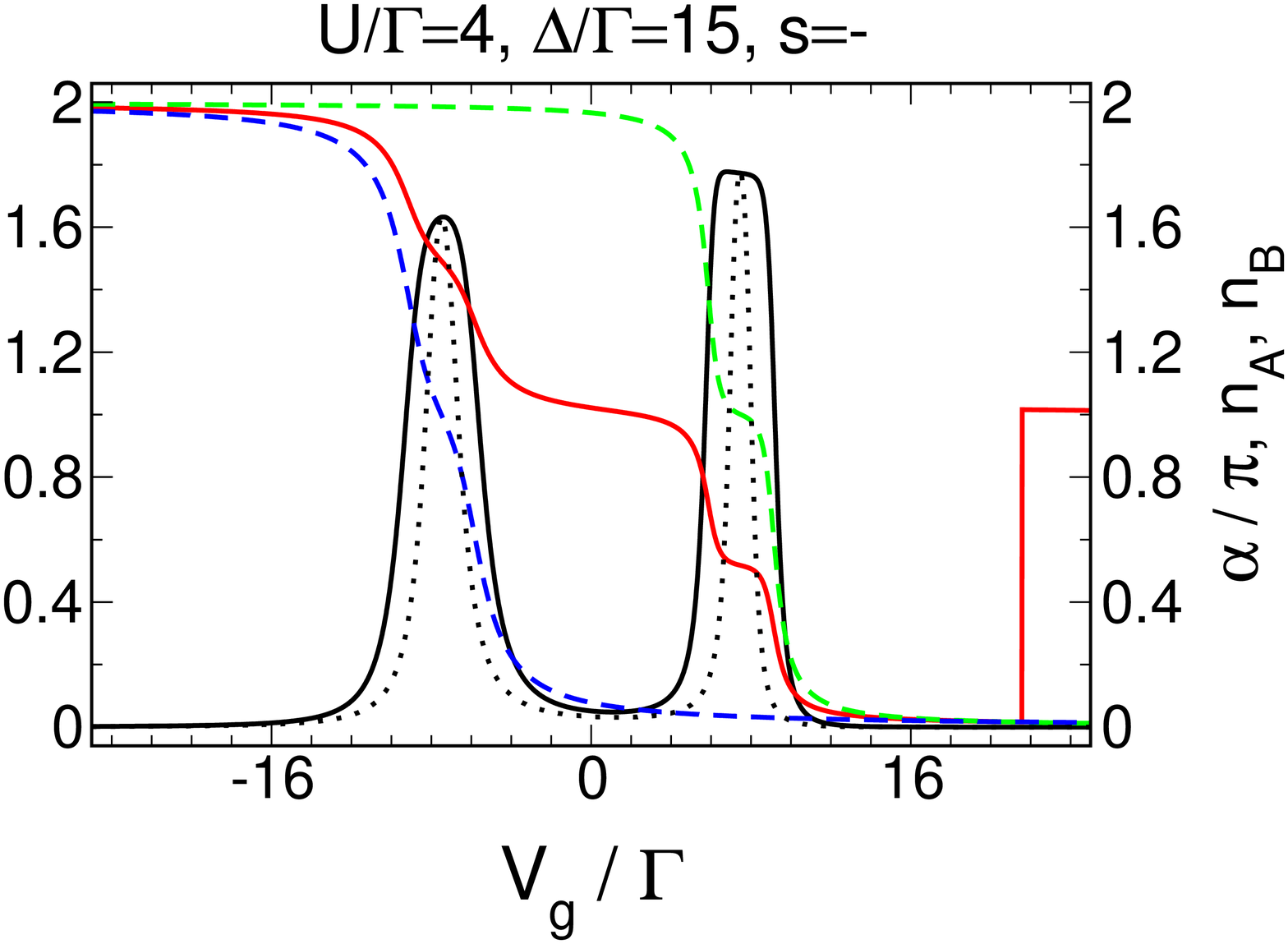}
        \caption{The same as Fig.~\ref{fig:MS.dd.loc_pd}, but for $s=-$ and purely local interactions $U/\Gamma=4.0$. In each plot, the conductance for $U/\Gamma=1.0$ but otherwise identical parameters is shown as well (black dotted lines). The crossover from small to large $\Delta$ is analogous to the spin-polarised case.}
\label{fig:MS.dd.loc_md}
\end{figure}

Relaxing the assumption of left-right symmetric hybridisations leads to a small but finite diagonal entry in the free propagator which stays small during the fRG flow. Hence one would not expect to find qualitatively different behaviour for such more general hybridisations, and this is indeed the case (Fig.~\ref{fig:MS.dd.loc_m}, left panel). The resonances prove to be robust against fairly large additional nearest-neighbour interactions and they only disappear if all interactions in the system are approximately of equal strength (Fig.~\ref{fig:MS.dd.loc_m}, right panel).

The transmission phase $\alpha$ jumps by $\pi$ at each of the three transmission zeros, but its change over each resonance is less than $\pi$. In the left-right symmetric case, the average level occupancies are exactly those of two single-level dots with hybridisations $\Gamma_{A,B}$, and again for arbitrary level-lead couplings the results are qualitatively the same (Fig.~\ref{fig:MS.dd.loc_md}, upper left panel).
\begin{figure}[t]	
	\centering
	      \includegraphics[width=0.475\textwidth,height=4.4cm,clip]{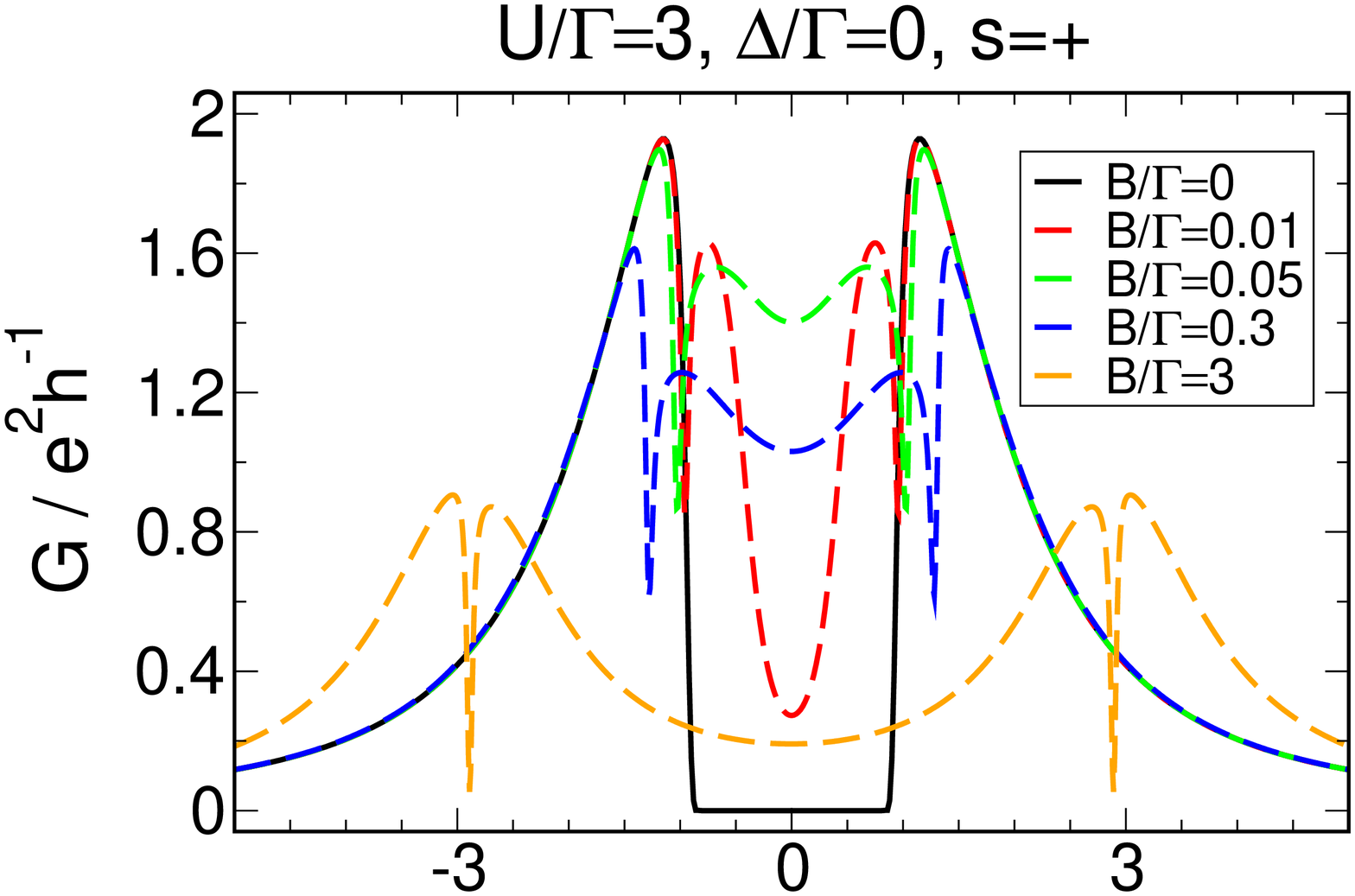}\hspace{0.035\textwidth}
        \includegraphics[width=0.475\textwidth,height=4.4cm,clip]{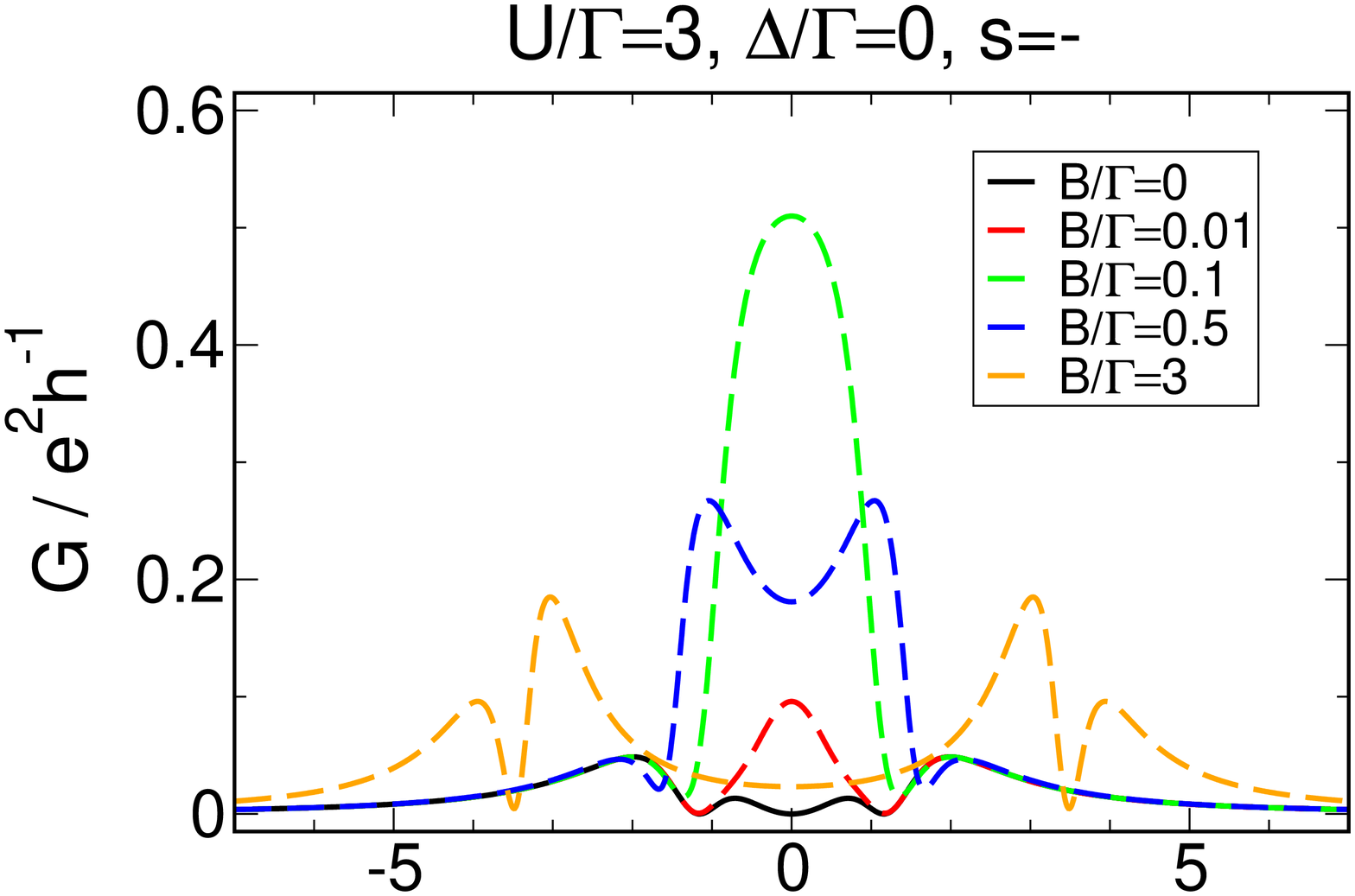}\vspace{0.3cm}
        \includegraphics[width=0.475\textwidth,height=5.2cm,clip]{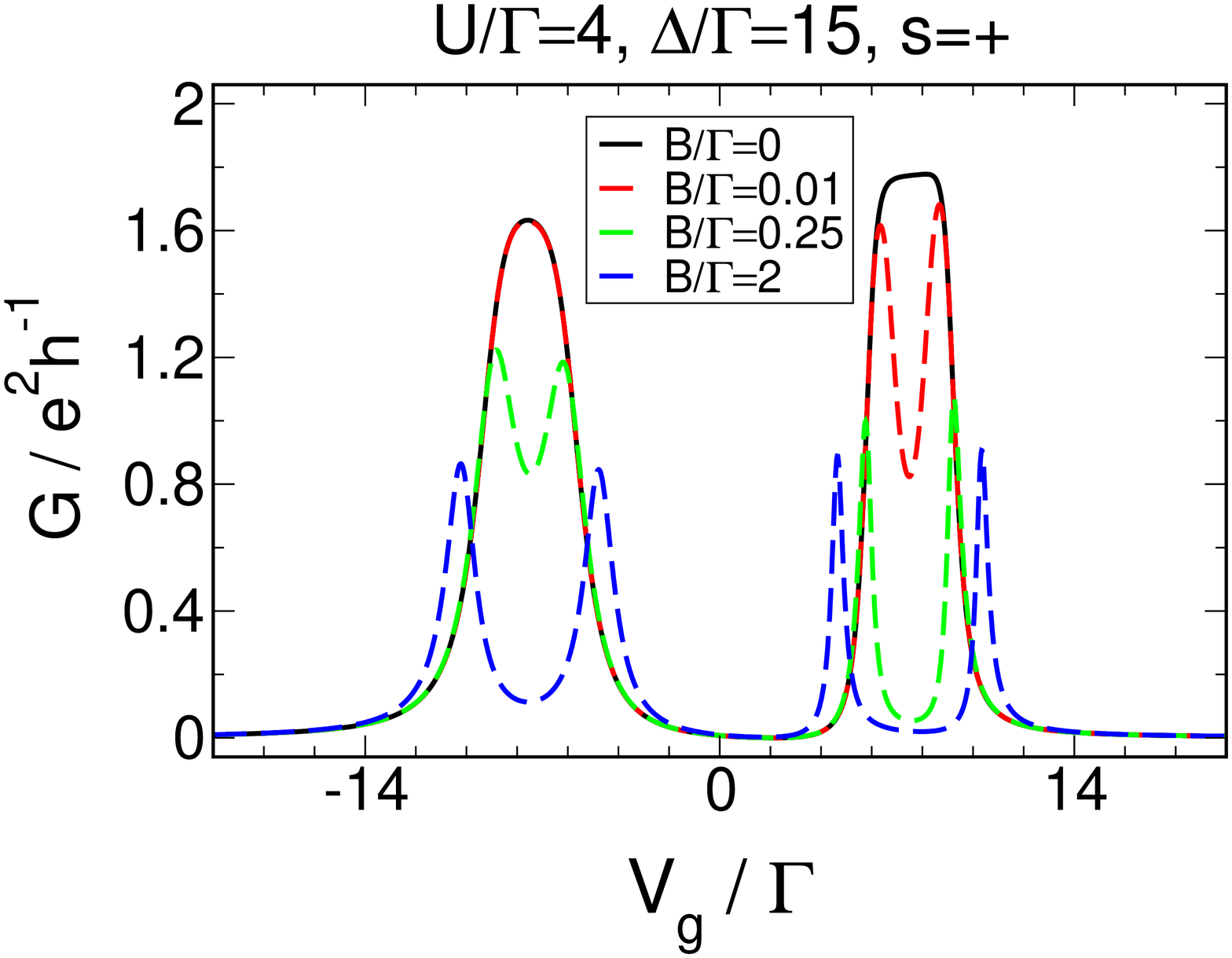}\hspace{0.035\textwidth}
        \includegraphics[width=0.475\textwidth,height=5.2cm,clip]{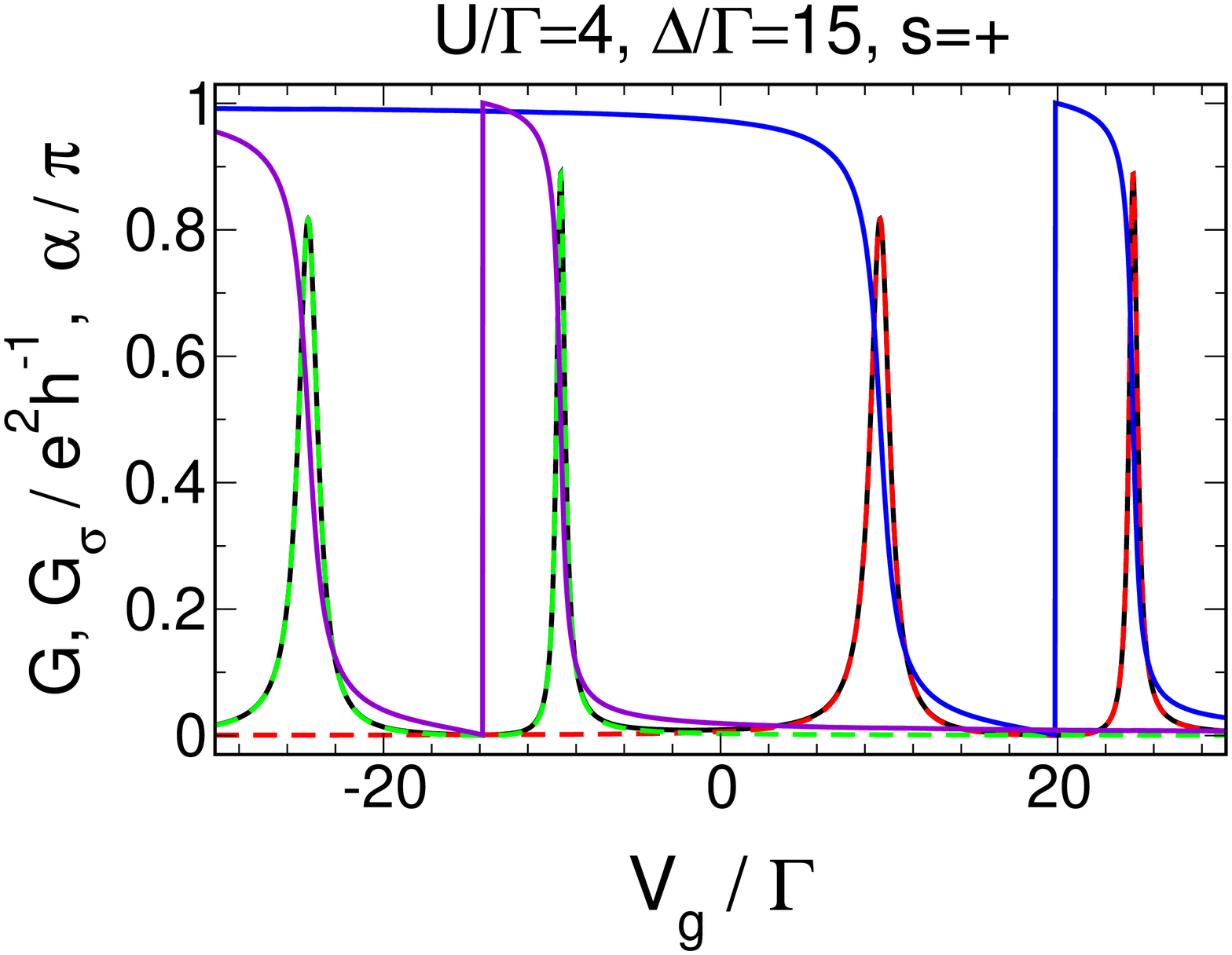}
        \caption{Conductance $G$ as a function of the gate voltage for a parallel double dot with generic level-lead hybridisations $\Gamma=\{0.1~0.2~0.5~0.2\}$ and purely local interactions $U$ for various magnetic fields. \textit{Lower panels:} Note that the Kondo temperatures for the two Kondo resonances determined from the corresponding single-level problems read $T_K^A/\Gamma=0.01$ and $T_K^B/\Gamma=0.25$. The right panel shows $G(V_g)$ for $B/\Gamma=30.0$ (black curve). The partial conductances (red: spin up electrons, green: spin down electrons) and the transmission phases ($\alpha_\uparrow$: blue, $\alpha_\downarrow$: violet) are presented as well.}
\label{fig:MS.dd.loc_b}
\end{figure}

As mentioned above, two of the three transmission zeros present if the interaction is large enough vanish if the level spacing becomes of order $U$. The third zero is shifted to larger $V_g$ and the corresponding resonance that is located at even larger gate voltages becomes vanishingly small. The peak that is moved outwards and gradually disappears is always the one one would associate with the more weakly coupled level (better: it is the right resonance that vanishes if $\Gamma_A$ is more weakly coupled). For further increase of the level spacing, the conductance in the region where originally the other three peaks were located evolves into one wide resonance which splits up and finally we recover two Kondo resonances for $\Delta\gg\Gamma$. This is completely similar to the $s=+$ case except for the fact that there is no transmission zero and corresponding jump by $\pi$ in between the peaks. If $U$ is so small that for degenerate levels only two resonances are observed, one of them splits up while the other is shifted outwards and becomes small if $\Delta$ is increased. One should note that this whole scenario is completely identical to the crossover regime of the spin-polarised double dot studied in Sec.~\ref{sec:OS.dd} (aside from the fact that it starts out with four instead of two resonances for $U$ sufficiently large).

The magnetic field dependence of the $G(V_g)$ curves is similar to the $s=+$ case. For degenerate levels, the conductance close to half-filling first increases and evolves into a maximum when a field is turned on (Fig.~\ref{fig:MS.dd.loc_b}, upper left panel). For left-right symmetric hybridisations, this can again be understood from the interfering single-levels dots. Since there the conductance at half filling is suppressed at an energy scale that exponentially depends on the strength of $\Gamma_A$ and $\Gamma_B$, one of the single-level $G(V_g)$ curves will fall off much earlier than the other one when a magnetic field is turned on. If $B$ exceeds a certain strength, the conductance at $V_g=0$ falls off again and for large $B$ we recover twice the spin-polarised, i.e. the noninteracting, structure. For equal strength of the interaction, the scales on which the conductance first increases and then begins to decrease again are of the same order of magnitude for both the $s=+$ and $s=-$ case (compare the upper panels in Fig.~\ref{fig:MS.dd.loc_b}). This suggests a small energy scale, or, in other words, a large effective interaction within the system. This is surprising since one would expect a large effective interaction to be recognised by a breakdown of the fRG scheme for fairly small $U$ compared to the `real' energy scales of the system. However, for $s=-$ no such breakdown was observed (see the next chapter for the clarification of the word `breakdown' in cases where no reference data known to be precise is available).

\subsubsection{Interactions Between All Electrons, s=+}
\begin{figure}[t]	
	\centering
        \includegraphics[width=0.475\textwidth,height=5.2cm,clip]{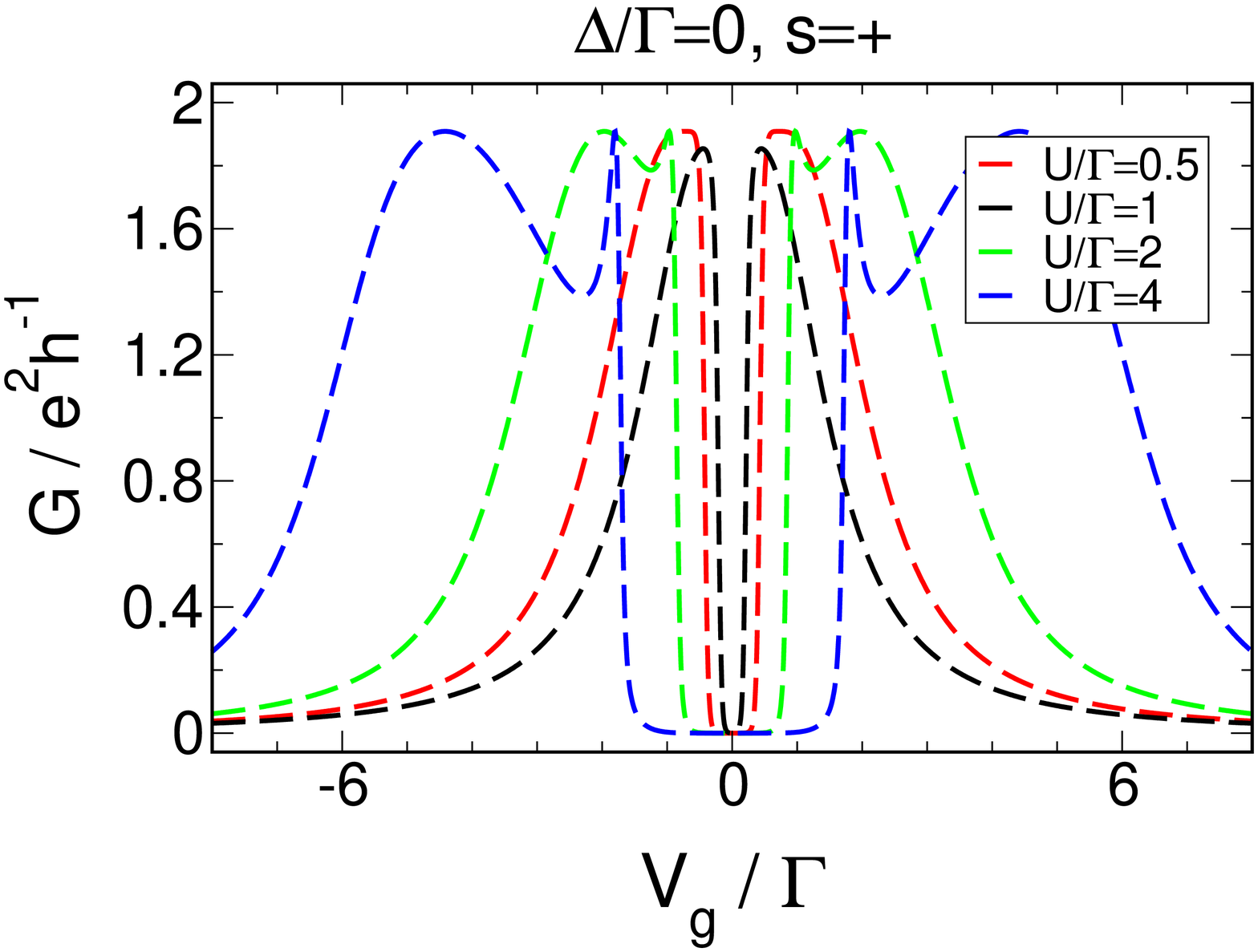}\hspace{0.035\textwidth}
        \includegraphics[width=0.475\textwidth,height=5.2cm,clip]{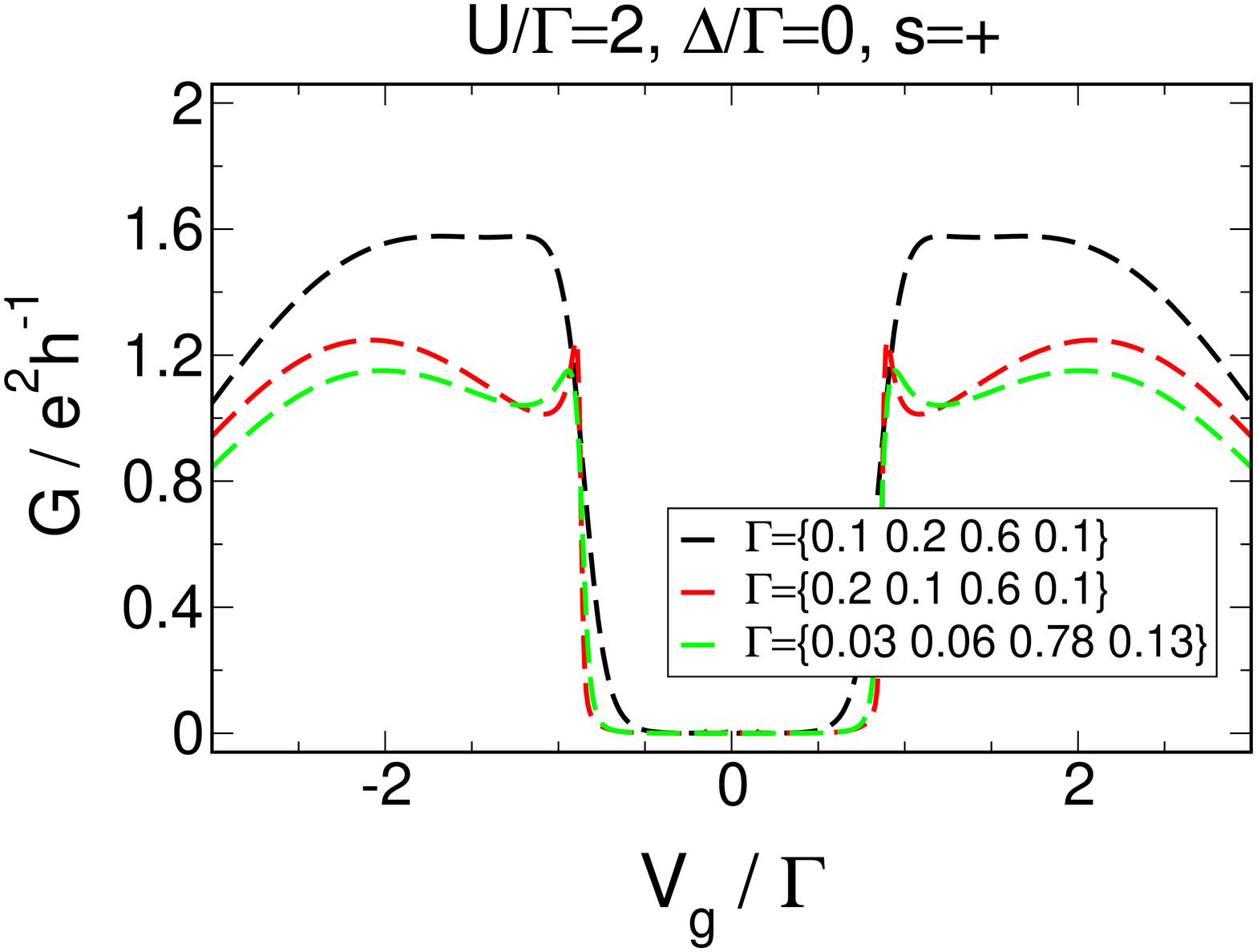}
        \caption{The appearance of the novel resonances for a parallel double dot with equal local and nearest-neighbour interactions $U$, degenerate levels, and $s=+$ in dependence of $U$ for fixed level-lead hybridisations $\Gamma=\{0.1~0.2~0.5~0.2\}$ (left panel) and in dependence of $\Gamma$ for fixed $U$ (right panel).}
\label{fig:MS.dd.all_pu}
\end{figure}

\begin{figure}[t]	
	\centering
        \includegraphics[width=0.475\textwidth,height=5.2cm,clip]{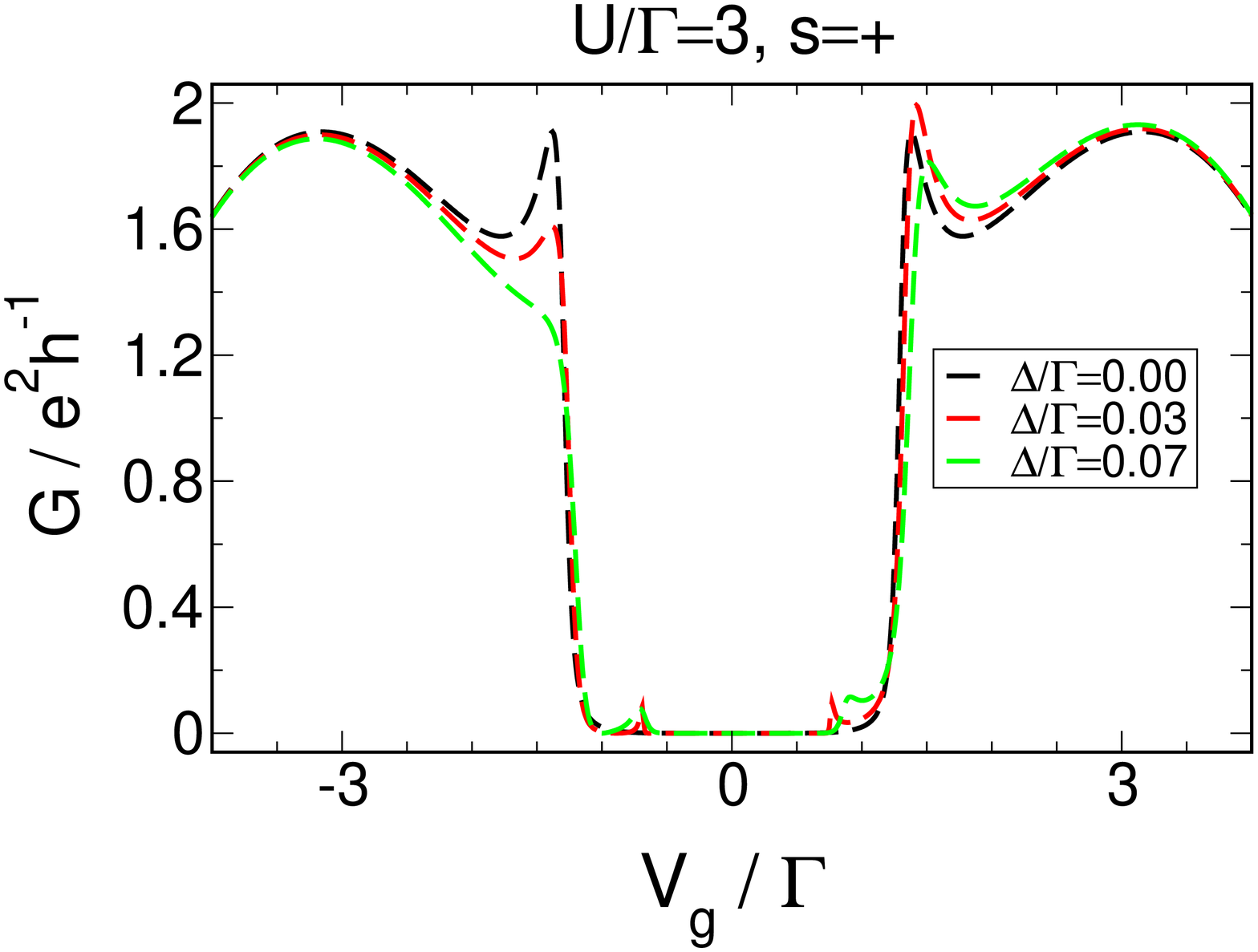}\hspace{0.035\textwidth}
        \includegraphics[width=0.475\textwidth,height=5.2cm,clip]{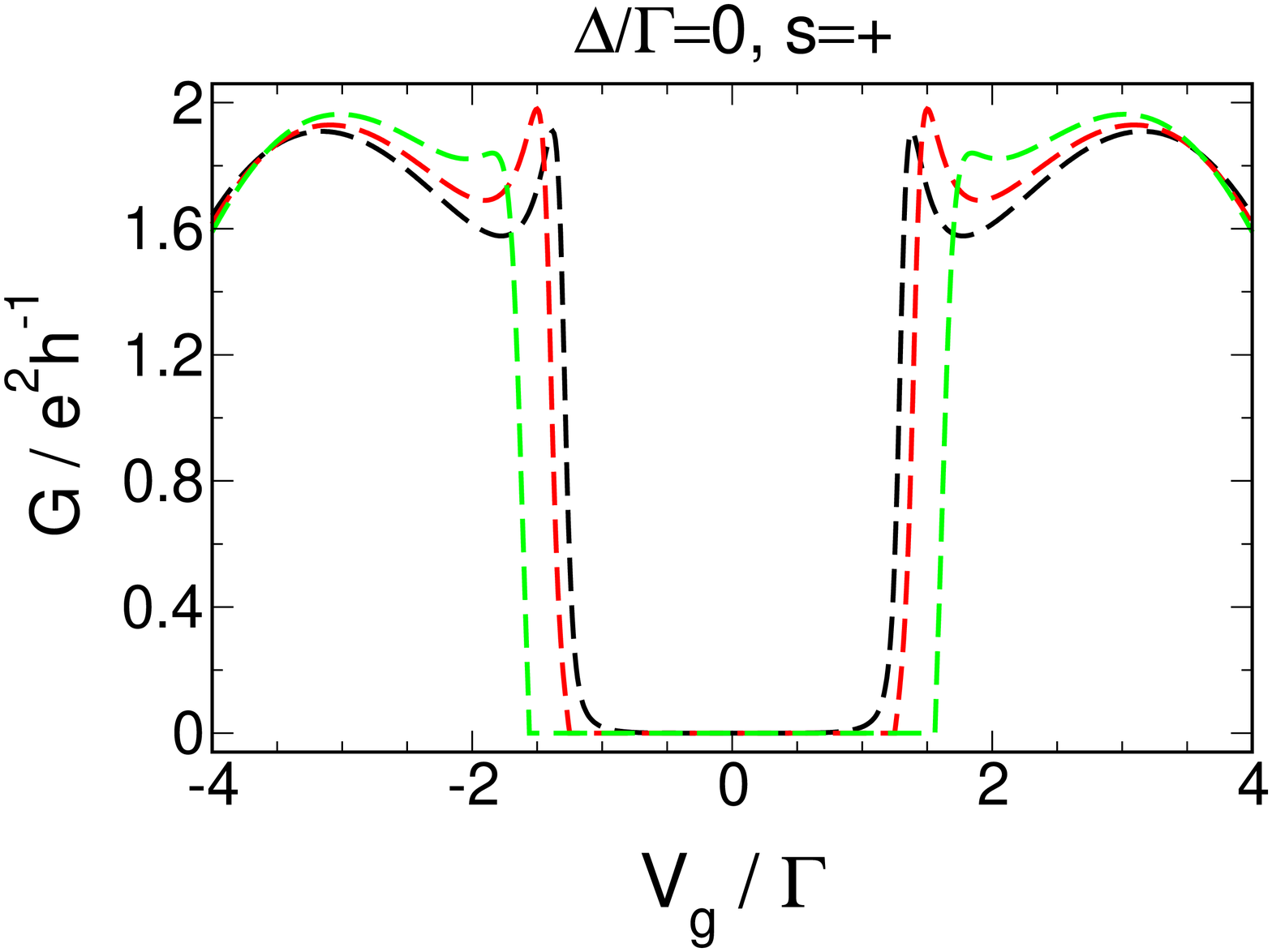}
        \caption{Stability of the new resonances against small level detunings (left panel) and asymmetric interactions (right panel). The hybridisations read $\Gamma=\{0.1~0.2~0.5~0.2\}$. In the left panel, the features of $G(V_g)$ in the valley are artifacts of the fRG approximation scheme (see next chapter). The parameters of the right panel are $U_{A,A}^{\sigma,\bar\sigma}/\Gamma=U_{B,B}^{\sigma,\bar\sigma}/\Gamma=U_{l,\bar l}^{\sigma,\sigma}/\Gamma=U_{l,\bar l}^{\sigma,\bar\sigma}/\Gamma=3.0$ (back), $U_{A,A}^{\sigma,\bar\sigma}/\Gamma=3.1$, $U_{B,B}^{\sigma,\bar\sigma}/\Gamma=2.9$, $U_{l,\bar l}^{\sigma,\sigma}/\Gamma=2.9$, $U_{l,\bar l}^{\sigma,\bar\sigma}/\Gamma=3.0$ (red) and $U_{A,A}^{\sigma,\bar\sigma}/\Gamma=3.2$, $U_{B,B}^{\sigma,\bar\sigma}/\Gamma=2.6$, $U_{l,\bar l}^{\sigma,\sigma}/\Gamma=2.7$, $U_{l,\bar l}^{\sigma,\bar\sigma}/\Gamma=3.2$ (green).}
\label{fig:MS.dd.all_ps}
\end{figure}

Next, we study the case where we have equal local and nearest neighbour interactions $U$ within the system. This should be an appropriate model for a single dot containing two levels. The easiest case is that with relative sign of the level-lead couplings $s=+$ since for $A$-$B$-symmetric hybridisations it can be mapped onto the side-coupled geometry with equal local and nearest neighbour interactions. For small level detunings $\Delta\ll\Gamma$ close to half-filling we find the well-known valley of width $U$ of suppressed transport associated with the second stage Kondo effect. For degenerate levels, by increasing the interaction strength the wings of the conductance peaks become flat and at a certain critical $U_c$ depending on the hybridisations two additional peaks split off so that we find a total of four transmission resonances located at $V_g\approx\pm U/2$ and $V_g\approx\pm U$. The additional peaks close to the valley of suppressed conductance become increasingly sharp when the interaction gets stronger. This whole scenario is quite similar to the appearance of additional resonances (the CIRs) within the spin-polarised model discussed in Sec.~\ref{sec:OS.dd}. The conductance valley at half-filling and the evolution of the four-peak structure when increasing the interaction strength is depicted in the left panel of Fig.~\ref{fig:MS.dd.all_pu}. The critical $U_c$ only depends on the difference between the left-right asymmetries of the different dots and on the bare $A$-$B$-asymmetry. The larger the former and the smaller the latter, the larger is $U_c$ (\ref{fig:MS.dd.all_pu}, right panel). In general, it was always found to be of the order of magnitude of $\Gamma$ for all systems considered. The additional resonances are robust against a very small level detuning $\Delta\ll\Gamma$ and against asymmetric interactions (Fig.~\ref{fig:MS.dd.all_ps}). The former explains why they were not observed when considering the side-coupled geometry since there we only focused on $t\lesssim\Gamma$.

\begin{figure}[t]	
	\centering
	      \includegraphics[width=0.495\textwidth,height=4.4cm,clip]{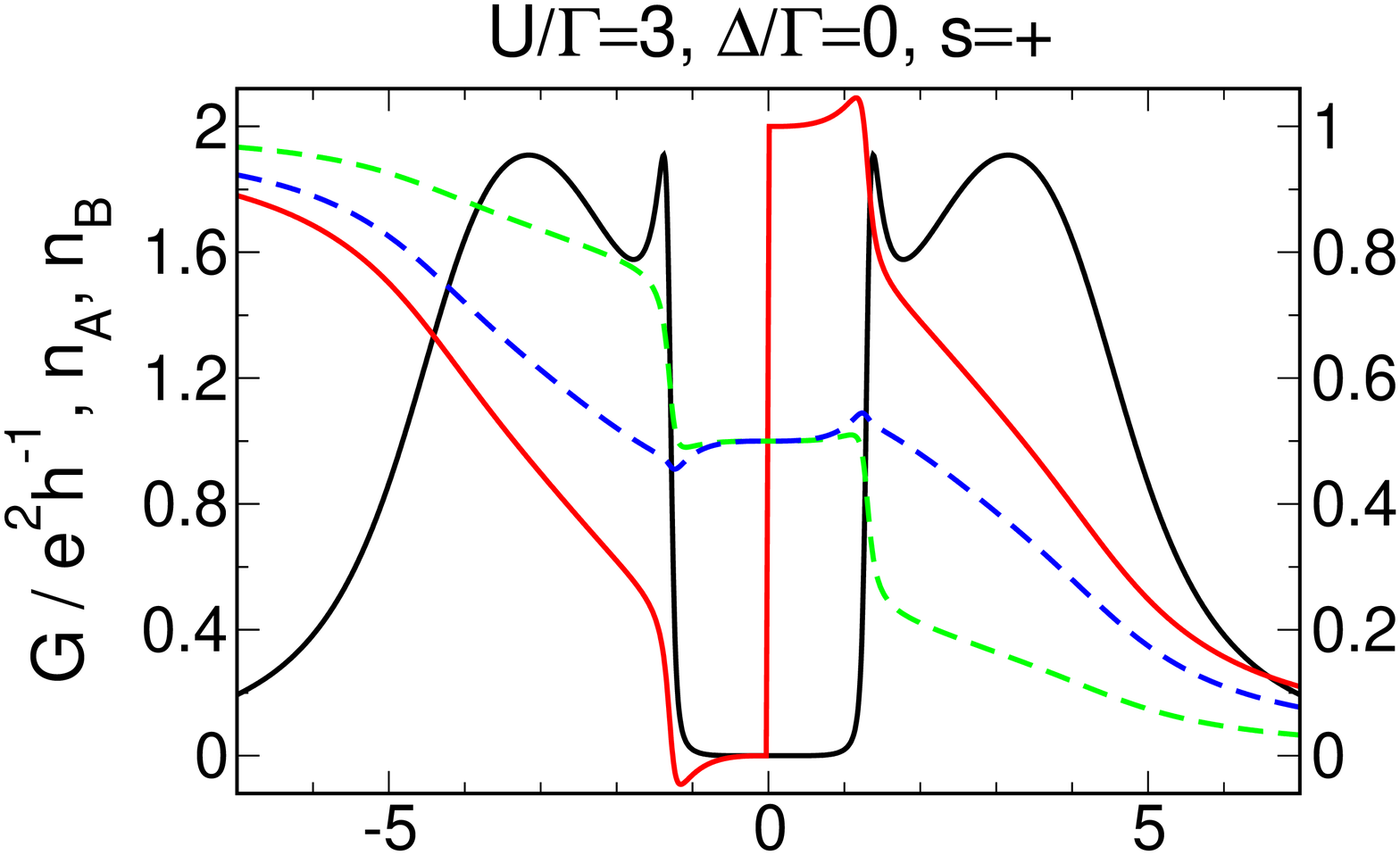}\hspace{0.015\textwidth}
        \includegraphics[width=0.475\textwidth,height=4.4cm,clip]{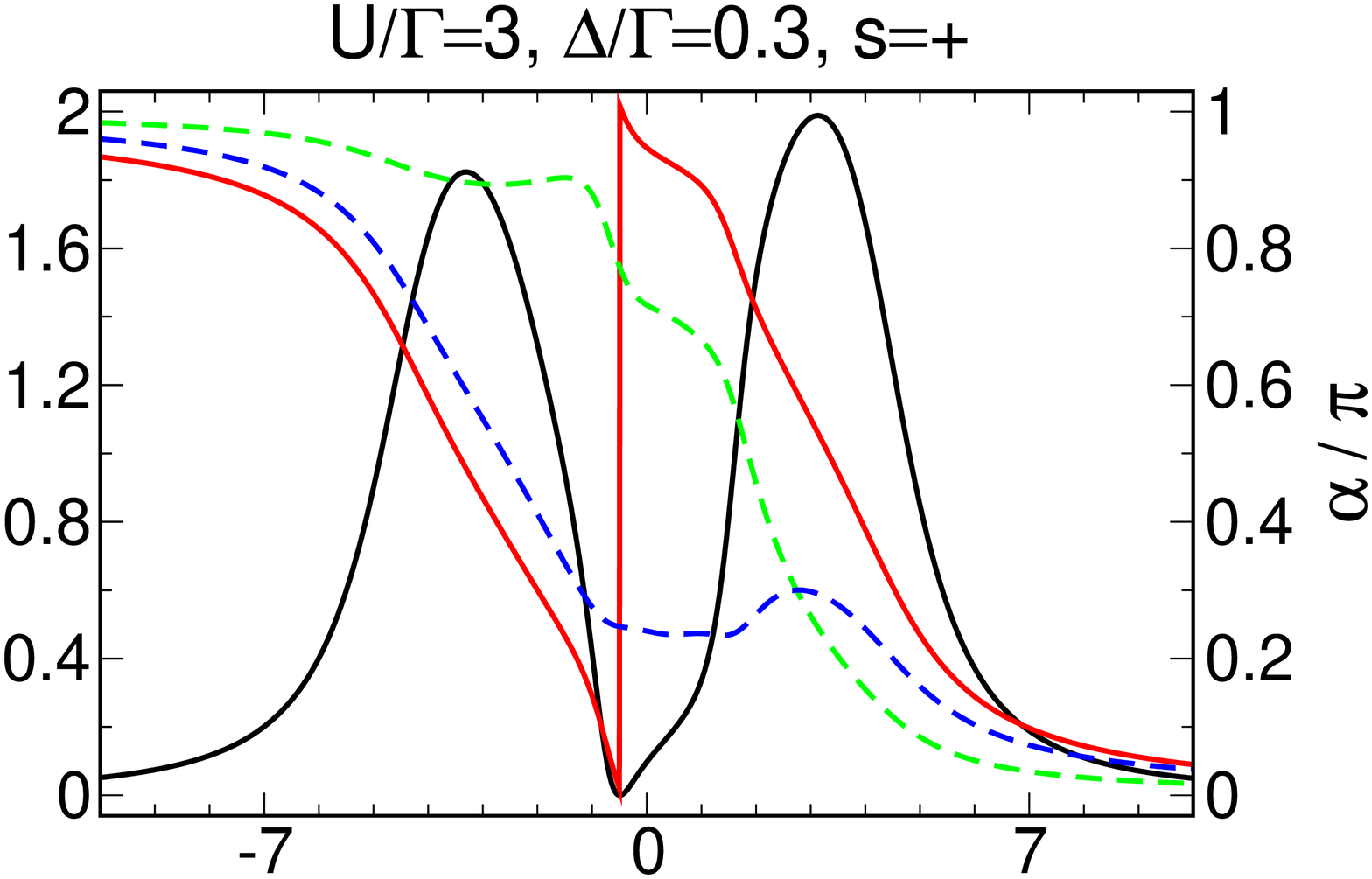}\vspace{0.3cm}
        \includegraphics[width=0.495\textwidth,height=5.2cm,clip]{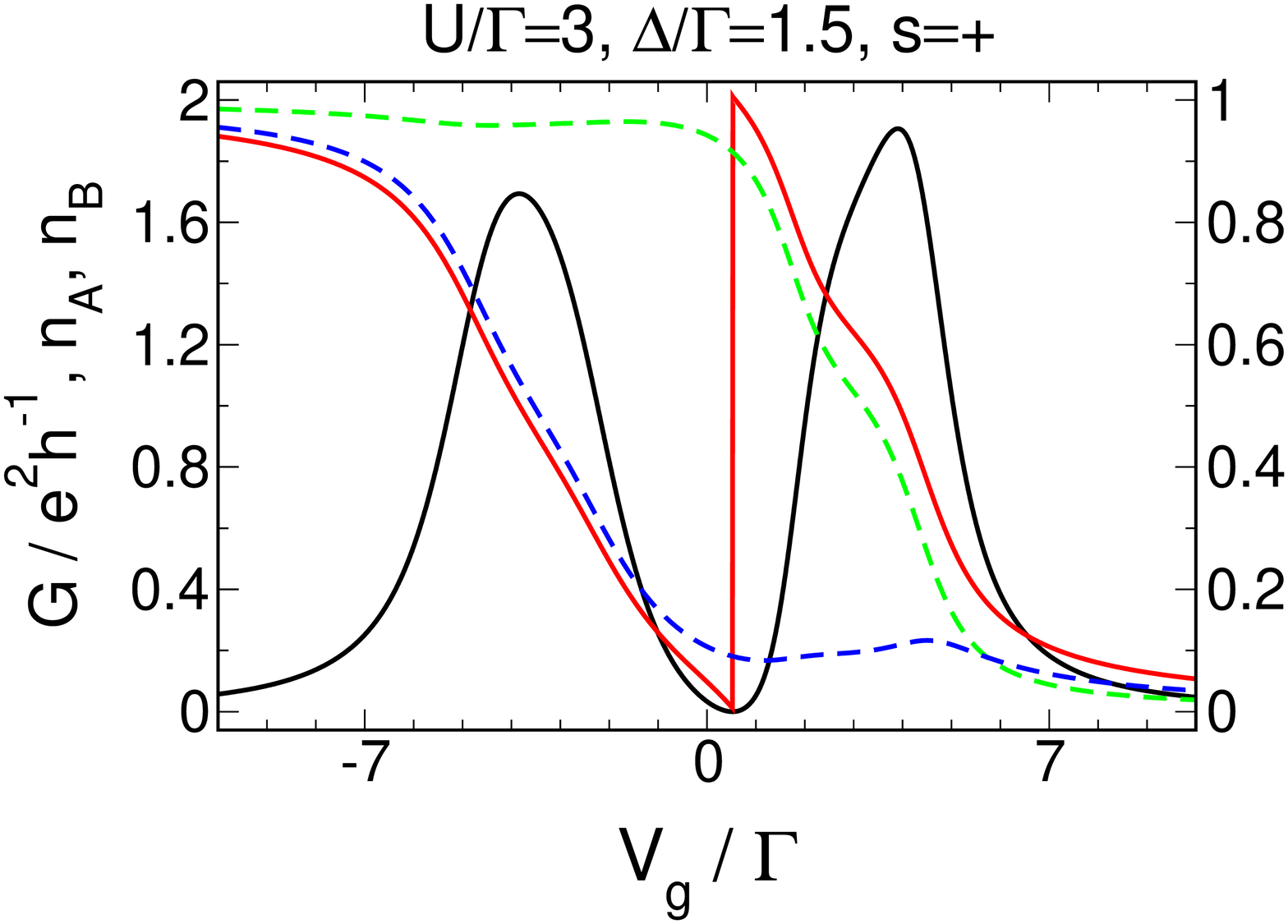}\hspace{0.015\textwidth}
        \includegraphics[width=0.475\textwidth,height=5.2cm,clip]{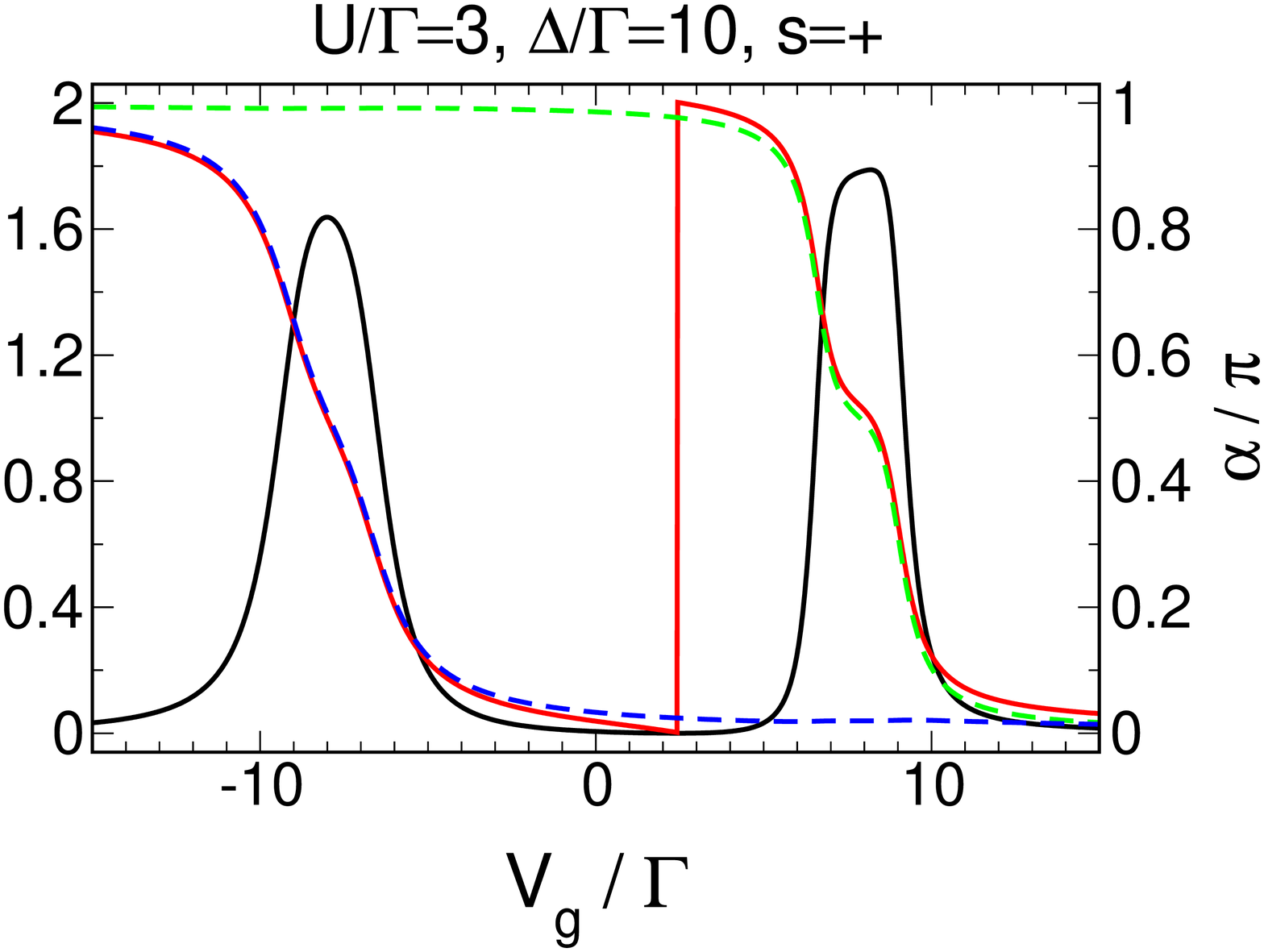}
        \caption{The same as Fig.~\ref{fig:MS.dd.loc_pd}, but with equal local and nearest-neighbour interactions.}
\label{fig:MS.dd.all_pd}\vspace{0.6cm}
	      \includegraphics[width=0.495\textwidth,height=4.4cm,clip]{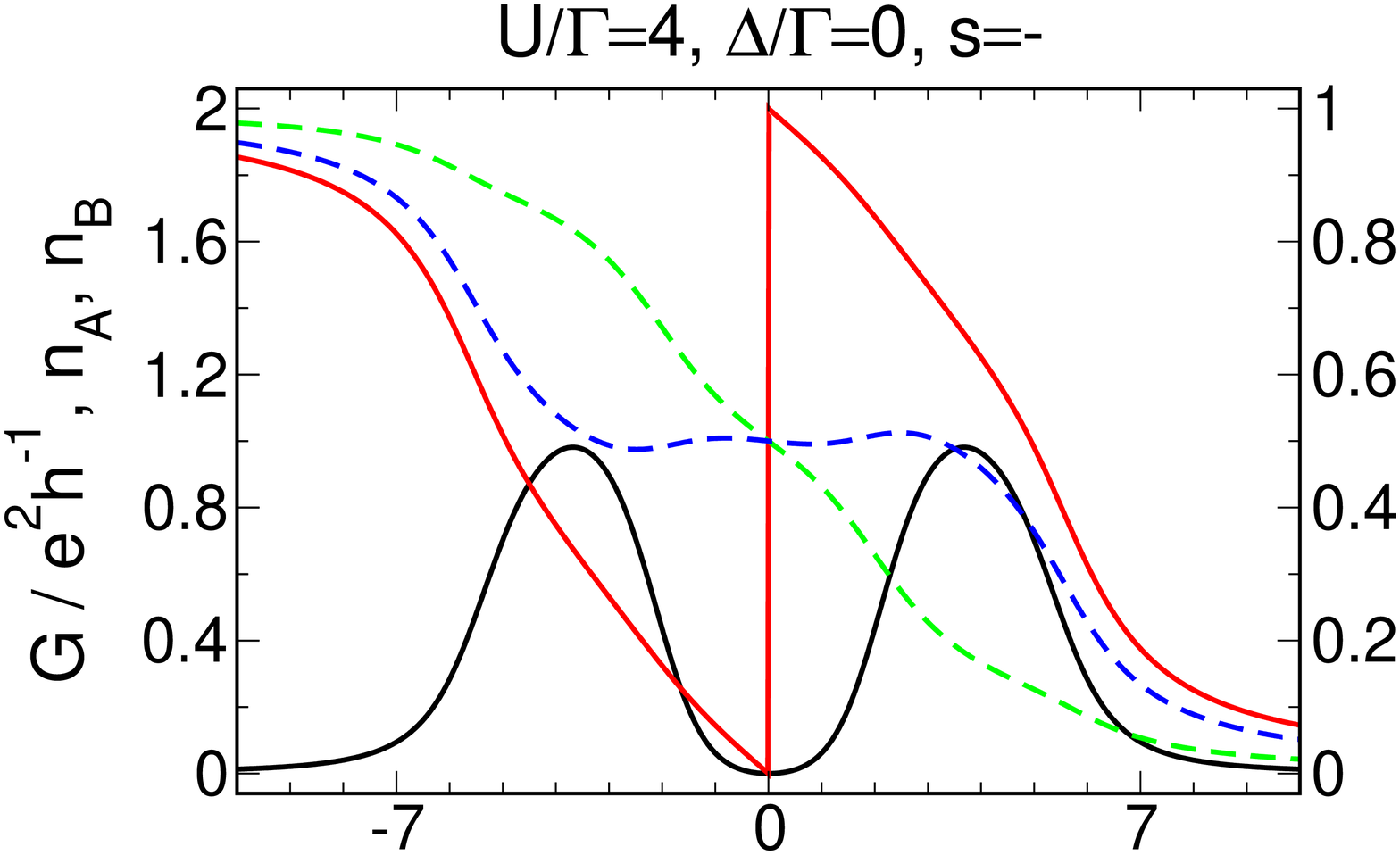}\hspace{0.015\textwidth}
        \includegraphics[width=0.475\textwidth,height=4.4cm,clip]{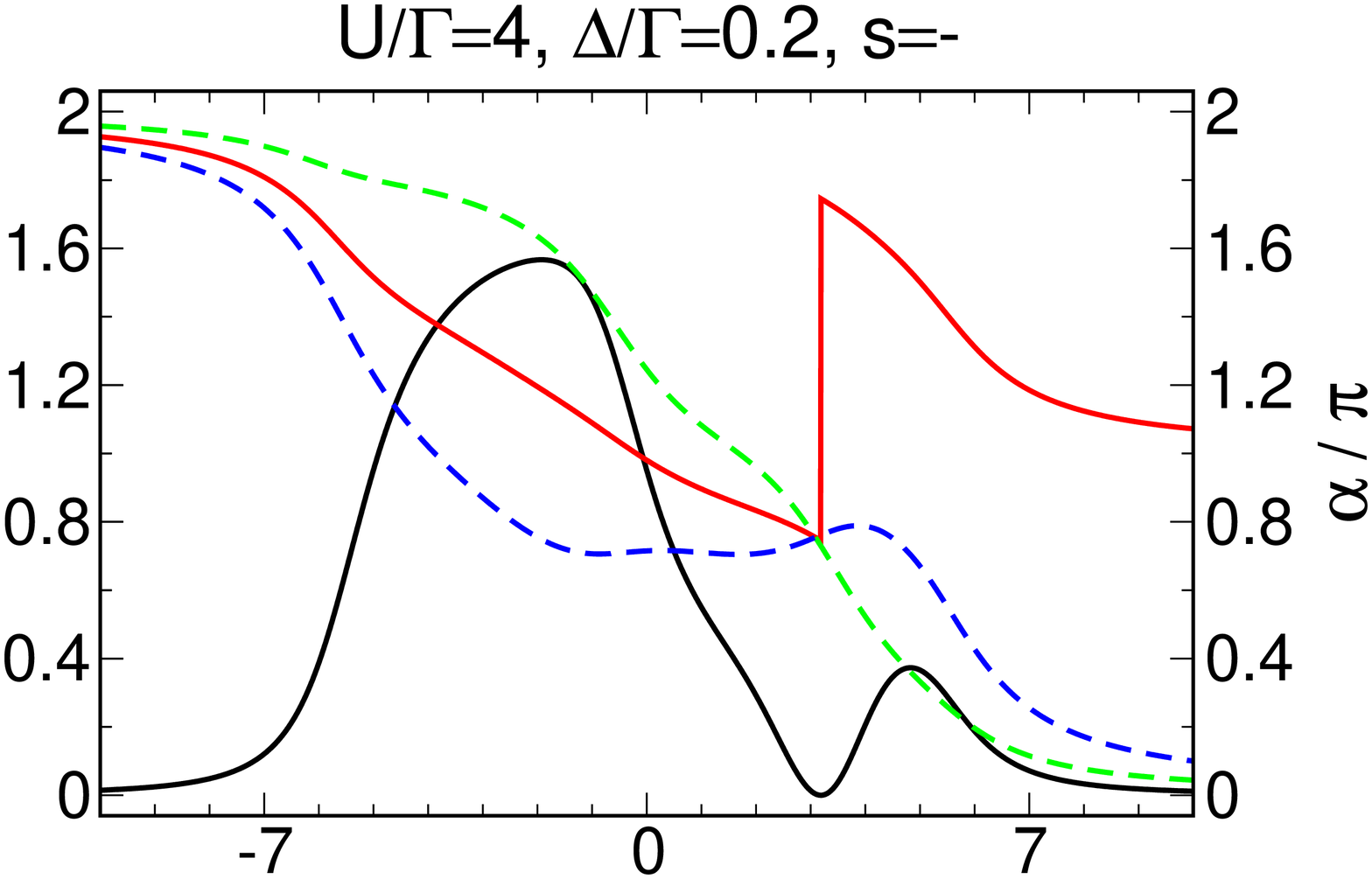}\vspace{0.3cm}
        \includegraphics[width=0.495\textwidth,height=5.2cm,clip]{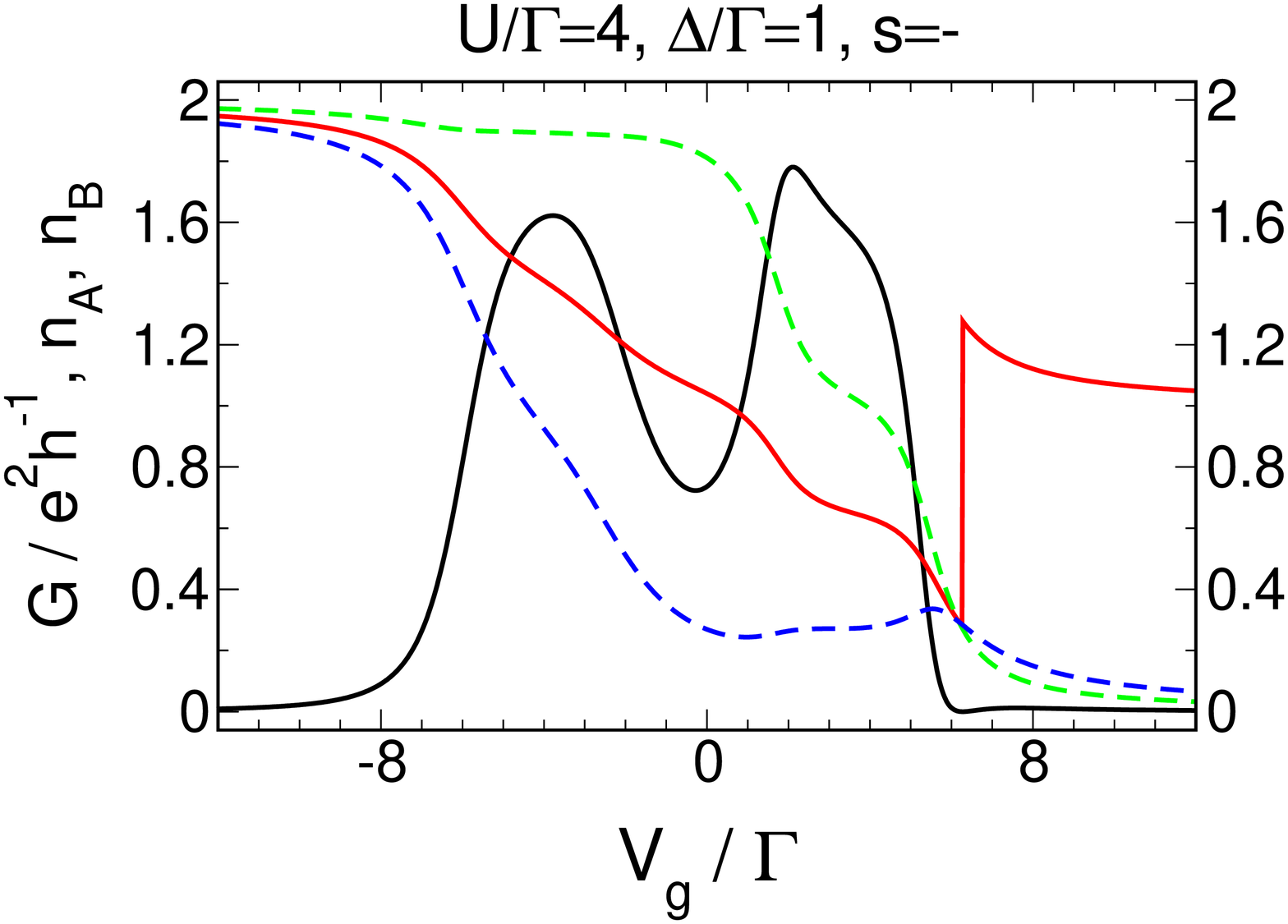}\hspace{0.015\textwidth}
        \includegraphics[width=0.475\textwidth,height=5.2cm,clip]{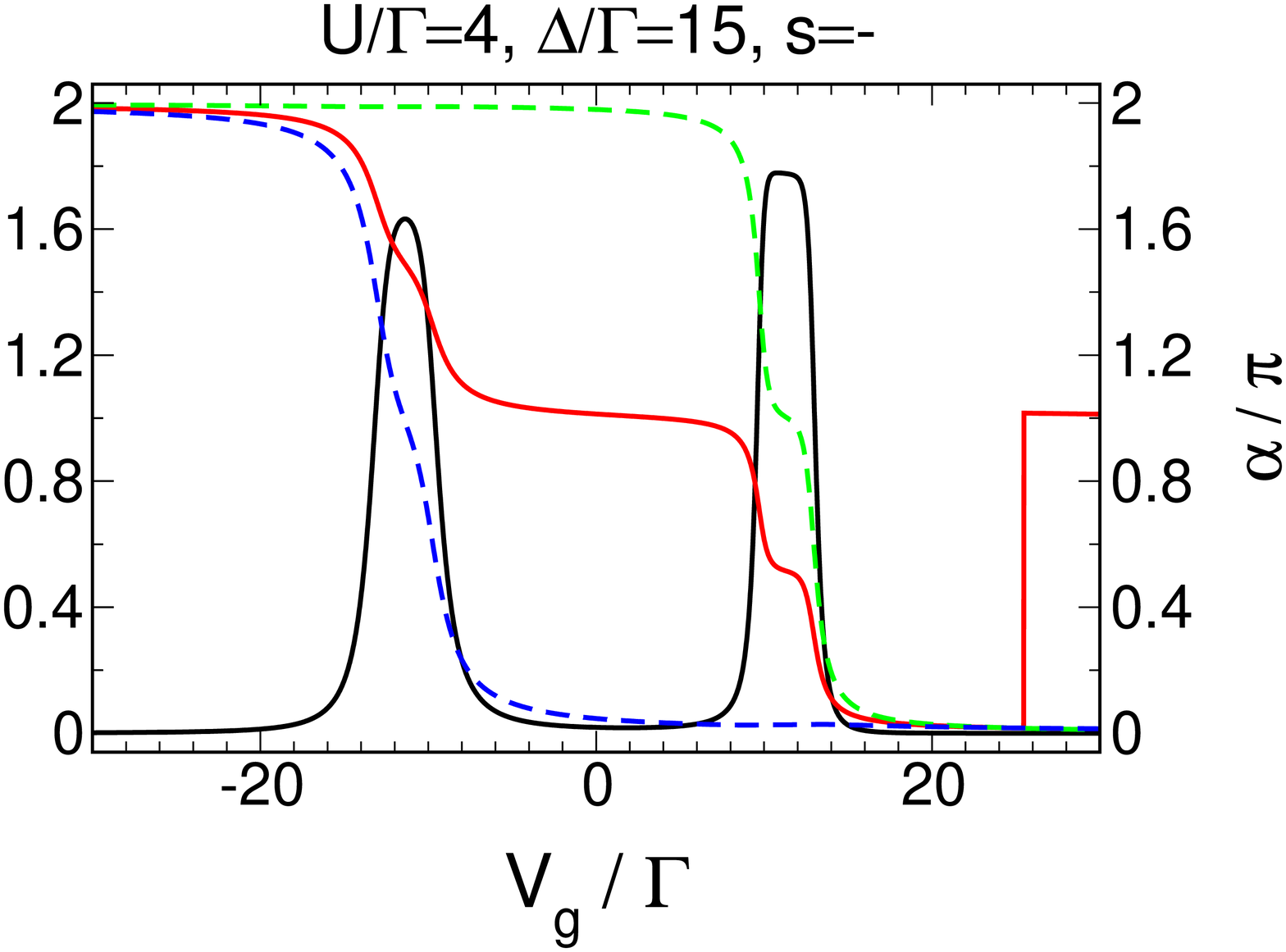}
        \caption{The same as Fig.~\ref{fig:MS.dd.loc_md}, but for equal local and nearest-neighbour interactions. The crossover from $\Delta\ll\Gamma$ to $\Delta\gg\Gamma$ is similar to the spin-polarised case.}
\label{fig:MS.dd.all_md}
\end{figure}

For $\Delta\ll\Gamma$, from large $|V_g|$ to the valley the transmission phase changes by $\pi$ over the resonances and jumps by $\pi$ close to $V_g=0$. If present, $\alpha$ evolves steeply when crossing one of the additional structures located at $V_g\approx\pm U/2$ which is again similar to the CIRs observed for the spin-polarised case. The average level occupancies both fall off with a slope determined by the corresponding $\Gamma_l$ when increasing $V_g\to -U/2$ and merge into a plateau of width $U$ with indications of a nonmonotonic dependence on the gate voltage. The details are shown in the upper left panel of Fig.~\ref{fig:MS.dd.all_pd}.

\afterpage{\clearpage}

Increasing the level spacing, only the outer resonances remain and their separation becomes larger, in complete analogy to the case where only local interactions are present. For $\Delta\gg\Gamma$ we recover the two Kondo resonances associated with the individual levels. Their position can be estimated by diagonalising the dot system decoupled from the leads. In this case, the lowest energies of states with one, two, three and four electrons read $V_g-\Delta/2-3U/2$, $2V_g-\Delta-2U$, $3V_g-\Delta/2-3U/2$, and $4V_g$, and hence at zero temperature the $V_g$-regions where the dot is occupied by one (three) electrons are given by $[(\Delta+U)/2,(\Delta+3U)/2]$ ($[-(\Delta+3U)/2,-(\Delta+U)/2]$). Therefore we expect two conductance peaks of width $U$ separated by $\Delta+2U$. The evolution from small to large $\Delta$ is shown in Fig.~\ref{fig:MS.dd.all_pd}.

The magnetic field dependence of the curves $G(V_g)$ is identical to the case with local correlations. The conductance at half filling first increases for fields much smaller than in the case considered before, which seems plausible because of the additional interaction terms. Further increase of $B$ destroys the first stage of the Kondo effect and $G$ decreases again. For large fields, we finally recover twice the spin-polarised structure in the lineshape of the conductance. In particular, we observe the correlation induced resonances in quantitative agreement with the spinless model if the interaction is large enough (Fig.~\ref{fig:MS.dd.all_b}). For large level spacings, the magnetic field dependence of the conductance is identical to the case where only local correlations are present.

\subsubsection{Interactions Between All Electrons, $s=-$}

\begin{figure}[t]	
	\centering
	      \includegraphics[width=0.475\textwidth,height=4.4cm,clip]{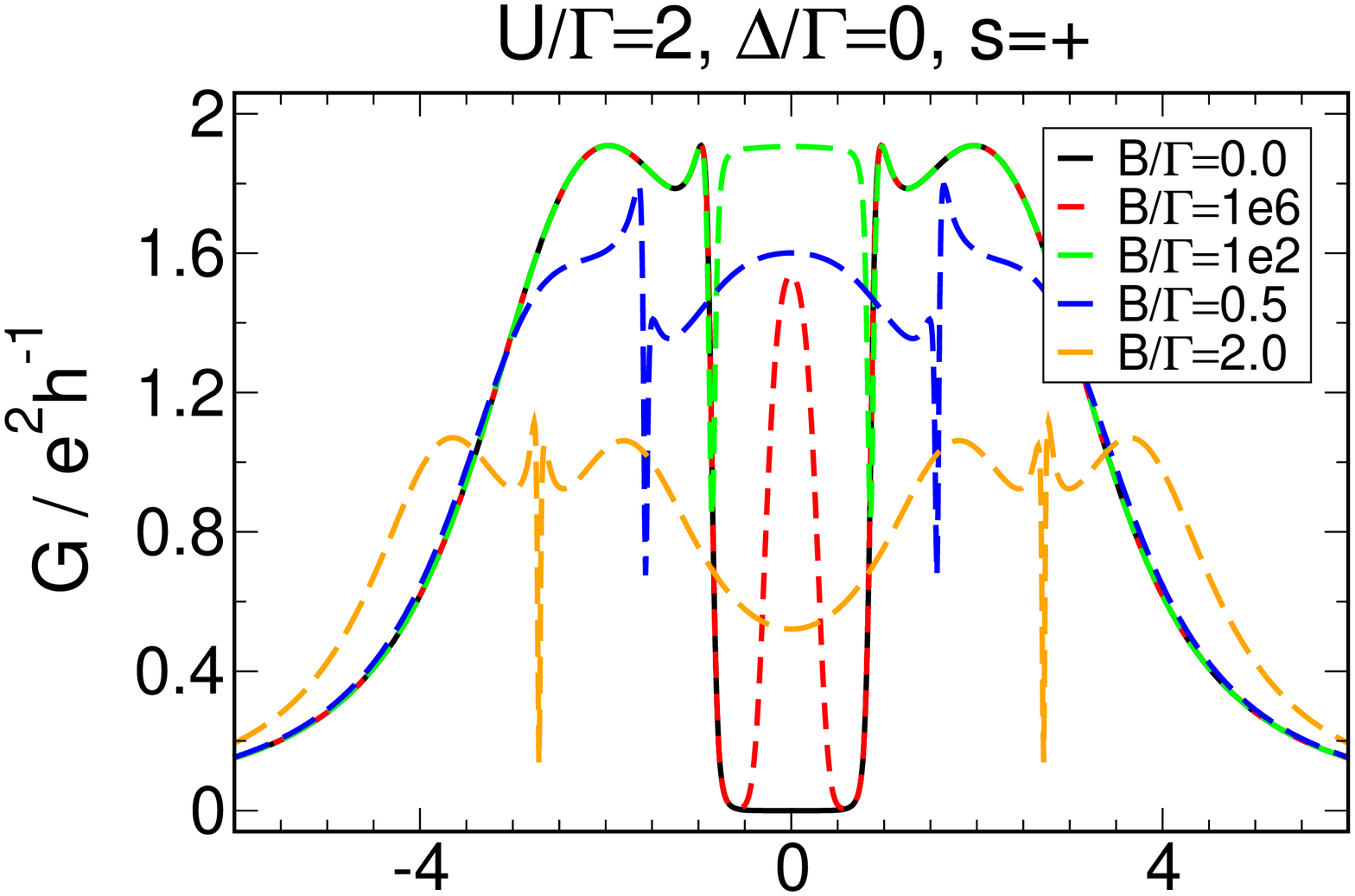}\hspace{0.035\textwidth}
        \includegraphics[width=0.475\textwidth,height=4.4cm,clip]{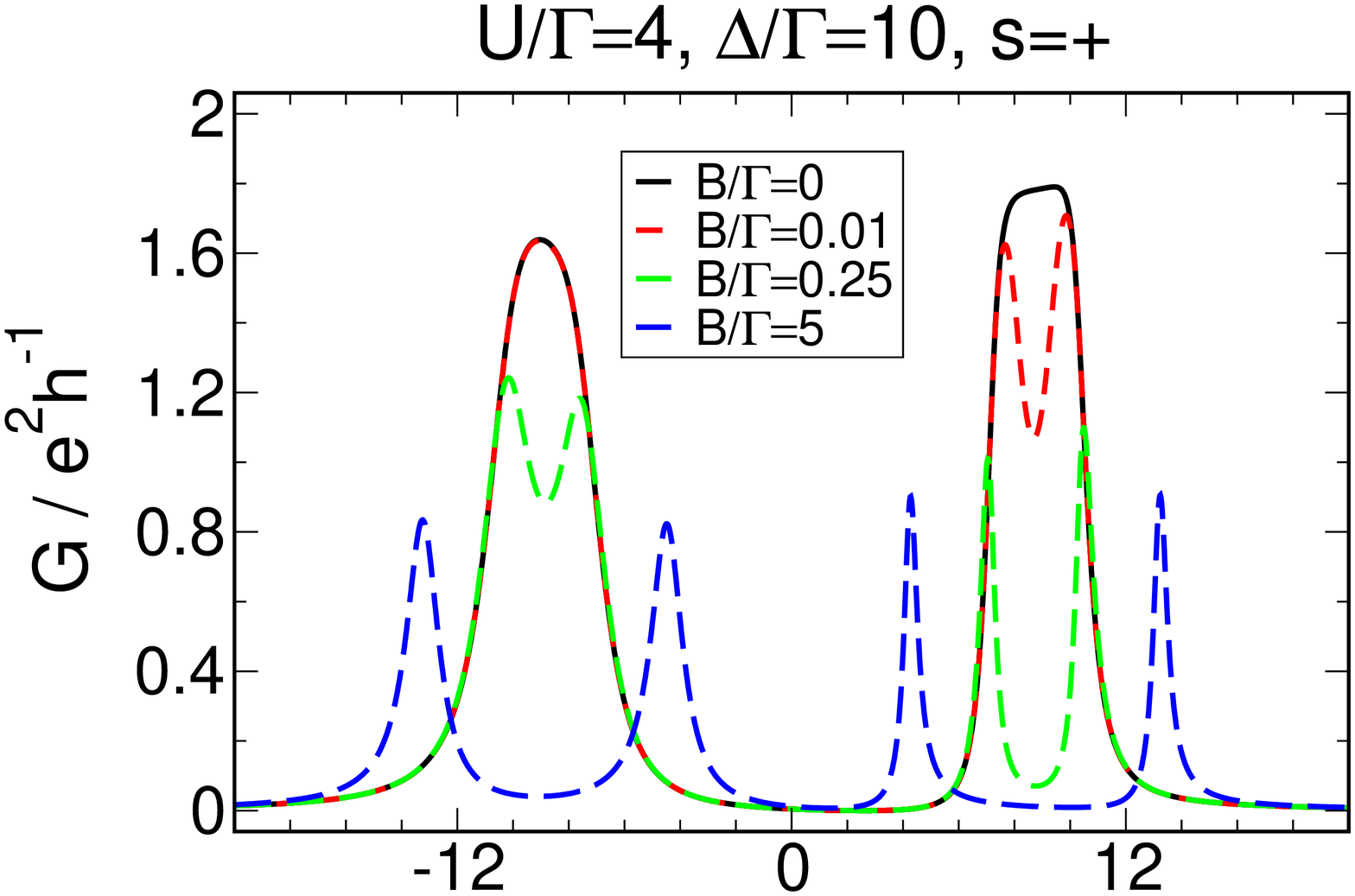}\vspace{0.3cm}
        \includegraphics[width=0.475\textwidth,height=5.2cm,clip]{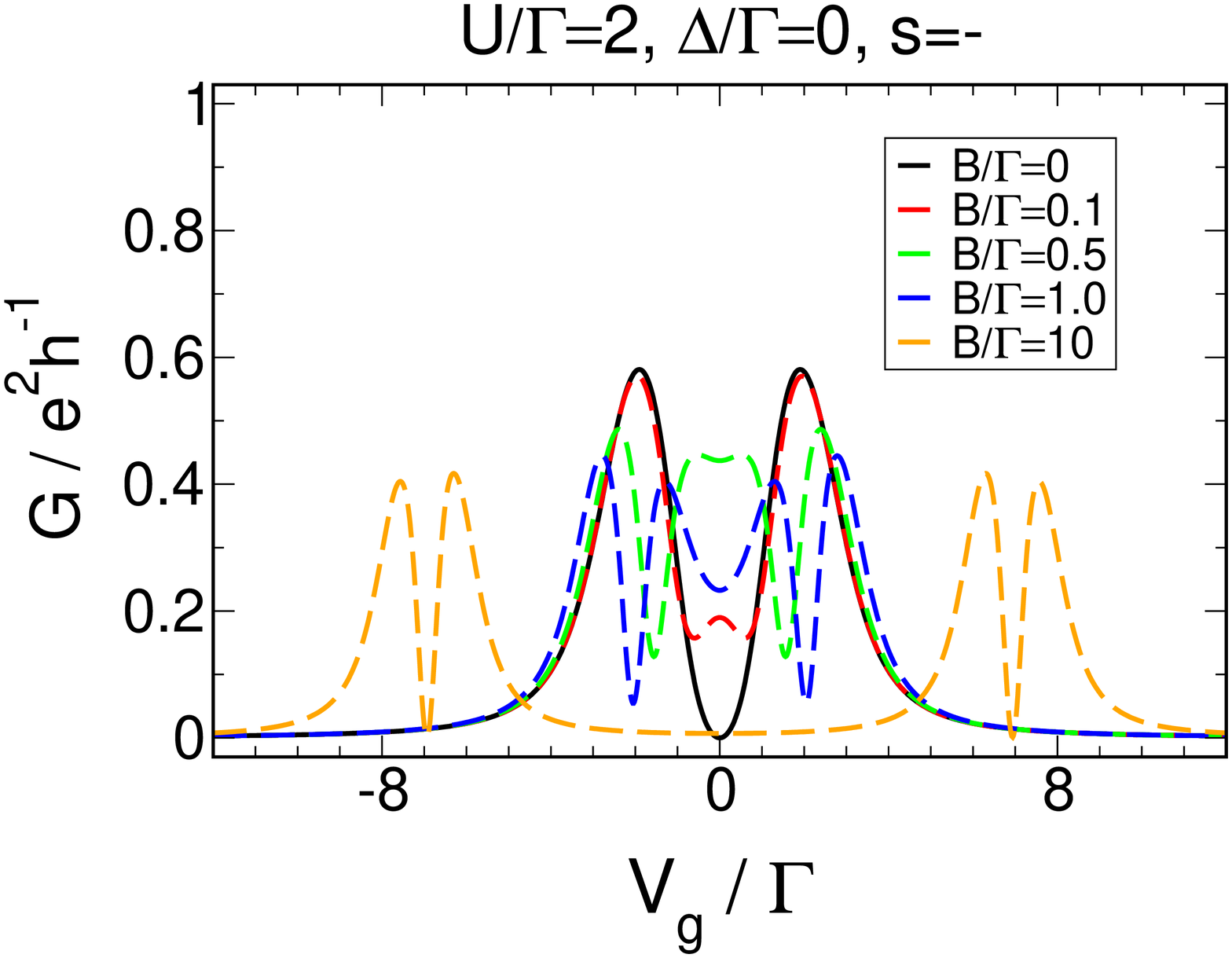}\hspace{0.035\textwidth}
        \includegraphics[width=0.475\textwidth,height=5.2cm,clip]{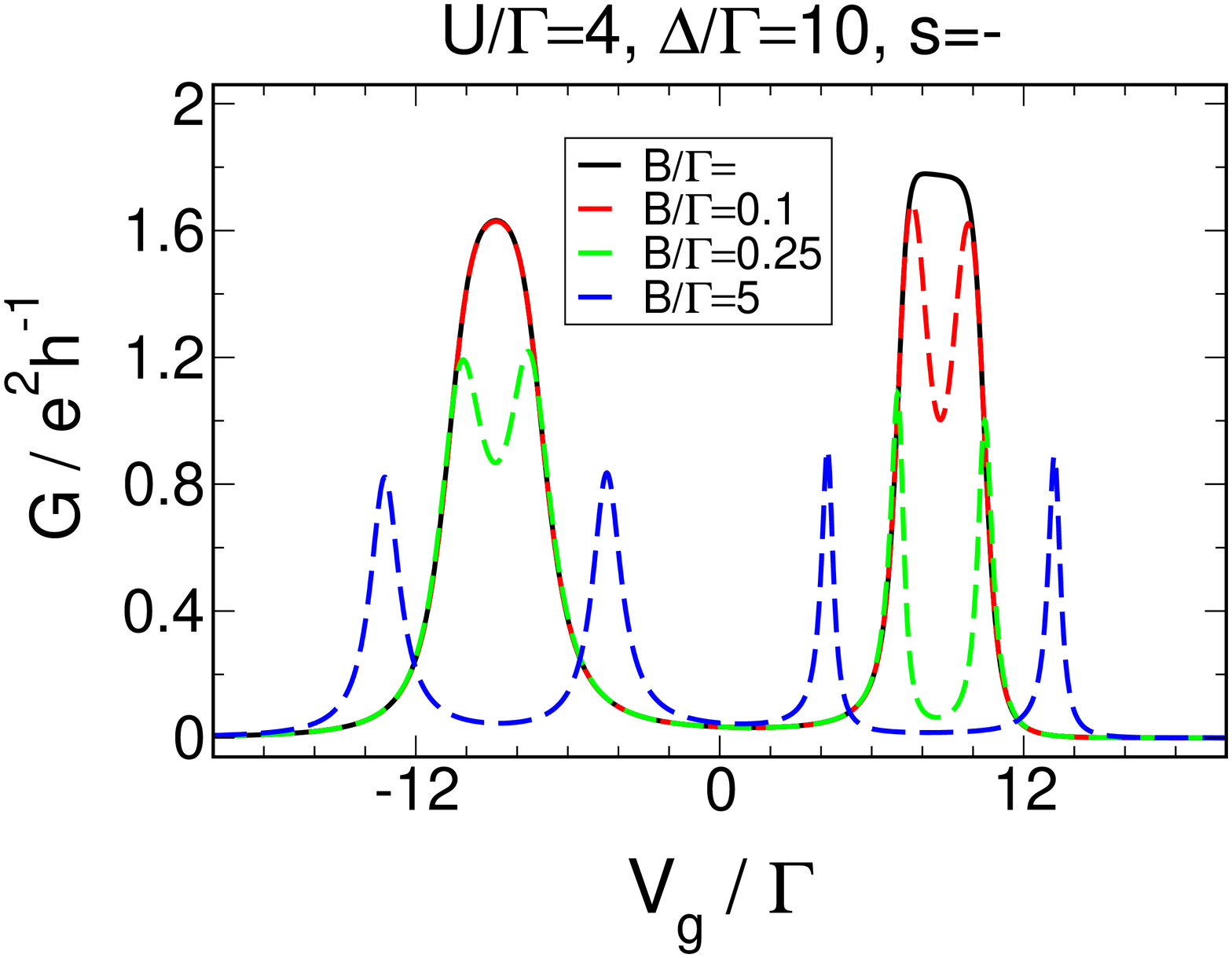}
        \caption{The same as Fig.~\ref{fig:MS.dd.loc_b}, but for equal local and nearest-neighbour interactions. Note that for the parameters of the upper left panel $U$ is large enough to recover the CIRs predicted by the spin-polarised model in the limit of large $B$.}
\label{fig:MS.dd.all_b}
\end{figure}

As described above, the additional resonances observed for $s=-$, degenerate levels and large enough local correlations (see Fig.~\ref{fig:MS.dd.loc_md}, upper left panel) gradually disappear and for equal local and nearest-neighbour interactions only two peaks located at $V_g\approx\pm U/2$ are found (Fig.~\ref{fig:MS.dd.all_md}, upper right panel). The transmission phase changes by $\pi$ over each of them and jumps by $\pi$ at the transmission zero in between. Surprisingly, the behaviour of the average level occupancies is very different from all previous situations. The occupation of the more strongly coupled level drops by one at each of the two resonances while being constant otherwise. The average occupancy of the more weakly coupled dot falls off continuously when the gate voltage is raised and the dots are heightened in energy.

Increasing the level spacing the evolution of the conductance is similar to the case with only local correlations. In particular, it is again analogous to the crossover regime of the spin-polarised case. The peak that `corresponds' to the more strongly coupled level splits up while the other is shifted outwards and gradually vanishes at a crossover scale depending on the dot parameters. For large $\Delta$, we find the two Kondo resonances of the individual levels separated by $\Delta+2U$ as for the $s=+$ case. In contrast to the latter, the phase evolves continuously in between the peaks (Fig.~\ref{fig:MS.dd.all_md}).

The magnetic field dependence of $G(V_g)$ for degenerate levels is puzzling as well. Its evolution with increasing field strength is identical to all other cases, but the conductance at half-filling is significantly raised at a scale much larger then in the situation where only local correlations are present (compare Fig.~\ref{fig:MS.dd.loc_b}, upper right panel to Fig.~\ref{fig:MS.dd.all_b}, lower left panel). For large magnetic fields, however, we recover twice the spin-polarised structure and in particular the CIRs if the interaction exceeds the critical strength depending on the hybridisations. For $\Delta\gg\Gamma$ we observe the usual splitting of the two single-level resonances.

A brief summary of the variety of phenomena arising for the different parameter regimes of the double dot (in particular for nearly degenerate levels) is given in Sec.~\ref{sec:MS.sum}.

\subsection{Parallel Triple Dots}\label{sec:MS.td}
\begin{figure}[t]	
	\centering
	      \includegraphics[width=0.495\textwidth,height=4.4cm,clip]{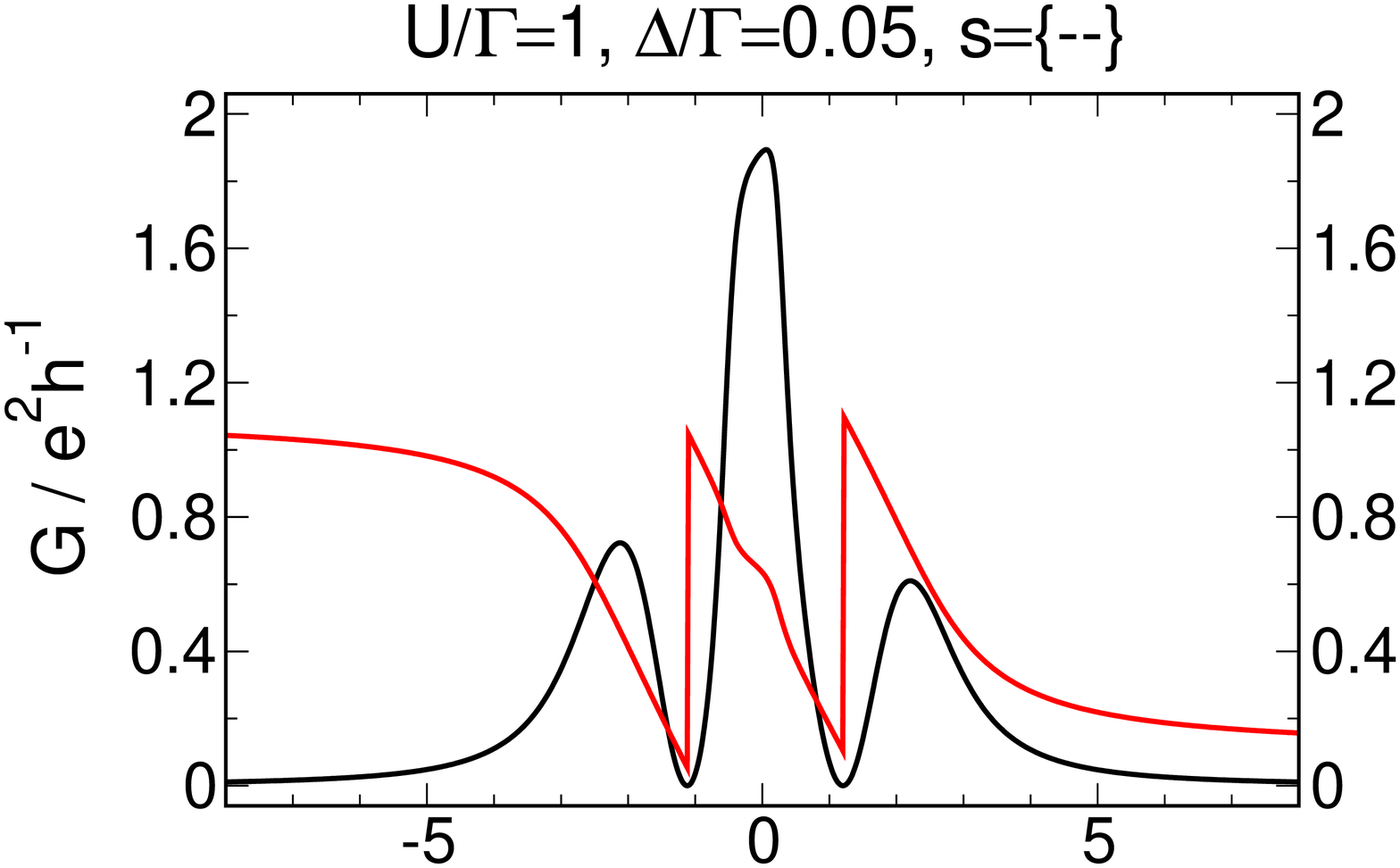}\hspace{0.015\textwidth}
        \includegraphics[width=0.475\textwidth,height=4.4cm,clip]{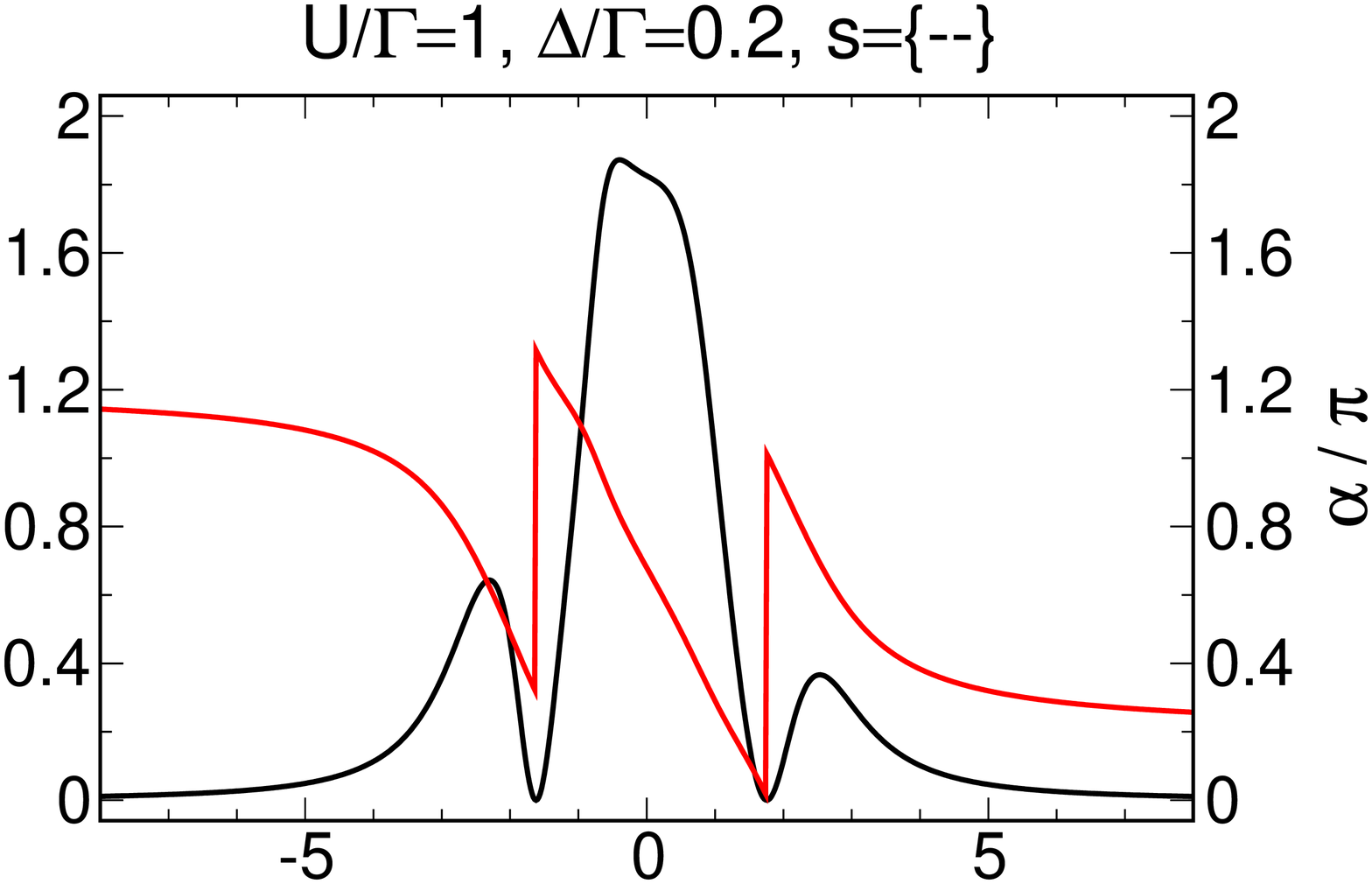}\vspace{0.3cm}
        \includegraphics[width=0.495\textwidth,height=5.2cm,clip]{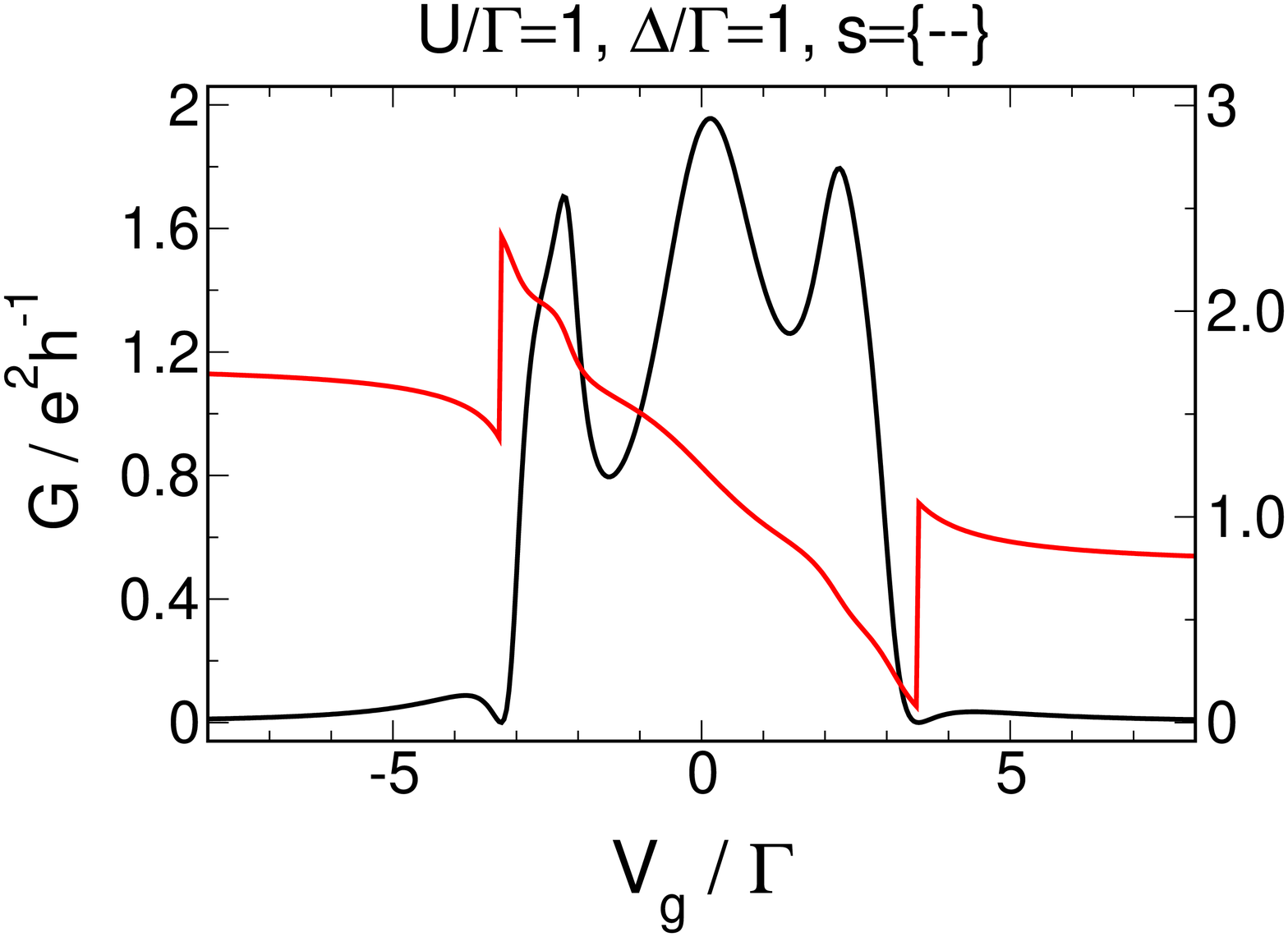}\hspace{0.015\textwidth}
        \includegraphics[width=0.475\textwidth,height=5.2cm,clip]{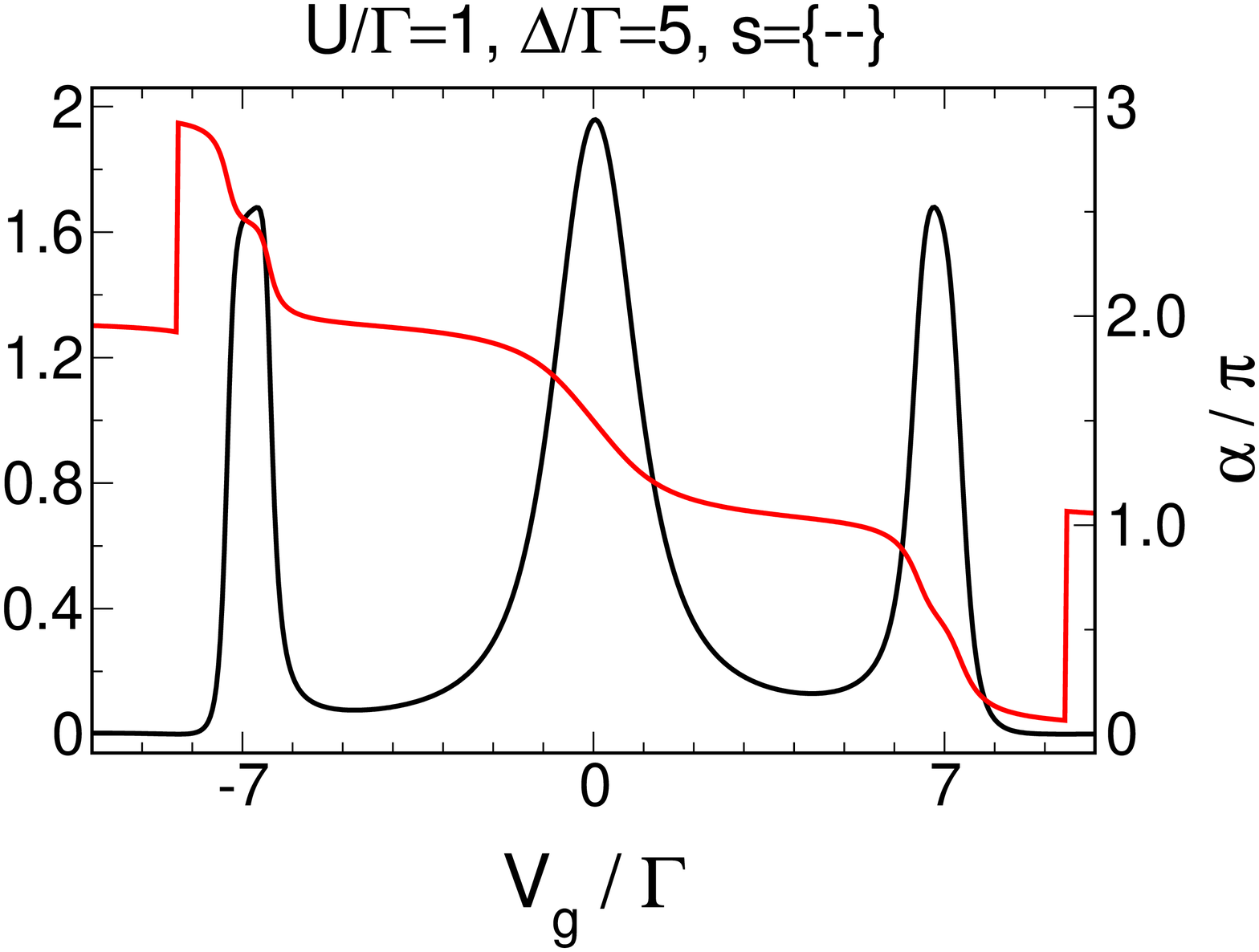}
        \caption{Conductance $G$ (black) and transmission phase $\alpha$ (red) as a function of the gate voltage for parallel triple dots with equal local- and nearest neighbour interactions and level-lead hybridisations from regime S, $\Gamma=\{0.06~0.14~0.3~0.4~0.07~0.03\}$. The crossover from $\Delta\ll\Gamma$ to $\Delta\gg\Gamma$ is illustrated by four different level spacings. It is analogous with the spin-polarised case. The smallest detuning is chosen such that the fRG results are still reliable.}
\label{fig:MS.td.delta}
\end{figure}

Finally, we will study the triple dot geometry similarly to the one of Sec.~\ref{sec:OS.td}, only that now we are accounting for the spin degree of freedom. The free propagator and the expression for the conductance follow similar to the spin-polarised case. We introduce equal local and nearest-neighbour interactions of strength $U$ and assume the level spacing $\Delta$ to be the same for all levels.

For large level spacings, the behaviour of the conductance is completely analogous to the double dot case. We observe three resonances of width $U$ (Fig.~\ref{fig:MS.td.delta}, lower right panel). The Kondo effect is active at each of them and the noninteracting Lorentzian lineshape is gradually transformed into a box-like structure depending on the size of the interaction compared to the corresponding hybridisations, $U/\Gamma_l$. The transmission phase changes by $\pi$ over each resonance with a lineshape similar to the single-level case. Transmission zeros and related phase jumps are located between those peaks associated to transport through levels whose relative couplings do not differ in sign.

Surprisingly, for $s=\{+-,--\}$ the dot enters a crossover regime completely similar to the spin-polarised case if the level spacing is lowered. In particular, some of the Kondo resonances merge while the height of others which are located outside the three-peak structure and are unobservably small for $\Delta\gg\Gamma$ gradually increases (Fig.~\ref{fig:MS.td.delta} shows an example for $s=\{--\}$). If one decreases the level spacing step by step, in the limit $\Delta\ll\Gamma$ one ends up with three resonances of comparable width and height with transmission zeros in between. The phase changes by $\pi$ over each peak and jumps by $\pi$ at the zeros (Fig.~\ref{fig:MS.td.delta}, upper left panel).

\begin{figure}[t]	
	\centering
        \includegraphics[width=0.475\textwidth,height=5.2cm,clip]{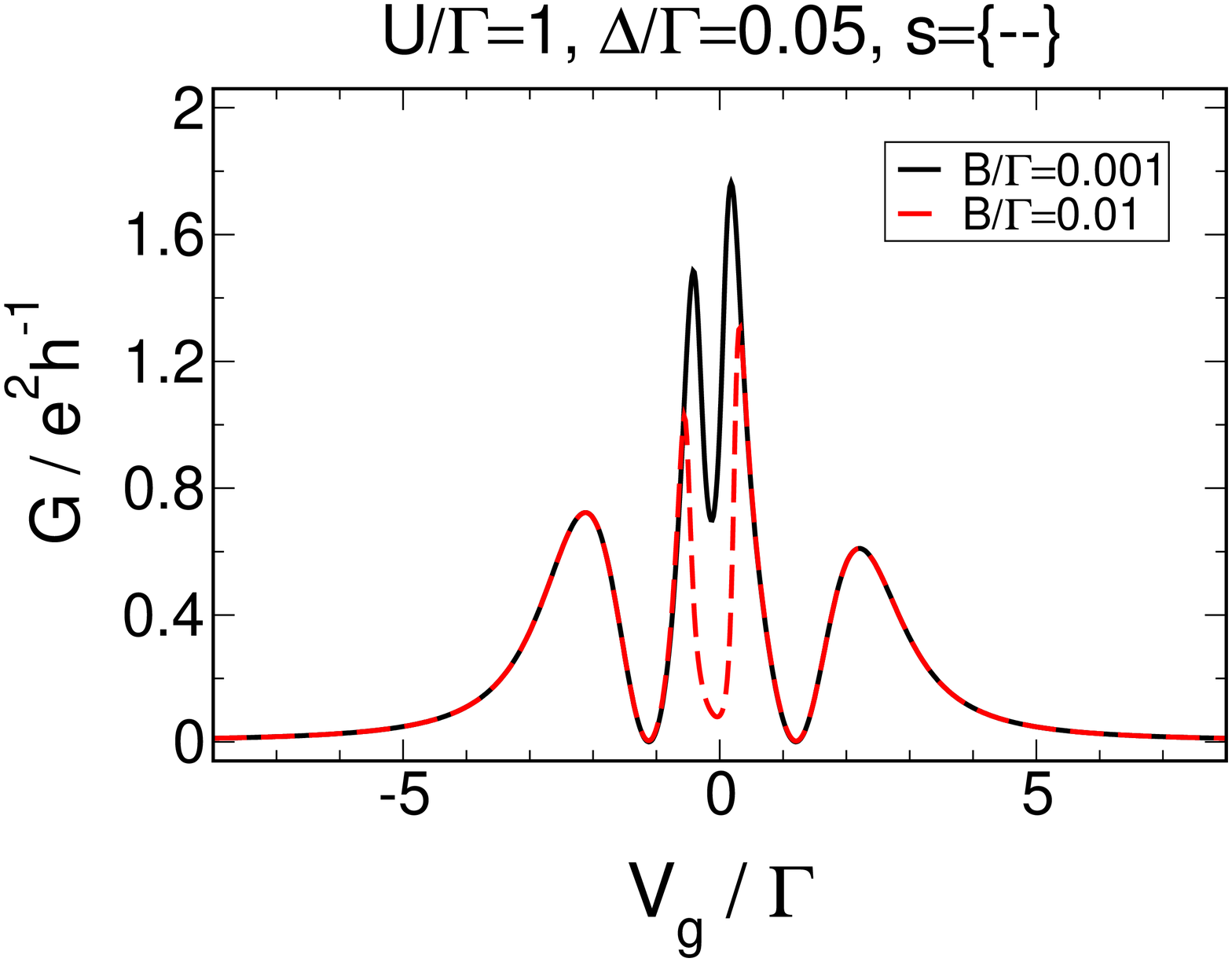}\hspace{0.035\textwidth}
        \includegraphics[width=0.475\textwidth,height=5.2cm,clip]{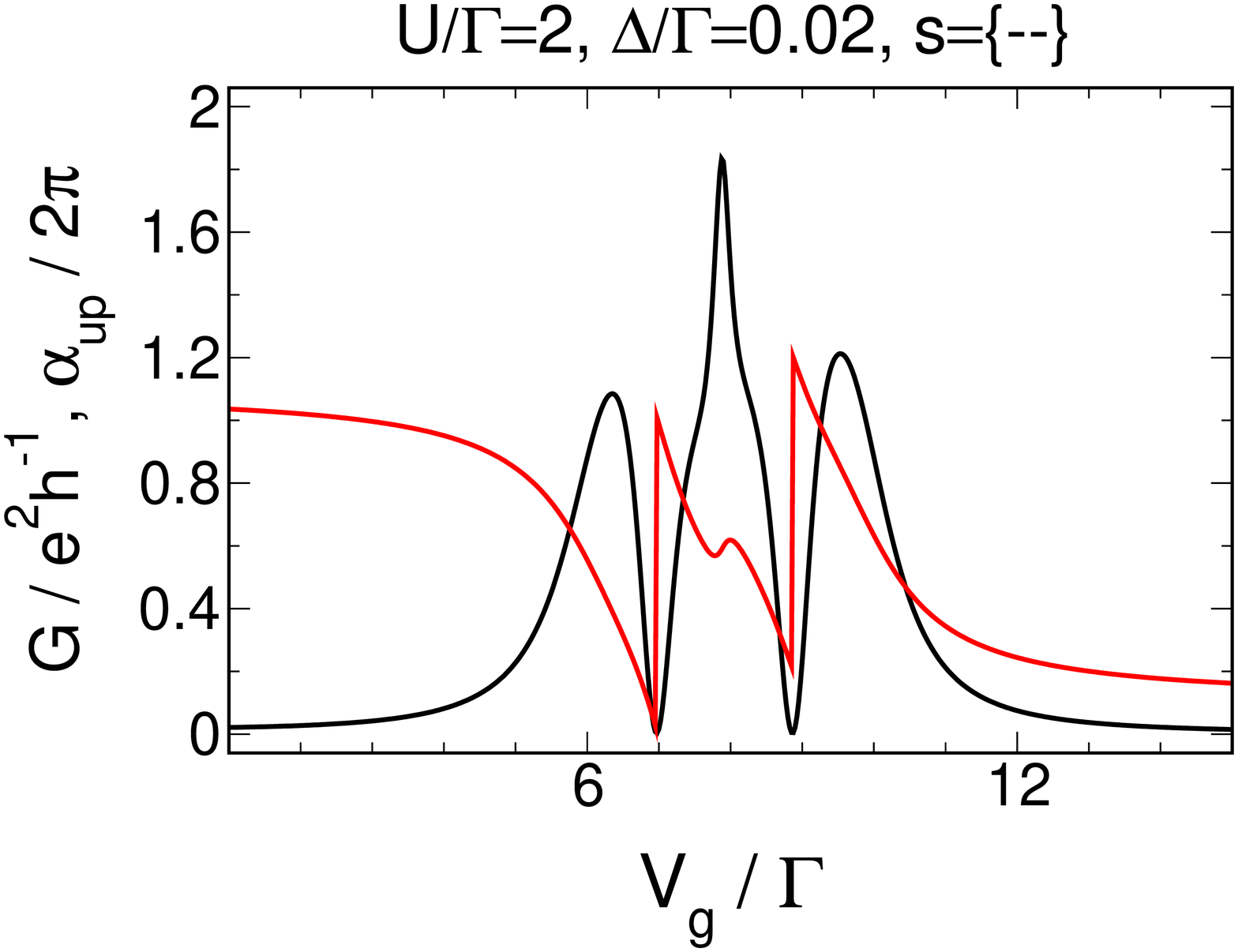}
        \caption{The same as Fig.~\ref{fig:MS.td.delta}, but for finite magnetic fields. The left panel suggests a large effective interaction within the system. The right panel shows that for large $B/\Gamma=5.0$ we recover the spin-polarised structures in $G(V_g)$ and $\alpha_\sigma(V_g)$.}
\label{fig:MS.td.b}
\end{figure}

In the spinful double dot case, we experienced problems with our fRG approximation for $s=+$ and small level spacings due to a strong effective correlation. It turns out that the same holds for the triple dot with $s=\{++\}$, and also for the other choices of the relative signs the fRG yields results that seriously deviate from exact NRG data, though at much smaller values of $\Delta$. However, despite the fact that for some choices of the hybridisations we are not able to perform reliable calculations for level spacings smaller than $\Delta_\tn{cross}$, it seems reasonable to assume that the behaviour of the conductance and the transmission phase in between the limits $\Delta\gg\Gamma$ to $\Delta\ll\Gamma$ is similar within the spin-polarised and the model containing spin in what concerns the number of peaks and transmission zeros.

The fact that we have to expect a strong correlation effect to be present for small level spacings is also signalled by the magnetic field dependence of $G(V_g)$. Applying very small fields $B\ll\Gamma$ leads to a significant change of the lineshape of the conductance (Fig.~\ref{fig:MS.td.b}, left panel). In analogy to the single-level case, the scale of magnetic fields destroying the Kondo resonance should be given by $\exp(-U/\Gamma)$. This is of the order of one for $U\approx\Gamma$, and hence the influence of small fields $B\ll\Gamma$ on $G(V_g)$ and the breakdown of the fRG for small level spacings could consistently be interpreted as some large effective interaction within the system. For the double dot with $s=+$, this large interaction was identified to be the second stage Kondo effect of the side-coupled geometry. Unfortunately, we are unable to present such an easy explanation here.

If we apply large magnetic fields to the system, we recover twice the spin-polarised structure in the lineshape of $G(V_g)$. In particular, we can quantitatively reproduce all the results of Sec.~\ref{sec:OS.td}, especially the three-peak structure of the conductance with the transmission zeros located in between for nearly degenerate levels (compare the right panel of Fig.~\ref{fig:MS.td.b} with the upper left panel of Fig.~\ref{fig:OS.td.pmpda}).

\subsection{Summary}\label{sec:MS.sum}

In the first section, we have presented how the fRG can be used to compute the most common signature of Kondo physics in zero temperature transport through quantum dots, the Kondo resonance exhibited by the conductance of a single impurity coupled to noninteracting leads. We have shown how the characteristic energy scale on which fluctuations destroy this resonance, the Kondo temperature, can be extracted from the linear-response conductance at $T=0$ by considering the magnetic field dependence of $G(V_g)$.

Next, we have turned to short Hubbard chains which are constructed by coupling $N$ of these single-level dots in series. In the limit of large nearest-neighbour hoppings, on which we mainly focused here, the conductance of these chains shows well-separated Kondo resonances each time they are occupied by an odd number of electrons. The most important physical observation was, however, the suppression of charge fluctuations exhibited if $N$ is an even number. Furthermore, a vanishing of transmission resonances was found for generic asymmetries in the hybridisations with the leads.

Next, we discussed transport through the side-coupled geometry where the size of the hopping $t$ between the embedded and the side-coupled dot determines two distinct regimes of different physics. For small $t$ the behaviour of the system is governed by a two-stage Kondo effect. In particular, due to the second stage Kondo effect taking place on the side-coupled dot the Kondo resonance that one would expect if only the embedded dot was present is transformed into a valley of width $U$ where the conductance is strongly suppressed due to the large $U/t^2$. On the other hand, two well-separated Kondo resonances attributed to individual Kondo effects of the bonding and antibonding states are observed for large $t$ (the molecular-orbital regime).

Finally, we put much effort in describing transport through the most complex geometry considered here, the spinful double dots. For nearly degenerate levels, the behaviour of $G(V_g)$ depends on the system parameters, in particular on the relative sign $s$ of the level-lead couplings. For $s=+$ the conductance is strongly suppressed in a region of width $U$ around $V_g=0$. This can be understood as the two-stage Kondo effect of the side-coupled geometry. In presence of equal interactions between all electrons, the `peaks' surrounding the conductance valley split up at a critical strength of $U$ such that for even larger interactions a total of four transmission resonances located at $V_g=\pm U$ and $V_g=\pm U/2$ are found. In both cases the conductance at half-filling is significantly influenced by the application of extremely small magnetic fields (compared to $\exp (-U/\Gamma)$). For $s=-$ and purely local interactions exceeding a critical scale of order $\Gamma$, $G(V_g)$ exhibits a four-peak structure of completely different nature. Within the fRG approximation scheme, this can be interpreted as interference between two independent single-level dots. If additional nearest-neighbour correlations are added to the system, two of the resonances gradually disappear. For purely local $U$, the application of a magnetic field of size comparable to the $s=+$ case has a strong impact on the conductance. Surprisingly this does no longer hold for equal interactions between all electrons in the dots. If the level spacing is increased above $\Delta_\tn{cross}<\Gamma$, the system enters a crossover regime completely similar to the spin-polarised case. The additional correlation induced features which are observed for $s=+$ and equal interactions between all electrons as well as for $s=-$ and purely local correlations vanish. For $s=-$, one of the remaining peaks splits up and the transmission zero is gradually moved outwards to larger $|V_g|$, while for $s=+$ only the lineshape of the two resonances changes. In all cases, one ends up with two well-separated Kondo peaks of strength $U/\Gamma_l$ associated with the individual levels in the limit $\Delta\gg\Gamma$. For $s=+$, a transmission zero (and a corresponding phase jump) is located in between, while $G$ stays finite and $\alpha$ evolves continuously for $s=-$.

The most important things one should learn from the discussion of the double dot geometry are the following. First, it excellently demonstrates how a two-particle interaction leads to complex structures (in particular to more than ordinary Coulomb blockade peaks) in the conductance. Even more, we demonstrated that the noninteracting limit does not exhibit the rich variety of different physics (and the same holds for the other geometries), stressing the importance of correlations. Second, one should note that in presence of large magnetic fields we precisely recover all the results of the spin-polarised model on a quantitative level, justifying that the latter is more than just a `toy model'. Finally it is important to realise that our findings for the case of almost equal interactions between all electrons (which is likely to be the experimentally realised situation) basically allow for an explanation of the `phase lapse' experiments in complete analogy to the discussion at the end of the last section. There is reason to believe that the same holds if more than two dots are considered (which was illustrated for $N=3$), but alternative methods are required to check this more systematically. 

\clearpage
\thispagestyle{empty}

\chapter{Comparison With Other Methods}

In this chapter, we will prove that the functional renormalization group indeed provides a powerfull tool to study transport through correlated quantum dot systems in the zero temperature limit. In order to do so, we will compare our $T=0$ results for almost every physical situation which has been previously investigated to numerical renormalization group (NRG, see \cite{nrg1,nrg2,nrg3}) data or, if the actual problem allows for it in some special cases, to the exact solution. As mentioned in the introduction, the NRG has been frequently used in recent papers as a reliable method non-perturbative in the two-particle interaction to study spinful dot systems. Only recently, also spinless systems where treated by mapping them back to ones with spin \cite{cir,theresanrg}. All data which has not been published so far has been checked against exact $U=0$ results ensuring that all NRG calculations presented here are very precise and therefore suited as a reference to compare the fRG against. In particular, for every system under consideration we will show up to which interaction strength we can still obtain reliable results by our second order fRG truncation scheme that neglects the frequency-dependence of the two-particle vertex as well as the flow of all higher-order vertex functions.

Next, we will establish a criterion to judge the quality of fRG results which is independent of the availability of reference data from other methods. In particular, we will argue that a breakdown of the usual second order approximation is always signalised by diverging components of the effective interaction at the end of the flow.

Finally, we will demonstrate that the fRG is superior to conventional Hartree-Fock calculations in describing the effects of the interaction between the electrons. For the simplest case of a spinless two-level dot, we will show that such a mean-field approach captures the Coulomb blockade physics, but it fails in correctly describing the phenomena connected with the appearance of the correlation induced resonances where the interaction is of particular importance. Furthermore, we will prove that taking into account quantum fluctuations remedies for some artifacts by which Hartree-Fock is plagued due to its mean-fieldish nature.

Unless stated otherwise, the fRG data presented in this chapter was always obtained by the usual second order truncation scheme that accounts for the flow of the self-energy and the two-particle vertex evaluated at zero external frequency, and we will generally speak of it as `the' fRG result, implying the linguistic sloppiness that in principle it is of course not forbidden to carry out arbitrary-order calculations. Furthermore, we will employ the same simplifications as in Chapter~\ref{sec:numericresults}. In particular, we will always perform the wide-band limit (which leads to energy-independent level-lead hybridisations) and set the chemical potential  of the noninteracting leads to zero. All data shown was obtained for $T=0$.

\section{NRG Calculations}

\subsection{Spin-Polarised Dots}

We will start comparing NRG calculations for parallel geometries of spinless quantum dots which we have investigated in great detail in Sec.~\ref{sec:OS}. The NRG data we use was provided by \cite{theresanrg} (see also \cite{phaselapsespaper}). Its reliability has been established by comparison with the exact $U=0$ case.

\subsubsection{Parallel Double Dots}
\begin{figure}[t]	
	\centering
        \vspace{-0.3cm}\includegraphics[width=0.475\textwidth,height=4.4cm,clip]{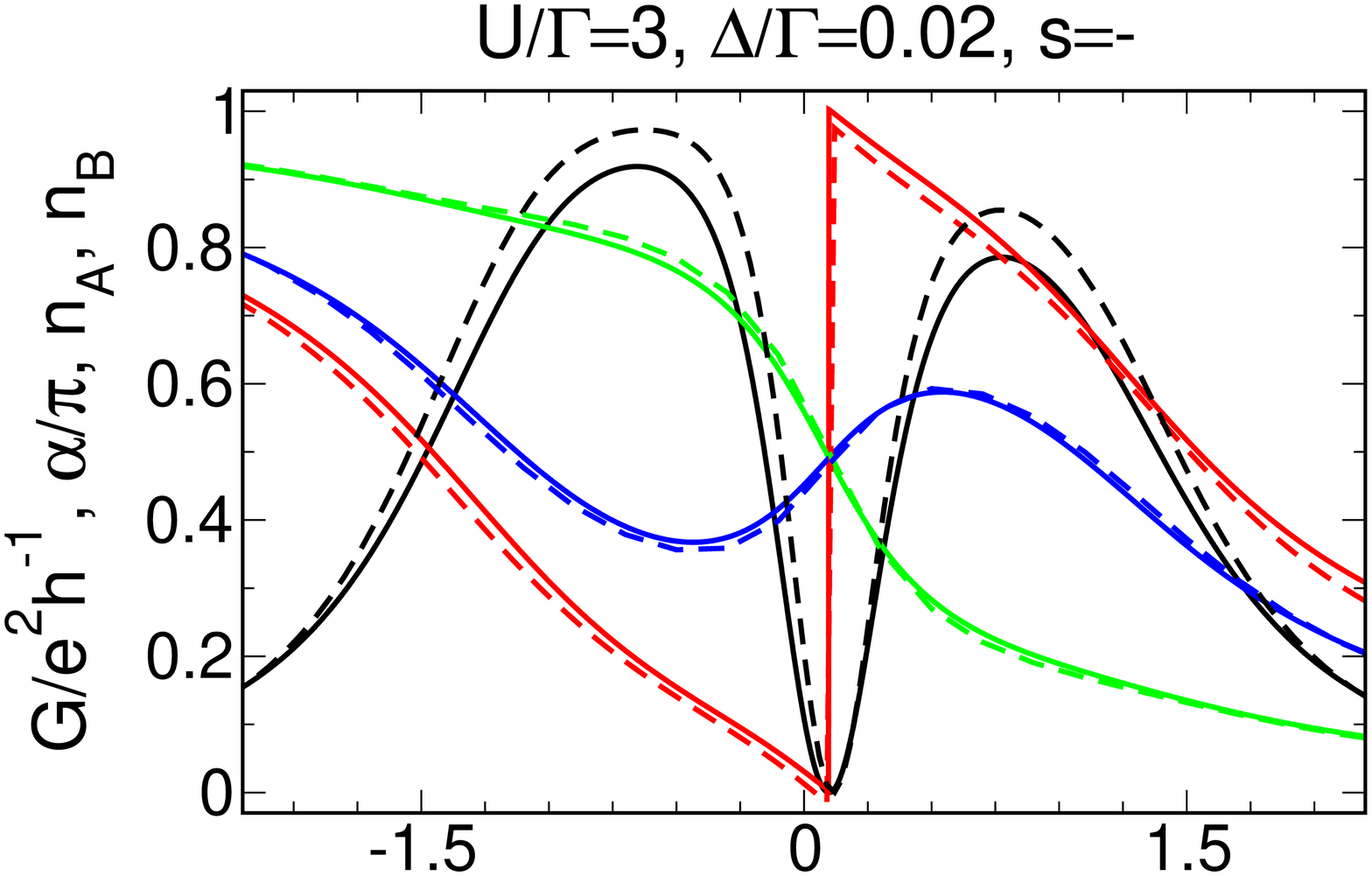}\hspace{0.035\textwidth}
        \includegraphics[width=0.475\textwidth,height=4.4cm,clip]{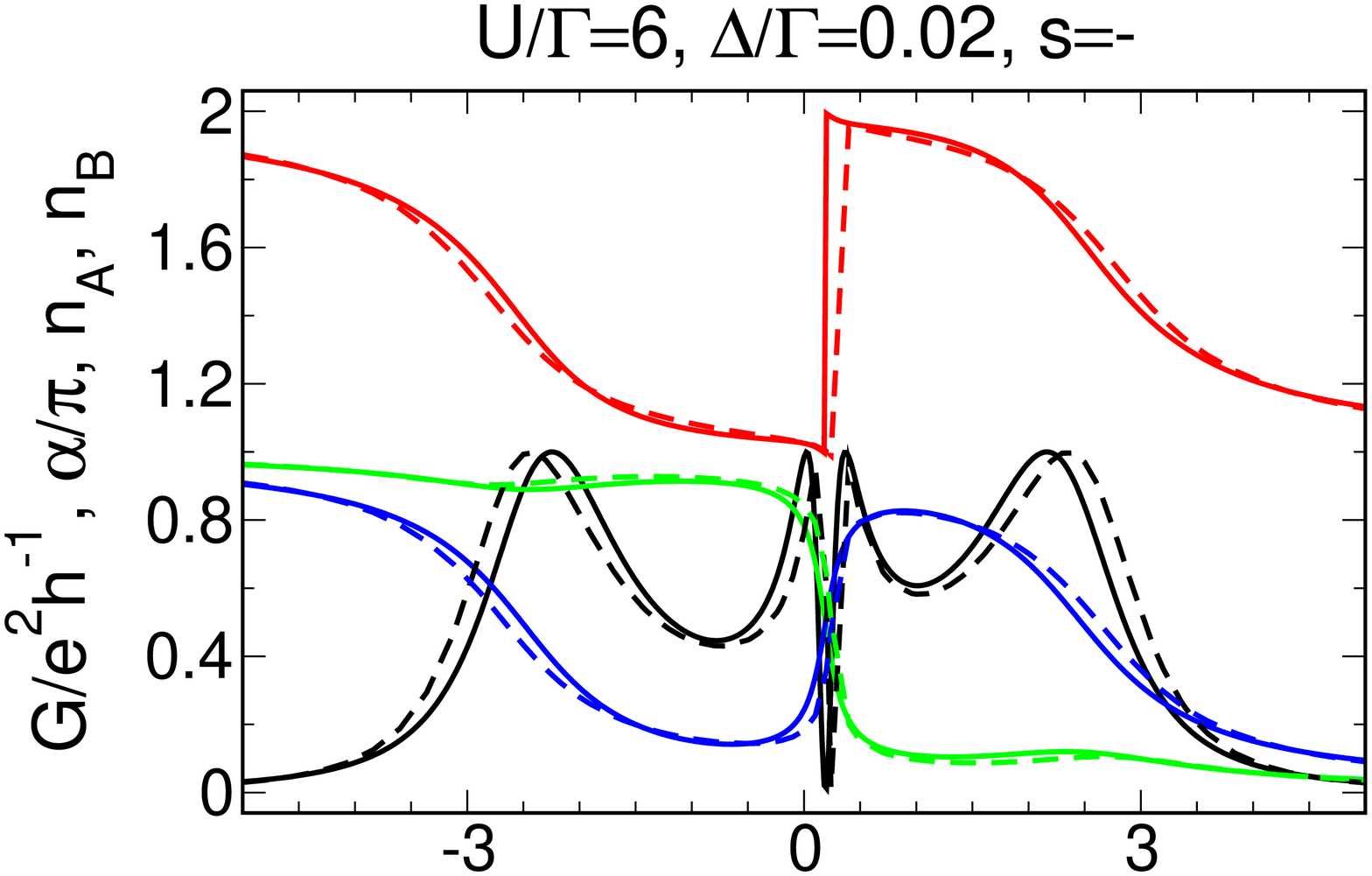}\vspace{0.2cm}
        \includegraphics[width=0.475\textwidth,height=4.4cm,clip]{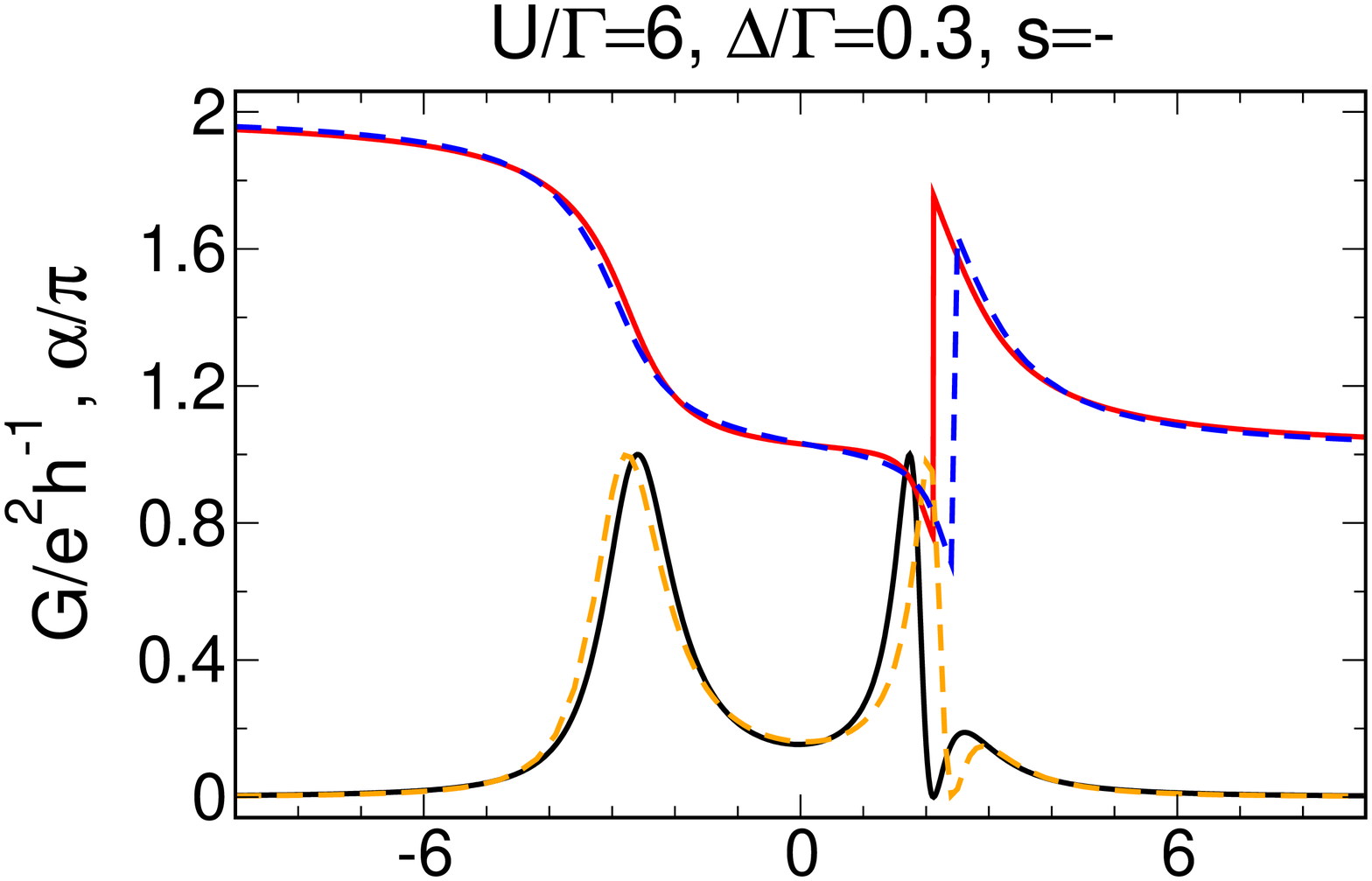}\hspace{0.035\textwidth}
        \includegraphics[width=0.475\textwidth,height=4.4cm,clip]{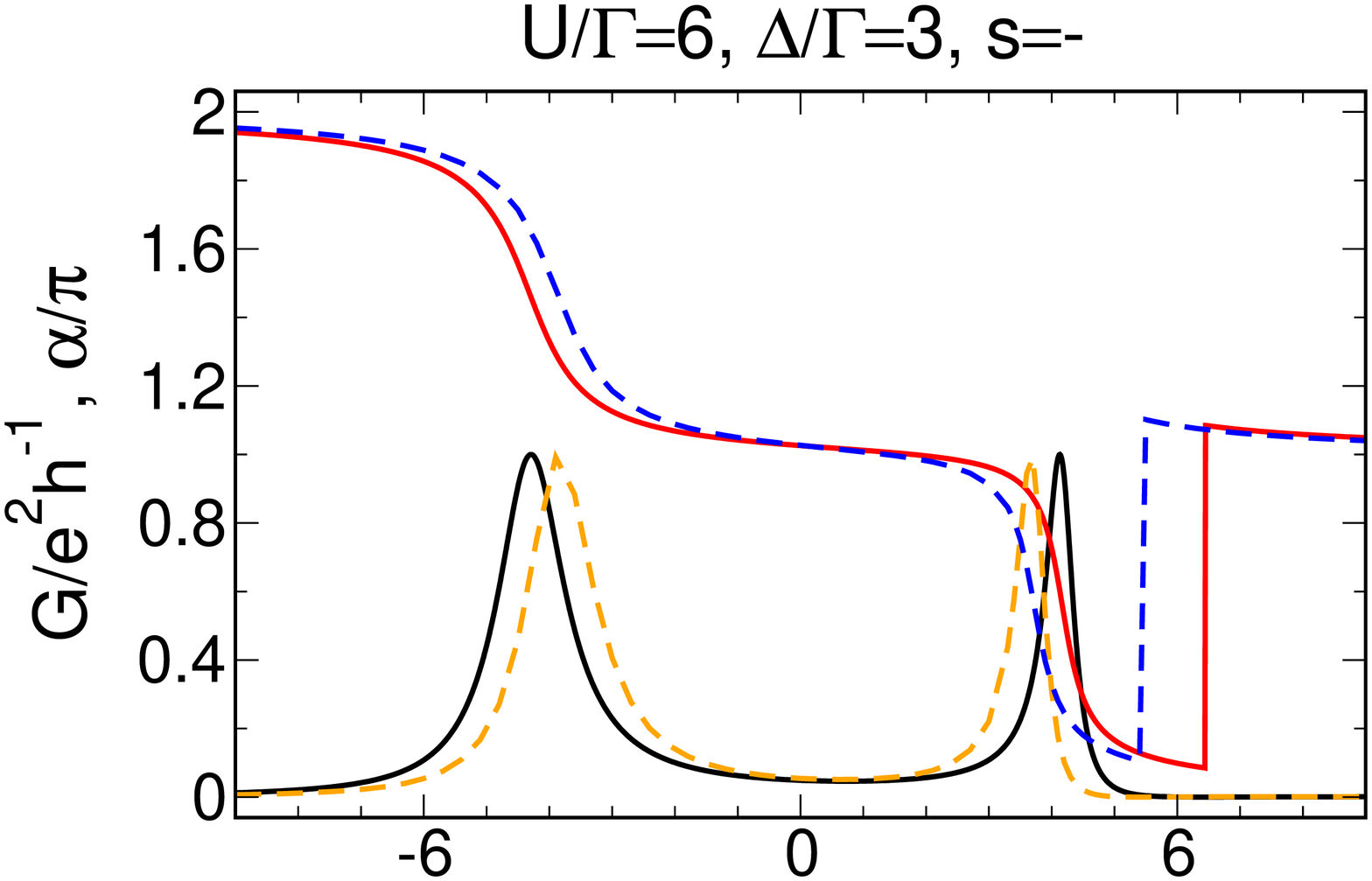}\vspace{0.2cm}
        \includegraphics[width=0.475\textwidth,height=5.2cm,clip]{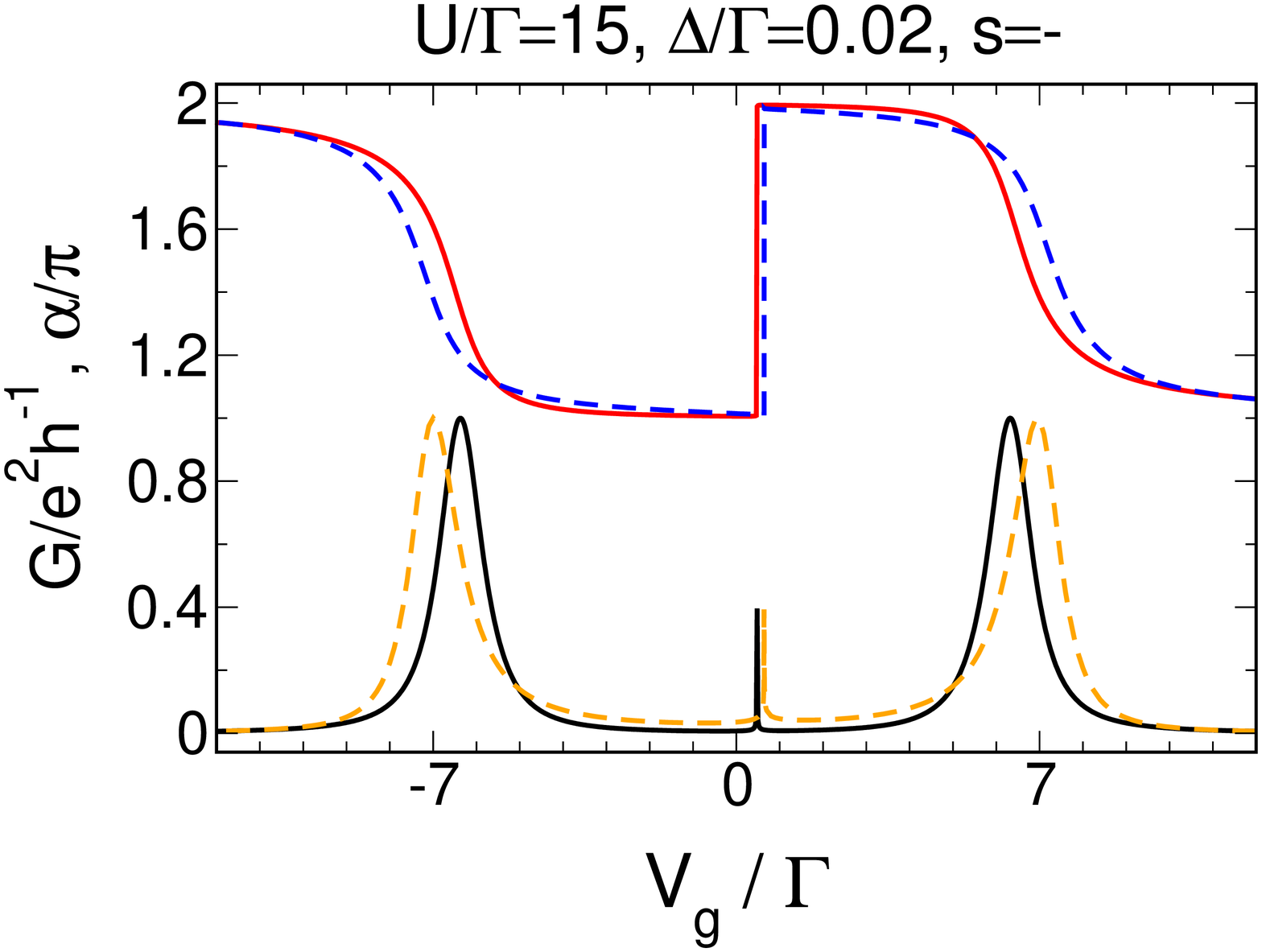}\hspace{0.035\textwidth}
        \includegraphics[width=0.475\textwidth,height=5.2cm,clip]{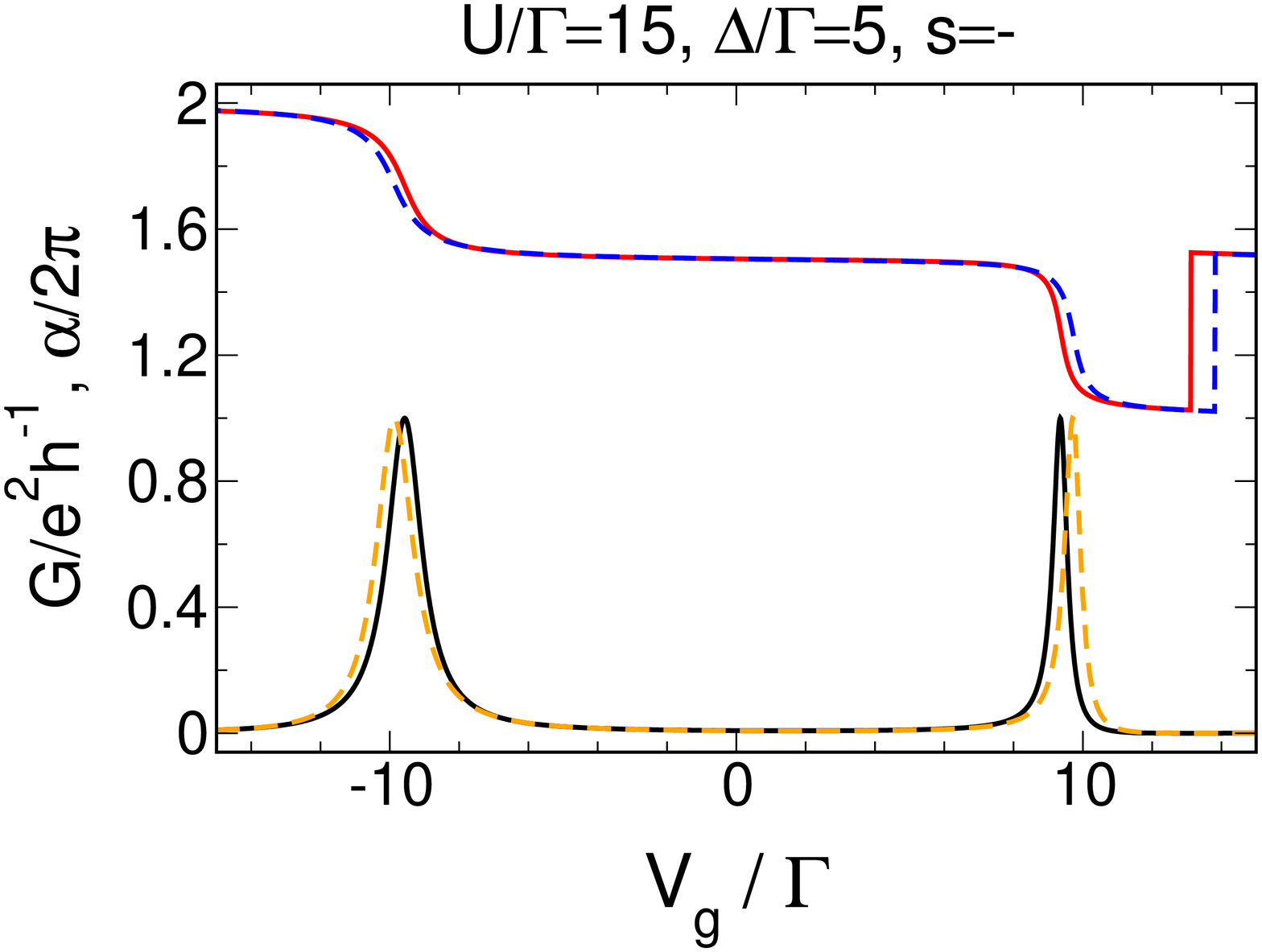}\vspace{-0.2cm}
        \caption{Comparison between fRG (solid lines) and NRG (dashed lines) data of the gate voltage dependence of the conductance $G$, transmission phase $\alpha$, and average level occupations (dot A: green, dot B: blue) of a spinless two-level dot with hybridisations $\Gamma=\{0.15~0.15~0.35~0.35\}$ for various parameters. The features in the lower left panel are remnants of the CIRs not observable at this scale. The NRG data was taken from \cite{theresanrg}.}
\label{fig:COMP.os.dd}\vspace{0.4cm}
        \includegraphics[width=0.475\textwidth,height=5.2cm,clip]{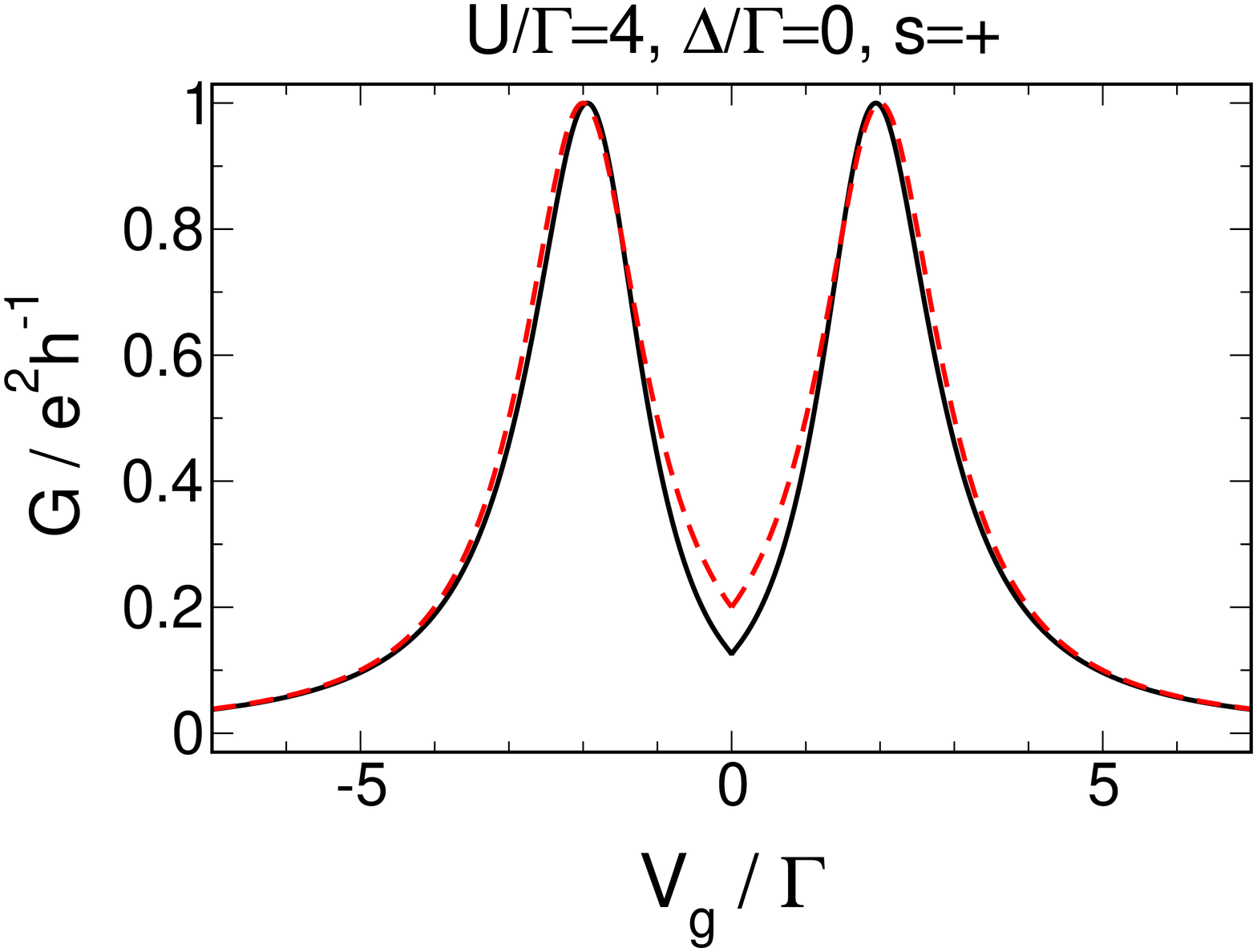}\hspace{0.035\textwidth}
        \includegraphics[width=0.475\textwidth,height=5.2cm,clip]{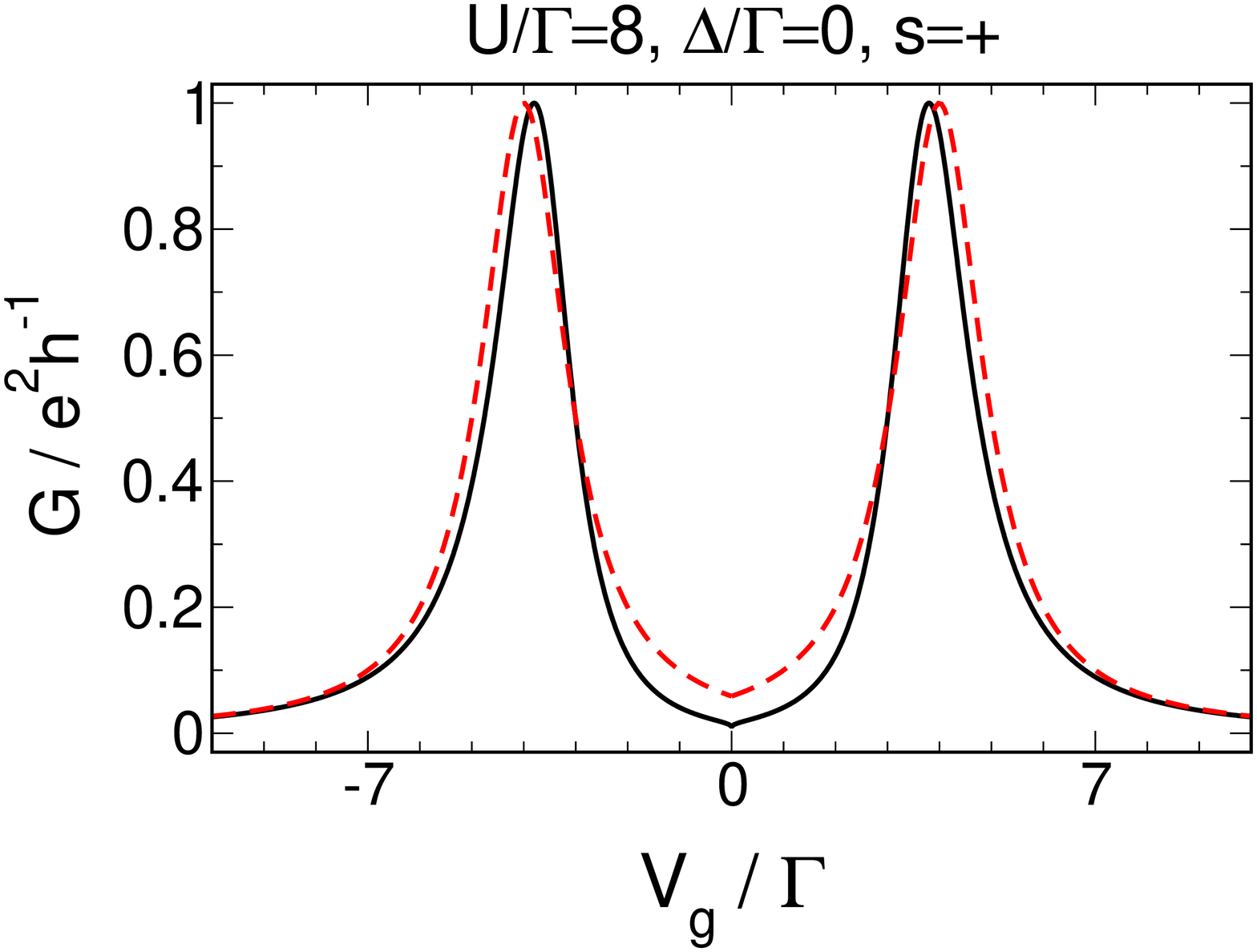}\vspace{-0.2cm}
        \caption{Comparison of the conductance $G$ calculated by fRG (black) with the exact solution (red) for a double dot with $\Gamma=\{0.25~0.25~0.25~0.25\}$ and degenerate levels. Note that due to the non-generic hybridisations no CIRs and transmission zero at $V_g=0$ are observed.}
\label{fig:COMP.os.dd_exact}
\end{figure}
\afterpage{\clearpage}

In Fig.~\ref{fig:COMP.os.dd}, we consider the parallel double dot with generic parameters $\{s,\Gamma\}$. For both small and large level spacings, the NRG and fRG curves for the conductance, the transmission phase and the average level occupancies as a function of the gate voltage are almost indistinguishable even for very large $U/\Gamma=15$. In particular, we can confirm the observed non-monotonic dependence of the occupancies on $V_g$ close to the particle-hole symmetric point. Even more important, the NRG calculations explicitly show that the additional sharp correlation induced resonances (CIRs) are no artifact of the fRG approximation (a much more detailed comparison between fRG and NRG in what concerns the CIRs can be found in \cite{cir}).

In fact, there is one special choice of parameters that allows for an exact solution of the interacting many-particle problem. Similar to Sec.~\ref{sec:MS.side}, the general double dot Hamiltonian with $A$-$B$-symmetric hybridisations and $s=+$,
\begin{equation*}
H =(V_g-U/2-\Delta/2)n_A + (V_g-U/2-\Delta/2)n_B
-\hspace{-0.2cm}\sum_{l=A,B\atop s=L,R}\left[t_sd_l^\dagger c_{0,s}+\tn{H.c.}\right] + Un_An_B,
\end{equation*}
can by introducing bonding and antibonding states, $|\pm\rangle:=\frac{1}{\sqrt{2}}\left(|A\rangle\pm|B\rangle\right)$, be mapped onto
\begin{equation*}
\tilde H= \sum_{l=+,-}(V_g-U/2)n_l +  \left[\frac{\Delta}{2}d_+^\dagger d_- -\sqrt{2}t_L d_+^\dagger c_{0,L} - t_R d_+^\dagger c_{0,R} + \tn{H.c.}\right]+ Un_+n_-.
\end{equation*}
For degenerate levels ($\Delta=0$), we have $[\tilde H,n_-]=0$, and hence the particle number operator $n_-$ is a conserved quantity. Therefore the conductance $G$ of the system is exactly given by
\begin{equation*}
G = \frac{4e^22t_L^22t_R^2}{h}\left|\frac{1}{V_g-U/2-2i[t_L^2+t_R^2]+f(V_g)U}\right|^2 ,
\end{equation*}
and particle-hole symmetry stipulates that the function $f(V_g)\in\{0,1\}$ which determines whether the antibonding state is occupied or not has to read $f(V_g)=0$ for $V_g>0$ and $f(V_g)=1$ otherwise. Therefore the conductance exhibits two Lorentzian peaks of width $2\Gamma$ separated by $U$ but no correlation induced resonances no matter how big the interaction is chosen. Therefore $A$-$B$-symmetric couplings together with $s=+$ do not represent the generic behaviour of the system (which can already be seen by considering the noninteracting case). This is the reason why this parameter set was completely ignored when discussing the double dot in Sec.\ref{sec:OS.dd}, but a without any effort exactly solvable model is always welcome as a comparison for any method whether it describes generic physics or not. It turns out that fRG results agree well with the exact ones (see Fig.~\ref{fig:COMP.os.dd_exact}).

\subsubsection{Parallel Dots with Four Levels}
\begin{figure}[t]	
	\centering
        \vspace{-0.3cm}\includegraphics[width=0.475\textwidth,height=4.4cm,clip]{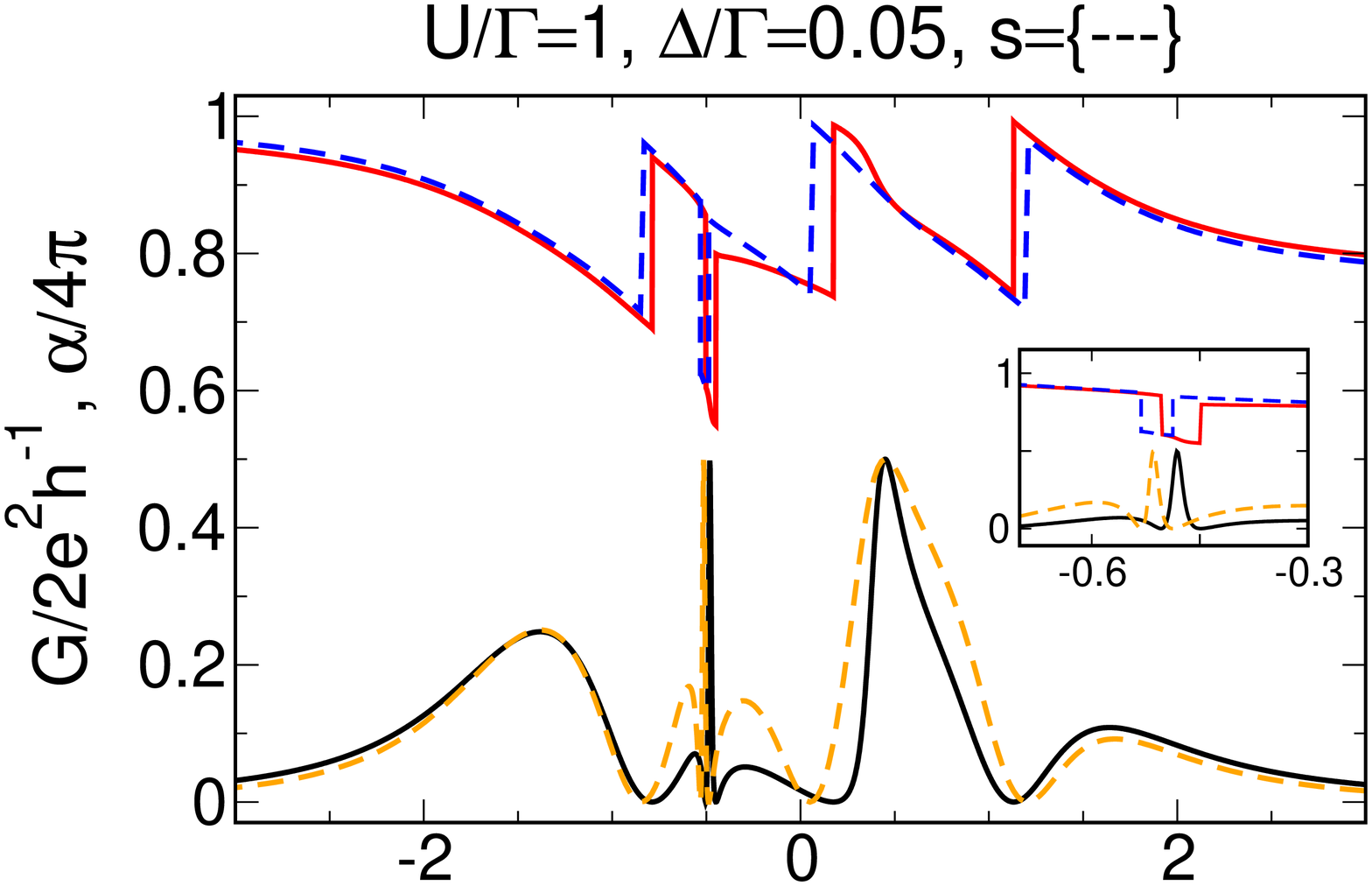}\hspace{0.035\textwidth}
        \includegraphics[width=0.475\textwidth,height=4.4cm,clip]{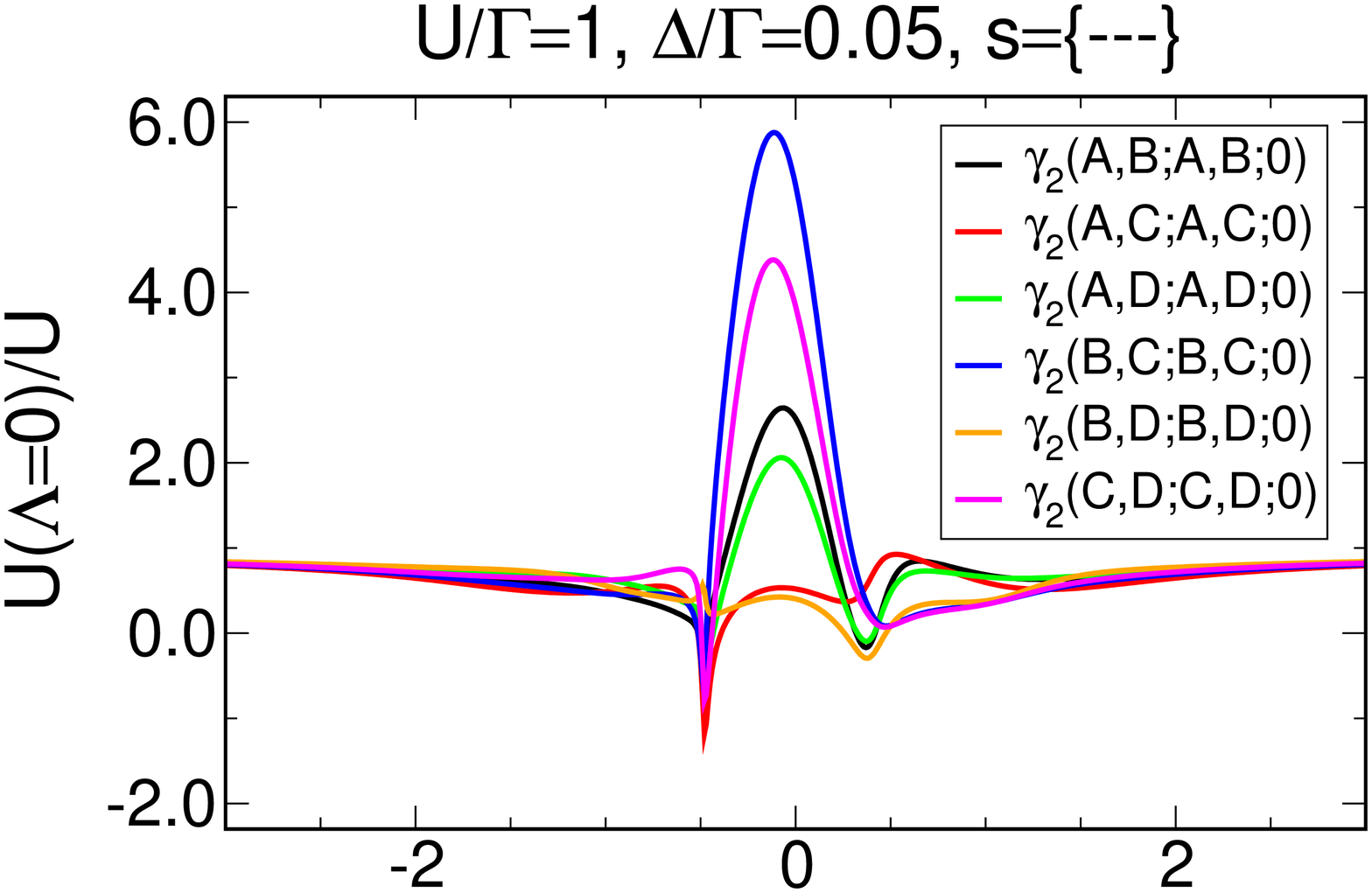}\vspace{0.3cm}
        \includegraphics[width=0.475\textwidth,height=5.2cm,clip]{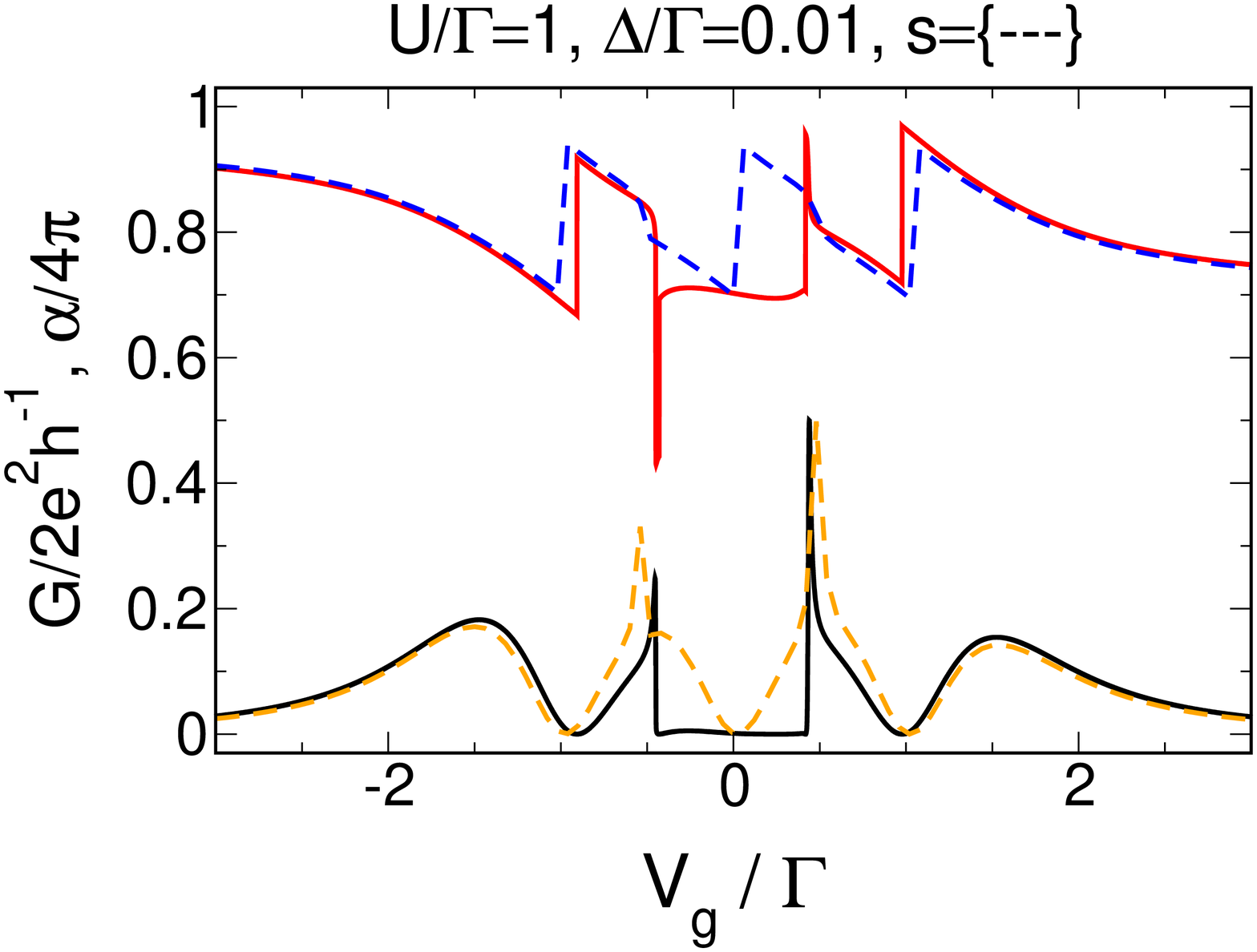}\hspace{0.035\textwidth}
        \includegraphics[width=0.475\textwidth,height=5.2cm,clip]{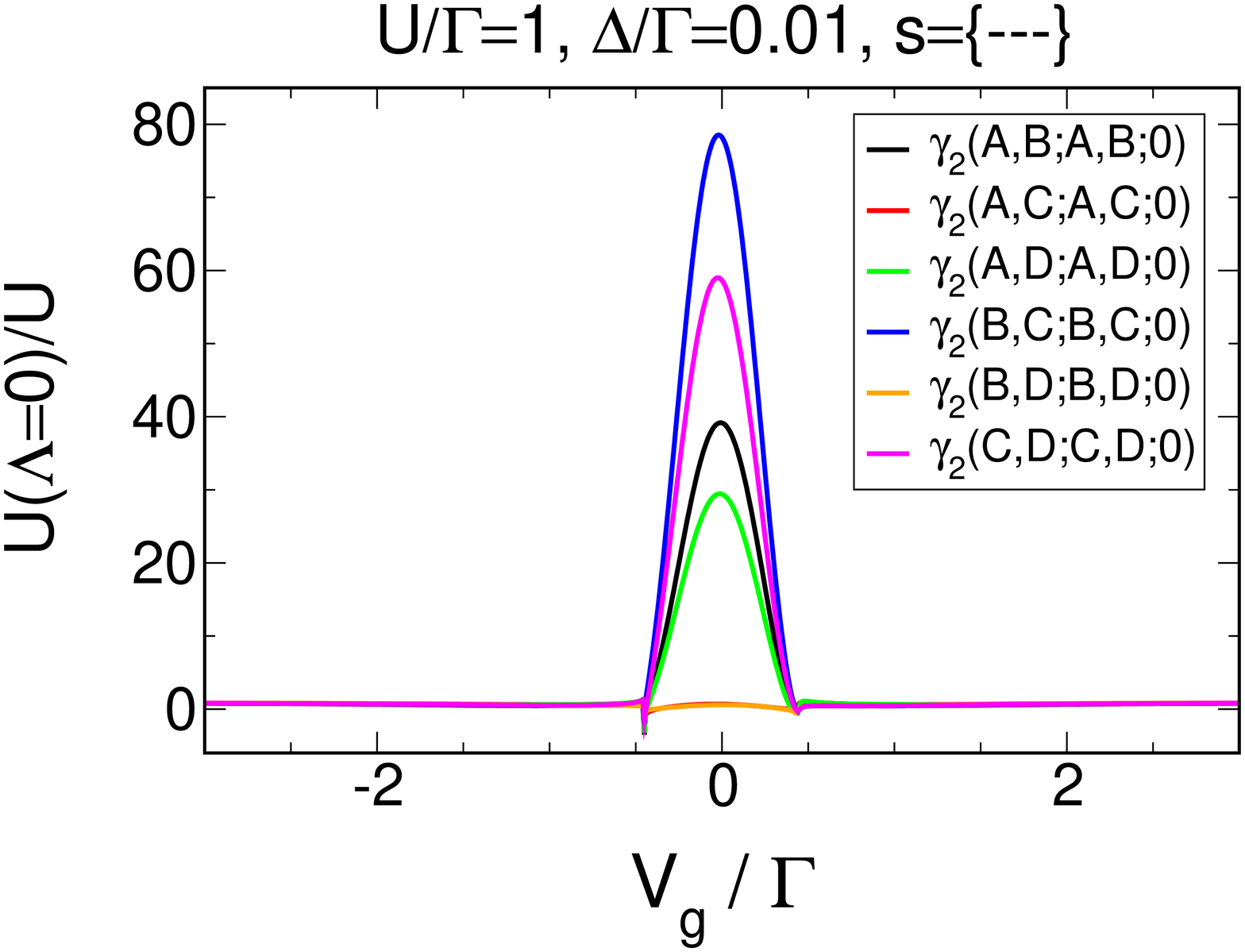}
        \caption{\textit{Left panels:} Comparison of fRG (solid lines) and NRG (dashed lines) calculations of the conductance $G$ (black, orange)  and transmission phase $\alpha$ (red, blue) as a function of the gate voltage of a spinless four-level dot with hybridisations $\Gamma=\{0.1~0.1~0.15~0.15~0.05~0.05~0.2~0.2\}$. The NRG data was taken from \cite{theresanrg}. \textit{Right panels:} Some components of the two-particle vertex at the end of the fRG flow.}
\label{fig:COMP.os.qd_a}\vspace{0.6cm}
        \includegraphics[width=0.475\textwidth,height=4.4cm,clip]{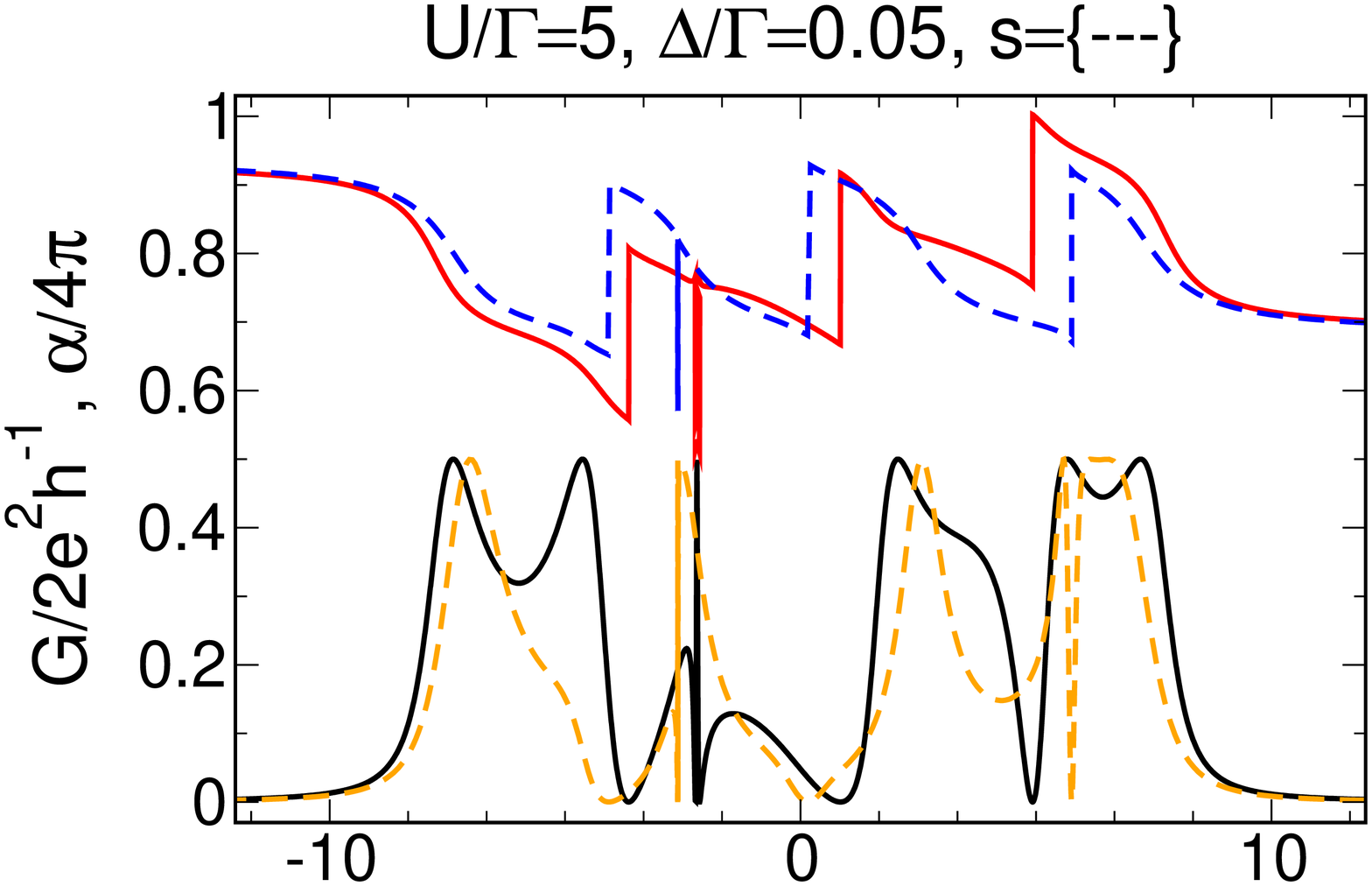}\hspace{0.035\textwidth}
        \includegraphics[width=0.475\textwidth,height=4.4cm,clip]{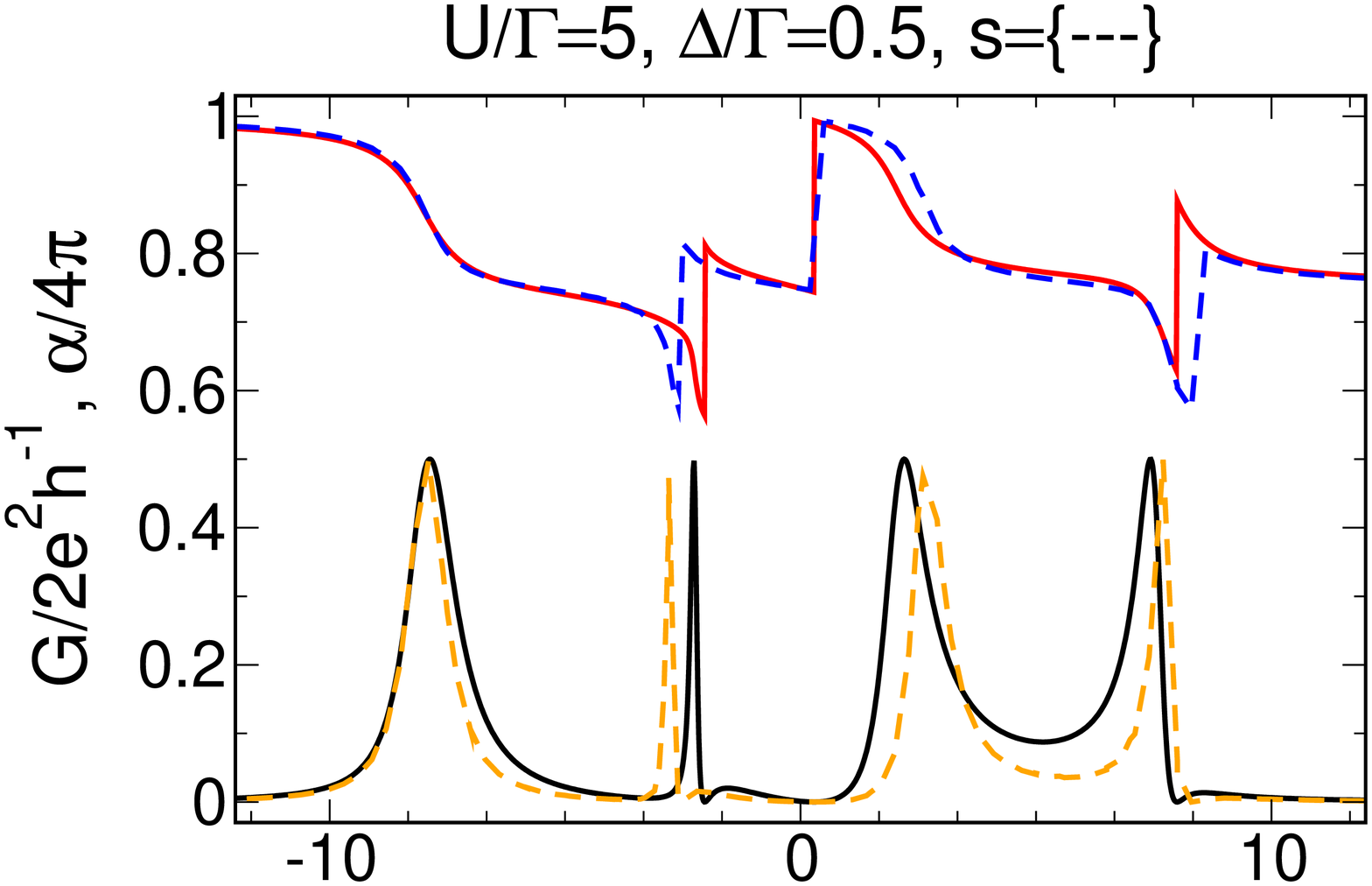}\vspace{0.3cm}
        \includegraphics[width=0.475\textwidth,height=5.2cm,clip]{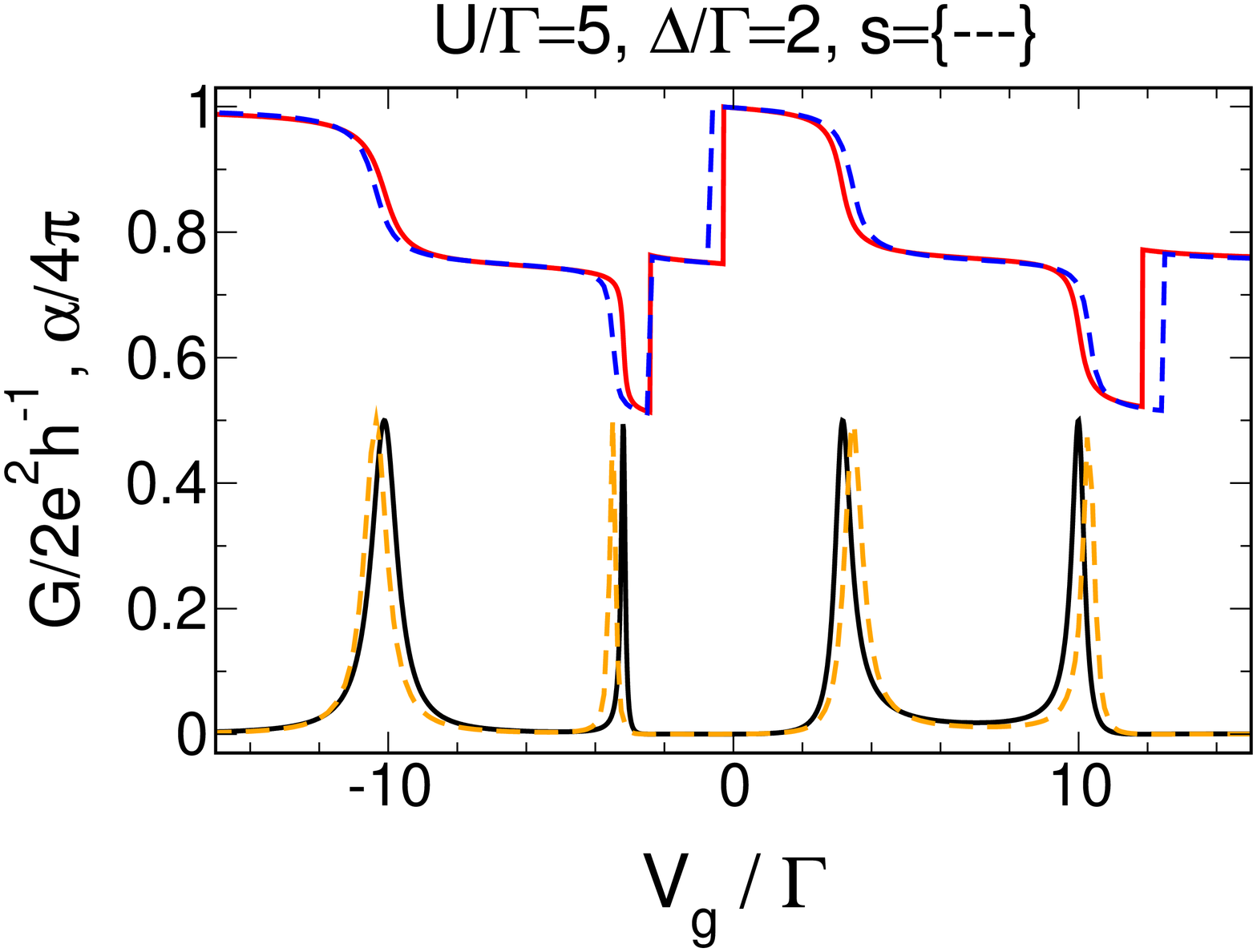}\hspace{0.035\textwidth}
        \includegraphics[width=0.475\textwidth,height=5.2cm,clip]{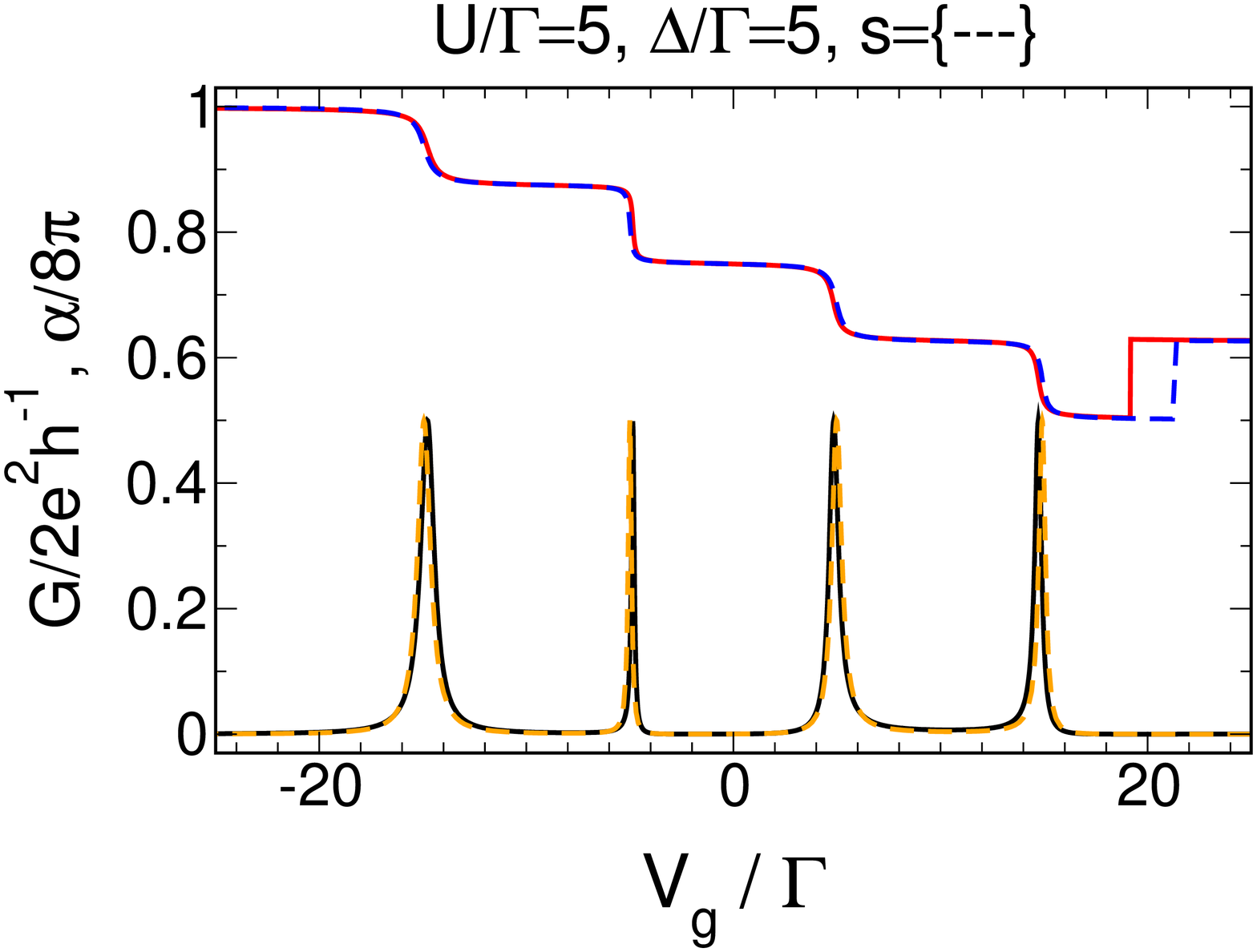}
        \caption{The same as the left panels of Fig.~\ref{fig:COMP.os.qd_a}, but for larger $U/\Gamma=5.0$.}\vspace{0.6cm}
\label{fig:COMP.os.qd_c}
\end{figure}
\afterpage{\clearpage}

Unfortunately, the triple dot geometry cannot be tackled by NRG (see Sec.~\ref{sec:COMP.summary} for more details), so that we turn to the four-level case next. Fig.~\ref{fig:COMP.os.qd_a} shows the comparison of the conductance $G(V_g)$ and the transmission phase $\alpha(V_g)$ computed by fRG and NRG for $U/\Gamma=1.0$, nearly degenerate levels, relative signs of the level-lead couplings $s=\{---\}$, and left-right symmetric hybridisations. The latter cause double phase lapses (DPLs) located near a very sharp resonance which appears in addition to the ordinary four Coulomb blockade peaks. Similar correlation induced phenomena were observed at the central resonances of the triple dot geometry. For $\Delta/\Gamma=0.05<\Delta_\tn{cross}/\Gamma$, it turns out that the results of both methods agree well. In particular, the very sharp DPLs are confirmed by NRG and are therefore no artifact of our truncation scheme. If the level spacing is further decreased, fRG and NRG results begin to deviate seriously for $V_g$ close to half-filling (see the comparison for $\Delta/\Gamma=0.01$). In particular, the number of transmission zeros computed by both methods is different and the conductance from fRG is much too strongly suppressed. This holds even when the interaction is gradually switched off, only that the detuning and the width of the affected region of $V_g$ where the fRG no longer returns reliable results decrease. Such behaviour was never observed for the double dot geometry where fRG and NRG agree extremely well even for degenerate levels, and unfortunately it lacks an easy explanation as a signature of some strong correlation effect with a large `effective' interaction.

The right panels of Fig.~\ref{fig:COMP.os.qd_a} show the effective interactions of the type $n_in_j$ at the end of the flow for the same four-level dot. Surprisingly, some of these interactions become attractive in some region of $V_g$, a phenomenon that we frequently observed for other geometries. More importantly, all $\gamma_2(i,j;i,j;\la)$ stay of order of their initial value $U$ if $\Delta$ is chosen such that fRG and NRG agree well. The same holds for all other components not forbidden by symmetries which are zero for $\la=\infty$ but in general of course generated by the flow. In contrast, some renormalized interaction terms grow large ($U(\la=0)/U\approx 100$) if the level spacing is so small that the fRG seriously deviates from the reliable NRG data. The evolution of the breakdown of the fRG approximation scheme in the limit $\Delta\ll\Gamma$ can be observed if the level spacing is successively lowered (see again Fig.~\ref{fig:COMP.os.qd_a}). For $\Delta/\Gamma=0.05$, close to $V_g=0$ some components of the two-particle vertex grow significantly larger than their initial value. Nevertheless, they still remain of roughly the same order and fRG and NRG only deviate quantitatively. On the other hand, for $\Delta/\Gamma=0.01$ some $\gamma_2$ become at least one order of magnitude larger than $U$ and the results of both methods strongly disagree. Alltogether, this will be the aforementioned self-consistent criterion to judge the trustworthiness of our results when no reference data is available: fRG calculations are reliable when no components of the two-particle vertex diverge.

If we increase the strength of the interaction (Fig.~\ref{fig:COMP.os.qd_c} shows $U/\Gamma=5.0$), the fRG and NRG calculations of the conductance deviate quantitatively for nearly degenerate levels ($\Delta/\Gamma=0.05$). In particular, the additional correlation induced structures are differently far evolved, while the number of transmission resonances and zeros and the transmission phase are still in good agreement. This becomes worse if the level spacing is lowered, and for $\Delta/\Gamma=0.01$ both methods do no longer yield the same number of transmission zeros. As before, this breakdown of our truncation scheme is signalised by some components of the two-particle vertex which  start to grow large if $\Delta$ is gradually decreased. On the other hand, the larger the level detuning becomes, the more $G(V_g)$ and $\alpha(V_g)$ agree, and for $\Delta\approx\Gamma$ the fRG and NRG results can barely be distinguished.

To summarise, one could say that the fRG performs extremely well for nearly degenerate levels for all parameters under consideration if the strength of the interaction is $U/\Gamma\approx 1.0$. Even for larger $U/\Gamma\approx 5.0$, it captures the right number of transmission resonances, but the NRG data is no longer matched quantitatively. This changes if the level spacing is increased, and for $\Delta\approx\Gamma$ the results of both methods can barely be distinguished. We encounter serious problems with our second order fRG approximation scheme if the level spacing becomes very small (typically $\Delta/\Gamma=0.01$ for $U/\Gamma\approx 1$).

\subsection{Spinful Dots}
\begin{figure}[t]	
	\centering
        \includegraphics[width=0.475\textwidth,height=4.8cm,clip]{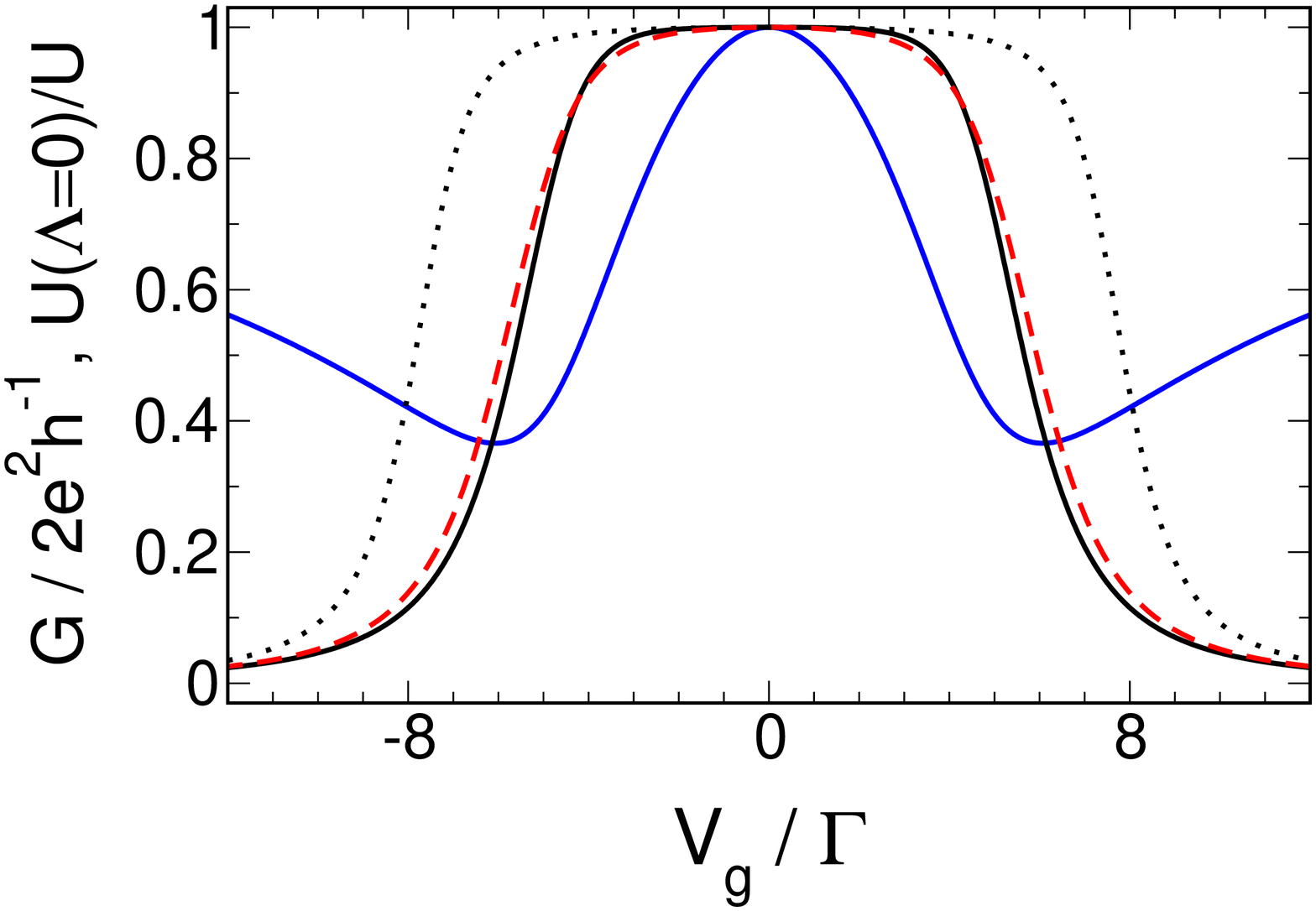}\hspace{0.035\textwidth}
        \includegraphics[width=0.475\textwidth,height=4.8cm,clip]{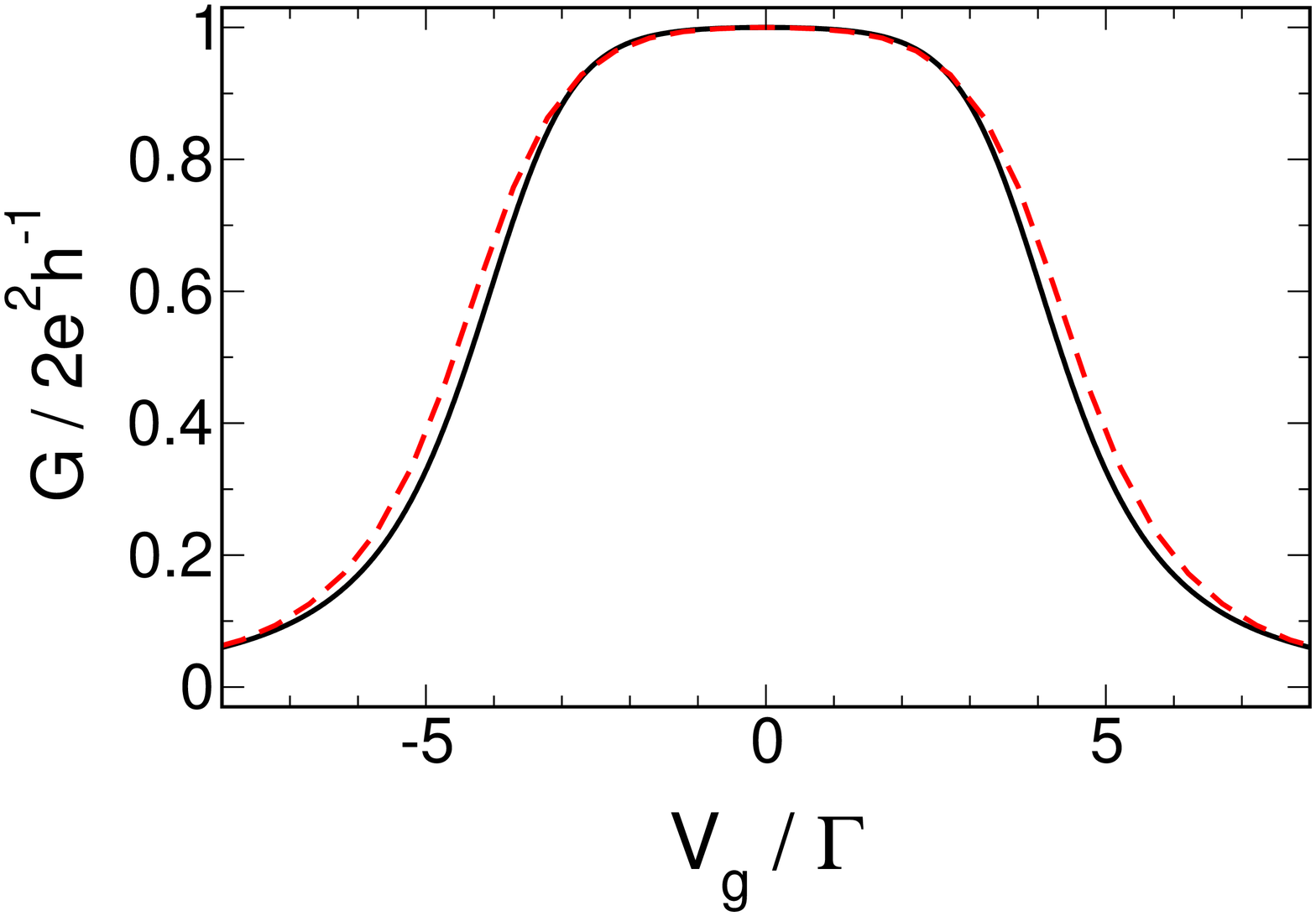}
        \caption{\textit{Left panel:} Gate voltage dependence of the conductance $G$ for a single-level dot with $U/\Gamma=4\pi$, $\Gamma_L/\Gamma=\Gamma_R/\Gamma=0.5$, and zero magnetic field obtained by fRG calculations (black solid line) and the exact Bethe ansatz solution (red line). The data for the latter was taken from \cite{kondonrg2}. First order fRG calculations (black dotted line) and the effective interaction at the end of the flow (blue line) are shown as well. \textit{Right panel:} The same for $U/\Gamma=3\pi$. The red line shows NRG data from \cite{kondonrg}.}
\label{fig:COMP.sd}
\end{figure}

In this section, we will compare fRG and NRG calculations for every geometry of spinful dots that was described in Sec.~\ref{sec:MS} in order to obtain a complete picture of how the fRG performs in the different physical situations.

\subsubsection{The Single-Level Case}

We will start with the simplest case of a single-level dot containing interacting spin up and down electrons. Surprisingly, this geometry allows for an analytic solution based on Bethe ansatz \cite{hewson}, and it shows that the fRG succeeds in describing the Kondo plateau which develops for large $U/\Gamma$ in $G(V_g)$ even on a quantitative level. In particular, for $U/\Gamma=4\pi$ (Fig.~\ref{fig:COMP.sd}, left panel), the fRG curve and the exact solution are barely distinguishable. The right panel of Fig.~\ref{fig:COMP.sd} shows another comparison for $U/\Gamma=3\pi$ with NRG data, which yields results practically identical to the exact ones on the scale of $U$. The agreement between fRG and the analytic Bethe ansatz solution becomes only sightly worse for the largest $U/\Gamma=25$ for which data for the latter is available in the literature. Consistent with the reliability of our results is that first and second order calculations are in good agreement, or, equivalently, that the renormalized interaction at the end of the flow does not diverge.

\begin{figure}[t]	
	\centering
        \includegraphics[width=0.475\textwidth,height=4.8cm,clip]{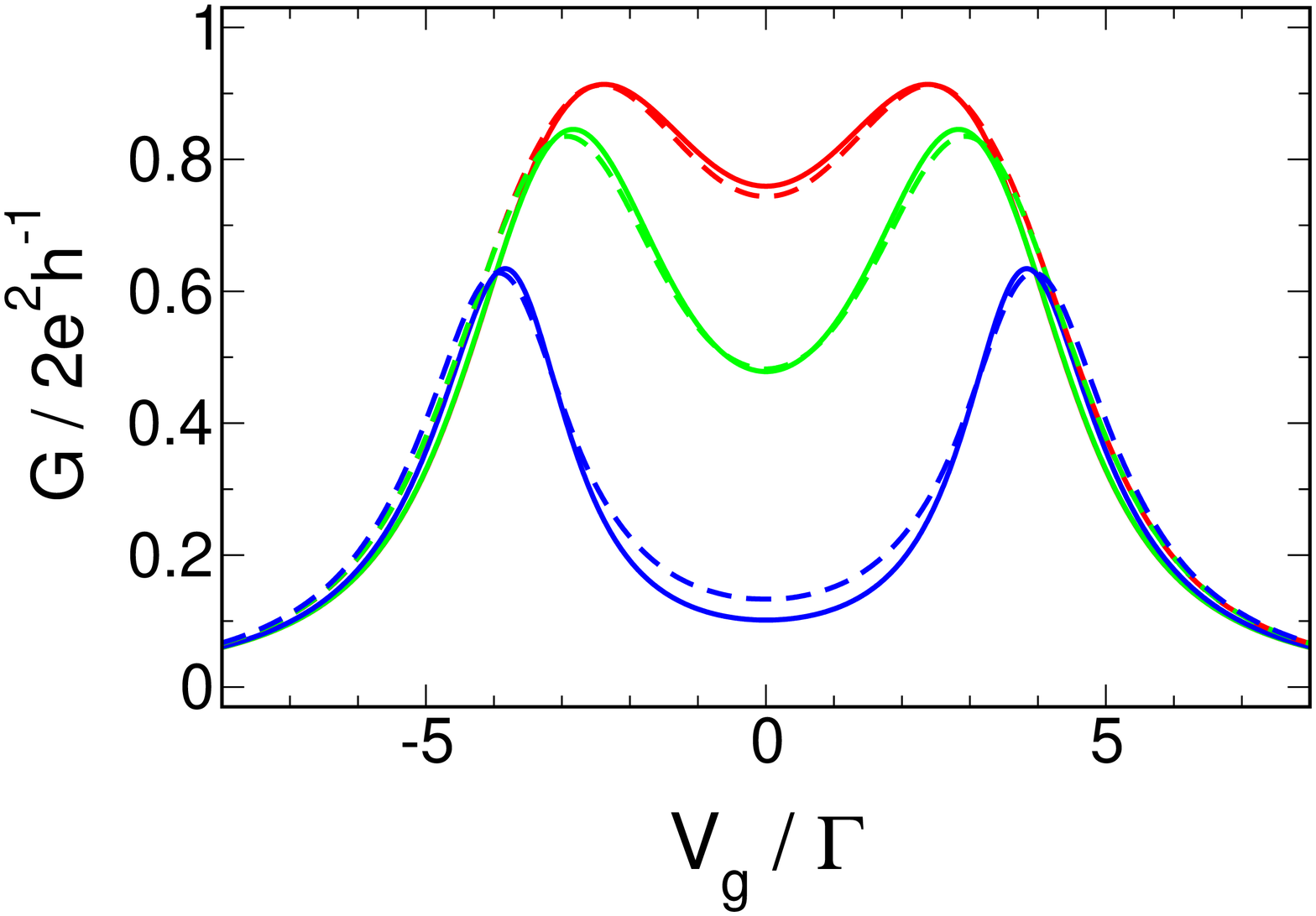}\hspace{0.035\textwidth}
        \includegraphics[width=0.475\textwidth,height=4.8cm,clip]{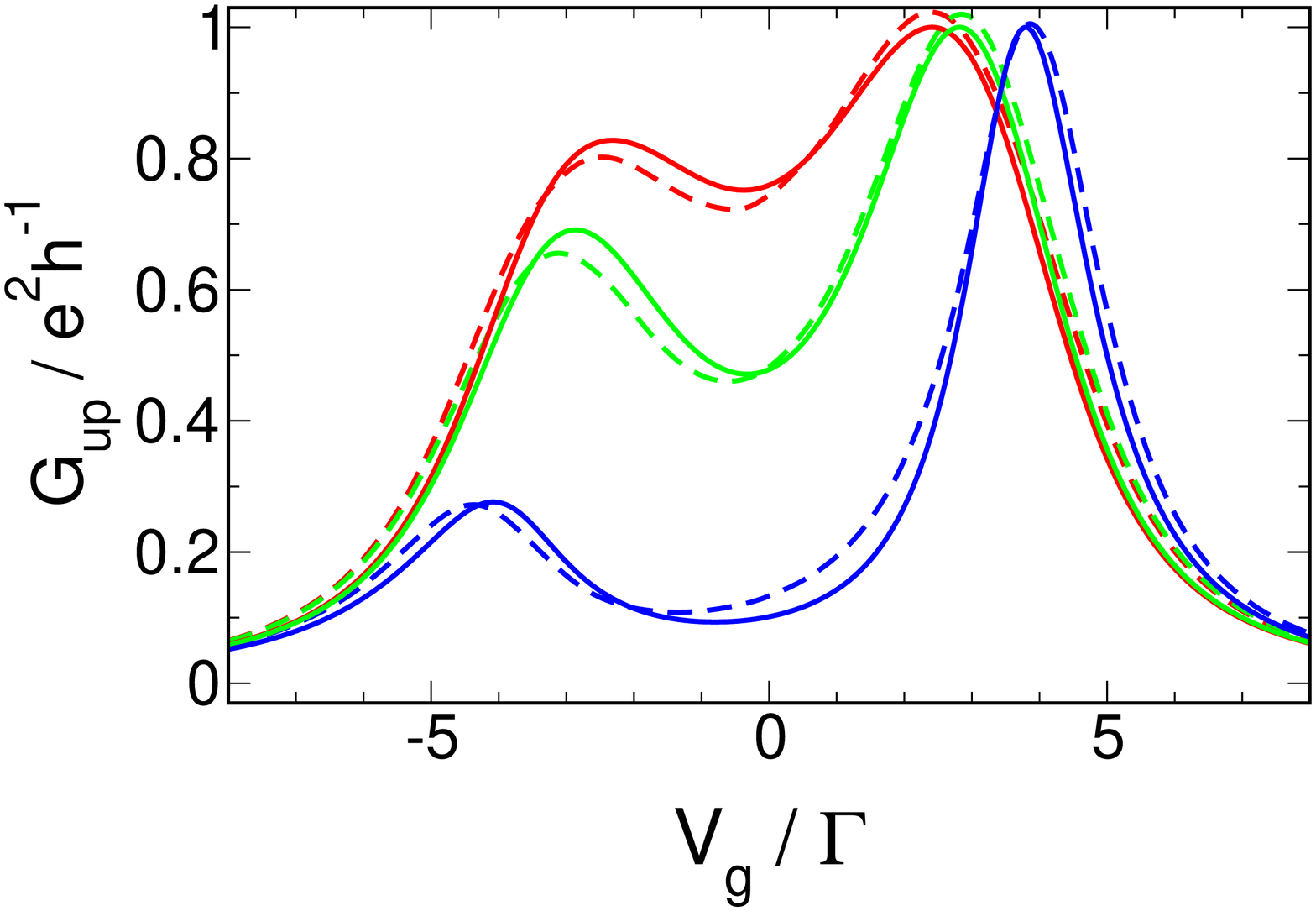}
        \caption{Conductance $G$ (left panel) and partial conductance $G_{\sigma=\tn{up}}$ (right panel) as a function of the gate voltage for a single-level dot with $U/\Gamma=3\pi$, $\Gamma_L/\Gamma=\Gamma_R/\Gamma=0.5$ for three different magnetic fields $B/T_K^\tn{NRG}=0.5$ (red), $B/T_K^\tn{NRG}=1.0$ (green), and $B/T_K^\tn{NRG}=5.0$ (blue). Each panel shows a comparison between fRG (solid lines) and NRG (dashed lines, taken from \cite{kondonrg2}) data. The Kondo temperature determined by NRG is $T_K^\tn{NRG}/\Gamma=0.116$.}
\label{fig:COMP.sd.b}
\end{figure}

Next, we consider how the fRG performs in presence of finite magnetic fields that lift the degeneracy between the electrons of both spin directions. For $B>0$, the Kondo resonance splits up at a scale determined by the Kondo temperature (which in an NRG approach is usually defined by the width of the sharp resonance at zero frequency in the spectral function), and it shows that the fRG excellently captures this behaviour (Fig.~\ref{fig:COMP.sd.b}). The same holds for the partial conductance of the spin up and spin down electrons. A point of particular importance is that the NRG calculations confirm that for $B\approx T_K^\tn{NRG}$ the conductance at half filling is indeed suppressed down to half the unitary limit, which justifies the reliability of our fRG definition of the Kondo `temperature' by the magnetic field dependence of $G(V_g)$. Due to the (practical) lack of NRG data for arbitrary interaction $U$, it seems meaningful to probe this reliability again by considering the dependence of $T_K^\tn{fRG}$ on the dot parameters $V_g$ and $U$. As shown in Sec.~\ref{sec:MS.sd}, it turns out that the numerically computed $T_K^\tn{fRG}(V_g,U)$ can indeed be fitted by a function consistent with the exact form of $T_K^\tn{exact}(V_g,U)$,
\begin{equation*}
f(U/\Gamma) = a(V_g)\exp\left[-\left|b(V_g)\frac{U}{\Gamma}-c(V_g)\frac{\Gamma}{U}\right|\right],
\end{equation*} 
with fit coefficients $a$, $b$, and $c$ that are to by determined as a function of the gate voltage. It proved that while $b\approx 0.32$ barely changes and is close to the exact value $\pi/8$, $c$ increases quadratically with $V_g$ while the dependence of $a$ on $U$ and the gate voltage is only weak. Alltogether, this shows that the fRG excellently describes the conductance for the single-level dot no matter how the parameters $U$ and $B$ are chosen.

\subsubsection{Linear Chains of Dots}
\begin{figure}[t]	
	\centering
        \includegraphics[width=0.475\textwidth,height=4.4cm,clip]{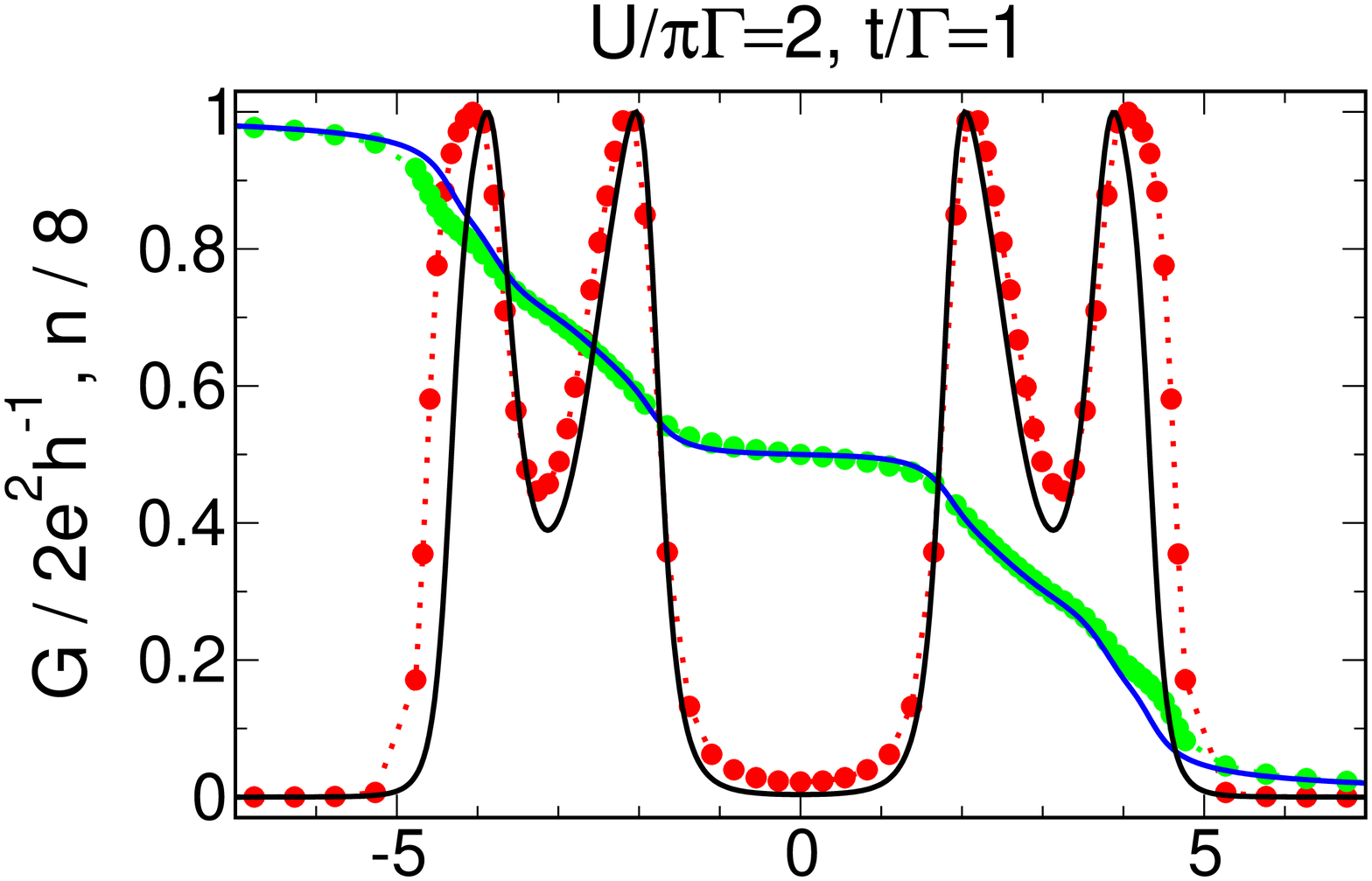}\hspace{0.035\textwidth}
        \includegraphics[width=0.475\textwidth,height=4.0cm,clip]{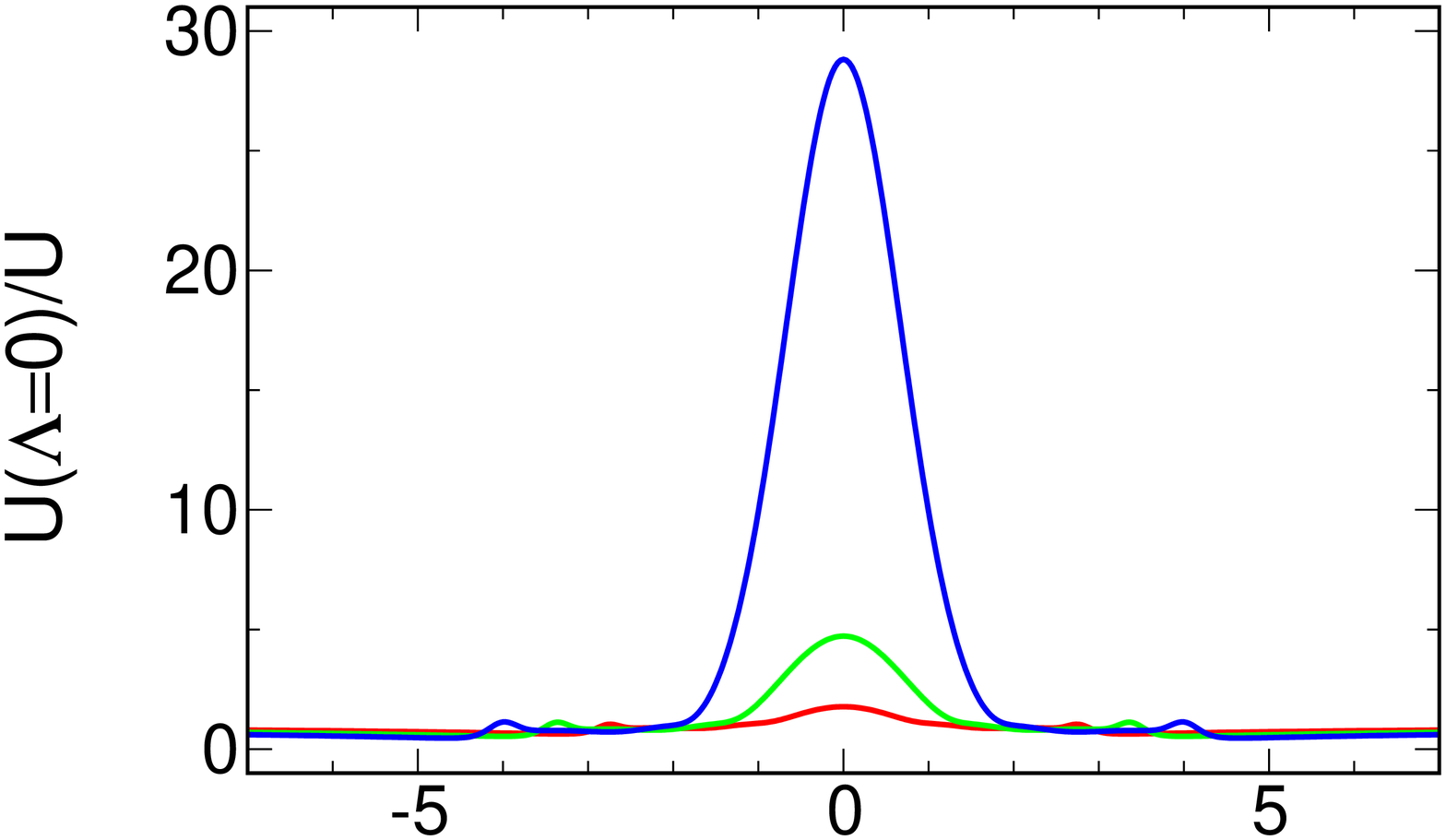}\vspace{0.3cm}
        \includegraphics[width=0.475\textwidth,height=5.2cm,clip]{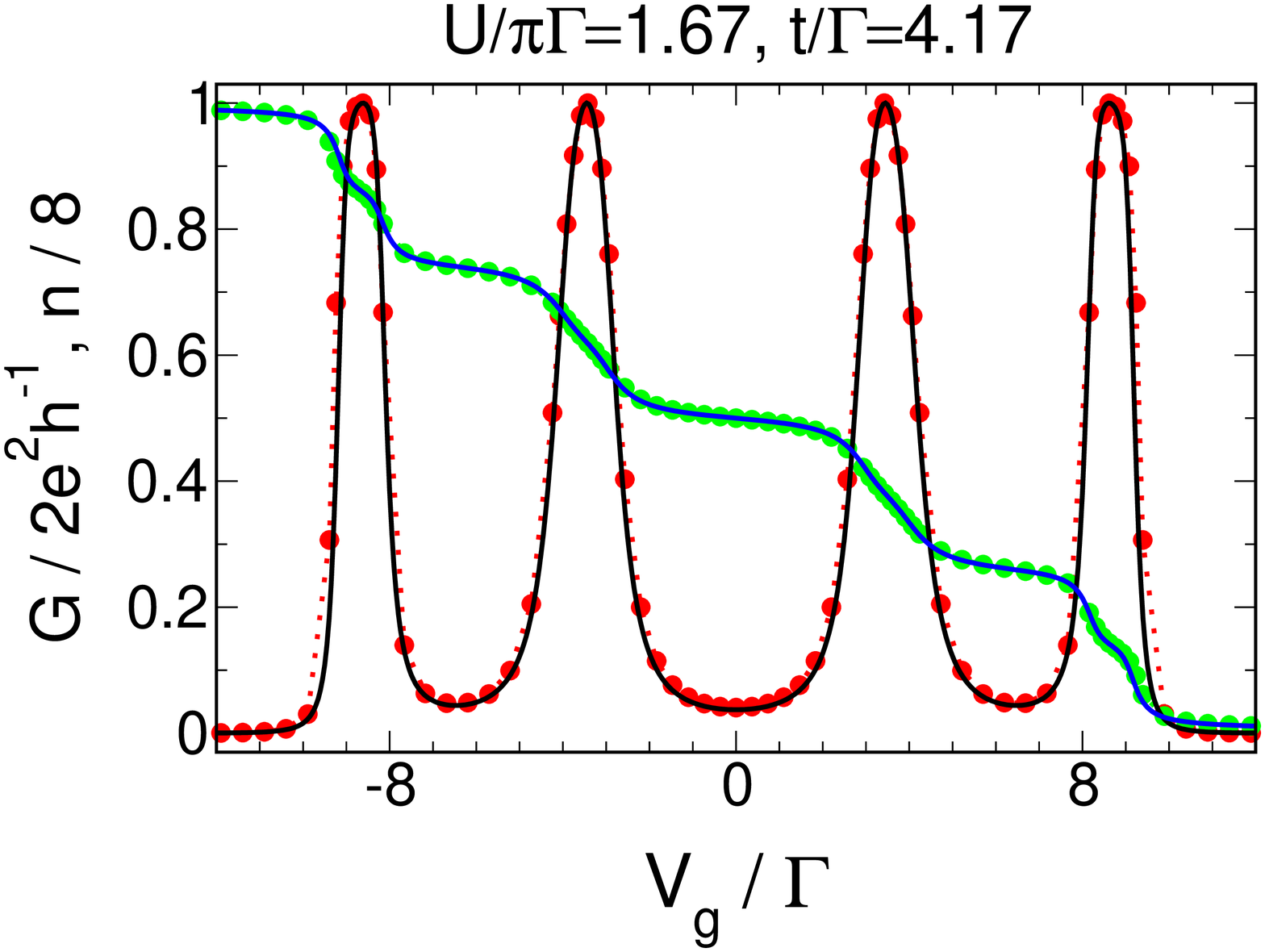}\hspace{0.035\textwidth}
        \includegraphics[width=0.475\textwidth,height=5.2cm,clip]{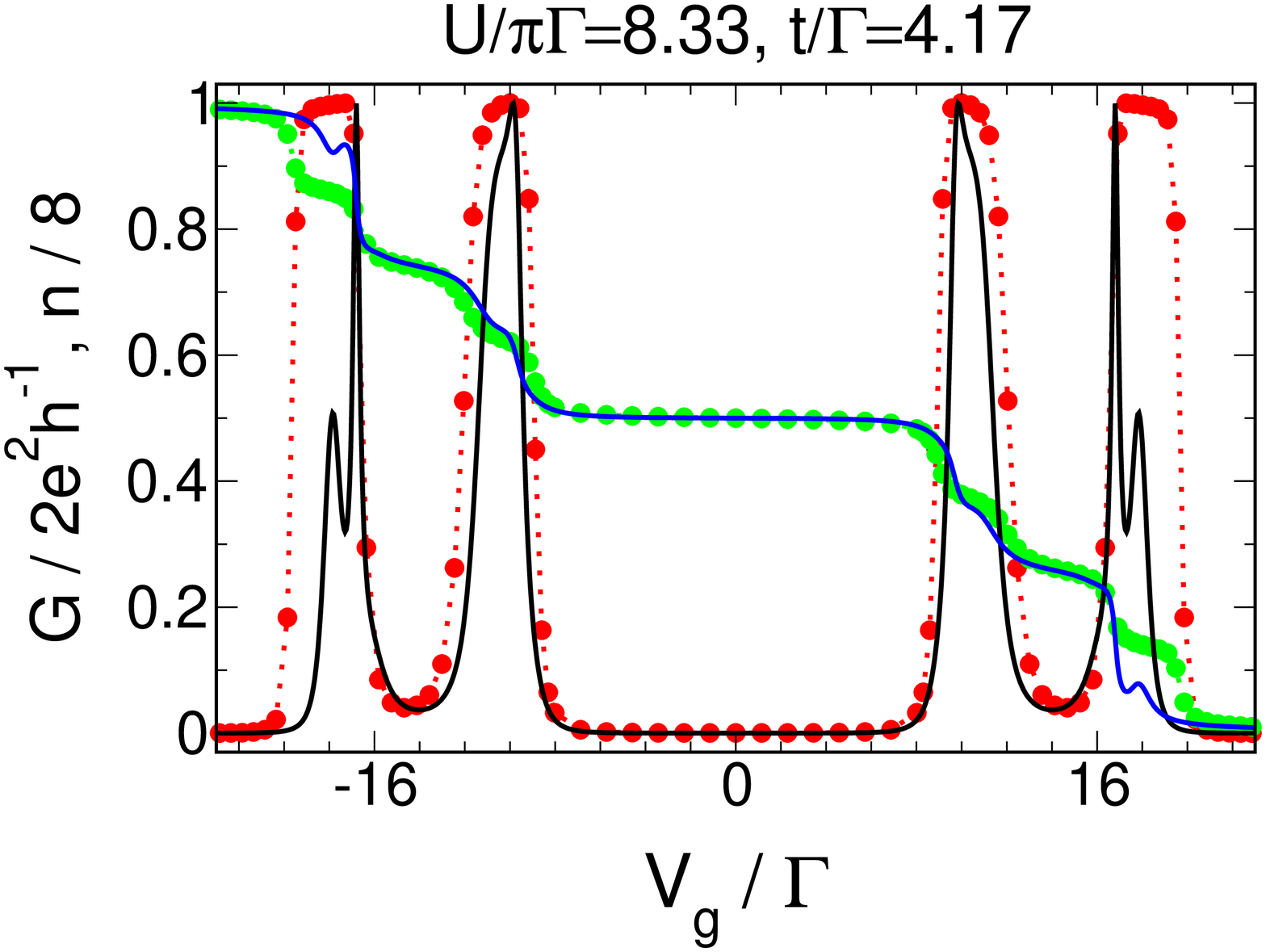}
        \caption{Comparison between fRG and NRG calculations of the conductance $G$ (fRG: black, NRG: red) and average occupation $n$ (fRG: blue, NRG: green) of a chain of four spin-degenerate dots with local interaction $U$, nearest-neighbour hopping $t$, and left-right symmetric couplings with the leads. Note that here $G$ and $n$ are not directly related. The NRG data was taken from \cite{chain4}. \textit{Upper right panel:} Effective local interaction on the first dot at the end of the fRG flow for $t/\Gamma=1.0$, $\Gamma_L=\Gamma_R$, $U/\Gamma=\pi$ (red), $U/\Gamma=1.5\pi$ (green), and $U/\Gamma=2\pi$ (blue).}
\label{fig:COMP.chain4}
\end{figure}

Next, we turn to the case of a few single-level dots coupled by a nearest-neighbour hopping $t$ to a short Hubbard chain. As explained in Sec.~\ref{sec:MS.chain}, the Kondo effect is active and the conductance for this geometry exhibits a resonance each time the dot region is filled by an odd number of electrons. For large $t$, on which we have mainly concentrated, the peaks are well-separated. If the chain comprises an even number of sites, there is a wide valley of small conductance around half-filling because charge fluctuations are strongly suppressed. On the other hand, half-filling corresponds to an odd number of electrons if the number of dots in the chain is odd, and hence the resonance at $V_g=0$ is much wider than the others in such cases.

By comparison to NRG data for the four-site chain provided by \cite{chain4}, it shows that our second order fRG truncation scheme quantitatively captures the physics, especially the wide conductance valley at half-filling, for interactions as large as $U/\Gamma=2\pi$ if the nearest-neighbour hopping is of size $t/\Gamma=1.0$ (Fig.~\ref{fig:COMP.chain4}, upper left panel). If $U$ is further increased, some components of the effective interaction at the end of the flow grow larger and larger for gate voltages on-resonance or close to $V_g=0$, signalising a breakdown of our approximation. This can already be observed for the aforementioned parameter set (Fig.~\ref{fig:COMP.chain4}, right panel): at half-filling, we find $U(\la=0)/U\approx 30$ and the conductance in the valley is significantly underestimated by fRG. The agreement between both methods improves if the hopping $t$ is increased. For example, for $t/\Gamma=4.167$ and $U/\Gamma=1.67\pi$, $G(V_g)$ and $n(V_g)$ are barely distinguishable on a scale $\Gamma$ (Fig.~\ref{fig:COMP.chain4}, lower left panel). Increasing the interaction to $U/\Gamma=8.33\pi$, some components of the two-particle vertex grow large and the fRG results seriously deviate from the reliable NRG data (Fig.~\ref{fig:COMP.chain4}, lower right panel).

\begin{figure}[t]	
	\centering
        \includegraphics[width=0.475\textwidth,height=5.2cm,clip]{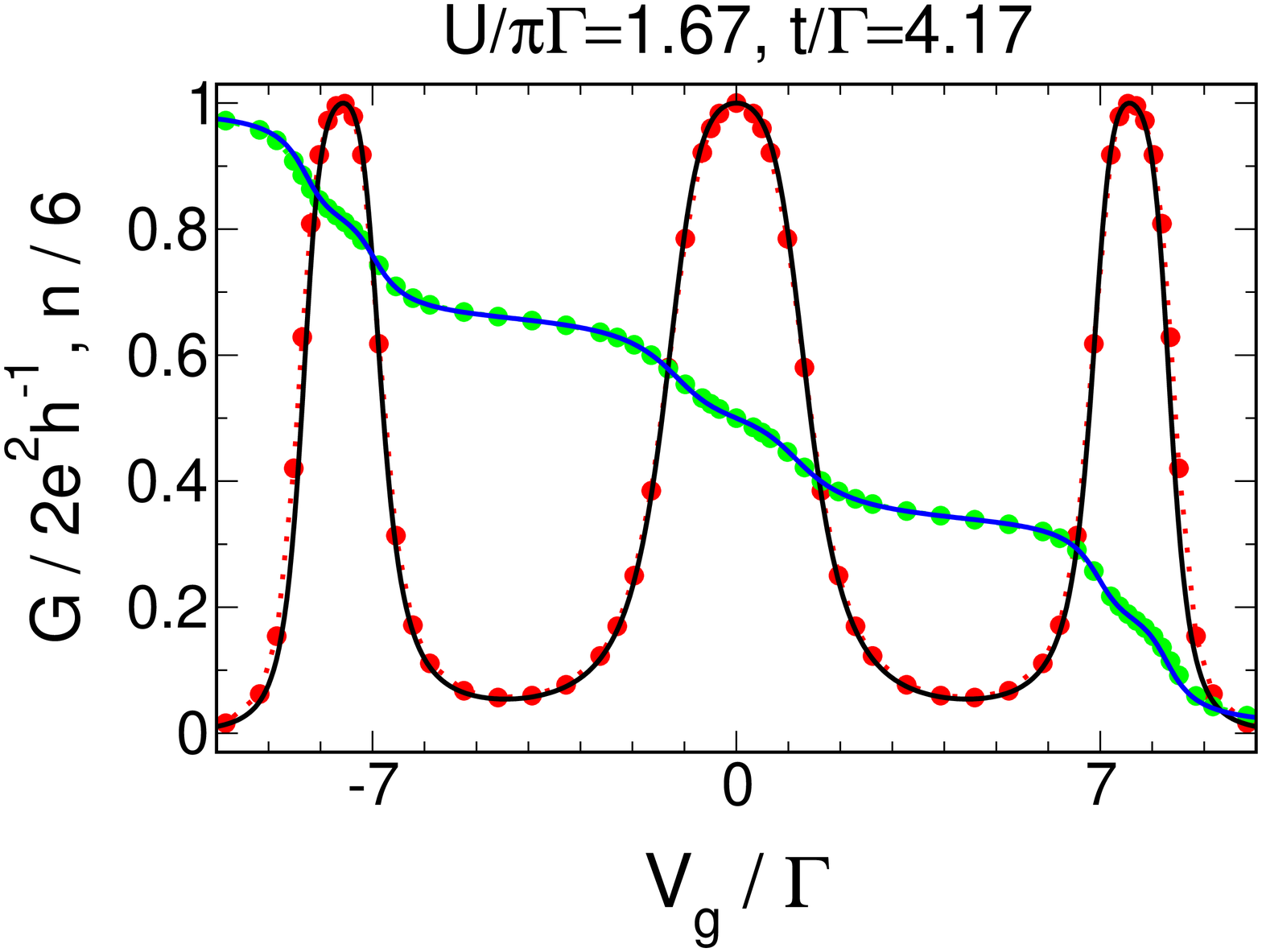}\hspace{0.035\textwidth}
        \includegraphics[width=0.475\textwidth,height=5.2cm,clip]{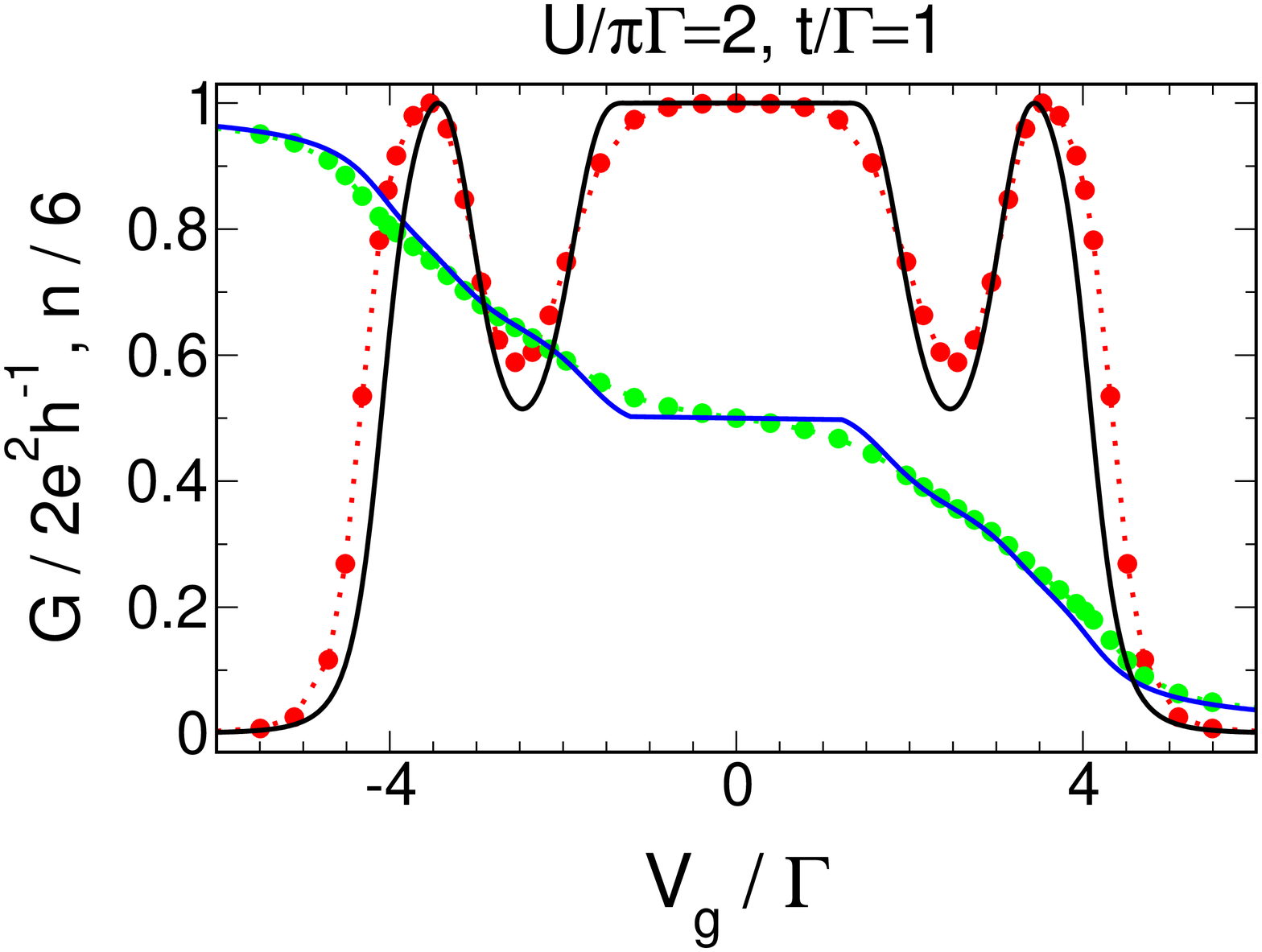}
        \caption{\textit{Left panel:}  The same as Fig.~\ref{fig:COMP.chain3}, but for a chain with three sites with NRG data from \cite{chain3a}.}
\label{fig:COMP.chain3}
\end{figure}

The picture when comparing fRG and NRG data (the latter taken from \cite{chain3a}) for a chain with three levels is basically the same. For $U/\Gamma=1.67\pi$ and $t/\Gamma=4.17$, the results from both methods can barely be distinguished (Fig.~\ref{fig:COMP.chain3}, left panel). The smaller the hopping between neighbouring sites becomes, the sooner the second order approximation scheme encounters problems for gate voltages on-resonance when $U$ is gradually increased. For example, for $U/\Gamma=2\pi$ and $t/\Gamma=1.0$, the fRG overestimates the suppression of charge fluctuation at half-filling (Fig.~\ref{fig:COMP.chain3}, right panel). Again, this breakdown is accompanied by diverging components of the two-particle vertex.

\subsubsection{The Side-Coupled Geometry}

Now, we turn the side-coupled geometry which was discussed in detail in Sec.~\ref{sec:MS.side}. We have argued that in presence of a local interaction $U$ two regimes with different physics arise in dependence of the inter-dot hopping $t$. Close to half-filling, either a local spin-singlet state is formed between the electrons on the two dots (large $t$), or the low-energy physics is dominated by a two-stage Kondo effect (small $t$). In both cases, the conductance at $V_g=0$ is heavily suppressed, and so at $T=0$ both phases can only be distinguished unambiguously by the magnetic field dependence of the conductance.

\begin{figure}
  \centering
        \includegraphics[width=0.475\textwidth,height=5.2cm,clip]{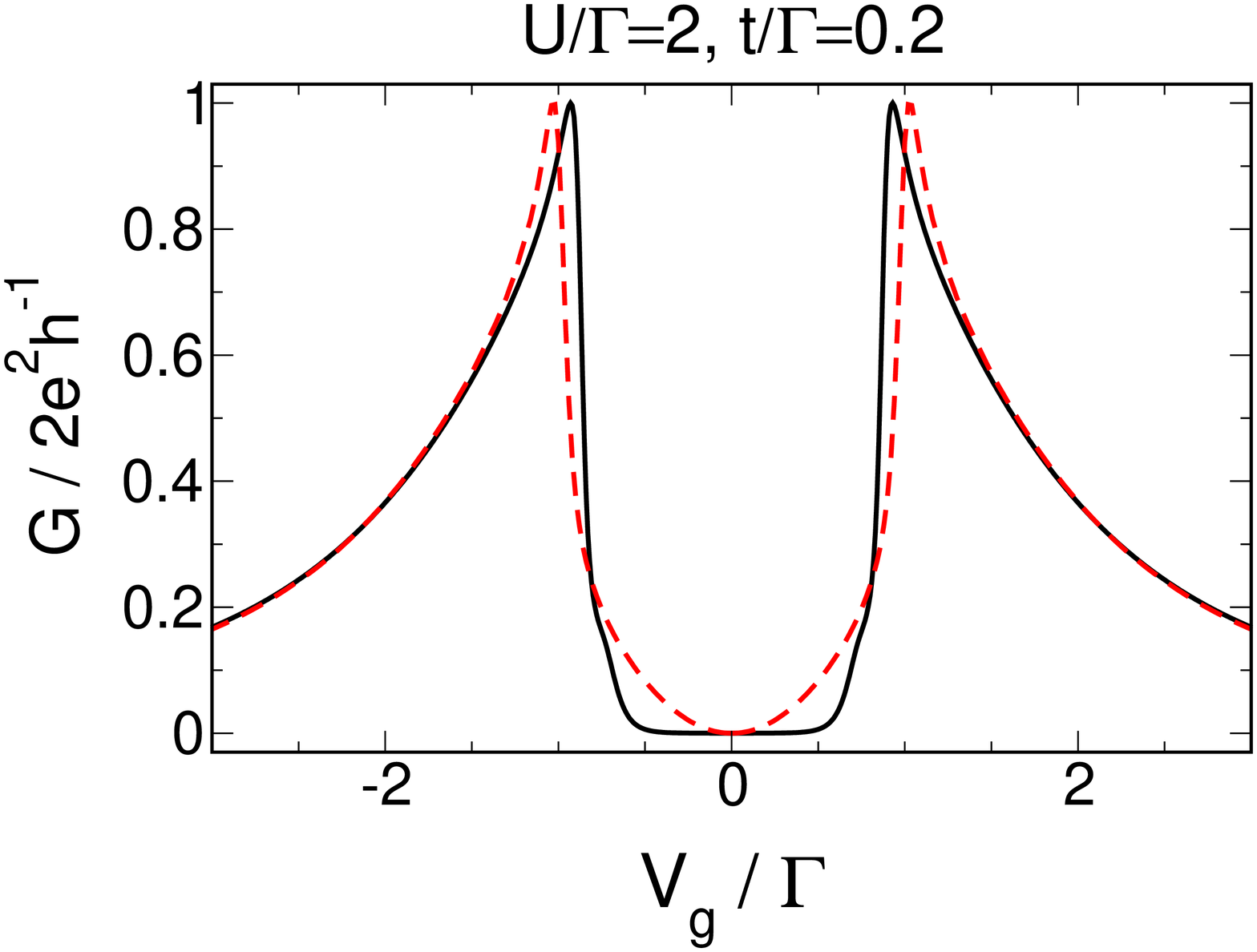}\hspace{0.035\textwidth}
        \includegraphics[width=0.475\textwidth,height=4.8cm,clip]{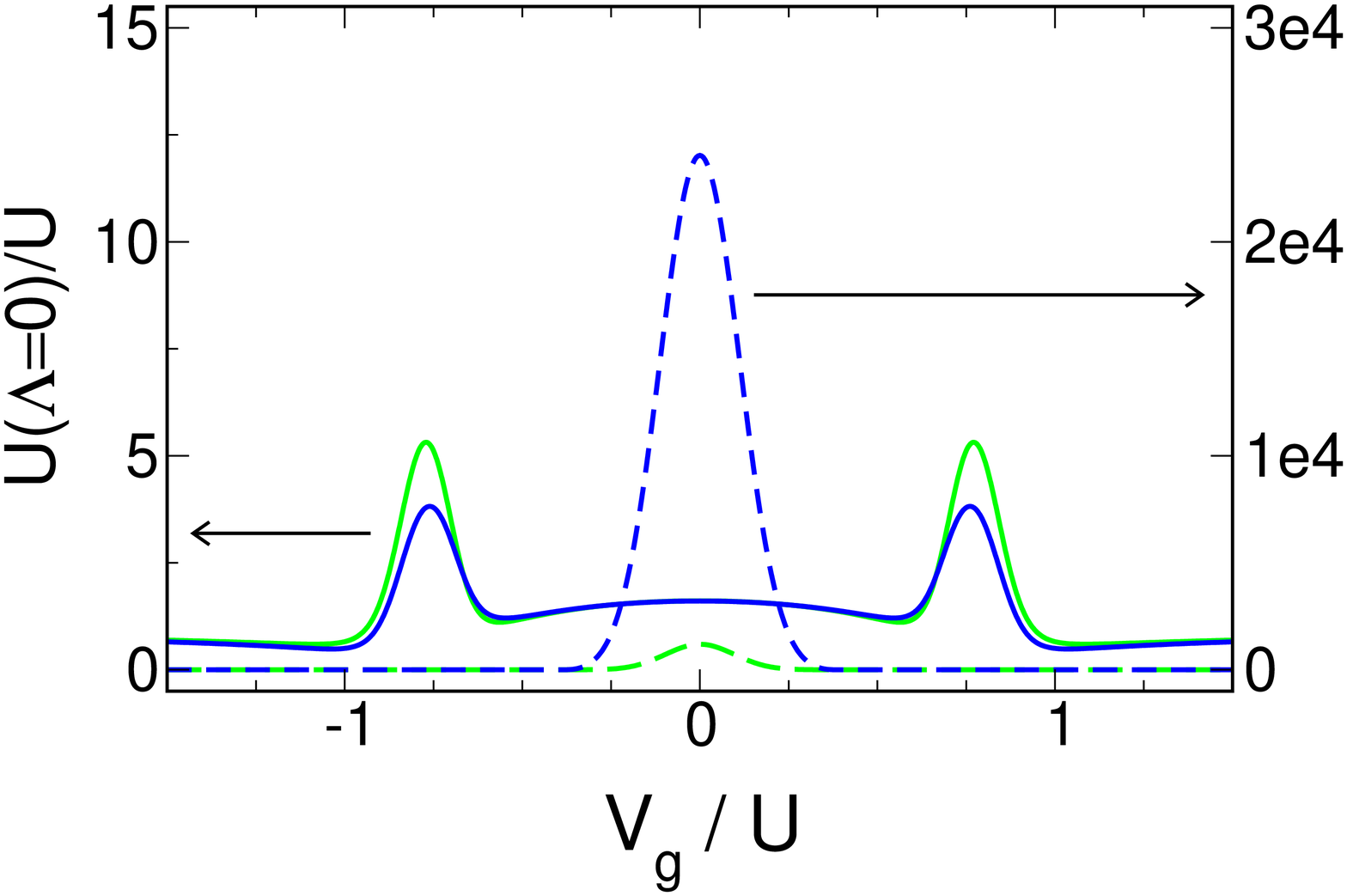}
        \caption{\textit{Left panel:} Comparison between NRG data from \cite{side2} and fRG for the conductance $G$ (fRG: black, NRG: red) of the side-coupled geometry in the two-stage Kondo regime. The level-lead couplings have been chosen left-right symmetric. \textit{Right panel:} Effective local interaction at the end of the fRG flow on the embedded dot (green) and on the side-coupled dot (blue) for the two different parameter sets $U/\Gamma=8.0$, $t/\Gamma=4.0$ (solid lines, left scale), and $U/\Gamma=2.0$, $t/\Gamma=0.2$ (dashed lines, right scale).}
\label{fig:COMP.side_a}
\end{figure}

In Fig.~\ref{fig:COMP.side_a}, we compare our second order fRG results with NRG calculations by \cite{side2}. For $U/\Gamma=2.0$ and $t/\Gamma=0.2$, both methods agree surprisingly well despite the large `effective' interaction $U/t^2=50$ (left panel). Nevertheless, one should note that the conductance close to half-filling is severely underestimated by fRG, and as usual this is signalised by large-growing components of the two-particle vertex (right panel, dashed lines). This becomes worse for smaller $t$, and the Fano-like feature in $G(V_g)$ observed by \cite{side3} for extremely large $U/t^2$ is out of reach within our approach. On the other hand, both methods agree in associating the aforementioned parameters to the two-stage Kondo regime. In particular, the evolution of $G(V_g)$ with small magnetic fields calculated by fRG and that with small temperatures obtained by NRG is qualitatively identical (compare the left panel of Fig.~\ref{fig:MS.side.b} with Fig.~6 in \cite{side2}). Moreover, the energy scale of the first stage Kondo effect for these parameters determined by fRG (by the field necessary to raise the conductance at $V_g=0$ to half the unitary limit), $(T_K^1)^\tn{fRG}\approx 1.16\Gamma$, is close to the NRG estimate, $(T_K^1)^\tn{NRG}\approx\Gamma$.

\begin{figure}
  \centering
        \includegraphics[width=0.475\textwidth,height=5.2cm,clip]{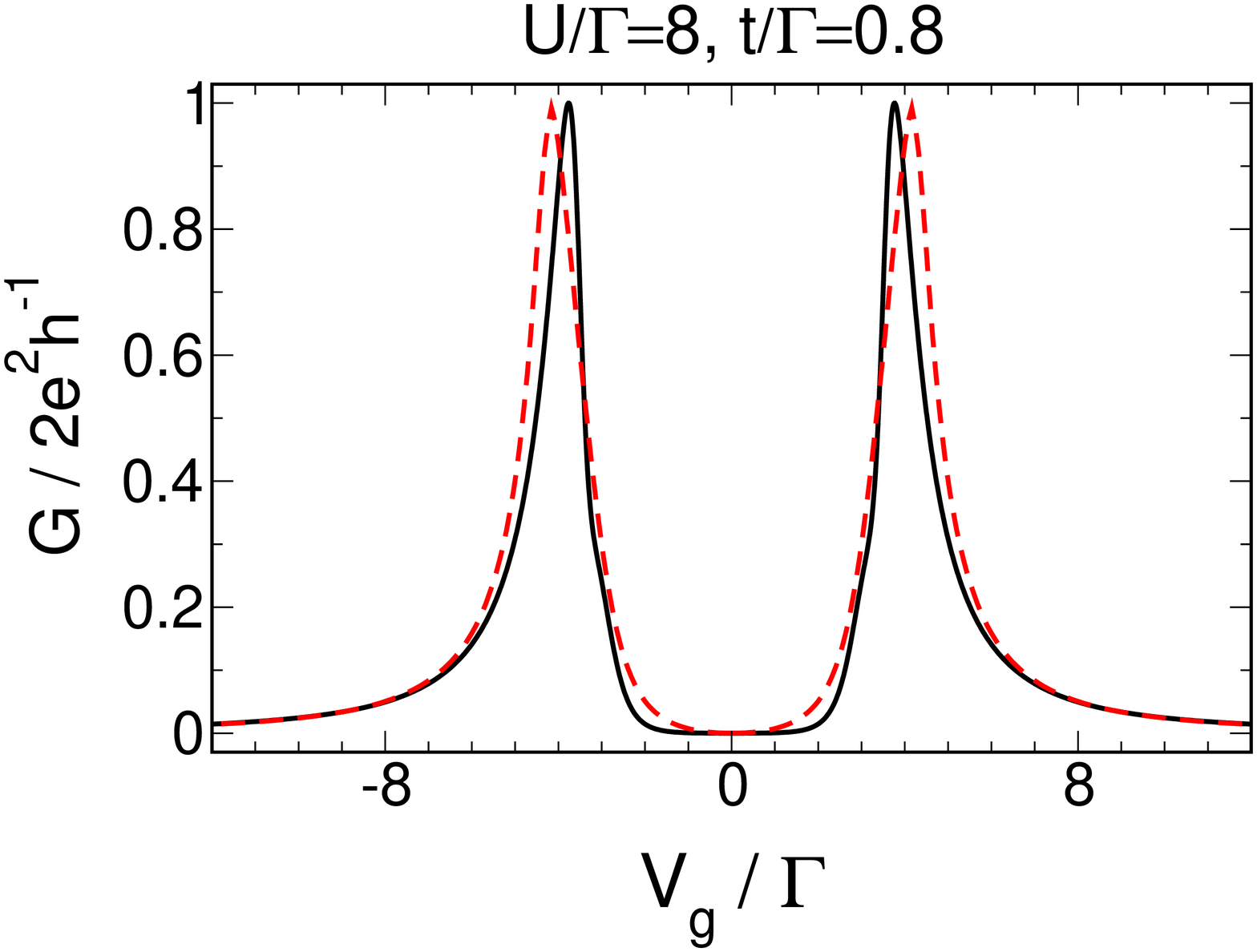}\hspace{0.035\textwidth}
        \includegraphics[width=0.475\textwidth,height=5.2cm,clip]{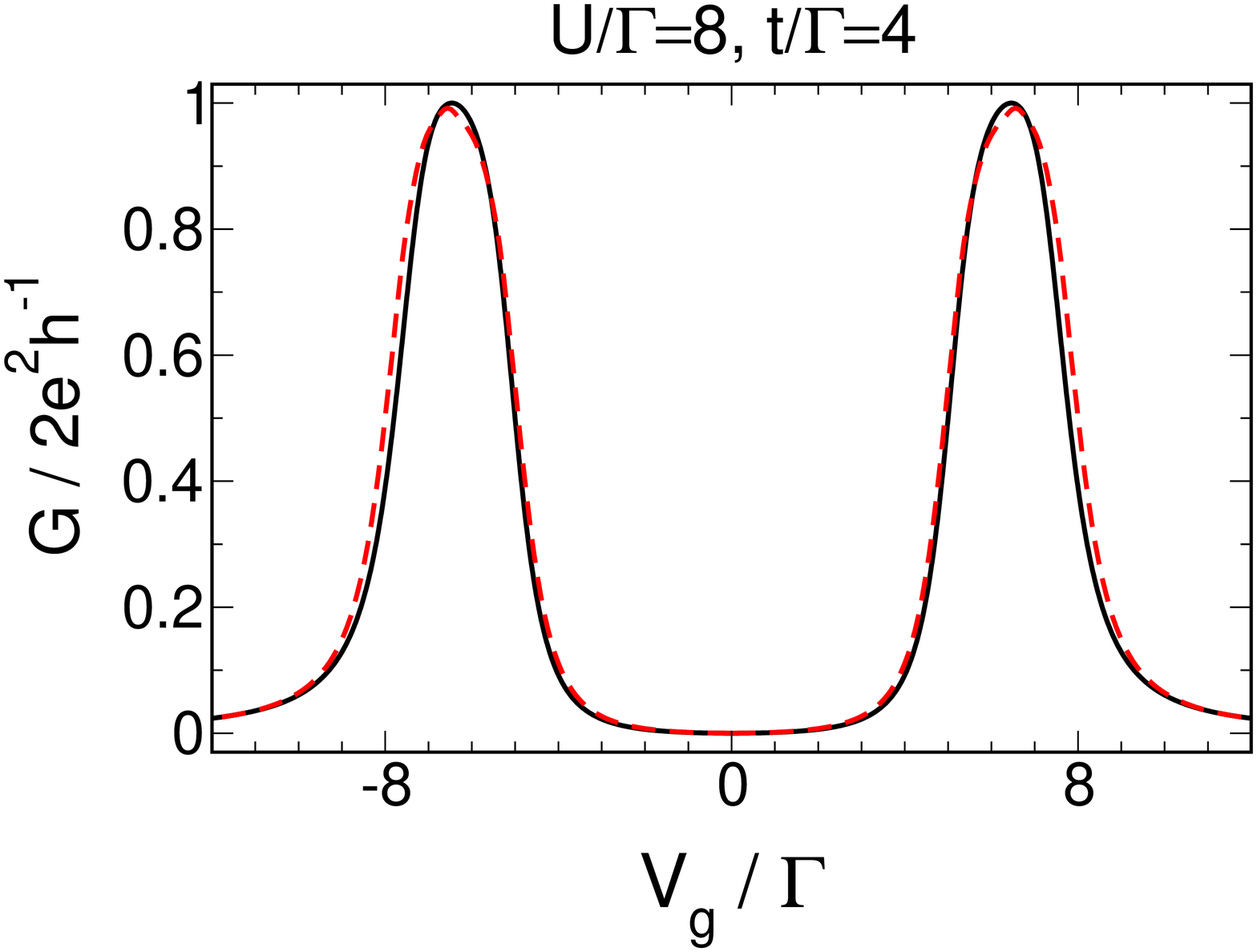}
        \caption{The same as in the left panel of Fig.~\ref{fig:COMP.side_a}, but for two different sets of parameters from the local spin-singlet phase.}
\label{fig:COMP.side_b}
\end{figure}

If we increase the level spacing, the agreement of the fRG calculations with the reliable NRG data becomes better. For $U/\Gamma=8.0$, $t/\Gamma=0.8$ (Fig.~\ref{fig:COMP.side_b}, left panel), the former still underestimates the suppression of $G$ close to $V_g=0$, but not as severely as for the parameters considered before, and consistently also the effective interaction does not grow that large ($U(\la=0)/U\approx500$ at $V_g=0$). Again, both methods agree in assigning the parameter set to the local spin-singlet phase. If we finally consider $U/\Gamma=8.0$, $t/\Gamma=4.0$ (Fig.~\ref{fig:COMP.side_b}, right panel), it proves hard to distinguish the fRG $G(V_g)$ curve from the precise NRG one, and $U(\la=0)$ stays always of order $U$ (right panel of Fig.~\ref{fig:COMP.side_a}, solid lines). Again, the physics is consistently identified to be dominated by the formation of the spin-singlet state, in particular the evolution of $G(V_g)$ with small temperatures and magnetic fields is similar (compare the right panel of Fig.~\ref{fig:MS.side.b} and Fig.~8 in \cite{side2}). The fRG computation of the energy scale of the Kondo effect active at each resonance (defined by the field required to suppress the conductance down to half the unitary limit), $(T_K)^\tn{fRG}\approx 0.16\Gamma$ is in good agreement with the NRG estimate, $(T_K)^\tn{NRG}\approx 0.156\Gamma$.

\subsubsection{Parallel Double Dots}

To the end of this section, we finally look whether the fRG performs similarly well in presence of Kondo physics that arises if the spinless double dot model considered in the previous section is extended by the spin degree of freedom. As reference data, we use very recent NRG calculations carried out by \cite{theresanrg}.

We start with describing the case where the level-lead couplings have the same relative sign, $s=+$. For nearly degenerate levels and even for small interactions $U\lesssim\Gamma$, the fRG significantly underestimates the conductance in a region of width $U$ around $V_g=0$, while still being in good agreement with NRG outside for large $U$ (see $U/\Gamma=3.0$ in the lower left panel of Fig.~\ref{fig:COMP.dd}). The same holds for the transmission phase and the average level occupancies which show an overpronounced plateau-like behaviour close to $V_g=0$. To understand this, it proves helpful to consider again the mapping of the parallel double dot back to the side-coupled geometry that was introduced in Sec.~\ref{sec:MS.side}. Strictly speaking, it requires $A$-$B$-symmetric hybridisations and the introduction of additional correlated hopping terms. However, we showed that the parallel dots with level spacing $\Delta$ can be understood as the side-coupled geometry with hopping $\Delta/2$ even if these assumptions are relaxed. In particular, we can easily interpret the double dot with nearly degenerate levels as the side-coupled dots being in the two-stage Kondo regime, and due to the very large $U/\Delta^2$, which is its relevant scale, it is not surprising that the second order fRG approximation fails to capture this effect quantitatively. Following this line of argumentation, it is also clear that the agreement between fRG and NRG improves if $\Delta$ is increased, and for $\Delta\approx\Gamma$ both results are practically identical on a scale $\Gamma$ (see $\Delta/\Gamma=5.0$ in the lower right panel of Fig.~\ref{fig:COMP.dd}).

Surprisingly, the fRG results compare well with the precise NRG data even for nearly degenerate levels if the relative sign of the level-lead couplings is $s=-$ (Fig.~\ref{fig:COMP.dd}, upper left panel). Consistent with this is the observation that in contrast to $s=+$ all components of the two-particle vertex at the end of the flow are of order of their initial value. The same holds if the level spacing is increased, and for $\Delta\approx\Gamma$ the curves produced by fRG and NRG are again barely distinguishable (Fig.~\ref{fig:COMP.dd}, upper right panel).

\section{Comparing fRG with fRG: Self-Consistency Checks}
\begin{figure}[t]	
	\centering
        \includegraphics[width=0.475\textwidth,height=4.4cm,clip]{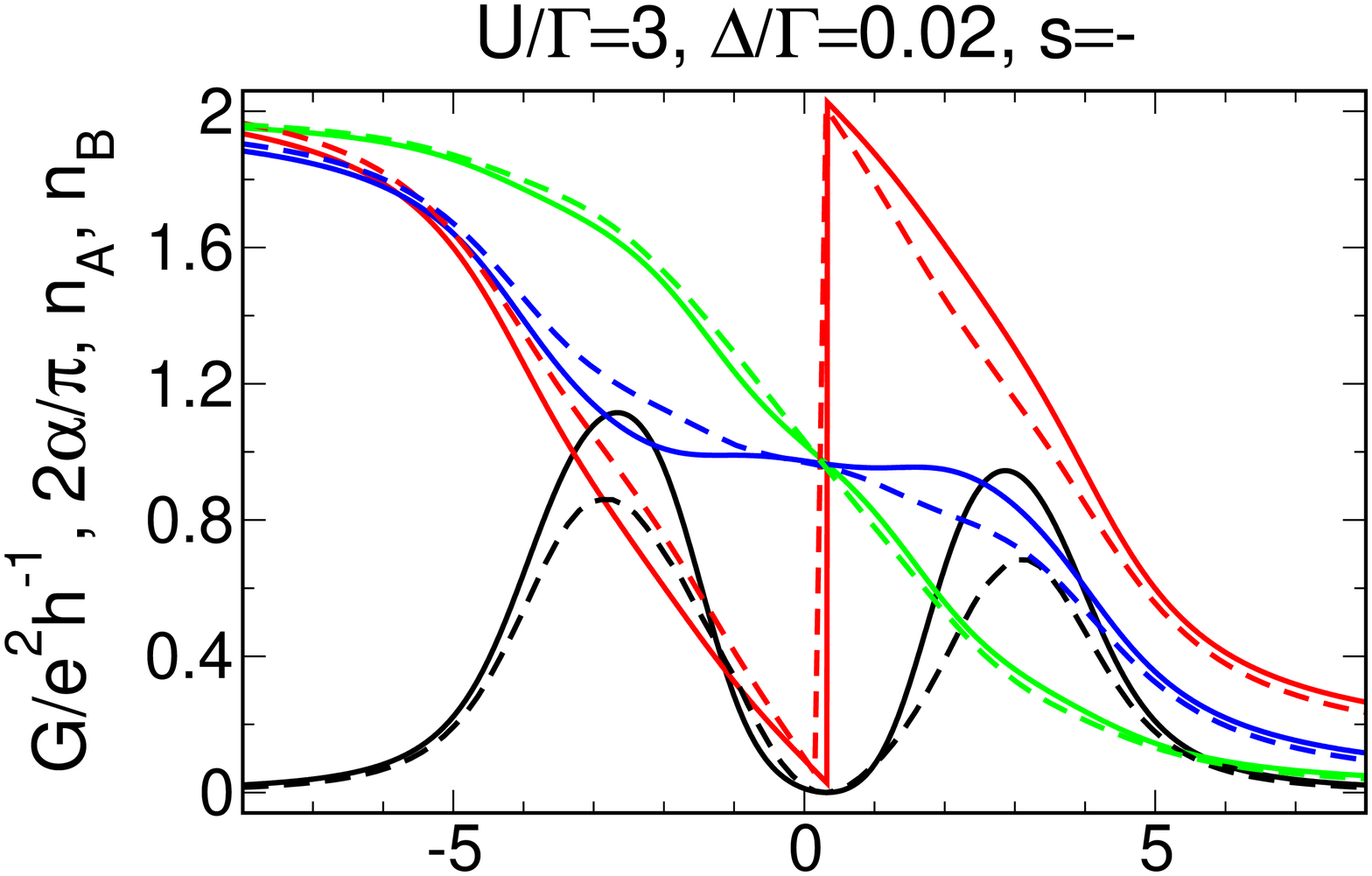}\hspace{0.035\textwidth}
        \includegraphics[width=0.475\textwidth,height=4.4cm,clip]{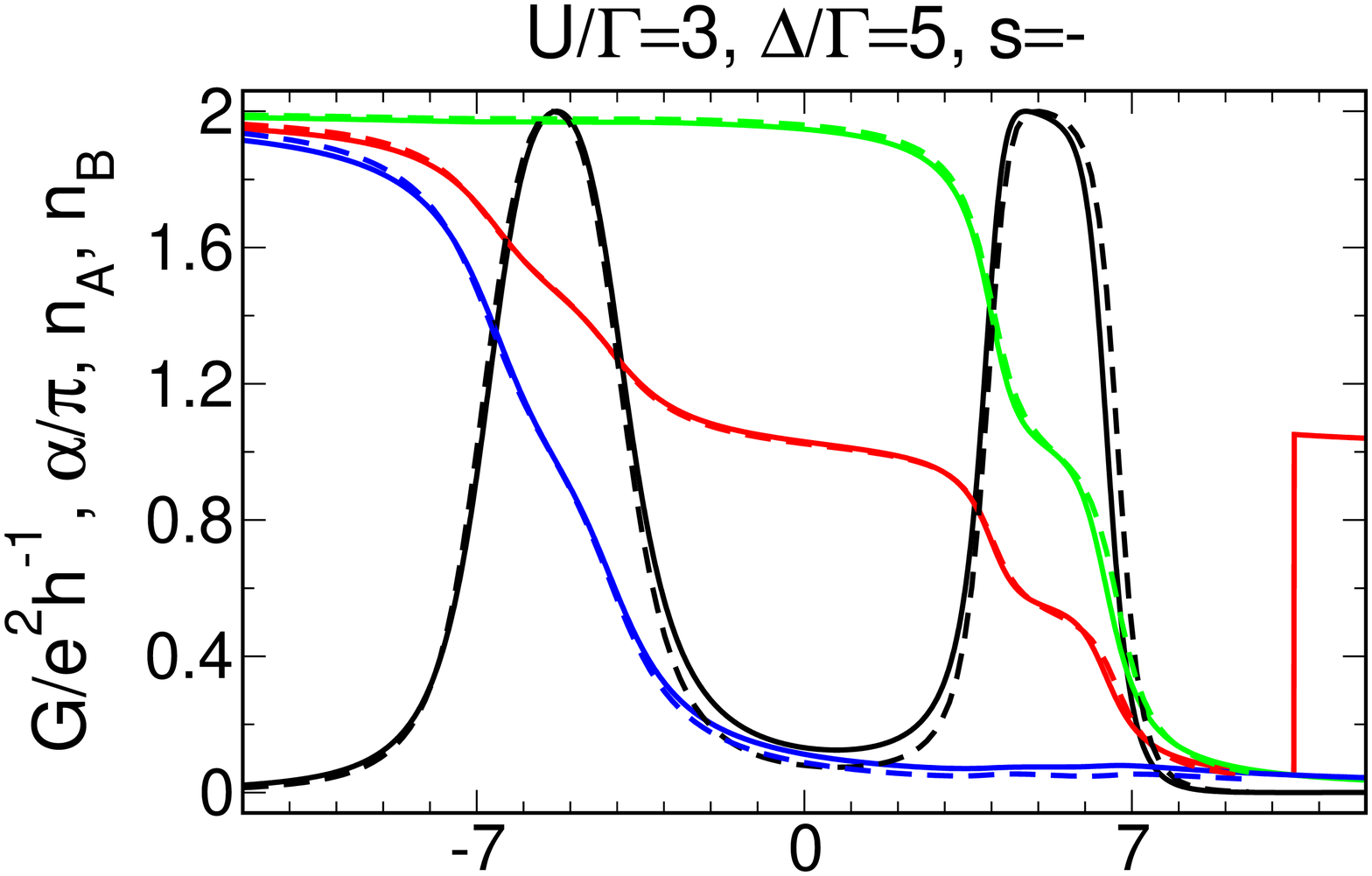}\vspace{0.3cm}
        \includegraphics[width=0.475\textwidth,height=5.2cm,clip]{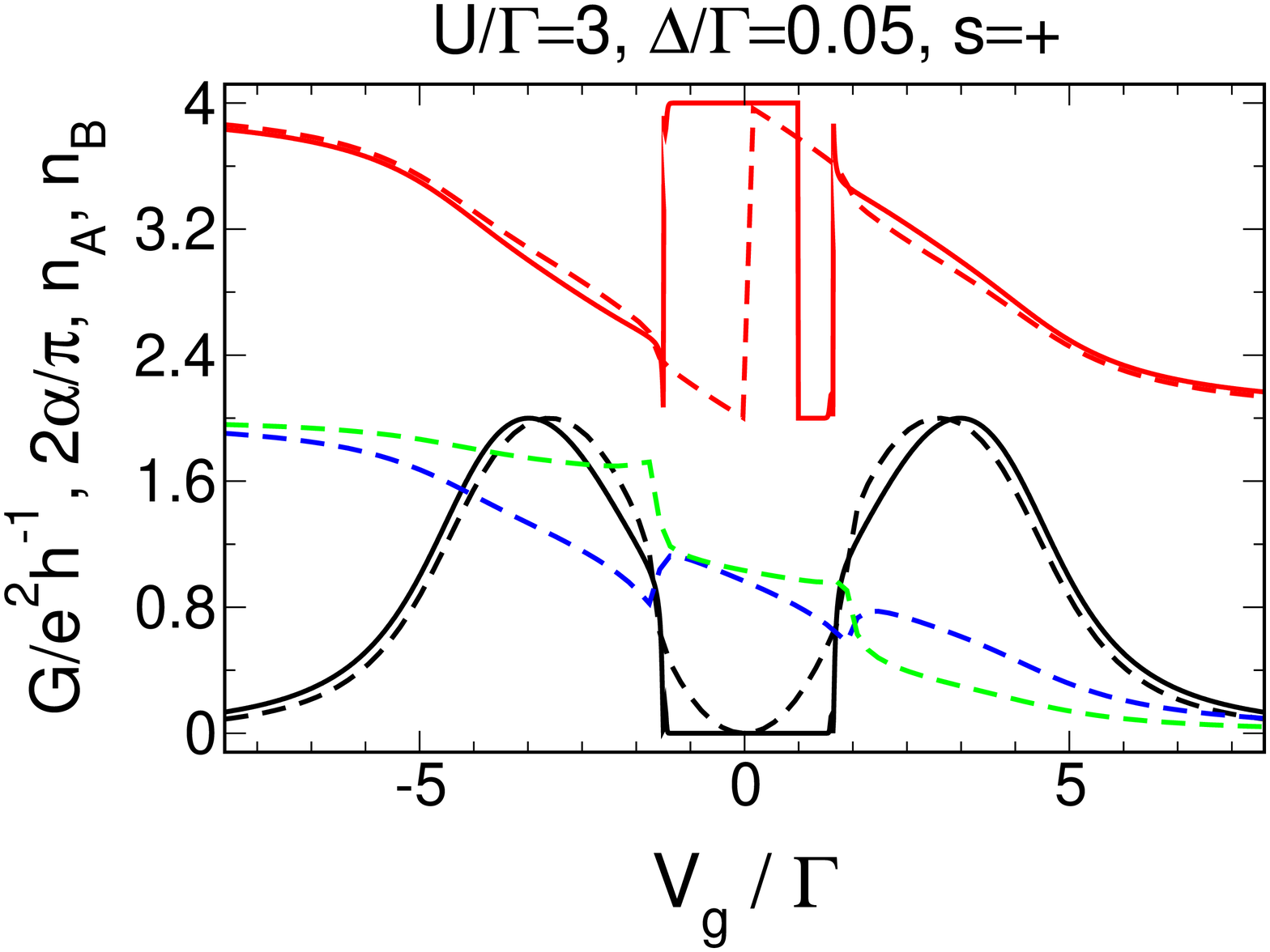}\hspace{0.035\textwidth}
        \includegraphics[width=0.475\textwidth,height=5.2cm,clip]{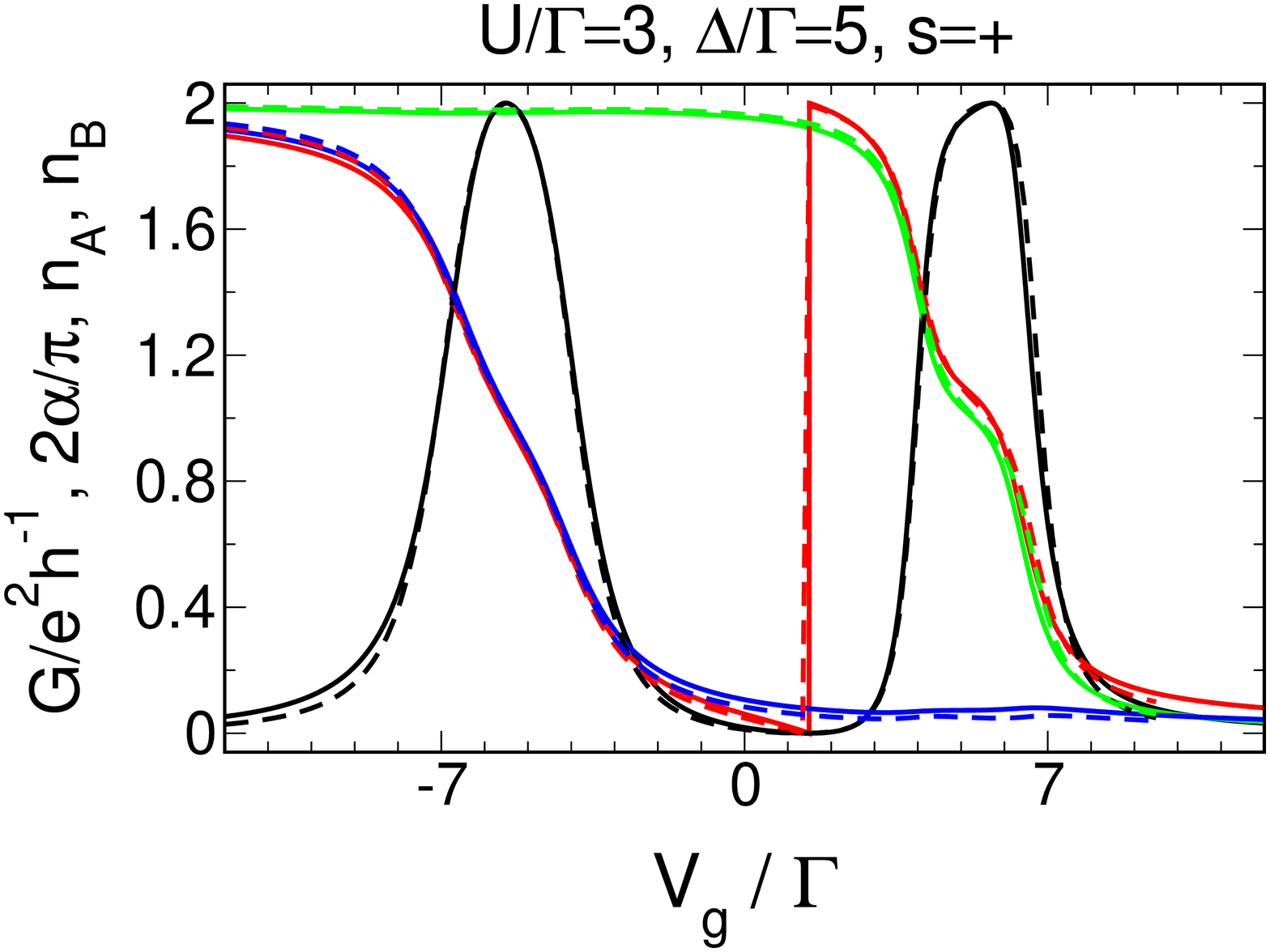}
        \caption{Comparison between fRG (solid lines) and NRG (dashed lines) data of the gate voltage dependence of the conductance $G$ (black), transmission phase $\alpha$ (red), and average level occupations (dot A: green, dot B: blue) of a spinful parallel two-level dot with equal local and nearest-neighbour interaction, and $\Gamma=\{0.15~0.15~0.35~0.35\}$ for four different parameter sets. The NRG data was taken from \cite{theresanrg}. Note that due to the non-generic symmetry of the hybridisations, $G(V_g)$ does not exhibit the additional resonances described in Sec.~\ref{sec:MS.dd} for $s=+$ and nearly degenerate levels. Furthermore, the average particle numbers computed by fRG are not shown for this parameter set because they exhibit numerically-caused oscillations in the $V_g$ regions where the second order approximation breaks down. Outside, they are in perfect agreement with the NRG data. Note that in the right panels fRG and NRG almost coincide rendering solid and dashed lines partly indistinguishable.}
\label{fig:COMP.dd}
\end{figure}

By considering many physical situations and parameter regions where the fRG performs differently well compared with precise NRG data and at the same time looking at the behaviour of the renormalized two-particle vertex at the end of the flow, we have motivated an inner-fRG criterion to judge the quality of our results when no reference data is available. More precisely, we observed that a breakdown of our second order approximation scheme was always accompanied by some components of $\gamma_2$ which grew at least one order of magnitude larger than their initial value, while remaining of order $U$ in situations where fRG and NRG agreed well. Equivalently, one would assume that fRG results are reliable if first and second order calculations, the former only accounting for the flow of the self-energy (see Sec.~\ref{sec:FRG.truncation}), yield qualitatively similar results.

\begin{figure}[t]	
	\centering
        \includegraphics[width=0.475\textwidth,height=4.0cm,clip]{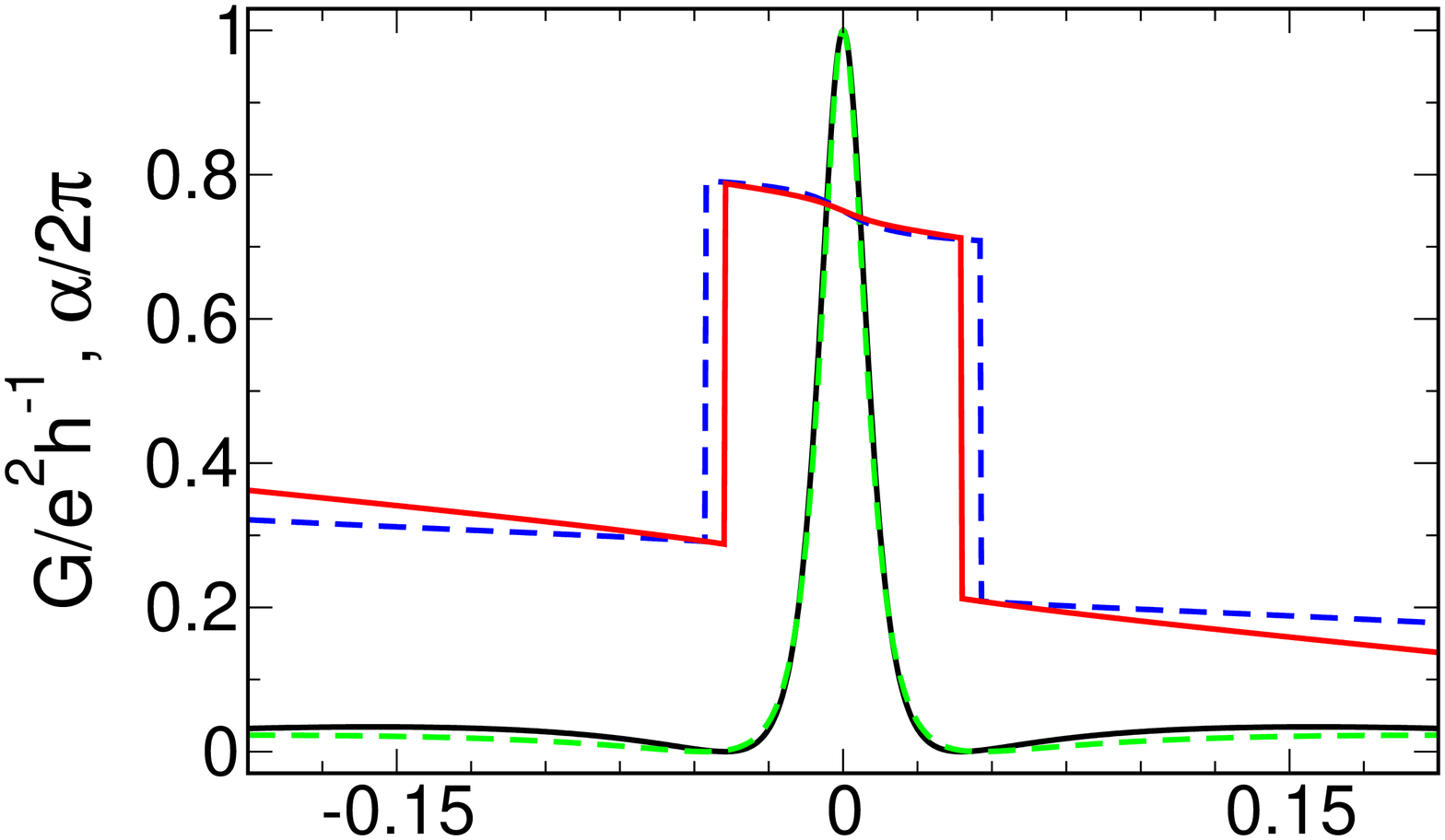}\hspace{0.035\textwidth}
        \includegraphics[width=0.475\textwidth,height=4.0cm,clip]{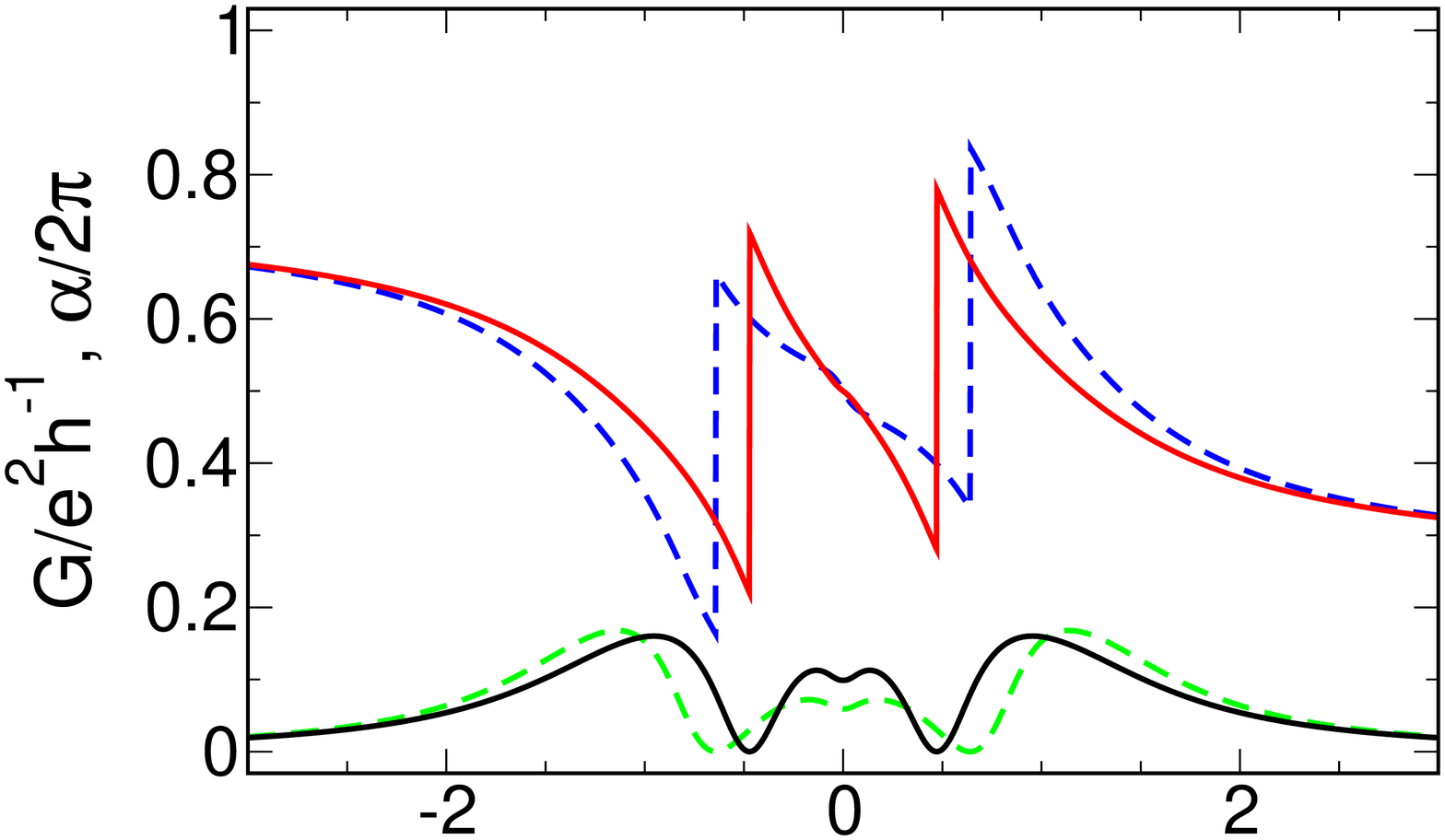}\vspace{0.3cm}
        \includegraphics[width=0.475\textwidth,height=4.8cm,clip]{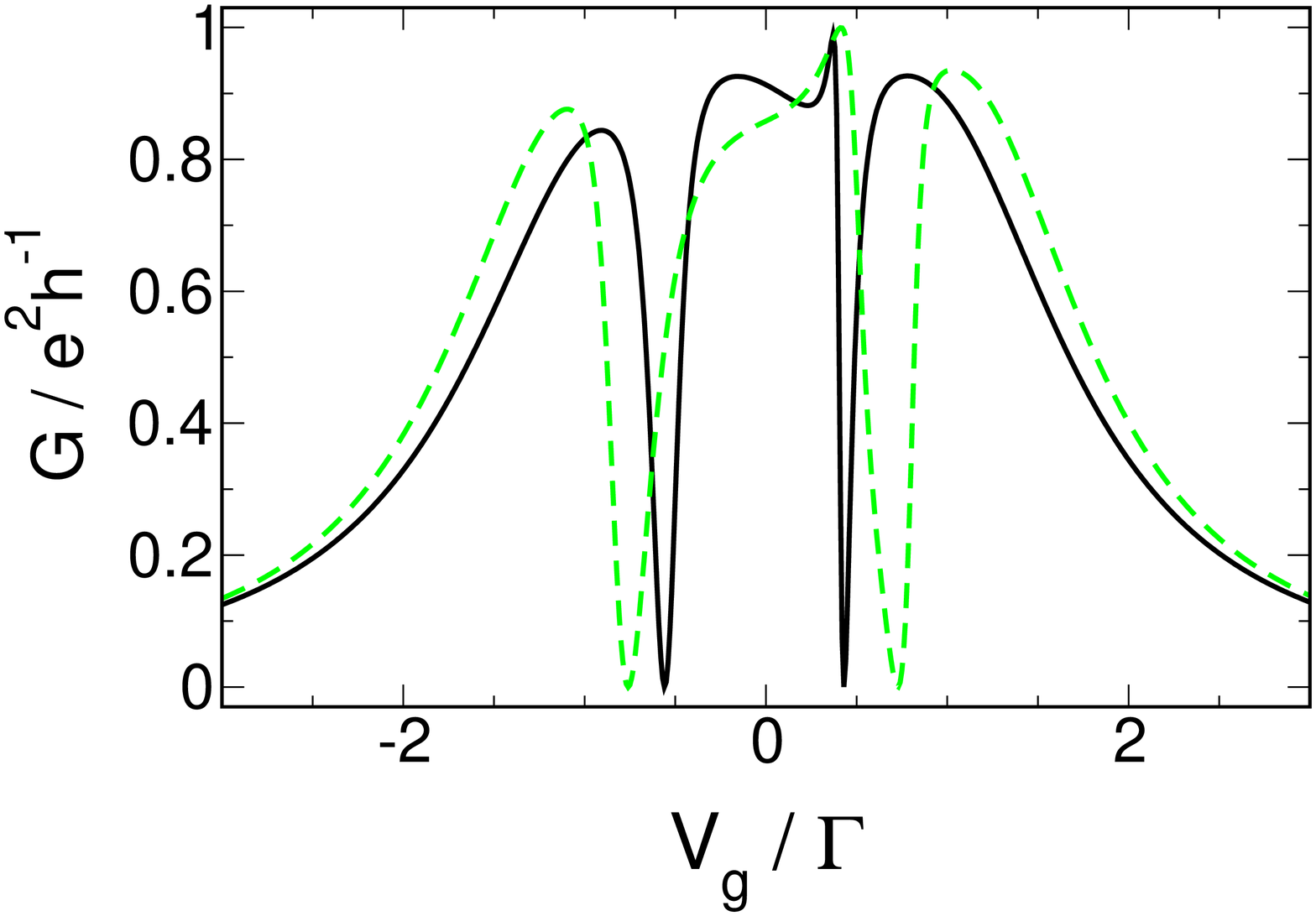}\hspace{0.035\textwidth}
        \includegraphics[width=0.475\textwidth,height=4.8cm,clip]{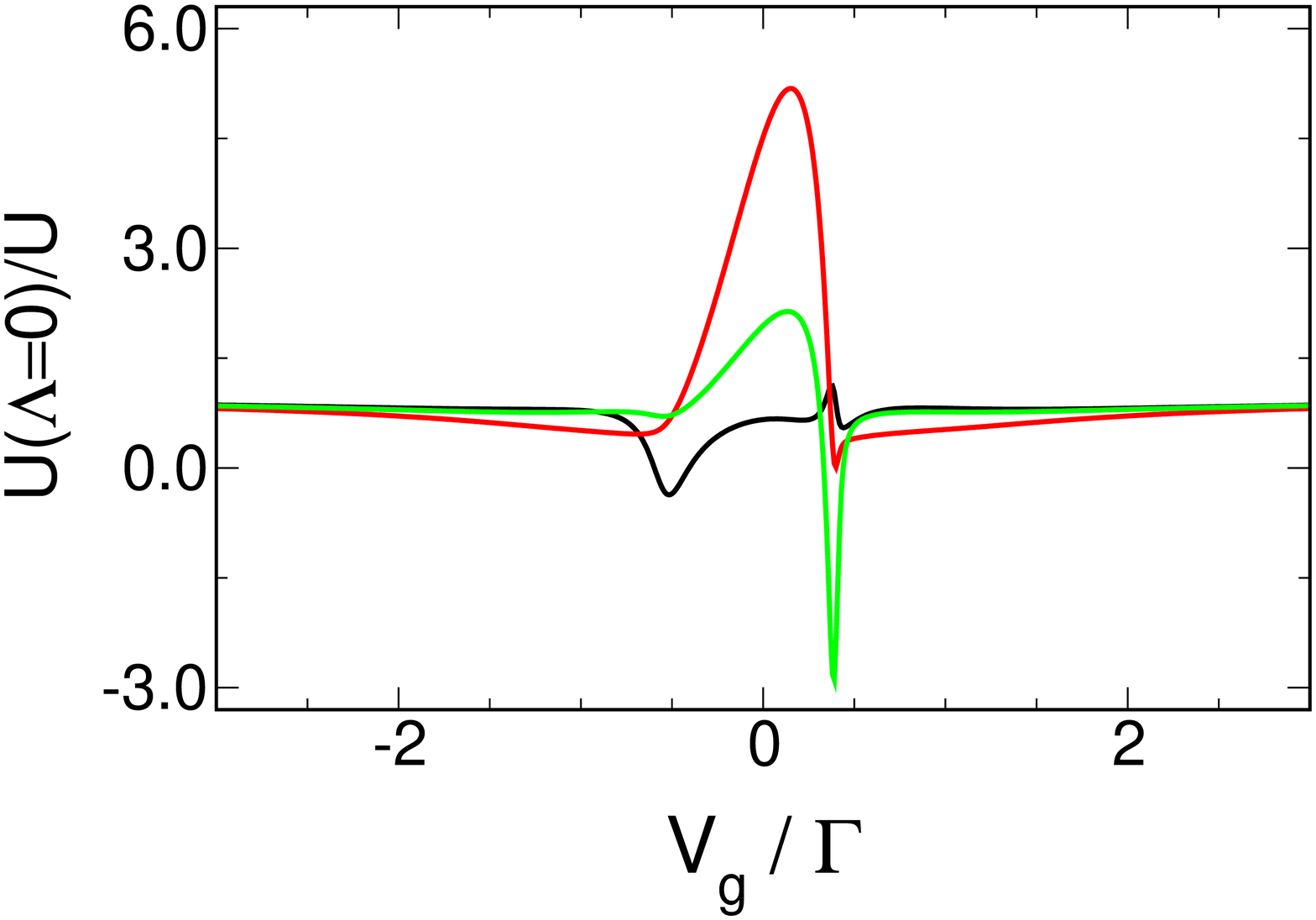}
        \caption{Comparison between first and second order fRG calculations of the conductance $G$ (second order: black, first order: green) and transmission phase $\alpha$ (second order: red, first order: blue) for different situations. \textit{Upper row:} Spinless triple dots, $U/\Gamma=1.0$, $\Delta/\Gamma=0.02$, $s=\{--\}$, $\Gamma=\{0.15~0.15~0.2~0.2~0.15~0.15\}$ (left panel), and $\Gamma=\{0.1~0.2~0.05~0.35~0.1~0.2\}$ (right panel). \textit{Lower row:} Spinless triple dots, $U/\Gamma=1.0$, $\Delta/\Gamma=0.05$, $s=\{+-\}$, $\Gamma=\{0.06~0.14~0.3~0.4~0.07~0.03\}$. The right panel shows the effective interactions at the end of the flow: $\gamma_2(A,B;A,B;0)$ (black), $\gamma_2(A,C;A,C;0)$ (red), $\gamma_2(B,C;B,C;0)$ (green).}
\label{fig:COMP.o1_1}
\end{figure}

We will now use this experience to explicitly demonstrate that our findings for the spinless triple dot geometry, which does not allow for an NRG reference calculation (see Sec.~\ref{sec:COMP.summary}), are indeed reliable. We start with nearly degenerate levels, relative signs of the couplings $s=\{--\}$, and left-right symmetric hybridisations with the leads. For this choice of parameters, $G(V_g)$ shows additional non-generic transmission zeros and associated double phase lapses (DPLs) sharply surrounding the correlation induced resonance close to $V_g=0$. Similar structures were observed for the four-level case and there confirmed by NRG. Here, it shows that first and second order fRG calculations are almost identical (and, equivalently, no components of the two-particle vertex grow large), and hence the appearance of the DPLs is obviously no artifact of our approximation (see Fig.~\ref{fig:COMP.o1_1}, upper left panel). The same holds for the fact that the additional zeros vanish if one lifts the non-generic left-right symmetry of the hybridisations in both the three- and four-level case (the upper right panel of Fig.~\ref{fig:COMP.o1_1} shows the comparison for the former). Finally, we prove that the correlation induced peaks, especially the ones exhibited at the central resonance, which appear for arbitrary choice of $s$ and $\{\Gamma\}$ can be considered as a reliable result of our calculations. As usual, it turns out that they show up consistently in first and second order, and all components of $\gamma_2$ stay of order of their initial value during the flow (Fig.~\ref{fig:COMP.o1_1}, lower panels).

\begin{figure}[t]	
	\centering
        \includegraphics[width=0.475\textwidth,height=4.8cm,clip]{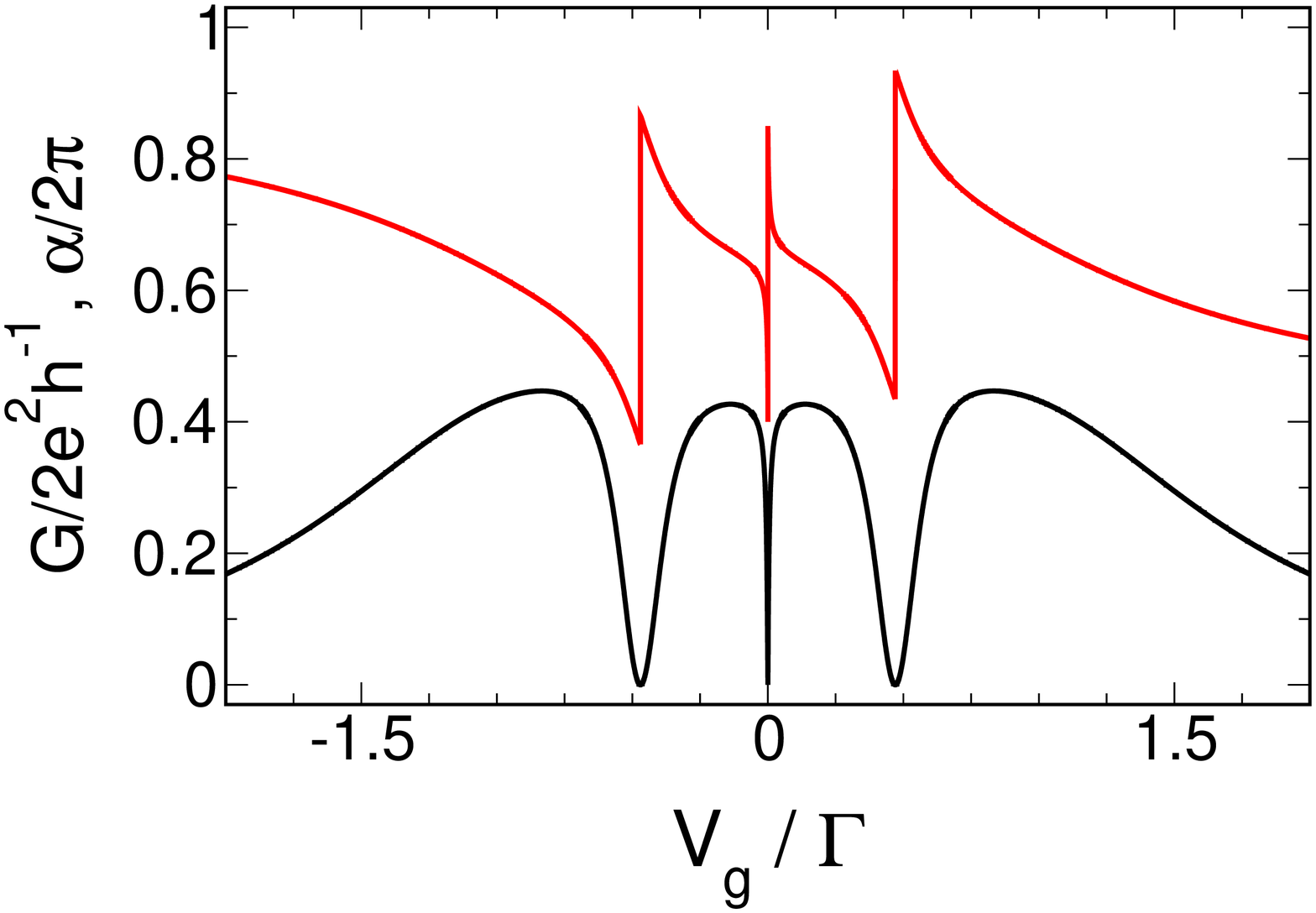}\hspace{0.035\textwidth}
        \includegraphics[width=0.475\textwidth,height=4.8cm,clip]{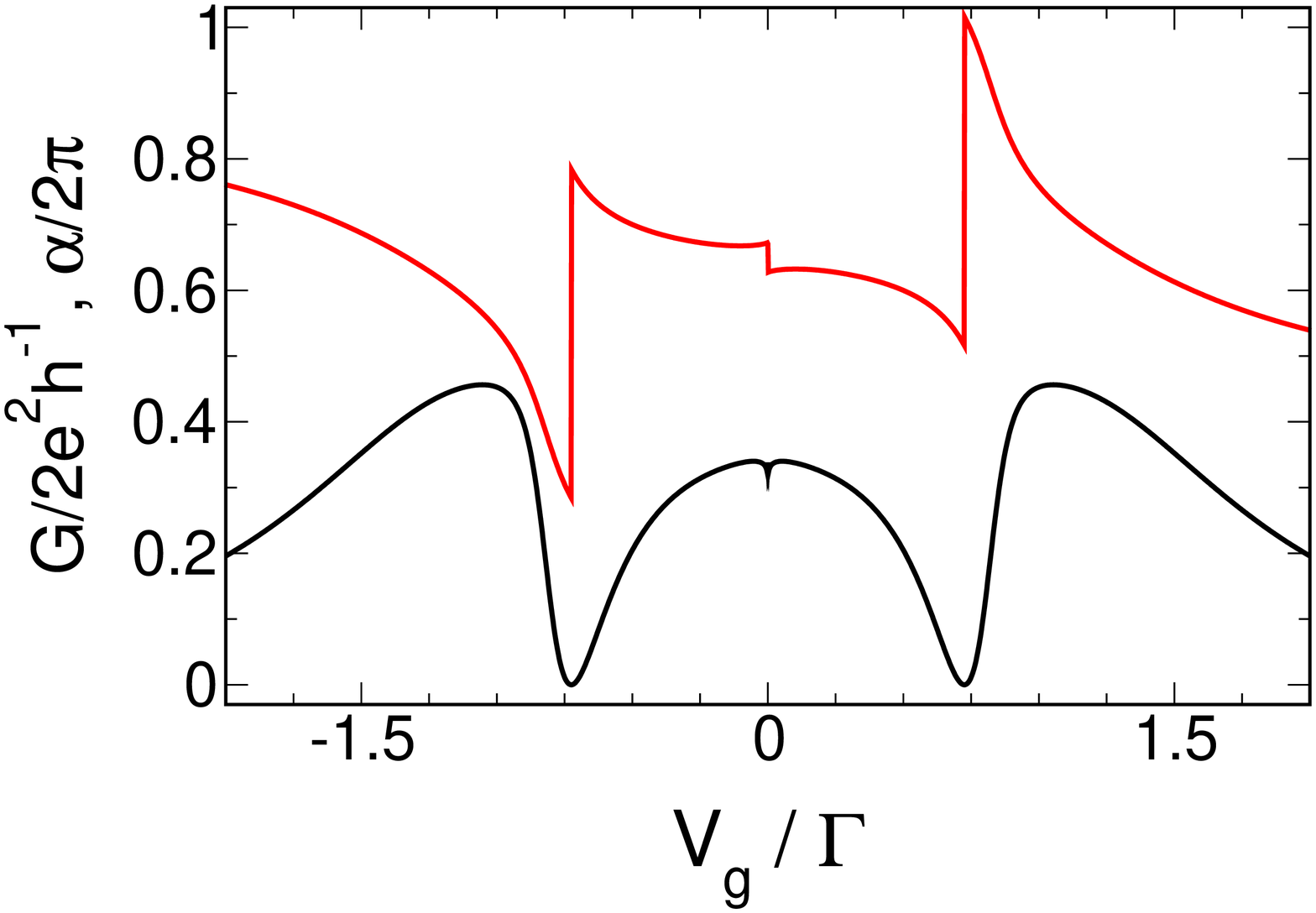}
        \caption{Comparison of second (left panel) and first (right panel) order fRG calculations of the conductance $G$ (black) and transmission phase $\alpha$ (red) for parallel triple dots with local interaction $U/\Gamma=1.0$, degenerate levels, $s=\{+-\}$, and hybridisations $\Gamma=\{0.06~0.14~0.3~0.4~0.07~0.03\}$. First and second order calculations differ qualitatively in what concerns the appearance of a transmission zero at $V_g=0$. Equivalently, some components of the two-particle vertex grow much larger than their initial value close to half-filling.}
\label{fig:COMP.o1_2}
\end{figure}

Next, we turn to the case of triple dots with completely degenerate levels, a situation which we carefully avoided in Sec.~\ref{sec:OS.td}. Here, the usual finite $U$ second order fRG calculations yield an additional transmission zero together with a $\pi$ phase jump at $V_g=0$ no matter how the relative signs of the level-lead couplings $s$ and the hybridisations are chosen (Fig.~\ref{fig:COMP.o1_2}, left panel). Unfortunately, the two-particle vertex grows large close to half-filling, and the first order truncation scheme yields a dip but no transmission zero at $V_g=0$, at least above the numerical limit $\delta V_g/\Gamma\approx 10^{-8}$, rendering the zero and the jump of $\alpha$ a questionable result (Fig.~\ref{fig:COMP.o1_2}, right panel). Fortunately, these problems vanish for generic $\Gamma_l^s$ if the degeneracy of the levels is lifted to $\Delta/\Gamma\approx 0.005$ for interactions of order $\Gamma$, and only the confirmed two transmission zeros remain; unfortunately, they appear even for small $U<\Gamma$.

When comparing with NRG results for the spinless four-level dot, we observed a similar breakdown of our approximation scheme for nearly degenerate levels, only that the scale $\Delta$ below which the fRG results ceased to be reliable as well the affected region of gate voltages around $V_g=0$ were larger. On the other hand, the method works well in the double dot case even if the level spacing is exactly zero. Alltogether, this leads to the conclusion that the fRG encounters serious problems in describing the correlations between electrons in nearly degenerate levels below a scale $\Delta$ which increases with the number of dots under consideration. For three, four, and six levels, we found this detuning to be $\Delta=0.005$, $\Delta=0.01$, and $\Delta=0.05$, respectively. Be that as it may; a plausible reason for a strong correlation effect that cancels the base of our fRG approximation (neglecting everything but the flow of the self-energy and the two-particle vertex evaluated at zero external frequency) remains to be found.

For one particular case, the self-consistency check can even be extended beyond the usual second order approximation scheme. As mentioned above, for spinful double dots with purely local interaction, relative sign of the level-lead couplings $s=+$, and left-right symmetric hybridisations, the free propagator is diagonal, and no off-diagonal elements can be generated by the flow. This implies that the problem separates into four independent flow equations corresponding to two single-dot problems, for which we know that fRG perfectly describes the physics on a quantitative level. Of course, this does not imply that same holds for the double dot case. Nevertheless, it was shown \cite{ralf} that using a truncation scheme that accounts for the full frequency dependence of the two-particle vertex leaves the single-level results in perfect agreement with the exact ones. Since accounting for frequencies does not spoil the split-up of the double dot problem for the aforementioned set of parameters, we can deduce that even if we took into care the flow of the full two-particle vertex, this would not lead to a quantitative improvement of the computed curves. As the influence of asymmetric hybridisations is only minor, we can expect the fRG results for the spinful double dot with local interactions and $s=-$ to be very reliable. This is an important observation since in the comparison with NRG data (Fig.~\ref{fig:COMP.dd}) the local- and nearest-neighbour interactions were assumed to be equal.

\section{Mean-Field Approaches}

Having established the power of fRG calculations, we will now compare them to a self-consistent Hartree-Fock approach (SCHFA) which treats the interaction between the electrons on a mean-field level. In particular, we refer to \cite{GG} (to be noted as GG from now on), who have extensively studied the spinless parallel double dot geometry by SCHFA.

GG start out from the the typical Hamiltonian describing the two-level dot,
\begin{equation*}\begin{split}
H = (V_g-U/2)n_A & +  (V_g-U/2+\Delta)n_B + Un_An_B \\ & - \left[c_{0,L}(t_A^Ld_A^\dagger+t_B^Ld_B^\dagger) + c_{0,R}(t_A^Rd_A^\dagger+t_B^Rd_B^\dagger) + \tn{H.c.}\right],
\end{split}\end{equation*}
which by sticking to a 3D subspace of all possible hybridisations defined by $\Gamma_A^L-\Gamma_A^R=\Gamma_B^R-\Gamma_B^L$ can be mapped onto
\begin{equation*}\begin{split}
\tilde H = (V_g-U/2)\tilde n_A & +  (V_g-U/2+\tilde\Delta)\tilde n_B + U\tilde n_A\tilde n_B - t \left[ \tilde d_A^\dagger \tilde d_B + \tilde d_B^\dagger \tilde d_A \right] \\ & - \left[c_{0,L}(t_A\tilde d_A^\dagger+t_B\tilde d_B^\dagger) + c_{0,R}(t_A\tilde d_A^\dagger+t_B\tilde d_B^\dagger) + \tn{H.c.}\right],
\end{split}\end{equation*}
with $\tilde\sigma=\tn{sgn}(t_At_B)=-1$. The key point of a mean-field treatment of the interaction is the replacement
\begin{equation*}\begin{split}
-\tilde n_A\tilde n_B = \tilde d_A^\dagger \tilde d_B^\dagger \tilde d_A \tilde d_B \rightarrow ~
& \tilde d_A^\dagger \tilde d_A \langle\tilde d_B^\dagger \tilde d_B\rangle +
\tilde d_B^\dagger \tilde d_B \langle\tilde d_A^\dagger \tilde d_A\rangle -
\tilde d_A^\dagger \tilde d_B \langle\tilde d_B^\dagger \tilde d_A\rangle \\ - &
\tilde d_B^\dagger \tilde d_A \langle\tilde d_A^\dagger \tilde d_B\rangle 
+ \langle\tilde d_A^\dagger \tilde d_A \rangle\langle\tilde d_B^\dagger \tilde d_B\rangle - 
|\langle\tilde d_a^\dagger \tilde d_B\rangle|^2,
\end{split}\end{equation*}
and the subsequent self-consistent determination of $\langle\tilde d_i^\dagger \tilde d_j\rangle$ within the (effective) noninteracting problem.

\begin{figure}[t]
  \centering
        \vspace{-0.5cm}\includegraphics[width=0.97\textwidth,clip]{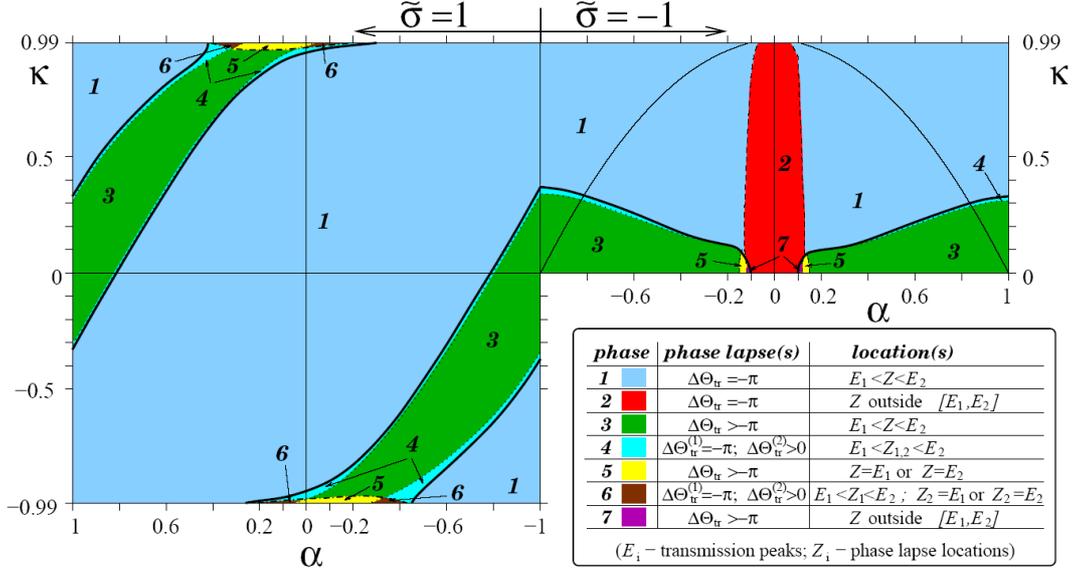}
        \caption{Phase diagram of spinless parallel double dots with parameters $U/\Gamma=6.4$, $\Delta/\Gamma=0.256$, $s=\tilde\sigma$, and left-right symmetric hybridisations. The inter-dot hopping $t$ is introduced as $t=-\kappa\Delta/2\sqrt{1-\kappa^2}$, and the couplings are determined by $\hat\alpha:=(|t_A|-|t_B|)/\sqrt{t_A^2+t_B^2}$. The different colours mark the different `phases' of the system which are characterised by the position of the transmission zero(s) $Z_i$ relative to the two transmission resonances $E_1$ and $E_2$ and the magnitude of the phase jump $|\Theta_\tn{tr}|$ at each zero. The diagram was obtained by \cite{GG} using a self-consistent Hartree-Fock calculation. (Reprinted by permission from the author.)}
\label{fig:COMP.gefen}
\end{figure}

Carrying out such calculations for fixed $U$ and $\tilde\Delta$ but variable $t$ and $A$-$B$-hybridisation asymmetry $\hat\alpha:=(|t_A|-|t_B|)/\sqrt{t_A^2+t_B^2}$,\footnote{We have introduced the hat in contrast to GG's notation to avoid notational confusion with the transmission phase.} GG always found two resonances located at $E_1$ and $E_2$ in the conductance $G(V_g)$, while the phase showed different behaviour in sharply separated regions of the parameter space $\{\hat\alpha,t\}$ (see Fig.~\ref{fig:COMP.gefen}). In particular, they distinguish three main phases where $\alpha(V_g)$ behaves differently: in phase 1 (2), there is a transmission zero and an associated phase jump (which GG call `lapse' because their gate voltage is defined with opposite sign) by $\pi$ in between (outside) the two resonances. Surprisingly, they observe jumps larger than $\pi$ located in between $E_1$ and $E_2$ for a significant region of $\{\hat\alpha,t\}$ (phase 3).

\begin{figure}[t]	
	\centering
        \includegraphics[width=0.495\textwidth,height=4.4cm,clip]{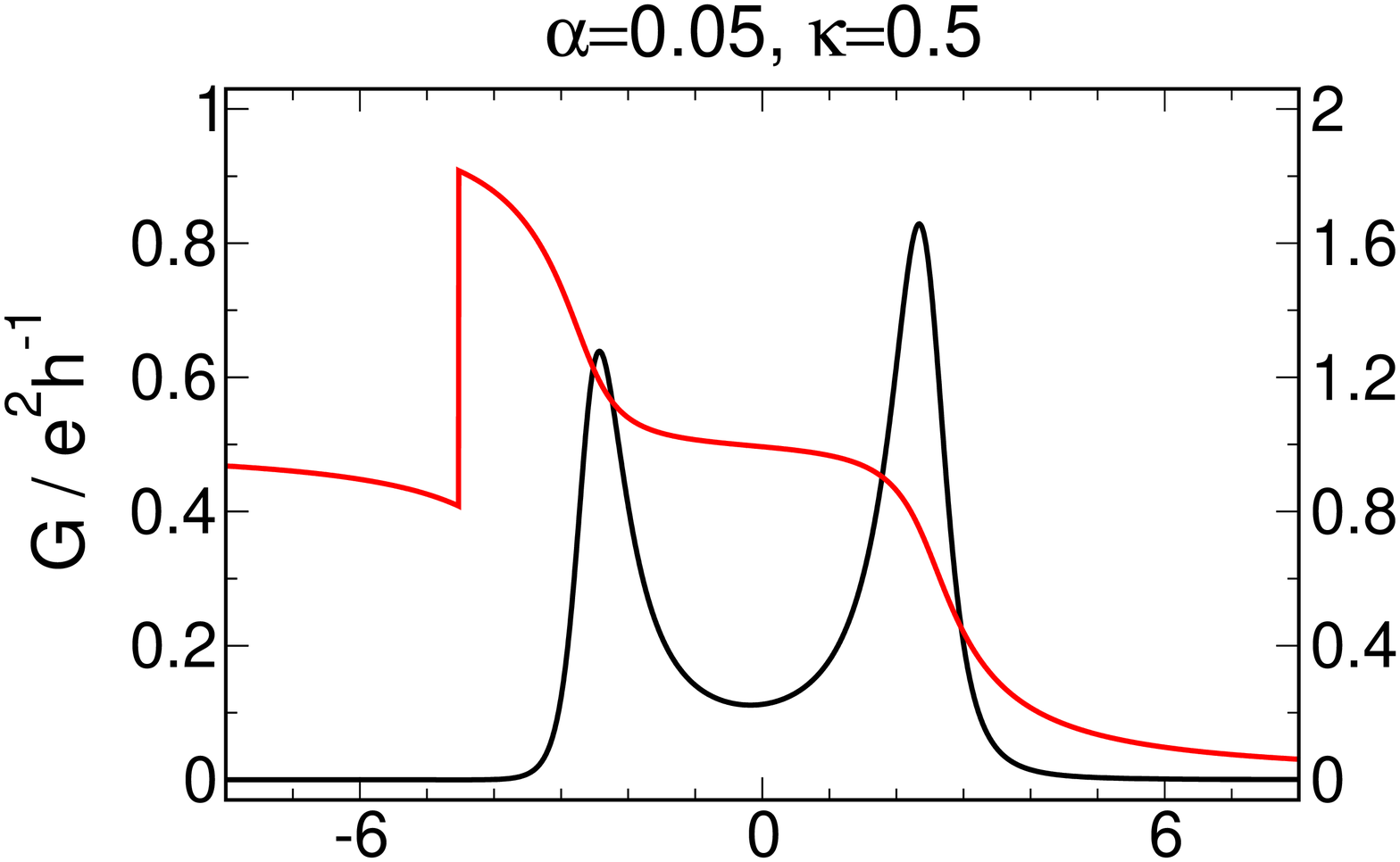}\hspace{0.015\textwidth}
        \includegraphics[width=0.475\textwidth,height=4.4cm,clip]{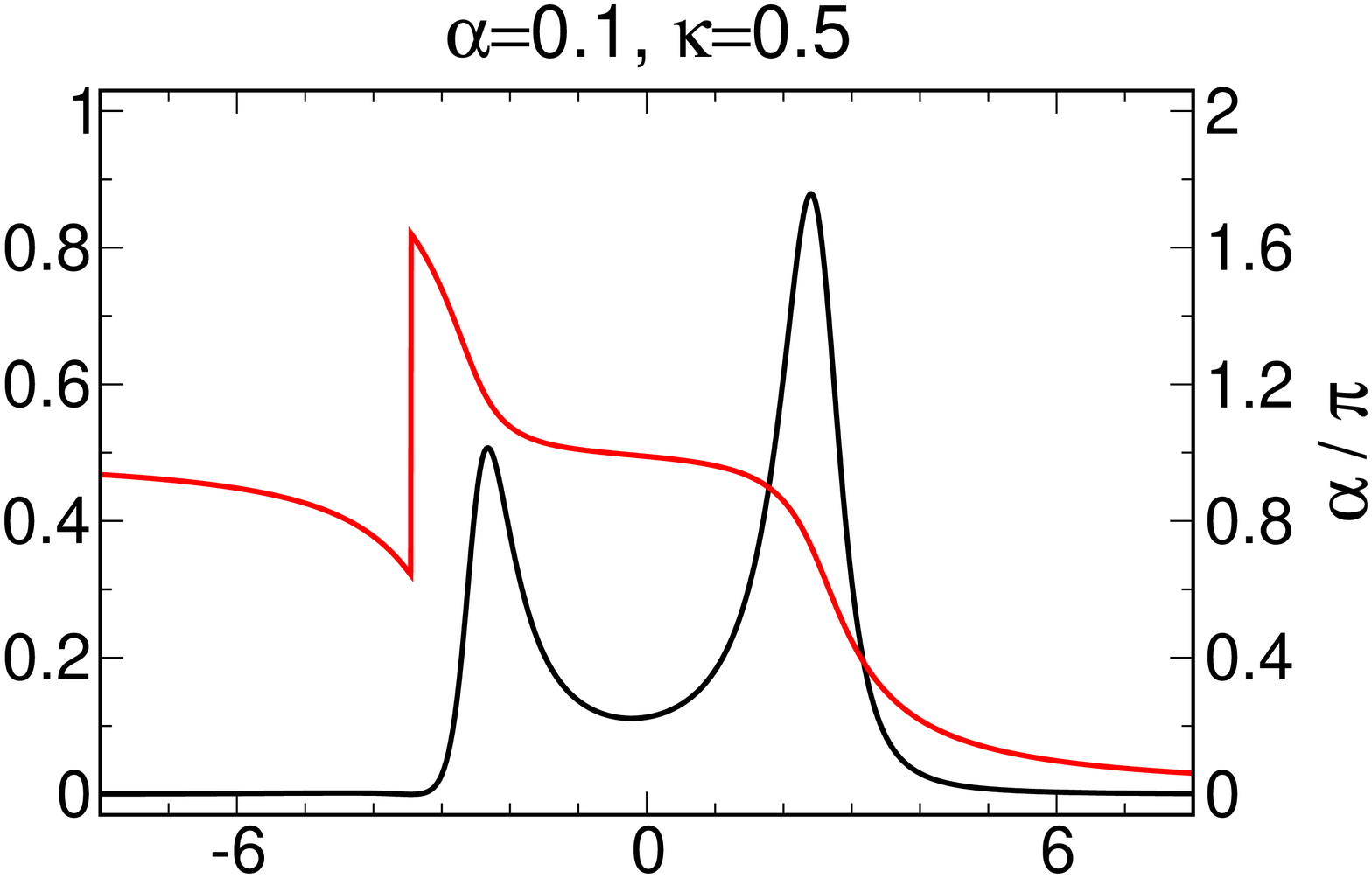}\vspace{0.3cm}
        \includegraphics[width=0.495\textwidth,height=5.2cm,clip]{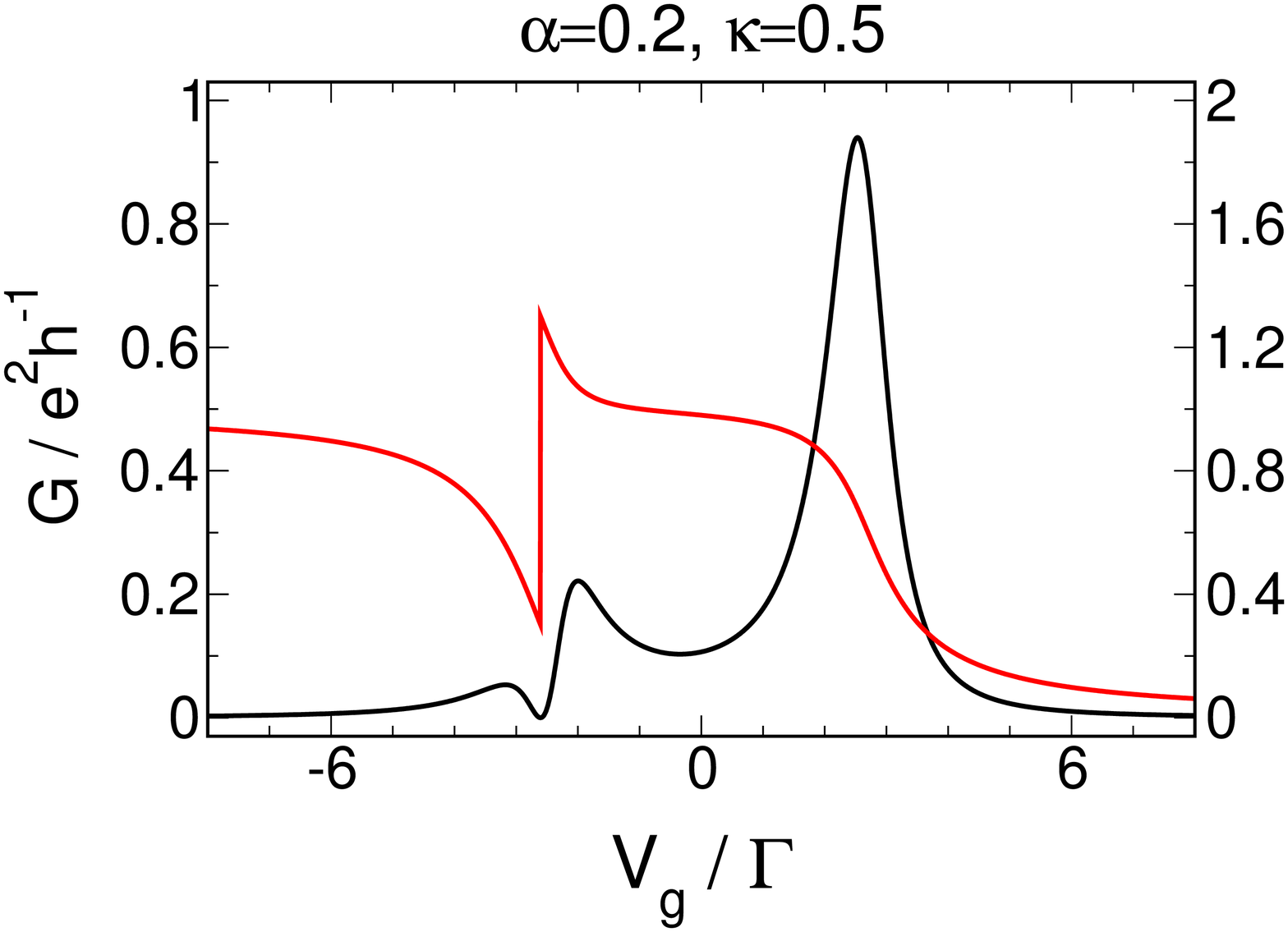}\hspace{0.015\textwidth}
        \includegraphics[width=0.475\textwidth,height=5.2cm,clip]{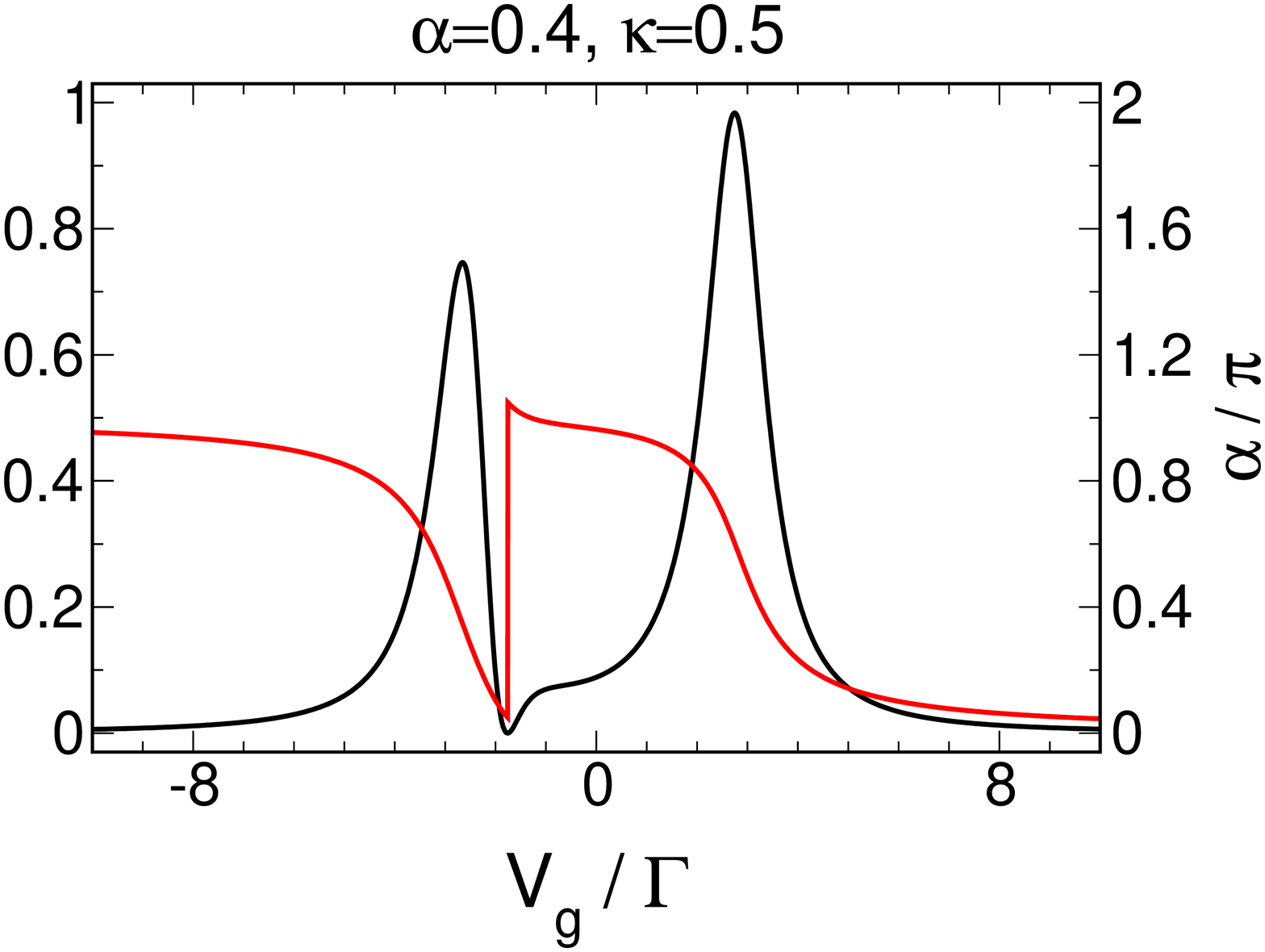}\vspace{0.8cm}
        \includegraphics[width=0.495\textwidth,height=4.4cm,clip]{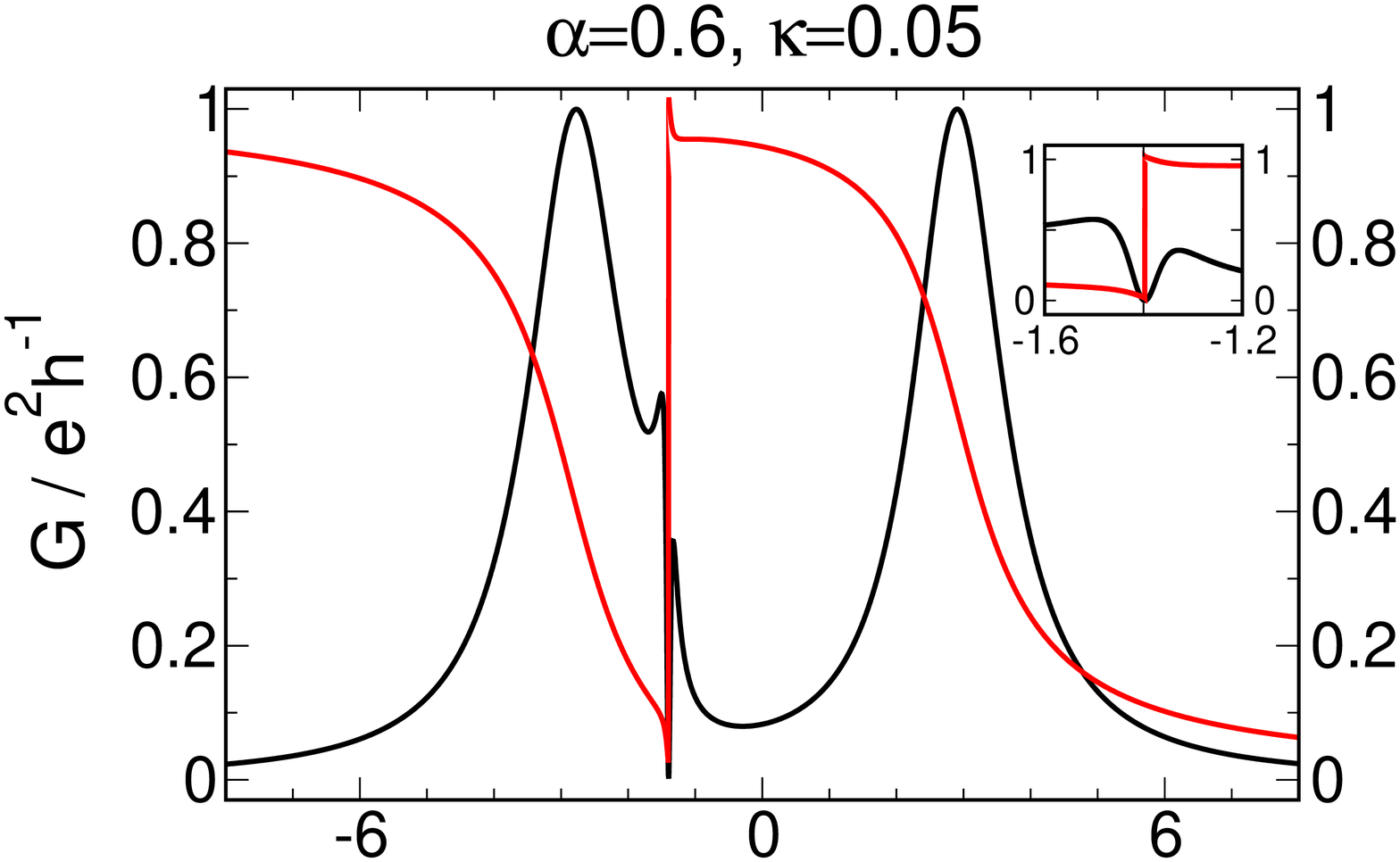}\hspace{0.015\textwidth}
        \includegraphics[width=0.475\textwidth,height=4.4cm,clip]{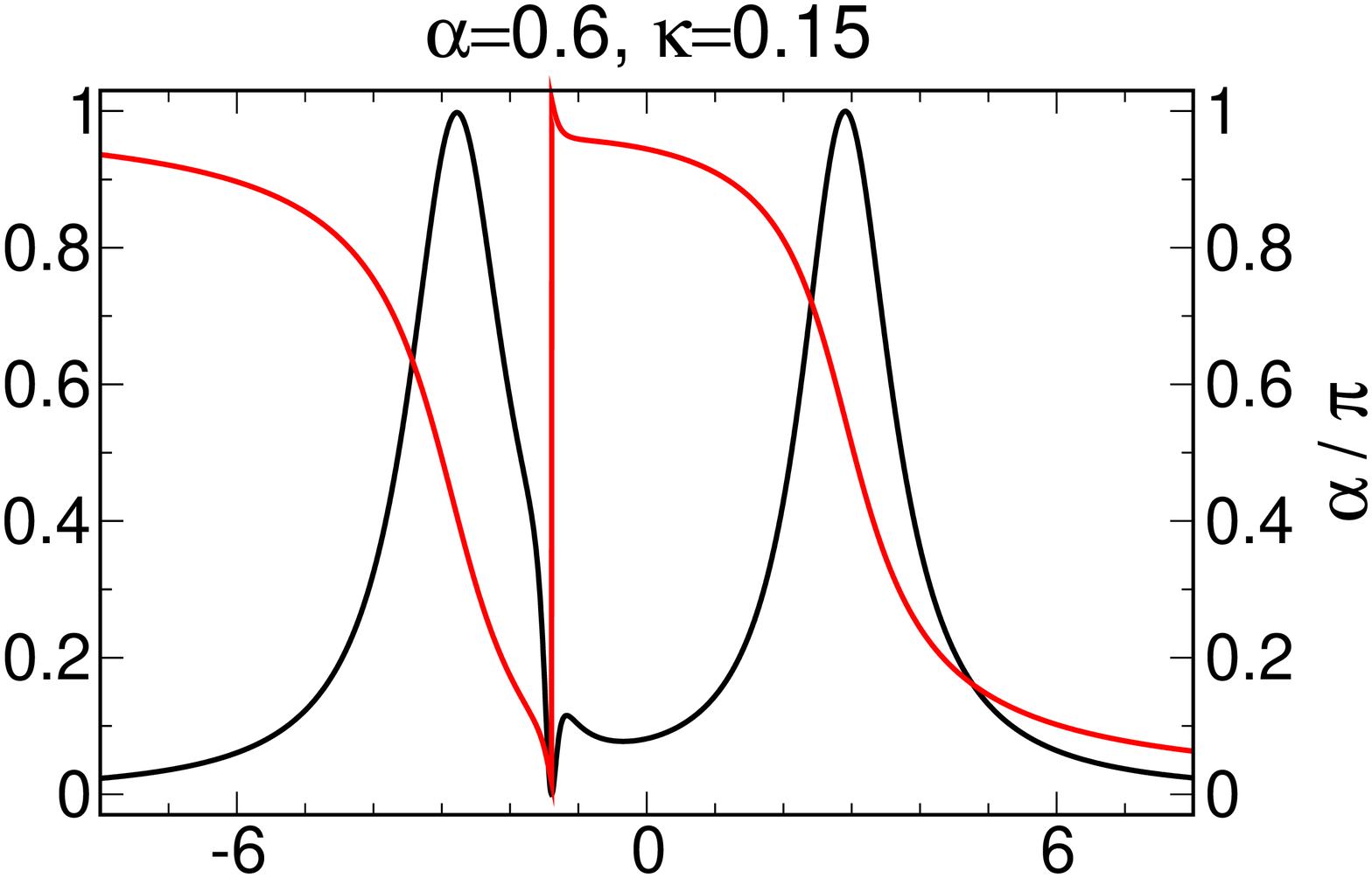}\vspace{0.3cm}
        \includegraphics[width=0.495\textwidth,height=5.2cm,clip]{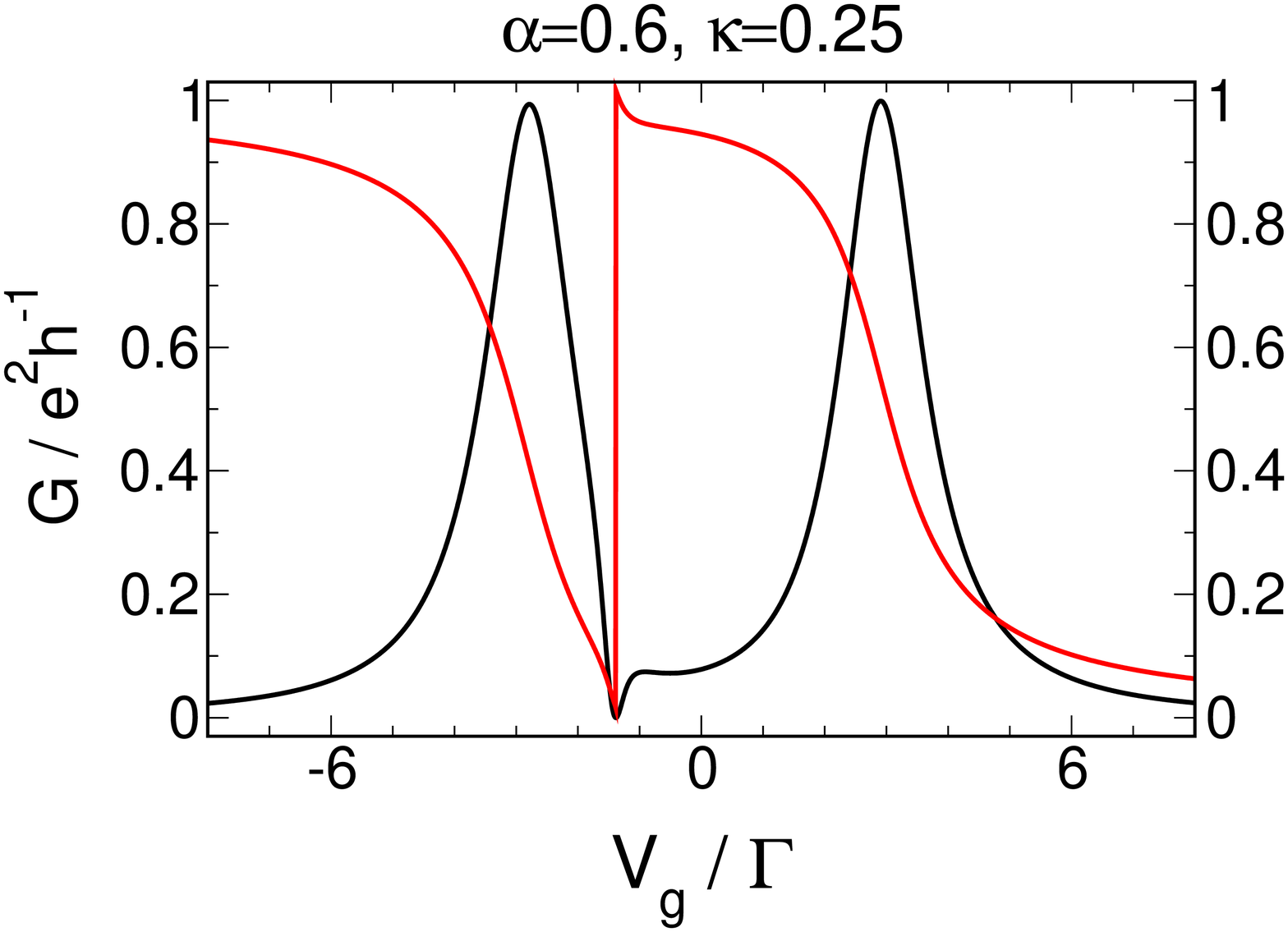}\hspace{0.015\textwidth}
        \includegraphics[width=0.475\textwidth,height=5.2cm,clip]{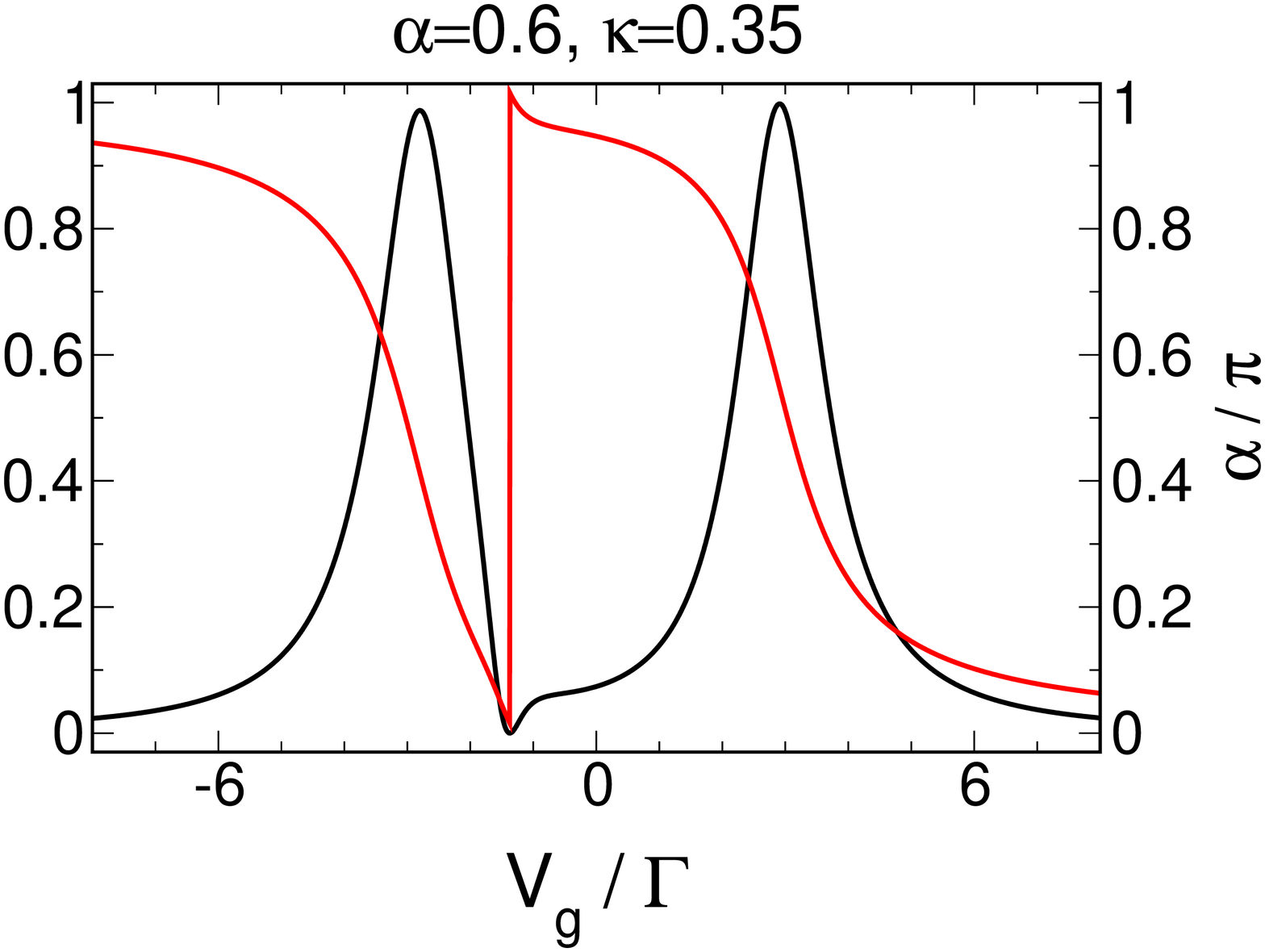}
        \caption{FRG calculation of the gate voltage dependence of the conductance $G$ (black) and transmission phase $\alpha$ (red) for spinless parallel double dots with parameters identical to those of the right panel ($s=\tilde\sigma=-1$) of Fig.~\ref{fig:COMP.gefen}. Note that the $\alpha$ in the caption of each plot is identical to the $\alpha$ introduced by GG and is called $\hat\alpha$ in the text.}
\label{fig:COMP.gefenfrg}
\end{figure}

If we carry out the usual second order fRG calculations for the dot Hamiltonian $\tilde H$,\footnote{In principal, it would of course be best to treat $H$ directly since this Hamiltonian contains the original dot parameters; nevertheless, we only want to demonstrate how the fRG performs in comparison with SCHFA, and thus we will focus on $\tilde H$ used by GG.} we can basically confirm GG's findings for the phase 1 and 2. For parameters associated with the former (latter), we find two resonances with a zero in between (outside). However, the transition between both regimes is smooth if quantum fluctuations are taken into account (see the upper panels of Fig.~\ref{fig:COMP.gefenfrg}).

For parameters from phase 3, going beyond the mean-field level has a more dramatic effect. In particular, the jump of the transmission phase at the zero in between the two resonances in this regime is precisely $\pi$ (and not larger than $\pi$). Furthermore, we observe additional correlation induced resonances described in Sec.~\ref{sec:OS.dd} in a region of $\{\hat\alpha,t\}$ that roughly coincides with GG's phase 3 which Hartree-Fock fails to capture. Hence correlations are of particular importance in this part of the parameter space, rendering it reasonable that a mean-field approach is plagued by artifacts like phase jumps larger than $\pi$. The transition from GG's phase 3 to phase 1 is shown in the lower panels of Fig.~\ref{fig:COMP.gefenfrg}.

\afterpage{\clearpage}

\section{Summary: Advantages and Disadvantages of the fRG}\label{sec:COMP.summary}

In this section, we have compared our $T=0$ fRG results to very precise NRG data from the literature. It proved that for every physical situation under consideration, the usual second order approximation scheme that only accounts for the flow of the self-energy and the two-particle vertex with external frequencies set to zero quantitatively agrees with NRG up to interaction strength where correlations between the electrons have strongly altered the $U=0$  physics. For large $U$ (where the meaning of `large' depends on the system under consideration), both methods begin to disagree as one would naturally expect because of the approximate nature of the fRG, so that it is no longer justified to neglect the frequency-dependence of $\gamma_2$ as well as all higher order functions. In some situations, the breakdown occurs at small $U$ compared to the usual energy scale set by the hybridisation strength. For models with spin, it proved possible to relate this to some strong correlation effect with a large `effective' interaction not obvious at first sight (for example the second stage Kondo effect in the side-coupled geometry; or the suppression of charge fluctuations for a half-filled linear chain). On the other hand, fRG encounters serious problems in describing even small interactions $U<\Gamma$ between the electrons on nearly degenerate ($\Delta\lll\Gamma$) spin-polarised parallel levels, and they become worse the more dots are considered. A reason (or a motivation) for this failure still needs to be found.

\begin{table}[t]
\renewcommand\tabularxcolumn[1]{m{#1}}
\begin{center}\begin{small}
\fbox{\begin{tabularx}{0.9\linewidth}{>{\vspace{0.1cm}\centering}m{0.33\linewidth}<{\vspace{0.1cm}}|>{\centering}X|>{\centering}X}
Geometry & Number of ODEs & Time required to compute 100 data points $G(V_g)$  \tabularnewline \hline
Spinless double dots & 4 & 0.21s \tabularnewline \hline
Spinless triple dots & 12 & 3.6s \tabularnewline \hline
Spinless four-level dots & 31 & 16.9s \tabularnewline \hline
Spinful single-level dot & 3 & 0.09s \tabularnewline \hline
Side-coupled geometry$\hspace{0.05cm}^3$ & 19 & 1.1s \tabularnewline \hline
Spinful double dots$\hspace{0.05cm}^3$ & 19 & 0.93s \tabularnewline \hline
Spinless double dots, finite T & 4 & 1.36s
\end{tabularx}}
\end{small}\end{center}
\caption{Typical times needed to numerically solve the second order fRG flow equations for various geometries on a 3GHz single-core engine.}
\label{tbl:COMP.times}
\end{table}

Alltogether, the comparison with NRG proved that fRG captures the physics in transport through quantum dots up to interaction strength where the latter is dominated by electron-electron correlations. In contrast to NRG, fRG has the following striking advantages:\footnotetext[3]{The computation time for these geometries is lower than that for the spinless triple dots (although we have to solve more independent equations) because all symmetries of the two-particle vertex were exploited. In particular, the right-hand side of the flow equations (\ref{eq:FRG.flowse3}, \ref{eq:FRG.flowww3}) was simplified analytically before implementation. This yields messy expressions in contrast to $N=1$ and $N=2$ (where the corresponding simplifications were given in the previous chapter).}
\begin{itemize}
\item
{\bfseries Flexibility:} fRG can describe any quantum dot geometry at will. In principal, it would even be possible to write a graphical user interface to build up geometries containing inter-dot hoppings, different level positions, arbitrary interaction terms, ..., automatically set up the corresponding flow equations, tell the computer to turn on a red flashing light if the two-particle vertex diverges, and provide a user who has never read a word about quantum many-particle systems with the conductance for the system. In contrast, NRG is limited to situations of high symmetry, and every different geometry needs individual implementation. For example, spinless parallel dots are treated by NRG by mapping them back onto systems with spin by an even-odd-transformation of the lead states. Unfortunately, this is only possible for left-right symmetric hybridisations, an even number of dots, and some special choices of the relative coupling signs.
\item
{\bfseries `Simpleness':} fRG is easy to implement. Having understood the formalism, writing down the flow equations for any situation is straight-forward (as stated above, write a graphical user interface...).
\item
{\bfseries Speed:} fRG is extremely fast and thus well-suited to scan wide regions of the parameter space of the system under consideration. Table~\ref{tbl:COMP.times} gives an impression of how long a mainstream personal computer with a 3GHz core needs to solve the set of coupled differential equations for typical parameters.
\item
{\bfseries Unarbitrarity:} The fRG scheme does not contain non-physical tunable parameters.
\end{itemize}

On the other hand, the disadvantages of the fRG are:

\begin{itemize}
\item
{\bfseries Approximate nature:} fRG still remains an approximate method. Therefore it fails to capture all effects which show up for extremely large (effective) interactions, such as the Fano-like behaviour observed for $U/t^2\gg1$ at the side-coupled geometry by \cite{side3}. Nevertheless, we showed that in most situations it works for $U$ strong enough to observe the physics dominated by correlations far beyond the mean-field level. Furthermore, there is reason to believe that the comparison between first and second order calculations provides a reliable criterion to judge the quality of fRG results in new situations where no reference data is available.
\item
{\bfseries Finite frequencies:} The calculation of finite-frequency properties by the simple second order fRG truncation scheme employed here is questionable.  For spinful models, in particular for the single-impurity case, we have argued that a frequency-independent self-energy leads to a spectral function which strongly differs from the exact one. Hence we cannot expect to obtain reliable $T>0$ results for such a situation (beyond the limit of very small temperatures), as it indeed shows by comparison with NRG. We can hope that things are different for spinless models, but this remains to be proven as soon as finite temperature reference data for the conductance of these systems to compare the fRG results against is available. Using a truncation scheme that accounts for the full frequency-dependence of the two-particle vertex seems, however, not very desirable since it would be accompanied with a vast increase of the computational resources required.
\end{itemize}

To summarise one could say that the fRG is a very promising tool to study zero temperature transport properties of quantum dots with local Coulomb correlations.

\newpage
\thispagestyle{empty}

\chapter{Summary \& Outlook}

In this thesis, we have studied zero temperature transport properties of few-level correlated quantum dots arranged in various geometries. The Hamiltonian that governs such systems consisted of several levels with hoppings and arbitrary interactions to model the geometry under consideration, and this `dot region' was connected to noninteracting tight-binding leads by tunnelling barriers. In order to tackle the two-particle interactions, we have employed the functional renormalization group. This method starts from an infinite set of flow equations for the one-particle irreducible vertex functions which exactly describe arbitrary electron-electron correlations. This set was truncated to render it solvable, leading to the truncation scheme used allover this thesis, accounting for the flow of the self-energy and the two-particle vertex evaluated at zero external frequency. Hence the fRG yields a frequency-independent self-energy which can be interpreted as a renormalization of the original single-particle energies of the system, and we showed that consistent with this approximation is to compute transport properties of quantum dots, in particular the linear-response conductance, with the effective noninteracting expressions.

With a tool at hand to tackle electron-electron correlations, we concentrated on two different physical problems. First, we studied the conductance $G$ and transmission phase $\alpha$ of parallel spin-polarised quantum dots as function of a gate voltage $V_g$ that simultaneously heightens the energies of all levels. The small computational effort required by fRG allowed us to systematically scan the vast parameter space of these systems, the special focus being on the influence of the single-particle level spacing $\Delta$. If the latter was assumed to be small compared to the total strength of the tunnelling barriers $\Gamma$, we found both the conductance and the transmission phase to show strikingly universal behaviour independent of all other parameters of the system, such as the individual level-lead couplings $\sqrt{\Gamma_l^s}$, their relative signs $s$, and the number $N$ of dots under consideration, assuming there is a two-particle interaction $U\gtrsim\Gamma$ in the system. In particular, we observed $N$ `Coulomb blockade' resonances of almost equal height, width of order $\Gamma$, and separation of order $U$. Additional structures (correlation induced features) showed up when the interaction strength was successively increased. The phase dropped by $\pi$ over each Coulomb peak and exhibited a $\pi$ jump at the $N-1$ transmission zeros located in between. In contrast, the behaviour of $G(V_g)$ and $\alpha(V_g)$ strongly depends on the dot parameters if the level spacing is increased. In the limit of large $\Delta$, we found $N$ Lorentzian resonances separated by $U+\Delta$ and the phase to drop by $\pi$ over each, but the number of transmission zeros (and associated jumps of $\alpha$) was observed to be governed by s. These findings might serve as an explanation of the transport properties of a quantum dot frequently measured in the experiment. While in some early ones \cite{phase3,phase2}, the transmission phase was observed to jump by $\pi$ in between all transmission resonances occurring if the energy of the dots was successively lowered (heightened), in contrast to the expected `mesoscopic' behaviour predicting $\alpha$ to exhibit jumps between some peaks while evolving continuously between others. Fortunately, in a very recent experiment \cite{phase1} also the occupation of the dot was measured, providing a clue to this puzzling behaviour. For dots occupied by only a few electrons ($\sim 5$; mesoscopic regime), the phase was indeed found to behave mesoscopically, in particular showing jumps only between some resonances which changed with the system under consideration. On the other hand, for dots with a fairly large occupation ($\sim 15$; universal regime), the universal behaviour of $\alpha$ jumping between all resonances independent of the actual experimental realisation was recovered. One major difference between both situations is the level spacing, which one would expect to decrease for the topmost filled levels. In this sense our calculations provide a natural explanation of the experiment: for nearly degenerate levels (those filled in the universal regime), we observed $N$ resonances with $N-1$ jumps of the transmission phase in between no matter how all other dot parameters are chosen while for large level spacings (associated with the mesoscopic regime) the behaviour of $\alpha$ depends on the relative signs of the level-lead couplings, which is determined by the parity of the underlying single-particle wavefunction and hence represents an experimentally uncontrolled quantity depending on the dot under consideration. Within the fRG approximation scheme, more insight into the behaviour of $G(V_g)$ and $\alpha(V_g)$ could be gained from diagonalising the effective dot Hamiltonian at the end of the flow. It turned out that for small $\Delta$ irrespective of $s$ the most strongly coupled effective level crosses the chemical potential upwards at each resonance, while close to the zeros it crosses another weakly coupled level (and crosses $\mu$ downwards), so that the transmission zero can be interpreted as arising from a Fano-antiresonance. Correlations enhance the asymmetries between the bare hybridisations such that this strongly coupled level is much broader than the others. In contrast, for large level spacings the effective level energies do not cross but depend linearly on the gate voltage (up to some shift by $U$ each time a level gets depleted), while their couplings are unchanged by the interaction. This explains the resemblance with the noninteracting case (only that the peak separation is now given by $U+\Delta$), and the behaviour of the transmission phase strongly depends on $s$.

Second, we studied various spinful geometries. We showed how the Kondo temperature, the energy scale on which the Kondo resonance exhibited by a single-level dot with local interaction is destroyed, can be extracted from $T=0$ calculations by the magnetic field dependence of the conductance. We investigated short Hubbard chains and the side-coupled geometry which have both been tackled before in NRG approaches, and we showed how the fRG can be used to scan the parameter space for regions of interesting physics. In particular, we observed a vanishing of transmission resonances for generic left-right asymmetric level-lead couplings in a three-site chain. We extensively studied the parallel double dot geometry which shows a variety of correlation induced features in different parameter regimes. We verified that if large magnetic fields are applied to the system all the results of the spin-polarised model are recovered. Importantly, our observations for parallel dot systems with equal interactions between all electrons allow for an interpretation of the `phase lapse experiments' almost analogously to the spin-polarised case. Unfortunately, the fRG encounters serious problems for double dots with nearly degenerate levels and $s=+$, and the same holds if more than two levels are considered. Therefore another method is needed to supplement the fRG results within this parameter region.

In the last chapter, we compared our fRG results with a very reliable method non-perturbative in the two-particle interaction, the numerical renormalization group (NRG). It turned out that our usual second order approximation scheme is in good quantitative agreement with NRG up to fairly large interactions, which is surprising for a method designed for small $U$. The word `fairly large' refers to the importance of the correlations: while for a spinful single-level impurity we can treat interactions as large as $U/\Gamma=30$, for the side-coupled geometry with small inter-dot hopping $t$ fRG begins to deviates from NRG at $U/\Gamma\approx 1$ due to the presence of a large effective interaction $U/t^2$ related to the second stage Kondo effect. To say it in other words, fRG still works well for parameters where correlations dominate the physics in the sense that all $U=0$ properties are significantly altered. Unfortunately, it is not always possible to relate the breakdown of the fRG approximation to an effective strong correlation effect. To be more precise, we encountered serious problems in the treatment of parallel spinless models with $\Delta\lll\Gamma$ which grew worse with the number of nearly degenerate levels under consideration. In contrast to NRG, fRG is very flexible, easy-to-implement, and computationally cheap, which renders it possible to perform exhaustive scans of the parameter space. At the same time, it allows for a treatment of the two-particle interactions far beyond the mean-field level, as was shown by comparison with Hartree-Fock results for a spinless double dot. Alltogether one could say that the fRG is a very promising tool to study zero temperature transport properties of quantum dots in presence of local Coulomb correlations.

There are a lot of things left to be done. First, one can of course apply the usual fRG approximation to any geometry at will. For example, we have started to describe the RKKY interaction induced between two single-level impurities with local interaction embedded in a noninteracting chain. Second, it would certainly prove an interesting challenge to supplement the explanation of the `phase lapse' experiments by calculations accounting for finite temperatures. To this end it would be interesting to compare the finite $T$ results for spinless models to a reference method of some kind (which might be a rate equation or equation-of-motion approach working at high $T$). Guided by the experience from spinful models where the simple frequency-independent fRG approximation computes a spectral function far from the exact one, one should suspect this comparison not to be very fortunate for large temperatures. Therefore: third, a reliable finite-temperature (or, more general, finite-frequency) extension of the fRG that refrains from taking into account the full frequency dependence of the two-particle vertex (since this would terribly spoil the advantage computationally cheapness) is needed.

\newpage
\thispagestyle{empty}

%


\chapter*{\vspace{-2cm}\centering\textsc{Thank You!}}\addcontentsline{toc}{chapter}{Thank You!}

Finally, I want to thank all those people without whom this work would not have been possible.
\\ \\
First and most of all, I want to thank my tutor Volker Meden for giving me the chance to write this diploma thesis on such an interesting topic under his supervision. I am grateful to him for teaching me countless things about physics and about how physics works, ... to cut a long story short: I could not have imagined a better tutor.
\\ \\
I am glad to have attended the very nice lecture series by Kurt Sch\"onhammer on various issues of quantum mechanics, and I want to thank him for the opportunity to write this thesis within his group.
\\ \\
Over the last year, I have had the opportunity for fruitful discussions with all members of my theory group in G\"ottingen and with people from outside, for which I am very thankful. In particular, I want to thank Jan von Delft for all the discussions about the `phase lapse problem'.
\\ \\
I thank Katharina Janzen and Patrick Pl\"otz for their useful comments on the drafts of this thesis.
\\ \\
I am grateful to Pablo S. Cornaglia, Theo Costi, Jan von Delft and Akira Oguri for providing their NRG and Bethe ansatz data. Very special thanks go out to Theresa Hecht from Munich for performing countless NRG calculations for my particular needs.
\\ \\
I thank Macmillan Publishers Ltd. (Nature Magazine) and Yuval Gefen for their permission to reprint some of their figures within this thesis.
\\ \\
Finally and very important to me, I am really glad to have worked in a group with such a nice atmosphere. In particular, I want to thank Jens, Magnus and Patrick who let me enjoy coming to office each day. I am indebted to many other friends from G\"ottingen and outside and in particular to my mother Marlies Karrasch without whom I would not have come so far.

\newpage
\thispagestyle{empty}

\end{document}